Université de Toulouse

# HABILITATION À DIRIGER DES RECHERCHES

Spécialité : Physique

présentée par

William KNAFO

# Magnetism in heavy-$f$-electron metals

Soutenue le 9 avril 2021 devant la commission d'examen :

| | |
|---|---|
| Gerry Lander | Président |
| Philippe Bourges | Rapporteur |
| Catherine Pépin | Rapporteur |
| Vladimir Sechovsky | Rapporteur |
| Pascale Foury-Leylekian | Examinateur |
| Michel Goiran | Examinateur |
| Jean-Pascal Rueff | Examinateur |

Laboratoire National des Champs Magnétiques Intenses
Centre National de la Recherche Scientifique,
Toulouse

Magnetism in heavy-$f$-electron metals

William Knafo

Laboratoire National des Champs Magnétiques Intenses
Centre National de la Recherche Scientifique,
Toulouse

# Contents

# Contents

# Contents

# Acknowledgments

I warmly thank Philippe Bourges, Catherine Pépin, and Vladimir Sechovsky for accepting to be 'Rapporteurs', Pascale Foury-Leylekian, Michel Goiran, and Jean-Pascal Rueff for accepting to be Members, and Gerry Lander for accepting to be the President of the Jury of my 'Habilitation à diriger des recherches'. I also thank Michel Goiran for accepting to be the 'Parrain' of this work and for his help in organizing the Zoom session for the defense. I thank Michèle Antonin and Eric Benoist from the University of Toulouse for their support and help concerning administrative questions.

I acknowledge the collaborators with whom I worked during the last twenty years on the fascinating thematics of research on heavy-fermion materials : Stéphane Raymond, Jacques Flouquet, Daniel Braithwaite, Dai Aoki, Louis-Pierre Regnault, Frédéric Bourdarot, Georg Knebel, Gérard Lapertot, Alexandre Pourret, Michal Vališka, Christoph Meingast, Thomas Wolf, Hilbert von Löhneysen, Frédéric Hardy, Niels van Dijk, Serdar Sakarya, Pascal Lejay, Pierre Haen, Françoise Lapierre, Bjorn Fåk, Fabienne Duc, Marc Nardone, Abdelaziz Zitouni, Jérôme Béard, Paul Frings, Harrison Rakoto, Gernot Scheerer, Kathrin Höfner, Tristan Thébault, Rikio Settai, Shuhei Kurahashi, Yusuke Hirose, Tatsuma Matsuda, Shingo Araki, Hiroyuki Nojiri, Keitaro Kuwahara, Karel Prokes, Eddy Lelièvre-Berna, Xavier Tonon, Jean-Pascal Rueff, Victor Balédent, Olivier Mathon, Sakura Pascarelli, but also colleagues with whom I had useful discussions and interactions: Claudine Lacroix, Mireille Lavagna, Catherine Pépin, Bernard Coqblin, Sébastien Burdin, Hisatomo Harima, Kasumasa Miyake, Charles Simon, Mucio Continentino, Valentin Taufour, Andres Santander-Syro, Yusei Shimizu, Fuminori Honda, Kamran Behnia, Ladislav Havela, Vladimir Sechovsky, Jiři Pospišil, Mladen Horvatic, Marc-Henri Julien, Shin Kambe, Yo Tokunaga, Kenji Ishida, John Mydosh, Jean-Pierre Sanchez, Pierre Dalmas de Réotier, Atsushi Miyake, Eric Ressouche, Gabriel Seyfarth, Koji Kaneko, and also those I may have forgotten here...

# Chapter 1

# Introduction

This document is a review on the quantum magnetic properties of heavy-fermion $f$-electron metals. It is a 'light' version of the manuscript prepared for my 'Habilitation à diriger des recherches', sent to the Jury on 29 October 2020 and defended on 9 April 2021. Small modifications were made to the manuscript on 20 July 2021. In this document, I present a personal view on the heavy-fermion problem, within a phenomenological approach guided by experiments.

This review presents a set of 'historical' works which established the ground bases of the thematic during the last decades. An exhaustive and systematic approach is privileged. After a general presentation in Chapter 2, the properties of heavy-fermion paramagnets, antiferromagnets, and ferromagnets are considered in Chapters 3, 4, and 5, respectively. Chapters 6 and 7 are dedicated to two specific compounds, $URu_2Si_2$ for which a 'hidden-order' phase constitutes a more-than-thirty-years-old unsolved mystery, and $UTe_2$, where multiple superconducting phases have been discovered in the last two years. Experiments performed using a panel of techniques ranging from microscopic (neutron scattering, NMR, etc.) to thermodynamic (specific heat, magnetization, etc.) and transport (electrical resistivity, etc.) probes, under extreme conditions of low temperatures, intense magnetic fields and high pressures, are reviewed. They show that magnetism plays a central role in the quantum critical properties of heavy-fermion systems. An emphasis is given to the intersite magnetic fluctuations, presented as the driving force for a heavy Fermi liquid, precursor of quantum magnetic criticality ending in magnetically-ordered phases. They are also suspected to drive an unconventional mechanism for superconductivity, which develops in the vicinity of quantum magnetic phase transitions induced under pressure or magnetic field. The appearance of magnetic fluctuations and ultimately magnetic order in heavy-fermion compounds occurs in a nearly-integer-valence regime, in which $f$ electrons have a dual itinerant-localized character. Fermi-surface and valence studies, which give complementary information about this duality, are also considered.

# Chapter 2

# From valence instabilities to heavy-fermion quantum magnetism

A review of the magnetic phases and their associated quantum instabilities in heavy-fermion $f$-electrons metals constitutes the main matter of this work. This first chapter progressively introduces the basic electronic properties of these systems, in which the magnetic properties are driven by $f$ electrons. It emphasizes that critical magnetic properties of heavy-fermion materials occur at the border of a valence instability, at which the electrons acquire a subtle dual itinerant-localized character. The Chapter is organized as follows. The interplay between magnetism and superconductivity in strongly-correlated-electron systems is presented in Section 2.1. Focus is made in Section 2.2 on the valence states in $f$-electron lanthanides and actinides, and in Section 2.3 on $f$-electron intermediate-valent and heavy-fermion systems. The relation between the electronic localization and the stabilization of magnetic exchange interactions is emphasized. Quantum magnetic criticality, which encompasses critical phenomena associated with the onset of long-range magnetic order, is presented in Section 2.4. The stabilization of unconventional superconductivity at the verge of magnetism is emphasized in Section 2.5.

## 2.1 Strongly-correlated-electron systems

Quantum magnetism of electrons is a central issue in condensed matter physics. It includes magnetically-ordered systems as ferromagnets, antiferromagnets, spin-density wave magnets, but also paramagnetic systems [Blundell 01, Schollwöck 04, Moriya 85]. Quantum magnets are magnets with strong quantum magnetic fluctuations, i.e., magnetic fluctuations persisting in the limit of zero temperature. These quantum magnetic fluctuations can be considered as precursors of long-range magnetic ordering and generally indicate nearby quantum magnetic phase transitions, i.e., magnetic phase transitions in the limit of zero temperature. Many quantum magnets are metallic, and some of them are also superconducting. Figure 2.1 presents schematically the overlap between quantum magnets, metals, and superconductors families. Left part of the diagram corresponds to insulators, where quantum magnetism is a property of localized electrons and often results from low-dimensional or frustrated magnetic exchange interactions. The right part of the diagram presents metals. Interestingly, quantum magnetism, but also unconventional superconductivity, is observed in metals in which electrons are close to a localized limit [Armitage 10, Johnston 10, Flouquet 05, von Löhneysen 07, Aoki 13, Pfleiderer 09]. The dual itinerant-localized nature of these strongly-correlated electrons is at the heart of a complex problem, where magnetism and superconductivity are intimately connected.

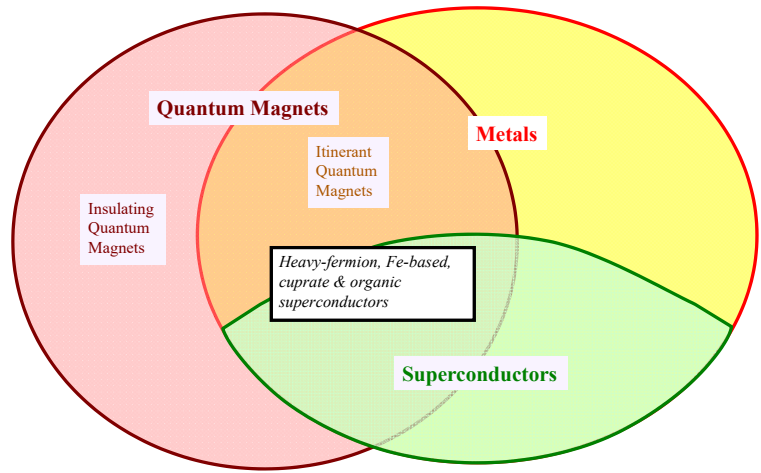

Figure 2.1: Schematic overlaps between quantum magnets, metals and superconductors.

The schematic generic phase diagrams, shown in Figure 2.2, illustrate the interplay between magnetism and superconductivity in heavy-fermion, iron-based pnictides and chalcogenides, and cuprate families. In these metallic strongly-correlated-electron materials, a superconducting phase often develops in the vicinity of a quantum magnetic phase transition, which can be adjusted by tuning pressure, chemical doping, or a magnetic field. The importance of quantum criticality, i.e., the critical properties associated with a quantum phase transition [Hertz 76], has been emphasized for a large variety of materials, ranging from high-temperature - cuprate [Valla 99] and iron-based [Shibauchi 14] - superconductors, heavy-fermion systems [Knafo 09a], to insulating low-dimensional quantum magnets [Coldea 10, Merchant 14]. In many of these systems, pressure (or doping) and magnetic field can destabilize a magnetically-ordered phase and lead to a critical non-Fermi-liquid behavior, that is a deviation from a 'standard' Fermi liquid metallic behavior [Stewart 01, Cooper 09]. Figure 2.3 summarizes schematically the interplay between the magnetic properties, the Fermi surface and superconductivity in strongly-correlated-electron materials. The magnetic properties and the Fermi surface are interdependent reflecting the itinerant-localized duality of the electrons, in relation in several compounds with indirect Rudermann-Kittel-Kasuya-Yosida (RKKY) magnetic exchange interactions [Rudermann 54, Kasuya 56, Yosida 57], nesting effects [Freeman 72], etc. In many systems, magnetism and superconductivity are presumably linked via a mechanism of superconductivity driven by magnetic fluctuations [Monthoux 07].

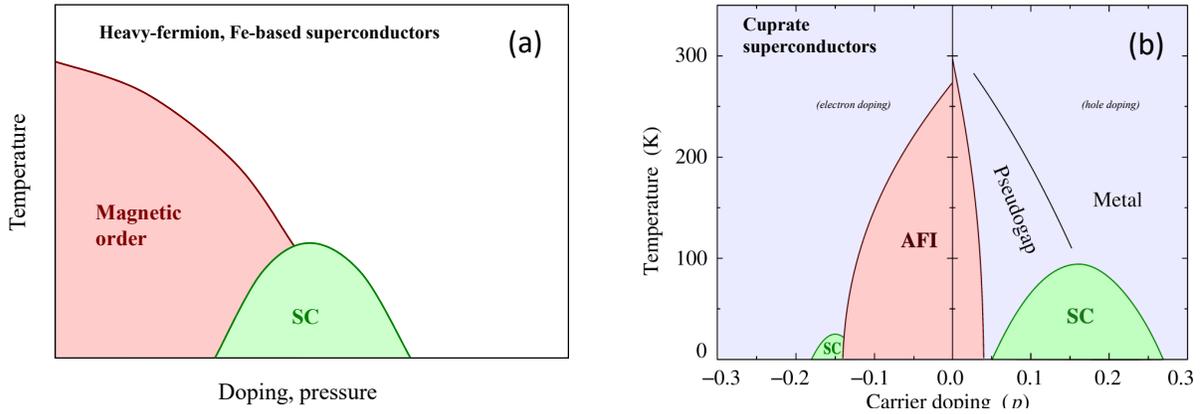

Figure 2.2: Typical pressure-temperature phase diagrams (a) of heavy-fermion and iron based, and (b) of cuprate superconductors [(b) adapted from [Peets 07]].

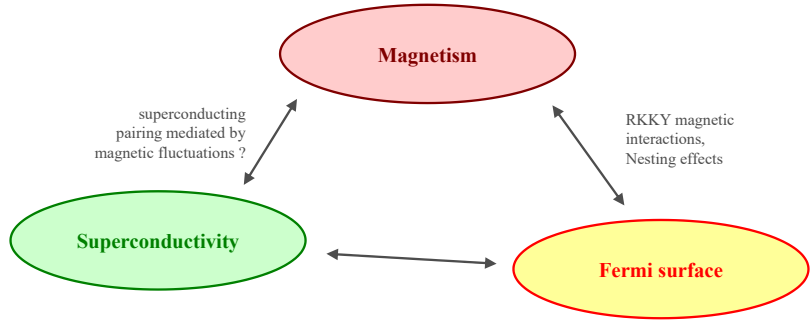

Figure 2.3: Basic relationship between quantum magnetism with superconductivity and Fermi surface.

Today, a challenge is to deepen our knowledge about the relationship between magnetism, the dual nature of electrons, and superconductivity. For that purpose, a first target is to investigate the electronic phase diagrams, with the aim to better understand the relationship between the different phases, the relevant energy scales, and the associated microscopic interactions. Tuning the properties of quantum magnets enables accessing quantum phase transitions between low-temperature phases and associated quantum critical behaviors. Multiple phases are observed, including magnetic order, hidden order, pressure- and field-induced superconductivity etc. A peculiarity of $f$-electron heavy-fermion systems is that their electronic energy scales are small enough to be easily tuned by chemical doping, pressure, and magnetic field, and high enough to be observable at temperatures offered by standard cryogenic [Stewart 01, Flouquet 05, von Löhneysen 07, Pfleiderer 09, Aoki 13]. For this reason, but also because of the richness of the observed quantum effects, they can be considered as textbook systems for the investigation of magnetism and superconductivity, and for the test of new concepts on a fascinating emerging quantum physics.

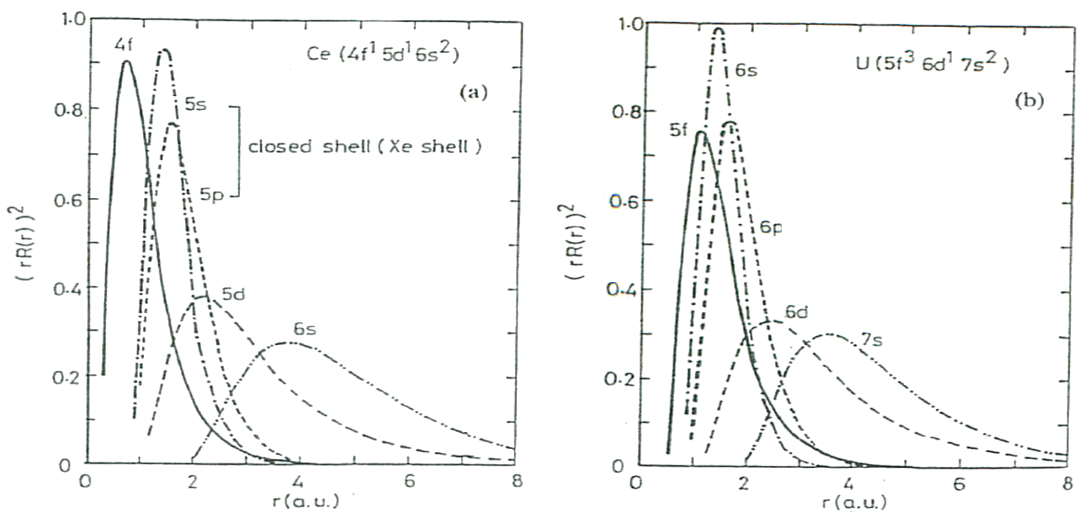

Figure 2.4: Electronic radial charge densities of last orbital shells in (a) cerium and (b) uranium (from [Ōnuki 91]).

## 2.2 The valence of $f$-electron systems

Basic properties of the electronic shells in lanthanide and actinide families of materials, which include most of heavy-fermion compounds, are presented here (see also [Ott 87, Moore 09, Arko 11, Bonnelle 15]).

### 2.2.1 Free-atom properties

In the lanthanides and actinides, the valence results from a subtle equilibrium between the electronic populations of quasi-localized $f$ orbitals and delocalized $s$ and $d$ orbitals. Table 2.3.4 lists the atomic number $Z$ and the electronic configuration of free atoms from the lanthanides and actinides, and of free-atoms from the $5d$ transition metals. Lanthanides are labeled here as the elements of atomic number $Z = 57$ (La) to $Z = 71$ (Lu), i.e., the fourteen elements from the $4f$ block and Lu, the first element of the $5d$ block in the periodic table. $5d$ transition metals are labeled here as the elements of atomic number $Z = 72$ (Hf) to $Z = 80$ (Hg) from the $5d$ block in the periodic table. Actinides are labeled here as the elements of atomic number $Z = 89$ (Ac) to $Z = 103$ (Lr), i.e., the fourteen elements from the $5f$ block and Lr, the first element of the $6d$ block in the periodic table. The electronic configuration of most of these elements follows the Madelung's rules, which consist in the successive filling of the subshells $6s$, $4f$, $5d$, $6p$, $7s$, $5f$, $6d$, $7p$ when $Z$ and, thus, the number of electrons are increased.

Outer-shell electrons which can participate to chemical bonding and electrical conduction processes are called valence electrons, and the valence $v$ of an atom is the number of its valence electrons. The $4f$ subshell in lanthanides and the $5f$ subshell in actinides lie close to the nucleus and do not contribute to valence bonding. The localized character of open $4f$ and $5f$ subshells in Ce and U atoms, respectively, is visible in the plot of the electron densities as function of distance to the atom nucleus, shown in Figure 2.4 [Ōnuki 91]. In the lanthanides, the valence corresponds to the number of electrons in the outer $6s$ and $5d$ subshells, and in the actinides it corresponds to the number of electrons in the outer $7s$ and $6d$ subshells.

Table 2.1: Atomic number $Z$ and electronic configuration of lanthanides, $5d$ transition metals, and actinides in free-atom form.

| Lanthanide | Z | Configuration | Transition metal ($5d$) | Z | Configuration | Actinide | Z | Configuration |
|---|---|---|---|---|---|---|---|---|
| La | 57 | [Xe]$5d^1 6s^2$ | Hf | 72 | [Xe]$4f^{14} 5d^2 6s^2$ | Ac | 89 | [Rn]$5d^1 7s^2$ |
| Ce | 58 | [Xe]$4f^1 5d^1 6s^2$ | Ta | 73 | [Xe]$4f^{14} 5d^3 6s^2$ | Th | 90 | [Rn]$6d^2 7s^2$ |
| Pr | 59 | [Xe]$4f^3 6s^2$ | W | 74 | [Xe]$4f^{14} 5d^4 6s^2$ | Pa | 91 | [Rn]$5f^2 6d^1 7s^2$ |
| Nd | 60 | [Xe]$4f^4 6s^2$ | Re | 75 | [Xe]$4f^{14} 5d^5 6s^2$ | U | 92 | [Rn]$5f^3 6d^1 7s^2$ |
| Pm | 61 | [Xe]$4f^5 6s^2$ | Os | 76 | [Xe]$4f^{14} 5d^6 6s^2$ | Np | 93 | [Rn]$5f^4 6d^1 7s^2$ |
| Sm | 62 | [Xe]$4f^6 6s^2$ | Ir | 77 | [Xe]$4f^{14} 5d^7 6s^2$ | Pu | 94 | [Rn]$5f^6 7s^2$ |
| Eu | 63 | [Xe]$4f^7 6s^2$ | Pt | 78 | [Xe]$4f^{14} 5d^9 6s^1$ | Am | 95 | [Rn]$5f^7 7s^2$ |
| Gd | 64 | [Xe]$4f^7 5d^1 6s^2$ | Au | 79 | [Xe]$4f^{14} 5d^{10} 6s^1$ | Cm | 96 | [Rn]$5f^7 6d^1 7s^2$ |
| Tb | 65 | [Xe]$4f^9 6s^2$ | Hg | 80 | [Xe]$4f^{14} 5d^{10} 6s^2$ | Bk | 97 | [Rn]$5f^9 7s^2$ |
| Dy | 66 | [Xe]$4f^{10} 6s^2$ | | | | Cf | 98 | [Rn]$5f^{10} 7s^2$ |
| Ho | 67 | [Xe]$4f^{11} 6s^2$ | | | | Es | 99 | [Rn]$5f^{11} 7s^2$ |
| Er | 68 | [Xe]$4f^{12} 6s^2$ | | | | Fm | 100 | [Rn]$5f^{12} 7s^2$ |
| Tm | 69 | [Xe]$4f^{13} 6s^2$ | | | | Md | 101 | [Rn]$5f^{13} 7s^2$ |
| Yb | 70 | [Xe]$4f^{14} 6s^2$ | | | | No | 102 | [Rn]$5f^{14} 7s^2$ |
| Lu | 71 | [Xe]$4f^{14} 5d^1 6s^2$ | | | | Lr | 103 | [Rn]$5f^{14} 6d^1 7s^2$ |

Following Madelung's rules, the configurations $4f^n 6s^2$ in the lanthanides and $5f^n 7s^2$ in the actinides are expected, implying that the elements from the $4f$ and $5f$ blocks are in valence $v = 2$. These rules are followed by a large part of $4f$- and $5f$-block elements in their free-atom state (Table 2.3.4). However, violations from Madelung's rules are observed. For three lanthanides (La, Ce, Gd), the $5d$ subshell 'attracts' one electron from the $4f$ subshell, and a configuration $4f^n 5d^1 6s^2$ corresponding to the valence $v = 3$ is achieved. For the five actinides (Ac, Pa, U, Np, Cm), the $6d$ subshell 'attracts' one electron from the $5f$ subshell, and a configuration $5f^n 6d^1 7s^2$ corresponding to the valence $v = 3$ is achieved. For the actinide Th, the $6d$ subshell 'attracts' two electrons from the $5f$ subshell, and the configuration $6d^2 7s^2$ corresponding to the valence $v = 4$ is achieved. These deviations result from the close energy levels of the subshells $4f$ and $5d$ in the lanthanides, and of the subshells $5f$ and $6d$ in the actinides.

### 2.2.2 Single-element metals

When metallic elements form a crystal, the valence electrons contribute to chemical bonding and to the conduction band, and modifications of the valence, in comparison with that of free atoms, are observed. In most of pure crystals, a compact arrangement of the atoms leads to a reduction of the atomic radius. In lanthanides and actinides crystals, this change of radius (in comparison with the free-ion state) can be accompanied by an increase of valence, part of the electrons from the inner subshells $4f$ and $5f$ moving to the outer subshells $5d$ and $6d$, respectively. These changes can be evidenced by comparing the evolution of the Wigner-Seitz radius $r_{WS}$, i.e., the mean radius of an atom in a metallic crystal, in single-element crystals made of $5d$ transition metals, $4f$ lanthanides and $5f$ actinides shown in Figure 2.5 [Arko 11]. The main features of Figure 2.5 are described below:

- $4f$ lanthanides: Most of the elements are in valence $v = 3$ with a configuration $4f^n 5d^1 6s^2$. The localized $4f$ subshell efficiently screens the nucleus charge, and the progressive filling of the localized $4f$ subshell when $Z$ is increased weakly affects $r_{WS}$, which saturates at $\simeq 2$ Å. A higher $r_{WS} \simeq 2.2 - 2.3$ Å in Yb and Eu indicates a smaller valence and a lesser attraction of the valence electrons by the nucleus. The configuration $4f^n 6s^2$ corresponding to valence $v = 2$ is favored by a half-filled ($n = 7$) $4f$ subshell in Eu and by a fully-filled ($n = 14$) $4f$ subshell in Yb.

- $5d$ transition metals: The $5d$ valence subshell is open. The number of $5d$ electrons and, thus, the valence increase with $Z$, and $r_{WS}$ shows a parabolic variation with $Z$ [Arko 11]. For the first elements of the $5d$ block, the decrease of $r_{WS}$ with increasing $Z$, from $r_{WS} \simeq 2.1$ Å for La to $r_{WS} \simeq 1.5$ Å for Os, is due to a partial screening of the nucleus charge by the outer-electron charges, leading to their stronger attraction by the nucleus and to the contraction of their orbitals. In the last elements of the line, the $5d$ subshell is more than half-filled and the screening is more efficient, leading to an increase of the radius with $Z$.

- $5f$ actinides: For small $Z$, a decrease of $r_{WS}$ with $Z$, from $r_{WS} \simeq 2.0$ Å for Th to

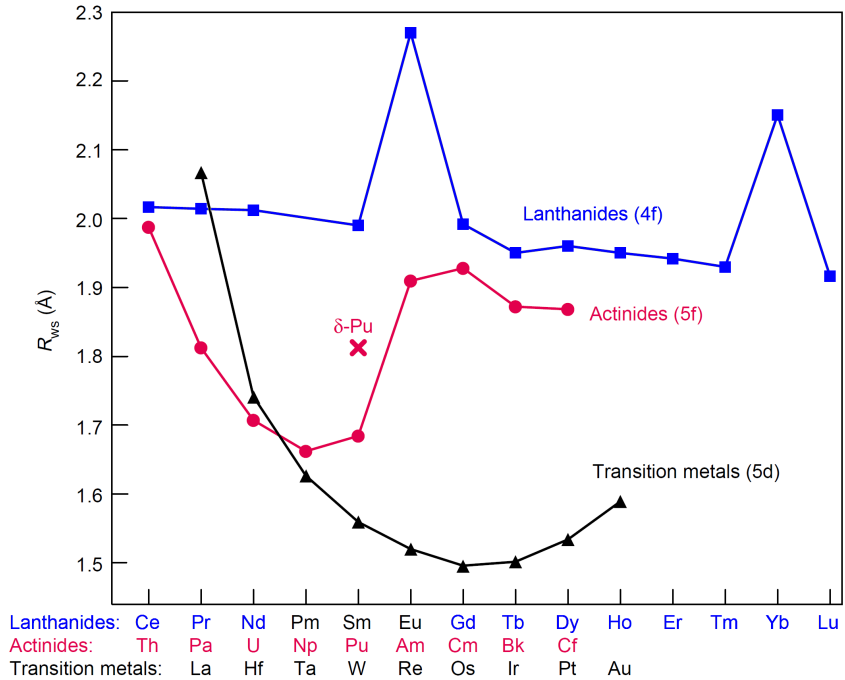

Figure 2.5: Wigner-Seitz radius of lanthanides, $5d$ transition metals, and actinides in single-element metals (from [Hills 00]).

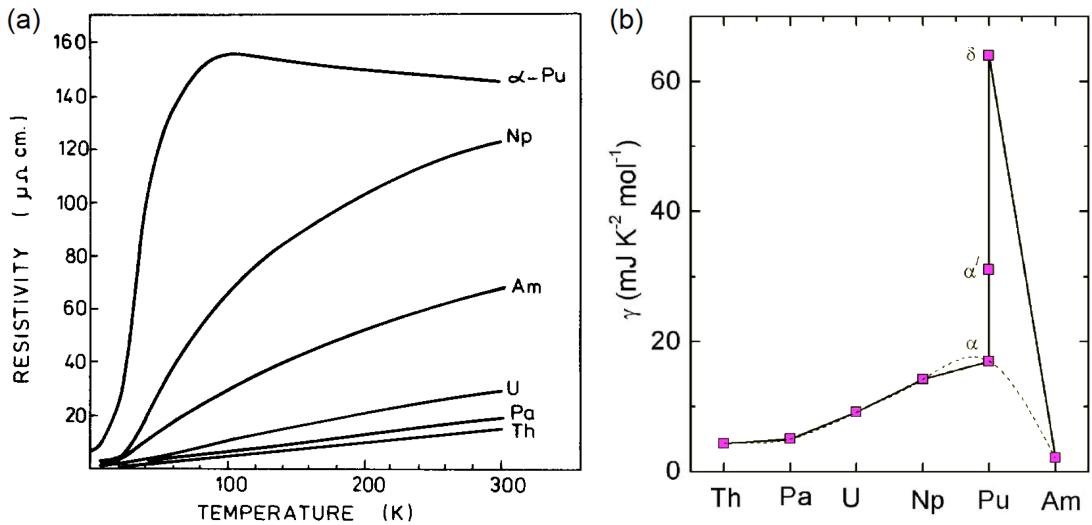

Figure 2.6: (a) Electrical resistivity versus temperature and (b) specific heat low-temperature Sommerfeld coefficient $\gamma$ of actinide-element metals (from [Müller 78, Lashley 05]).

$r_{WS} \simeq 1.65$ Å for Np, is similar to that observed for $5d$ elements. The number of itinerant electrons from the $6d$ subshell and thus, the valence, increase with $Z$. The lighter actinides in their single-element metallic form are in an itinerant limit similar to that of $5d$ transition metals. A step-like increase of $r_{WS}$ observed for larger $Z$ elements (Pu-Am) indicates a sudden reduction of valence due to the filling of the localized $5f$ band. From Am to Cf, an almost constant radius $r_{WS} \simeq 1.9$ Å close to that found in most lanthanides is the indication of a valence reaching, or almost reaching, the value $v = 3$. By analogy with the lanthanides, the heavier actinides in their single-element metallic form can be considered in a localized, or nearly-localized, limit.

In their single-element crystalline state, the situation of $5f$ actinides is intermediate between that of the $4f$ lanthanides and $5d$ transition metals. $5d$ transition metals correspond to a limit with itinerant electrons, their physics being controlled by the outer and delocalized $5d$ electrons. Oppositely, most of the lanthanides have the same number of valence electrons (valence $v = 3$), and their electronic properties mainly depend on their numbers of localized $4f$ electrons. Most of $4f$ lanthanides, but also $5f$ actinides, can be considered in their 'localized' limit when they are in valence $v = 3$. Deviations towards higher valences, due to the additional filling of itinerant $5d$ or $6d$ subshells, are commonly identified as signatures of electronic itinerancy. In some systems, as Yb, Sm and Eu, where the valences $v = 2$ and 3 are in competition, valence $v = 2$ is generally considered as the localized limit, while valence $v = 3$ is considered as an itinerant configuration.

In both their free-ion (Table 2.3.4) and single-element crystalline (Figure 2.5) states, the $5f$ actinides have higher valences than the $4f$ lanthanides. This is due to a $5f$ subshell which is less localized than the $4f$ subshell, resulting in closer $5f$ and $6d$ subshells and, thus, in a transfer of electrons to the valence band which is more favored in the actinides than in the lanthanides (see Figure 2.4 [Ōnuki 91]).

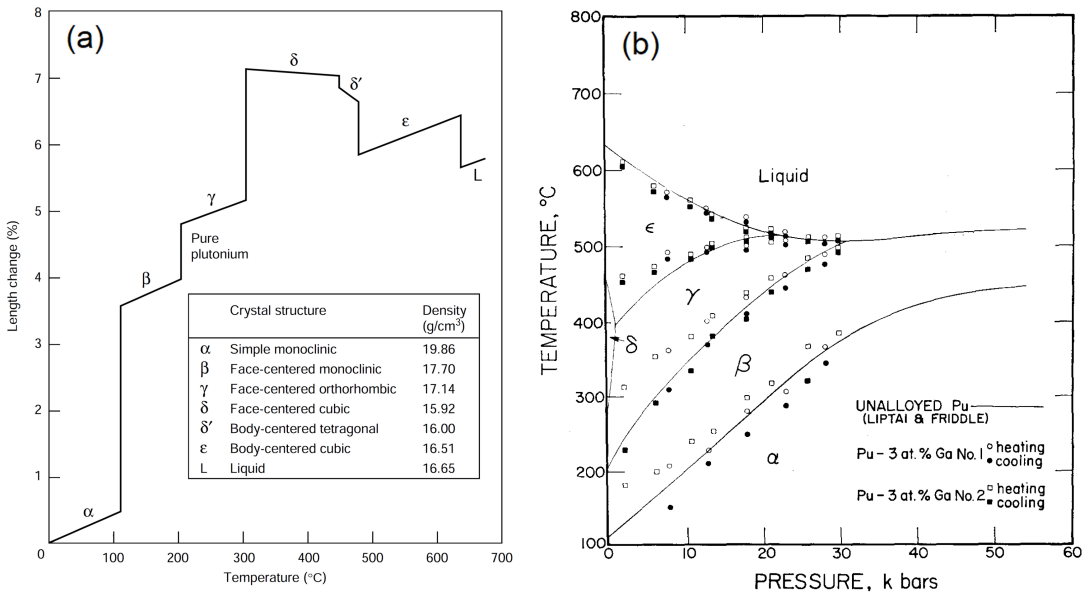

Figure 2.7: (a) Length versus temperature and (b) pressure-temperature phase diagram of Pu (from [Liptai 67, Boring 00]).

Figure 2.6 presents (a) the electrical resistivity $\rho$ versus temperature [Müller 78] and (b) the low-temperature Sommerfeld coefficient $\gamma = C_p/T$, where $C_p$ is the specific heat and $T$ is the temperature, [Lashley 05] of pure-actinide metals. Increase of $\rho$ confirms the progressive localization of outer electrons when $Z$ increases [Müller 78]. Figure 2.6 further shows the peculiar properties of Pu in the actinide series: this element presents the maximal electrical resistivity, the maximal low-temperature slope of the resistivity [Müller 78], and the maximal Sommerfeld specific-heat coefficient $\gamma$ [Lashley 05]. These features are related with the proximity of Pu metal from a valence instability, as indicated by the variation of the Wigner-Seitz radius as function of $Z$ in actinide metals, shown in Figure 2.5. In relation with this nearby valence instability, Pu presents a rich phase diagram, where multiple phases are successively stabilized when the temperature is decreased [Hecker 00]. These transitions are accompanied by strong lattice variations related with valences changes. The strongest contraction indicates that valence reaches its maximal value in the low-temperature phase $\alpha$ (see Figure 2.7(a) [Boring 00]). Figure 2.7(b) further shows that, by directly acting on the contraction of the crystal, pressure permits to tune the valence instability. The itinerant phase $\alpha$ is stabilized under pressure, as shown by the increase of its transition temperature and by the disappearance of the phase $\delta$ [Liptai 67]. Intermediate valences $v = 2.72$ (corresponding a number $n_f \simeq 5.28$ of $5f$ electrons) in the phase $\delta$ and $v = 2.84$ (corresponding a number $n_f \simeq 5.16$ of $5f$ electrons) in the phase $\alpha$ were estimated in [Booth 12].

Other $f$-block elements in their pure crystalline state, as Ce, U, and Np, are close to a valence instability. Transitions between allotropic states are reported in these systems when the temperature is lowered, a smaller metallic radii indicating higher valences at low temperature [Koskenmaki 78, Boring 00, Grenthe 11, Yoshida 11]. The case of Ce will be considered with more details in Section 2.3.

Table 2.2: Selection of valence configurations in lanthanides and actinides.

| Atom | Inner shells | Valence shells | Valence $v$ | $f$-shell | | |
|------|-------------|----------------|-------------|-----------|-----------|-----------|
| | | | | $S$ | $L$ | $J$ |
| Ce | [Xe]4f$^2$ | 6s$^2$ | 2 | 1 | 5 | 4 |
| | [Xe]4f$^1$ | 5d$^1$6s$^2$ | 3 | 1/2 | 3 | 5/2 |
| | [Xe] | 5d$^2$6s$^2$ | 4 | - | - | - |
| Sm | [Xe]4f$^6$ | 6s$^2$ | 2 | 3 | 3 | 0 |
| | [Xe]4f$^5$ | 5d$^1$6s$^2$ | 3 | 5/2 | 5 | 5/2 |
| Eu | [Xe]4f$^7$ | 6s$^2$ | 2 | 7/2 | 0 | 7/2 |
| | [Xe]4f$^6$ | 5d$^1$6s$^2$ | 3 | 3 | 3 | 0 |
| Yb | [Xe]4f$^{14}$ | 6s$^2$ | 2 | - | - | - |
| | [Xe]4f$^{13}$ | 5d$^1$6s$^2$ | 3 | 1/2 | 3 | 7/2 |
| U | [Rn]5f$^3$ | 6d$^1$7s$^2$ | 3 | 3/2 | 6 | 9/2 |
| | [Rn]5f$^2$ | 6d$^2$7s$^2$ | 4 | 1 | 5 | 4 |
| | [Rn]5f$^1$ | 6d$^3$7s$^2$ | 5 | 1/2 | 3 | 5/2 |
| Pu | [Rn]5f$^6$ | 7s$^2$ | 2 | 3 | 3 | 0 |
| | [Rn]5f$^5$ | 6d$^1$7s$^2$ | 3 | 5/2 | 5 | 5/2 |
| | [Rn]5f$^4$ | 6d$^2$7s$^2$ | 4 | 2 | 6 | 4 |

### 2.2.3 Compounds and alloys

In lanthanide and actinide compounds and alloys, which include a large number of inter-metallic materials, the last $f$ and $d$ orbitals are modified by the environment of other atoms (metals or ligands), and deviations of the valence states, in comparison with the pure-elements metals, are observed. Experimentally, several valence configurations have been identified or proposed for a large series of compounds. Table 2.2 summarizes different integer-valence configurations of Ce, Sm, Eu, Yb, U and Pu in compounds where valence transitions and/or heavy-fermion behaviors have been reported or suspected [Varma 76, Lawrence 81, Rueff 06, Booth 12, Janoschek 15]. In these systems, an intermediate valence with a non-integer value of valence is generally observed, which can be interpreted as the signature of valence fluctuations between states of integer valence. In Yb, Sm and Eu based materials, the valence states $v = 2, 3$ are the most favorable, and deviations to valences higher than 2 are considered as the signature of itinerancy. In Ce compounds, the valence states $v = 3, 4$ are generally more favorable, and deviations to valences higher than 3 are the signature of itinerancy. An intermediate valence in pure Ce metal was shown to result from fluctuations within three states of integer valences $v = 2$, 3, and 4, the weight of valence 2 being smaller than that of the two other valences [Rueff 06] (see Section 2.3.1). As well, intermediate valence in Pu metal was related to fluctuations between the configurations of integer valences $v = 2, 3$ and 4 and intermediate valence in some U compounds was related to fluctuations between the configurations of integer valences

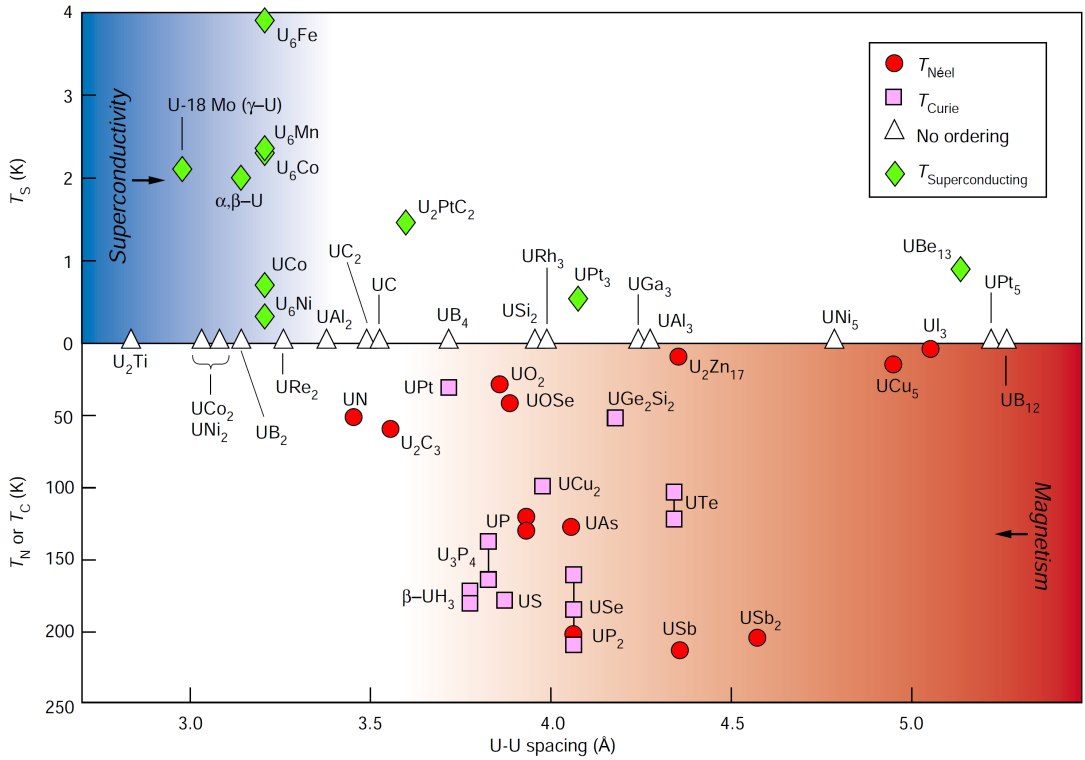

Figure 2.8: Hill plot of transition temperatures versus the minimal $d_{U-U}$ distance for a selection of U-based materials (from [Hill 70, Boring 00]).

$v = 3$, 4 and 5 (see below) [Booth 12, Janoschek 15].

As discussed above for single-element metals, the valence of elements is intimately related with their ionic radius. The Hill plot in Figure 2.8 shows that, in the case of U-based compounds, a paramagnetic ground state and sometimes a superconducting phase are generally stabilized for small distances $d_{U-U}$ between U atoms, while magnetically-ordered states are favored for larger distances $d_{U-U}$ [Hill 70, Boring 00]. A change of the degree of overlap between $5f$ subshells of U neighbors has been emphasized to explain this diagram [Hill 70, Arko 11]. However, the critical distance $d_{U-U}^c \simeq 3.5$ Å separating the two regimes corresponds to twice the critical value of the Wigner-Seitz radius $r_{WS}^c \simeq 1.7 - 1.9$ Å reached in the actinides Pu and Am lying at the valence instability (see Figure 2.5). A change of valence and, thus, of the itinerant or localized character of the electrons, in relation with a change of the atomic radius, may occur at the Hill limit. The characteristics of the two regimes separated by the Hill limit are summarized below:

- In dense U-based materials, small interatomic U-U distances and, thus, small U-atom radius, favor an itinerant limit ending in the stabilization of a paramagnetic ground state, and sometimes of superconductivity. A larger valence due to the delocalization of $5f$ electrons into the conduction band is expected in these materials.

- In materials with a higher number of non-U atoms, large interatomic U-U distances generally coincide with the stabilization of long-range magnetic ordering. These compounds are close to the localized limit, for which a smaller valence related to a higher number of localized $5f$ electrons is expected. Ultimately, when the interatomic distance becomes bigger ($\gtrsim 5$ Å), magnetic exchange interactions collapse and a paramagnetic ground state is established.

However, the definition of the localized and itinerant limits is not trivial. In [Booth 12], an intermediate valence $v = 3.29$ (corresponding a number $n_f \simeq 2.71$ of $5f$ electrons) was extracted in the antiferromagnet UCd$_{11}$, for which the large interatomic distance $d_{U-U} = 6.56$ Å [Fisk 84] would suggest a localized limit, while an intermediate valence $v = 4.08$ (corresponding a number $n_f \simeq 1.92$ of $5f$ electrons) was extracted for the paramagnetic UCoGa$_5$, where a smaller interatomic distance $d_{U-U} = 4.24$ Å is the indication of a more itinerant case [Moreno 05].

The Hill plot contains all elements of heavy-fermion quantum criticality, which will be addressed in the next Sections and Chapters. The delocalization of the $f$ electrons is generally described as resulting from a hybridization between $f$ localized orbitals and $s - d$ valence orbitals contributing to the conduction band (see the Anderson [Anderson 61] and Kondo [Kondo 64] models in Subsection 2.3). For a smaller atomic radius, the spatial overlap between these orbitals (see Figure 2.4) may increase, so that a stronger hybridization drives a compensation of localized magnetic moment and, thus, the stabilization of paramagnetism. In the localized limit, the stabilization of a magnetically-ordered ground state is the consequence of magnetic exchange interactions between $f$ atoms moments. These interactions are generally described by indirect RKKY exchange [Rudermann 54, Kasuya 56, Yosida 57], or by more subtle anisotropic-hybridization-mediated exchange [Cooper 85] (see Subsection 2.4).

The observation of superconductivity arising in itinerant systems close to their localized regime, i.e., superconductivity preferentially developing in bad metals rather than in good metals, could naively sound as a contradiction. However, the proximity of magnetic order, favored in the more-localized side of the phase diagrams, is accompanied by the development of precursors magnetic fluctuations [Hertz 76, Moriya 85, Millis 93]. While phonons were proposed to drive conventional superconductivity within Bardeen-Cooper-Schrieffer theory [Bardeen 57], enhanced quantum magnetic fluctuations at a magnetic phase transition are suspected to drive an unconventional mechanism for superconductivity in many strongly-correlated-electron systems [Monthoux 07]. Superconductivity in heavy-fermion systems will be introduced in Section 2.5.

Finally, beyond the Hill plot, more subtle approaches are needed to describe real three-dimensional compounds. In many $f$-electron compounds the physical properties are anisotropic and a single parameter, as the smaller distance between $f$ atoms, is not enough for a complete description, as shown by the large scattering of data points in the Hill plot (see Figure 2.8). Anisotropies in the lattice structure imply that interatomic distances are non-equivalent along different lattice directions, and the presence of $d$-metals and $p$-ligands can modify the orbitals of the $f$-atom (via electronic bonding, crystalline electric field etc.). Such anisotropic effects play a role in the degree of localization of the electrons, in the magnetic anisotropy and in the magnetic exchange mechanisms. They may be captured properly for a microscopic modeling of the electronic properties.

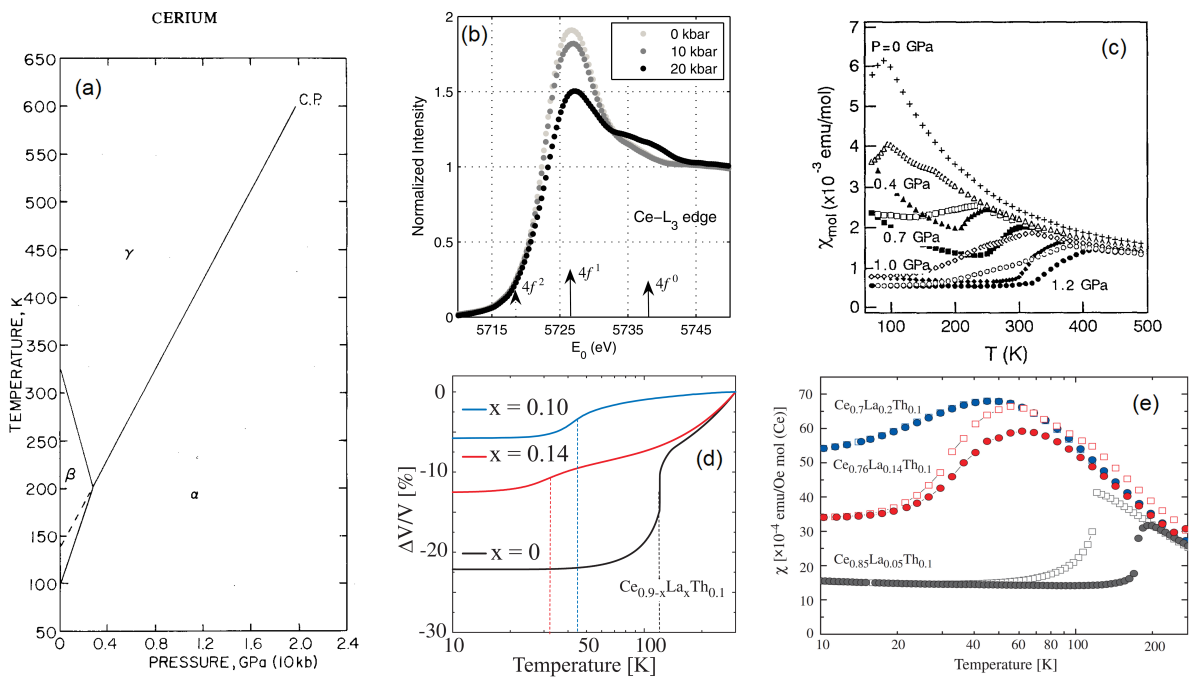

Figure 2.9: (a) Pressure-temperature phase diagram (from [Koskenmaki 78]), (b) pressure dependence of x-ray absorption spectroscopy spectra (from [Rueff 06]) and (c) pressure dependence of the magnetic susceptibility of cerium (from [Naka 94]). Temperature dependence of (d) length and (e) magnetic susceptibility of $Ce_{0.9-x}La_xTh_{0.1}$ (from [Lashley 06]).

## 2.3   From intermediate valence to heavy-fermion physics

This Section introduces how the progressive localization of $f$ electrons can induce a crossover from intermediate-valence into heavy-fermion physics. Experimental signatures of the strongly-renormalized Fermi-liquid regime reported in these systems are presented. Hybridization and Kondo-effect models developed for the description of these systems are introduced. This Section ends with electronic structure and Fermi surface measurements and their comparison with band calculations.

### 2.3.1   Intermediate-valence systems

Intermediate valences have been observed in many $f$ electrons systems (see Section 2.2.3 and in [Varma 76, Lawrence 81, Rueff 06, Booth 12, Janoschek 15]). Here, we focus on the particular case of metallic cerium and its alloys, which corresponds to a textbook example of tunable valence instability ending in a heavy-fermion limit. Figure 2.9 presents properties of Ce metal under pressure or doped with La and Th. At ambient pressure, Ce undergoes two successive structural transitions between the phases $\gamma$ and $\beta$ at $T_\beta \simeq 320$ K and between the phases $\beta$ and $\alpha$ at $T_\alpha \simeq 100$ K [Koskenmaki 78]. The phases $\gamma$ and $\alpha$ are isostructural, having a fcc crystal structure, and the phase $\beta$ has a dhcp crystal structure. While a 1 % volume change, from 34.4 to 34.8 Å$^3$/Ce, is observed at the transition $\gamma \rightarrow \beta$, a large 18 % volume change, from 34.8 to 28.5 Å$^3$/Ce, indicates that the transition $\beta \rightarrow \alpha$ is accompanied by a change of valence [Koskenmaki 78].

The phase diagram in Figure 2.9(a) shows that the phase $\alpha$ is stabilized under pressure, its critical temperature $T_\alpha$ increasing almost linearly, ending in a critical point at $T = 600$ K and $p \simeq 2$ GPa (from [Koskenmaki 78]). The temperature $T_\beta$ decreases under pressure, and the phase $\beta$ vanishes at the pressure $p = 0.3$ GPa, above which the transition line $T_\alpha$ directly separates the isostructural phases $\gamma$ and $\alpha$. Although a smaller number of phases is observed, the phase diagram of Ce under pressure strikingly resembles that of Pu (see Figure 2.7(b)). In spite of an open $5f$ subshell expected to be less localized than an open $4f$ subshell (see Figure 2.4), this indicates similarities between the physics of lanthanides and actinides.

Figure 2.9(b) presents resonant inelastic x-ray scattering experiments performed on Ce under pressure at $T = 300$ K (from [Rueff 06]). Analysis of these data led to intermediate valences $v = 3.03$ (corresponding a number $n_f \simeq 0.97$ of $4f$ electrons) in the phase $\gamma$ and $v = 3.19$ (corresponding a number $n_f \simeq 0.81$ of $4f$ electrons) in the phase $\alpha$ [Rueff 06], confirming that $T_\alpha$ is a valence transition. Figure 2.9(c) shows that the magnetic susceptibility $\chi$ follows a Curie-type behavior at high temperatures, without a clear anomaly at the temperature $T_\beta$. A step-like decrease of $\chi$ is observed at the transition temperature $T_\alpha$, at which a strong hysteresis occurs [Naka 94]. Under pressure, the increase of $T_\alpha$ induces a deviation from the Curie-like behavior at smaller values of $\chi$, which almost saturates at low temperature. Magnetic susceptibility measurements indicate that the valence transition to the state $\alpha$ is accompanied by a deep change in the magnetic properties, illustrating an intimate relationship between the valence and magnetism.

Alternatively to pressure, chemical doping permits to tune the valence instability of Ce. Figures 2.9(d,e) present the variations with temperature of the volume and magnetic susceptibility, respectively, of doped Ce$_{0.9-x}$La$_x$Th$_{0.1}$ alloys [Lashley 06]. While Th-doping has an effect similar to pressure, leading to the enhancement of $T_\alpha$ up to $\simeq 150$ K in Ce$_{0.9}$Th$_{0.1}$, La-doping has the opposite effect and is similar to a negative pressure, leading to a decrease of $T_\alpha$ to less than 50 K in Ce$_{0.9-x}$La$_x$Th$_{0.1}$ with $x = 0.1, 0.14$. As well as pressure, Th-doping leads to the disappearance of the phase $\beta$, but the combination of Th- and La- doping permits to stabilize a direct $\gamma \rightarrow \alpha$ transition at low temperature, without the intermediate phase $\beta$. While a clear first-order transition is observed at $T_\alpha$ in Ce$_{0.9}$Th$_{0.1}$, being associated by step-like decreases by $\simeq 15$ % of the volume and by more than half of the magnetic susceptibility at low temperature, the anomaly broadens and transforms into a crossover, rather than a well-defined phase transition, when $T_\alpha$ decreases in Ce$_{0.8}$La$_{0.1}$Th$_{0.1}$ and Ce$_{0.76}$La$_{0.14}$Th$_{0.1}$ [see Figures 2.9(d,e)]. This crossover is associated with a broad maximum at the temperature $T_\chi^{max}$, followed by a low-temperature saturation of the magnetic susceptibility, which are the signatures of a strongly-renormalized Fermi liquid state, i.e., a heavy-fermion behavior. The crossover to a heavy-fermion regime coincides with a smaller volume variation, which indicates a smaller valence variation than that reported in pure Ce or in Ce$_{0.9}$Th$_{0.1}$.

While the physics of intermediate-valent systems is driven by an increase of electrons itineracy, heavy-fermion behavior develops at the onset of an almost temperature-independent va-lence, presumably close to an integer value corresponding to a localized-electrons, or nearly-localized-electrons, limit. Intermediate-valent compounds are associated with larger energy scales than heavy-fermion compounds. Typically, valence transitions (as in Ce [Koskenmaki 78]) or crossovers (as in CeSn$_3$ [Béal-Monod 80]) are observed at temperatures $T > 50$ K, while a heavy Fermi liquid develops at temperatures $T < 50$ K, and often at temperatures smaller than a few kelvins. The main signatures of heavy-fermion physics in $f$-electron systems and their description by hybridization models are presented in Sections 2.3.2, 2.3.3, and 2.3.4.

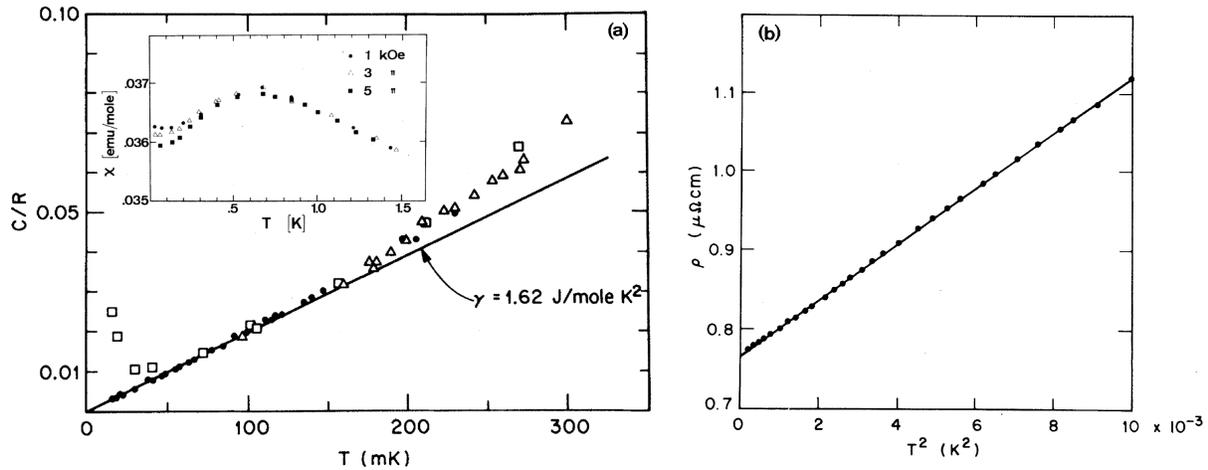

Figure 2.10: (a) Heat capacity (main panel), magnetic susceptibility (Inset) versus temperature, and (b) electrical resistivity versus square of temperature of CeAl$_3$ (from [Andres 75]).

### 2.3.2 Heavy Fermi liquid

A Fermi liquid corresponds to a renormalized state of matter, where the electrons acquire an enhanced effective mass $m^*$. The mass enhancement, in comparison with the free-electron mass, is driven by the electronic interactions. Basic properties of a Fermi liquid are:

- a linear temperature-variation of the specific heat $C_p$, corresponding to a constant Sommerfeld coefficient $\gamma = C_p/T \propto m^*$, where $T$ is the temperature,

- a saturating magnetic susceptibility $\chi \propto m^*$,

- a quadratic temperature-variation of the electrical resistivity $\rho = \rho_0 + AT^2$, where $\rho_0$ is the residual resistivity and $A \propto m^{*2}$ is the quadratic coefficient.

In 1975, a heavy-fermion behavior was identified for the first time in a material, the paramagnet CeAl$_3$, by Andres *et al* [Andres 75]. A Fermi liquid associated with an effective mass, that is three orders of magnitude larger than the free electron mass, was evidenced. Plots of the specific heat and magnetic susceptibility of CeAl$_3$ versus temperature and of its electrical resistivity versus the square of the temperature are presented in Figure 2.10. They show a large enhancement of the sommerfeld coefficient $\gamma \simeq 1.6$ J·mol$^{-1}$·K$^{-2}$ for $T \leq 200$ mK, a saturation of the magnetic susceptibility to $\chi \simeq 36 \cdot 10^{-3}$ emu·mol$^{-1}$ for $T \leq T_\chi^{max} \simeq 500$ mK, and a quadratic temperature dependence of the electrical resistivity with the coefficient $A = 35$ $\mu\Omega$·K$^{-2}$ for $T \leq 100$ mK. Instead of a transition, where sharp anomalies are observed in all physical properties at a unique temperature, a broad crossover occurs at the progressive onset of the Fermi-liquid regime, and different temperature scales, rather than a single one, correspond to its signatures in the different physical quantities (see also the prototypical heavy-fermion paramagnet CeRu$_2$Si$_2$, where quadratic electrical resistivity is reported for $T < T^* \simeq 500$ mK [Daou 06] while the magnetic susceptibility nearly saturates for $T < T_\chi^{max} \simeq 10$ K [Fisher 91]). The Fermi-liquid regime, delimited by the upper temperature scale $T_\chi^{max}$, established in paramagnets will be labeled here as a correlated paramagnetic (CPM) regime. A Fermi liquid, i.e., a regime of electronic correlations, can also be identified at low temperatures in the magnetically-ordered phases of heavy-fermion systems (see Section 2.4).

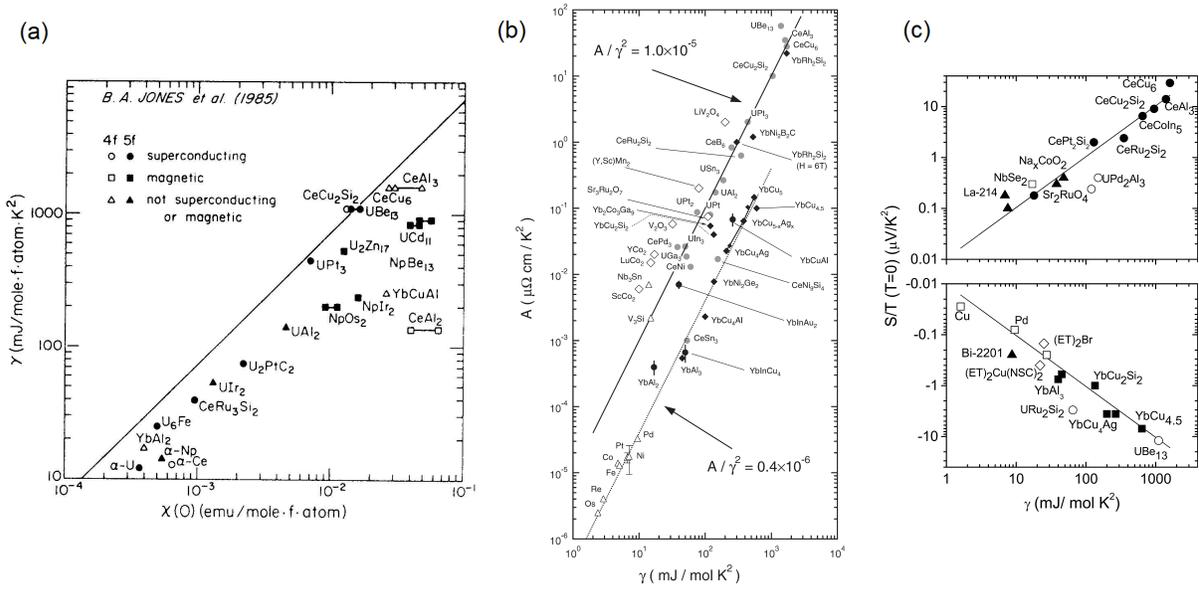

Figure 2.11: (a) Low-temperature Sommerfeld heat-capacity coefficient $\gamma = C/T$ versus the low-temperature magnetic susceptibility $\chi$ (from [Lee 86]), (b) quadratic coefficient $A$ in the temperature dependence of the electrical resistivity versus the Sommerfeld coefficient $\gamma$ (from [Tsujii 03]), and (c) low-temperature Seebeck thermo-electric power coefficient $S$ versus the Sommerfeld coefficient $\gamma$ (from [Behnia 04]) for various Fermi-liquid systems.

Following the discovery of a heavy-fermion behavior in CeAl$_3$, a large number of heavy-fermion materials have been identified. Extensive reviews can be found in [Stewart 84, Ott 87, Fisk 95, Radousky 00, Flouquet 05]. Figure 2.11 shows the comparison of the low-temperature physical properties for a large number of systems. It permits confirming the pertinence to describe them, within a first approximation, as a Fermi liquid governed by a single parameter, which can be the effective mass $m^*$:

- Figure 2.11(a) presents a plot in logarithmic scales of the low-temperature Sommerfeld

heat-capacity coefficient $\gamma = C/T$ versus low-temperature magnetic susceptibility $\chi$ for a large number of heavy-fermion materials [Lee 86]. This plot shows that $\gamma \propto \chi$, and equivalently that the ratio $R_W = \gamma/\chi$ introduced theoretically by Wilson [Wilson 75] is nearly constant and, thus, that $\gamma$ and $\chi$ are controlled by a single parameter, within first approximation.

- Figure 2.11(b) presents a plot in logarithmic scales of the quadratic coefficient $A$ in the temperature dependence of the electrical resistivity versus the Sommerfeld coefficient $\gamma$ for a large number of Fermi-liquid materials (from [Tsujii 03]). This plot demonstrates that $A \propto \gamma^2$ and equivalently that the Kadowaki-Woods ratio $R_{KW} = A/\gamma^2$ [Kadowaki 86] is nearly constant. $\sqrt{A}$ and $\gamma$ are, thus, controlled by a single parameter within first approximation.

- Figure 2.11(c) shows that the Seebeck thermo-electric power coefficient $S$ divided by the temperature scales with the Sommerfeld coefficient $\gamma$ for a large number of Fermi-liquid materials [Sakurai 01, Behnia 04]. As well as $\gamma$ and $\chi$, $S/T$ is, thus, controlled by the effective mass too. However, in relation with the sign of the Seebeck coefficient $S$, either a positive or a negative coefficient characterizes the proportionality of $S/T$ and $\gamma$. An electron contribution is expected to lead to a negative Seebeck coefficient, while a hole contribution is expected to lead to a positive Seebeck coefficient. A positive coefficient is observed for Ce compounds and some U compounds while a negative coefficient is observed for Yb compounds and some U compounds. Anisotropic Seebeck coefficients with a sign varying with the electrical current direction have also been reported in heavy-fermion magnets (see for instance [Boukahil 14, Palacio 13, Palacio 15]). The relation between the sign of $S$ and the nature of the electronic correlations driving the Fermi liquid regime still needs to be clarified.

In a Fermi liquid and within first approximation, we have the relation $\gamma \propto \chi \propto \sqrt{A} \propto S/T \propto m^*$ between several physical quantities. Other quantities, not considered here, are also the signature of a Fermi liquid and scale together, being controlled by the same parameter $m^*$ as well. In Section 2.4.3 and Chapter 3, the relation between the effective mass $m^*$ with i) the energy scale of a microscopic phenomenon, the intersite magnetic correlations developing in the correlated paramagnetic regime, and ii) the temperature boundary $T_\chi^{max}$ and the field boundary $H_m$ of this regime, will be emphasized.

### 2.3.3 Hybridization and Kondo effect

A hybridization between the $f$ electrons and the conduction electrons is a ground basis for modeling $f$-electron intermediate-valent and heavy-fermion systems. It describes the interaction between $f$ electrons and the electrons from the conduction band, which partly originates from the $s$ and $d$ valence shells of $f$ atoms, and the deviations from integer valences induced in these systems. Detailed reviews about hybridization models and theories can be found in [Lawrence 81, Lee 86, Ott 87, Hewson 93, Tsunetsugu 97, Arko 99, Kuramoto 00, Riseborough 16]. Historically, a $s - d$ exchange between the spin of localized $d$-electrons and that of conduction $s$-electrons was first introduced by Zener to describe the magnetic properties of transition metals [Zener 51]. Ten years later, Anderson proposed a model for localized magnetic impurities in metals where, in addition to a hybridization interaction, he introduced a Coulomb repulsion between the localized spins from $d$-shell electrons [Anderson 61]. Soon after, Kondo used the $s - d$ hybridization model to describe the scattering of conduction electrons by localized magnetic impurities [Kondo 64]. He proposed an explanation for the minimum, followed by an increase at low temperature, of the electrical resistivity observed in alloys containing diluted magnetic impurities (see Figure 2.12(a)). The Kondo effect was born. Its Hamiltonian can be written in the simple form:

$$H_K = J_K \mathbf{S} \cdot \mathbf{s}, \tag{2.1}$$

where $J_K > 0$ is an antiferromagnetic coupling between a localized magnetic spin $\mathbf{S}$ and the spin $\mathbf{s}$ of an itinerant electron. The Kondo effect leads to the formation of a singlet ground state, corresponding to a localized magnetic moment screened by the spin of the conduction electrons [Yosida 66]. Schrieffer and Wolff further showed that the $s - d$ Kondo model corresponds to a limiting case of the Anderson model [Schieffer 66]. In 1969, Coqblin and Schrieffer extended the Kondo model to localized electrons carrying a magnetic moment of quantum number $J = L \pm S$, for $f$-electron systems where spin-orbit interaction is active [Coqblin 69]. They extracted the Kondo temperature, i.e., the characteristic temperature scale of the Kondo effect:

$$T_K \sim D \exp\left(-\frac{1}{N\rho(E_F)J_K}\right), \tag{2.2}$$

where $D$ is the energy width of the conduction band, $\rho(E_F)$ the density of states of conduction electrons at the Fermi surface, $J_K$ is the Kondo exchange interaction, and $N$ is the multiplicity of the localized-electron ground state. Soon after, Cornut and Coqblin generalized this model to $f$-electron systems where a crystal-field lifts the degeneracy of the spin-orbit ground state levels [Cornut 72] (see also Section 2.4.1.1). Nozières further showed that a low-temperature Fermi-liquid behavior can result from the $s - d$ Kondo model [Nozières 74], while Wilson highlighted that the low-temperature magnetic susceptibility and specific heat are related by a constant ratio [Wilson 75]. In single-impurity models, the Kondo temperature is the temperature scale of the Fermi-liquid regime. Within certain sets of parameters, such models can lead to a broad maximum of the magnetic susceptibility, at a temperature $T_\chi^{max} \sim T_K$, accompanying the onset of a Fermi liquid (see for instance [Newns 80, Rajan 83, Cox 86, Cox 87, Zwicknagl 90, Hewson 93]). $T_K$ is inversely proportional to the effective mass $m^*$ of the Fermi liquid, i.e., $T_K \propto 1/m^*$, and the low-temperature magnetic susceptibility $\chi$ and heat-capacity $C_p$ are related to $T_K$ (see [Tsvelick 82, Riseborough 16]) by:

$$\chi \propto \frac{C_p}{T} \propto \frac{1}{T_K}. \qquad (2.3)$$

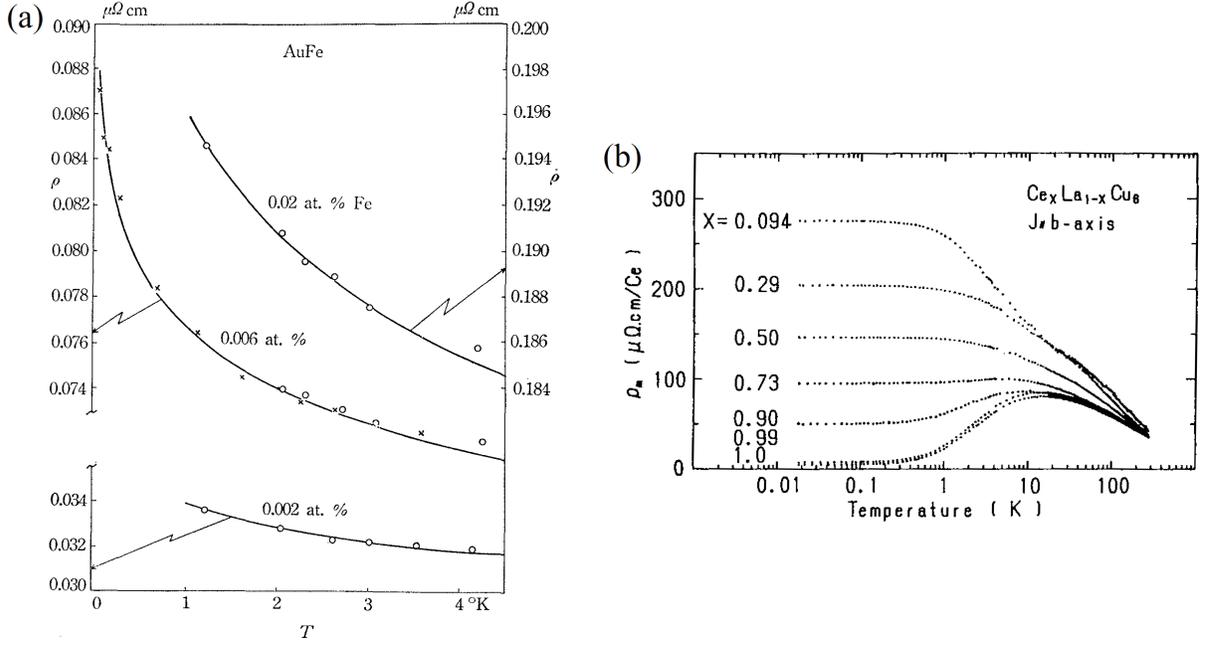

Figure 2.12: Electrical resistivity versus temperature (a) of dilute AuFe alloys and their fit by the Kondo model (from [Kondo 64]), and (b) of $Ce_xLa_{1-x}Cu_6$ alloys (from [Ōnuki 87]).

Samples with a Fermi-liquid regime driven by a large effective mass $m^*$ are those labeled as heavy-fermion systems. A small temperature $T_\chi^{max} \leq 50$ K marks the onset of the heavy-fermion correlated regime and can be identified as the Fermi-liquid temperature. Intermediate-valent systems are the place of a stronger hybridization, associated with higher energy scales and, thus, with higher temperature scales than those of heavy-fermion systems. In $CeSn_3$, a broad maximum in the magnetic susceptibility is observed at a high-temperature $T_\chi^{max} \simeq 150$ K [Lawrence 79, Béal-Monod 80], indicating a progressive crossover to an intermediate valent-regime similar to the heavy-fermion regime, and suggesting the achievement of a moderately-enhanced effective mass. However, other intermediate-valent systems, as Ce [Koskenmaki 78], undergo a first-order valence transition leading to a very different feature, a sudden fall, of the magnetic susceptibility at low temperature (see Section 2.3.1).

Figure 2.12(b) presents the electrical resistivity versus temperature of $Ce_xLa_{1-x}Cu_6$ alloys [Ōnuki 87]. Compounds with small Ce contents $x$ correspond to a situation with diluted and non-interacting magnetic impurities, whose electrical resistivity increases at low temperature and can be described by a single-impurity Kondo effect [Kondo 64]. For higher Ce contents $x$, the magnetic impurities form a dense lattice where intersite interactions develop, leading to a maximum in the electrical resistivity at the temperature $T_\rho^{max}$. In these so-called Kondo-lattice compounds, a Fermi-liquid quadratic variation of the electrical resistivity occurs at temperatures $T \ll T_\rho^{max}$. The pioneering theoretical works developed for diluted magnetic impurities in a metal were adapted for such periodic arrangement of localized magnetic moments, leading to periodic Anderson models (see for instance [Leder 78]) and Kondo-lattice models (see for instance [Doniach 77]). These lattice models aim describing real crystalline systems, where intersite effects as magnetic exchange interactions lead to long-range magnetic order.

However, due to the complexity of the heavy-fermion many-body problem, a quantitative description of the electronic properties and of their microscopic origin is often lacking. Difficulty results from the presence of multiple magnetic interactions paths, single-ion magnetic anisotropy and anisotropic magnetic exchanges, which add to the dual localized-itinerant character of the electrons. A challenge for theories is to describe quantitatively the magnetic fluctuations, which play a central role in the critical properties of heavy-fermion systems. Alternatively to Kondo and Anderson models of hybridization effects, magnetic-fluctuations theories assumed the presence of magnetic interactions (see Section 2.4.3.2). Magnetic-fluctuations models allow a qualitative description of the quantum critical magnetic properties when these interactions can be tuned [Hertz 76, Millis 93, Moriya 85]. The question of quantum magnetic phase transitions and magnetic ordering in heavy-fermion Kondo lattices will be considered in Section 2.4.

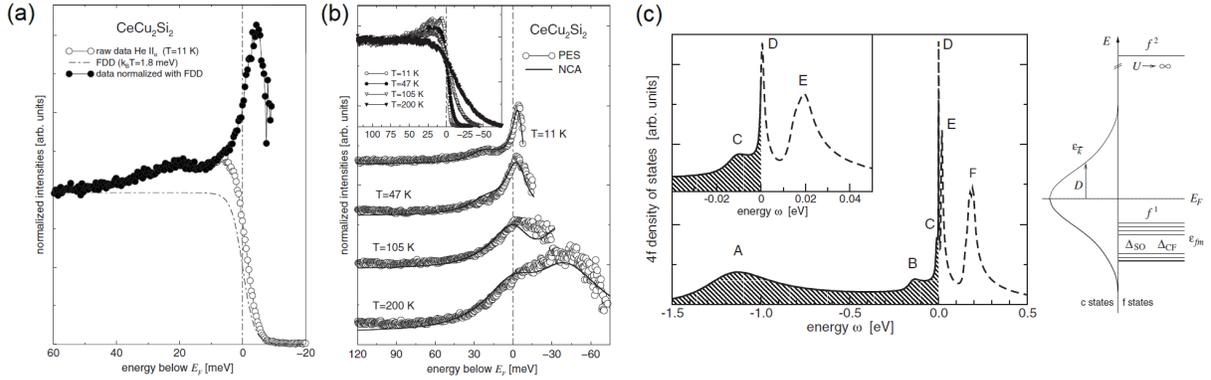

Figure 2.13: (a) Photoemission spectrum (open circles: raw data, closed circles: normalized data) at $T = 11$ K, and (b) photoemission spectra at temperatures from 11 to 200 K of CeCu$_2$Si$_2$. (c) Calculation of $4f$ spectrum of CeCu$_2$Si$_2$ based on single-impurity Anderson model (A: ionization peak, B: Kondo resonance, C,E: crystal field, and B,D: spin-orbit satellites) and sketch of the energy-levels scheme for conduction $c$ electrons and $f$ electrons before their hybridization (from [Ehm 07]).

### 2.3.4 Electronic structures and Fermi surfaces

$f$-electrons heavy-fermion systems are metallic and are composed of multiple electronic bands, to which $f$ electrons generally contribute. Photoemission spectroscopy and quantum oscillation techniques permit to investigate their electronic structure and Fermi surface, i.e., the surface in the reciprocal space corresponding to conduction electrons at the Fermi energy. Here, an introduction to the electronic structure and Fermi surface of heavy-fermion systems is given. Complementary information about photoemission spectroscopy and band-structure calculations can be found in [Arko 99, Ehm 07, Moore 09, Fujimori 16, Zwicknagl 16], and reviews about the study of Fermi surfaces by quantum oscillations techniques can be found in [Ōnuki 95, Settai 07, Ōnuki 12].

Photo-emission spectroscopy measurements indicate that lanthanide [Ehm 07] and actinide [Moore 09, Fujimori 16] heavy-fermion compounds present a high density of electronic states close to the Fermi level. Two examples of such studies are presented below:

- In Figure 2.13(a), a normalized photoemission spectrum, measured on the heavy-fermion

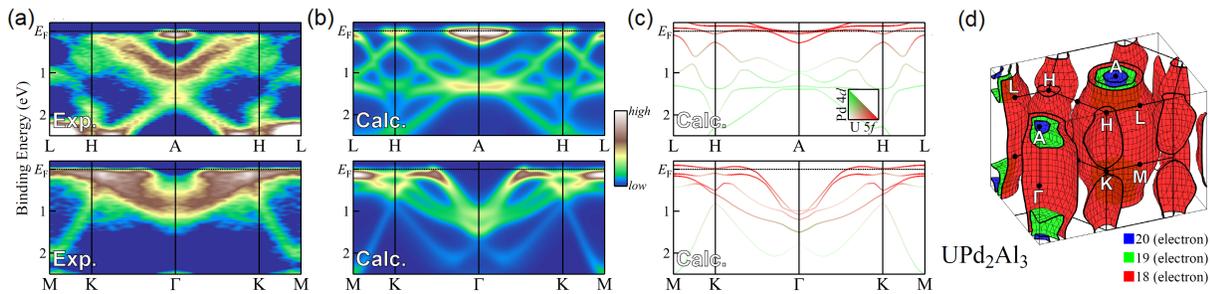

Figure 2.14: ARPES spectra of UPd$_2$Al$_3$ (a) measured at $T = 20$ K and (b) simulated by band-structure calculation, along the A-H-L and Γ-K-M high-symmetry lines. (c) Calculated energy band dispersions, where the color coding represents the contributions of U 5$f$ and Pd 4$d$ states. (d) three-dimensional shape of calculated Fermi surface in UPd$_2$Al$_3$ (from [Fujimori 16]).

paramagnet CeCu$_2$Si$_2$ at the temperature $T = 11$ K, shows the density of electronic states and reveals the presence of a resonance just above the Fermi energy [Ehm 07]. Figure 2.13(b) further shows that, when the temperature is raised, the resonance becomes broader and progressively vanishes. A description by a band-structure calculation, assuming a Kondo-type hybridization between localized $f$ electrons and conduction electrons, is presented in Figure 2.13(c). This single-impurity Anderson model captures the main features of the experimental spectrum: the Kondo resonance close to the Fermi level and satellites related with crystal-field and spin-orbit states.

- Angle-resolved photo-emission spectroscopy (ARPES) enables extracting wavevector **k**-dependent photo-emission spectra. Figure 2.14(a) shows ARPES data collected on the heavy-fermion compound UPd$_2$Al$_3$ in its paramagnetic phase, at a temperature $T = 20$ K higher than the Néel temperature $T_N = 14$ K of this antiferromagnet (from [Fujimori 16]). These features are compared to the results from band calculations assuming itinerant $f$ electrons, shown in Figures 2.14(b,c). The theoretical spectrum presents weakly-dispersive bands with strong 5$f$-character near the Fermi surface, and strongly-dispersive bands with a dominant 4$d$-character far below the Fermi level. Figure 2.14(d) shows the Fermi surface, i.e., points in the reciprocal space at which the Fermi energy is reached, of three electronic bands deduced from these calculations.

While ARPES measurements can be only performed at ambient pressure and zero magnetic field, Fermi surfaces can be studied under pressure and magnetic field by quantum-oscillation techniques, via the measurement of Shubnikov-de-Haas (SdH) and de-Haas-van-Alphen (dHvA) effects. Figure 2.15(a) presents an example of dHvA quantum oscillations observed in the magnetization of UPd$_2$Al$_3$, and their Fourier transform (from [Haga 99]). Peaks at a frequency $F$ in the Fourier transform are related to a section of extremal area $A \propto F$ (maximum or minimum), perpendicular to the magnetic-field direction, of the Fermi surface. The angular dependence of the dHvA different frequencies, shown in Figure 2.15(b) for UPd$_2$Al$_3$, are compared with the results from band-structure calculations, shown in Figure 2.15(c), based on an itinerant-5$f$-electron model [Ōnuki 07]. An analysis of similar quantum-oscillation data in UPd$_2$Al$_3$ was also done using a model considering dual itinerant-localized 5$f$ electrons by Zwicknagl [Zwicknagl 16].

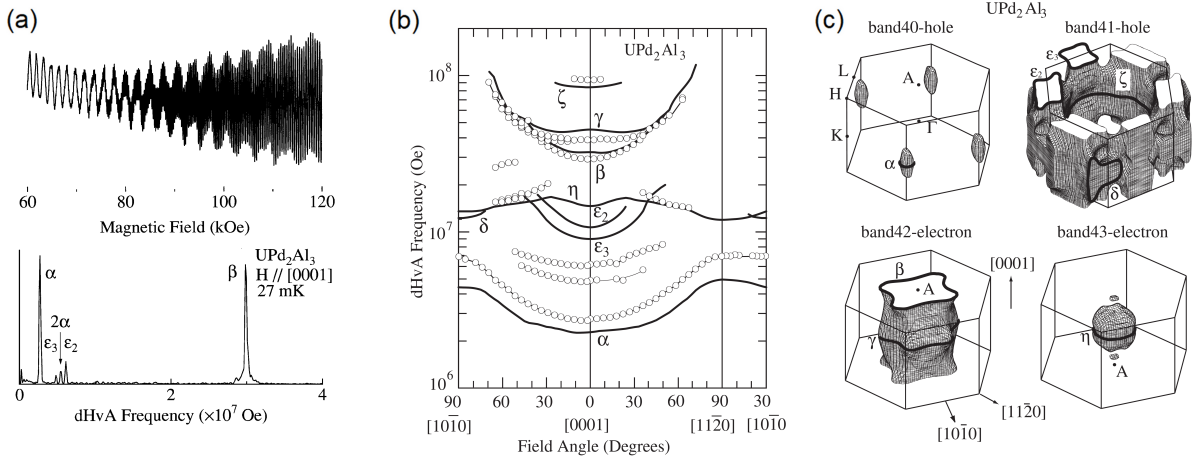

Figure 2.15: (a) de Haas-van Alphen quantum oscillations in the magnetization and their Fourier transformation, (b) angular dependence of the de Haas-van Alphen frequencies, where the thick solid lines are the results of band calculations, and (c) three-dimensional view of calculated Fermi-surface sheets of UPd$_2$Al$_3$ (from [Haga 99, Ōnuki 07]).

Quantum oscillations measurements permit to access different information about the electronic properties:

- *Delocalization of f electrons.* The comparison with isostructural compounds whose last $f$ subshell is empty (or full) can indicate the pertinence of a localized- or itinerant description of $f$-electrons in lanthanide or actinide compounds (see Table). For instance, the Fermi surface of Ce compounds in a localized limit [Xe]4$f^1$5$d^1$6$s^2$ is generally similar to that from their La-based parents of electronic configuration [Xe]5$d^1$6$s^2$. In the localized antiferromagnet CeIn$_3$ [Ebihara 93], the Fermi surface resembles that of LaIn$_3$ [Umehara 91]. As well, the Fermi surface of Ce compounds in an itinerant limit [Xe]4f$^{1-\delta}$5d$^{1+\delta}$6s$^2$ can be similar to that from their Th-based parents of electronic configuration [Rn]6$d^2$7$s^2$. In CeIn$_3$ in its paramagnetic phase induced under pressure, the Fermi surface [Settai 05] resembles that of ThIn$_3$, which indicates that CeIn$_3$ is in an itinerant limit under pressure [Matsuda 05].

- *Cyclotron masses.* By analyzing the temperature-dependence of quantum-oscillations measurements, one can extract the cyclotron mass $m^*_{c,i}$ of an orbit from a Fermi-surface band $i$. In a Fermi-liquid regime driven by itinerant electrons, the effective mass $m^*$ is proportional to the density of states at the Fermi level, and it can be related to the cyclotron

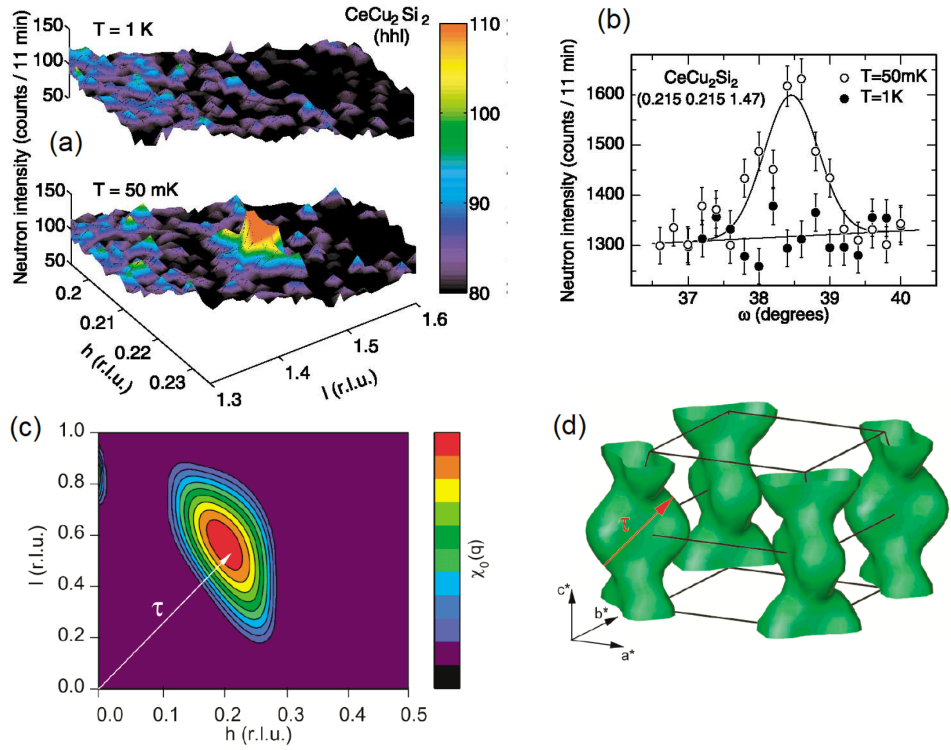

Figure 2.16: (a) Mapping of magnetic neutron-diffraction intensity in the $(hhl)$ plane of the reciprocal lattice, (b) plot of the intensity versus $w$ angle (rocking curve of the sample) around the Bragg peak at the wavevector $\tau = (0.215, 0.215, 0.53)$, at the temperatures $T = 50$ mK and 1 K, and plot of the low-temperature static susceptibility $\chi_0$ in the $(hhl)$ plane estimated from the previous neutron measurements, for CeCu$_2$Si$_2$ A-phase sample. (d) View of a Fermi surface sheet calculated by a renormalized-band method (from [Stockert 04]).

masses by [Ōnuki 95, Aoki 00]:

$$m^* \propto \sum_i m^*_{b,i} \propto \sum_i m^*_{c,i} a_i, \qquad (2.4)$$

where $m^*_{b,i}$ is the mass of the band $i$, and where $a_i$ is a constant depending on the topology of the band. Electronic correlations lead to an enhancement of the effective mass by $(1+\lambda)$, where $\lambda$ is the enhancement factor. One can write $m^* = (1+\lambda)m_t$, where $m_t$ is the effective mass of a system without correlations. Equivalently, one can write $m^*_{c,i} = (1+\lambda_i)m^*_{c,i,t}$, where $m^*_{c,i,t}$ is the cyclotron mass of an orbital $i$ in a system without correlations and $\lambda_i$ is the enhancement factor specific to the orbital $i$. In a situation with enhanced electronic correlations on hot spots or hot lines from the Fermi surface, $\lambda_i$ can be different from $\lambda$. Otherwise, one can assume $\lambda_i = \lambda$. Determining fully a Fermi surface and, thus, extracting the full effective mass from the sum rule in Equation (2.4), is a difficult experimental task. A high number of Fermi-surface sheets is often present, and high band-cyclotron masses, in relation with a strong heavy-fermion character, request very-low-temperature measurements, sometimes beneath standard experimental accessible temperatures.

- *Nesting.* Assuming that itinerant electrons drive the magnetic properties, a nesting of the Fermi surface, i.e., the presence of two parallel parts of the Fermi surface connected with a vector $\mathbf{k}$, can be the signature of magnetic ordering with wavevector $\mathbf{k}$. A nesting effect was identified by Stockert *et al* in $CeCu_2Si_2$ in its A-phase, where antiferromagnetism is stabilized at temperatures below $T_N = 0.8$ K [Stockert 04]. Figure 2.16(a-c) shows different plots of neutron-diffraction measurements, in which a Bragg peak emerges at the incommensurate wavevector $\tau = (0.215, 0.215, 0.53)$, and Figure 2.16(d) presents a view of a Fermi-surface sheet calculated by a renormalized-band method, in which $\tau$ is identified as a nesting vector.

In heavy-fermion systems, the magnetic properties are driven by nearly-localized $f$-electrons, which also contribute to the Fermi surface. Magnetic fluctuations are at the heart of the heavy Fermi-liquid properties induced in the proximity of quantum magnetic phase transitions, at which long-range magnetic order develops (see Section 2.4). However, band calculations generally describe itinerant $f$ electron systems but ignore their magnetic properties. For instance, the identification or proposition of nesting effects remains rare. A challenge for a quantitative modeling of heavy-fermion quantum critical magnets would be to describe the interplay between the magnetic and the band properties. For that purpose, new generations of band calculations, in which the effects of intersite magnetic correlations between nearly-localized $f$-electrons would be considered, are needed.

## 2.4 Quantum magnetic properties

We have seen that the localization of $f$-electrons drives a progressive change from intermediate-valence to heavy-fermion physics, and ultimately to long-range magnetic order. Here, basic magnetic properties of localized or nearly-localized $f$ electronic states are presented. Within a competition with the Kondo effect, magnetic anisotropy and magnetic exchange interactions induce long-range magnetically-ordered phases in these systems. Quantum magnetic phase transitions are induced and are accompanied by quantum critical properties. Models based on magnetic fluctuations and describing quantum magnetic phase transitions are introduced. In complement, reviews about the quantum critical properties of heavy-fermion magnetic systems can be found in [Stewart 84, Stewart 01, Flouquet 05, von Löhneysen 07, Gegenwart 08, Knebel 11a, Aoki 13].

### 2.4.1 Ingredients for magnetism

#### 2.4.1.1 Crystal field and single-ion magnetic anisotropy

In $f$-electron compounds, spin-orbit coupling leads to a ground state of quantum number $J = L \pm S$, where $L$ is the total orbital momentum and $S$ the total spin of electrons from the incompletely-filled $f$ subshell. The spin-orbit ground-state is a multiplet of degeneracy $N = 2J+1$, composed of states $|J, M_J\rangle$, noted $|M_J\rangle$ in the following. $M_J$ differs by an integer number from $J$ and takes values $-J \leq M_J \leq J$ [Cohen-Tannoudji 06]. Spin-orbit electronic states carry a magnetic moment $\mu = gJ\mu_B$, where

$$g = 3/2 + ((S(S+1) - L(L-1))/(2J(J+1)))$$ (2.5)

is the Landé factor and $\mu_B$ the Bohr magneton. A crystalline electric field is induced by the local environment of the $f$ atoms and drives a modification of the $f$ orbitals [Lea 62, Walter 84, Walter 86]. It ends in a splitting of the spin-orbit multiplet into crystal-field levels, which consist in linear combinations of the $|M_J\rangle$ states. In relation with the lattice local environment, a crystal-field ground state presents anisotropic magnetic properties. Ultimately, magnetic exchange between $f$-electron magnetic moments (see Section 2.4) can lead to magnetic ordering, i.e., the stabilization of a ground state with magnetic moments with a wavevector **k**, equivalent to a splitting of the crystal-field ground state.

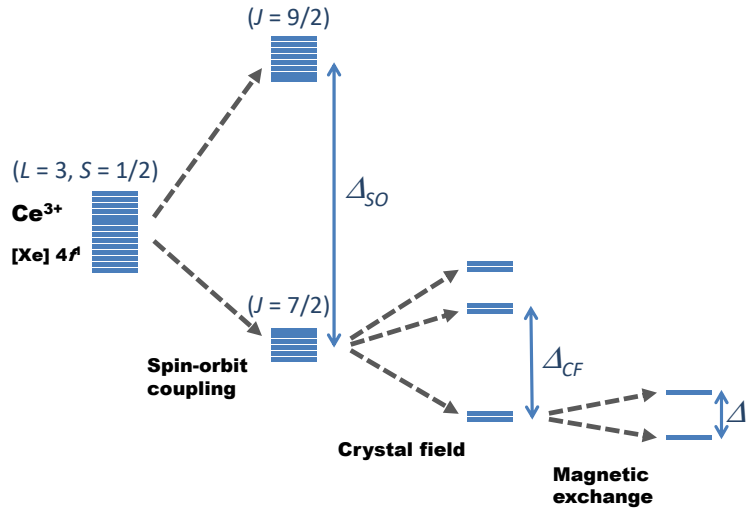

Figure 2.17: Schematic view of Ce$^{3+}$ $f$-electron-levels splitting by spin-orbit, crystal field, and magnetic exchange.

Figure 2.17 shows schematically the effects of spin-orbit coupling, crystal field, and magnetic exchange interaction in a Ce compound. This hierarchy of energy scales is illustrated in Figure 2.18 by inelastic neutron scattering spectra obtained on the heavy-fermion antiferromagnet CeIn$_3$, which presents a spin-orbit excitation at the energy $\Delta_{SO} \simeq 250 - 300$ meV, a crystal-field excitation at the energy $\Delta_{CF} \simeq 10 - 15$ meV, and a dispersive (wavevector-**k** dependent) spin-wave excitation, i.e., a signature of the magnetically-ordered state in this weakly anisotropic (cubic) system, at energies $\Delta < 3$ meV [Murani 90, Murani 93, Knafo 03].

In compounds with a small Kondo temperature $T_K < \Delta_{CF}$, i.e., close to a localized limit, the Kondo effect leads to a broadening of the $f$ electron levels by an energy bandwidth $\Delta \sim$ $T_K$ (see CeIn$_3$ in Figure 2.18(b) [Knafo 03]). In compounds with a large Kondo temperature $T_K > \Delta_{CF}$, the full spin-orbit multiplet, of degeneracy $N = 2J + 1$, instead of the crystal-field ground state with reduced degeneracy, is involved in the Kondo hybridization with the conduction electrons [Hanzawa 85]. The collapse of crystal-field splitting has been observed by inelastic-neutron-scattering experiments in CeIn$_{3-x}$Sn$_x$ samples with Sn-contents $x > 1.5$, for which a high Kondo temperature is achieved [Murani 93].

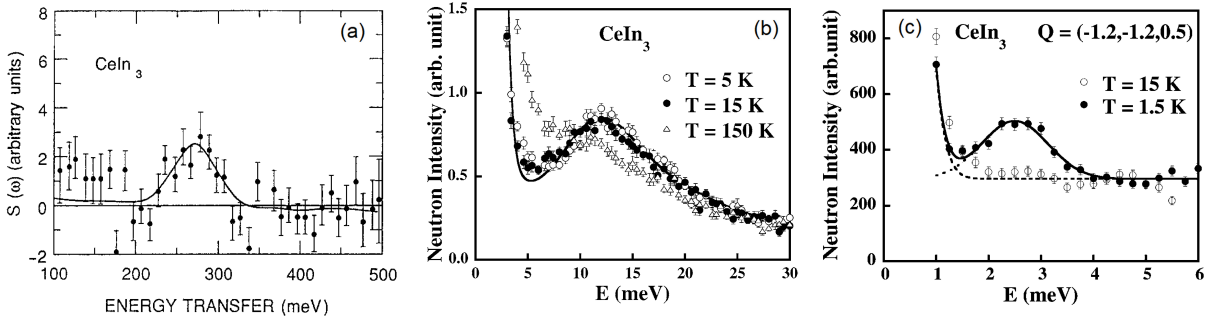

Figure 2.18: (a) Spin orbit (from [Murani 90]), (b) crystal-field, and (c) spin-wave (from [Knafo 03]) excitations observed by inelastic neutron scattering in CeIn$_3$.

Assuming localized $f$ electrons, one can calculate the magnetic susceptibility using a Van Vleck-type formula:

$$\chi^i = \frac{N_A(g\mu_B)^2}{Z} \sum_n (\beta \sum_{m,\epsilon_n=\epsilon_m} \mid \langle m \mid J_i \mid n \rangle \mid^2 e^{-\beta\epsilon_n}$$
$$+ \sum_{m,\epsilon_n \neq \epsilon_m} \mid \langle m \mid J_i \mid n \rangle \mid^2 \frac{e^{-\beta\epsilon_m} - e^{-\beta\epsilon_n}}{\epsilon_n - \epsilon_m}), \qquad (2.6)$$

where $N_A$ is the Avogadro number, $Z$ is the partition function, $\beta = 1/k_BT$, $i$ is the magnetic field direction, $n$ and $m$ are the indices of two crystal-field levels, and $\epsilon_n$ and $\epsilon_m$ are their respective energies [van Vleck 32, Aviani 01]. As first approximation, a fit to the high-temperature magnetic susceptibility by a Van-Vleck formula can give valuable information about the crystal-field states and related single-ion magnetic anisotropy. Figure 2.19(a) shows that, in the tetragonal paramagnet CeRu$_2$Si$_2$, a strong Ising magnetic anisotropy, revealed by the hierarchy $\chi_c \gg \chi_a$ in the magnetic susceptibilities, is a consequence of an almost pure $\mid \pm 5/2 \rangle$ groundstate [Haen 87a, Haen 92, Dakin 92].

In most heavy-fermion and intermediate-valent materials, Kondo hybridization develops at low-temperature and leads to deviations from Van-Vleck behavior. 'Kondo-modified' Van-Vleck formula can be used to describe the anisotropy of the magnetic susceptibility. By considering a single-impurity Kondo effect, Aviani $et$ $al$ analyzed the magnetic susceptibility of diluted-$f$-atom tetragonal paramagnets Ce$_x$La$_{1-x}$Cu$_{2.05}$Si$_2$ of low density of Ce atoms (contents $x \ll 1$) [Aviani 01]. Figure 2.20 shows the best fit to the magnetic susceptibility data of Ce$_{0.06}$La$_{0.94}$Cu$_{2.05}$Si$_2$, which is consistent with a ground-state doublet $\mid\Gamma_7^1 >= \eta \mid \pm 5/2 \rangle +$ $\sqrt{1 - \eta^2} \mid \mp 3/2\rangle$, with $\eta = 0.816$, and an excited-state quartet composed of $\mid \Gamma_7^2 \rangle = -\sqrt{1 - \eta^2} \mid \pm 5/2\rangle + \eta \mid \mp 3/2\rangle$ and $\mid \Gamma_6 \rangle = \mid \pm 1/2\rangle$ at the energy $\Delta_{CF} \simeq 30$ meV.

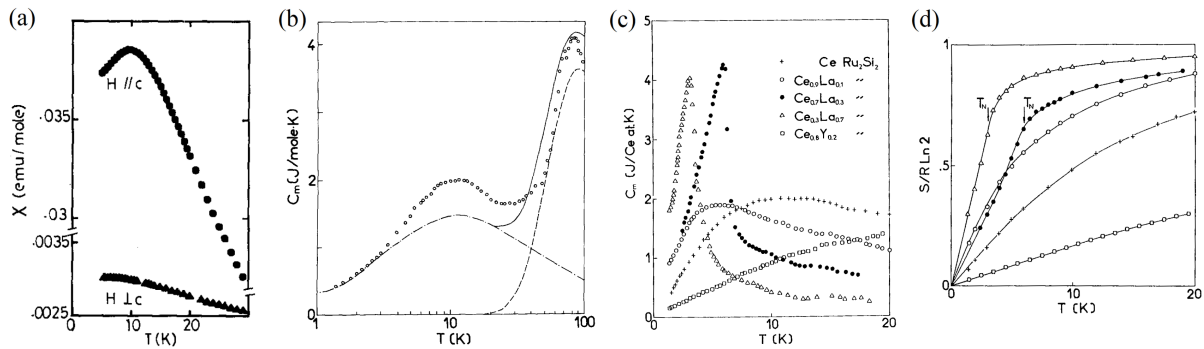

Figure 2.19: Magnetic susceptibility versus temperature of CeRu$_2$Si$_2$ for **H** $\parallel$ **c** and **H** $\perp$ **c** (from [Haen 87a]). Magnetic heat capacity (b) of CeRu$_2$Si$_2$ versus temperature up to 200 K (from [Besnus 85]), and (c) of a selection of Y- and La-doped CeRu$_2$Si$_2$ alloys versus temperature up to 20 K. (d) Magnetic entropy of Y- and La-doped CeRu$_2$Si$_2$ alloys versus temperature up to 20 K (from [Besnus 87]).

In complement to the analysis of magnetic susceptibility data, information about the electronic ground state of $f$-electron systems and their crystal-field levels can be extracted from heat-capacity versus temperature measurements. Once the non-magnetic lattice background has been subtracted, the magnetic part of the heat capacity $C_p^m$ can be related with the magnetic entropy $S_m$ by the relation:

$$S_m = \int \frac{C_p^m}{T} \partial T. \qquad (2.7)$$

Figure 2.19(b) shows the heat capacity versus temperature of the paramagnet CeRu$_2$Si$_2$. Its analysis confirms that the ground state is an electronic doublet, and that the first excited crystal-field state is a doublet with a gap $\Delta_{CF} \simeq 20$ meV [Besnus 85]. Figures 2.19(c-d) present the heat capacity and magnetic entropy, respectively, extracted for Y- and La-doped CeRu$_2$Si$_2$ alloys, whose ground state is either a Fermi-liquid paramagnet or an antiferromagnet [Besnus 87]. For all samples, the entropy at $T = 20$ K is smaller than the entropy $R\ln 2$ expected for a doublet, $R\ln 2$ being almost reached at temperatures above the Néel temperature (noted $T_m$ in the graphs) of the antiferromagnets, which corresponds to a nearly-localized $f$-electrons limit. The consideration of the electronic entropy permits to conclude that the low-temperature magnetic properties of most heavy-fermion systems, including CeRu$_2$Si$_2$ and its alloys, can be considered at first approximation as those of a $f$-electron doublet.

### 2.4.1.2 Magnetic exchange interactions

In Kondo-lattice compounds, Rudermann-Kittel-Kasuya-Yosida (RKKY) magnetic exchange interactions [Rudermann 54, Kasuya 56, Yosida 57], i.e., an indirect exchange between $f$-electron localized magnetic moments mediated by the conduction electrons, generally develop at low temperatures. They lead to magnetic fluctuations and/or long-range magnetic order, implying deviations from Van-Vleck paramagnetic susceptibility. The RKKY Hamiltonian can be written in the form:

$$H_{RKKY} = \sum_{i,j} J_{RKKY}(\mathbf{r}_{ij}) \mu_i \cdot \mu_j, \qquad (2.8)$$

where $J_{RKKY}(\mathbf{r}_{ij})$ is the RKKY interaction between magnetic moments $\mu_i$ and $\mu_j$ carried on two neighboring atoms $i$ and $j$, respectively, distant by $\mathbf{r}_{ij}$. The RKKY exchange can be expressed as:

$$J_{RKKY}(\mathbf{r}_{ij}) = C(\mathbf{r}_{ij}) \rho(E_F) J_K^2, \qquad (2.9)$$

where $J_K$ is the single-site Kondo exchange defined earlier (see Equation 2.8) and $\rho(E_F)$ is the density of states at the Fermi energy $E_F$. The sign of the factor $C(\mathbf{r}_{ij})$, and thus of the RKKY interaction, results from Friedel oscillations of the Coulomb potential created by a charged impurity (see Figure 2.21) [Friedel 58, Villain 16]. Depending on the distance between two magnetic atoms, either a ferromagnetic or an antiferromagnetic coupling can be induced by the RKKY interaction.

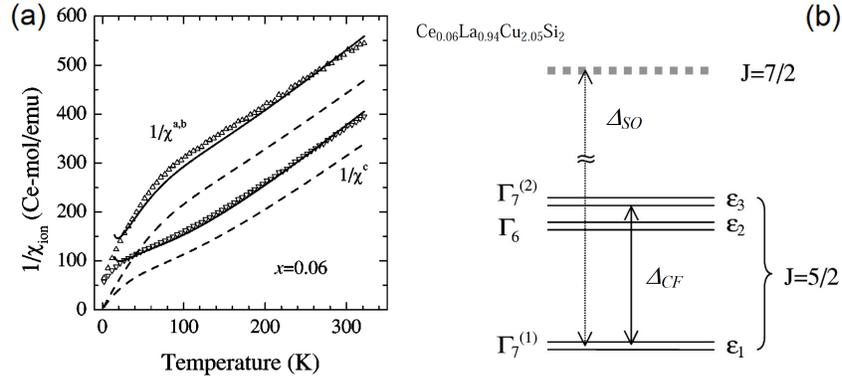

Figure 2.20: (a) Inverse of the magnetic susceptibility versus temperature and its fits following van-Vleck formula (dashed lines) and single-ion-Kondo-van-Vleck formula (full lines), and (b) crystal-field states determined by single-ion-Kondo-van-Vleck fit in $Ce_{0.06}La_{0.94}Cu_{2.05}Si_2$ (from [Aviani 01]).

The combination of magnetic exchange interactions with single-ion magnetic anisotropy generally ends with the development of long-range magnetic order, the magnetic moments being aligned along the high-temperature easy magnetic axis fixed by the crystal field (see for instance the Ising system $Ce_{1-x}La_xRu_2Si_2$ [Haen 87a, Quezel 88, Fisher 91]). However, anisotropy resulting from the magnetic exchange interactions can also compete with the single-ion crystal-field magnetic anisotropy, leading to long-range magnetic order with moments perpendicular to the high-temperature easy-magnetic axis. Such competition is observed in several systems, as $CeRhIn_5$, where long-range antiferromagnetic order sets in with moments $\perp$ **c**, **c** being the high-temperature easy magnetic axis fixed by the crystal field [Hegger 00, Bao 00]. Theories beyond the RKKY model, as the anisotropic hybridization models developed by Cooper *et al* [Cooper 85], can describe anisotropic magnetic interactions (see also [Sechovsky 98, Arko 99]). In addition, other mechanisms, as direct exchange interactions between overlapping $f$ orbitals [Sechovsky 98] or indirect superexchange interactions in systems with ligand atoms [Anderson 59](for instance $UO_2$ [Caciuffo 11]), can be operative in $f$ electron systems.

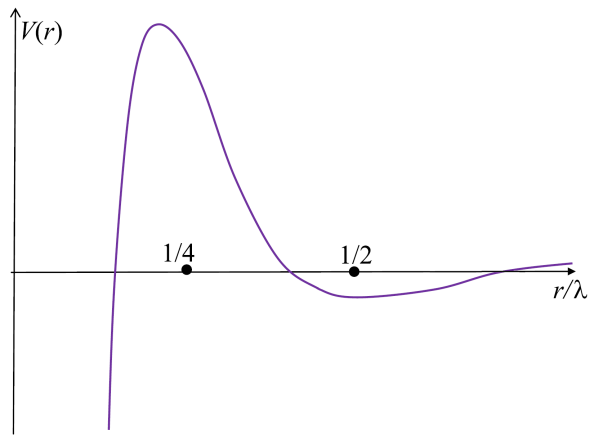

Figure 2.21: Friedel oscillations of the Coulomb potential as function of the distance $r$ from a charged impurity (from [Villain 16]).

### 2.4.2 Quantum magnetic phase transitions

#### 2.4.2.1 Long-range magnetic order

We have seen in Sections 2.2.3 and 2.3 that heavy-fermion states and long-range magnetic order phases develop in the limit of localized $f$ electrons in lanthanides and actinides compounds. A reduced strength of the Kondo hybridization in materials within localized limit allows magnetic exchange interactions to operate, leading in many compounds to the onset of long-range magnetic order. Most of heavy-fermion systems are in the vicinity of a quantum magnetic phase transition, at which their magnetic exchange interactions and magnetic anisotropy are strong enough to induce long-range magnetic ordering.

Doniach showed that, for a one-dimensional lattice [Doniach 77], a quantum magnetic phase transition can result from the competition between the Kondo hybridization [Kondo 64] and indirect RKKY magnetic exchange interaction [Rudermann 54, Kasuya 56, Yosida 57]. Figure 2.22 presents the variations of the temperatures scales $T_K \propto \exp(-1/(N\rho(E_F)J_K))$ characteristic of the Kondo effect (see Equation 2.2) and $T_{RKKY} \propto J_{RKKY} = C\rho(E_F)J_K^2$ characteristic of the RKKY interaction (see Equation 2.9) as function of a tuning parameter $\delta$. The parameter $\delta$ is equal to $\rho(E_F)J_K$, where $J_K$ is the Kondo coupling between localized and conduction electrons, and $\rho(E_F)$ the density of conduction electrons at the Fermi energy $E_F$. For a large coupling $\delta > \delta_c$, the Kondo effect dominates and a paramagnetic state is reached. For a small coupling $\delta < \delta_c$, the RKKY interaction dominates and a magnetically-ordered state is reached. A quantum phase transition, i.e., a phase transition in the limit of zero-temperature, between these two phases is induced at the critical coupling value $\delta_c$, at which the two energy scales are comparable. A superconducting phase, not considered in the original work from Doniach but observed at the quantum magnetic phase transition of many heavy-fermion compounds (see Section 2.5), is also indicated in Figure 2.22.

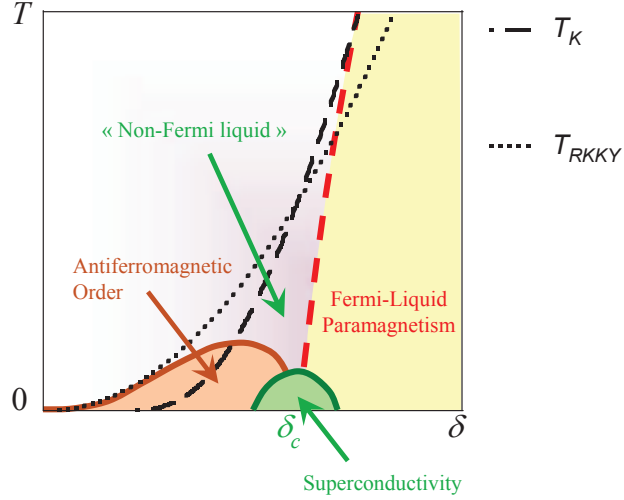

Figure 2.22: Doniach's phase diagram. The short dashed line is the RKKY temperature $T_{RKKY}$, at which magnetic order would occur in the absence of Kondo screening. The long-dashed line is the Kondo temperature $T_K$, below which $f$-electron moments are screened by the conduction electrons. For $T_{RKKY} = T_K$, magnetic ordering temperature is driven to zero, leading to a quantum phase transition between an antiferromagnetic and a paramagnetic Fermi-liquid ground states. In many heavy-fermion systems, a superconducting pocket develops in the vicinity of the quantum phase transition (from [Aoki 13]).

CeIn$_{3-x}$Sn$_x$ constitutes a textbook example of a progressive change from intermediate-valence to Kondo physics, and ultimately to long-range magnetic order, in agreement with the picture from Doniach [Doniach 77]. Figures 2.23(a-b) present magnetic susceptibility $\chi$ versus temperature data of CeIn$_{3-x}$Sn$_x$ compounds. In pure CeSn$_3$ a broad maximum of $\chi$ at $T_\chi^{max} \simeq 150$ K has been interpreted as the signature of a valence crossover towards low-temperature intermediate-valence state [Béal-Monod 80]. At room temperature, Vegard's law, i.e., a linear variation with doping $x$ of the lattice parameter $a$, is followed for $x \leq 2.5$. For $x > 2.5$, a downwards deviation from Vegard's law can be considered as a precursor of the valence crossover occurring at lower temperatures, presumably for $T < T_\chi^{max}$ [Béal-Monod 80]. In CeIn$_{3-x}$Sn$_x$ alloys, $T_\chi^{max}$ decreases when the Sn-content $x$ is decreased. The enhancement of $\chi$ at low temperature indicates the onset of a heavy-fermion regime for $0.5 \leq x \leq 1.5$. At lower Sn-dopings $x < x_c \simeq 0.4$, long-range antiferromagnetic order develops at low temperature, with a maximal Néel ordering temperature $T_N = 10$ K in pure $CeIn_3$ [Lawrence 79]. The higher low-temperature magnetic susceptibility $\chi \simeq 15 \cdot 10^{-3}$ emu/mol, and thus, the heavier effective mass, are observed at the critical concentration $x_c$, where they are a factor 7.5 higher than in the intermediate-valent compound $CeSn_3$. This maximum of the effective mass is a signature of quantum magnetic criticality, which will be addressed in Section 2.4.2.2.

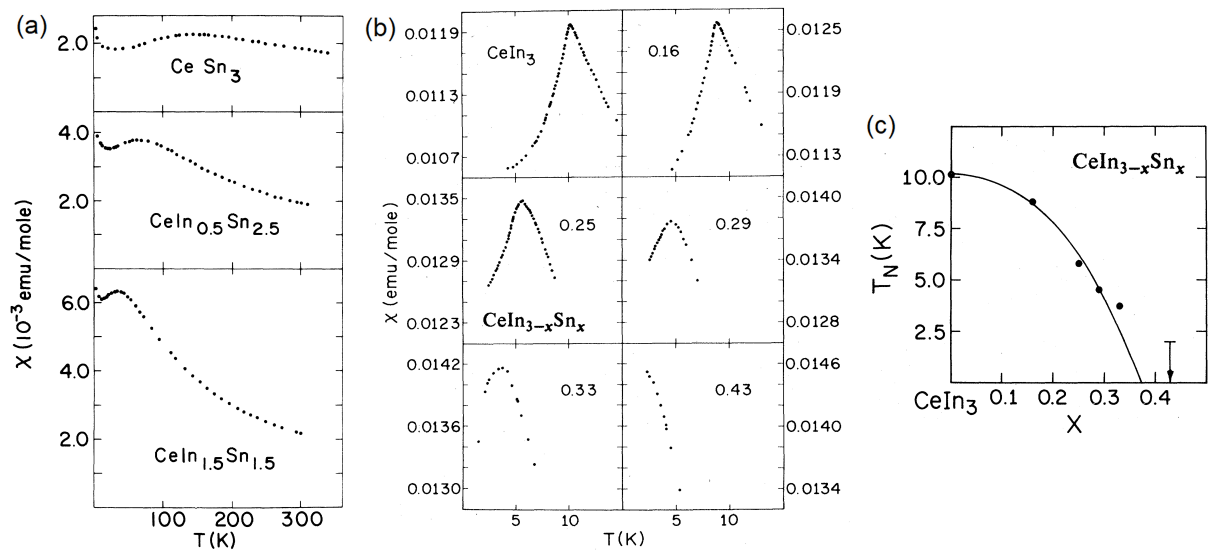

Figure 2.23: Magnetic susceptibility versus temperature of $CeIn_{3-x}Sn_x$ (a) for $x \geq 1.5$ (from [Béal-Monod 80]), (b) for $x < 0.5$ (from [Lawrence 79]) and (c) Néel temperature versus doping $x$ (from [Lawrence 79]).

Alternatively to chemical doping, pressure allows tuning quantum magnetic phase transitions in heavy-fermion compounds. Contrary to doping, pressure does not affect the sample quality and opens the route to phenomena only observable in high-quality samples, as unconventional superconductivity or de Haas - van Alphen and Shunikov-de Haas quantum oscillations. Complementarily, a magnetic field permits to continuously tune the magnetic ground state of these systems, ending in a polarized paramagnetic regime at high magnetic fields, where the electronic correlations have been (partly or totally) quenched. Figure 2.24 presents the characteristic three-dimensional (3D) magnetic field - pressure - temperature phase diagram of a heavy-fermion Ising (easy-axis-anisotropic) antiferromagnet in a magnetic field applied along its easy magnetic axis [Knafo 17]. Starting from an antiferromagnetic (AF) ground state at zero-field and ambient pressure, a correlated paramagnetic (CPM) regime is generally established under pressure at temperatures $T < T_\chi^{max}$. The magnetic susceptibility nearly saturates and indicates the onset of a heavy Fermi liquid in the CPM regime. Both AF phase and CPM regime are destabilized by a magnetic field, at a critical field $H_c$ and at a metamagnetic, or pseudo-metamagnetic, field $H_m$, respectively. The superconducting phase, which develops at the quantum magnetic phase transition of many heavy-fermion compounds, is also represented

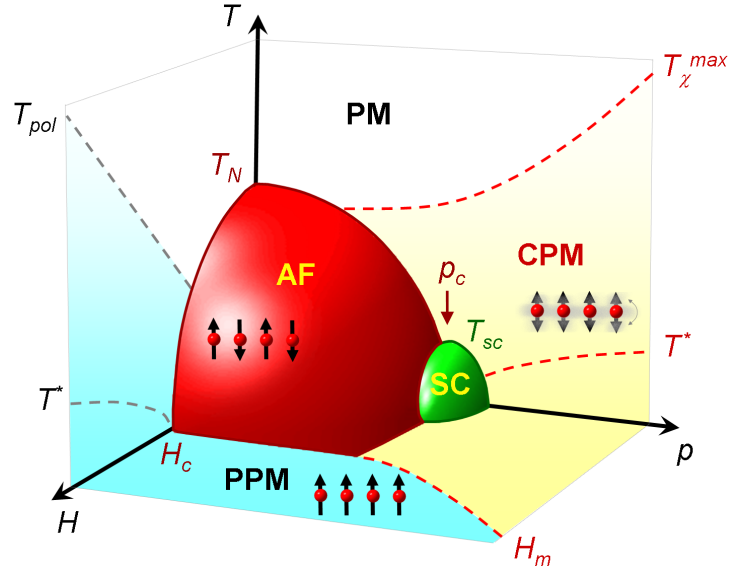

Figure 2.24: Schematic magnetic field - pressure - temperature phase diagram of a heavy-fermion Ising antiferromagnet in a magnetic field applied along its easy magnetic axis. AF, SC, CPM and PPM denote the antiferromagnetic and superconducting phases, the correlated paramagnetic and polarized paramagnetic regimes, respectively. Sketches of antiferromagnetic moments, polarized moments, and fluctuating moments (assuming Ising magnetic anisotropy) are presented for the AF phase and the PPM and CPM regimes, respectively. The parameters $T_N$, $T_\chi^{max}$, $T_{pol}$, $T_{sc}$, $p_c$, $H_c$, and $H_m$ presented in this phase diagram are defined in the text (from [Knafo 17]).

in this 3D phase diagram.

Beyond the case of Ising antiferromagnets, a large variety of magnetically-ordered phases is observed in heavy-fermion materials, from ferromagnetism to antiferromagnetism, including collinear and non-collinear structures, with Ising (easy-axis) or XY (easy-plane) magnetic anisotropies, spin-density waves with incommensurate wavevectors $\mathbf{k}$, single or multi-$\mathbf{k}$ structures etc. Competing magnetic interactions can also lead to magnetic frustration in these systems. The proximity to a quantum magnetic phase transition is indicated by reduced long-range ordered magnetic moments, associated with the presence of a non-Fermi liquid critical regime, which will be introduced in Section 2.4.2.2, and strong magnetic fluctuations, which will be introduced in Section 2.4.3.

### 2.4.2.2 Quantum critical properties

Deviations from a Fermi-liquid behavior, i.e., critical power or logarithmic laws in different physical quantities, are observed in the vicinity of a quantum magnetic phase transition in heavy-fermion compounds. In such low-temperature non-Fermi liquid regime, the electrical resistivity varies as $\rho(T) = \rho_0 + AT^n$, with $1 < n < 1.5$, the magnetic susceptibility can vary as $\chi(T) \propto \chi_0(1 - c(T/T_0)^{1/2})$, and the specific heat can vary as $C/T \propto -\ln T$. An extensive review of non-Fermi-liquid phenomena is presented in [Stewart 01].

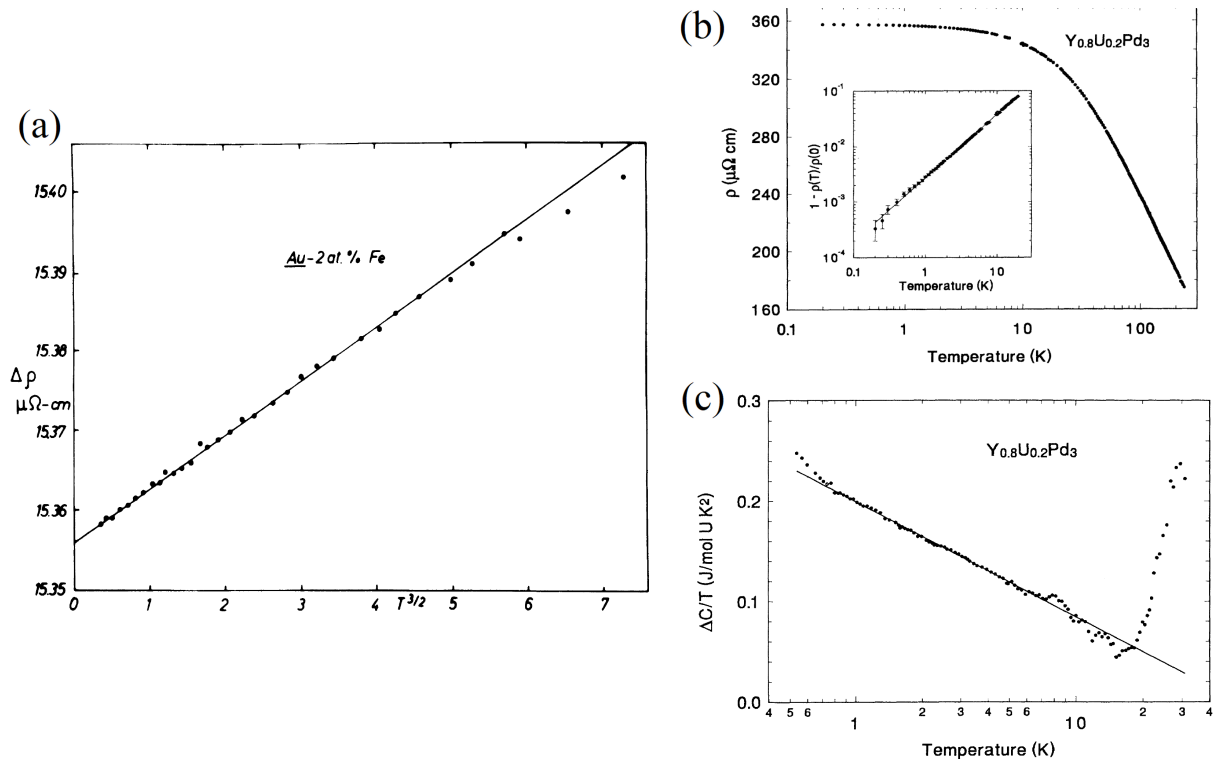

Figure 2.25: Non-Fermi liquid behavior (a) in the electrical resistivity of 2%-Fe AuFe alloys [Mydosh 73], and in (b) the electrical resistivity and (c) the heat capacity of $Y_{0.8}U_{0.2}Pd_3$ (from [Seaman 91]).

Figure 2.25 presents the first experimental reports of a non-Fermi-liquid behavior, (a) in an Au-Fe alloy of 2%-Fe content, which is a quantum spin system close to a ferromagnetic phase transition [Mydosh 73], and (b-c) in $Y_{0.8}U_{0.2}Pd_3$, a diluted U-alloy [Seaman 91]. While a low-temperature $T^{3/2}$ increase of the electrical resistivity was reported for the Au-Fe alloys (see Figure 2.25(a)), a linear decrease of the resistivity with temperature was reported for $Y_{0.8}U_{0.2}Pd_3$, in a low-temperature regime where a logarithmic variation of the heat capacity was also identified (see Figure 2.25(b)).

In the dense Kondo lattice $CeCu_{6-x}Au_x$, a quantum magnetic phase transition is induced at the critical Au-doping $x = 0.1$ [Schlager 93]. Paramagnetism sets in down to the lowest temperatures for $x < x_c$ and long-range antiferromagnetism is established for $x > x_c$. Figure 2.26 presents the heat capacity divided by the temperature (a) of $CeCu_{6-x}Au_x$ compounds, (b) of $CeCu_{5.8}Au_{0.2}$ under pressure, and (c) of $CeCu_{5.8}Au_{0.2}$ in a magnetic field $\mathbf{H} \parallel \mathbf{c}$ [von Löhneysen 96]. A second-order-like step in the heat capacity indicates the onset of long-range magnetic order at the Néel temperature $T_N$ for $x > x_c$, at ambient pressure and zero-field. Similar signatures of antiferromagnetic ordering occur at $T_N$ for $p < p_c \simeq 3$ kbar and

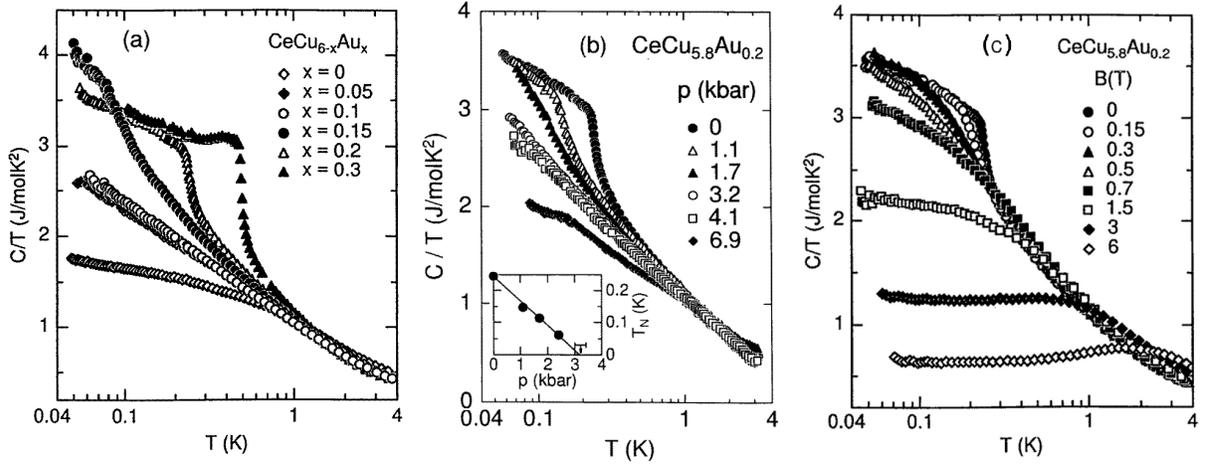

Figure 2.26: Non-Fermi liquid behavior in the heat capacity of (a) $CeCu_{6-x}Au_x$, (b) $CeCu_{5.8}Au_{0.2}$ under pressure, and (c) $CeCu_{5.8}Au_{0.2}$ in a magnetic field $\mathbf{H} \parallel \mathbf{c}$ (from [von Löhneysen 96]).

$\mu_0 H < \mu_0 H_c \simeq 0.5$ T in the antiferromagnet $CeCu_{5.8}Au_{0.2}$. At the critical doping $x_c$, but also at the critical pressure $p_c$ and at the critical field $H_c$, a non-Fermi liquid behavior is evidenced by a logarithmic variation of the heat capacity divided by the temperature $C_p/T$, ending in the maximal value of the Sommerfeld coefficient $\gamma =_{T\to 0} C_p/T$ in the limit of zero-temperature, and thus of the effective mass $m^*$.

Figure 2.27(a) presents the magnetic-field-temperature phase diagram of the heavy-fermion weakly-anisotropic antiferromagnet $YbRh_2Si_2$ in a magnetic field $\mathbf{H} \parallel \mathbf{c}$ [Gegenwart 02]. Small energy scales lead to a Néel temperature $T_N = 70$ mK, and to a vanishing of the antiferromagnetic phase in magnetic fields $\mu_0 H > \mu_0 H_c \simeq 0.65$ T. Figure 2.27(b) presents the electrical resistivity versus temperature of $YbRh_2Si_2$ in a magnetic field $\mathbf{H} \parallel \mathbf{c}$, emphasizing that a quadratic Fermi-liquid $T^2$ variation of the low-temperature electrical resistivity occurs in both the antiferromagnetic phase for $H < H_c$ and in the high-field polarized paramagnetic phase, for $H > H_c$. At the critical field $H_c$, a linear increase of the electrical resistivity is the signature of a non-Fermi liquid. Such $T$-linear variation of $\rho$ is also observed at fields $H \neq H_c$ for $T > T_N, T_\rho^*$, where $T_\rho^*$ the temperature below which a Fermi-liquid $T^2$ law is found. Figure (c) superposes a color plot of the $n$ coefficient, extracted from $\rho = \rho_0 + AT^n$ on the $(H, T)$ phase diagram of $YbRh_2Si_2$, emphasizing the coincidence of the non-Fermi liquid with the quantum magnetic phase transition induced by a magnetic field [Custers 03].

## 2.4.3 Quantum magnetic fluctuations

### 2.4.3.1 Phenomenology

Magnetic fluctuations, i.e., fluctuations of the magnetic moments, play a crucial role in the quantum critical properties of heavy-fermion lanthanide and actinide materials, in both their paramagnetic and magnetically-ordered phases. Figure 2.28(b) shows schematically that a magnetic moment can be decomposed into a static ordered part and fluctuating parts, which can be longitudinal or transverse. While paramagnets are only the place of magnetic fluctuations, long-range-ordered magnets consist in a superposition of magnetic fluctuations and magnetically-ordered moments. In heavy-fermion long-range-magnetically-ordered compounds, a reduced low-temperature magnetic moment often results from the presence of magnetic fluctuations, which grow in the proximity of a quantum magnetic phase transition. As well, thermal magnetic fluctuations are responsible for a reduction of the ordered moment in magnets at temperatures $T \lesssim T_{C,N}$. The 'total' magnetic moment is fixed by the electronic ground-state properties, which result from spin-orbit and crystal-field effects. It is smaller than the effective magnetic moment $\mu_{eff} = g\sqrt{J(J+1)}\mu_B$, which can be determined from Curie-Weiss behaviors in the high-temperature magnetic susceptibility. Here, the generic term 'magnetic fluctuations' will be preferred to 'spin fluctuations', since $f$-electron magnetism results from the combination of spin and orbit contributions. One note that the crystal-field often leads to an $f$-electron doublet ground-state (within a localized description), whose properties can be considered as those of a pseudo-spin 1/2 (see Section 2.4.1.1). In a paramagnet, magnetic fluctuations are also sometimes labeled as paramagnons, by analogy with the magnons, i.e., the transverse magnetic excitations, observed in weakly-anisotropic long-range-ordered magnets.

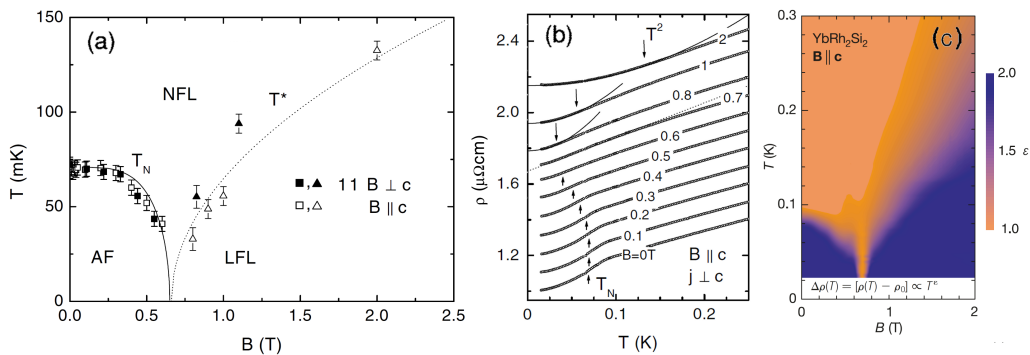

Figure 2.27: (a) Magnetic-field-temperature phase diagram (from [Gegenwart 02]), (b) electrical resistivity $\rho$ versus temperature (from [Gegenwart 02]), and (c) evolution of the exponent $\varepsilon$ from a non-Fermi fit $\rho = \rho_0 + B \cdot T^\varepsilon$ (from [Custers 03]) of the resistivity of $YbRh_2Si_2$ in a magnetic field $\mathbf{H} \parallel \mathbf{c}$.

Figures 2.28(a,c-d) shows sketches of the magnetic moments in an antiferromagnet and in a paramagnet with antiferromagnetic fluctuations (example of antiferromagnetic moments with wavevector $\mathbf{k} = (1/2, 0, 0)$ and Ising magnetic anisotropy). While antiferromagnetism is a static long-range order of the magnetic moments, antiferromagnetic fluctuations correspond to a short-range order, in which the magnetic moments fluctuate with time. At a given time $t$, islands of short-range antiferromagnetic order can be defined in the crystal, having finite correlation lengths $\xi_i$ of a few interatomic distances, and a finite correlation time $\tau$, which corresponds to the mean time duration of the islands. Apart from the islands, the magnetic moments have

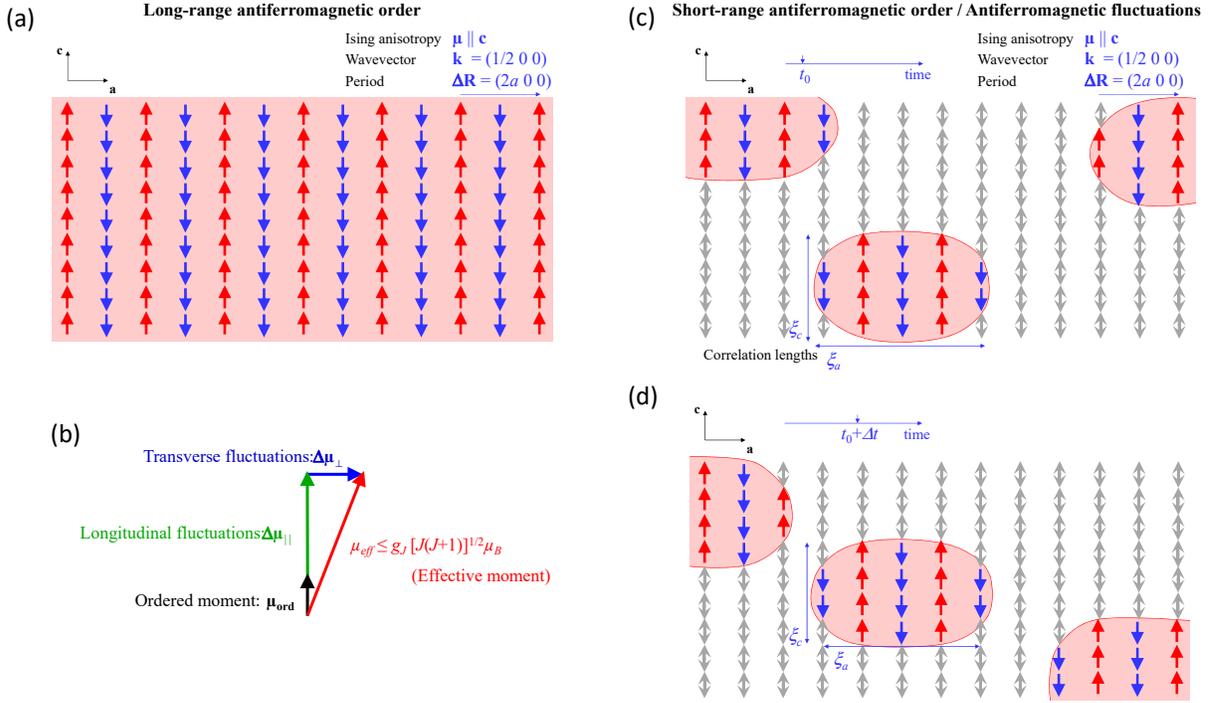

Figure 2.28: Sketch of a two-dimensional lattice of magnetic moments in (a) a long-range antiferromagnetically ordered phase and a short-range antiferromagnetically ordered regime, i.e., with antiferromagnetic fluctuations, at two different times (c) $t_0$ and (d) $t_0 + \Delta t$. These magnetic arrangements correspond to an Ising magnetic anisotropy (moments aligned along the easy magnetic axis $\mathbf{c}$ and a wavevector $\mathbf{k} = (1/2, 0, 0)$. (b) Schematic representation of a magnetic moment, including long-range ordered and transverse/longitudinal fluctuating components.

random and non-correlated directions (here up or down). At a time $t + \Delta t$, with $\Delta t > \tau$, the antiferromagnetic islands have moved to other places of the crystal.

The dynamical magnetic susceptibility $\chi_{\alpha\beta}(\mathbf{q}, E)$, where $\mathbf{q}$ is the wavevector and $E = \bar{h}\omega$ is the energy, is a complex tensor permitting to describe the magnetic fluctuations. It can be defined as:

$$M_\alpha(\mathbf{q}, E) = \sum_\beta \frac{\chi_{\alpha\beta}(\mathbf{q}, E)}{H_\beta(\mathbf{q}, E)}, \tag{2.10}$$

where $\alpha, \beta = x, y, z$ and $\mathbf{M}(\mathbf{q}, E)$ is the magnetization induced by a magnetic field $\mathbf{H}(\mathbf{q}, E)$. Alternatively, it can be expressed as [Kuramoto 00]:

$$\chi_{\alpha\beta}(\mathbf{q}, E) = \frac{1}{\mu_B^2} \int_0^\infty e^{(iEt/\bar{h})} \langle [\mu_\alpha(\mathbf{q}, t), \mu_\beta(-\mathbf{q}, 0)] \rangle dt, \tag{2.11}$$

where $\mu_\alpha(\mathbf{q}, t) = \sum_\mathbf{r} \mu_\alpha(\mathbf{r}, t) e^{-i\mathbf{q}\mathbf{r}}$ is the Fourier transform of the $\alpha$ component of the magnetic moment $\mu(\mathbf{r}, t)$ at the position $\mathbf{r}$ and time $t$. For simplicity the indexes $\alpha$ and $\beta$ are abandoned in the following, and formulas will be expressed for the isotropic case.

The dynamical magnetic susceptibility can be decomposed into a real and an imaginary components defined by $\chi(\mathbf{q}, E) = \chi'(\mathbf{q}, E) + i\chi''(\mathbf{q}, E)$. It can be probed by different experimental methods, as:

- measurement of the static susceptibility $\chi$, which is related to the dynamical susceptibility by:

$$\chi = M/H = \chi'(\mathbf{q} = 0, E = 0) = \chi'(\mathbf{q} = 0),\quad (2.12)$$

where $M$ is the magnetization and $H$ is the magnetic field,

- inelastic neutron scattering, which probes the wavevector and energy dependence of the magnetic susceptibility. Neutron scattered intensity can be expressed as:

$$S(\mathbf{q}, E) \propto \frac{1}{1 - e^{-E/k_B T}} \chi''(\mathbf{q}, E).\quad (2.13)$$

In the following, we will use the following convention for neutron-scattering experiments, where the magnetic signal emerges at a neutron momentum transfer $\mathbf{Q}$, related with the wavevector $\mathbf{q}$ of the sample reciprocal lattice by:

$$\mathbf{Q} = \tau + \mathbf{q},\quad (2.14)$$

where $\tau$ corresponds to the position of a non-magnetic nuclear Bragg peak. The wavevectors characteristic of intersite magnetic correlations, i.e., at which larger neutron scattered intensity emerges from a $\mathbf{q}$-independent background, will be noted $\mathbf{k}_i$.

- nuclear magnetic resonance (NMR), which allows accessing several magnetic quantities [Alloul 16], as the magnetic Knight shift:

$$K \propto \chi_p,\quad (2.15)$$

where $\chi_p$ is the Pauli susceptibility of conduction electrons, which can coincide with that of itinerant $f$-electrons, as in heavy-fermion systems, and the spin-lattice relaxation rate $1/T_1$, which is linked to the magnetic susceptibility by:

$$\frac{1}{T_1 T} \propto \sum_{\mathbf{q}, E \to 0} |A(\mathbf{q})|^2 \frac{\chi''(\mathbf{q}, E)}{E},\quad (2.16)$$

where $A(\mathbf{q})$ is a $\mathbf{q}$-dependent hyperfine coupling. NMR experiments are performed in a magnetic field, and the NMR spin-lattice relaxation rate $1/T_1$ probes the magnetic fluctuations components perpendicular to the magnetic field. The longitudinal component of the magnetic fluctuations, i.e., the component parallel to the magnetic field, can be accessed by studying the NMR spin-spin relaxation rate $1/T_2$.

In paramagnets close to a quantum magnetic phase transition, magnetic fluctuations generally correspond to damped quasi-elastic excitations. They can be described using random-phase approximation theories, in which the dynamical magnetic susceptibility is given by:

$$\chi(\mathbf{q}, E) = \frac{\chi'(\mathbf{q})}{1 - iE/\Gamma(\mathbf{q})}, \tag{2.17}$$

where $\chi'(\mathbf{q})$ is the real part of the static susceptibility and $\Gamma(\mathbf{q})$ is the relaxation rate at wavevector $\mathbf{q}$. $\chi'(\mathbf{q})$ probes the intensity of the magnetic fluctuations and can be expressed as the integral:

$$\chi'(\mathbf{q}) = \frac{1}{\pi} \int_{-\infty}^{\infty} \frac{\chi''(\mathbf{q}, E)}{E} dE. \tag{2.18}$$

Equation 2.17 implies:

- for inelastic neutron scattering:

$$S(\mathbf{q}, E) \propto \frac{1}{1 - e^{-E/k_B T}} \frac{E\chi'(\mathbf{q})/\Gamma(\mathbf{q})}{1 + (E/\Gamma(\mathbf{q}))^2}, \tag{2.19}$$

- for nuclear magnetic resonance:

$$\frac{1}{T_1 T} \propto \sum_{\mathbf{q}} |A(\mathbf{q})|^2 \frac{\chi'(\mathbf{q})}{\Gamma(\mathbf{q})}. \tag{2.20}$$

With the aim to describe the magnetic exchange interactions, the susceptibility can be expressed as:

$$\chi(\mathbf{q}, E) = \frac{\chi_0(E)}{1 - J(\mathbf{q})\chi_0(E)}, \tag{2.21}$$

where $\chi_0(E)$ is a local $\mathbf{q}$-independent magnetic susceptibility and $J(\mathbf{q}) = \sum_{\mathbf{r}} J(\mathbf{r})e^{-i\mathbf{q}\mathbf{r}}$ is the magnetic exchange expressed in the reciprocal space, i.e., the Fourier transform of intersite magnetic exchanges $J(\mathbf{r})$ in the crystal, $\mathbf{r}$ being the distance between two magnetic atoms. In a material with intersite magnetic fluctuations of wavevector $\mathbf{k}$, $\chi(\mathbf{q}, E)$ is $\mathbf{q}$-dependant and $|J(\mathbf{q})|$ is maximum for $\mathbf{q} = \mathbf{k}$. Equivalently, $\chi'(\mathbf{q})$ and $\Gamma(\mathbf{q})$ are maximum and minimum, respectively, at the wavevector $\mathbf{k}$. The relaxation time is inversely proportional to the relaxation rate and is given by $\tau(\mathbf{k}) = \bar{h}/\Gamma(\mathbf{k})$. As well, the correlation length is inversely proportional to the width $\kappa(\mathbf{k})$ in $\mathbf{q}$-space of the fluctuations spectrum, and is given by $\xi(\mathbf{k}) = 1/\kappa(\mathbf{k})$.

The effects of the onset of ferromagnetic fluctuations, with the magnetic wavevector $\mathbf{k}_{FM} = 0$, and antiferromagnetic fluctuations, with a wavevector $\mathbf{k}_{AF} \neq 0$, on the wavevector- and temperature-variations of $\chi'(\mathbf{q})$ and $\Gamma(\mathbf{q})$ are summarized in Figure 2.29. Figure 2.29(a,d) show that the onset of low-temperature intersite magnetic correlations leads to the enhancement of $\chi'(\mathbf{q})$ at the magnetic wavevector $\mathbf{k}_{FM}$ or $\mathbf{k}_{AF}$, respectively, and to its reduction at other wavevectors, in comparison with a local $\mathbf{q}$-independent magnetic susceptibility $\chi'_0$ corresponding to $J(\mathbf{q}) = 0$. Figure 2.29(b,e) show the effects of intersite magnetic fluctuations on the temperature dependence of $\chi'(\mathbf{q})$. In the case of ferromagnetic fluctuations, the bulk magnetic susceptibility $\chi = \chi'(\mathbf{k}_{FM} = 0)$ (see Equation 2.12) presents at low temperature an upwards deviation from the high-temperature behavior (generally of Curie-Weiss form). Oppositely, a downwards deviation of $\chi$ occurs at low temperatures in the case of antiferromagnetic fluctuations. This can result in a broad maximum in the magnetic susceptibility, at a temperature $T_\chi^{max} \sim T^*$, where $T^*$ is the characteristic temperature of intersite magnetic fluctuations. Figures 2.29(c,f) further show that $\Gamma(\mathbf{q})$ at the magnetic wavevector $\mathbf{k}_{FM}$ or $\mathbf{k}_{AF}$, respectively, is reduced at temperatures below the temperature scale $T^*$.

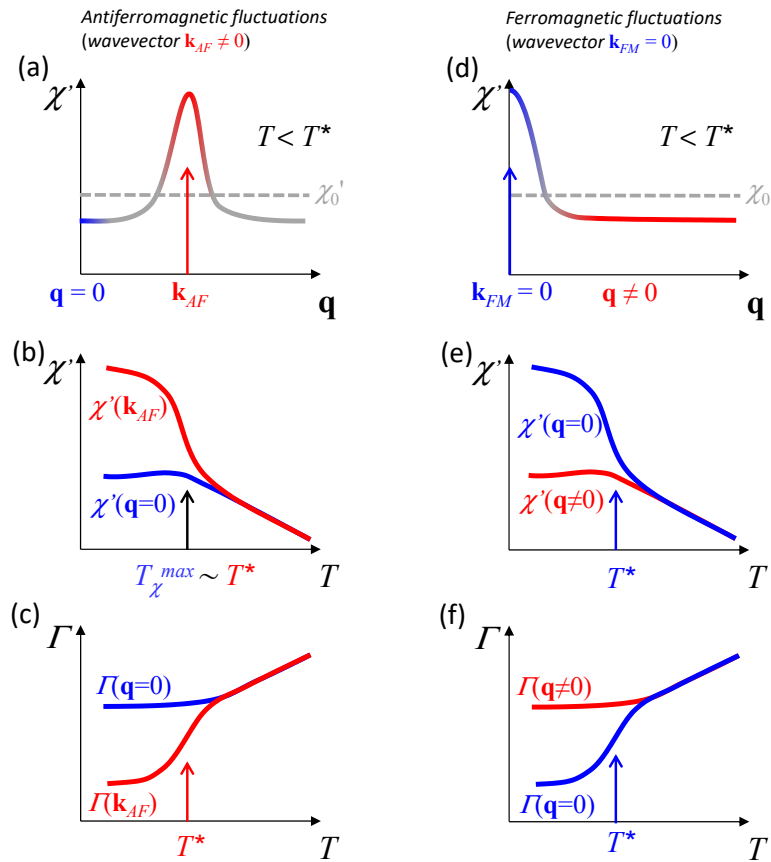

Figure 2.29: (a) Wavevector-dependence of $\chi'(\mathbf{q})$, (b) temperature-dependence of $\chi'(\mathbf{k}_{FM} = 0)$, and (c) temperature-dependence of $\chi'(\mathbf{k}_{FM})$ in the case of ferromagnetic fluctuations. (d) Wavevector-dependence of $\chi'(\mathbf{q})$, (e) temperature-dependence of $\chi'(\mathbf{k}_{AF})$, and (f) temperature-dependence of $\chi'(\mathbf{k}_{AF})$ in the case of antiferromagnetic fluctuations. The graphs in panels (b,c,e,f) are in a log-log scale.

A well-documented case presenting low-energy quantum magnetic fluctuations is the heavy-fermion paramagnet CeCu$_6$. Figures 2.30(b-c) present inelastic-neutron-scattering measurements performed on this compound [Regnault 87, Rossat-Mignod 88]. At the low temperature $T = 100$ mK, damped magnetic fluctuations are peaked at the wavevector $\mathbf{k} = (0, 0, 0.9)$. Figure 2.30(e) shows the temperature-variation of the inter-site and single-site relaxation rates $\Gamma_{is}$ and $\Gamma_{ss}$, respectively. $\Gamma_{is}$ and $\Gamma_{ss}$ were extracted from an analysis of the energy dependence of the spectra, using two contributions, a quasielastic term for the single-site contribution and a weakly-inelastic term for the intersite contribution, by Rossat-Mignod *et al* [Rossat-Mignod 88]. The intersite relaxation rate $\Gamma_{is}$ saturates to a value $\simeq 0.2$ meV $\simeq 2.5$ K (1 meV $\leftrightarrow$ 11.6 K) for $T < T^* \simeq 1$ K, and it increases significantly for $T > T^*$. The conver-gence of $\Gamma_{is}$ and $\Gamma_{ss}$ for $T \gtrsim 2$ K indicates the loss of the intersite correlations. Figure 2.30(f) also shows that the correlation lengths saturate for $T < T^*$ to values $\xi_c \simeq 4$ Å and $\xi_a \simeq 9$ Å, and strongly decrease at temperatures $T > T^*$. The temperature scale $T^*$ coincides with the onset of a low-temperature linear-$T$ Korringa variation of the NMR spin-lattice relaxation rate $1/T_1$ [Kitaoka 87] shown in Figure 2.30(g). The temperature scale $T^* \simeq 1$ K of the intersite correlations also drives the Fermi-liquid regime formation indicated by a downward deviation from Curie-Weiss behavior of the magnetic susceptibility in a field along the easy magnetic axis **c** shown in Figure 2.30(a) [Schröder 92], and by the nearly-saturation of the heat capacity divided by temperature shown in Figure 2.30(d) [von Löhneysen 01].

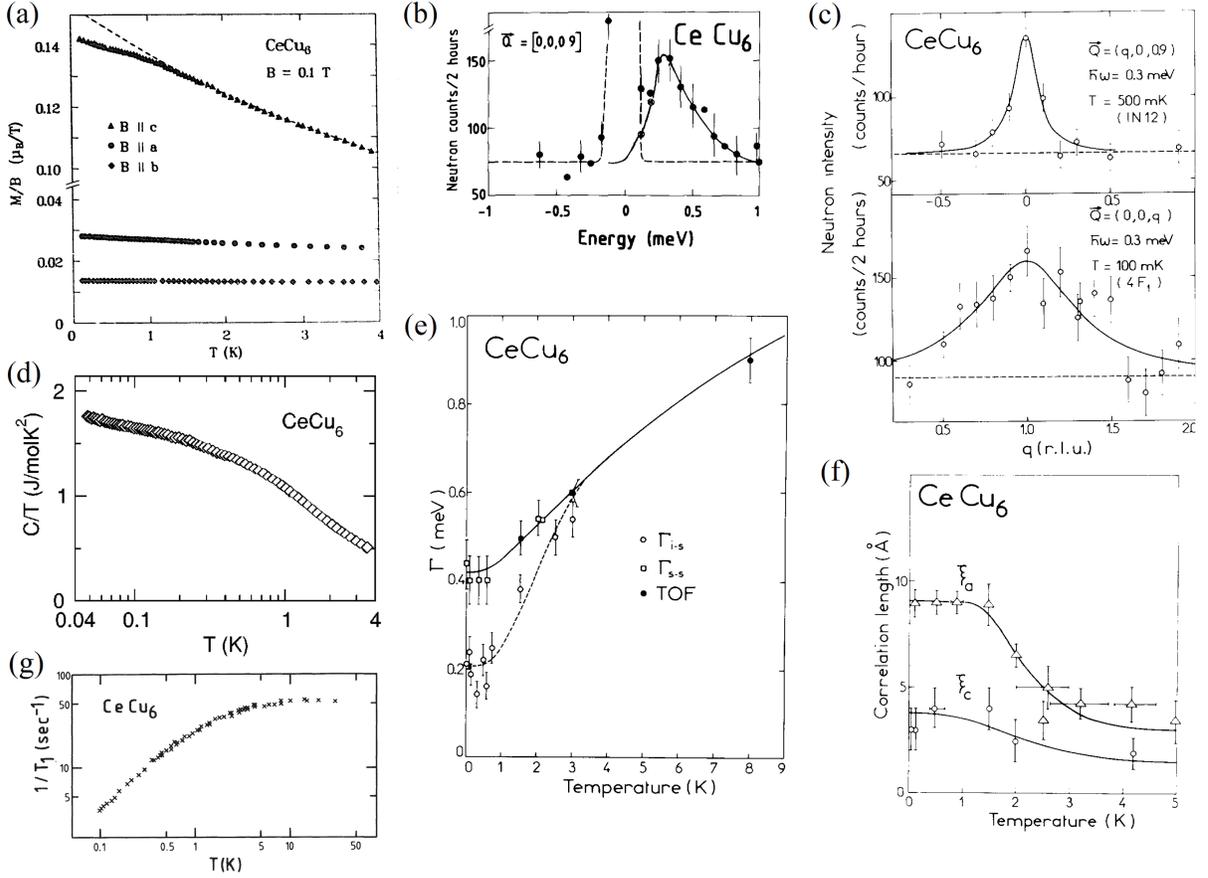

Figure 2.30: (a) Magnetic susceptibility versus temperature, for $\mathbf{H} \parallel \mathbf{a}, \mathbf{b}, \mathbf{c}$ [Schröder 92], (b) neutron scattering intensity versus energy transfer at the momentum transfer $Q = (0, 0, 0.9)$ and $T = 100$ mK [Regnault 87], (c) neutron scattering intensity versus $Q_h$ and $Q_l$ around the momentum transfer $Q = (0, 0, 0.9)$, at the energy transfer $E = \hbar\omega = 0.3$ meV and $T = 100$ mK [Rossat-Mignod 88], (d) specific heat divided by temperature versus temperature [von Löhneysen 01], (e) magnetic correlation lengths $\xi_a$ and $\xi_c$ versus temperature, (f) inter-site and single-site relaxation rates $\Gamma_{is}$ and $\Gamma_{ss}$ versus temperature [Rossat-Mignod 88], and NMR spin-lattice relaxation rate $1/T_1$ versus temperature [Kitaoka 87] measured in CeCu$_6$.

### 2.4.3.2 Theories

A series of theoretical works focused on the relation between intersite magnetic fluctuations and the formation of a Fermi liquid in paramagnets close to a quantum magnetic phase transition. The driving role played by magnetic fluctuations in the development of a Fermi-liquid regime in the magnetic susceptibility has been emphasized in the early works of Doniach and Engelsberg [Doniach 66], Schrieffer and Berk [Schrieffer 67], and Béal-Monod *et al* [Béal-Monod 68] in the late sixties. In parallel, Mills and Lederer described how, in a metal, the magnetic fluctuations of localized moments can lead to a Fermi-liquid $T^2$ behavior in the electrical resistivity via a $s - d$ exchange [Mills 66]. Kaiser and Doniach further demonstrated that a linear $T$ behavior in the electrical resistivity is expected at temperatures just above the Fermi-liquid temperature $T^*$ [Kaiser 70]. This linear behavior recalls the non-Fermi-liquid behaviors $\rho \sim T^\varepsilon$ observed in heavy-fermion systems in the vicinity of their quantum magnetic phase transition, at which the Fermi-liquid regime collapses (see Section 2.4.2.2). By considering the temperature dependence of the magnetic susceptibility of nearly-magnetically-ordered metals, Jullien *et al* described the electrical resistivity up to temperatures well above the characteristic temperature of the magnetic fluctuations [Jullien 74]. They succeeded to obtain a plateau, and within certain sets of parameters, a maximum of the electrical resistivity versus temperature, indicating that the electrical resistivity of metals close to a magnetic phase transition, as those of heavy-fermion magnets, may be controlled by intersite magnetic fluctuations. Lopes *et al* found out that, at low temperatures, the relaxation rate of the magnetic fluctuations, which can be extracted by inelastic neutron scattering, is inversely proportional to the characteristic temperature scale of the magnetic fluctuations, noted here $T^*$, so that $\Gamma \sim 1/T^*$ [Lopes 83]. Béal-Monod and Lawrence emphasized that, in the intermediate-valence compounds CeIn$_{3-x}$Sn$_x$, a Fermi-liquid regime of magnetic susceptibility $\chi \sim 1/T^*$, reached at temperatures $T < T_\chi^{max}$, results from the onset of magnetic-fluctuations of temperature scale $T^* \sim T_\chi^{max}$ [Lawrence 79, Béal-Monod 80, Lawrence 81]. While intermediate-valence systems have a large $T^*$, heavy-fermion systems have a small $T^*$ leading to intense low-energy magnetic fluctuations, in relation with a large effective mass. Shiba showed that a Korringa law of the NMR relaxation rate:

$$\frac{1}{T_1 T} \propto \frac{1}{T^{*2}} \propto m^{*2} \tag{2.22}$$

is followed in the Fermi-liquid regime of dilute Kondo-alloys [Shiba 75] (see also [Kitaoka 87]). By considering $\mathbf{q}$-dependent magnetic fluctuations, Kuramoto obtained the relation:

$$\chi'(\mathbf{q})\Gamma(\mathbf{q}) = \frac{2C}{\pi}, \tag{2.23}$$

where $C = \mu_{eff}^2$ is the Curie constant and $\mu_{eff}$ the effective moment of the $f$-electrons [Kuramoto 87]. This relation describes the saturation of the relaxation rate $\Gamma(\mathbf{q})$ and of the susceptibility $\chi'(\mathbf{q})$ in a low-temperature Fermi-liquid regime. Combined with Equation 2.20, we extract the relation:

$$\frac{1}{T_1 T} \propto \sum_{\mathbf{q}} \frac{|A(\mathbf{q})|^2}{\Gamma(\mathbf{q})^2} \tag{2.24}$$

which describes the Korringa law for a heavy Fermi liquid. By considering the entropy gained by the establishment of short-range order, Edwards and Lonzarich expressed the Sommer-feld coefficient of the heat capacity as function of the relaxation rate of magnetic fluctuations [Edwards 92]:

$$\gamma = \frac{C_p}{T} \sim \sum_{\mathbf{q}} \frac{1}{\Gamma(\mathbf{q})} \sim \sum_{\mathbf{q}} \chi'(\mathbf{q}). \tag{2.25}$$

Finally, one obtain that magnetic fluctuations with wavevector $\mathbf{k}$ can drive to a Fermi-liquid regime of effective mass:

$$m^* \sim \frac{1}{\Gamma(\mathbf{k})} \sim \frac{C_p}{T} \sim \sqrt{\frac{1}{T_1 T}}, \tag{2.26}$$

developing at temperatures smaller than the temperature $T^* \simeq \Gamma(\mathbf{k})$ characteristic of intersite magnetic fluctuations.

Magnetic-fluctuations theories were adapted by Hertz, Millis, and Moriya to describe quantum magnetic phase transitions, i.e., a magnetic phase transition occurring in the limit of $T \rightarrow 0$ when a non-thermal tuning parameter $\delta$ is adjusted [Hertz 76, Millis 93, Moriya 95]. In quantum-phase-transition models, a transition temperature characteristic of the ordered phase can be tuned, ending in a quantum critical point at a critical value $\delta_c$ of the tuning parameter [Sondhi 97, Vojta 03, von Löhneysen 07] (see Figure 2.31). Fluctuations of the order parameter increase progressively on approaching the critical point, and a correlation length varying as $\xi \sim (\delta_c - \delta)^{-\nu}$ and a correlation time varying as $\tau \sim \xi^z \sim (\delta_c - \delta)^{-\nu z}$ are associated with these fluctuations (see Figure 2.28). At the quantum phase transition, $\xi$ and $\tau$ diverge, or almost diverge, and short-range fluctuating order is transformed into long-range static order. In the theories of Hertz, Millis, and Moriya, the order parameter is magnetic (either ferromagnetic or antiferromagnetic), and its fluctuations diverge at the quantum critical point, which is the endpoint for $T \rightarrow 0$ of a transition temperature $T_x$ (Curie temperature $T_C$ for a ferromagnet and Néel temperature $T_N$ for an antiferromagnet) [Hertz 76, Millis 93, Moriya 95]. While thermal critical properties are observed in the physical quantities on approaching $T_x$, quantum critical properties occur near a quantum phase transition, either when $\delta$ is adjusted close to $\delta_c$ in the limit $T \to 0$, or when $T$ is increased at $\delta = \delta_c$. The disordered state corresponds to a Fermi liquid of high effective mass $m^*$ driven by the fluctuations of the magnetic order parameter, which is established at temperatures $T < T^*$. In the ordered state, the presence of fluctuations near the critical point leads to reduced magnetically-ordered moments. In these models, the Fermi-liquid paramagnetic regime vanishes at the quantum critical point, where $T^* \to 0$, ending in a divergence of $m^*$ controlled by the divergence of the magnetic correlations. Figure 2.32 presents schematically the temperature-variation of the relaxation rate $\Gamma$ of intersite magnetic fluctuations expected for a low-temperature Fermi-liquid ground state, where $\Gamma$ saturates, and for a quantum critical case, where $\Gamma \to 0$ with $T \to 0$, following a quantum critical behavior $\Gamma \sim T^\beta$. In the case of an antiferromagnet phase transition with three-dimensional exchange, the exponent $\beta = 3/2$ is expected [Continentino 01]. The temperature dependence of these critical fluctuations drives a non-Fermi-liquid regime where critical power or logarithmic laws characterize the different measurable physical quantities (heat capacity, magnetization etc.). However, the Hertz-Millis-Moriya models failed in describing quantitatively the non-Fermi-liquid behaviors reported in heavy-fermion systems [Stewart 01]. Critical fluctuations of the order parameter have been observed at the quantum phase transition of the heavy-fermion critical antiferromagnet $Ce_{1-x}La_xRu_2Si_2$, validating at least qualitatively, and for this system, the pertinence of a Hertz-Millis-Moriya picture [Knafo 09a]. Alternative models, based on a local scenario of quantum criticality driven by $\mathbf{q}$-independent phenomena, such as a Kondo breakdown at the quantum critical point, have been proposed [Si 01, Coleman 02]. An experimental and consensual verification that a local quantum magnetic criticality scenario is applicable for a given heavy-fermion system constitutes an experimental challenge [Schröder 00, Knafo 05].

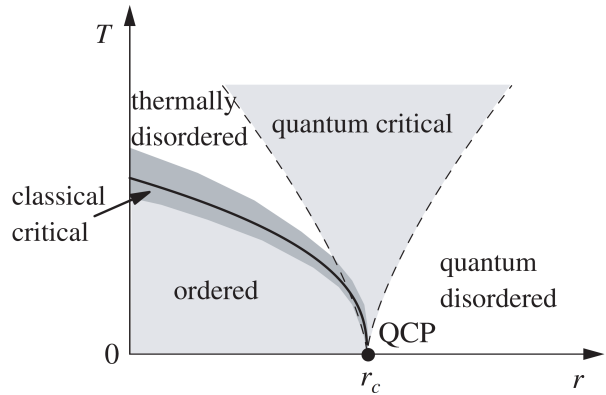

Figure 2.31: Generic quantum critical phase diagram, where $r$ is a tuning parameter and $T$ the temperature. The quantum critical point (QCP) is defined at the zero-temperature extrapolation, at $r = r_c$, of the transition temperature associated with the ordered phase (from [Vojta 03]).

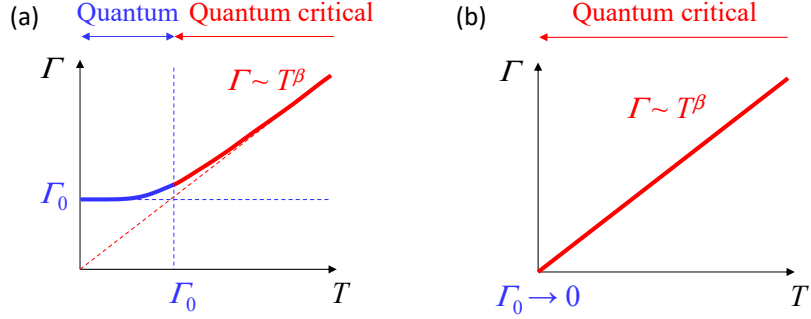

Figure 2.32: Log-log plot of the temperature-variation of the relaxation rate of intersite magnetic fluctuations expected (a) for a compound reaching a Fermi-liquid regime at low temperature and (b) for a quantum critical compound.

## 2.5   Unconventional superconductivity

The Section presents a focus on unconventional superconductivity, which develops near quantum magnetic phase transitions in heavy-fermion magnets. The role played by the magnetic fluctuations, the observation of a magnetic resonance in the superconducting state, and the challenge to determine the superconducting order parameter are emphasized. Further reviews about the interplay between magnetism and superconductivity in heavy-fermion compounds can be found in [Ōnuki 04, Pfleiderer 09, Knebel 11a, Aoki 13, Griveau 14, White 15, Aoki 19a].

### 2.5.1   Superconductivity at a quantum magnetic phase transition

The first report of bulk superconductivity in a heavy-fermion material, CeCu$_2$Si$_2$, was made in 1979 by Steglich *et al* [Steglich 79]. Clear signatures in the electrical resistivity, magnetic susceptibility, and heat capacity versus temperature shown in Figures 2.33(a-b) show the onset of superconductivity at the temperature $T_{sc} = 0.6$ K. Figure 2.33(c) focuses on the electronic heat capacity $C_p^{el}$ (after substraction of the phonon background) divided by temperature versus temperature in magnetic fields **H** ∥ **a** up to the superconducting field $\mu_0 H_{sc} \simeq 1.9$ T [Kittaka 14]. At zero field, a step-like variation of $C_p^{el}/T$ at $T_{sc}$ is followed by a strong decrease of $C_p^{el}/T$ at low temperature ending with $C_p^{el}/T \rightarrow 0$ for $T \rightarrow 0$. At $H_{sc}$, $C_p^{el}/T$ almost saturates to a large Sommerfeld coefficient $\gamma \simeq 0.8 - 0.9$ J/molK$^2$ for $T < 1$ K, indicating that a heavy-fermion ground state is restored. The conservation of the electronic entropy is accompanied by an isosbestic point, where all curves cross at $T_x \simeq 0.3$ K. It indicates that the $4f$ electrons driving the heavy-fermion behavior at $H_{sc}$ are also those becoming superconducting at zero magnetic field. In their ground state, $4f$ electrons are in a crystal-field doublet $|\Gamma_7^1> = \eta \mid \pm 5/2 \rangle + \sqrt{1 - \eta^2} \mid \mp 3/2 \rangle$, with $\eta = 0.83$, determined by Ōnuki *et al* from a fit to the magnetic susceptibility [Ōnuki 85]. A similar ground state was found for the diluted compound Ce$_{0.06}$La$_{0.94}$Cu$_{2.05}$Si$_2$ [Aviani 01] (see Section 2.4.1.1).

The discovery of superconductivity in a Ce-based heavy-fermion material almost coincided

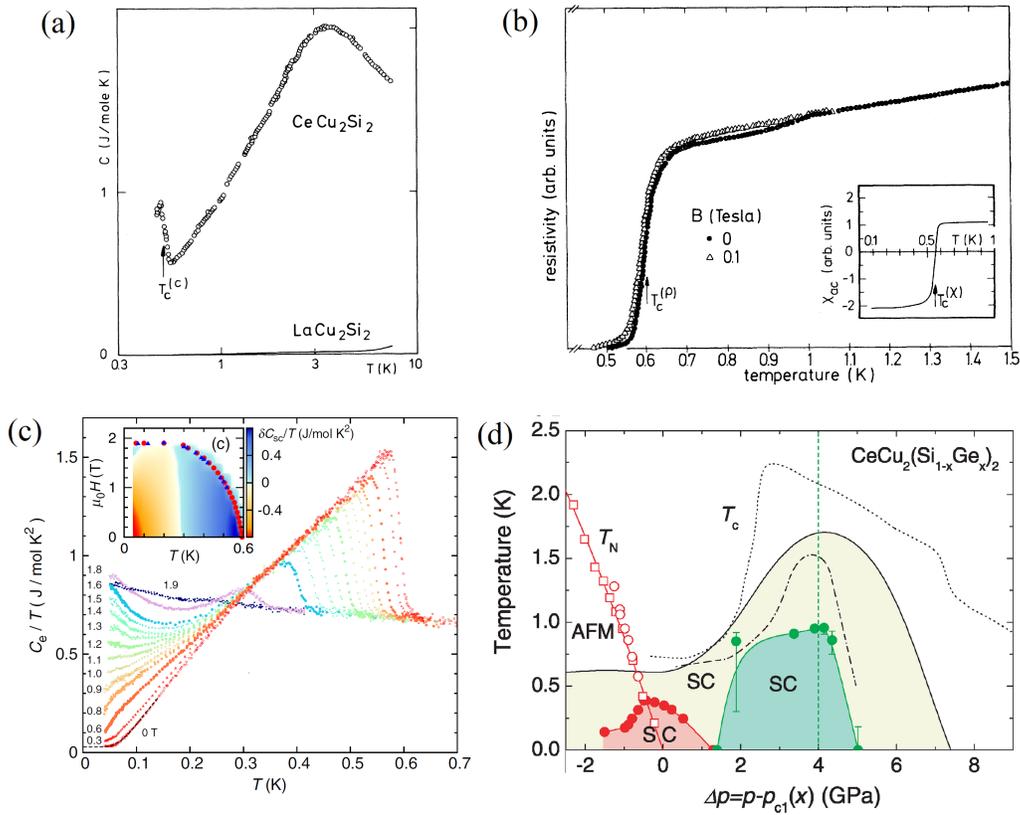

Figure 2.33: (a) Heat capacity versus temperature (from [Steglich 79]), (b) electronic heat capacity divided by temperature versus temperature and magnetic-field-temperature phase diagram (Inset) in magnetic fields $\mathbf{H} \parallel \mathbf{a}$ (from [Kittaka 14]), (c) electrical resistivity and magnetic susceptibility (Inset) (from [Steglich 79]) of $CeCu_2Si_2$, and (d) pressure-temperature phase diagram of $CeCu_2(Si_{1-x}Ge_x)_2$ (from [Yuan 03])

with that of superconductivity in U-based heavy-fermion compounds. Even before the report of superconductivity in $CeCu_2Si_2$ [Steglich 79], signatures of heavy-fermion superconductivity could be found in a study of $UBe_{13}$ published in 1975 by Bucher *et al.* [Bucher 75]. In this work, a low-temperature increase of the heat capacity was reported, but it was not recognized as the onset of a heavy-fermion behavior (the first identification of heavy-fermion behavior came a few months later with the $CeAl_3$ case [Andres 75]), and superconductivity was observed, but it was interpreted as resulting from 'non-intrinsic' filamentary effects. Heavy Fermi liquid and superconductivity were finally identified as intrinsic properties of $UBe_{13}$ in 1983 by Ott *et al* [Ott 83]. Soon after, superconductivity was discovered in $UPt_3$ [Stewart 84b], $URu_2Si_2$ [Palstra 85, Schlabitz 86], $UNi_2Al_3$ [Geibel 91a] and $UPd_2Al_3$ [Geibel 91b]. Although the role of magnetic fluctuations was sometimes suspected (see for instance [Stewart 84b]), the first evidence of superconductivity induced in the vicinity of a quantum magnetic transition was found by Jaccard *et al* in a study of the antiferromagnet $CeCu_2Ge_2$ under pressure [Jaccard 92]. In this work, the similarity between CeCu$_2$Ge$_2$ under high pressure and CeCu$_2$Si$_2$ at ambient pressure was highlighted. Small variations of stoichiometry in CeCu$_2$Si$_2$ compounds were found to allow tuning the magnetic phase transition, leading to the paramagnetic and superconducting phase in 'S'-type crystals (see above) and to an antiferromagnetic phase in 'A'-type crystals [Stockert 04] (see Section 2.3.4). Ge-doping on the Si site also permits to stabilize long-range magnetic order. When combined with pressure, it allows continuously tuning the quantum phase transition, leading to the pressure-temperature phase diagram shown in Figure 2.33(d). In addition to the superconducting phase observed near the critical pressure $p_c$ at the magnetic phase transition, a second superconducting phase is stabilized at a pressure $p_v > p_c$. While the role of antiferromagnetic fluctuations was emphasized for the stabilization of the first superconducting phase near $p_c$ [Stockert 11] (see Section 2.5.2), valence fluctuations were proposed to drive the superconducting phase near $p_v$ [Holmes 04]. However, a smooth variation of valence was reported in the vicinity of $p_v$ [Rueff 11].

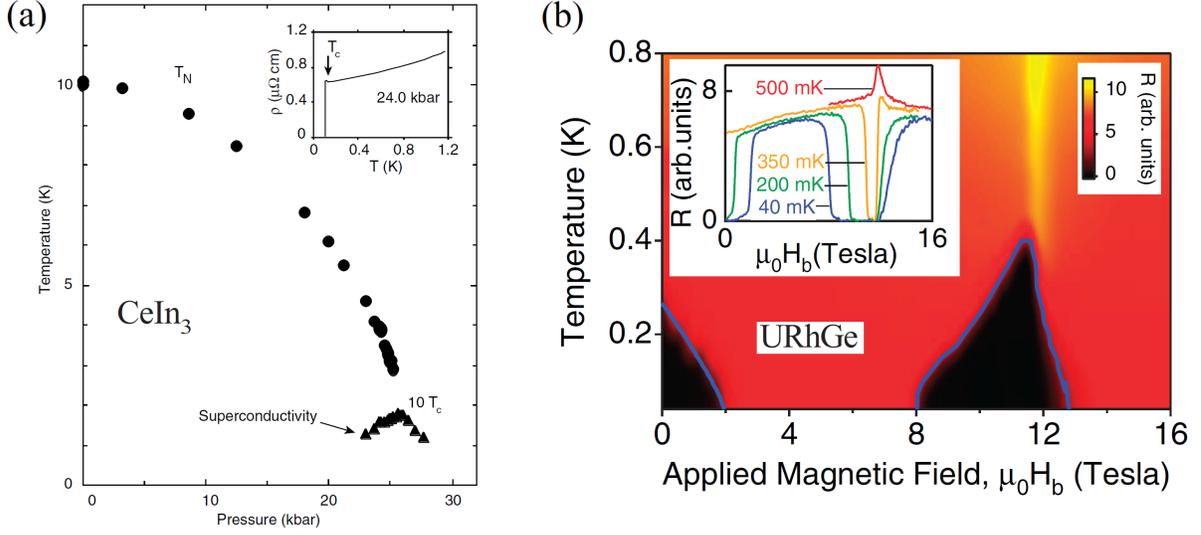

Figure 2.34: (a) Pressure-temperature phase diagram of CeIn$_3$ (from [Mathur 98]) and (b) magnetic field - temperature phase diagram of URhGe in a magnetic field **H** ∥ **b** (from [Lévy 05]). The insets in the two panels show resistivity versus temperature for each compound.

Over the last decades, the number of known heavy-fermion superconductors increased substantially [Pfleiderer 09]. Contrary to the early-discovered case of CeCu$_2$Si$_2$, most of heavy-fermion superconductors present a single superconducting phase glued to a quantum magnetic phase transition, generally of antiferromagnetic nature. Figure 2.34(a) shows the pressure-temperature phase diagram of CeIn$_3$, which is a prototypical example of superconductivity induced at an antiferromagnetic phase transition, for which the critical role of antiferromagnetic fluctuations for the formation of the superconducting Cooper pairs has been proposed [Mathur 98]. Although a large number of Ce and U superconducting compounds have been discovered, their critical temperatures are limited to less than 3.5 K for Ce superconductors (maximum observed in $CeNiC_2$ under pressure [Kittaka 14]) and less than 3.9 K for U super-conductors (maximum observed in $U_6Fe$ [Chandrasekhar 58]). Oppositely, much fewer Np and Pu superconductors were found, mainly due to safety and experimental constraints in relation with their radioactivity [Griveau 14], but higher superconducting temperatures were found, with a maximum $T_{sc} = 4.9$ K in the single Np superconductor $NpPd_5Al_2$ [Aoki 07] and a maximum of 18.5 K in Pu compounds (maximum observed in $PuCoGa_5$ [Sarrao 02]). The proximity of Np and U compounds from a valence instability, as indicated by the plot of the Wigner-Seitz radius of the actinides in single-element metals (see Figure 2.5) [Hills 00], may be related with the larger energy scales associated with superconductivity in these materials.

Superconductivity has also been observed in a few U-based ferromagnetic heavy-fermion materials: $UGe_2$ [Saxena 00] and UIr [Akazawa 04] under pressure, in the vicinity of the collapse of their ferromagnetic transition line, and in URhGe [Aoki 01] and UCoGe [Huy 07] at ambient pressure. In these ferromagnets, the strong exchange field and the suspected role of ferromagnetic fluctuations indicate that a spin-triplet superconducting order parameter with equal-spin pairing may be realized [Aoki 19a]. On the contrary, antiferromagnetic fluctuations are suspected to be the 'glue' for superconductivity and to lead to a singlet order parameter in most of unconventional superconductors. NMR experiments on UCoGe highlighted the role of magnetic fluctuations and brought microscopic support for a triplet-state scenario of superconductivity [Hattori 14, Manago 19a].

A peculiar property of the U-based ferromagnetic superconductors is the possibility to induce, or reinforce, superconductivity near a metamagnetic transition driven by a magnetic field [Sheikin 01, Lévy 05, Aoki 09a]. This is perhaps the most striking manifestation of magnetically-mediated superconducting pairing. Figure 2.34(b) presents the magnetic-field-temperature phase diagram of URhGe in a magnetic field applied along its hard magnetic axis **b**, showing that beyond the superconducting critical field $\mu_0 H_{sc} = 2$ T delimiting the superconducting ground-state phase, a second superconducting phase is induced under in a magnetic field between 8 and more than 12 T [Lévy 05]. This field-induced superconducting phase coincides with a metamagnetic transition occurring at $\mu_0 H_m = 12$ T. The metamagnetic transition is associated with a sudden rotation of the ferromagnetic moments $\mu_{FM}$ from their easy magnetic axis direction **c** for $H < H_m$ towards the direction **b**, along which the magnetic field is applied, for $H > H_m$. Critical magnetic fluctuations, presumably of ferromagnetic nature, and whose signature was evidenced by electrical resistivity [Miyake 08] and nuclear magnetic resonance (NMR) [Tokunaga 15] measurements, are suspected to drive an unconventional mechanism of superconductivity at the metamagnetic transition of URhGe.

Recently, the route to magnetic-field-induced superconductivity in non-ferromagnetic compounds was opened by the discovery of superconductivity a paramagnetic compound, $UTe_2$. Two superconducting phases develop in the vicinity of a metamagnetic transition induced by a magnetic field applied either along the hard magnetic axis **b** or along a peculiar hard direction tilted by $\simeq 25°$ from **b** in the (**b**, **c**) plane [Knebel 19, Ran 19b, Knafo 2x]. These last findings may boost the search for superconductivity induced by a magnetic field in other compounds.

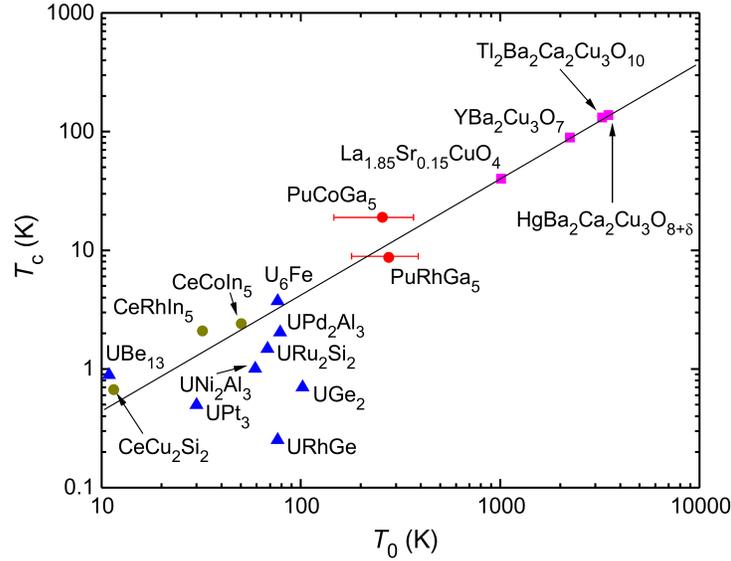

Figure 2.35: Superconducting critical temperature versus characteristic magnetic-fluctuations temperature $T_0$ for several heavy-fermion and cuprate superconductors (from [White 15]).

### 2.5.2 Magnetic fluctuations and resonance

From the original Hill plots, suggestion was made that magnetism and superconductivity may be antagonist in $f$-electron systems [Hill 70] (see Figure 2.8 in Section 2.2.3). Hill noticed that magnetism is mainly observed for localized $f$-electrons stabilized in compounds with large interatomic distances, while superconductivity is mainly observed for itinerant $f$-electrons stabilized in compounds with small interatomic distances. However, the tuning by pressure or chemical doping of heavy-fermion materials showed that superconductivity is often induced at a quantum magnetic phase transition (see Section 2.5.1). Superconductivity in these materials may, thus, result from the magnetic properties, rather than being in competition with them. As shown in Sections 2.3.2 and 2.4.3, the Fermi-liquid regime of heavy-fermion paramagnets is controlled by intersite magnetic fluctuations, which become maximum at a quantum phase transition (see the $Ce_{1-x}La_xRu_2Si_2$ quantum critical case [Knafo 09a], or the Hertz-Millis-Moriya models [Hertz 76, Millis 93, Moriya 95]). Knowing that $f$-electron superconductivity develops from a high-temperature heavy-fermion state (see Section 2.5.1), magnetic fluctuations have been proposed to drive the mechanism leading to the formation of superconducting Cooper pairs. The role of magnetic fluctuations in unconventional heavy-fermion, but also iron-based and cuprate superconductors, rather than that of phonons as in conventional superconductors, has been emphasized [Monthoux 07, Hinks 09, Pines 13], and magnetically-mediated-superconductivity theories have been developed [Monthoux 99, Monthoux 01, Moriya 00].

Figure 2.35 presents a plot of the superconducting temperature $T_{sc}$ versus an estimate of the characteristic temperature $T^*$ (noted $T_c$ and $T_0$, respectively, in the graph) of the magnetic-fluctuations in the non-superconducting 'normal' state of several heavy-fermion and cuprate superconductors [White 15]. This plot highlights that a higher superconducting temperature is reached in systems where the magnetic fluctuations have a higher energy scale, confirming that they may play a role in the mechanism for superconductivity.

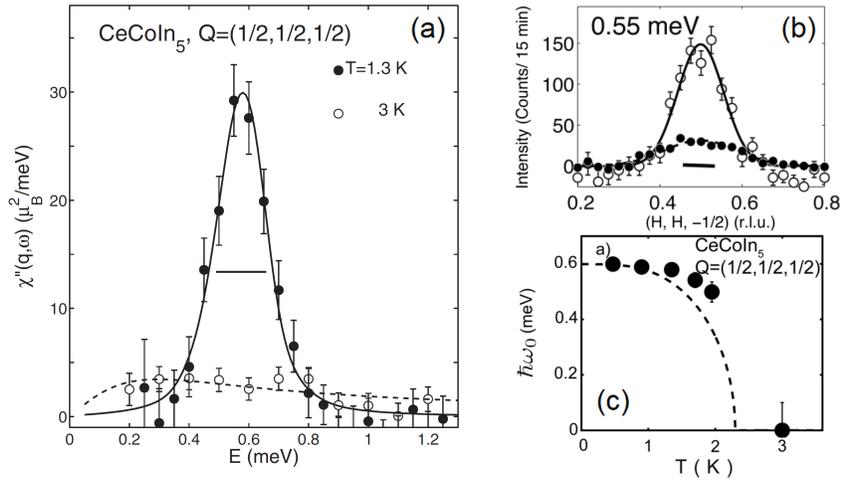

Figure 2.36: (a) Energy dependence, (b) wavevector dependence, and (c) temperature dependence of the gap of the magnetic resonance at the wavevector $\mathbf{q} = (0.5\ 0.5\ 0.5)$ observed in the superconducting phase of $CeCoIn_5$ by inelastic neutron scattering (from [Stock 08]).

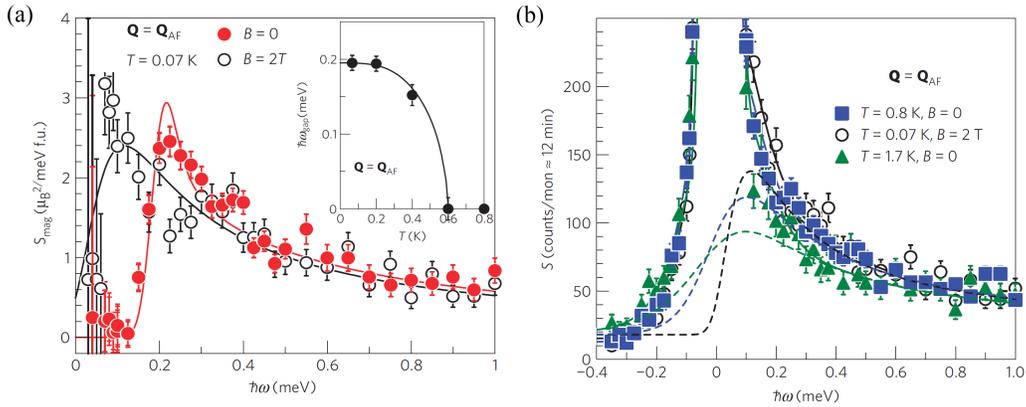

Figure 2.37: Inelastic neutron scattering versus energy at the antiferromagnetic wavevector $\mathbf{k} = (0.215\ 0.215\ 0.542)$, (a) at $T = 70$ mK and $\mu_0 H = 0$ and 2 T, and (b) at $T = 800$ mK/1.7 K and $\mu_0 H = 0$ and 2 T, and at $T = 70$ mK and $\mu_0 H = 2$ T, of $CeCu_2Si_2$. Inset of (a) shows the variation of the gap at the antiferromagnetic wavevector as function of temperature (from [Stockert 11]).

In 1991, Rossat-Mignod $et\ al$ discovered a magnetic resonance mode in the superconducting phase of the high-temperature-superconductivity cuprate system $YBa_2Cu_3O_{6+x}$ by inelastic neutron scattering [Rossat-Mignod 91]. Similar observation of magnetic resonances resulting from a sudden gapping of the magnetic fluctuations spectra were observed in the supercon-ducting state of other heavy-fermion, iron-based and cuprate compounds [Yu 09]. In particular, Stock *et al* observed the first magnetic resonance in the superconducting state of a heavy-fermion material, the paramagnet CeCoIn$_5$, at the wavevector $\mathbf{q} = (0.5\ 0.5\ 0.5)$ characteristic of antiferromagnetic correlations. As shown in Figure 2.36, the resonance corresponds to a sharp inelastic peak in energy and wavevector scans. The resonance energy, i.e., the gap value, reaches $E_r = 0.6$ meV at low temperature, and suddenly vanishes at temperatures higher than the superconducting temperature $T_{sc} = 2.3$ K. Figure 2.37 presents inelastic neutron scattering performed on CeCu$_2$Si$_2$. The superconducting state is also accompanied by a magnetic resonance. It mainly consists in a gapping of the antiferromagnetic quasi-elastic peak observed in the normal non-superconducting phase, at zero-field at temperatures $T > T_{sc} = 600$ mK, or at low-temperature combined with a magnetic field $\mu_0 H = 2$ T $> \mu_0 H sc$.

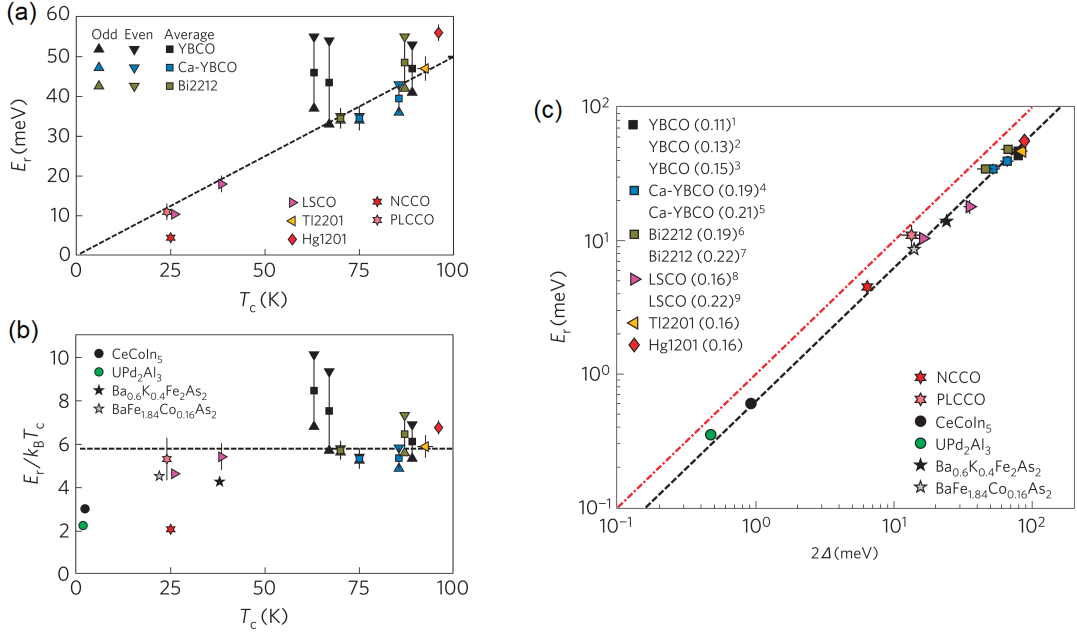

Figure 2.38: (a) Gap of the magnetic resonance versus superconducting temperature (b) gap of the magnetic resonance divide by the superconducting temperature versus superconducting temperature, and (c) gap of the magnetic resonance versus the superconducting gap in cuprate, heavy-fermion and iron-based superconductors (from [Yu 09]).

Figure 2.38 presents a comparison between the energy $E_r$ of the superconducting resonance determined by neutron scattering, the superconducting temperature $T_{sc}$, and the superconducting pairing gap $\Delta$ determined from different sets of experimental techniques (photo-emission, tunneling, thermal conductivity, etc.), for a large number of heavy-fermion, iron-based and cuprate superconductors [Yu 09]. A universal relation $E_r \sim T_{sc} \sim \Delta$ is followed for these systems where $T_{sc}$ varies over more than two orders of magnitude. This relation suggests that these different techniques may probe the same electronic gap, and it confirms the importance of magnetism for superconductivity in strongly-correlated-electron unconventional superconduc-tors.

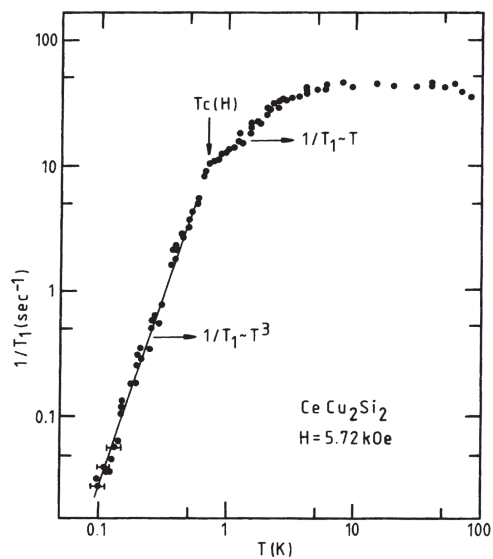

Figure 2.39: NMR relaxation rate $1/T1$ versus temperature $T$ of superconducting CeCu$_2$Si$_2$ (from [Kitaoka 87]).

The modification of the magnetic excitation spectrum in the superconducting state is also visible in the temperature variation of the NMR relaxation rate $1/T_1$. Figure 2.39 presents the temperature variation of $1/T_1$ in the heavy-fermion paramagnetic superconductor CeCu$_2$Si$_2$ [Kitaoka 87], where a Korringa linear law indicates a Fermi-liquid regime at temperature $T_{sc} = 600$ mK $< T < 5$ K, which is replaced by a $T^3$ variation in the superconducting phase for $T < T_{sc}$. Such $T^3$ variation, observed in many unconventional superconductors, will be discussed in Section 2.5.3

## 2.5.3 Superconducting order parameter

A challenge is to describe the unconventional superconducting state of heavy-fermion materials. As discussed in Section 2.5.2, superconductivity and, thus, the superconducting order parameter and the associated gapped-excitation spectrum, result from a pairing mechanism, which is presumably driven by magnetic fluctuations. While a conventional phonon-mechanism of superconductivity drives to a $s$-wave order parameter associated with an isotropic superconducting gap, unconventional magnetic-fluctuation-mechanism of superconductivity can drive to $p$-, $d$- or $f$-wave order parameters associated with an anisotropic superconducting gap, which cancels along line- and/or point- nodes of the Fermi surface (see Figure 2.40 [Yonezawa 16]). NMR Knight-shift measurements are generally used, as a local probe of the Pauli paramagnetic susceptibility, to determine if the superconducting pairing is of spin-singlet or spin-triplet nature. In the case of spin-singlet pairing, the total spin of a Cooper pair is equal to $S = 0$ and, for all magnetic field directions, the associated Pauli susceptibility strongly decreases when the temperature is lowered below the superconducting transition temperature $T_{sc}$. Spin-singlet superconducting states are either $s$- or $d$-wave superconductors. In the case of spin-triplet pairing, the total spin of a Cooper pair is equal to $S = 1$ and the associated Pauli susceptibility is temperature-independent for a magnetic field $\mathbf{H} \perp \mathbf{d}$, and it decreases with the temperature for a magnetic field $\mathbf{H} \parallel \mathbf{d}$, where the vector $\mathbf{d}$ is characteristic of the superconducting order. Figure 2.41 shows the temperature dependence of the NMR Knight shift of heavy-fermion materials, either identified as spin-singlet superconductors (UPd$_2$Al$_3$ and URu$_2$Si$_2$) or spin-triplet superconductors (UNi$_2$Al$_3$, UPt$_3$, and UTe$_2$) [Kotegawa 19]. A large upper critical field, relatively with the superconducting temperature and observed for all field directions, is also generally interpreted as the signature of a spin-triplet state, for which no Pauli limitation is expected for particular directions of the magnetic field. This is for instance the case of the ferromagnetic superconductors URhGe and UCoGe [Aoki 19a].

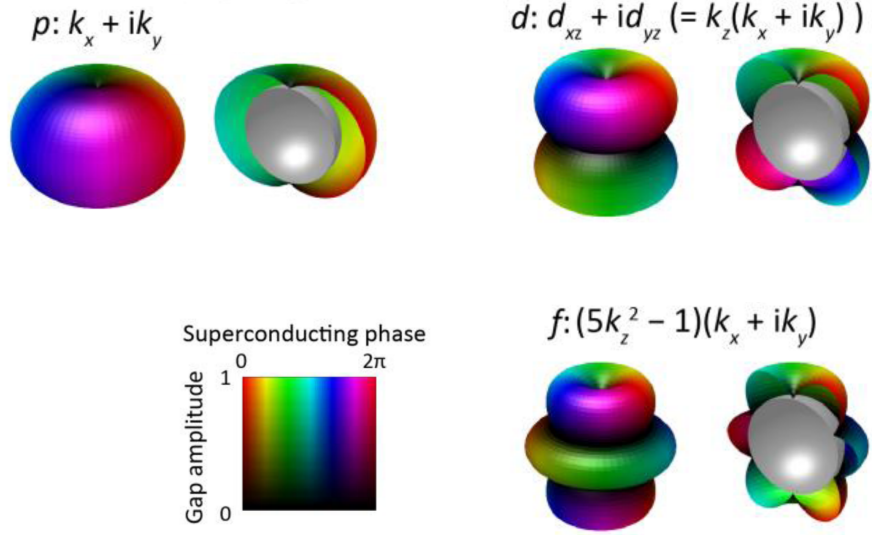

Figure 2.40: Schematics of superconducting $p$-, $d$-, and $f$-wave states. The colors represent the superconducting gap amplitude and phase as shown in the bottom-left panel, with black corresponding to gap nodes. Cross-sections are shown, in which the normal-state Fermi surface is in grey (from [Yonezawa 16]).

In single-band superconductors in zero-magnetic field or in the limit of zero-magnetic field, the following variations of the heat-capacity and NMR relaxation rate are expected [Sigrist 91]:

- $C_p \propto T$ and $1/T_1 \propto T$ for a gapless quasielastic excitation (Fermi liquid),

- $C_p \propto T^2$ and $1/T_1 \propto T^3$ for a line-node gapped excitation,

- $C_p \propto T^3$ and $1/T_1 \propto T^5$ for a point-node gapped excitation.

A $T^3$ variation of the NMR relaxation rate is observed in many heavy-fermion superconductors and is generally interpreted as the signature of a gapped excitation with line nodes [Sigrist 91, Kuramoto 00]. However, a full superconducting gap without line node has been proposed from heat capacity, thermal conductivity, and penetration depth measurements on $CeCu_2Si_2$ by Yamashita *et al* [Yamashita 17], which would imply that the NMR relation rate varying as $1/T_1 \propto T^3$ [Kitaoka 87, Ishida 99] is not a signature of line nodes in the superconducting gap. Angle-resolved thermal-conductivity and heat-capacity, where the angle is related with the magnetic-field direction, can also be used to study the anisotropy of the superconducting gap [Matsuda 06, Sakakibara 16]. Figure 2.42 presents angle-resolved thermal-conductivity measurements performed on CeCoIn$_5$, which were interpreted as the signature of a fourfold-symmetry superconducting gap.

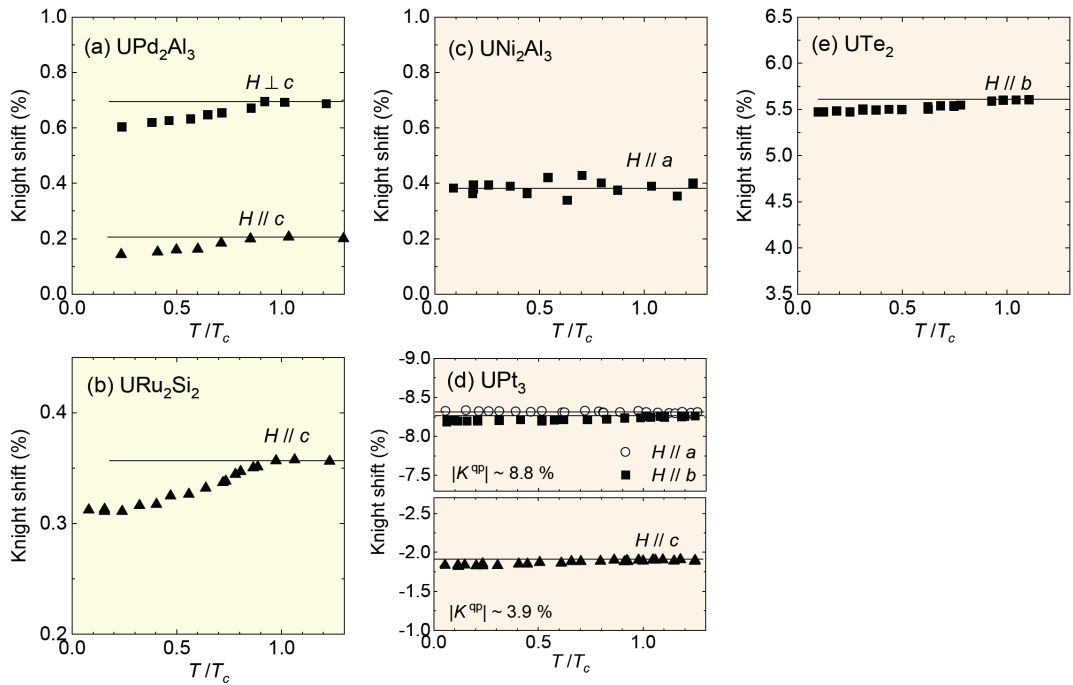

Figure 2.41: NMR Knight shift versus normalized temperature $T/T_{sc}$ in the superconducting state of (a) $UPd_2Al_3$, (b) $URu_2Si_2$, (c) $UNi_2Al_3$, (d) $UPt_3$, and (e) $UTe_2$ (from [Kotegawa 19]).

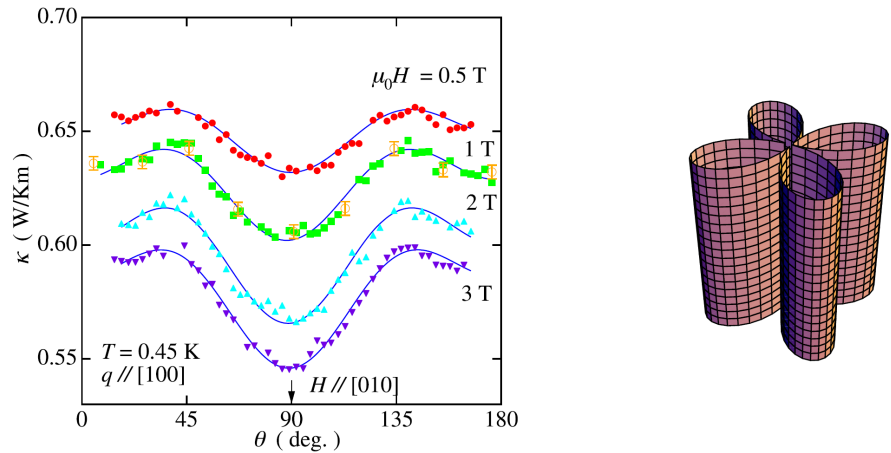

Figure 2.42: Angular dependence of the thermal conductivity of $CeCoIn_5$, indicating a fourfold symmetry of the superconducting gap (from [Knebel 11a, Izawa 01]).

Since magnetic fluctuations are suspected to be the glue for unconventional superconductivity in heavy-fermion materials, it is important to determine how the magnetic properties evolve when tuning parameters are adjusted. Extracting the electronic phase diagrams of these systems under a various set of tuning parameters (mainly chemical doping, pressure, magnetic field) and modeling magnetic fluctuations and long-range orders are prerequisites to describe the superconducting order parameter and the related gapped excitations.

# Chapter 3

# Correlated paramagnetism

This Chapter focuses on the properties of heavy-fermion paramagnets. The correlated paramagnetic ground state of these materials is presented in Section 3.1. Signatures of the electronic correlations in the Fermi-liquid regime and the role played by intersite magnetic fluctuations is highlighted. The effects of a magnetic field applied on these correlated paramagnets is presented in Section 3.2. The physical properties at a metamagnetic transition, or pseudo-metamagnetic crossover, to a high-field polarized paramagnetic regime are detailed. The enhancement of the effective mass, the role played by the magnetic fluctuations, and Fermi-surface reconstructions at the metamagnetic transition are emphasized.

## 3.1 Heavy-fermion ground state

Signatures of the electronic correlations developing at low-temperature in various physical properties of heavy-fermion paramagnets are presented in Section 3.1.1. In Section 3.1.2, intersite magnetic fluctuations are identified as the basic phenomenon driving the correlated paramagnetic (CPM) regime of these materials. Section 3.1.1 constitutes an extension of Section 2.3.2 and Section 3.1.2 is an extension to Section 2.4.3.1.

### 3.1.1 Bulk properties

Figures 3.1(a-b) present magnetic-susceptibility $\chi$ and heat-capacity divided by temperature $C_p/T$ versus temperature curves for a selection of heavy-fermion paramagnets [Settai 07, von Löhneysen 00, von Löhneysen 01, Torikachvili 07, Raymond 10]. Below a temperature $T^*$ characteristic of a heavy Fermi-liquid regime, both quantities tend to saturate to a high value $\propto m^*$. The quantum critical alloy $CeCu_{5.9}Au_{0.1}$, and to a lesser degree its parent $CeCu_6$, are characterized by large values of $\chi$ and $C_p/T$, indicating the heavy character of their electrons [von Löhneysen 00, von Löhneysen 01]. With a Sommerfeld heat-capacity coefficient $\gamma \simeq 8$ J/molK$^2$, $YbCo_2Zn_{20}$ is the 'heaviest' heavy-fermion paramagnet known so far [Torikachvili 07]. Oppositely, lighter masses characterize the intermediate-valent systems $CeRu_2$ and $CeSn_3$, indicating that they are far from a quantum magnetic phase transition [Béal-Monod 80, Raymond 10].

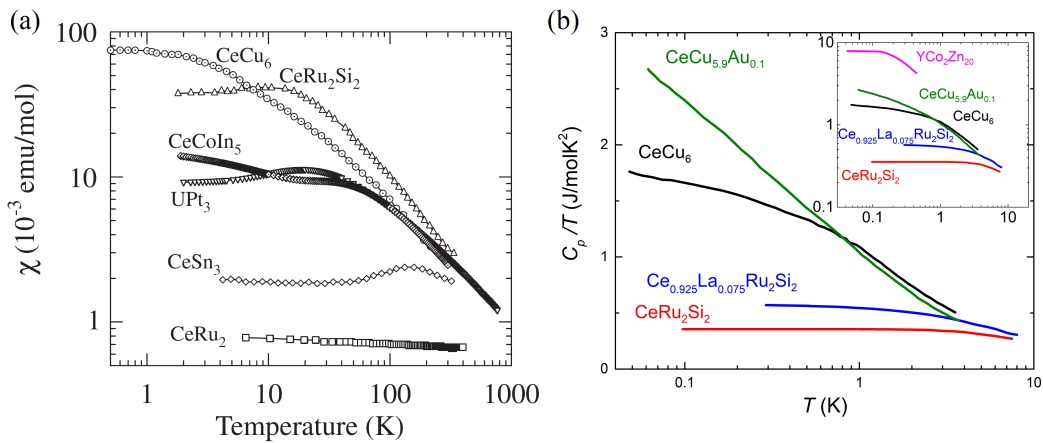

Figure 3.1: (a) Magnetic susceptibility versus temperature in a magnetic field applied along the easy magnetic axis (from [Settai 07]) and (b) heat capacity divided by temperature versus temperature for a selection of heavy-fermion paramagnets (data from [von Löhneysen 00, von Löhneysen 01, Torikachvili 07, Raymond 10]). For $CeRu_2Si_2$ and $Ce_{0.925}La_{0.075}Ru_2Si_2$, the non-electronic background has been subtracted to the heat capacity [Raymond 10].

A comparison of these two graphs indicates that, in the low-temperature Fermi-liquid regime, $\chi$ and $C_p/T$ are not strictly proportional. This deviation from the Wilson law can be qualitatively explained assuming that antiferromagnetic correlations drive the Fermi-liquid regime. Knowing that the presence of magnetic fluctuations peaked at an antiferromagnetic wavevector $\mathbf{k}_{AF} \neq 0$ implies $\chi = \chi'(\mathbf{q} = 0) \ll \chi'(\mathbf{k}_{AF})$ (see Equation 2.12) and $C_p/T \sim \sum_{\mathbf{q}} \chi'(\mathbf{q}) \propto \chi'(\mathbf{k}_{AF})$ (see Equation 2.25), a stronger Wilson ratio $R_W = C_p/T\chi$ is expected in systems subject to strong antiferromagnetic correlations. Such upwards deviation of the Wilson ratio is observed in heavy-fermion compounds with high $\chi$ and $C_p/T$ values (see Figure 2.11(a)).

Complementarily to Figure 2.19 where heat capacity and entropy data were shown, Figure 3.2 presents a large set of physical quantities measured on the heavy-fermion paramagnet $CeRu_2Si_2$. Figure 3.2(a) shows that the electrical resistivity $\rho$ of $CeRu_2Si_2$ is much larger than in non-$4f$ reference compound $LaRu_2Si_2$, over a wide range of temperatures up to 300 K, indicating the presence of electronic correlations [Haen 87a]. Similarly to the heat-capacity $C_p$ data (Figure 2.19(a)), the thermoelectric power $S$ shown in Figure 3.2(b) is anisotropic and reveals at least two electronic contributions, the first one weighting at temperatures $T < 50$ K and the second one weighting at temperatures $T > 50$ K [Amato 89]. The low-temperature contribution to the thermoelectric power is characteristic of the Fermi-liquid regime, and its positive sign highlights the role played by positive-charge carriers. The role of holes is confirmed by the positive Hall coefficient maximum developing in the same temperature range, and also controlled by the Fermi-liquid regime (see Figure 3.2(c)) [Aoki 04]. Figure 3.2(d) shows that large and anisotropic thermal-expansion coefficients $\alpha_i = 1/L_i \partial L_i/\partial T$, where $L_i$ is the sample length along the direction $i$, are also observed in the Fermi-liquid regime [Lacerda 89]. The Inset of Figure 3.2(d) shows that La-doping leads to a decrease of the temperature scale $T^*$ of the

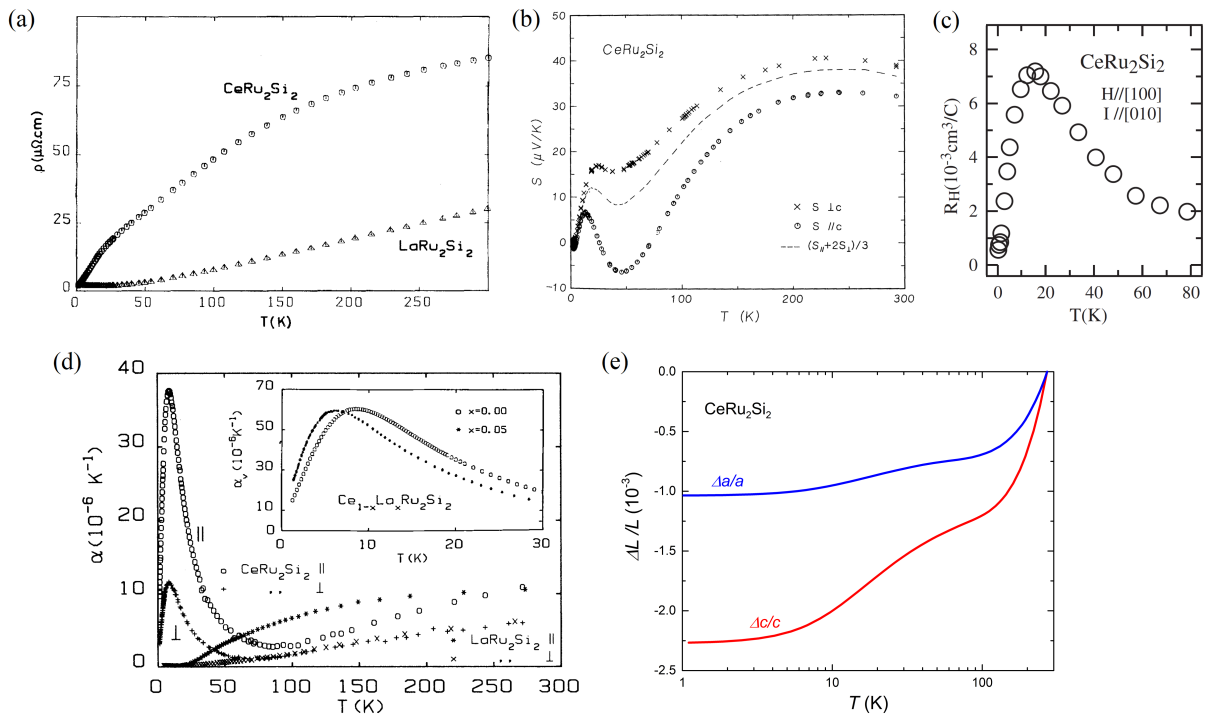

Figure 3.2: (a) Electrical resistivity versus temperature of CeRu$_2$Si$_2$ and LaRu$_2$Si$_2$ (from [Haen 87a]), (b) thermoelectric power versus temperature of CeRu$_2$Si$_2$ (from [Amato 89]), (c) Hall resistivity versus temperature of CeRu$_2$Si$_2$ (from [Aoki 04]), (d) thermal expansion coefficients $\alpha_a$ and $\alpha_c$ versus temperature of CeRu$_2$Si$_2$ and LaRu$_2$Si$_2$ (from [Lacerda 89]) and, in the Inset,volume thermal expansion coefficient $\alpha_v$ versus temperature of CeRu$_2$Si$_2$ and Ce$_{0.95}$La$_{0.05}$Ru$_2$Si$_2$ (from [de Visser 90]), and (e) variation of the lattice parameters in relative lattice units $\Delta a/a$ and $\Delta c/c$ versus temperature of CeRu$_2$Si$_2$ (adapted from [Lacerda 89]).

Fermi liquid (which can be estimated at the maximum of $\alpha_i$). Figure 3.2(e) further shows the lattice parameters variations $\Delta a/a$ and $\Delta c/c$ extracted from the thermal expansion measurements. The establishment of the Fermi-liquid regime corresponds to a low-temperature volume collapse $\Delta V/V = 2\Delta a/a + \Delta c/c \sim 0.1 - 0.2$ %. By comparison with the volume collapses of a few % reported in intermediate-valent systems (see Section 2.3.1), we can conclude that the setup of the heavy-fermion regime in CeRu$_2$Si$_2$ corresponds to a tiny variation of valence, confirming the closeness of this system to a localized limit.

### 3.1.2 Intersite magnetic fluctuations

The role of magnetic fluctuations has been highlighted in Section 2.4.3.1, where their characteristic energy scale $\Gamma(\mathbf{k}) \sim 1/\chi'(\mathbf{k})$ extracted from quasielastic neutron scattering spectra was shown to drive a Fermi-liquid regime. In Figure 2.30, magnetic fluctuations spectra observed by inelastic neutron scattering on CeCu$_6$ were considered. Figure 3.3 shows similar spec-tra obtained for three other heavy-fermion paramagnets. CeCu$_{5.9}$Au$_{0.1}$ is at a quantum magnetic phase transition, as indicated by its almost-diverging heat capacity (see Figures 2.26 and 3.1), and is characterized by quasi-two-dimensional magnetic fluctuations [Stockert 98]. The magnetic fluctuations are peaked along rod-like features in the $(\mathbf{a}^*, \mathbf{c}^*)$ plane of the reciprocal space. Figure 3.3(a) shows that the magnetic fluctuations at a wavevector characteristic of these correlations are enhanced at low energies $E < 0.5$ meV and low-temperatures $T < 1$ K. These correlations presumably contribute to the critical low-temperature enhancement of the heat-capacity [Stockert 98]. Similarly, Figures 3.3(b-c) show that low-energy intersite magnetic fluctuations, identified at wavevectors $\mathbf{k}$ characteristic of antiferromagnetic correlations, develop at low temperature in the heavy-fermion paramagnets CeNi$_2$Ge$_2$ [Kadowaki 03] and YbAgGe [Fåk 05], respectively, indicating nearby antiferromagnetic phase transitions from these two systems.

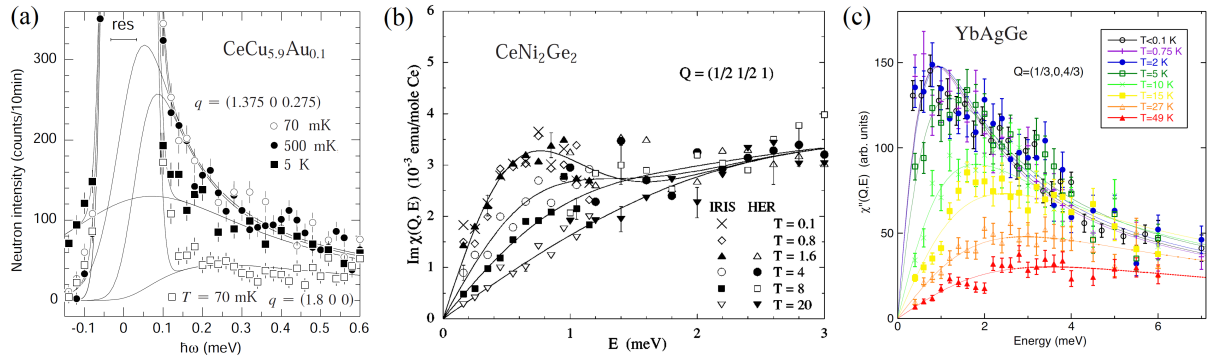

Figure 3.3: Inelastic neutron scattering spectra (a) of CeCu$_{5.9}$Au$_{0.1}$ at temperatures from 70 mK to 5 K, at the momentum transfer $\mathbf{Q} = (1.375, 0, 0.275)$ characteristic of the antiferromagnetic correlations and $\mathbf{Q} = (1.8, 0, 0)$ characteristic of the single-site background (from [Stockert 98]), (b) of CeNi$_2$Ge$_2$ at temperatures from 100 mK to 20 K, at the momentum transfer $\mathbf{Q} = (0.5, 0.5, 1)$ characteristic of the antiferromagnetic correlations (from [Kadowaki 03]), and (c) of YbAgGe at temperatures from 100 mK to 49 K, at the momentum transfer $\mathbf{Q} = (1/3, 0, 4/3)$ characteristic of the antiferromagnetic correlations (from [Fåk 05]).

In the following, a focus is made on CeRu$_2$Si$_2$ and its parent compounds, whose magnetic-fluctuations spectra have been intensively studied over several decades. Figure 3.4 summarizes different sets of inelastic-neutron-scattering experiments, illustrating that RKKY interactions lead to enhanced quasielastic, or broad and weakly-inelastic, dynamical magnetic susceptibility at the incommensurate wavevectors $\mathbf{k}_1 = (0.31, 0, 0)$, $\mathbf{k}_2 = (0.31, 0.31, 0)$, and $\mathbf{k}_3 = (0, 0, 0.35)$ [Rossat-Mignod 88, Sato 99, Kadowaki 04]. Interestingly, the intensities and relaxation rates of the spectra at these three wavevectors are similar. These magnetic fluctuations correspond to a short-range antiferromagnetic order (of spin-density wave-type) indicating the proximity of quantum magnetic phase transitions associated with the setup of long-range magnetic order with one of these wavevectors. Chemical doping (with La, Ge, or Rh) and/or magnetic field tunings can favor one particular interaction and establish antiferromagnetic long-range ordering with either $\mathbf{k}_1$, $\mathbf{k}_2$, or $\mathbf{k}_3$ wavevectors [Quezel 88, Haen 02,

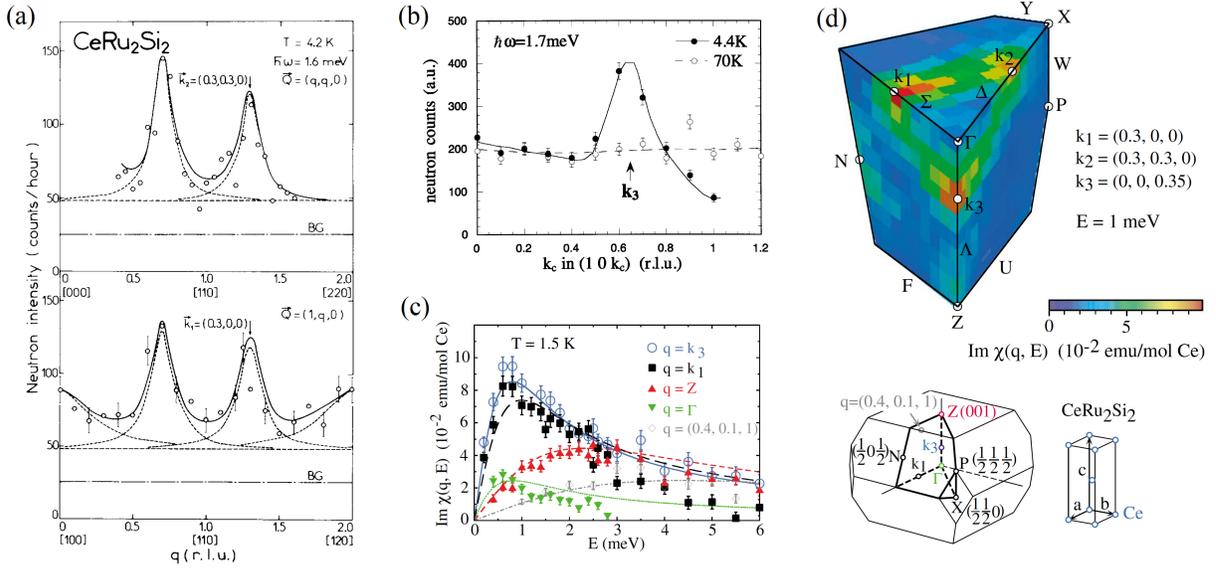

Figure 3.4: Wavevector-scans of inelastic neutron scattering spectra of CeRu$_2$Si$_2$ at an energy transfer $E = 1.6 - 1.7$ meV and a temperature $T \simeq 4.2$ K (a) along $(q, q, 0)$ and $(q, 0, 0)$ directions (from [Rossat-Mignod 88]), and (b) $(0, 0, q)$ direction (from [Sato 99]). (c) Comparison of inelastic neutron scattering energy spectra measured on CeRu$_2$Si$_2$ at $T = 1.5$ K at various wavevectors and (d) three-dimensional representation in the reciprocal space of the imaginary part of the magnetic susceptibility extracted from these data (top) with representation of the Brillouin zone and related elementary lattice unit cell (bottom) (from [Kadowaki 04]).

Mignot 91a, Watanabe 03]. Upon doping by La, a quantum phase transition to an antiferromagnetic state is governed by antiferromagnetic fluctuations with wavevector $\mathbf{k}_1$, i.e., fluctuations of the antiferromagnetic order parameter, which are peaked at the transition [Knafo 09a] (see Chapter 4). At wavevectors sufficiently far from $\mathbf{k}_1$, $\mathbf{k}_2$, and $\mathbf{k}_3$, magnetic fluctuations persist and can be considered as the signature of a local (or single-site) Kondo effect [Rossat-Mignod 88, Kadowaki 04, Raymond 07a, Knafo 09a].

Figures 3.5(a-b) present inelastic-neutron-scattering spectra measured at different temperatures on Ce$_{0.925}$La$_{0.075}$Ru$_2$Si$_2$, which lies at a quantum magnetic phase transition induced by La-doping [Knafo 04]. Measurements at the momentum transfers $\mathbf{Q}_1 = (0.69, 1, 0)$, corresponding to the magnetic wavevector $\mathbf{k}_1 = (0.31, 0, 0)$, and $\mathbf{Q}_0 = (0.44, 1, 0)$, corresponding to a wavevector $\mathbf{q}_0 = (0.56, 0, 0)$ representative of the local single-site signal, are shown. A higher intensity at the wavevector $\mathbf{k}_1$ indicates the development of low-energy correlations at low-temperature. Figures 3.5(c-d) show the temperature variation of the static susceptibility and of the relaxation rate extracted from fits to the data by a quasielastic Lorentzian term (see Equation 2.19). $\chi'(\mathbf{q}, T)$ increases and $\Gamma(\mathbf{q}, T)$ decreases when the temperature is decreased, before saturating for $T < T_0 \simeq 18$ K at the wavevector $\mathbf{q}_0$ and for $T < T_1 \simeq 2.5$ K at the wavevector $\mathbf{k}_1$. The saturation of the relaxation rate $\Gamma(\mathbf{k}_1, T \to 0) \simeq k_B T_1$ for $T < T_1$ is identified as the microscopic origin of the low-temperature Fermi-liquid. In CeRu$_2$Si$_2$, larger temperature scales

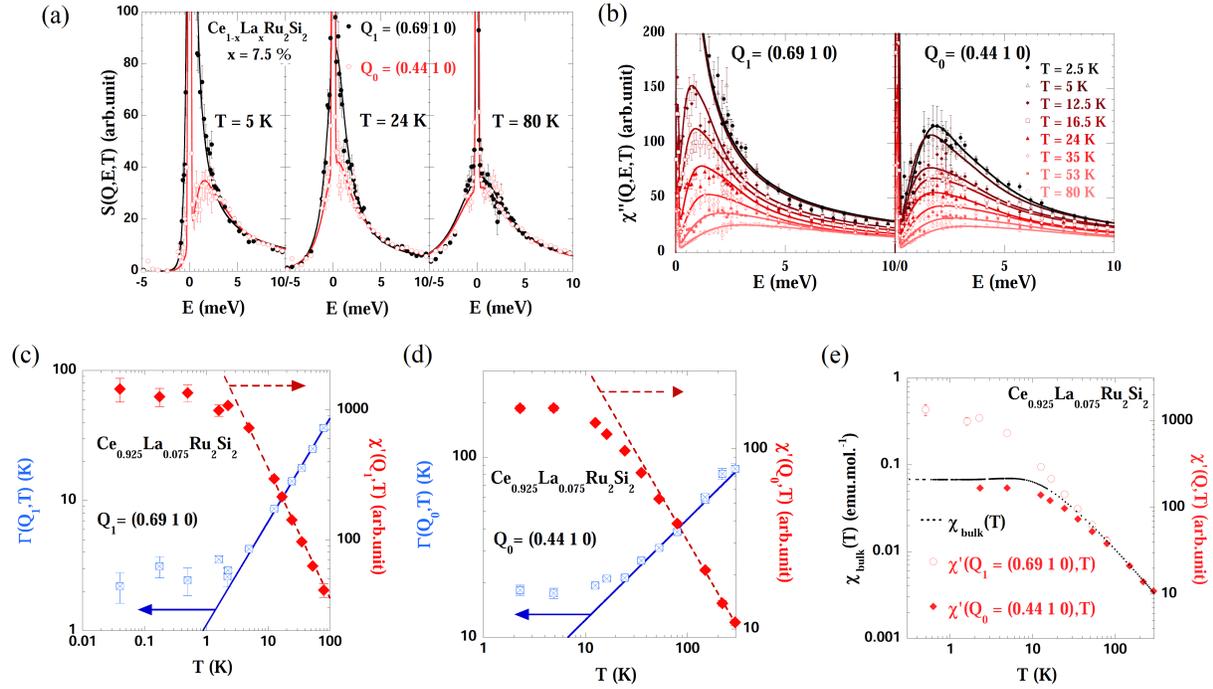

Figure 3.5: (a) Inelastic neutron scattering spectra at the momentum transfers $\mathbf{Q_1} = (0.69, 1, 0)$ and $\mathbf{Q_0} = (0.44, 1, 0)$, at temperature $T = 5$, 24 and 80 K, (b) corresponding dynamical susceptibility extracted at temperatures $2.5 \leq T \leq 80$ K, temperature variations of the extracted static susceptibility and relaxation rate (c) at $\mathbf{Q_1} = (0.69, 1, 0)$ and (d) $\mathbf{Q_0} = (0.44, 1, 0)$, and (e) comparison between the temperature dependences of the bulk susceptibility with the static susceptibilities measured at these two momentum transfers, in $Ce_{0.925}La_{0.075}Ru_2Si_2$ (from [Knafo 04, Raymond 07a]).

$T_1 = 9$ K and $T_0 = 22$ K are the signature of weaker correlations, i.e., less intense magnetic fluctuations, in relation with a more-distant quantum phase transition [Raymond 07a, Knafo 09a]. However, no divergence of the correlations, predicted theoretically for $T \rightarrow 0$, is observed for the quantum critical compound $Ce_{1-x}La_xRu_2Si_2$ (see Section 2.4.3.2). For $T > T_1$, power laws are observed in the $T$-dependence of $\chi'(\mathbf{k_1}, T)$ and $\Gamma(\mathbf{k_1}, T)$:

$$\chi'(\mathbf{k_1}, T) = C_1/T^{\alpha_1} \quad \text{and} \quad \Gamma(\mathbf{k_1}, T) = a_1 T^{\beta_1} \tag{3.1}$$

with $\alpha_1 = 1 \pm 0.05$ and $\beta_1 = 0.8 \pm 0.05$. As well, for $T > T_0$, power laws are observed in the $T$-dependence of $\chi'(\mathbf{q_0}, T)$ and $\Gamma(\mathbf{q_0}, T)$:

$$\chi'(\mathbf{q_0}, T) = C_0/T^{\alpha_0} \quad \text{and} \quad \Gamma(\mathbf{q_0}, T) = a_0 T^{\beta_0} \tag{3.2}$$

with $\alpha_0 = 1 \pm 0.1$ and $\beta_0 = 0.6 \pm 0.2$. Similar $\sqrt{T}$-variations of the relaxation rate were observed for other heavy-fermion compounds (see for instance [Lopes 83, Loidl 89]). While magnetic fluctuations models are clearly needed to describe intersite magnetic effects, as those with wavevector $\mathbf{k_1}$ in $Ce_{0.925}La_{0.075}Ru_2Si_2$, impurity models may be sufficient to describe single-site magnetic effects, as those with wavevector $\mathbf{q_0}$ in $Ce_{0.925}La_{0.075}Ru_2Si_2$. Interestingly, Kondo-impurity [Bickers 85] or Anderson-impurity [Kuroda 92] models, but also a magnetic-fluctuations model [Lopes 83], permitted to extract a relaxation rate varying as $\Gamma \sim \sqrt{T}$ at temperatures higher than the Fermi-liquid temperature.

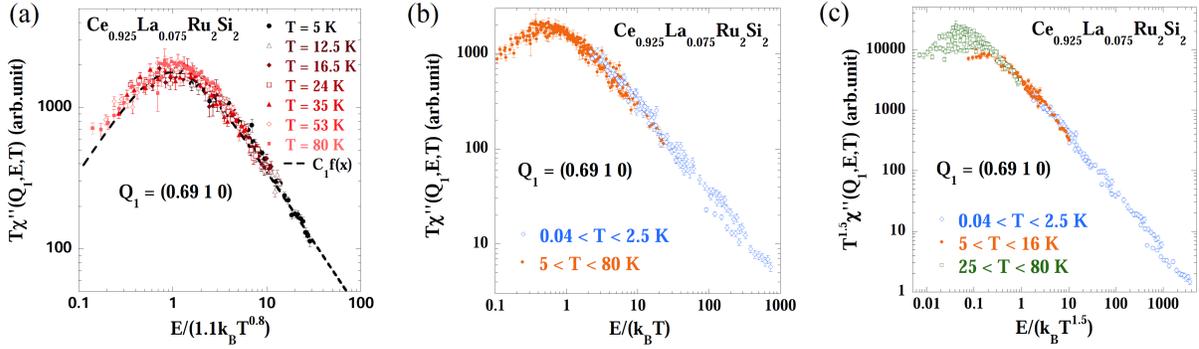

Figure 3.6: Scaling plots $T^{\alpha}\chi''(\mathbf{Q_1}, E, T)$ versus $T/E^{\beta}$ of neutron dynamical susceptibility data measured at $\mathbf{Q_1} = (0.69, 1, 0)$ in $Ce_{0.925}La_{0.075}Ru_2Si_2$, (a) with $\alpha = 1$ and $\beta = 0.8$ as determined from the temperature dependence of $\chi'(\mathbf{Q_1}, T)$ and $\Gamma'(\mathbf{Q_1}, T)$, respectively, at temperatures $T > T_1 = 2.5$ K (intrinsic scaling behavior), and (b) with $\alpha = 1$ and $\beta = 1$ and (c) with $\alpha = 3/2$ and $\beta = 3/2$ plotted for all investigated temperatures 100 mK $\leq T \leq 80$ K (artificial or non-intrinsic scalings) (from [Knafo 04, Knafo 05]).

Figure 3.5(e) compares the temperature dependence of the static susceptibilities $\chi'(\mathbf{q_0}, T)$ and $\chi'(\mathbf{k_1}, T)$ with the bulk susceptibility $\chi(T) = M(T)/H = \chi'(\mathbf{q} = 0, T)$. At high-temperatures, the absence of intersite magnetic correlations leads to an almost $\mathbf{q}$-independent signal. At low temperatures, a strong enhancement of $\chi'(\mathbf{k_1}, T)$ is due to the development of antiferromagnetic correlations. The temperature variations of $\chi(T)$ and $\chi'(\mathbf{q_0}, T)$ are similar, indicating the absence, or a small amplitude, of intersite correlations at the corresponding wavevectors $\mathbf{q} = 0$ and $\mathbf{q_0}$. The decrease of the magnetic susceptibility $\chi(T)$ below the temperature $T_\chi^{max} \simeq 5$ K is identified as a consequence of the onset of intersite magnetic correlation. It will be discussed in Section 3.2.

Following Equations 2.13, 2.19 and 3.2, one can rewrite, for $T > T_1$, the dynamical susceptibility at wavevector $\mathbf{k_1}$ as:

$$T\chi''(\mathbf{k_1}, E, T) = C_1 f\left(\frac{E}{a_1 T^{0.8}}\right),$$ (3.3)

where $f(x) = x/(1 + x^2)$ with $x = E/(a_1 T^{0.8})$. Figure 3.6(a) shows a scaling plot of $T\chi''(\mathbf{k_1}, E, T)$ versus $E/T^{0.8}$ from energy spectra measured by inelastic neutron scattering at the wavevector $\mathbf{k_1}$ and temperatures $T > T_1$. In this graph, the collapse of all data on a single curve governed by Equation 3.3 is a consequence from the observed $T$-power-law variations of $\chi'(\mathbf{k_1}, T)$ and $\Gamma(\mathbf{k_1}, T)$.

Scaling plots of the dynamical susceptibility $T^\alpha \chi''(E, T)$ versus $E/T^\beta$ were made for other heavy-fermion paramagnets, as $UCu_{5-x}Pd_x$ [Aronson 95], $CeCu_{5.9}Au_{0.1}$ [Schröder 98], and $Ce(Ru_{0.24}Fe_{0.76})_2Ge_2$ [Montfrooij 03], and were highlighted as the signature of unconventional quantum criticality followed down to $T \to 0$. However, such graphs have to be interpreted carefully. Figures 3.6(b-c) show two 'artificial' scaling plots $T\chi''(\mathbf{k}_1, E, T)$ versus $E/T$ and $T^{3/2}\chi''(\mathbf{k}_1, E, T)$ versus $E/T^{3/2}$ of the dynamical susceptibility of $Ce_{1-x}La_xRu_2Si_2$, made in a large range of temperatures $100\ mK \leq T \leq 80\ K$ [Knafo 05]. Data from the quantum regime at temperatures $T < T_1$, for which a $E/T^\beta$ scaling is not relevant due to temperature-independent fluctuations, also collapse on these scaling plots. This apparent contradiction can be simply explained. An appropriate manipulation of temperature-independent fluctuations spectra (equivalent to a translation of data in the scaling plots in log-log scale) can allow all data collapsing on a single line, without needing that quantum criticality is followed down to $T \to 0$. Indeed, at temperatures $T < T_1$, the imaginary part of the dynamical susceptibility is temperature-independent and can be expressed as:

$$\chi''(\mathbf{Q}_1, E, T) = \chi'(\mathbf{Q}_1, 0) \frac{E/\Gamma(\mathbf{Q}_1, 0)}{1 + (E/\Gamma(\mathbf{Q}_1, 0))^2}, \tag{3.4}$$

so that, at energies $E \gg \Gamma(\mathbf{k}_1, T \to 0)$ corresponding to the collected data, one obtain:

$$\chi''(\mathbf{k}_1, E, T) \sim E^{-1}, \tag{3.5}$$

ending in the trivial form:

$$T^\alpha \chi''(\mathbf{Q}_1, E, T) \sim \left( \frac{E}{T^\alpha} \right)^{-1}, \tag{3.6}$$

Equation 3.6 drives the artificial scaling plotted in Figures 3.6(b-c). Other kinds of artificial scaling plots can be obtained from a temperature-independent Fermi-liquid regime and will depend on the fluctuations spectrum and on the energy range of the measured data. To avoid dealing with artificial $E/T^\beta$ scaling graphs and, thus, ending with non-relevant physical information, a preliminary analysis of the temperature-dependence of the energy spectra is mandatory.

## 3.2 Magnetic-field-induced phenomena

Metamagnetic transitions induced in heavy-fermion paramagnets under a magnetic field are presented. After an introduction to their signatures in thermodynamic and transport properties in Section 3.2.1, focus will be given to the magnetic fluctuations in Section 3.2.2 and to the Fermi surface in Section 3.2.3.

### 3.2.1 Signatures of metamagnetism

Metamagnetism labels first-order magnetic transitions induced in strongly-anisotropic magnets by a magnetic field [Stryjewski 77]. The metamagnetic field, noted $H_m$, can be defined at a large and sudden jump $\Delta M$ of the magnetization. Generally, metamagnetism is observed in antiferromagnets of Ising uniaxial magnetic anisotropy, and it corresponds to a sudden alignment of the magnetic moments along the magnetic-field direction. The high-magnetic-field regime beyond the metamagnetic field is a polarized paramagnet, where intersite magnetic interactions have been released by the magnetic field.

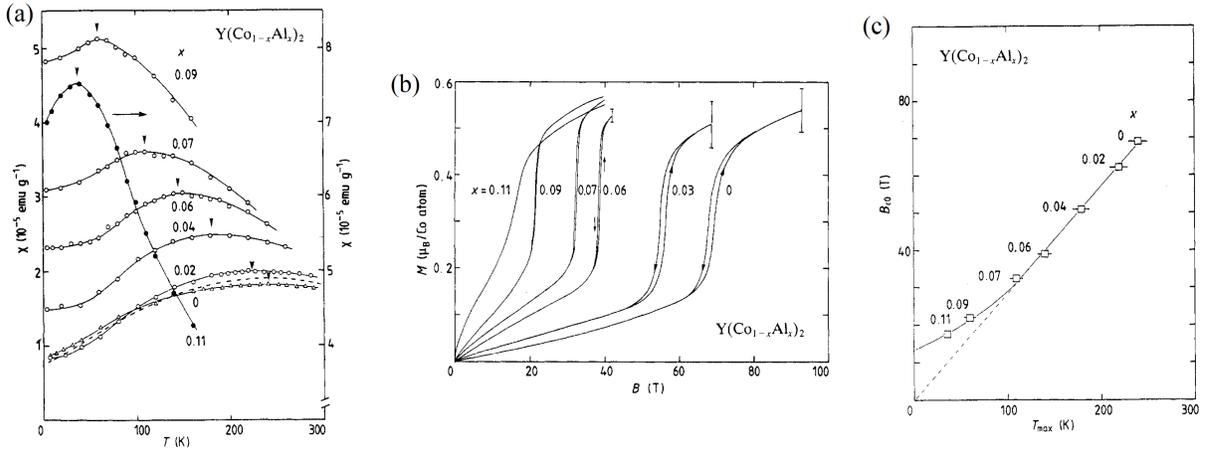

Figure 3.7: (a) Temperature-dependence of the magnetic susceptibility, (b) magnetic-field-dependence of the magnetization, and (c) plot of the metamagnetic field (noted $H_c$ here) versus the temperature at the maximum of the magnetic susceptibility (noted here $T_{max}$) of $Y(Co_{1-x}Al_x)_2$ (from [Sakakibara 90a]).

In anisotropic paramagnets close to a magnetic phase transition, i.e., nearly-ferromagnets or nearly-antiferromagnets, metamagnetism can also occur. Figures 3.7(a-b) present the magnetic susceptibility versus temperature and the magnetization versus magnetic field, respectively, of nearly ferromagnetic 'Laves-phase' compounds $Y(Co_{1-x}Al_x)_2$ [Sakakibara 90a]. A broad maximum in the magnetic susceptibility at a temperature $T_\chi^{max}$ ranging from $\simeq 40$ K for $x = 0.11$ to $\simeq 250$ K for $x = 0$ and a metamagnetic transition at a magnetic field $\mu_0 H_m$ ranging from $\simeq 20$ T for $x = 0.11$ to $\simeq 70$ T for $x = 0$ are observed. The empirical relation $H_m \propto T_\chi^{max}$ is emphasized by the plot of $H_m$ versus $T_\chi^{max}$ shown in Figure 3.7(c), indicating that both quantities are controlled by a unique parameter.

From the thermodynamic Maxwell relation:

$$\frac{\partial M}{\partial T} = \frac{\partial S}{\partial H},$$ (3.7)

the entropy $S$ increases with the field $H$ at temperatures $T < T_\chi^{max}$ where $\partial M/\partial T > 0$, and it decreases with $H$ at temperatures $T > T_\chi^{max}$ where $\partial M/\partial T < 0$ [Sakakibara 90b]. This implies that the entropy is maximal at the metamagnetic field $H_m$, above which $T_\chi^{max}$ vanishes (see for instance Figure 3.8(e)). Although they did not observe signatures of magnetic fluctuations, Sakakibara et al. proposed that the increase of entropy is controlled by an enhancement of the magnetic fluctuations induced by the magnetic field [Sakakibara 90b].

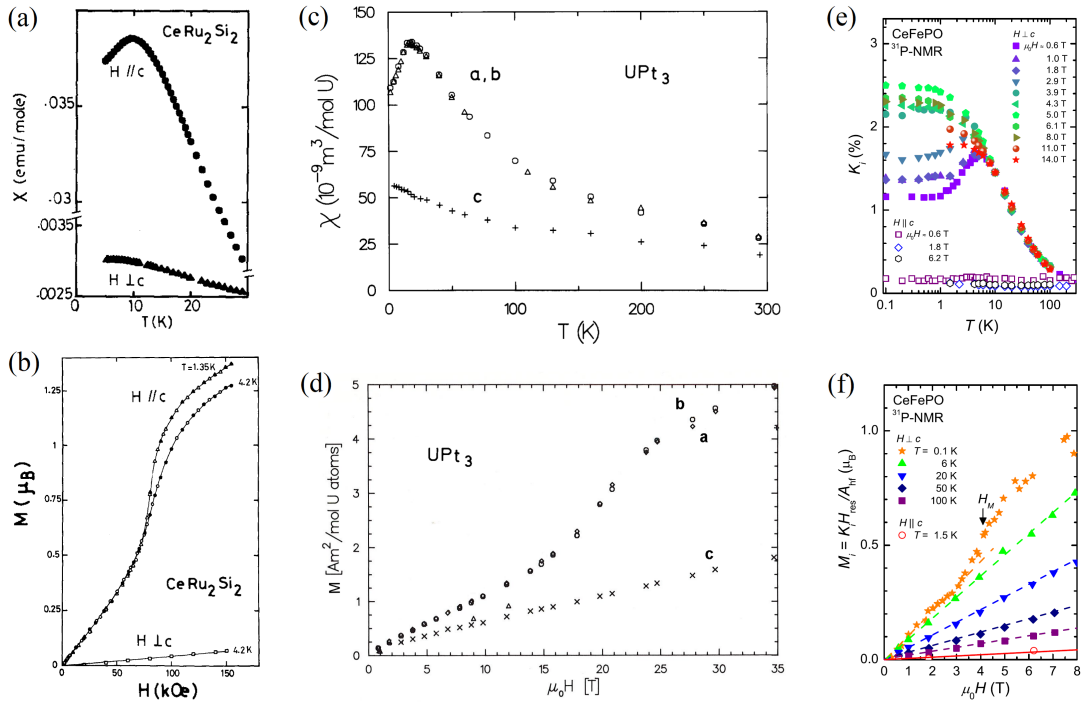

Figure 3.8: (a) Temperature-dependence of the magnetic susceptibility and (b) magnetic-field-dependence of the magnetization of $CeRu_2Si_2$ in magnetic fields $\mathbf{H} \parallel \mathbf{c}$ and $\mathbf{H} \perp \mathbf{c}$ (from [Haen 87a]). (c) Temperature-dependence of the magnetic susceptibility and (d) magnetic-field-dependence of the magnetization of $UPt_3$ in magnetic fields $\mathbf{H} \parallel \mathbf{c}$ and $\mathbf{H} \perp \mathbf{c}$ (from [Frings 84, Frings 85, de Visser 87a]). (e) Temperature-dependence of the NMR Knight shift $K_i$ and (f) magnetic-field dependence of the associated local magnetization $M_i$, where $i$ represents the magnetic field direction, of CeFePO in magnetic fields $\mathbf{H} \parallel \mathbf{c}$ and $\mathbf{H} \perp \mathbf{c}$ (from [Kitagawa 11]).

Metamagnetism, sometimes called pseudo-metamagnetism (when a crossover rather than a first-order transition occurs), is also observed in heavy-fermion anisotropic paramagnets. Figure 3.8 presents magnetic susceptibility versus temperature and magnetization versus magnetic field, measured either by a macroscopic magnetization probe or by a microscopic NMR probe (via the NMR Knightshift, see Equation 2.15 in Section 2.4.3.1), of the heavy-fermion paramagnets $CeRu_2Si_2$ [Haen 87a], $UPt_3$ [Frings 84, Frings 85, de Visser 87a], and CeFePO [Kitagawa 11]. A metamagnetic transition is induced at a magnetic field $\mu_0 H_m$ applied along the easy magnetic axis (or easy magnetic plane), ranging from 4 T in CeFePO, 7.8 T in $CeRu_2Si_2$ to 20 T in $UPt_3$. The metamagnetic transition is in relation with a maximum observed at the temperature $T_\chi^{max}$ in the low-field magnetic susceptibility, which ranges from 5 K in CeFePO, 9 K in $CeRu_2Si_2$ to 20 K in $UPt_3$. The relation between $T_\chi^{max}$ and $H_m$, which both delimitate a correlated paramagnetic (CPM) regime, will be discussed in Section 3.3.

Figure 3.9 presents a large set of physical quantities measured in $CeRu_2Si_2$ under a magnetic field applied along its easy magnetic axis $\mathbf{c}$. Figures 3.9(a-b) shows magnetostriction coef-ficients $\lambda_i = 1/L_i \partial L_i / \partial \mu_0 H$, where $L_i$ is the sample length along the direction $i$ ($i = a, c$ or $V$ for volume) versus magnetic field at $T = 1.3$ K [Lacerda 89] and the volume thermal-expansion coefficient $\alpha_V = 1/L_V \partial L_V / \partial T$ versus temperature at different field values [Paulsen 90]. For $\mu_0 H < \mu_0 H_m = 7.8$ T, the thermal-expansion coefficients present a broad maximum at the temperature $T_\alpha^{max}$ characteristic of the establishment of the CPM regime. For $H > H_m$, the low-temperature thermal expansion becomes negative, a broad minimum at the temperature $T_\alpha^{min}$ being a signature of the boundary of the PPM regime. From Ehrenfest/Clapeyron relationships, the positive anomalies in $\alpha_V$ indicate that $\partial T_\alpha^{max} / \partial p > 0$, while the negative anomalies in $\alpha_V$ indicate that $\partial T_\alpha^{min} / \partial p < 0$, where $p$ is the hydrostatic pressure, confirming that $T_\alpha^{max}$ and $T_\alpha^{min}$ are the boundaries of two different regimes. Figure 3.9(d) presents the temperature-magnetic-field phase diagram extracted from these thermal-expansion measurements, emphasizing the borderlines of the CPM and PPM regimes, separated by $H_m$ at low-temperature [Holtmeier 95]. A sharp and positive enhancement of $\lambda_i$ at $H_m$ is observed for all directions, indicating a sudden expansion of the lattice at the metamagnetic transition [Lacerda 89]. Figure 3.9(c) presents the magnetic-field-variation of the lattice parameters $\Delta a/a$ and $\Delta c/c$ extracted from the $\lambda_i(H)$ curves. The magnetic field leads to a low-temperature volume expansion $\Delta V/V = 2\Delta a/a + \Delta c/c \sim 0.1\,\%$, which cancels the contraction of same order of magnitude observed at temperatures $T < T_\chi^{max}$ (see Figure 3.8(e) in Section 3.1.1). These dilatometry data indicate that, while the establishment of the the CPM regime below $T_\chi^{max}$ is accompanied by a tiny increase of valence indicating a 'moderate' itinerant character of the electrons, the transition at $H_m$ to the polarized paramagnetic regime (PPM) is accompanied by reduction of the valence, indicating that the electrons are back to a localized limit similar to that at low-field and high temperature. Figure 3.9(e) presents a measurement of valence performed by x-ray absorption spectroscopy under pulsed magnetic field [Matsuda 12]. A small valence change $\Delta v = -0.005$ is induced at $H_m$, in agreement with the conclusions reached by considering the length variations. Figure 3.9(f) shows electrical-resistivity $\rho_{xx}$ and Hall-resistivity $\rho_H = \rho_{xy}$ measurements, at temperatures from $< 10$ mK to 6 K [Daou 06]. The metamagnetic transition leads to strong enhancements of $\rho_{xx}$ and $\rho_H$ at high temperature, which vanish at low-temperatures into small step-like variations of $\rho_{xx}$ and $\rho_H$. Daou *et al* interpreted the absence of strong discontinuity of $\rho_{xx}$ and $\rho_H$ at $H_m$ as the indication that the $f$ electrons do not become suddenly localized and that the Fermi surface is continuously modified at $H_m$ [Daou 06]. In Section 3.2.3, Fermi-surface measurements at the metamagnetic transition of CeRu$_2$Si$_2$ will be presented and discussed in light of other experiments, as those presented here and magnetic fluctuations studies presented in Section 3.2.2.

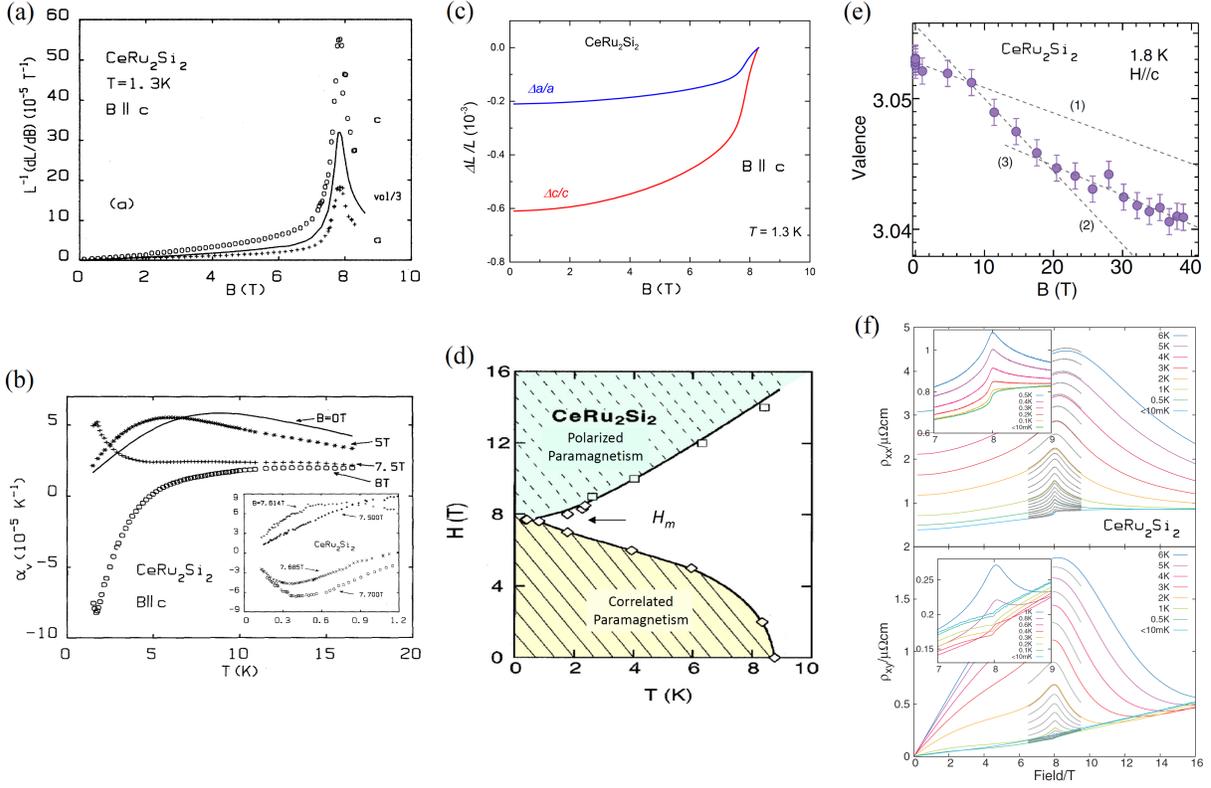

Figure 3.9: (a) Magnetostriction coefficients $\lambda_i$ ($i = a, c$ or volume) versus magnetic field, at $T = 1.3$ K (from [Lacerda 89]), (b) volume thermal expansion coefficient $\alpha_V$ versus temperature at different magnetic-field values (from [Paulsen 90]), (c) relative lengths variations $\Delta L_i / L_i$ ($i = a, c$ or volume) versus magnetic field at $T = 1.3$ K (extracted from [Lacerda 89]), (d) magnetic-field-temperature phase diagram extracted from the thermal expansion data (from [Holtmeier 95]), (e) variation of valence versus magnetic field extracted from x-ray absorption spectroscopy at $T = 1.8$ K (from [Matsuda 12]), and (f) electrical resistivity and Hall resistivity versus magnetic field, at different temperatures from $< 10$ mK to 6 K (from [Daou 06]), of $CeRu_2Si_2$ in a magnetic field $\mathbf{H} \parallel c$.

### 3.2.2 Effective mass and magnetic fluctuations

Figures 3.10(a-b) present the variation of the heat-capacity Sommerfeld coefficient $\gamma =_{T\to 0} C_p/T$ of UPt$_3$ in a magnetic field $\mathbf{H} \parallel \mathbf{a}$ (from [van der Meulen 90]) and (b) CeRu$_2$Si$_2$ in a magnetic field $\mathbf{H} \parallel \mathbf{c}$ (from [van der Meulen 91]), respectively. For both compounds, a large enhancement of $\gamma$ indicates the enhancement of the effective mass and, thus, of the low-temperature magnetic fluctuations, at the metamagnetic field $H_m$. Complementarily to Figure 3.10(b), Figure 3.10(c) shows the variation with magnetic field of the square root $\sqrt{A}$ of the quadratic coefficient from the electrical resistivity of CeRu$_2$Si$_2$ at different pressures. At ambient pressure, the field variations of $\sqrt{A}$ and $\gamma$ are quite similar, indicating an almost field-independent Kadowaki-Woods coefficient $R_{KW} = A/\gamma^2$ (see Section 2.3.2). Inset of Figure 3.10(a) also shows that the Kadowaki-Woods coefficient $R_{KW}$ is also almost field-independent when metamagnetism is crossed in UPt$_3$. Figure 3.10(c) shows that under pressure, smaller maximum values of $A$ are observed at the metamagnetic field $H_m$ of CeRu$_2$Si$_2$ [Aoki 11a]. The metamagnetic field $H_m$, as well as the temperature $T_\chi^{max}$ at the maximum of the magnetic susceptibility, increase linearly with $p$, which reinforces the correlated paramagnetic regime of CeRu$_2$Si$_2$ [Haen 87b, Mignot 88]. The proportionality of $H_m$ and $T_\chi^{max}$ with the low-temperature susceptibility $1/\chi(T \to 0)$ and the coefficient $1/\sqrt{A}(H = 0)$ was evidenced

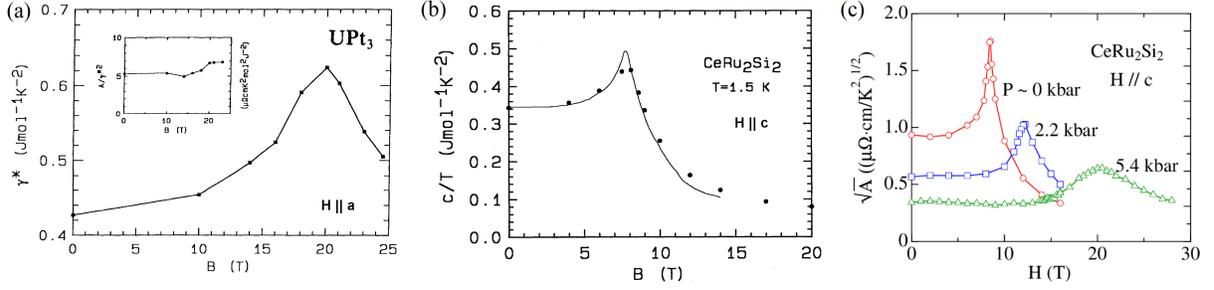

Figure 3.10: Magnetic-field variation of heat-capacity Sommerfeld coefficient $\gamma$ of (a) UPt$_3$ in a magnetic field $\mathbf{H} \parallel \mathbf{a}$ (from [van der Meulen 90]) and (b) CeRu$_2$Si$_2$ in a magnetic field $\mathbf{H} \parallel \mathbf{c}$ (from [van der Meulen 91]), and magnetic-field variation of the square root $\sqrt{A}$ of the electrical resistivity quadratic coefficient of CeRu$_2$Si$_2$ in a magnetic field $\mathbf{H} \parallel \mathbf{c}$, at pressures $p = 0$, 2.2, and 5.4 kbar (from [Aoki 11a]).

under pressure [Mignot 88], indicating that, within first approximation, all of these quantities are controlled by a unique parameter, the effective mass $m^*$:

$$H_m \propto T_\chi^{max} \propto 1/\chi(T \to 0) \propto 1/\sqrt{A} \propto m^*. \qquad (3.8)$$

This confirms, thus, the Fermi-liquid nature of the CPM regime in CeRu$_2$Si$_2$ under pressure.

The enhancement of the effective mass at the metamagnetic transition of heavy-fermion paramagnets indicates that critical magnetic fluctuations may operate. Figure 3.11 presents measurements of the NMR relaxation rate $1/T_1T$ of CeRu$_2$Si$_2$ and CeFePO through their metamagnetic transition [Ishida 98, Kitagawa 11]. Figures 3.11(a-b) show that $1/T_1T$ saturates at temperatures $T \lesssim T_\chi^{max}$ for $H < H_m$ and passes through a maximum for $T \to 0$ and $H = H_m$ [Ishida 98]. For $H > H_m$, the low-temperature ground state is a polarized paramagnetic regime established below a temperature scale $T_{PPM}$, which can be identified here at a maximum of $1/T_1T$, and which increases with $H$. Figures 3.11(c-d) show similar variations of $1/T_1T$, which is largely enhanced at the metamagnetic field $H_m$ for $T \to 0$, in CeFePO. Contrary to CeRu$_2$Si$_2$ where a saturation of $1/T_1T$ was observed, a maximum of $1/T_1T$ occurs at the temperature $T \simeq T_\chi^{max}$ for $H < H_m$ [Kitagawa 11]. These variations of the NMR relation rate $1/T_1T$ indicate that magnetic fluctuations, here summed over all reciprocal space (see Equation 2.20 in Section 2.4.3.1), are peaked at the metamagnetic field $H_m$, but also at the temperature $T_{PPM}$ characteristic of the high-field PPM regime, and that they saturate for $T \lesssim T_\chi^{max}$, with (CeFePO) or without (CeRu$_2$Si$_2$) presenting a maximum at $T_\chi^{max}$.

NMR relaxation-rate measurements do not allow extracting the wavevector-dependence of the magnetic fluctuations and to access to energy spectra (see Equation 2.20 in Section 2.4.3.1). For this purpose, inelastic neutron scattering constitutes a unique tool. Figure 3.12 presents inelastic neutron scattering measurements performed on CeRu$_2$Si$_2$ under a magnetic field $\mathbf{H} \parallel \mathbf{c}$ [Rossat-Mignod 88, Flouquet 02, Flouquet 04]. Figures 3.12(a-b) show energy $E$- and momentum transfer $\mathbf{Q}$- scans, respectively, of low-temperature neutron-scattered intensity at and near the momentum transfer $\mathbf{Q}_2 = (0.7, 0.7, 0)$ corresponding to wavevector $\mathbf{k}_2 \simeq (0.3, 0.3, 0)$, at which antiferromagnetic fluctuations are peaked at zero-field [Rossat-Mignod 88, Flouquet 02] (see. Figure 3.8 in Section 3.1.2). A loss of magnetic correlations with wavevector $\mathbf{k}_2$ is shown by the reduction of scattered intensity combined with a loss of the wavevector-dependence of the signal in magnetic fields $H > H_m$. Figure 3.12(c) shows a plot of neutron scattering intensity, measured at wavevector $\mathbf{k}_2$, at the energy $E = 1.6$ meV and temperature $T = 1.5$ K, versus magnetic field, indicating that the correlations with wavevector $\mathbf{k}_2$ vanish at the metamagnetic field $H_m$ [Rossat-Mignod 88]. Figure 3.12(d-top) shows that, as well as those with wavevector $\mathbf{k}_2$, the magnetic fluctuations with wavevector $\mathbf{k}_1 = (0.31, 0, 0)$, which also develop in the CPM regime of $CeRu_2Si_2$ (see. Figure 3.4 in Section 3.1.2), vanish at $H_m$ [Flouquet 04]. The collapse of antiferromagnetic fluctuations at $H_m$ confirms that the CPM regime, which corresponds to a low-field heavy Fermi liquid, is controlled by these intersite antiferromagnetic correlations. This picture is confirmed by the observation that the temperature boundary $T_\chi^{max} = 10$ K of the CPM regime is almost equal to the relaxation rate $\Gamma_1 = 9$ K of the magnetic fluctuations with wavevector $\mathbf{k}_1$ [Raymond 07a, Knafo 09a], both being also related to $H_m$, the high-field boundary of the CPM regime, by the empirical relation 1 K $\leftrightarrow$ 1 T [Aoki 13] (see Section 3.3).

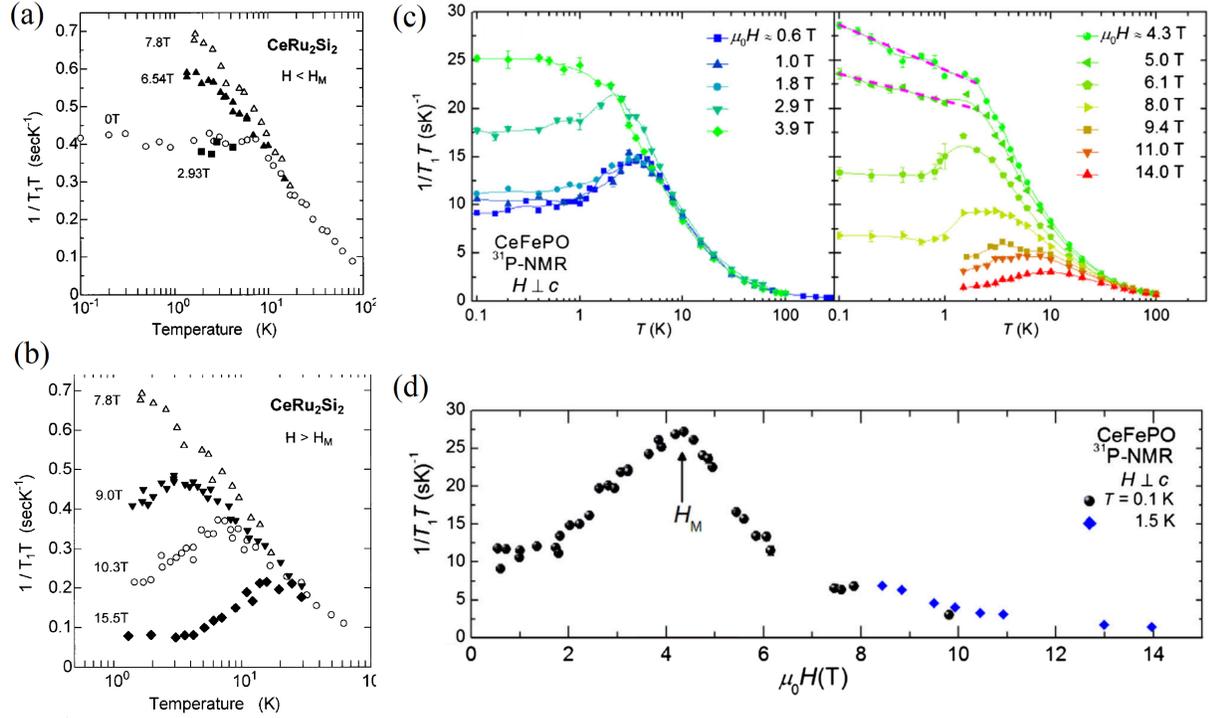

Figure 3.11: Temperature-variation of the NMR relaxation rate $1/T_1T$ of $CeRu_2Si_2$ in magnetic fields (a) $H < H_m$ and (b) $H > H_m$ applied along **c** (from [Ishida 98]), and (c) of CeFePO in magnetic fields $\mathbf{H} \perp \mathbf{c}$. (d) Low-temperature magnetic-field variation of the NMR relaxation rate $1/T_1T$ of CeFePO in magnetic fields $\mathbf{H} \perp \mathbf{c}$ (from [Kitagawa 11]).

Figure 3.12(d-bottom) presents the magnetic-field-variation of the neutron-scattering intensity, at the energy $E = 0.4$ meV and temperature $T = 170$ mK, at momentum trans-

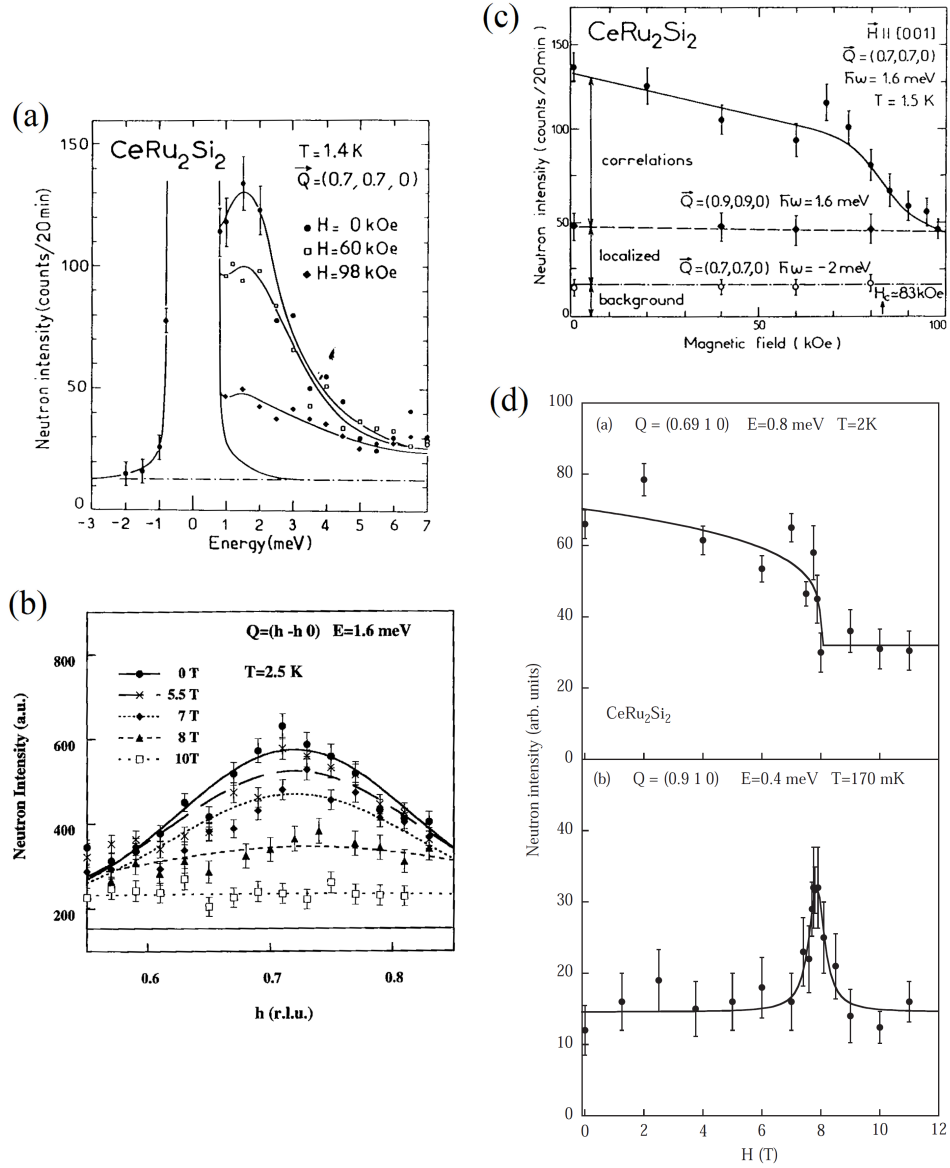

Figure 3.12: Neutron scattering intensity on CeRu$_2$Si$_2$ in magnetic fields $\mathbf{H} \perp \mathbf{c}$ (a) versus energy, at the momentum transfer $\mathbf{Q}_2 = (0.7, 0.7, 0)$ (corresponding to wavevector $\mathbf{k}_2$) at different magnetic fields and at the temperature $T = 1.4$ K (from [Rossat-Mignod 88]), (b) versus $h$, at the momentum $\mathbf{Q}(h, h, 0)$ at the energy transfer $E = 1.6$ meV, at different magnetic fields and at the temperature $T = 2.5$ K (from [Flouquet 02]), versus magnetic field (c) at $\mathbf{Q}_2 = (0.7, 0.7, 0)$ (corresponding to wavevector $\mathbf{k}_2$) at the energy transfer $E = 1.6$ meV and the temperature $T = 1.5$ K (from [Rossat-Mignod 88]), (d-top) at $\mathbf{Q}_2 = (0.69, 0, 0)$ (corresponding to wavevector $\mathbf{k}_1$) at the energy transfer $E = 0.8$ meV and the temperature $T = 2$ K, and (d-bottom) at $\mathbf{Q} = (0.9, 1, 0)$ (near the ferromagnetic wavevector $\mathbf{k} = 0$) at the energy transfer $E = 0.4$ meV and the temperature $T = 170$ mK (from [Flouquet 04]).

fer $\mathbf{Q} = (0.9, 1, 0)$, which probes the ferromagnetic fluctuations near the nuclear position $\tau = (1, 1, 0)$, corresponding to the ferromagnetic wavevector $\mathbf{k} = 0$ [Flouquet 04]. A sharp enhancement of the scattered intensity indicates that ferromagnetic fluctuations develop in a narrow magnetic-field window around $H_m$. These ferromagnetic fluctuations, first observed by Sato *et al* [Sato 01], are identified as the critical magnetic fluctuations of the metamagnetic transition, beyond which a polarized paramagnetic regime, where the magnetic moments are aligned parallel to the field direction, is stabilized. The enhancement of the effective mass $m^*$ at $H_m$ (see for instance Equation 3.8) may result from the combined decrease of antiferromagnetic fluctuations, which control the effective mass at low fields, and increase of ferromagnetic fluctuations. In magnetic fields $H > H_m$, the fall of the effective mass and the disappearance of antiferromagnetic and ferromagnetic fluctuations indicates a regime where intersite magnetic fluctuations have vanished. Interestingly, the increase of $A$ under magnetic field is such that the ratio $A(H_m)/A(H = 0)$ is independent of pressure [Aoki 11a]. Assuming that $A(H = 0)$ is controlled by antiferromagnet fluctuations, and that $A(H_m)$ is controlled by ferromagnet fluctuations, this indicates that stronger antiferromagnetic fluctuations in the CPM regime lead to stronger ferromagnetic fluctuations at $H_m$, suggesting a transfer of spectral weight from antiferromagnetic fluctuations into ferromagnetic fluctuations at $H_m$.

As well as the quench of antiferromagnetic long-range order into a polarized paramagnetic regime drives the metamagnetic transition in antiferromagnets [Stryjewski 77], the quench of antiferromagnetic short-range order, i.e., the antiferromagnetic fluctuations (see Figure 2.28 in Section 2.4.3.1), drives the metamagnetic transition from correlated paramagnetism to polarized paramagnetism in CeRu$_2$Si$_2$. The survey of intersite magnetic fluctuations of CeRu$_2$Si$_2$ should be extended to other heavy-fermion paramagnets to check whether the physics revealed at its metamagnetic transition is representative of a general behavior.

### 3.2.3   Fermi surface

Fermi-surface reconstructions generally accompany metamagnetism in heavy-fermion compounds. Figure 3.13 presents a synthesis of Fermi-surface studies performed on CeRu$_2$Si$_2$ across its metamagnetic transition (see [Aoki 14] for a review about Fermi surface properties of CeRu$_2$Si$_2$ and its alloys). Figure 3.13(a) presents the angular-dependence of the de Haas-van Alphen (dHvA) frequencies extracted along different field directions, resulting from Fermi surfaces for $H < H_m$ and $H > H_m$ [Takashita 96]. Figure 3.13(c) shows thermoelectrical-power measurements, which changes sign at $H_m$ and presents quantum oscillations on both sides of $H_m$ [Boukahil 14]. Focus on the quantum-oscillation signal for $H < H_m$ and $H > H_m$ is made in Figure 3.13(d). The resulting Fourier-transform spectra are presented in Figure 3.13(e), indicating a change of the quantum-oscillations frequencies, and thus, of the Fermi surface, at $H_m$. The magnetic-field dependence of the frequencies extracted from dHvA measurements in a magnetic field $\mathbf{H} \parallel \mathbf{c}$ is shown in Figure 3.13(f) [Takashita 96]. Progressive changes of the Fermi surface lead to continuous variations of the frequencies on both sides of $H_m$. A Fermi-surface reconstruction leads to sudden changes of the dHvA frequencies at $H_m$. Figure 3.13(b) presents three-dimensional views of the Fermi sheets of CeRu$_2$Si$_2$ [Yamagami 93] and LaRu$_2$Si$_2$ [Yamagami 92, Settai 95], which can be considered as a non-4$f$ reference for CeRu$_2$Si$_2$, at low magnetic field. The difference between these two Fermi surfaces was proposed to result from the presence of itinerant $f$-electrons in CeRu$_2$Si$_2$ [Yamagami 93].

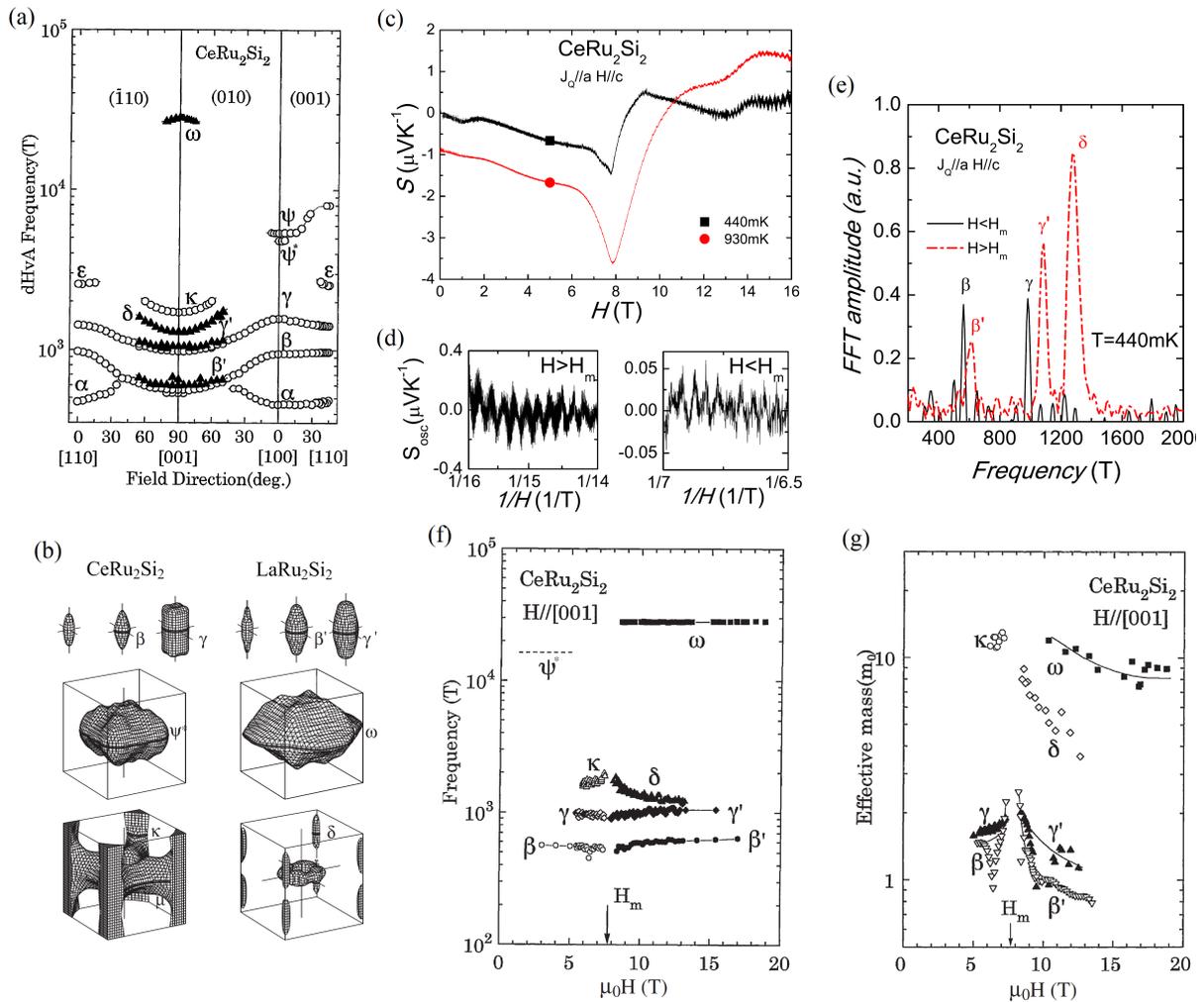

Figure 3.13: (a) dHvA frequencies versus magnetic field direction for $H < H_m$ (open symbols) and $H > H_m$ (full symbols) CeRu$_2$Si$_2$ (from [Takashita 96]), (b) three-dimensional views of Fermi surface sheets of CeRu$_2$Si$_2$ and LaRu$_2$Si$_2$ measured by dHvA technique (from [Yamagami 92, Yamagami 93]), (c) field-variation of low-temperature thermoelectric power $S$ for $\mathbf{H} \parallel \mathbf{c}$ and a current $\mathbf{J} \perp \mathbf{c}$, with (d) a focus on the quantum oscillations of $S$ versus $1/H$ for $H > H_m$ (left) and $H < H_m$ (right) and (e) corresponding Fourier transformations (from [Boukahil 14]), and magnetic-field dependence (f) of the dHvA frequencies and (g) of the cyclotron masses of the observed sheets of CeRu$_2$Si$_2$ in a magnetic field $\mathbf{H} \parallel \mathbf{c}$ (from [Takashita 96]).

Takashita *et al* suggested that the similarity between the Fermi surface of CeRu$_2$Si$_2$ in magnetic fields $H > H_m$ and that of LaRu$_2$Si$_2$ at low fields is an indication of electronic local-ization in the PPM regime of CeRu$_2$Si$_2$ [Takashita 96]. However, from electrical-resistivity and Hall-resistivity measurements, Daou *et al* proposed that a continuous change occurs at $H_m$, both regimes being controlled by itinerant-$f$ electrons [Daou 06] (see Figure 3.9(f) in Section 3.2.1). The observation by inelastic neutron scattering of a remaining wavevector-**q**-independent magnetic-fluctuations signal in magnetic fields $H > H_m$ [Rossat-Mignod 88] (see Figure 3.12(f) in Section 3.2.2) can also be interpreted as the signature of a single-site Kondo effect implying itinerant $4f$-electrons.

Figure 3.13(g) presents the magnetic-field variation of the cyclotron masses $m_c^i$ of the dHvA frequencies $i$ observed in CeRu$_2$Si$_2$ [Takashita 96]. At zero-field, the masses of these bands account for a large part of the effective mass $m^*$ estimated from the heat capacity Sommerfeld coefficient [Tautz 95] (see Equation 2.4 in Section 2.3.4). Maxima of the cyclotron masses $m_c^i$ are observed in the vicinity of the metamagnetic transition, indicating that the carriers from these Fermi-surface bands contribute to the enhancement of $m^*$, leading to a maximum of the heat-capacity at $H_m$ [van der Meulen 90] (see Figures 3.9(c,e) in Section 3.2.1). This suggests that the magnetic fluctuations, which were also found to drive the enhancement of the effective mass on both sides of the metamagnetic transition, may be a property of itinerant $4f$ electrons. However, the heavy-fermion limit in Ce-compounds corresponds to a nearly-localized $4f$-electrons state, as indicated by a valence close to 3 [Matsuda 12] and by the small lattice changes [Lacerda 89, Paulsen 90] reported in CeRu$_2$Si$_2$ under varying temperature and magnetic field (see Figure 3.10(a) in Section 3.2.2). The dual itinerant/localized nature of the $4f$ electrons emphasized here for CeRu$_2$Si$_2$ constitutes a subtle issue in most of heavy-fermion compounds, and its understanding will be a key for a thoughtful description of their properties.

## 3.3 Synthesis

A synthesis of the magnetic properties of the correlated paramagnetic regime of heavy-fermion systems is made here. Table 3.1 summarizes the characteristics of several heavy-fermion paramagnets undergoing a field-induced transition to a polarized paramagnetic regime. In this Table, the metamagnetic field $H_m$, the temperature $T_\chi^{max}$ at the maximum of the susceptibility, the corresponding field direction, and the relaxation rate $\Gamma$ of the magnetic fluctuations are listed. Figure 3.14 presents in a log-log scale a $H_m$ versus $T_\chi^{max}$ plot of these data. It extends to a higher number ($\simeq 30$) of heavy-fermion materials the similar plots already presented in Refs. [Inoue 01, Fukuhara 96, Sugiyama 02, Õnuki 04, Takeuchi 10, Hirose 11], and is restricted here to paramagnets [Aoki 13]. A similar $H_c$ versus $T_N$ plot will be shown for heavy-fermion anti-ferromagnets in Chapter 4. A striking feature of Figure 3.14 is that, for most of heavy-fermion paramagnets, $H_m$ and $T_\chi^{max}$ are related by a simple correspondence 1 T $\leftrightarrow$ 1 K. Although $H_m$ and $T_\chi^{max}$ vary by more than two decades, from 0.57 T and 0.32 K, respectively, in YbCo$_2$Zn$_{20}$ (**H** $\parallel$ [110]) to > 50 T and > 50 K, respectively, in CeRh$_2$Si$_2$ and URu$_2$Si$_2$ under pressure (**H** $\parallel$ **c**), most of the compounds considered here are characterized by the empirical rule:

$$H_m \propto T_\chi^{max} \tag{3.9}$$

which indicates that, at first approximation, both quantities are controlled by a single energy scale. UCoAl constitutes an exception to this rule: in a magnetic field applied along its easy magnetic axis **c**, it is characterized by small values of $H_m$ in comparison with the values of $T_\chi^{max}$. This deviation may result from the proximity of UCoAl to a ferromagnetic phase transition (see Section 5.3.1.2).

Table 3.1: Temperature of the maximum (or kink) in the magnetic susceptibility, magnetic field of the transition, field direction, and inelastic-neutron-scattering relaxation rate in various heavy-fermion paramagnets.

| Material | $T_\chi^{max}$ (K) | $H_m$ (T) | **H** ∥ | Relaxation rate $\Gamma$ (K) | |
|---|---|---|---|---|---|
| CeRu$_2$Si$_2$ | 10 | 7.8 | **c** | 10 ($T \to 0$) | ♯a |
| CeRu$_2$Si$_2$ ($p = 1 \to 2$ kbar) | 11.7 → 14 | 10.3 → 13 | **c** | - | ♯b |
| Ce$_{1-x}$Y$_x$Ru$_2$Si$_2$ ($x = 1.5 \to 10$ %) | 12.5 → 25 | 9.5 → 19.3 | **c** | - | ♯c |
| Ce$_{1-x}$La$_x$Ru$_2$Si$_2$ ($x = 3 \to 7.5$ %) | 7.4 → 5 | 6.2 → 4 | **c** | - → 2.5 ($T \to 0$) | ♯d |
| CeRu$_2$(Si$_{1-x}$Ge$_x$)$_2$ ($x = 3.5 \to 7$ %) | 6 → 3.8 | 6 → 4.2 | **c** | - | ♯e |
| CeFe$_2$Ge$_2$ | (kink at $\simeq$ 20 K) | 30 | **c** | - | ♯f |
| CeNi$_2$Ge$_2$ | 28 | 42 (powder) | **c** | 40 ($T \to 0$) | ♯g |
| CeIrIn$_5$ | (kink at $\simeq$ 30 K) | 30-40 | **c** | 14 ($\langle \mathbf{q} \rangle, T = 8$ K) | ♯h |
| CeCu$_6$ | (kink at $\simeq 1 - 1.5$ K) | 1.7 | **c** | 2.5 ($T \to 0$) | ♯i |
| CeTiGe | 25 | 12.5 | (polycrystal) | - | ♯j |
| CeFePO | 5 | 4 | **c** | - | ♯k |
| CeRh$_2$Si$_2$ ($p = 0.85 \to 1.2$ GPa) | 42 → 58 | 40.2 → 51.6 | **c** | - | ♯l |
| YbIr$_2$Zn$_{20}$ | 7.4 | 9.7 (12) | [100] ([110]) | - | ♯m |
| YbRh$_2$Zn$_{20}$ | 5.3 | 6.4 | [100] | - | ♯n |
| YbCo$_2$Zn$_{20}$ | 0.32 | 0.6 (0.57) | [100] ([110]) | - | ♯o |
| YbCu$_{5-x}$Ag$_x$ ($x = 0 \to 1$) | 17 → 40 | 10 → 40 | (cubic) | - | ♯p |
| UPt$_3$ | 20 | 20 | $\perp$ **c** | 30 | ♯q |
| USn$_3$ | (kink at $\simeq 15 - 20$ K) | 30 | **a** | 60 ($\langle \mathbf{q} \rangle$) | ♯r |
| UTe$_2$ ($p = 1$ bar → 0.6 GPa) | 35 → 27 | 35 → 25 | **b** | - | ♯s |
| URu$_2$Si$_2$ ($p = 1$ bar → 1.6 GPa) | 55 → 72 | 36.5 → 41.3 | **c** | 50 → - ($T \geq T_0$) | ♯t |
| UIr$_2$Zn$_{20}$ | 5.2 | 2 | **c** | - | ♯u |
| UCo$_2$Zn$_{20}$ | 8.5 | 7.5 | **c** | - | ♯u |
| UCoAl ($p = 1$ bar → 1.2 GPa) | 20 → 28 | 0.65 → 3.6 | **c** | - | ♯v |

♯a [Paulsen 90, Fisher 91, Raymond 07a, Knafo 09a], ♯b [Mignot 88], ♯c [Park 95, Haen 95],
♯d [Fisher 91, Knafo 04, Matsumoto 08, Matsumoto 10], ♯e [Matsumoto 11], ♯f [Ebihara 95, Sugawara 99], ♯g [Fukuhara 96, Fåk 00],
♯h [Takeuchi 01, Palm 03, Willers 10], ♯i [Rossat-Mignod 88, Schröder 92, von Löhneysen 93], ♯j [Deppe 12], ♯k [Kitagawa 11],
♯l [Muramatsu 99, Knafo 17], ♯m [Takeuchi 10], ♯n [Hirose 11], ♯o [Hirose 11, Takeuchi 11], ♯p [Sarrao 99, Tsujii 01],
♯q [Frings 84, Frings 85, de Visser 87a, Bernhoeft 95], ♯r [Loewenhaupt 90, Sugiyama 02], ♯s [Knafo 19a, Miyake 19, Knebel 20],
♯t [Broholm 91, Inoue 01, Pfleiderer 06, Scheerer 12, Knafo 20a], ♯u [Hirose 15], ♯u [Mushnikov 99].

The magnetic fluctuations have been investigated by neutron scattering in the zero-field ground state of several heavy-fermion paramagnets. Table 3.1 lists the values of the low-temperature relaxation rate $\Gamma = \Gamma(\mathbf{k})$ of antiferromagnetic fluctuations (of wavevector $\mathbf{k}$) for the systems Ce$_{1-x}$La$_x$Ru$_2$Si$_2$ with $x \leq 7.5$% [Knafo 04, Knafo 09a], CeNi$_2$Ge$_2$ [Fåk 00], CeCu$_6$ [Rossat-Mignod 88], UPt$_3$ [Bernhoeft 95] and URu$_2$Si$_2$ [Broholm 91] (for CeNi$_2$Ge$_2$ a low-energy-excitation reported with an energy scale of 0.6 meV ↔ 6 K and identified as the origin of a low-temperature non-Fermi-liquid upturn of the heat capacity [Kadowaki 03] is not considered here). The relationship:

$$\Gamma \sim T_\chi^{max} \tag{3.10}$$

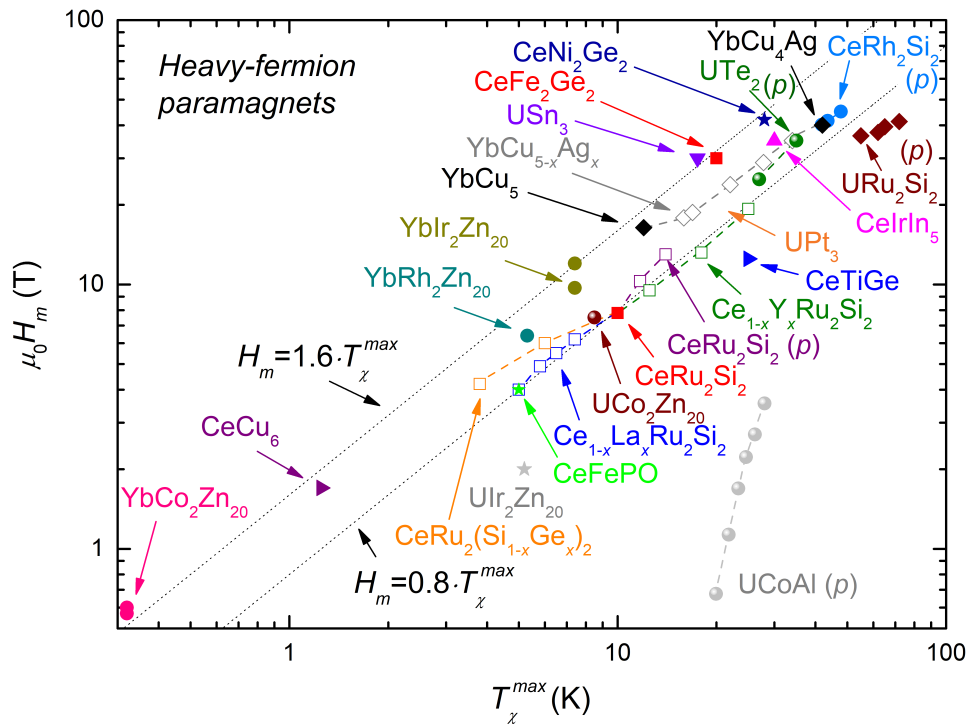

Figure 3.14: Metamagnetic field as a function of the temperature at the maximum (or kink) of the magnetic susceptibility for various heavy-fermion paramagnets (data from [Paulsen 90, Fisher 91, Mignot 88, Park 95, Haen 95, Fisher 91, Knafo 04, Matsumoto 08, Matsumoto 10, Matsumoto 11, Ebihara 95, Sugawara 99, Fukuhara 96, Takeuchi 01, Palm 03, Schröder 92, von Löhneysen 93, Deppe 12, Kitagawa 11, Muramatsu99, Knafo 17, Takeuchi 10, Hirose 11, Takeuchi 11, Sarrao 99, Tsujii 01, Frings 84, Frings 85, de Visser 87a, Sugiyama 02, Knafo 19a, Miyake 19, Inoue 01, Pfleiderer 06, Scheerer 12, Knafo 20a]).

indicates that $T_\chi^{max}$ is a temperature characteristic of the antiferromagnetic fluctuations. In most of the systems considered here [Knafo 04, Knafo 09a, Fåk 00, Rossat-Mignod 88, Broholm 91], the antiferromagnetic fluctuations progressively set up when the temperature is decreased for $T > T_\chi^{max}$, as indicated by the strong decrease of $\Gamma$, and they saturate for $T \lesssim T_\chi^{max}$, where $\Gamma$ reaches its minimal value (see Figure 2.32 in Section 2.4.3.2). The low-temperature fall of $\Gamma(\mathbf{k})$ is generally accompanied by an increase of the static susceptibility $\chi'(\mathbf{k}) \simeq 1/\Gamma(\mathbf{k})$ (see [Kuramoto 87, Knafo 09a]). At temperatures below the Fermi-liquid temperature $T^*$, while ferromagnetic fluctuations with wavevector $\mathbf{k} = 0$ would lead to an enhancement of the static susceptibility $\chi = M/H = \chi'(\mathbf{q} = 0)$ (see Equation 2.12 in Section 2.4.3.1), antiferromagnetic fluctuations with wavevector $\mathbf{k} \neq 0$ lead to an enhancement of the staggered susceptibility $\chi'(\mathbf{k})$. The enhancement of $\chi'(\mathbf{k})$ is accompanied by a downwards deviation of $\chi'(\mathbf{q})$ at wavevectors $\mathbf{q}$ characteristic of single-site effects, i.e., distant from $\mathbf{k}$, and thus, generally by a downward deviation of $\chi = \chi'(\mathbf{q} = 0)$ ending in a susceptibility maximum at a tem-perature $T_\chi^{max} \simeq T^*$ (see Figure 2.29 in Section 2.4.3.1). In CeIrIn$_5$ [Willers 10] and USn$_3$ [Loewenhaupt 90], data were collected and summed over a large volume of the reciprocal space. For these two systems, the extracted relaxation rate, corresponding to that of an 'averaged' wavevector $\langle \mathbf{q} \rangle$, has also a similar value than $T_\chi^{max}$ (see Table 3.1), suggesting that a larger contribution from intersite magnetic fluctuations than from the local single-site fluctuations. As emphasized in [Scheerer 12], the case of URu$_2$Si$_2$ above its hidden-order temperature $T_0 = 17.5$ K presents similarities with the situation reported for CeRu$_2$Si$_2$, since the extrapolation of neutron data (measured up to 17 T) is compatible with a loss of antiferromagnetic correlations above $\mu_0 H_m \simeq 35$ T [Bourdarot 03] (see also Chapter 6).

In Section 3.2.2, the CPM regime delimited by $T_\chi^{max}$ and $H_m$ in CeRu$_2$Si$_2$ was shown to be associated with antiferromagnetic correlations, which suddenly collapse at the metamagnetic boundary $H_m$. When crossing $T_\chi^{max}$ or $H_m$, no drastic change of local ($\mathbf{q}$-independent) Kondo magnetic fluctuations has been observed [Raymond 99a, Raymond 07a, Knafo 09a], which indicates no sudden modification of the single-site Kondo effect and, thus, no Kondo breakdown at $T_\chi^{max}$ or $H_m$. The empirical laws from Equations 3.9 and 3.10 followed by a large number of materials permit to suggest that, by analogy with the textbook CeRu$_2$Si$_2$ case, the CPM regime delimited by $T_\chi^{max}$ and $H_m$ in heavy-fermion paramagnets may be controlled by short-range antiferromagnetic correlations, which progressively vanish at temperatures $T > T_\chi^{max}$ under zero magnetic field, and suddenly vanish under magnetic fields $H > H_m$ at low temperatures. This picture is compatible with models presented in Section 2.4.3.2, in which a Fermi-liquid is controlled by quantum magnetic fluctuations. One can further extend Equation 2.26 to the more general formula:

$$m^* \sim \frac{1}{\Gamma(\mathbf{k})} \sim \frac{C_p}{T} \sim \sqrt{\frac{1}{T_1 T}} \sim \frac{1}{T_\chi^{max}} \sim \frac{1}{H_m}, \tag{3.11}$$

confirming the pertinence of a Fermi-liquid picture based on intersite magnetic fluctuations to describe the CPM regime of heavy-fermion paramagnets.

Alternatively to the phenomenological description proposed here, where intersite magnetic fluctuations drive the CPM regime and, thus, control its boundaries $T_\chi^{max}$ and $H_m$ in heavy-fermion paramagnets, a picture where $T_\chi^{max}$ may be related with the onset of a Kondo hybridization between $f$- and conduction electrons has been proposed (see for instance [Inoue 01, Sugiyama 02, Ōnuki 04, Takeuchi 10, Hirose 11]). In this picture, the maximum of the magnetic susceptibility corresponds to a crossover from a high-temperature regime where the $f$-electrons are localized to a low-temperature regime where they are itinerant. This hypothesis is compatible with the assumption proposed in Ref. [Kitagawa 11] that the metamagnetic field $H_m$ in CeFePO is associated with a breakdown of the Kondo effect, or similarly with the proposition that $H_m$ in CeRu$_2$Si$_2$ is driven by a reconstruction, from a Fermi surface with itinerant $4f$ electrons for $H < H_m$ to a Fermi surface with localized $4f$ electrons for $H > H_m$ [Takashita 96, Aoki 14]. These hypotheses assume that $T_\chi^{max}$ and $H_m$ are respectively related to the temperature- and magnetic field-induced destruction of the Kondo hybridization, as expected from a single-impurity Kondo model [Rajan 82, Rajan 83]. They also imply that a drastic reduction of $\mathbf{q}$-independent Kondo magnetic fluctuations should occur for $T > T_\chi^{max}$ and $H > H_m$, respectively.

Beyond these pictures, theories based on a single-site hypothesis [Wohlfarth 62, Evans 92, Hanzawa 97, Ōno 98], as well as theories considering intersite magnetic correlations [Konno 91, Yamada 93, Takahashi 98, Satoh 98], have been developed to describe metamagnetism in heavy-fermion paramagnets. However, a complete theory describing all properties of heavy-fermion paramagnets in their CPM regime and at its boundaries, i.e., the interplay between antiferromagnetic and ferromagnetic fluctuations, their relations with the CPM boundaries $T_\chi^{max}$ and $H_m$, with the effective mass $m^*$, but also with the Fermi surface, the cyclotron masses $m_i^c$ of the observed orbits, and the effective masses $m_b^*$ of each band would be needed for a deeper understanding of heavy-fermion physics.

# Chapter 4

# Antiferromagnetism

This Chapter focuses on antiferromagnets and nearly-antiferromagnets. After an introduction to the basic properties of heavy-fermion antiferromagnets (Section 4.1), the characteristics of quantum antiferromagnetic phase transitions (Section 4.2) and the effects of a magnetic field on heavy-fermion antiferromagnets (Section 4.3) are presented. Complementarily, a general introduction to antiferromagnetism can be found in [Chikazumi 97].

## 4.1 Antiferromagnetic ground state

Basic properties of heavy-fermion antiferromagnets are presented here. Bulk thermodynamic properties (Section 4.1.1) and the microscopic signatures of long-range magnetic order and intersite magnetic fluctuations developing in the antiferromagnetic ground state of these systems (Section 4.1.2) are displayed for a selection of compounds.

### 4.1.1 Bulk properties

Figure 4.1 shows the magnetic susceptibility [Settai 07] and heat capacity divided by temperature [Rosch 97, Graf 98, Sheikin 02, Custers 03, Raymond 10] versus temperature for a selection of heavy-fermion antiferromagnets, for which the Néel transition temperature $T_N$ covers a broad range of values, from 80 mK in YbRh$_2$Si$_2$ [Rosch 97] to 36.5 K in CeRh$_2$Si$_2$ [Graf 98]. Figure 4.1(a) compares the magnetic susceptibility $\chi$ of several antiferromagnets in a low magnetic field applied along an easy magnetic direction, i.e., a direction along which the antiferromagnetic moments are aligned. In these systems, long-range magnetic ordering leads to a kink or a sharp maximum in $\chi$ at $T_N$. $\chi$ is reduced at temperatures $T < T_N$ for all considered compounds. The strongest reduction of $\chi$ at low-temperature, where it reaches a value comparable to its room-temperature value, is observed in CeRh$_2$Si$_2$, the antiferromagnet with the highest $T_N$ of the selection presented here. In several materials, as CeRhIn$_5$, CeIrSi$_3$, CePt$_3$Si, but also CeRh$_2$Si$_2$, a kink in $\chi$ is observed at $T_N$ and a broad maximum in $\chi$ is observed at a temperature $T_\chi^{max} > T_N$. This maximum is the signature of a high-temperature correlated-paramagnetic regime similar to that occurring in quantum correlated paramagnets (in which the CPM regime develops at temperatures down to $T \to 0$, see Section 3). Figure 4.1(b) compares the electronic heat capacity divided by temperature of several antiferromagnets at zero magnetic field [Rosch 97, Graf 98, Sheikin 02, Custers 03, Raymond 10]. In decreasing temperatures, a sharp step-like increase of $C_p/T$ is observed at $T_N$ for all compounds, indicating the second-order nature of the antiferromagnetic transition. It is followed either by a decrease (YbRh$_2$Si$_2$, CePd$_2$Si$_2$, and CeRh$_2$Si$_2$) or by a saturation (CeCu$_{5.8}$Au$_{0.2}$ and Ce$_{0.87}$La$_{0.13}$Ru$_2$Si$_2$) of $C_p/T$ at temperatures $T < T_N$. For $T \to 0$, a saturation of the Sommerfeld coefficient $\gamma =_{T \to 0} C_p/T$ indicates the presence of quantum magnetic fluctuations. These fluctuations are more intense, as shown by a larger $\gamma$, in compounds for which $T_N$ is smaller. They indicate the proximity of a quantum magnetic phase transition, at which long-range antiferromagnetic order is destabilized. From the integration of $C_p/T$ versus $T$ data, a magnetic entropy $S_m \leq R\ln 2$ can be extracted for these systems, whose low-temperature electronic properties are, thus, those of a $f$-electron doublet.

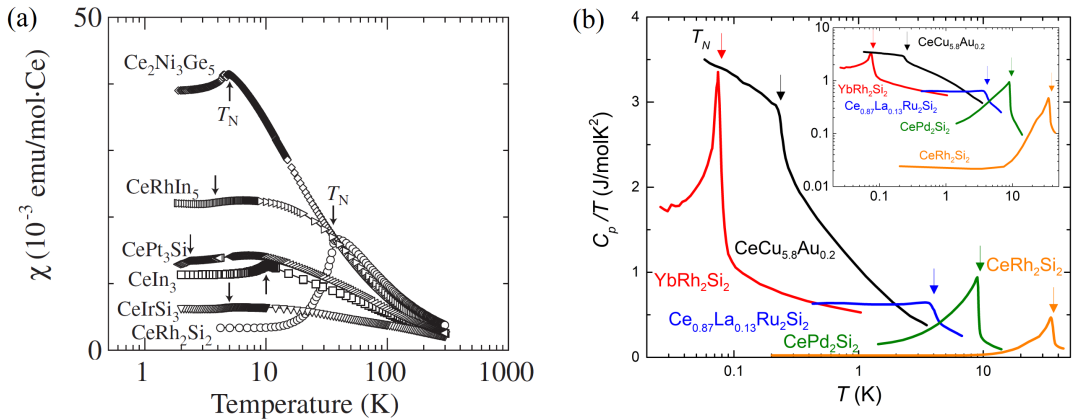

Figure 4.1: (a) Magnetic susceptibility versus temperature in a magnetic field applied along the easy magnetic axis (from [Settai 07]) and (b) heat capacity divided by temperature versus temperature for a selection of heavy-fermion antiferromagnets (data from [Rosch 97, Graf 98, Sheikin 02, Custers 03, Raymond 10]). The non-electronic background has been subtracted to the heat capacity.

Magnetic anisotropy plays an important role in heavy-fermion antiferromagnets (see Section 2.4.1.1). Figure 4.2 shows the magnetic susceptibility versus temperature measured with a magnetic field applied along different crystallographic directions for three antiferromagnets: CeRh$_2$Si$_2$ [Settai 97], CeRhIn$_5$ [Takeuchi 01], and CePd$_2$Si$_2$ [van Dijk 00].

- Figure 4.2(a) shows that CeRh$_2$Si$_2$ has a strong Ising magnetic anisotropy, revealed by $\chi_c(T = 300\,\text{K})/\chi_a(T = 300\,\text{K}) > 2$ and $\chi_c(T_N)/\chi_a(T_N) > 5$ [Settai 97]. The high-temperature anisotropy is compatible with an almost pure $|\pm 5/2\rangle$ groundstate, i.e., weakly-mixed with $|\pm 3/2\rangle$. For $T < T_N$, while $\chi_a$ is almost unchanged, $\chi_c$ is strongly reduced when $T$ is decreased, ending in $\chi_c \gtrsim \chi_a$ for $T < 10\,\text{K}$, and suggesting the alignment of antiferromagnetic moments $\mu_{AF} \parallel \mathbf{c}$.

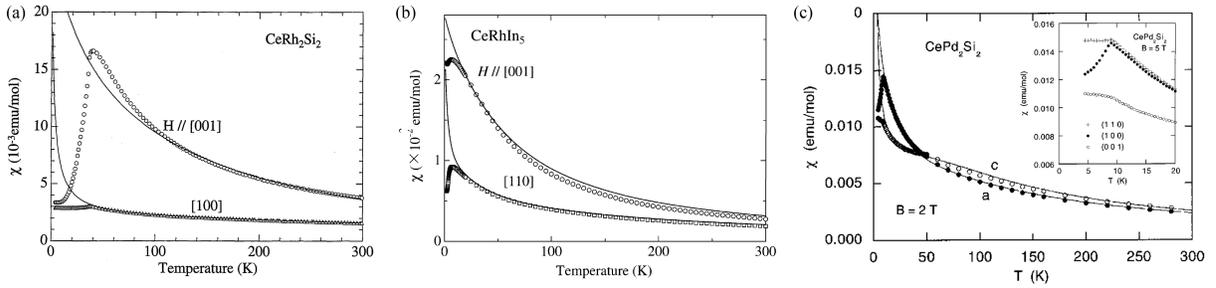

Figure 4.2: Magnetic susceptibility versus temperature, for different magnetic-field directions, of (a) CeRh$_2$Si$_2$ (from [Settai 97]), (b) CeRhIn$_5$ (from [Takeuchi 01]), and (c) CePd$_2$Si$_2$ (from [van Dijk 00]).

- Figure 4.2(b) shows that CeRhIn$_5$ has a Ising magnetic anisotropy, revealed by $\chi_c(T = 300 \text{ K})/\chi_a(T = 300 \text{ K}) > 1.3$ and $\chi_c(T_N)/\chi_{[110]}(T_N) > 2$, i.e., weaker than that of CeRh$_2$Si$_2$ [Takeuchi 01]. The crystal-field groundstate of CeRhIn$_5$ may consist in a state made of mixed $|\pm5/2\rangle$ and $|\pm3/2\rangle$ terms. The reduction of $\chi_{[110]}$ and the small variation of $\chi_c$ at temperatures $T < T_N$ are the indication for antiferromagnetic moments $\mu_{AF} \parallel$ [110].

- Figure 4.2(c) shows that CePd$_2$Si$_2$ has a weak Ising magnetic anisotropy at high temperature, with $\chi_c(T = 300 \text{ K})/\chi_a(T = 300 \text{ K}) \gtrsim 1$ and $\chi_c(T = 100 \text{ K})/\chi_a(T = 100 \text{ K}) \gtrsim 1.1$ [van Dijk 00]. This highlights that the ground state of CePd$_2$Si$_2$ strongly deviates from pure $|\pm5/2\rangle$, implying a large $|\pm3/2\rangle$ term. A crossing of $\chi_c$ and $\chi_a$ at $T \simeq 50$ K, which presumably results from the onset of anisotropic magnetic correlations, leads to a reversal of the magnetic anisotropy ending with $\chi_c(T_N)/\chi_a(T_N) \simeq 0.7$. The large decrease of $\chi_a(T)$ and the saturation of $\chi_c(T)$ and $\chi_{110}(T)$ for $T < T_N$ indicate antiferromagnetic moments $\mu_{AF} \parallel$ a.

High-temperature fits to the magnetic susceptibility led to quasi-pure $|\pm5/2\rangle$ ground-state doublets in CeRh$_2$Si$_2$ [Settai 97], CeRhIn$_5$ [Takeuchi 01], and to a ground-state doublet dominated by $|\pm3/2\rangle$ in CePd$_2$Si$_2$ [van Dijk 00]. However, different electronic schemes were extracted from inelastic neutron scattering and x-ray absorption spectroscopy studies for these three systems at low temperature [Steeman 90, Christianson 04, Hansmann 08, Willers 12]. A change of the magnetic anisotropy driven by the onset of magnetic exchange interactions may be responsible for a different electronic ground state at low temperature.

### 4.1.2 Magnetic order and fluctuations

Neutron diffraction, i.e., elastic neutron scattering, permits to probe long-range magnetic ordering at a given wavevector **k**. Figure 4.3 presents neutron-diffraction experiments performed on a selection of heavy-fermion antiferromagnets having a strong (Ce$_{0.87}$La$_{0.13}$Ru$_2$Si$_2$ and CeRh$_2$Si$_2$), intermediate (CeRhIn$_5$), or small (CePd$_2$Si$_2$) magnetic anisotropy, and for which antiferromagnetic order is established with the moments aligned either $\parallel \mathbf{c}$ or $\perp \mathbf{c}$, and with commensurate or incommensurate wavevectors. Magnetic structures associated with an incommensurate wavevector generally correspond to a spin-density-wave, while magnetic structures with a commensurate wavevector $\mathbf{k} \neq 0$ (and sometimes $\mathbf{k} = 0$, in the case of a lattice unit cell containing two magnetic ions) correspond to an antiferromagnetic order. In the following we will label by antiferromagnetism both kinds of staggered magnetic order, and their characteristic magnetic fluctuations, i.e., those corresponding to short-range magnetic order with the same wavevector and ordering direction, will be called antiferromagnetic fluctuations.

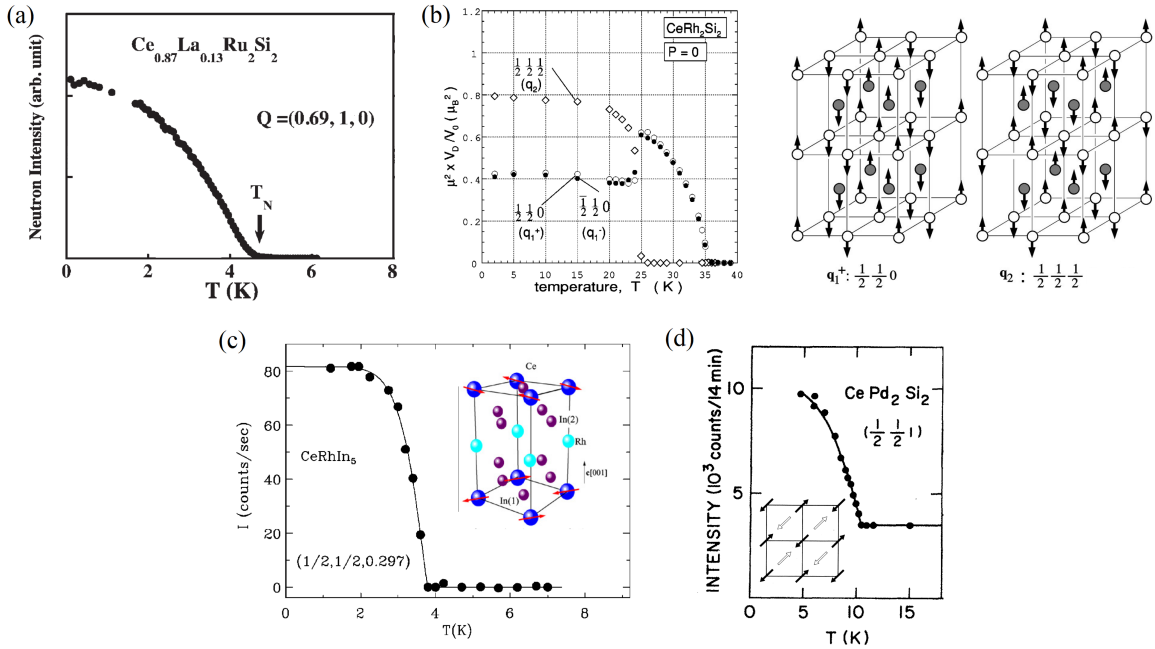

Figure 4.3: Neutron diffracted intensity at the magnetic Bragg peaks developing below the Néel temperature $T_N$ of the heavy-fermion antiferromagnets (a) $Ce_{0.87}La_{0.13}Ru_2Si_2$ (from [Raymond 10]), (b) $CeRh_2Si_2$ (from [Kawarazaki 00]), (c) $CeRhIn_5$ (from [Bao 00]), and (d) $CePd_2Si_2$ (from [Grier 84]). Sketchs of the magnetic structure corresponding to the different wavevector-contributions are shown in the right panel of (b), as inset of (c), and a projection of the magnetic structure in the plane $\perp \mathbf{c}$ is shown as inset of (d).

- Figure 4.3(a) shows the temperature-dependence of the neutron diffracted intensity on $Ce_{0.87}La_{0.13}Ru_2Si_2$, at the magnetic Bragg peak of momentum transfer $\mathbf{Q}_1 = (0.69, 1, 0)$ corresponding to the incommensurate wavevector $\mathbf{k}_1 = (0.31, 0, 0)$ [Raymond 10] (see Equation 2.14 in Section 2.4.3.1). From the increase of scattered intensity $\Delta I(\mathbf{Q}_1) \propto \mu(\mathbf{k}_1)^2$ below the Néel temperature $T_N = 4.4$ K, one can extract the value of the antiferromagnetic moment $\mu(\mathbf{k}_1) \simeq 1\mu_B /$ Ce stabilized along the easy magnetic axis $\mathbf{c}$ at low temperature [Quezel 88]. In this strongly anisotropic system, for which $\chi_c(T_N)/\chi_a(T_N) \simeq 50$ [Haen 92], the magnetic order direction corresponds to the easy magnetic axis fixed

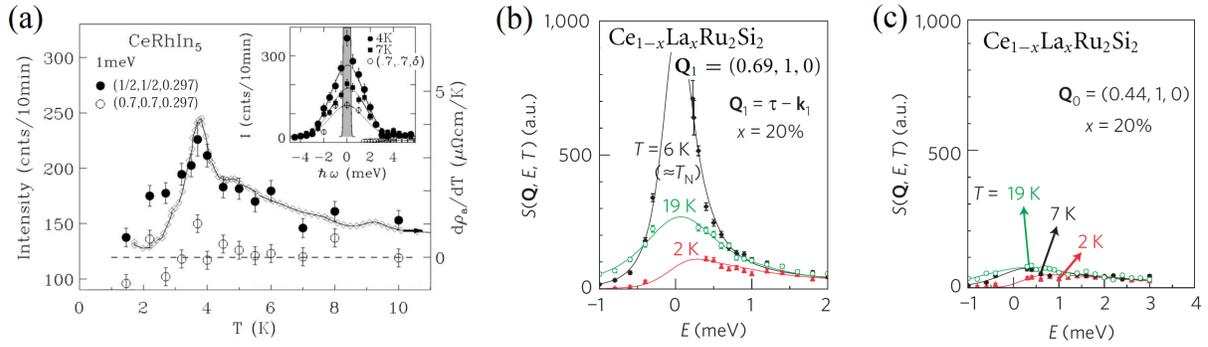

Figure 4.4: Temperature-dependence of magnetic fluctuations probed by neutron inelastic scattering in heavy-fermion antiferromagnets. (a) Temperature dependence of neutron intensity at the energy $E = 1$ meV in CeRhIn$_5$, at the antiferromagnetic momentum transfer $\mathbf{Q} = (0.5, 0.5, 0.297)$ and at the momentum transfer $\mathbf{Q} = (0.7, 0.7, 0.297)$ representative of the $\mathbf{Q}$-independent background, compared with the temperature-derivative $\partial\rho/\partial T$ of the electrical resistivity, the Inset showing the energy spectra at these two momentum transfers, for different temperatures (from [Bao 02]). Energy spectra in Ce$_{0.8}$La$_{0.2}$Ru$_2$Si$_2$, at different temperatures, (b) at the antiferromagnetic momentum transfer $\mathbf{Q}_1 = (0.69, 1, 0)$ and (c) at the momentum transfer $\mathbf{Q}_0 = (0.44, 1, 0)$ representative of $Q$-independent local magnetic fluctuations (from [Knafo 09a]).

from the crystal-field.

- The left part of Figure 4.3(b) presents the temperature-dependence of the neutron diffracted intensity on CeRh$_2$Si$_2$, at magnetic Bragg peaks of commensurate wavevectors $\mathbf{q}_1^{\pm} = (\pm 1/2, 1/2, 0)$ and $\mathbf{q}_2 = (1/2, 1/2, 1/2)$ [Kawarazaki 00]. CeRh$_2$Si$_2$ presents two antiferromagnetic phases: AF1 for $T_{1,2} = 26.5$ K $< T < T_N = 36.5$ K, where long-range magnetic order was identified as composed of two single-$\mathbf{k}$ domains of wavevectors $\mathbf{q}_1^{\pm}$, and AF2 for $T < T_{1,2}$, where long-range magnetic order was identified as a multiple-$\mathbf{k}$ phase, made of the superposition of oscillating magnetic moments with wavevectors $\mathbf{q}_1^{\pm}$ and $\pm\mathbf{q}_2$ [Kawarazaki 00]. In both phases, the magnetic moments are aligned along the direction $\mathbf{c}$, which corresponds to the easy magnetic axis fixed from the crystal-field. The right part of Figure 4.3(b) shows sketches of the magnetic-moment components with wavevectors $\mathbf{q}_1^+$ and $\mathbf{q}_2$.

- Figures 4.3(c-d) present the neutron diffracted intensities and sketches of the magnetic structure, for CeRhIn$_5$ with the incommensurate wavevector $\mathbf{k} = (1/2, 1/2, 0.297)$ [Bao 00] and CePd$_2$Si$_2$ with the commensurate wavevector $\mathbf{k} = (1/2, 1/2, 0)$ [Grier 84], respectively. For both systems, magnetic moments $\mu_{AF} \perp \mathbf{c}$ are ordered within a direction perpendicular to the high-temperature easy magnetic axis $\mathbf{c}$ driven by crystal field effects. This indicates that the magnetic-ordering direction results from another mechanism, presumably an anisotropic magnetic exchange, developing at low-temperature.

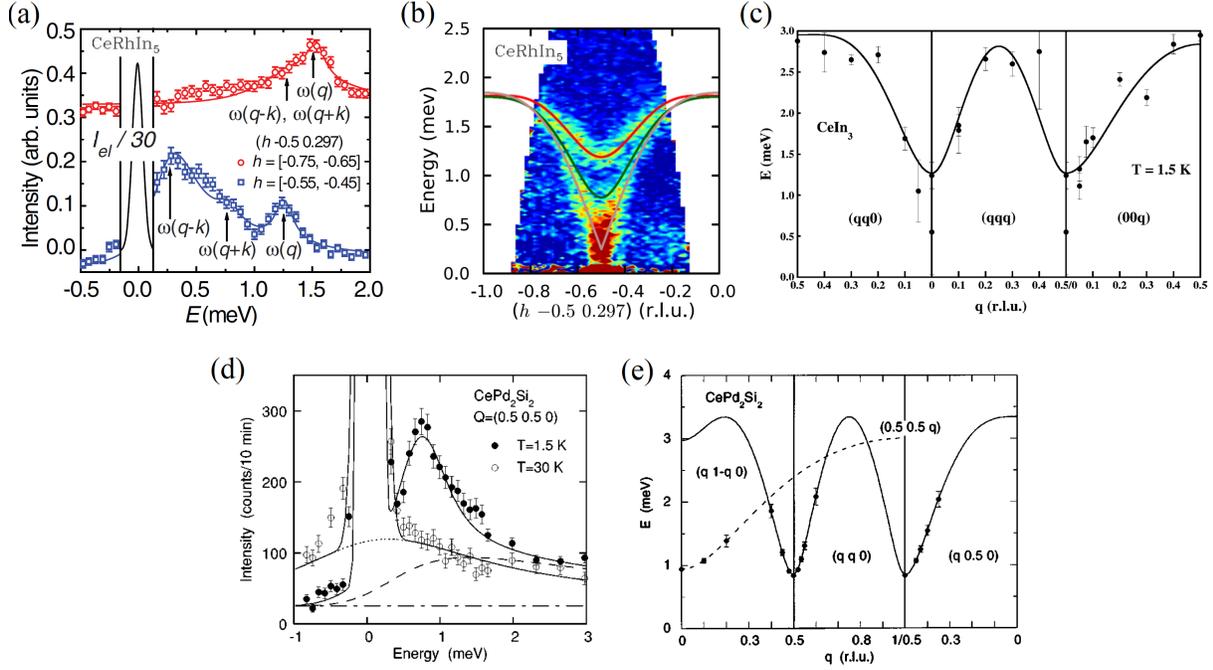

Figure 4.5: Wavevector-dispersion of magnetic fluctuations probed by neutron inelastic scattering in heavy-fermion weakly-anisotropic antiferromagnets. (a) Energy spectra at two momentum transfers close to the antiferromagnetic Bragg peak in CeRhIn$_5$ and (b) wavevector-dispersion of the observed spin-wave excitation (from [Das 14]). (c) Wavevector-dispersion of the observed spin-wave excitation observed in CeIn$_3$ (from [Knafo 03]). (d) Energy spectra in CePd$_2$Si$_2$ at different temperatures at the antiferromagnetic momentum transfer $\mathbf{Q} = (0.5, 0.5, 0)$ and (e) wavevector-dispersion of the associated spin-wave excitation (from [van Dijk 00]).

Heavy-fermion antiferromagnets are generally the place of intense magnetic fluctuations indicating their proximity from a thermal or quantum magnetic transition. Figure 4.4 presents the temperature-dependence of magnetic fluctuations spectra obtained in two antiferromagnets, highlighting the enhancement of the critical magnetic fluctuations at the Néel temperature $T_N$, onset of antiferromagnetism.

- Figure 4.4(a) shows the temperature-dependence of the inelastic neutron intensity at the antiferromagnetic momentum transfer $\mathbf{Q} = (0.5, 0.5, 0.297)$, measured at an energy transfer $E = 1$ meV [Bao 02]. This signal corresponds to the fluctuations of the order parameter of the antiferromagnetic phase, and it is peaked at $T_N$. Such enhancement of the fluctuations of the order parameter is expected at a second-order phase transition. For $T > T_N$, it results from the increase of the antiferromagnetic correlation length, i.e., the typical size of the short-range order, when the temperature is decreased. At $T = T_N$, the divergence of the correlation length leads to the appearance of long-range and static magnetic order (see Figures 2.28(b-d) in Section 2.4.3.1). For $T < T_N$, the magnetic fluc-tuations also grow when the temperature increases and approaches $T_N$, in relation with the progressive reduction of the ordered magnetic moment.

- Figure 4.4(b) presents the magnetic excitations spectra obtained by inelastic neutron scattering on the antiferromagnet $Ce_{0.8}La_{0.2}Ru_2Si_2$ at its antiferromagnetic wavevector $\mathbf{k}_1 = (0.31, 0, 0)$, at different temperatures [Knafo 09a]. Here again, the fluctuations of the antiferromagnetic order parameter are peaked at $T_N = 6$ K, indicating their critical role for the transition. Figure 4.4(c) presents the neutron-scattering spectra obtained on $Ce_{0.8}La_{0.2}Ru_2Si_2$ at the wavevector $\mathbf{q}_0 = (0.56, 0, 0)$ far from the antiferromagnetic wavevector and representative of a $\mathbf{q}$-independent local magnetic signal. The magnetic fluctuations with wavevector $\mathbf{q}_0$ are not peaked at $T_N$, showing that local magnetic fluctuations do not play a critical role for the transition, as expected within standard phase-transition theories [Knafo 09a].

In strongly anisotropic antiferromagnets, as $Ce_{0.8}La_{0.2}Ru_2Si_2$ (see Figure 4.4(b)) [Knafo 09a], transverse magnetic fluctuations are forbidden, and longitudinal magnetic fluctuations are associated with damped quasielastic, or nearly-quasielastic, magnetic excitations spectra. In weakly-anisotropic antiferromagnets, transverse magnetic fluctuations, historically called spin waves, can develop at temperatures $T < T_N$. Spin-waves present a dispersive wavevector-$\mathbf{k}$-dependent energy gap, whose minimum is observed at the wavevector corresponding to the antiferromagnetic order. Figure 4.5 focuses on the experimental observation of spin-waves in heavy-fermion weakly-anisotropic antiferromagnets.

- Figures 4.5(a-b) show the inelastic neutron scattering spectra obtained on $CeRhIn_5$ at two wavevectors $\mathbf{q} \simeq (-0.6, -0.5, 0.297)$ and $\mathbf{k}_{AF} \simeq (-0.45, -0.5, 0.297)$, and the corresponding wavevector-energy-dispersion, plotted simultaneously in a color plot where the intensity of the magnetic fluctuations is mapped out [Das 14] (see also [Stock 15]).

- While $CeRhIn_5$ has a tetragonal structure and is an Ising-high-temperature paramagnet becoming a weak-XY-antiferromagnet at low-temperature, $CeIn_3$ has a weak anisotropy simply resulting from its cubic structure. Complementarily to Figure 2.18(c) presented in Section 2.4.1.1, Figure 4.5(c) shows the spin-wave dispersion extracted from inelastic-neutron-scattering experiments on $CeIn_3$, where magnetic order develops with the wavevector $\mathbf{k}_{AF} = (1/2, 1/2, 1/2)$ [Knafo 03]. The similarity between the magnetic excitations of the cubic antiferromagnet $CeIn_3$ and the weakly-anisotropic antiferromagnet $CePd_2Si_2$, whose spectrum at the antiferromagnetic wavevector $\mathbf{k}_{AF} = (1/2, 1/2, 0)$ and associated spin-wave dispersion curve are shown in Figures 4.5(d,e), respectively, can be emphasized [Knafo 03, van Dijk 00]. Both compounds possess a similar Néel temperature $T_N = 10$ K, a similar ordered magnetic moment $\mu_{AF} = 0.5 - 0.6 \ \mu_B/$ Ce, a superconducting phase induced at a similar critical pressure $p_c \simeq 25 - 30$ kbar, but a Sommerfeld coefficient, a superconducting temperature, and a crystal-field gap twice bigger in $CePd_2Si_2$ than in $CeIn_3$. This suggests a possible relation between the strength of planar magnetic anisotropy, the development of enhanced magnetic fluctuations, and that of superconductivity in quantum-critical heavy-fermion systems.

## 4.2 Quantum antiferromagnetic phase transitions

Quantum phase transitions, i.e., transitions in the limit of $T \to 0$, between an antiferromagnetic and a paramagnetic phase can be induced by pressure or chemical doping in most of heavy-fermion systems. After an introduction to their basic thermodynamic and electrical-transport properties, Section 4.2.1 presents the electronic phase diagrams of these quantum critical systems, where the destabilization of long-range magnetic order often coincides with the appearance of a superconducting phase. In Section 4.2.2, the quantum-critical role of intersite magnetic fluctuations, which are enhanced at the quantum antiferromagnetic phase transition, is emphasized. Section 4.2.3 shows that Fermi-surface reconstructions coincide with the destabilization of antiferromagnetism in these systems. Finally, Section 4.2.4 shows that smooth valence variations can be observed at the quantum antiferromagnetic phase transition of heavy-fermion systems.

### 4.2.1 Bulk properties of phase diagrams

Heavy-fermion systems undergo quantum antiferromagnetic phase transitions which can be tuned by chemical doping or under pressure. Their magnetic phase diagrams show multiple magnetic phases, and sometimes superconductivity developing in the vicinity of the magnetic phase transition. In this Section, the bulk properties and the typical electronic phase diagrams of a selection of heavy-fermion systems crossing a quantum antiferromagnetic phase transition are presented.

Complementarily to Figures 2.19(c) (heat capacity of $Ce_{1-x}La_xRu_2Si_2$) and 2.26(a-b) (heat capacity of $CeCu_{6-x}Au_x$) shown in Sections 2.4.1.1 and 2.4.2.2, respectively, Figure 4.6 shows a set of basic properties reported on heavy-fermion quantum critical antiferromagnets.

- Figure 4.6(a) presents the heat capacity divided by temperature of $Ce_{1-x}La_xRu_2Si_2$ alloys, emphasizing a critical enhancement of $\gamma =_{T \to 0} C_p/T$ at $x_c = 7.5$ [Raymond 10]. For $x < x_c$, a Fermi-liquid correlated-paramagnetic regime is established at low temperatures. For $x > x_c$, a large second-order-like anomaly in the heat capacity at $T_N = 4.4$ K for $x = 13\%$ and $T_N = 6$ K for $x = 20\%$ is the signature of long-range magnetic order. A second transition, also associated with a second-order-like anomaly, and whose nature has not been clearly identifier so far, occurs at a temperature $T_L < T_N$ [Regnault 90, Jacoud 92, Raymond 10]. Figure 4.6(b) shows that a kink in the magnetic susceptibility is observed at $T_N$ for $x > x_c$, in addition to the broad maximum of the magnetic susceptibility at $T_\chi^{max}$, onset of the CPM regime [Fisher 91]. The decoupling of $T_N$ and $T_\chi^{max}$ is larger for $x$ near $x_c$, i.e., in the vicinity of the quantum phase transition.

- Figure 4.6(c) presents electrical resistivity $\rho$ versus temperature curves of $CePd_2Si_2$ under pressure, emphasizing that a non-Fermi-liquid regime $\rho \sim T^{1.2}$ occurs near the critical pressure $p \simeq 2.5 - 3$ GPa at which antiferromagnetism collapses and superconductivity appears [Grosche 96].

- Figure 4.6(d) shows the evolution under pressure of the heat-capacity divided by temperature versus temperature of $CeRhIn_5$ [Knebel 04]. Under pressure, the collapse of the large second-order-like step anomaly in $C_p$ at the Néel temperature $T_N$ is replaced by another large and second-order anomaly, at the onset of superconductivity at the temperature $T_{sc}$. At the critical pressure $p_c \simeq 2-2.5$ GPa, $T_{sc}$ is maximal and $C_p/T$ saturates to a high value indicative of a heavy-fermion regime with strong magnetic fluctuations.

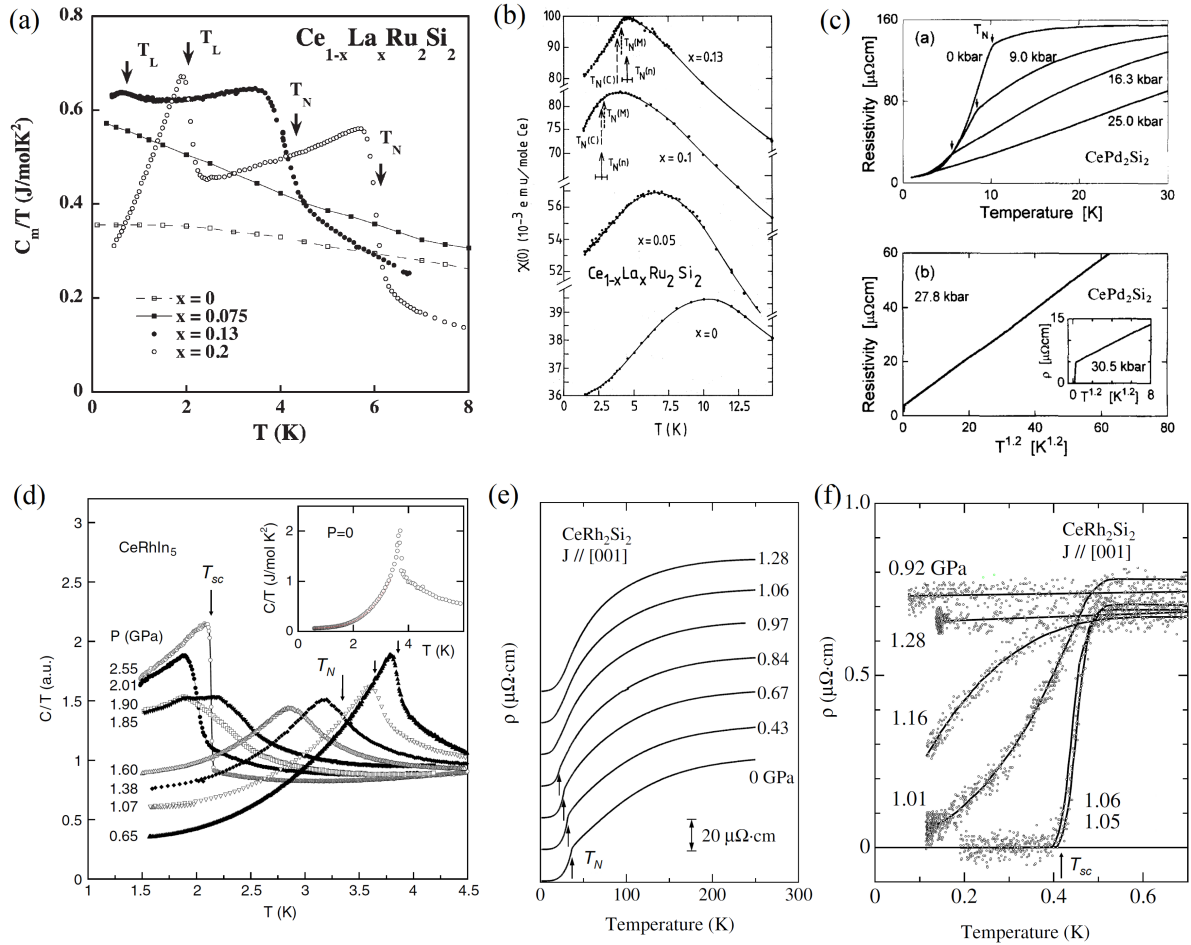

Figure 4.6: (a) Heat capacity divided by temperature (from [Raymond 10]), and (b) low-temperature magnetic susceptibility for **H**||**c** (from [Fisher 91]), versus temperature of $Ce_{1-x}La_xRu_2Si_2$ alloys. (c) Electrical resistivity versus temperature (top) and versus $T^{1.2}$ (bottom) of $CePd_2Si_2$ under pressure (from [Grosche 96]). (d) Heat capacity divided by temperature versus temperature of $CeRhIn_5$ at different pressures (from [Knebel 04]). Electrical resistivity versus temperature of $CeRh_2Si_2$ under pressure, (e) at temperatures up to 300 K, (f) at temperatures up to 0.7 K (from [Araki 02b]).

- Figure 4.6(e-f) show the electrical resistivity versus temperature of $CeRh_2Si_2$ under pressure, emphasizing the collapse at $p_c = 1$ GPa of the Néel temperature $T_N$, and of the development of a superconducting phase near $p_c$ [Araki 02b]. In this material, a Fermi-liquid $T^2$ variation of the electrical resistivity is observed for all pressures, including in the vicinity of $p_c$ at temperatures $T > T_{sc}$.

Signatures of magnetic fluctuations developing at temperatures $T > T_N$ can be seen in the heat-capacity data, where a large $C_p/T$ term precedes the establishment of long-range magnetic order (Figures 4.6(a,d)), and in the electrical resistivity, where a large resistive contribution is present (Figures 4.6(c,e)).

Figure 4.7 presents the phase diagrams of a selection of heavy-fermion antiferromagnets at their quantum antiferromagnetic phase transition. Figure 4.7(a) shows the chemical doping-temperature phase diagram of $Ce_{1-x}La_xRu_2Si_2$, where antiferromagnetism is induced with wavevector $\mathbf{k}_1 = (0.31, 0, 0)$ for $x > x_c = 7.5\%$ [Quezel 88]. The right scale of the graph indicates the magnitude of the low-temperature ordered antiferromagnetic moment, which reaches $\simeq 1.2\ \mu_B$/Ce in $Ce_{0.8}La_{0.2}Ru_2Si_2$, for which $T_N = 6$ K. Figures 4.7(b-f) presents the pressure-temperature phase diagrams of the antiferromagnets $CePd_2Si_2$ [Mathur 98], $CeRh_2Si_2$ [Araki 02b, Muramatsu99, Knafo 17], $CeIn_3$, and $CeRhIn_5$, and of the paramagnet $CeCoIn_5$ [Ōnuki 04], respectively. For the four antiferromagnets, long-range magnetic ordering disappears at pressures $p$ higher than a critical pressure $p_c$ ranging from 1 GPa in $CeRh_2Si_2$ to $2.5 - 3$ GPa in $CePd_2Si_2$, $CeRh_2Si_2$, and $CeRhIn_5$. Figure 4.7(c) shows that, for $CeRh_2Si_2$, the transition temperature $T_{1,2} < T_N$, at which a change of the antiferromagnetic structure was reported, vanishes a a pressure $p_{1,2} \simeq 0.5 - 0.6$ GPa $< p_c$ (see Figure 4.3(b) in Section 4.1.2) [Kawarazaki 00]. The phase diagram of $CeRh_2Si_2$ also includes the variation of the high-temperature scale $T_\chi^{max}$ at the onset of the correlated paramagnetic regime, showing that, for this system, the CPM regime precedes the establishment of antiferromagnetic order for $p < p_c$ and it becomes the low-temperature ground state for $p > p_c$ [Muramatsu99, Knafo 17]. The decoupling of $T_\chi^{max}$ and $T_N$ increases when $p_c$ is approached. Such decoupling is observed in a large number of heavy-fermion antiferromagnets, as $Ce_{1-x}La_xRu_2Si_2$ (see Figure 4.6(b)), where it indicates the closeness of a quantum magnetic phase transition.

Superconductivity is induced in a narrow pocket in the vicinity of the quantum magnetic phase transition of the four antiferromagnets considered in Figures 4.7(b-e). Due to the similarity of their superconducting phase boundary, $CeCoIn_5$ at ambient pressure is often considered as equivalent, or at least very similar, to its isostructural parent $CeRhIn_5$ under a pressure $p \gtrsim p_c$ [Ōnuki 04, Shishido 05, Settai 07]. The role of magnetic anisotropy for the enhancement of superconductivity was suggested by a comparison between the cubic $CeIn_3$ and the tetragonal $CePd_2Si_2$ (see discussion in Section 4.1.2 [Knafo 03]). One can also compare $CeIn_3$ and its tetragonal parents $CeRhIn_5$ and $CeCoIn_5$, built from $CeIn_3$ and $TIn_2$ (T=Rh,Co) alternating layers stacked along the direction c. The anisotropy of their magnetic properties [Hegger 00, Petrovic 01], but also that of their Fermi surfaces [Ōnuki 04, Shishido 05, Settai 07], may play a role for the enhancement of their superconducting temperature up to $T_{sc} \simeq 2.2$ K, in comparison with the maximal superconducting temperature $T_{sc} \simeq 200$ mK reported for the isotropic $CeIn_3$ compound [Walker 97].

Figure 4.8(a) presents the temperature-volume phase diagrams of $CeCu_2Si_2$, $CeCu_2Ge_2$, and $CeAu_2Si_2$ measured by electrical resistivity under pressure. The similarities of three phase diagrams suggest that $CeCu_2Si_2$ at ambient pressure is equivalent to $CeCu_2Ge_2$ and $CeAu_2Si_2$ at their quantum magnetic phase transition, where antiferromagnetic order collapses and superconductivity develops [Ren 14]. In an earlier study, Yuan et al [Yuan 03] established the

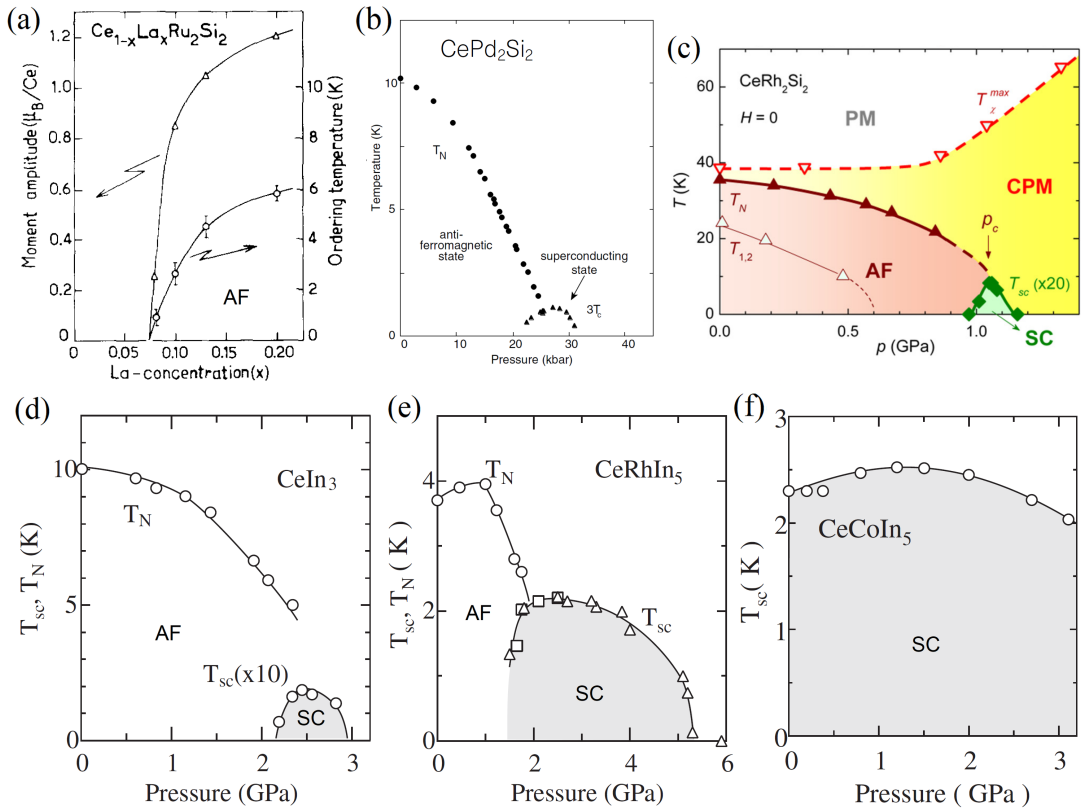

Figure 4.7: (a) Chemical doping-temperature phase diagram and variation with doping of the ordered antiferromagnetic moment $Ce_{1-x}La_xRu_2Si_2$ alloys (from [Quezel 88]), and pressure-temperature phase diagrams of (b) $CePd_2Si_2$ [Mathur 98], (c) $CeRh_2Si_2$ (from [Araki 02b, Muramatsu99, Knafo 17]), (d) $CeIn_3$, (e) $CeRhIn_5$, and (f) $CeCoIn_5$ (from [Ōnuki 04]).

chemical-doping-temperature phase diagram of $CeCu_2(Si_{1-x}Ge_x)_2$ (see Figure 2.33(c) in Section 2.5.1). They observed two separated superconducting domes in this system, the first one being connected with the fall of the antiferromagnetic phase transition $T_N$ and the second one being induced by a valence crossover to an intermediate-valent regime (see also [Holmes 04]). However, a x-ray-absorption study showed that valence in $CeCu_2Si_2$ varies smoothly under pressure, without any anomaly indicative of a clear valence crossover [Rueff 11]. As shown in Figures 4.8(b-d), two broad maxima in the electrical resistivity of these systems are observed at the temperatures $T_1^{max}$ and $T_2^{max}$ [Jaccard 98, Holmes 04, Ren 16]. While the maximum at $T_1^{max}$ can be ascribed to the electronic correlations of the $4f$-electron-doublet ground state, the maximum at $T_2^{max}$ is presumably driven by correlations associated with the first excited crystal-field state. At a pressure $p^* > p_c$, where $p_c$ is the antiferromagnetic critical pressure, the two maxima merge into a single maximum, indicating a loss of the low-energy crystal-field degeneracy, presumably due to the enhancement of the Kondo hybridization (see Section 2.4.1.1.1). The fact that $T_{sc}$ is maximum indicates that the merging of the two maxima is related with the mechanism driving to superconductivity at $p^*$.

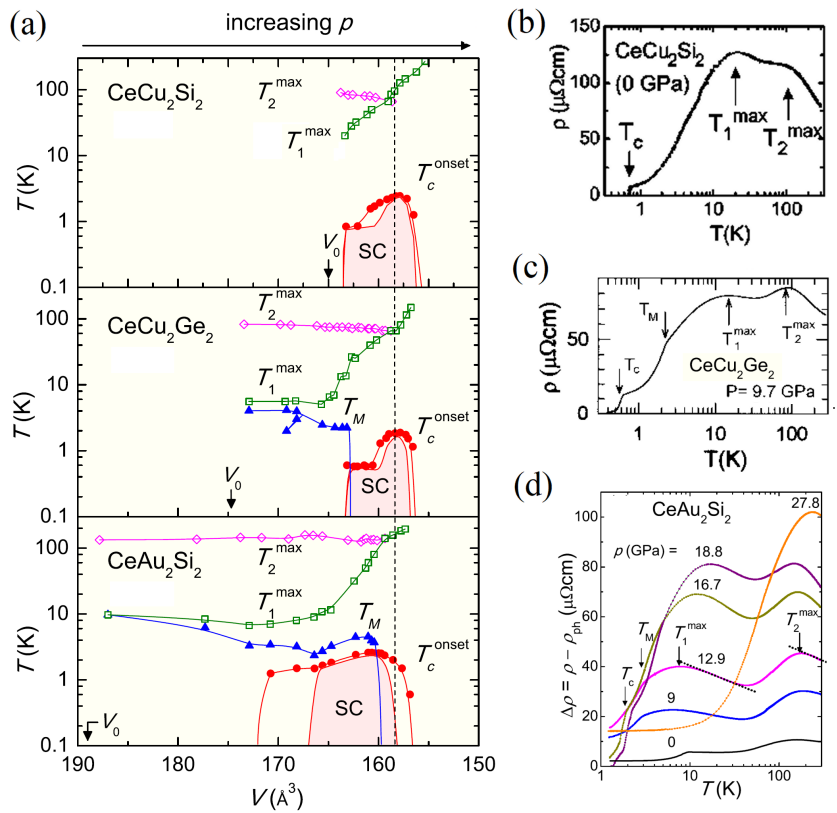

Figure 4.8: (a) Temperature-volume phase diagrams of $CeCu_2Si_2$ (top), $CeCu_2Ge_2$ (middle), and $CeAu_2Si_2$ (bottom) extracted from experiments under pressure (from [Ren 14]), electrical resistivity versus temperature of (b) $CeCu_2Si_2$ at ambient pressure (from [Holmes 04]) and (c) $CeCu_2Ge_2$ at $p = 9.7$ GPa (from [Jaccard 98]), and (d) non-phononic contribution to the electrical resistivity versus temperature of $CeAu_2Si_2$ at different pressures (from [Ren 16]).

Similar merging at a pressure $p^*$ of two high-temperature maxima in the electrical resistivity, have been observed in other heavy-fermion compounds, as $CeRhIn_5$ [Ren 17] and $CePt_2In_7$ [Sidorov 13], where superconductivity develops at an antiferromagnetic quantum phase transition at a pressure $p_c$ coinciding with $p^*$. Understanding the respective direct or indirect roles played by the Kondo hybridization, the valence fluctuations, the crystal-field effects and the low-temperature magnetic fluctuations for the establishment of superconductivity at $p_c$ and $p^*$ constitute a challenging and, so far, unsolved issue.

## 4.2.2 Critical magnetic fluctuations

As well as intersite magnetic fluctuations were emphasized in Sections 3.1.2 and 3.2.2 as a driving force for the Fermi-liquid regime established in heavy-fermion correlated paramagnets, we show here that they play a critical role at the quantum magnetic phase transition of heavy-fermion antiferromagnets. In this Section, studies of magnetic fluctuations by macroscopic and microscopic experiments performed on a selection of textbook systems are presented.

A first focus is given on the signatures of critical magnetic fluctuations induced in the strongly-anisotropic Ising system $CeRh_2Si_2$ and in the isotropic cubic system $CeIn_3$. Figures 4.9(a-b) present the pressure variation of the quadratic coefficient $A$ of the electrical resistivity [Araki 02b] and of the Sommerfeld coefficient $\gamma$ of the heat capacity [Graf 97], respectively, of $CeRh_2Si_2$. Both quantities are maximum, indicating an enhancement of the effective mass, and thus of the critical magnetic fluctuations, at the critical pressure $p_c$. Contrary to $CeRh_2Si_2$, where a Fermi-liquid variation $\rho = \rho_0 + AT^2$ is observed at all pressures, a non-fermi-liquid regime with a variation $\rho = \rho_0 + AT^n$, with $n \simeq 1.5$, is observed in the vicinity of the critical pressure of $CeIn_3$. Figures 4.9(c-d) show the variations with pressure of the $T$-power-law coefficient $n$, and of the coefficient $A$ extracted for $CeIn_3$ [Knebel 01]. While $n$ deviates from 2 in the vicinity of $p_c$ and reaches its minimal value $n \simeq 1.5$ at $p_c$, the coefficient $A$ is found to be maximum, indicating an enhanced effective mass, at $p_c$. This is confirmed by an NMR study of the relaxation rate $1/T_1T$, which saturates at low-temperatures in the high-pressure regime and whose low-temperature value is maximum at $p_c$ [Kawasaki 08]. Knowing that a Fermi liquid driven by magnetic fluctuations leads to a NMR relaxation rate $1/T_1T \propto m^{*2}$ (see Equations 2.22 and 2.26 introduced in Section 2.4.3.2), the low-temperature enhancement of $1/T_1T$ indicates that critical magnetic fluctuations drive the enhancement of the effective mass of $CeIn_3$ at $p_c$.

Inelastic neutron scattering permits to extract the wavevector-dependence of the magnetic fluctuations and is perfectly adapted to identify the microscopic nature of critical magnetic fluctuations at a heavy-fermion quantum phase transition. Figure 4.10(a-b) presents inelastic-neutron-scattering spectra measured on $Ce_{1-x}La_xRu_2Si_2$ at low temperature ($T \simeq 2$ K), at the wavevector $\mathbf{k}_1 = (0.31, 0, 0)$ characteristic of the long-range magnetic order established for $x > x_c$, and at a wavevector $\mathbf{q}_0 = (0.56, 0, 0)$ characteristic of a $\mathbf{q}$-independent local magnetic contribution, respectively [Knafo 09a]. An enhancement of the intersite magnetic fluctuations, which correspond to short-range antiferromagnetic order, with wavevector $\mathbf{k}_1$, is visible at the critical content $x_c$, while the variation of the local magnetic fluctuations, probed at $\mathbf{q}_0$, shows peculiar feature at $x_c$. Figure 4.10(c) summarizes the quantum critical properties of $Ce_{1-x}La_xRu_2Si_2$, by showing the doping-variation of the antiferromagnetically-ordered moment with wavevector $\mathbf{k}_1$ (top) and the doping-temperature magnetic phase diagram of the system. The temperature scales $T_1$ and $T_0$ plotted in this phase diagram are defined as the low-temperature linewidths of the quasielastic spectra at the wavevectors $\mathbf{k}_1$ and $\mathbf{q}_0$, respectively: $T_1 = \Gamma(\mathbf{k}_1, T \to 0)$ and $T_0 = \Gamma(\mathbf{k}_0, T \to 0)$. The monotonous variation of $T_0$ indicates that the local magnetic fluctuations do not drive the quantum magnetic phase transition. Oppositely, the minimum value of $T_1$ at the critical concentration $x_c$ indicates that the intersite magnetic fluctuations with wavevector $\mathbf{k}_1$ are critical. These fluctuations are identified as the driving force of the Fermi-liquid regime established at temperatures $T < T_1$, in the paramagnetic phase for $x \leq x_c$, and $T_1 = T^* \simeq 1/m^*$ is identified as the characteristic temperature of the Fermi-liquid regime (see Equation 2.26 in Section 2.4.3.2).

Figure 4.11 presents further details about the temperature- and doping-variations of the magnetic fluctuations measured by inelastic neutron scattering on $Ce_{1-x}La_xRu_2Si_2$ [Knafo 09a]. Using a quasielastic Lorentzian lineshape (see Equation 2.19 in Section 2.4.3.1), the static sus-ceptibility $\chi'(\mathbf{q}, T)$ and the relaxation rate $\Gamma(\mathbf{q}, T)$ have been extracted for the two wavevectors $\mathbf{k}_1$ characteristic of the magnetic-order-parameter fluctuations and $\mathbf{q}_0$ characteristic of the local single-site magnetic fluctuations, for a large range of temperatures and chemical doping covering the phase diagram presented in Figure 4.10(c). Figures 4.11(d,e) show a continuous variation of $\chi'(\mathbf{q}_0, T)$ and $\Gamma(\mathbf{q}_0, T)$ when crossing the antiferromagnetic boundary $T_N$ or $x_c$, which indicates that local magnetic fluctuations at wavevector $\mathbf{q}_0$ do not play a critical role. Oppositely, Figure 4.11(a) shows that $\chi'(\mathbf{k}_1, T)$ presents well-defined maxima in $T$-scans at $T_N = 4.4$ K for $x = 13\%$ and $T_N = 6$ K for $x = 20\%$, and at $x_c$ in a $x$-scan in the limit of low-temperatures. Figure 4.11(b) shows that $\Gamma(\mathbf{k}_1, T)$ presents well-defined minima in $T$-scans at $T_N = 4.4$ K for $x = 13\%$ and $T_N = 6$ K for $x = 20\%$, and at $x_c$ in a $x$-scan in the limit of low-temperatures. These maxima in the static susceptibility (plotted with the colored scale in the phase diagram in Figure 4.10(c)) combined with the minima in the relaxation rate indicate that the magnetic fluctuations with wavevector $\mathbf{k}_1$ are maximum and, thus, drive the critical properties, via a thermal transition at $T_N$ or a quantum transition at $x_c$ in the limit of $T \rightarrow 0$. As expected from the magnetic-fluctuations Fermi-liquid theory developed by Kuramoto, Figures 4.11(c,f) show that the product $\chi'(\mathbf{q}, T)\Gamma(\mathbf{q}, T)$ is independent of $\mathbf{q}$ for the two wavevectors considered here (see Equation 2.23 in Section 2.4.3.2) [Kuramoto 87]. This product is also found to be temperature-independent, which suggests a possible extension of the theory of Kuramoto to finite temperatures.

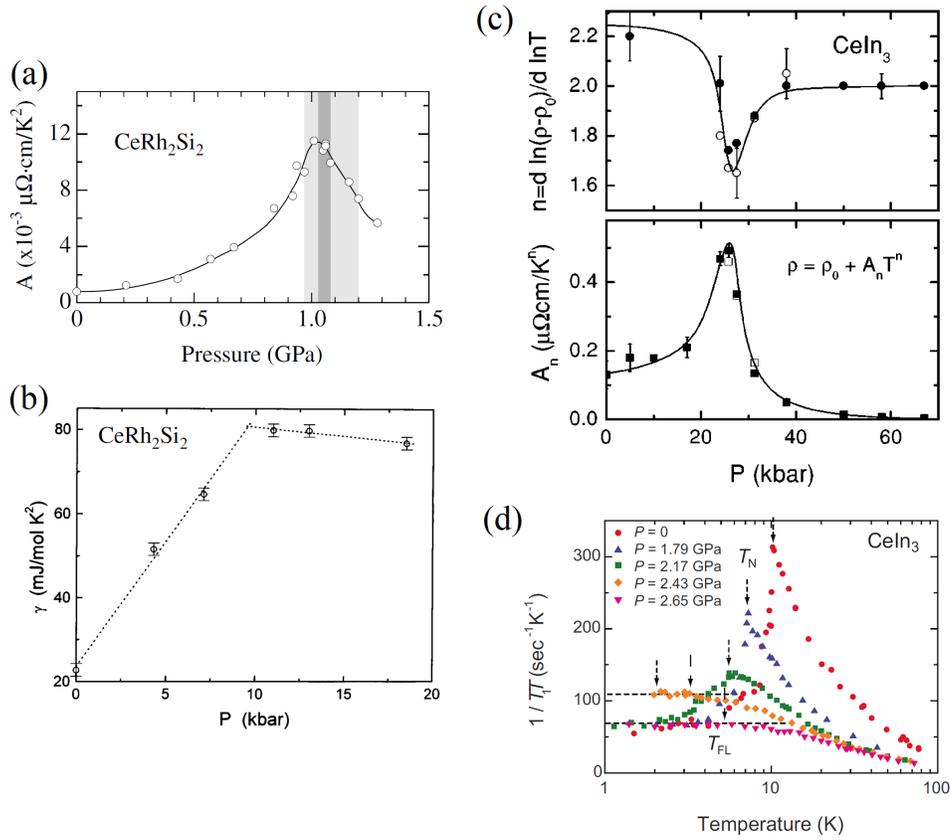

Figure 4.9: (a) Pressure-variation of (a) the quadratic coefficient $A$ in the temperature-dependence of the electrical resistivity (from [Araki 02b]) and (b) of the Sommerfeld heat-capacity coefficient $\gamma$ of $CeRh_2Si_2$ (from [Graf 97]). (c) Pressure variation of the power-law coefficient $n$ in the temperature-dependence of the electrical resistivity and of the coefficient $A$, extracted assuming a $T^n$ law (open squares: $n \neq 2$, closed squares: $n = 2$) (from [Knebel 01]), and temperature-dependence of the NMR relaxation rate $1/T_1T$ measured at different pressures in $CeIn_3$ (from [Kawasaki 08]).

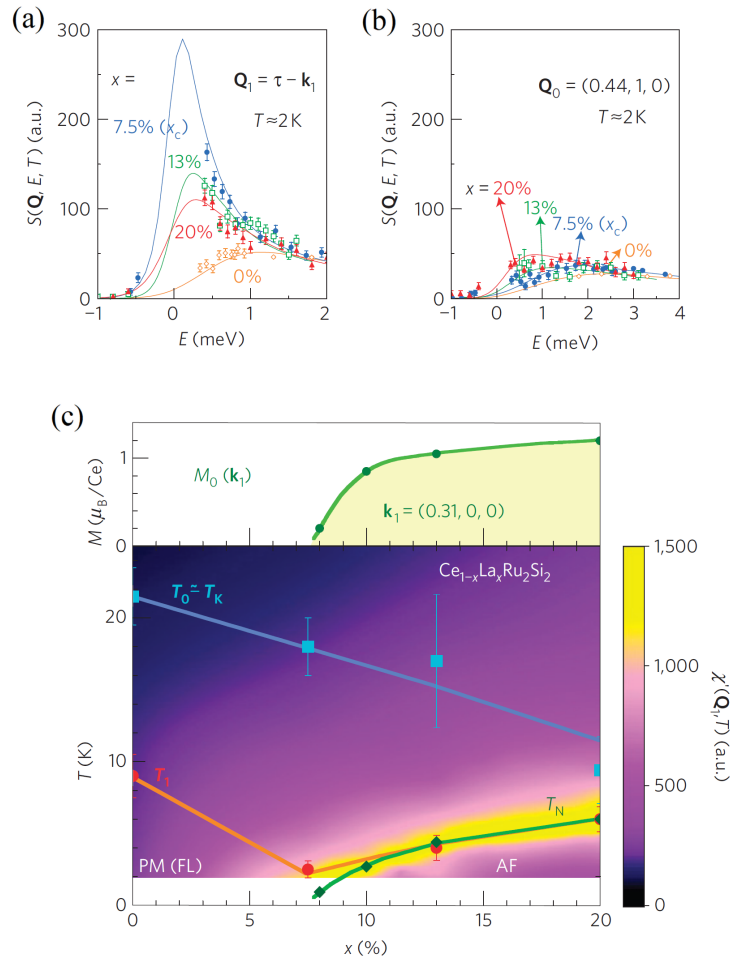

Figure 4.10: Energy spectra measured on $Ce_{1-x}La_xRu_2Si_2$ at $T \simeq 2$ K for different lanthanum contents $x$, (b) at the antiferromagnetic momentum transfer $\mathbf{Q}_1 = (0.69, 1, 0)$ and (c) at the momentum transfer $\mathbf{Q}_0 = (0.44, 1, 0)$ representative of $Q$-independent local magnetic fluctuations. (c) Doping-temperature magnetic phase diagram of $Ce_{1-x}La_xRu_2Si_2$, where the temperature scales $T_1$ and $T_0$ correspond to the low-temperature linewidths of the spectra measures at the momentum transfers $\mathbf{Q}_1$ and $\mathbf{Q}_0$ (from [Knafo 09a]).

The extensive study of the wavevector-, temperature- and doping-dependence of the mag-netic fluctuations in $Ce_{1-x}La_xRu_2Si_2$ permitted to validate a conventional scenario of quantum criticality driven by order-parameter fluctuations [Knafo 09a, Hertz 76, Millis 93, Moriya 95]. Similar work should be repeated to determine if conventional order-parameter-fluctuations models are the ground basis for the description of quantum criticality in other heavy-fermion systems. Alternative local models of quantum criticality were proposed by Si *et al* [Si 01] and Coleman *et al* [Coleman 02], following the observation of an 'anomalous' critical scaling of the dynamical magnetic susceptibility extracted from neutron-scattering experiments on $CeCu_{6-x}Au_x$ [Schröder 00]. However, such scaling plots are not a definitive proof for quantum criticality (see discussion and Figure 3.6 in Section 3.1.2 ) [Knafo 05]. The 'local' scenario is based on the hypothesis that quantum criticality is driven by a single-site local magnetic effect, implying that the magnetic fluctuations at all wavevectors of the reciprocal space are critical. To verify this hypothesis, a demonstration that the magnetic fluctuations at all wavevectors of the reciprocal space are maximum at the quantum magnetic phase transition (critical doping or pressure), in the limit of zero-temperature, is needed. Oppositely, the observation of a monotonous variation of the magnetic fluctuations at a wavevector characteristic of the single-site magnetic fluctuations, as at $\mathbf{q}_0$ for $Ce_{1-x}La_xRu_2Si_2$, is sufficient to invalidate a local scenario of quantum magnetic criticality.

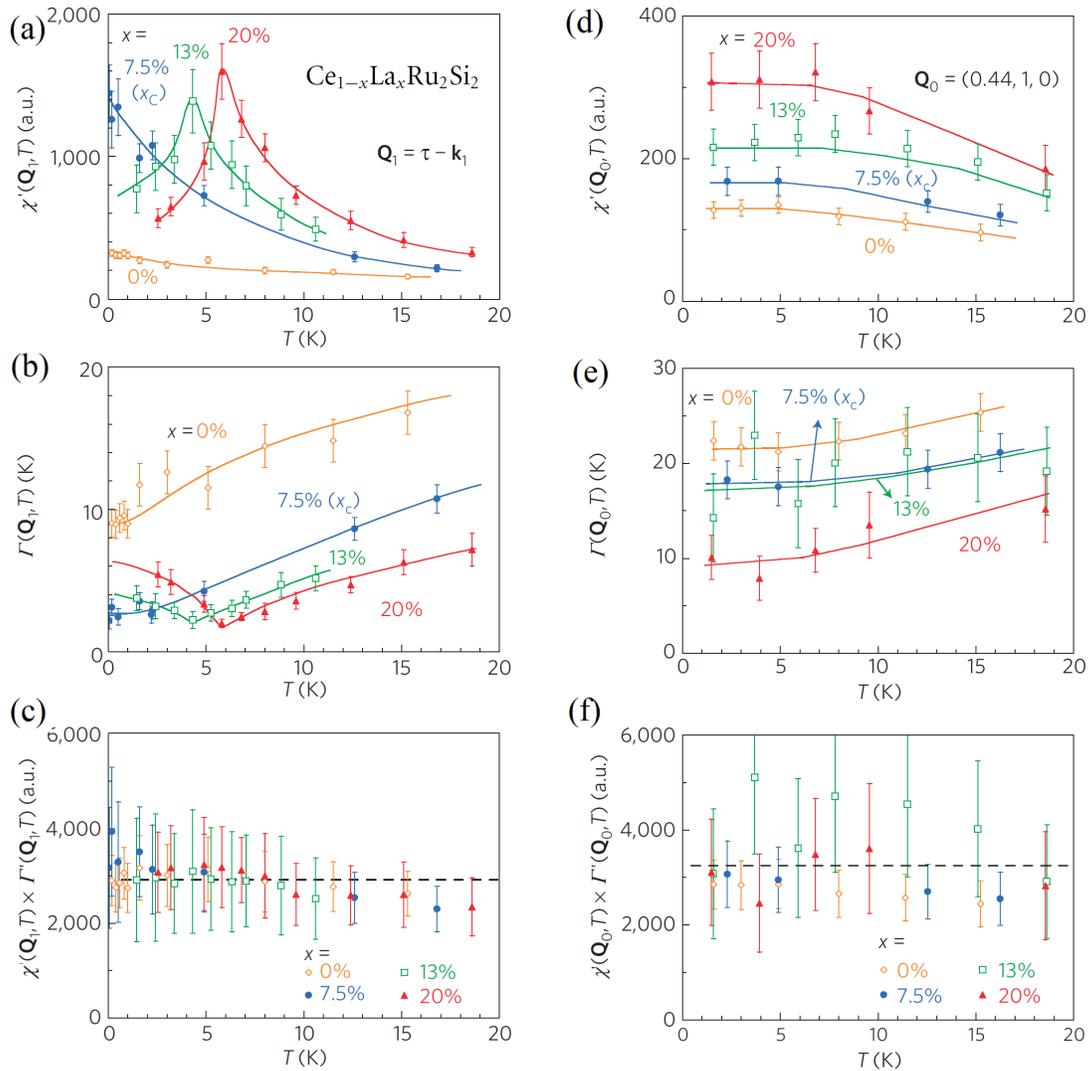

Figure 4.11: Temperature-dependence of the parameters extracted from a fit by a quasielastic lorentzian lineshape to the inelastic neutron scattering spectra measured on $Ce_{1-x}La_xRu_2Si_2$ alloys: (a) the static susceptibility $\chi'(\mathbf{Q}_1)$, (b) the linewidth $\Gamma(\mathbf{Q}_1)$ measured at the momentum transfer $\mathbf{Q}_1$, and (c) their product, and (d) the static susceptibility $\chi'(\mathbf{Q}_0)$, (e) the linewidth $\Gamma(\mathbf{Q}_0)$ measured at the momentum transfer $\mathbf{Q}_0$, and (f) their product (from [Knafo 09a]).

### 4.2.3 Fermi surface

In parallel to the study of intersite magnetic fluctuations, which play a critical role at the metamagnetic transition of heavy-fermion paramagnets (see Section 3.2.2) and at the quantum antiferromagnetic phase transition of heavy-fermion antiferromagnets (see Section 4.2.2), a large experimental and theoretical effort has been devoted to the investigation of the Fermi surfaces. In Section 3.2.3, it has been shown that the metamagnetic transition of heavy-fermion Ce-based paramagnets is accompanied by a Fermi-surface reconstruction. A Fermi surface with itinerant $4f$ electrons was identified in the correlated paramagnetic regime stabilized for $H < H_m$ and a Fermi surface with localized $4f$ electrons was proposed for the polarized paramagnetic regime reached for $H > H_m$. In this Section, focus is given to the Fermi-surface investigation by dHvA technique of heavy-fermion antiferromagnets at their pressure-induced quantum magnetic phase transition. Fermi-surface studies of two textbook systems, $CeRh_2Si_2$ and $CeRhIn_5$, are presented here. They show that Fermi-surface reconstructions accompany the quantum magnetic phase transition and indicate a tendency towards electronic localization in the antiferromagnetic phase.

Figure 4.12 summarizes a study performed by Araki *et al* of the Fermi surface of $CeRh_2Si_2$ under pressure [Araki 01, Araki 02a]. At low-temperature, this antiferromagnet is characterized by two pressure-induced phase transitions, at $p_{1,2} \simeq 0.5 - 0.6$ GPa where the antiferromagnetic structure is modified, and at $p_c \simeq 1$ GPa where antiferromagnetism vanishes, being replaced by a correlated paramagnetic regime for $p > p_c$ (see Figure 4.7(c) in Section 4.2.1). Figures 4.12(a-b) show the dHvA oscillating signal measured in a magnetic field $\mathbf{H} \parallel \mathbf{c}$, at a low temperature $T = 28.5$ mK and ambient pressure, and the Fourier-transform spectra of these oscillations measured at different pressures, respectively. Significant changes of the spectra are induced under pressure. They are summarized in Figure 4.12(c), where the observed dHvA frequencies are plotted as function of pressure. At the pressure $p_{1,2}$, the orbits $o$, $\kappa$, and $\xi$ of low frequencies $0.5 \lesssim F \lesssim 2$ kT (1 kT=$10^7$ Oe) disappear and the orbits $p$, $q$, and $r$ of higher frequencies $1.5 \lesssim F \lesssim 6$ kT appear. At the pressure $p_c$, the five observed orbits $d$, $p$, $q$, $\pi$ and $r$ of frequencies $1.5 \lesssim F \lesssim 6.5$ kT are replaced by three orbits $C$, $A$, and $B$, of higher frequencies $4 \lesssim F \lesssim 8$ kT. Knowing that dHvA frequencies are proportional to extremal areas of Fermi-surface orbits perpendicular to the magnetic-field direction [Ōnuki 95], this progressive increase of the dHvA frequencies under pressure suggests the increase of the Fermi-surface volume un-der pressure. Araki *et al* emphasized that the Fermi surface of $CeRh_2Si_2$ in its low-pressure antiferromagnetic phase is similar to that of the non-$4f$ reference compound $LaRh_2Si_2$, suggesting that $CeRh_2Si_2$ in its antiferromagnetic phase has localized $4f$ electrons [Araki 01, Araki 02a]. They also proposed that $4f$-itinerant electrons drive the physics of $CeRh_2Si_2$ in its high-pressure correlated-paramagnetic regime. Figure 4.12(d) shows that the Fermi-surface reconstructions at $p_{1,2}$ and $p_c$ are accompanied by a modification of the orbit cyclotron masses $m_c^i$, which smoothly increase at pressures $p < 0.8$ GPa (from $m_c^i \simeq (2-6)m_0$ at $p = 0$ to $m_c^i \simeq (3-8)m_0$ at $p_{1,2}$, where $m_0$ is the free-electron mass), before suddenly increasing on approaching $p_c$ ($m_c^i > 25m_0$ at $p \gtrsim p_c$) and then smoothly decreasing for $p > p_c$. A heavy-fermion behavior can be, thus, identified as driven by itinerant electrons in the CPM regime of $CeRh_2Si_2$.

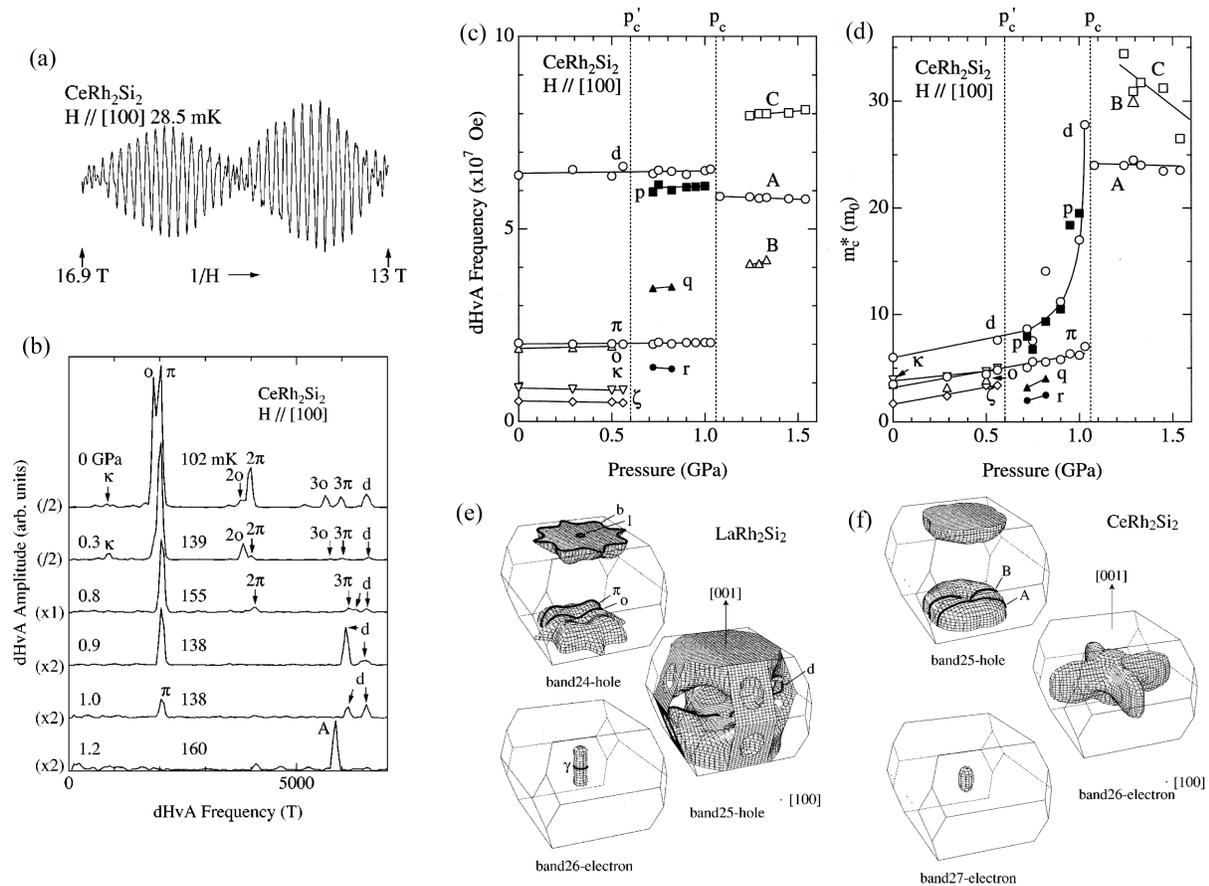

Figure 4.12: Fermi-surface reconstruction at the pressure-induced magnetic phase transition of $CeRh_2Si_2$ evidenced by dHvA measurements. (a) dHvA signal measured for $\mu_0\mathbf{H} \parallel \mathbf{c}$ varying from 13 to 17 T, at $T \simeq 30$ mK, (b) Fourier transform of the dHvA signal at different pressures, pressure-dependence of (d) the dHvA frequencies and (e) the cyclotron mass of the observed orbits, and three-dimensional sketches of the Fermi surface of $CeRh_2Si_2$ assuming itinerant $4f$ electrons and that of the non-$4f$ reference $LaRh_2Si_2$ (from [Araki 01, Araki 02a]).

Figure 4.13 summarizes a dHvA study of the Fermi surface of $CeRhIn_5$ under pressure per-formed by Shishido *et al* [Shishido 05]. Figure 4.13(a) presents quantum oscillations measured on CeRhIn$_5$ under a magnetic field $\mathbf{H} \parallel \mathbf{c}$, at a low temperature $T = 80$ mK and ambient pressure. The similarity between the Fourier transform of the oscillations with that of the non-4$f$ reference compound LaRhIn$_5$ was emphasized. It was interpreted as the indication that the Fermi surface of CeRhIn$_5$ at ambient pressure is very close to that of LaRhIn$_5$ and, thus, as a signature of localized-$4f$ electrons in the antiferromagnetic phase of CeRhIn$_5$ at low pressure. Figure 4.13(b) presents quantum oscillations measured on CeRhIn$_5$ under a magnetic field $\mathbf{H} \parallel \mathbf{c}$, at a low temperature $T = 80$ mK and at a pressure $p = 2.9$ GPa $> p_c$. The Fourier transform of the oscillations is very similar with that of the paramagnet CeCoIn$_5$, indicating that CeRhIn$_5$ in its paramagnetic regime induced under pressure can be considered, within first approximation, as equivalent to CeCoIn$_5$. As well as the Fermi surface of CeCoIn$_5$ can be calculated assuming itinerant $4f$ electrons, the electronic properties of CeRhIn$_5$ in its high-pressure paramagnetic regime were proposed to be driven by itinerant $4f$ electrons. Figure 4.13(c) shows the pressure-dependence of the dHvA frequencies, and of their cyclotron masses, extracted for CeRhIn$_5$. The frequencies are almost constant in the low-pressure antiferromagnetic phase, they suddenly change at $p_c$, and they also remain almost constant in the high-pressure paramagnetic regime. On the contrary to the frequencies, the cyclotron masses strongly vary in the antiferromagnetic and paramagnetic phases, from $\simeq (5-10)m_0$ at ambient pressure to $\simeq (20-60)m_0$ for $p \lesssim p_c$ and $\simeq (20-30)m_0$ for $p \simeq 3$ GPa.

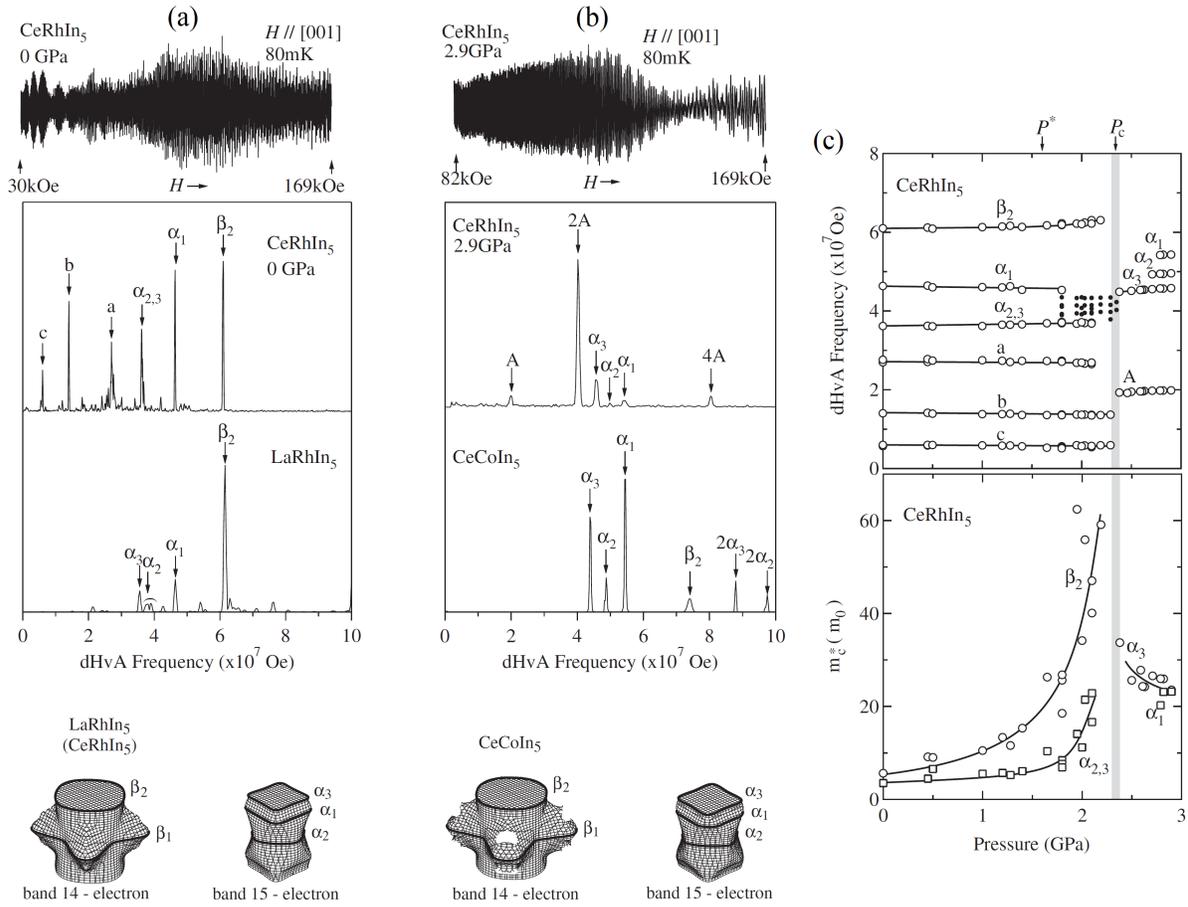

Figure 4.13: Fermi surface reconstruction at the pressure-induced magnetic phase transition of CeRhIn$_5$ evidenced by dHvA measurements. (a) dHvA signal measured on CeRhIn$_5$ at $p = 0$ GPa for $\mu_0\mathbf{H} \parallel \mathbf{c}$ varying from 3 to 17 T, at $T \simeq 80$ mK (top), its Fourier-transform spectrum compared with the Fourier-transform spectrum of dHvA oscillations in LaRhIn$_5$ (middle), and three-dimensional sketch of the Fermi surface bands 14 and 15 of LaRhIn$_5$ from theoretical calculations (bottom). (b) dHvA signal measured on CeRhIn$_5$ at $p = 2.9$ GPa for $\mu_0\mathbf{H} \parallel \mathbf{c}$ varying from 8 to 17 T, at $T \simeq 80$ mK (top), its Fourier-transform spectrum compared with the Fourier-transform spectrum of dHvA oscillations in CeCoIn$_5$ (middle), and three-dimensional sketch of the Fermi surface bands 14 and 15 of CeCoIn$_5$ from theoretical calculations (bottom). (c) Pressure-dependence of the dHvA frequencies (top) and of the associated cyclotron masses (bottom) extracted for CeRhIn$_5$ under pressure with $\mathbf{H} \parallel \mathbf{c}$ (from [Shishido 05]).

To summarize, as well as metamagnetism in the paramagnet CeRu$_2$Si$_2$ was suggested to be driven by an itinerant-to-localized change of the nature of $4f$ electrons (see Section 3.2.3) [Takashita 96, Aoki 14], the antiferromagnetic-to-paramagnetic quantum phase transitions in CeRh$_2$Si$_2$ [Araki 01, Araki 02a] and CeRhIn$_5$ [Shishido 05] were suggested to be induced by a localized-to-itinerant change of $4f$ electrons. Further, the cyclotron masses of the observed orbits were found to be enhanced at both field-and-pressure-induced transitions. Their relation with the enhancement of the effective mass by bulk measurements indicates that the magnetic fluctuations, also identified to drive the enhancement of the effective mass in several quantum critical heavy-fermion systems (see Sections 3.2.2 and 4.2.2), may, thus, be a property of itinerant $f$ electrons.

### 4.2.4 Valence

In Ce compounds, the main integer valence states $v = 3$ and 4 correspond to $n_{4f} = 1$ and 0, respectively, $4f$ electron per Ce atom (see Table 2.2 in Section 2.2.2). Pressure (or appropriate chemical doping) induces a delocalization of the $4f$ electrons into the valence band, leading to deviation from the integer-valent state $v = 3$ with $n_{4f} = 1$, corresponding to the localized limit, to a state of intermediate valence $v = 3 + \delta$ with $n_{4f} = (1 - \delta)$, corresponding to an itinerant case. Ultimately, the limit $n_{4f} = 0$ corresponds to a non-magnetic ground state. The situation in Yb compounds is more subtle. The main integer valence states $v = 2$ and 3 correspond to $n_{4f} = 14$ and 13, respectively, $4f$ electrons per Yb atom. Contrary to Ce compounds, the loss of $4f$ electrons, which is induced under pressure, leads to a magnetic state with $n_{4f} = 13$ in Yb compounds, whose non-hybridized integer-valent state $v = 2$ corresponds to a non-magnetic ground state with $n_{4f} = 14$.

Heavy-fermion compounds are magnetically-ordered, or nearly-ordered, systems, implying that their valence must be compatible with a magnetic ground state. The Hill plot shown in Figure 2.8 of Section 2.2.3 supports the picture that the onset of magnetism and nearby heavy-fermion paramagnetism, in $f$-electron systems occurs in a limit of nearly-localized $f$-electrons with a nearly-integer valence state. In Ce- and Yb-based heavy-fermion compounds, magnetic properties generally develop for a small deviation $\delta \ll 1$ to integer valence, i.e., for a valence $v = 3 + \delta$ with $n_{4f} = (1 - \delta)$ in Ce-systems and a valence $v = 3 - \delta$ with $n_{4f} = (13 + \delta)$ in Yb-systems.

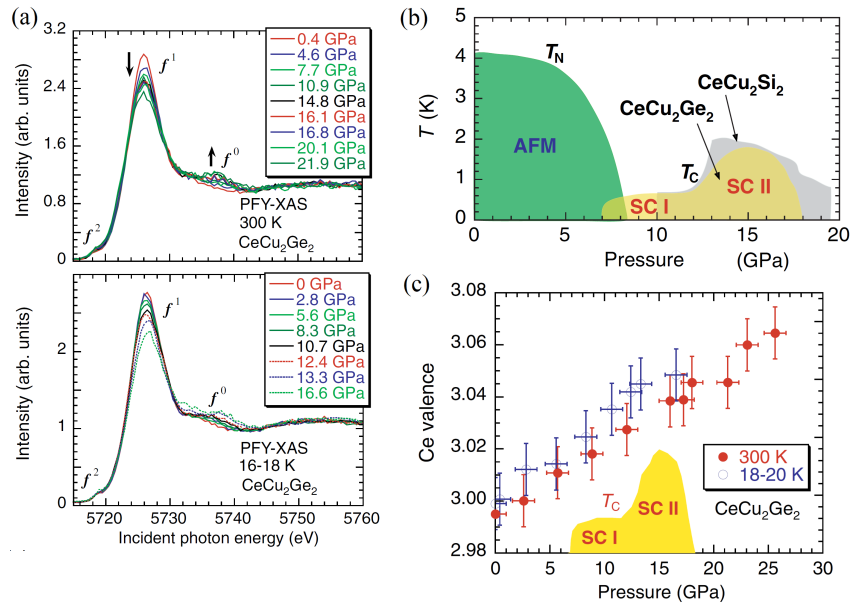

Figure 4.14: (a) X-ray absorption spectroscopy spectra measured at $T = 300$ K (top) and $T = 18$ K (bottom) at different pressures, (b) pressure-temperature phase diagram, and (c) valence variation extracted from the x-ray absorption spectroscopy spectra in CeCu$_2$Ge$_2$ (from [Yamaoka 14]).

Complementarily to the study of magnetic fluctuations and Fermi surfaces, the study of the valence configuration can, thus, offer valuable information about the electronic states involved in quantum critical phenomena. In this Section, investigations of valence by x-ray absorption spectroscopy on two heavy-fermion compounds at their quantum magnetic phase transition, CeCu$_2$Ge$_2$ and YbNi$_3$Ga$_9$ under pressure, are presented. They permit to confirm that a magnetic heavy-fermion state develop in systems of nearly-integer valence, and that the onset of magnetism is related with a smoothed variation of valence.

- Figure 4.14 presents a study of valence by x-ray absorption spectroscopy in the heavy-fermion antiferromagnet CeCu$_2$Ge$_2$ under pressure [Yamaoka 14]. The spectra measured at two temperatures $T = 300$ K and $T = 16 - 18$ K ($> T_N$) are shown in Figure 4.14(a), the pressure-temperature phase diagram is recalled in Figure 4.14(b) , and the variation of valence extracted from these spectra is shown in Figure 4.14(c). A small variation of valence $\Delta v \simeq 1\%$ is observed between $T = 300$ K and $T = 18$ K, suggesting that variations of valence in the antiferromagnetic phase may not be significant. Under pressure, a smooth variation of valence is reported, the valence varying by $\Delta v \simeq 6\%$ between ambient pressure and $p = 2.5$ GPa. No anomaly is observed at the pressure $p_c = 8$ GPa where antiferromagnetism collapses, or at the pressure $p^* \simeq 15$ GPa where the superconducting temperature is maximum (see Figure 2.33 in Section 2.5.1).

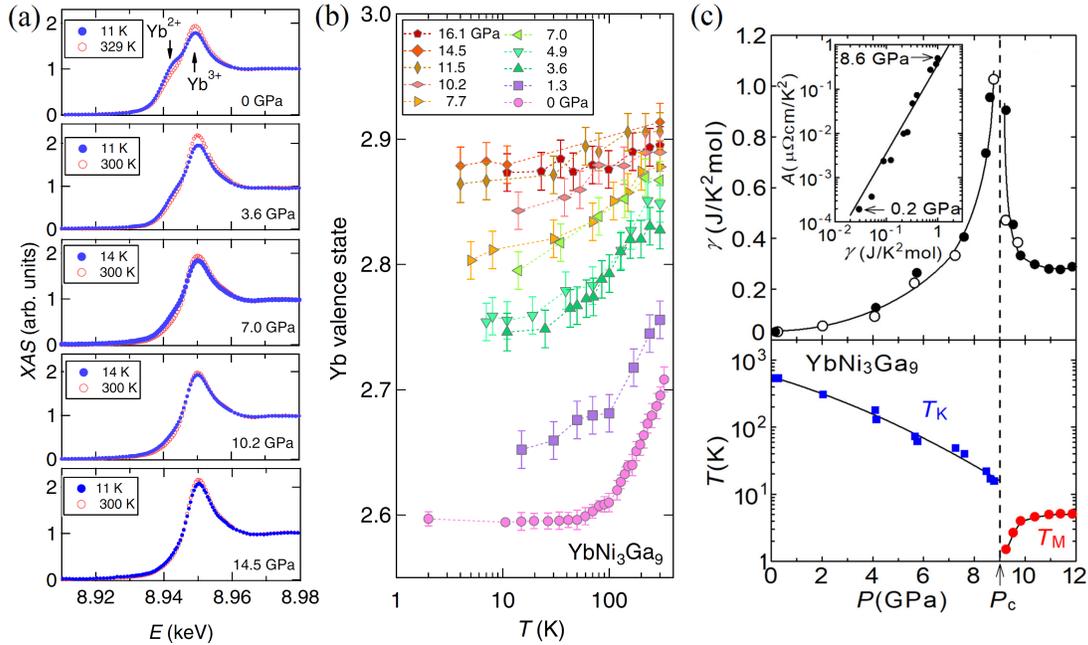

Figure 4.15: (a) X-ray absorption spectroscopy spectra measured at $T = 11 - 14$ K and $T = 300 - 329$ K at different pressures, (b) temperature-variation of valence at different pressures (from [Matsubayashi 15]), (c) heat-capacity Sommerfeld coefficient $\gamma$ versus pressure (top), plot of quadratic resistivity coefficient $A$ versus $\gamma$ (inset of top), and pressure-temperature phase diagram (bottom) of $YbNi_3Ga_9$ (from [Umeo 18]).

- Figure 4.15 presents a study of valence by x-ray absorption spectroscopy in $YbNi_3Ga_9$ under pressure [Matsubayashi 15]. The spectra measured at two temperatures $T = 300 - 329$ K and $T = 11 - 14$ K are presented in Figure 4.15(a), the temperature-variation of valence extracted from these spectra is shown in Figure 4.15(b) [Matsubayashi 15], the variations of the Fermi-liquid Sommerfeld coefficient $\gamma$ and resistivity coefficient $A$, and the pressure-temperature phase diagram are given in Figure 4.15(c) [Umeo 18]. At ambient pressure, $YbNi_3Ga_9$ is an intermediate-valent system whose valence, estimated at $v = 2.6$ at low-temperature, strongly increases with temperature, reaching the value $v = 2.7$ at $T = 300$ K. Under pressure, the low-temperature valence increases smoothly, before saturating to a value $v \simeq 2.85 - 2.9$ for $p > 10$ GPa. This saturation nearly coincides with the onset of a magnetically-ordered phase, presumably of antiferromagnetic nature [Matsubayashi 15], under pressures $p > p_c \simeq 9$ GPa. The large enhancement of the $\gamma$ and $A$ Fermi-liquid coefficients at pressures $p \sim p_c$ is observed when the valence becomes nearly-integer.

The smooth variations of valence in $CeCu_2Ge_2$ and $YbNi_3Ga_9$ under pressure indicate a progressive modification of the Kondo hybridization at the critical pressures $p_c$ or $p^*$, ending in a magnetically-ordered phase corresponding to a nearly-integer-valence limit. This contrasts with Kondo-breakdown-type scenarios, where a sudden change from localized to itinerant $f$-electron is expected to occur at the quantum magnetic phase transition [Si 01, Coleman 02]. In YbNi$_3$Ga$_9$, the enhancement at $p \sim p_c$ of the $\gamma$ and $A$ Fermi-liquid coefficients, and of the associated critical magnetic fluctuations, is observed once $n_{4f} \simeq 13$. This confirms that a nearly-integer-valence state constitutes a favorable condition for the development of magnetic correlations, and ultimately of long-range magnetism.

## 4.3 Magnetic-field-induced phenomena

As well as chemical doping and pressure, a magnetic field can induce a quantum phase transition between an antiferromagnetic and a paramagnetic ground states. Once high enough, a magnetic field ultimately leads to a polarized paramagnetic regime, where most of the electronic correlations have been squeezed out and all magnetic moments have been polarized. In relation with their magnetic anisotropy, different polarization processes are observed in anisotropic heavy-fermion antiferromagnets. After a presentation to the magnetic-field-temperature phase diagrams of heavy-fermion antiferromagnets and to their basic thermodynamic and electrical transport properties in Section 4.3.1, the effects of a magnetic field to their microscopic electronic properties, i.e., the antiferromagnetic order, the magnetic fluctuations, the Fermi surface, and the valence, are detailed in Section 4.3.2. In complement to this Section, an introduction to magnetic models can be found in [de Jongh 01] and an introduction to metamagnetic transitions in strongly-anisotropic antiferromagnets can be found in [Stryjewski 77].

### 4.3.1 Magnetic-moment reorientations

The signatures in basic thermodynamic and electrical-transport properties of the magnetic-field polarization of heavy-fermion antiferromagnets are presented here. First, a general introduction to the effects of a magnetic field on 'model' antiferromagnets is given in Section 4.3.1.1. Focus is then made on the high-magnetic-field properties of different anisotropy classes of heavy-fermion antiferromagnets. While most of them have a strong magnetic anisotropy, some of them also have a weak magnetic anisotropy. Section 4.3.1.2 presents the properties of strongly-Ising uniaxial antiferromagnets and Section 4.3.1.3 presents the properties of strongly-XY planar antiferromagnets. Finally, Section 4.3.1.4 presents the properties of weakly-Ising uniaxial antiferromagnets and Section 4.3.1.4 presents the properties of weakly-XY planar antiferromagnets.

#### 4.3.1.1 Basic properties

Figure 4.16 presents schematically the high-magnetic-field properties of 'simple' antiferromagnets. The indexes $\parallel$ and $\perp$ indicate the directions of the magnetic field $\mathbf{H} \parallel \mu_{AF}$ and $\mathbf{H} \perp \mu_{AF}$, respectively, relatively to the ordered antiferromagnetic moments $\mu_{AF}$. A three-dimensional magnetic exchange is assumed, and the cases of a strong and a weak Ising magnetic anisotropy, with or without magnetic domains, are considered:

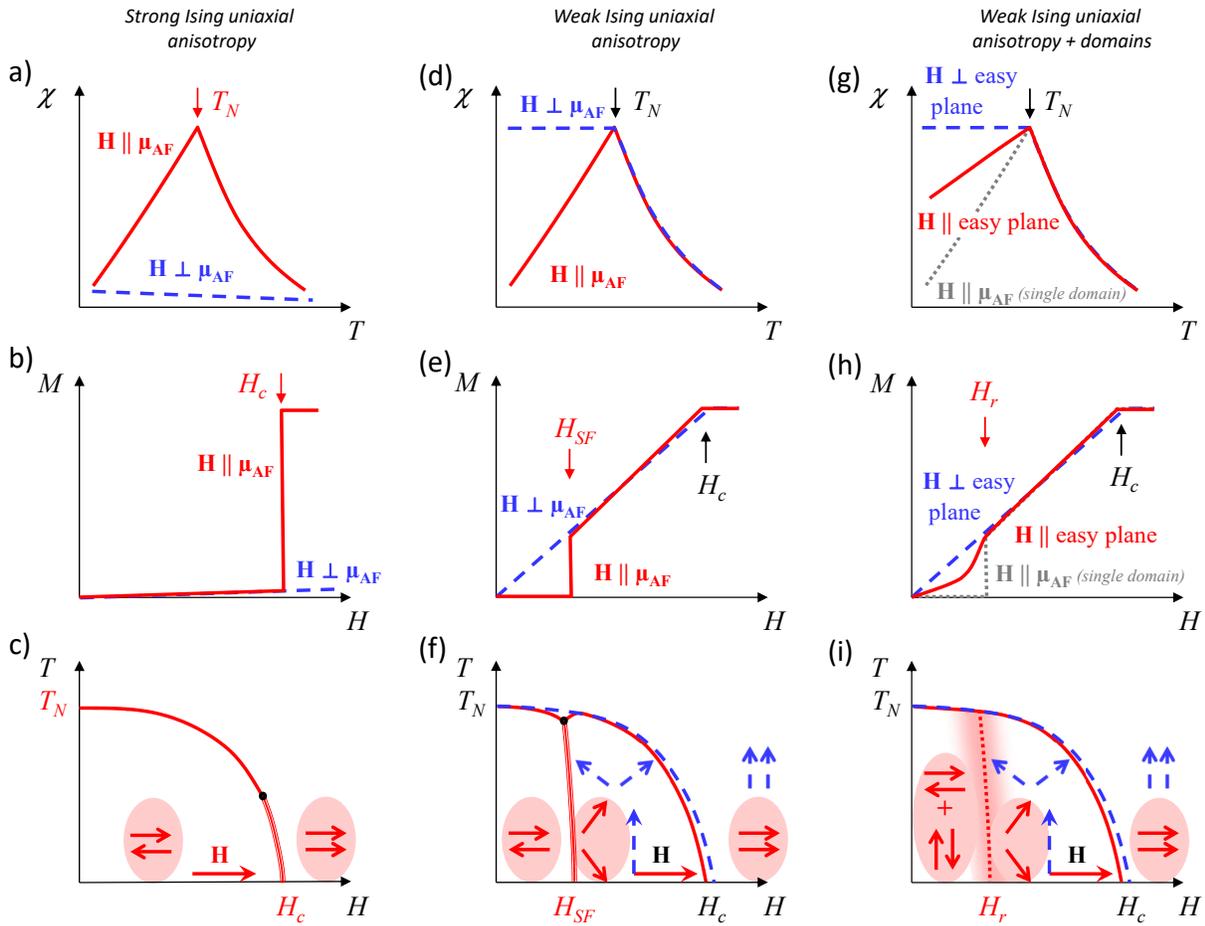

Figure 4.16: Schematic behavior expected for antiferromagnets in a magnetic field applied parallel or perpendicular to the antiferromagnetic moments direction. (a) Low-field magnetic susceptibility versus temperature, (b) low-temperature magnetization versus magnetic field, and (c) magnetic-field-temperature phase diagram of antiferromagnets with strong Ising uniaxial anisotropy. (d) Low-field magnetic susceptibility versus temperature, (e) low-temperature magnetization versus magnetic field, and (f) magnetic-field-temperature phase diagram of antiferromagnets with weak Ising uniaxial anisotropy. (g) Low-field magnetic susceptibility versus temperature, (h) low-temperature magnetization versus magnetic field, and (i) magnetic-field-temperature phase diagram of antiferromagnets with weak Ising uniaxial anisotropy and magnetic domains (due to several equivalent easy-axis directions).

- Figures 4.16(a-c) show the magnetic susceptibility, magnetization and magnetic-field-temperature phase diagram, respectively, of antiferromagnets with a strong Ising magnetic anisotropy and magnetic moments $\mu_{AF} \parallel$ **c**. As well as magnetic fluctuations perpendicular to the direction **c**, the high-field alignment of the magnetic moments perpendicular to **c** are forbidden. The high-temperature magnetic susceptibility is strongly anisotropic, with $\chi_{\parallel} \gg \chi_{\perp}$, and long-range magnetic order leads to a sharp second-order peak in $\chi_{\parallel}$ at $T_N$, with a strong reduction of $\chi_{\parallel}$ at low-temperatures. The high-magnetic-field magnetization processes are also strongly anisotropic. While no field-induced transition can be generally observed for $\mathbf{H} \perp \mathbf{c}$, a sharp first-order metamagnetic transition with a step-like variation $\Delta M$ of the magnetization is induced at the critical field $H_c$ delimiting the antiferromagnetic phase. The phase diagram shows a second-order antiferromagnetic boundary at high-temperature, which becomes first-order at low-temperature. The first-order and second-order lines are related by a tricritical point [Stryjewski 77].

- Figures 4.16(d-f) show the magnetic susceptibility, magnetization and magnetic-field-temperature phase diagram, respectively, of antiferromagnets with a weak Ising magnetic anisotropy. The high-temperature magnetic susceptibility is almost isotropic, with $\chi_{\parallel} \sim \chi_{\perp}$, and long-range magnetic order leads to a sharp second-order peak in $\chi_{\parallel}$ at $T_N$, with a strong reduction of $\chi_{\parallel}$ at low-temperatures. The transverse susceptibility saturates to a value $\chi_{\perp} > \chi_{\parallel}$ at temperatures $T < T_N$. In a magnetic field $\mathbf{H} \parallel \mu_{AF}$, a spin-flop transition is induced at $H_{SF}$. A first-order step-like anomaly in the magnetization is observed at $H_{SF}$. For $H_{SF} < H < H_c$, canted moments are composed of an antiferromagnetic component perpendicular to the field-direction and of a polarized component parallel to the field-direction, and $M$ increases almost linearly with $H$. A second-order kink in $M(H)$ occurs at the antiferromagnetic phase boundary $H_c$.

- Figures 4.16(g-i) show the magnetic susceptibility, magnetization and magnetic-field-temperature phase diagram, respectively, of antiferromagnets in which a small Ising magnetic anisotropy leads to the formation of magnetic domains. Such XY-type antiferromagnetism can be observed in tetragonal or hexagonal structures with, respectively, two or three equivalent easy magnetic axes $\perp \mathbf{c}$. Due to an averaging over the domains, the magnetic susceptibility is nearly isotropic in the easy plane, reaching a value larger than the intrinsic magnetic susceptibility $\chi_{\parallel}$ of an isolated domain. For $H_r < H < H_c$ a canted antiferromagnetic state with an antiferromagnetic component perpendicular to the magnetic-field direction, and for which $M$ increases linearly with $H$, is stabilized. The reorientation field $H_r$ corresponds to a crossover governed by the progressive reorientation from a multi-domain low-field regime to a single-domain high-field state. A second-order kink in $M(H)$ occurs at the antiferromagnetic phase boundary delimited by the critical field $H_c$.

In weakly-anisotropic antiferromagnets, a magnetic field induces an effective transverse magnetic anisotropy, ending in the stabilization of a canted state whose antiferromagnetic components are perpendicular to the field direction. In systems subject to strong magnetic fluctuations, magnetic-field-induced anisotropy can lead to an increase of the Néel temperature $T_N$ [Knafo 07, Knafo 08]).

The prototypical cases presented in Figure 4.16 do not cover all possible cases, but they indicate some general features expected for Ising and XY antiferromagnets. In heavy-fermion systems, deviations from these simple pictures can be driven by a large variety of magnetic exchange interactions and anisotropies:

- multiple exchange interactions, along different directions, possibly associated with different magnetic anisotropies. Such interactions are generally in competition, due to different wavevectors, geometrical frustration, etc.,

- a competition between the magnetic anisotropy imposed by the crystal field and that from the magnetic-exchange interaction. Long-range magnetic order can also develop at low temperature, with the magnetic moments aligned along a hard magnetic direction of the high-temperature magnetic susceptibility, controlled by crystal-field effects,

- a competition between magnetic exchange interactions and the Kondo effect, ending in a reduction of the magnetic ordering temperature and of the ordered magnetic moments, and in an increase of the magnetic fluctuations. In the vicinity of a quantum magnetic phase transition, such magnetic fluctuations induce a broad maximum in the magnetic susceptibility at a temperature $T_\chi^{max} > T_N$ and metamagnetism at a magnetic field $H_m > H_c$. Their progressive quench in a field is associated with a large initial slope of $M(H)$. The quench of remaining magnetic fluctuations can also induce a large slope of $M(H)$ for $H > H_c$.

- although most heavy-fermion system are the place of three-dimensional magnetic exchange interactions, low-dimensional (one-dimensional, two-dimensional, magnetic ladders etc.) magnetic exchange can lead to magnetic fluctuations, which can also result in a broad maximum in the magnetic susceptibility,

- a magnetic field can also lead to a modification of the magnetic wavevector of antiferromagnetic order, either progressively, or suddenly. In some compounds a magnetic field transforms a spin-density-wave of incommensurate wavevector $\mathbf{k}$ into antiferromagnetism with a commensurate wavevector $\mathbf{k}'$ (close to $\mathbf{k}$). It can also transform a multi-$\mathbf{k}$ to a single-$\mathbf{k}$ (single-domain) magnetic structure, or a single-$\mathbf{k}$ structure made of domains to a single domain. Finally, a magnetic field can induce magnetic phases with modified wavevectors.

#### 4.3.1.2  Ising anisotropy

Here, a focus is made on heavy-fermion antiferromagnets with a strong Ising magnetic anisotropy. The properties of heavy-fermion antiferromagnets with a weak Ising anisotropy will be considered in Section 4.3.1.4.

Figure 4.17 presents the magnetization versus magnetic field of two prototypical antiferromagnetic systems with a strong Ising uniaxial anisotropy, $Ce_{1-x}La_xRu_2Si_2$ (see Figure 3.8(a) in Section 3.2.1 and Figure 4.2.1 in Section 4.6) and $CeRh_2Si_2$ (see Figure 4.1.1(b) in Section 4.1), in a magnetic field applied along their easy magnetic axis $\mathbf{c}$. In both cases, a sharp first-order transition is induced at the antiferromagnetic boundary $H_c$ to a polarized paramagnetic regime with a large magnetization $M \simeq 1.5\mu_B/Ce$. However, instead of a single transition, as expected in a simple case (see Figures 4.16(a-c) in Section 4.3.1.1), these systems show a two-step metamagnetic behavior.

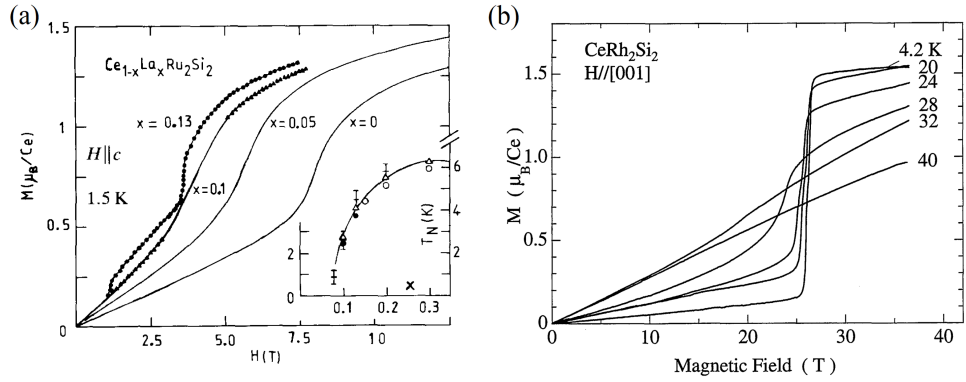

Figure 4.17: Magnetization versus magnetic field (a) of $Ce_{1-x}La_xRu_2Si_2$ in a magnetic field $\mathbf{H} \parallel \mathbf{c}$ at $T = 1.5$ K, the inset of (a) showing the doping-$x$-variation of the Néel temperature (from [Haen 90]), and (b) of $CeRh_2Si_2$ in a magnetic field $\mathbf{H} \parallel \mathbf{c}$ at different temperatures (from [Settai 97]).

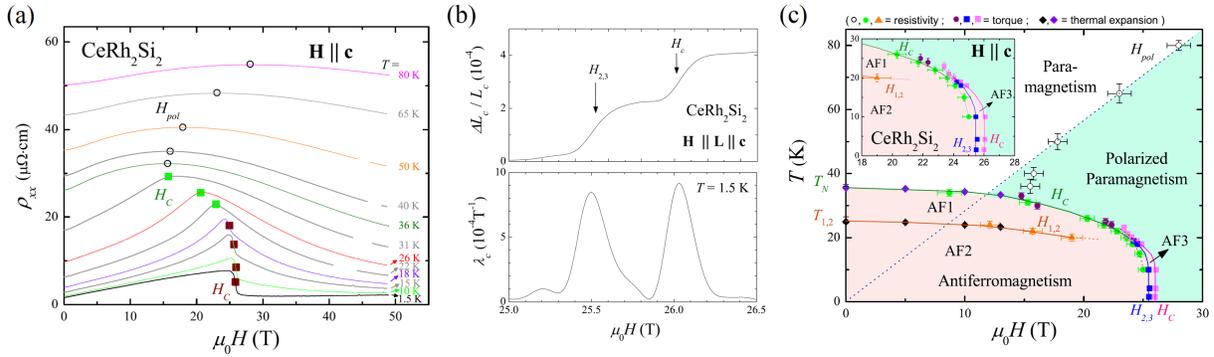

Figure 4.18: (a) Electrical resistivity $\rho$ versus magnetic field at different temperatures, (b) low-temperature relative length variation $\Delta L_c/L_c$ (top) and magnetostriction coefficient $\lambda_c = \partial(\Delta L_c/L_c)/\partial(\mu_0 H)$ (bottom) versus magnetic field, and (c) magnetic-field-temperature phase diagram of $CeRh_2Si_2$ in a magnetic field $\mathbf{H} \parallel \mathbf{c}$ (from [Knafo 10]).

Figure 4.17(a) presents the magnetization of $Ce_{1-x}La_xRu_2Si_2$ in the limit of low-temperatures ($T = 1.5$ K). The metamagnetic transition at $H_m$ in the correlated paramagnetic compounds with $x \leq x_c = 7.5$ % is replaced by a sharper metamagnetic transition at $H_c$ in the antiferromagnetic compounds with $x > x_c$ [Haen 90]. For $x = 13$ %, two first-order transitions are reported at $\mu_0 H_a = 1.1$ T, corresponding to a change of the antiferromagnetic structure, and at $\mu_0 H_c = 3.6$ T, corresponding to the antiferromagnetic phase boundary (see Figure 4.26 in Section 4.3.2.1). Figure 4.17(b) shows the magnetization versus magnetic field of $CeRh_2Si_2$ at different temperatures, indicating a sharp two-steps metamagnetic process at the characteristic field $\mu_0 H_{2,3} = 25.5$ T and at the antiferromagnetic boundary $\mu_0 H_c = 26$ T [Settai 97]. These transitions vanish when the temperature is raised above $T_N = 36.5$ K.

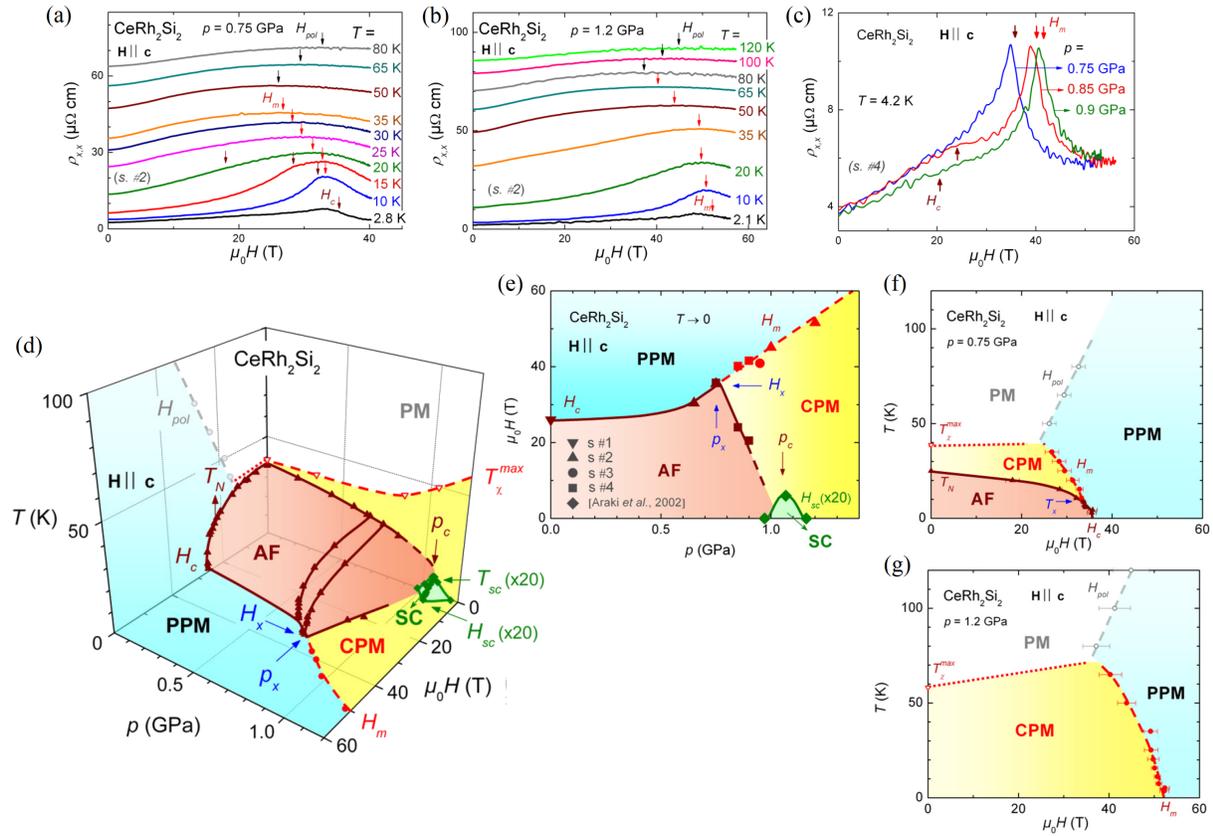

Figure 4.19: Electrical resistivity $\rho$ versus magnetic field at different temperatures (a) at the pressure $p = 0.75$ GPa, (b) at the pressure $p = 1.2$ GPa, and (c) at $T = 4.2$ K and various pressures, three-dimensional magnetic-field-pressure-temperature phase diagram, and two dimensional (e) pressure-magnetic phase diagram in the limit of $T \rightarrow 0$, (f-g) magnetic-field-temperature phase diagrams at the pressures $p = 0.75$ and 1.2 GPa, respectively, of $CeRh_2Si_2$ in a magnetic field $\mathbf{H} \parallel \mathbf{c}$ (from [Knafo 17]).

Figure 4.18 focuses on the properties of $CeRh_2Si_2$ in a magnetic field $\mathbf{H} \parallel \mathbf{c}$ [Knafo 10]. Its electrical resistivity $\rho$ versus magnetic field is presented in Figure 4.18(a), indicating that the sharp first-order transition at low temperature, characterized by a sudden fall of $\rho$ at $H_c$, is replaced at high temperature by a second-order transition, characterized by a kink in $\rho$. At temperatures $T > T_N$, the anomaly at $H_c$ has vanished and has been replaced by a broad maximum at $H_{pol}$ characteristic of a crossover to the polarized paramagnetic regime. Figure 4.18(b) shows that the two-steps metamagnetic process is visible in the length variation $\Delta L_c/L_c$ measured along $c$, which increases by $2 \cdot 10^{-4}$ at each transition, and in its field derivative, the magnetostriction coefficient $\lambda_c$ which shows two sharp maxima at $H_{2,3}$ and $H_c$. The small length variations indicate that the valence is almost unaffected by these field-induced transitions. Figure 4.18(c) summarizes the magnetic-field-temperature phase diagram of $CeRh_2Si_2$ in a magnetic field $\mathbf{H} \parallel \mathbf{c}$. It emphasizes the presence of three antiferromagnetic phases, AF1, AF2, and AF3, whose boundaries merge at the tetracritical point (24 T,20 K). As well as AF1 and AF2 correspond to two different magnetic structures (see Figure 4.3 in Section 4.1.2) [Kawarazaki 00], the transition between AF2 and AF3 at $H_{2,3}$ corresponds to the field-stabilization of a third magnetic structure [Kumar unp.].

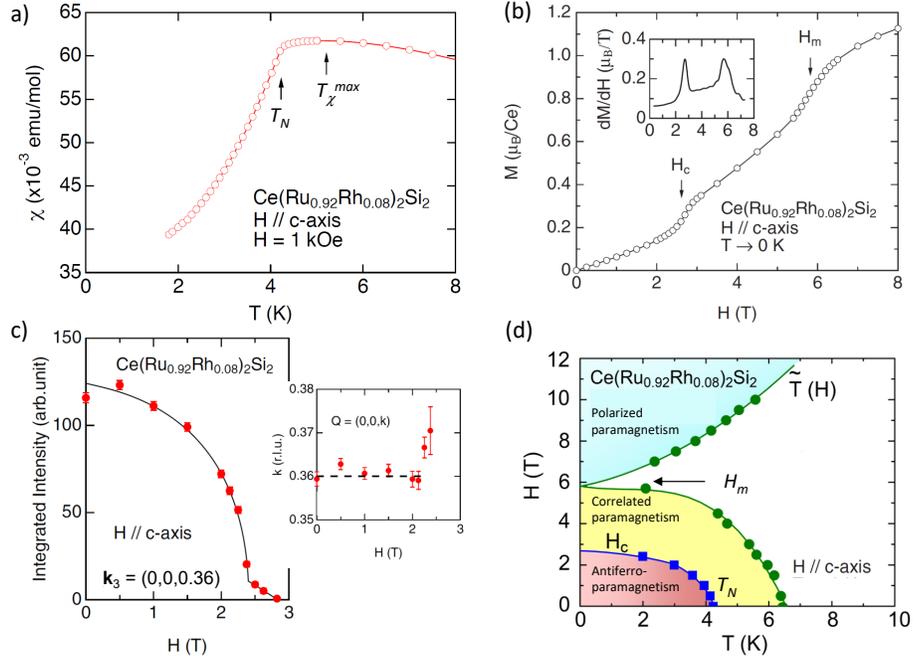

Figure 4.20: (a) Low-field magnetic susceptibility versus temperature, (b) low-temperature magnetization versus magnetic field, (c) neutron diffracted intensity at the wavevector $k_3 = (0, 0, 0.36)$ and wavevector-component $\delta$ of $k = (0, 0, \delta)$ versus magnetic field, and (d) temperature-magnetic-field phase diagram of Ce(Ru$_{0.92}$Rh$_{0.08}$)$_2$Si$_2$ in a magnetic field $\mathbf{H} \parallel \mathbf{c}$ (from [Aoki 12a, Aoki 13]).

In combination with a magnetic field, the application of pressure permits to continuously tune the phase diagram of a heavy-fermion antiferromagnet in three dimensions. Figure 4.19 summarizes such study performed on CeRh$_2$Si$_2$ in a magnetic field $\mathbf{H} \parallel \mathbf{c}$ [Knafo 17]. Figures 4.19(a-c) present electrical resistivity versus magnetic field data measured at different sets of temperatures and pressures, emphasizing the signatures of the phase boundaries $H_c$, $H_m$, and $H_{pol}$ characteristic of the boundaries of the AF, CPM and PPM phases, respectively. Two-dimensional $(p, H)$ and $(H, T)$ and a three-dimensional $(H, p, T)$ phase diagrams constructed from these data are shown in Figures 4.19(d-f). At $p = 0.75$ GPa $< p_c$, $H_c$ and $H_m$ are decoupled for $T > T_x \simeq 10$ K and they converge for $T < T_x$. Under pressure, a decoupling of $T_\chi^{max}$, which increases, and $T_N$, which decreases, is observed. A low-temperature decoupling of $H_c$ and $H_m$ results from the high-temperature decoupling of $T_\chi^{max}$ and $T_N$. It occurs in the vicinity of the quantum magnetic phase transition for $p_x = 0.75$ K $< p \lesssim p_c$. At these pressures, a sequence AF $\rightarrow$ CPM $\rightarrow$ PPM can be induced by a magnetic field.

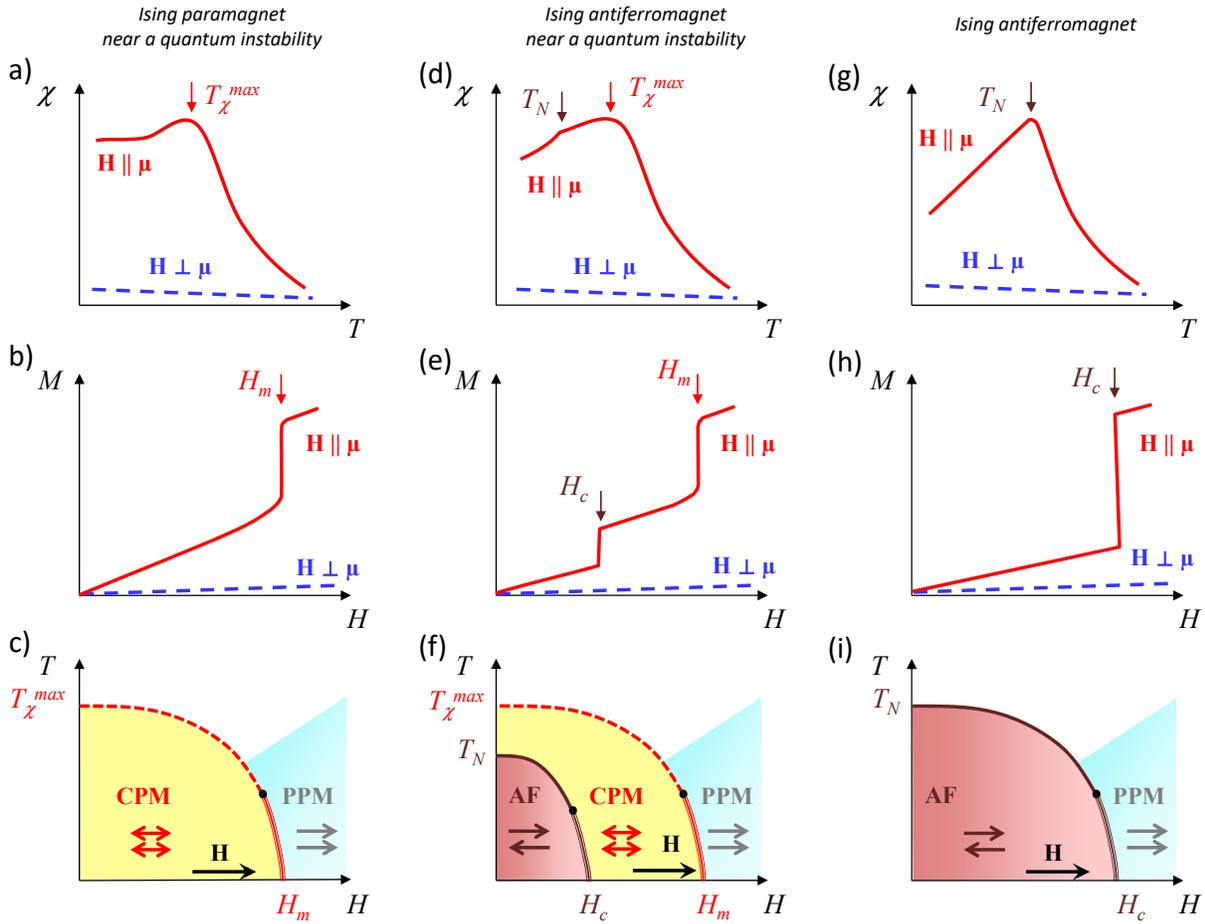

Figure 4.21: Schematic behavior expected for Ising heavy-fermion magnets close to a quantum magnetic phase transition, in a magnetic field applied parallel or perpendicular to the antiferromagnetic moments direction. (a) Low-field magnetic susceptibility versus temperature, (b) low-temperature magnetization versus magnetic field, and (c) magnetic-field-temperature phase diagram of Ising paramagnets near a quantum phase transition. (d) Low-field magnetic susceptibility versus temperature, (e) low-temperature magnetization versus magnetic field, and (f) magnetic-field-temperature phase diagram of Ising antiferromagnets very close to a quantum phase transition. (g) Low-field magnetic susceptibility versus temperature, (h) low-temperature magnetization versus magnetic field, and (i) magnetic-field-temperature phase diagram of Ising antiferromagnets near a quantum phase transition.

A similar low-temperature decoupling of $H_c$ and $H_m$ was observed in Ce(Ru$_{0.92}$Rh$_{0.08}$)$_2$Si$_2$, which is also a strongly Ising anisotropic antiferromagnet near a quantum magnetic phase transition, in a magnetic field $\mathbf{H} \parallel \mathbf{c}$ [Aoki 12a]. As shown in Figure 4.20, $T_\chi^{max}$ and $T_N$ are decoupled and long-range antiferromagnetic order with wavevector $\mathbf{k}_3 = (0, 0, 0.36)$ vanishes at the antiferromagnetic boundary $\mu_0 H_c = 3$ T. Interestingly, a small modification of the magnetic wavevector is observed when approaching $H_c$. By analogy with the high-field physics of the parent paramagnetic compound CeRu$_2$Si$_2$, one can speculate that the CPM regime stabilized at low-temperature for $\mu_0 H_c < \mu_0 H < \mu_0 H_m \simeq 6$ T also corresponds to a heavy-fermion regime driven by intersite magnetic fluctuations, which collapse for $H > H_m$.

Similarly to the paramagnetic case, where the low-temperature magnetic susceptibility saturates in a Fermi-liquid regime controlled by intersite magnetic fluctuations, the proximity of antiferromagnets to their quantum magnetic phase transition is accompanied by low-temperature quantum magnetic fluctuations, which drive the magnetic susceptibility to large low-temperature values. The related linear increase of $M(H)$ for $H < H_c$ can, thus, been interpreted as the signature of the progressive field-induced quench of these intersite magnetic fluctuations into a polarized magnetic moment. The step-like variation $\Delta M$ of the magnetization at $H_c$ is driven by the sudden alignment of the antiferromagnetic moments. In Ce(Ru$_{0.92}$Rh$_{0.08}$)$_2$Si$_2$ and Ce$_{0.87}$La$_{0.13}$Ru$_2$Si$_2$, the large slope in $M(H)$ indicates intense quantum magnetic fluctuations for $H < H_c$, while the small jump $\Delta M$ at $H_c$ indicates a small ordered antiferromagnetic moment in the AF phase, i.e., the characteristics of a close vicinity of a quantum magnetic phase transition. Oppositely, in CeRh$_2$Si$_2$, the small slope in $M(H)$ indicates weak quantum magnetic fluctuations for $H < H_c$, while the large jump $\Delta M$ at $H_c$ indicates a large ordered antiferromagnetic moment in the AF phase, also compatible with the large ordering temperature $T_N$ of this system.

Figure 4.21 summarizes schematically the properties of heavy-fermion magnets with strong Ising anisotropy in a magnetic field. It emphasizes the relation between $T_\chi^{max}$ and $H_m$ in paramagnets close to an antiferromagnetic phase transition (Figures 4.21(a-c)), the decoupling of $T_\chi^{max}$ and $T_N$ driving that of $H_m$ and $H_c$ in antiferromagnets very close to the quantum phase transition (Figures 4.21(d-f)), and the relation between $T_N$ and $H_c$ in antiferromagnets more distant from the quantum phase transition (Figures 4.21(g-i)). At low-temperatures, first-order transitions are often observed. When the temperature is increased, the first-order transition at $H_m$ transforms, via a critical endpoint, into a crossover delimiting the CPM regime and that the first-order transition at $H_c$ transforms, via a tricritical point, into a second-order phase transition delimiting the AF phase [Stryjewski 77]. In the Figure, schemes of the magnetic moments indicate their main state in the different phases, from long-range antiferromagnetic order in the AF phase, short-range magnetic fluctuations in the CPM regime, to polarized moments with field in the PPM regime. For simplicity, the multiple antiferromagnetic phases reported experimentally are not been represented in this Figure.

### 4.3.1.3  XY anisotropy

The effects of a magnetic field on heavy-fermion antiferromagnets with a XY planar magnetic anisotropy are presented here. In these materials, long-range antiferromagnetic order is associated with the formation of antiferromagnetic domains, in which the magnetic moments are aligned along one of the $\geq 2$ equivalent easy magnetic directions in the easy plane. A focus is made on the well-documented U(Pd$_{1-x}$Ni$_x$)$_2$Al$_3$ alloys, in which the XY magnetic anisotropy is strong. The properties of heavy-fermion antiferromagnets with a weak XY-anisotropy will be considered in Section 4.3.1.5.

Figure 4.22 presents the magnetic properties of UPd$_2$Al$_3$ in magnetic fields $\mathbf{H} \perp \mathbf{c}$ and

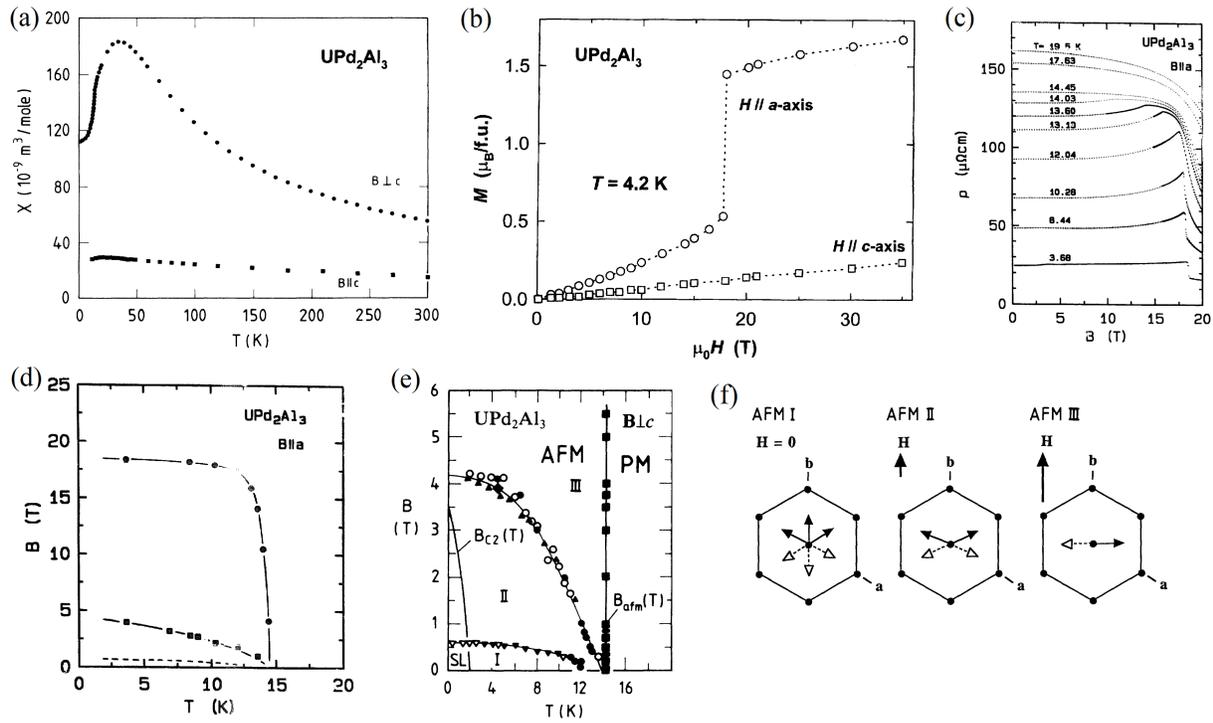

Figure 4.22: (a) Magnetic susceptibility versus temperature (from [Grauel 92]), (b) magnetization versus magnetic field at $T = 4.2$ K (from [de Visser 92, Sechovsky 98]), (c) electrical resistivity versus magnetic field at various temperatures (from [de Visser 94]) of $UPd_2Al_3$ in magnetic fields $\mathbf{H} \parallel \mathbf{a}, \mathbf{c}$. Magnetic-field-temperature phase diagram of $UPd_2Al_3$ in a magnetic field $\mu_0\mathbf{H} \parallel \mathbf{a}$ up to (d) 25 T (from [de Visser 94]) and (e) 6 T (from [Grauel 92]). (f) Schematic representation of the magnetic domain selection processes deduced from neutron-diffraction experiment in a magnetic field $\mathbf{H} \parallel \mathbf{a}$ (from [Kita 94]).

$\mathbf{H} \parallel \mathbf{c}$. Figure 4.22(a) shows that the magnetic susceptibility is strongly anisotropic. The low-temperature hierarchy $\chi_a \gg \chi_c$ indicates the strong XY planar magnetic anisotropy, which progressively develops when the temperature is decreased from 300 K to 35 K (here $\chi_a$ labels the magnetic susceptibility measured in a magnetic field $\mathbf{H} \perp \mathbf{c}$) [Grauel 92]. A broad maximum of $\chi_a$ at $T_\chi^{max} = 35$ K is compatible with a progressive onset of antiferromagnetic fluctuations with moments $\mu \perp \mathbf{c}$. The kink at the Néel temperature $T_N = 14.5$ K, followed by a large low-temperature decrease of $\chi_a$, suggests the onset of long-range antiferromagnetism with moments $\mu_{AF} \perp \mathbf{c}$. Neutron-diffraction experiments confirmed this picture and indicated that antiferromagnetism is established with the commensurate wavevector $\mathbf{k} = (0, 0, 1/2)$ [Krimmel 92]. Contrary to $\chi_a$, $\chi_c$ shows little anomalies at $T_\chi^{max}$ and $T_N$ [Grauel 92]. Figure 4.22(b) presents the high-field magnetization $M$ of $UPd_2Al_3$ in magnetic fields $\mathbf{H} \parallel \mathbf{a}, \mathbf{c}$ [de Visser 92, Sechovsky 98] (see also [Sakon 02]). $M$ increases almost linearly, with a slope corresponding to the low-temperature magnetic susceptibility, for $\mathbf{H} \parallel \mathbf{c}$. For $\mathbf{H} \parallel \mathbf{a}$, a larger initial slope of $M(H)$ is followed by a first-order metamagnetic transition, with a sudden magnetization variation $\Delta M \simeq 0.8 \mu_B$/ U, observed at the antiferromagnetic boundary $\mu_0 H_c = 18.5$ T. Figure 4.22(c) shows the field-variation of the electrical resistivity $\rho$ of UPd$_2$Al$_3$ in magnetic fields $\mathbf{H} \parallel \mathbf{a}$ for a large set of temperatures, indicating a sudden step-like decrease of $\rho$ at $H_c$ at low-temperatures. $H_c$ decreases with increasing temperatures, before collapsing at temperatures $T > T_N$ [de Visser 94].

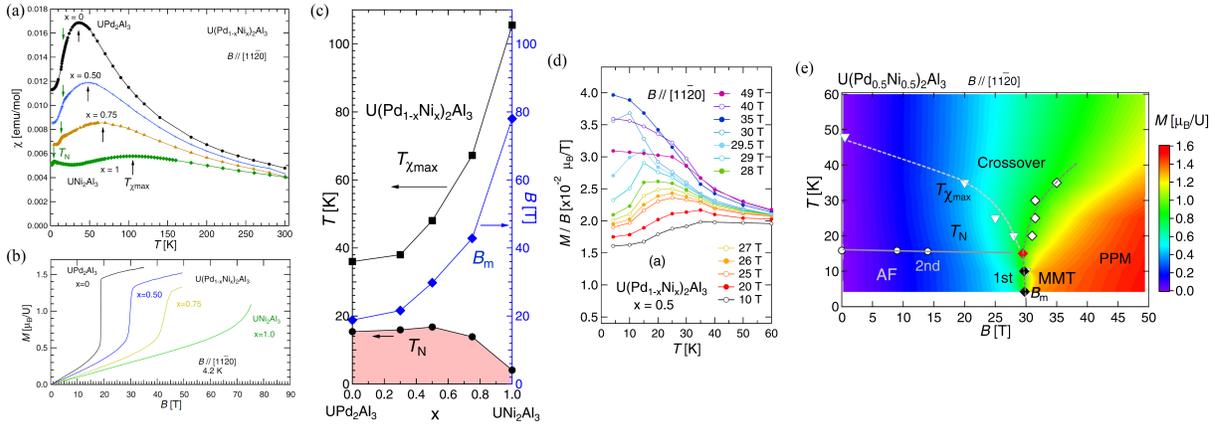

Figure 4.23: (a) Magnetic susceptibility versus temperature, (b) magnetization versus magnetic field at $T = 4.2$ K, (c) doping-temperature and magnetic-field-doping phase diagrams of U(Pd$_{1-x}$Ni$_x$)$_2$Al$_3$ alloys, (d) Magnetic susceptibility versus temperature at different magnetic fields and (e) magnetic-field-temperature phase diagram of U(Pd$_{0.5}$Ni$_{0.5}$)$_2$Al$_3$ in a magnetic field $\mathbf{H} \parallel [11\overline{2}0] \perp \mathbf{c}$ (from [Mochizuki 19].

The temperature-magnetic-field phase diagram of UPd$_2$Al$_3$ in magnetic fields $\mathbf{H} \perp \mathbf{c}$ is presented in Figure 4.22(d-e). In addition to the antiferromagnetic phase boundary, delimited by $H_c$ and $T_N$, two transition lines at $\simeq 0.6$ and 4.2 T are observed inside the antiferromagnetic phase $\mathbf{H} \perp \mathbf{c}$ [Grauel 92]. Kita *et al* showed from neutron diffraction experiments that these low-field lines are driven by magnetic-domain reorientation processes, ending in the high-field selection of domains where canted magnetic moments have an antiferromagnetic component $\perp \mathbf{H}$ (see Figure 4.22(f)) [Kita 94]. This indicates that a magnetic field applied in the easy plane leads to an effective transverse magnetic anisotropy. Field-induced domain selection is not a specificity of heavy-fermion antiferromagnets and was observed in broad range of antiferromagnets, as in oxides (with a reversible process, as in the hexagonal BaNi$_2$V$_2$O$_8$ [Knafo 07]) or iron-based compounds (with a reversible or irreversible process, as in Ba(Fe$_{1-x}$Co$_x$)$_2$As$_2$ [Chu 10, Fisher 11, Ruff 12], EuFe$_2$As$_2$ [Zapf 14], Fe$_{1.1}$Te [Knafo 13, Fabrèges 17]). As shown in [Tanatar 10, Chu 10, Fisher 11, Ruff 12, Zapf 14, Fabrèges 17], anisotropic susceptibilities and electrical resistivities can be revealed by a domain selection induced under a magnetic field or a uniaxial pressure. Similarly to these well-documented cases, the anisotropy of the in-plane electrical resistivity observed in UPd$_2$Al$_3$ in magnetic fields $\mu_0 \mathbf{H} \perp \mathbf{c}$ of 7 T, i.e., beyond the domain selection field [Köhler 93], may be an intrinsic property of magnetic domains.

Similarly to Ising antiferromagnets (see the Ce(Ru$_{0.92}$Rh$_{0.08}$)$_2$Si$_2$ case in Section 4.3.1.2), UPd$_2$Al$_3$ is characterized by a decoupling of $T_\chi^{max}$ and $T_N$, shown in Figure 4.22. Figure 4.23 presents the evolution of this decoupling and of its relationship with magnetic-field induced phenomena in U(Pd$_{1-x}$Ni$_x$)$_2$Al$_3$ alloys in a magnetic field $\mathbf{H} \parallel [11\overline{2}0]$ ($\perp \mathbf{c}$) [Mochidzuki 19]. Figure 4.23(a) shows the magnetic susceptibility versus temperature of U(Pd$_{1-x}$Ni$_x$)$_2$Al$_3$ compounds, emphasizing that, while $T_\chi^{max}$ increases from 35 to 105 K, $T_N$ decreases from 14.5 to 4.5 K when Ni-doping varies from $x = 0$ to 1. The decrease of $T_N$ is accompanied by a change of magnetic structure, from a commensurate wavevector $\mathbf{k} = (0, 0, 1/2)$ for UPd$_2$Al$_3$ [Krimmel 92] to an incommensurate wavevector $\mathbf{k} = (1/2 \pm \delta, 0, 1/2)$, with $\delta = 0.11$, for UNi$_2$Al$_3$ [Schröder 94]. As in UPd$_2$Al$_3$ [Kita 94], the antiferromagnetic moments in UNi$_2$Al$_3$ lie in the hexagonal plane and a domain reorientation occurs in a magnetic field $\mathbf{H} \perp \mathbf{c}$, ending in a canted phase with antiferromagnetic moments $\mu_{AF} \perp (\mathbf{H}, \mathbf{c})$ [Lussier 97]. Figure 4.23(b) presents low-temperature magnetization versus magnetic field measurements on U(Pd$_{1-x}$Ni$_x$)$_2$Al$_3$. A metamagnetic transition characterizes the borderline of antiferromagnetism at a critical field $H_c$ varying from 19 to 78 T for Ni-doping $x = 0$ to 1. The sharpness of the transition and the amplitude of the magnetization jump $\Delta M$ at $H_c$ decreases with $x$. Figures 4.23(c-d) show the magnetization divided by magnetic field versus temperature for different field values and the magnetic-field-temperature phase diagram, respectively, of U(Pd$_{0.5}$Ni$_{0.5}$)$_2$Al$_3$. They emphasize that the collapse of $T_\chi^{max}$ is linked with the antiferromagnetic boundary. As reported for pure UPd$_2$Al$_3$ [de Visser 94], $T_N$ in U(Pd$_{0.5}$Ni$_{0.5}$)$_2$Al$_3$ is almost field-independent up to $H_c$. Mochidzuki $et$ $al$ find that $T_\chi^{max} \propto H_c = H_m$, implying that the value of the AF boundary $H_c$ is fixed by the high-temperature boundary $T_\chi^{max}$ of the CPM regime, with no direct relation with the AF temperature boundary $T_N$. Contrary to strongly-Ising antiferromagnets (see Section 4.3.1.2), the low-field decoupling of $T_\chi^{max}$ and $T_N$ in the strongly-XY antiferromagnets U(Pd$_{1-x}$Ni$_x$)$_2$Al$_3$ does not drive a low-temperature decoupling of $H_m$ and $H_c$. This difference may originate from the nature of magnetic anisotropy.

Experiments on other XY antiferromagnets and modelings of their magnetic properties are needed to determine whether the behavior reported for U(Pd$_{1-x}$Ni$_x$)$_2$Al$_3$ is representative, or not, of this class of materials.

### 4.3.1.4  Weak Ising anisotropy

A few heavy-fermion magnets are characterized by a weak Ising magnetic anisotropy. Figure 4.24 summarizes the properties of one of them, the antiferromagnet YbNiSi$_3$ in magnetic fields applied either parallel or perpendicular to its easy magnetic axis $\mathbf{b}$. Figures 4.24(a-b) present plots of the magnetic susceptibility and of its inverse, respectively, versus temperature [Avila 04]. They show a small high-temperature Ising anisotropy, with a decrease of $\chi_b$ measured for $\mathbf{H} \parallel \mathbf{b}$, and a saturation of $\chi_\perp$ measured for $\mathbf{H} \perp \mathbf{b}$, at temperatures $T < T_N$. This indicates that long-range magnetic ordering is associated with antiferromagnetic moments $\mu_{AF} \parallel \mathbf{b}$. Figures 4.24(c-d) show electrical resistivity versus magnetic field $\mathbf{H} \parallel \mathbf{b}$ curves measured at different temperatures [Grube 06], and magnetization versus magnetic field curves measured at low-temperature for $\mathbf{H} \parallel \mathbf{b}$ and $\mathbf{H} \perp \mathbf{b}$ [Avila 04], respectively. For $\mathbf{H} \parallel \mathbf{b}$, a spin-flop transition at the magnetic field $\mu_0 H_{sf} = 1.6$ T induces step-like anomaly in $\rho(H)$ and $M(H)$. No spin-flop transition is observed for $\mathbf{H} \perp \mathbf{b}$. The antiferromagnetic phase boundary leads to a kink-like anomalies at the critical fields $\mu_0 H_c^\parallel \simeq 8.3$ T for $\mathbf{H} \parallel \mathbf{b}$ and $\mu_0 H_c^\perp \simeq 9.5$ T for $\mathbf{H} \perp \mathbf{b}$. These phase transitions are summarized in the two temperature-magnetic-field phase diagrams presented in Figure 4.24(e) [Grube 07]. The properties of YbNiSi$_3$ correspond exactly to those expected for a weakly-anisotropic Ising antiferromagnet (see Figures 4.16(d-f) in Section 4.3.1.1).

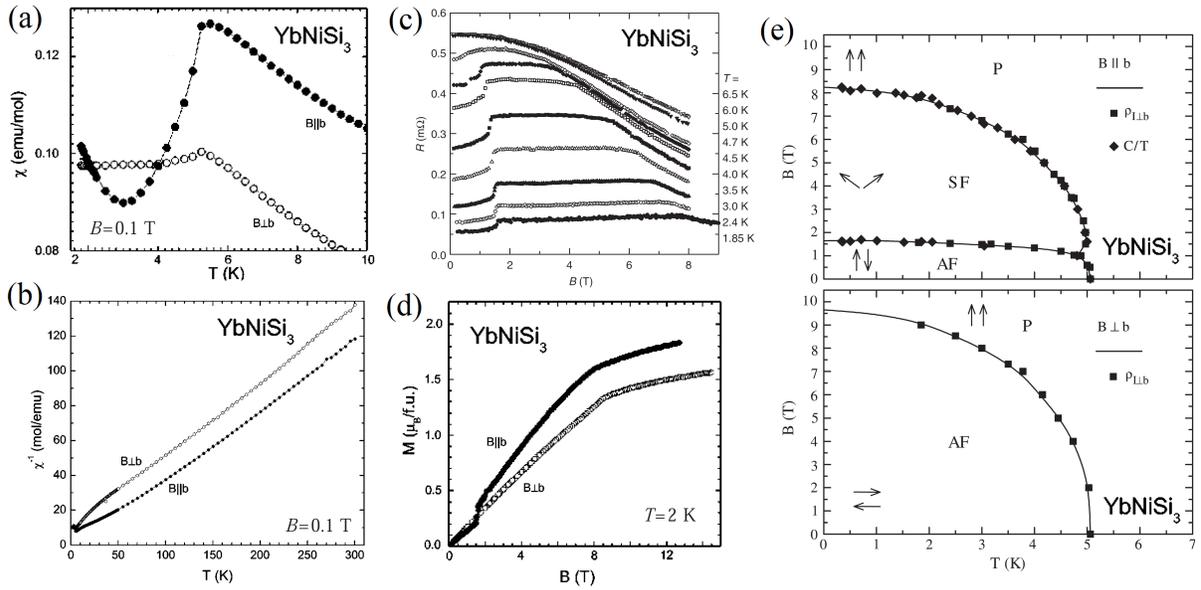

Figure 4.24: (a) Magnetic susceptibility versus temperature and (b) inverse magnetic susceptibility versus temperature for $\mu_0 \mathbf{H} \parallel \mathbf{b}$ and $\mu_0 \mathbf{H} \perp \mathbf{b}$ (from [Avila 04]), (c) electrical resistivity versus magnetic field, at different temperatures, for $\mu_0 \mathbf{H} \parallel \mathbf{b}$ (from [Grube 06]), (d) magnetization versus magnetic field at $T = 2$ K for $\mu_0 \mathbf{H} \parallel \mathbf{b}$ and $\mu_0 \mathbf{H} \perp \mathbf{b}$ (from [Avila 04]), and (e) temperature-magnetic-field phase diagrams for $\mu_0 \mathbf{H} \parallel \mathbf{b}$ and $\mu_0 \mathbf{H} \perp \mathbf{b}$ of YbNiSi$_3$ (from [Grube 07]).

### 4.3.1.5 Weak XY anisotropy

As well as the weakly-Ising heavy-fermion antiferromagnets, there are a few weakly-XY heavy-fermion antiferromagnets. Here, the high-magnetic-field properties of one of them, the antiferromagnet CeRhIn$_5$ are considered. This system belongs to the family of '115' heavy-fermion superconductors, for which amongst the highest superconducting temperatures $T_{sc}$ up to 2.3 K in Ce-based system have been reported [Sarrao 07]. While its magnetic susceptibilities $\chi_c > \chi_a$ show a large Ising magnetic anisotropy (see Figure 4.2(b) in Section 4.1.1) [Settai 97], long-range magnetic order sets in at temperatures $T < T_N = 3.8$ K with antiferromagnetic moments $\mu_{AF} \perp \mathbf{c}$, indicating a XY magnetic anisotropy, and with the incommensurate wavevector $\mathbf{k} = (1/2, 1/2, 0.297)$ (see Figure 4.3(c) in Section 4.1.2) [Bao 00]. This XY-planar magnetic anisotropy presumably results from magnetic exchange interactions, which develop at low temperatures and compete with the Ising magnetic anisotropy driven by high-temperature crystal-field effects. Figure 4.25(b) shows the magnetic-field-temperature phase diagrams determined from heat-capacity measurements for $\mathbf{H} \parallel \mathbf{a}, \mathbf{c}$ [Jiao 15]. The weak anisotropy is shown by the similar critical fields $H_c^a \simeq H_c^c \simeq 50$ T found for the two field-direction $\mathbf{H} \parallel \mathbf{a}, \mathbf{c}$, respectively. Another indication for a weak magnetic anisotropy is the observation of transverse spin-wave excitations by inelastic neutron scattering (see Figures 4.5(a-b) in Section 4.1.2) [Das 14, Stock 15]. Figure 4.25 (a) shows the magnetic-field variation of the magnetization measured at $T = 1.3$ K for $\mathbf{H} \parallel \mathbf{c}$ and $\mathbf{H} \perp \mathbf{c}$ [Takeuchi 01]. For $\mathbf{H} \parallel \mathbf{c}$, no phase transition is observed and $M(H)$ increases almost linearly. A reduction of the slope indicates an approaching saturation, with $M > 1.5$ $\mu_B$ for $\mu_0 H > 50$ T. For $\mathbf{H} \perp \mathbf{c}$, a small step-like increase of $M(H)$ is observed at $\mu_0 H^* \simeq 2$ T and a kink in $M(H)$ at $\mu_0 H_c \simeq 50$ T indicates the boundary of the antiferromagnetic phase.

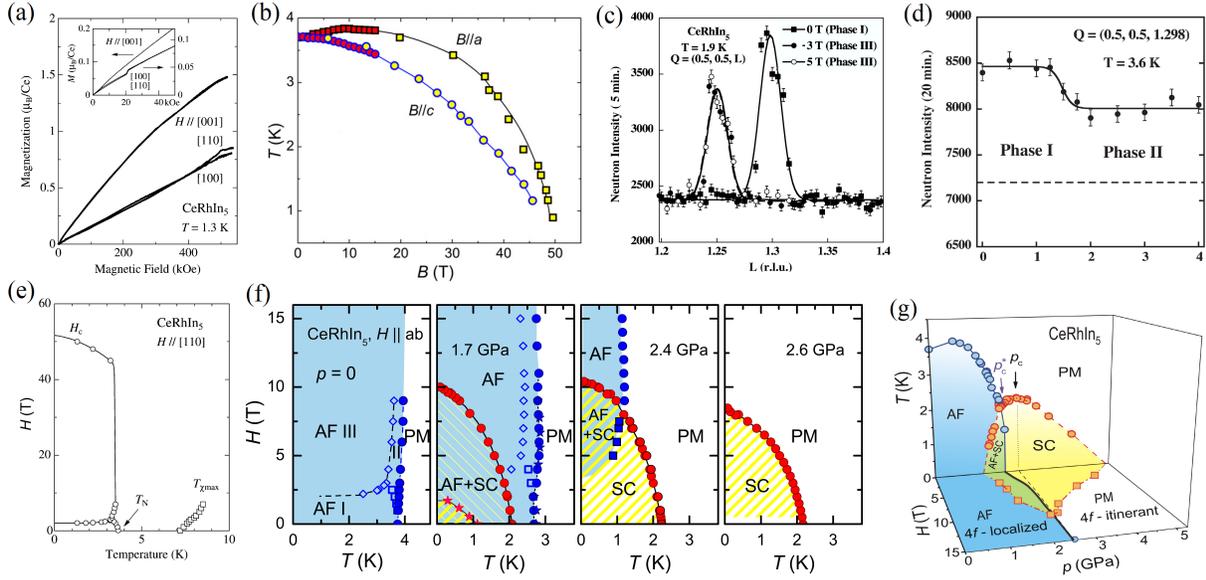

Figure 4.25: (a) Magnetization versus magnetic field at $T = 1.3$ K for $\mathbf{H} \parallel [100]$, [110], and [001] (from [Takeuchi 01]), magnetic-field-temperature phase diagrams for $\mathbf{H} \parallel \mathbf{a}, \mathbf{c}$ (from [Jiao 15]), (c) $(1/2, 1/2, Q_l)$ scans of neutron diffracted intensity at $T = 1.9$ K in a magnetic field $\mathbf{H} \parallel [1\bar{1}0]$, (d) magnetic-field dependence of neutron diffracted intensity at $T = 3.6$ K and momentum transfer $\mathbf{Q} = (1/2, 1/2, 1.298)$ (from [Raymond 07b]), (e) temperature-magnetic-field phase diagram for $\mu_0\mathbf{H} \parallel [110]$ up to 60 T (from [Takeuchi 01]), (f) temperature-magnetic-field phase diagram for $\mu_0\mathbf{H} \perp \mathbf{c}$ up to 16 T under different pressures (from [Knebel 11a]), and (g) three-dimensional pressure-magnetic-field-temperature phase diagram for $\mu_0\mathbf{H} \perp \mathbf{c}$ up to 15 T (from [Knebel 11b]) of CeRhIn$_5$.

The temperature-magnetic-field phase diagram deduced from magnetization measurements with $\mathbf{H} \perp \mathbf{c}$ is shown in Figure 4.25 (e) [Takeuchi 01] and in the left panel of Figure 4.25 (f) [Knebel 11a]. The zero-field antiferromagnetic phase (phase I) corresponds to a non-collinear helicoidal order with wavevector $\mathbf{k}_{inc} = (1/2, 1/2, 0.298)$, where the magnetic moments $\mu_{AF} \perp \mathbf{c}$ are ordered ferromagnetically within planes $\perp \mathbf{c}$ and spiral transversely along the direction **c** [Bao 00, Curro 00]. As shown in Figure 4.25 (c), the phase for $H > H^*$ (phase III) is characterized by a small modification of the magnetic wavevector, which becomes commensurate and equals $\mathbf{k}_c = (1/2, 1/2, 1/4)$, and by a transverse alignment of the antiferromagnetic moments $\mu_{AF} \perp (\mathbf{c}, \mathbf{H})$ [Raymond 07b]. Similarly to a spin-flop transition or to a domain-alignment expected for weakly-Ising- or XY- anisotropic antiferromagnets, respectively (see Figures 4.16(d-f) and (g-i) in Section 4.3.1.1), the transition at $H^*$ results from an effective transverse anisotropy induced by a magnetic field $\mathbf{H} \perp \mathbf{c}$. In a field tilted by $\simeq 20°$ towards the direction $\mathbf{c}$, a transition at $\mu_0 H^* = 28$ T was reported from electrical-resistivity measurements. A peculiar anisotropic planar electrical resistivity observed for $H^* < H < H_c \simeq 50$ T was interpreted as the signature of a nematic state [Ronning 17]. Alternatively to a nematic scenario, a magnetic-moment reorientation, in relation with an effective transverse magnetic anisotropy induced by the magnetic field, could drive the anisotropy of the in-plane resistivity (see for instance [Köhler 93, Tanatar 10, Chu 10, Fisher 11, Zapf 14]. Figure 4.25 (d) further shows the presence, at temperatures $T \lesssim T_N$, of a high-field phase (phase II) with the incommensurate wavevector $\mathbf{k}_{inc}$, with an antiferromagnetic moment smaller than that of the zero-field phase I.

Figures 4.25 (f-g) present the temperature-magnetic-field phase diagrams at different pressures, and the three-dimensional pressure-magnetic-field-temperature phase diagram of CeRhIn$_5$ in a magnetic field $\mathbf{H} \perp \mathbf{c}$ [Knebel 11a, Knebel 11b]. They emphasize that a superconducting phase develops in the vicinity of the quantum magnetic phase transition induced in this system under pressure. Interestingly, a magnetic-field-induced reentrance of the antiferromagnetic phase is observed at pressures $p \gtrsim p_c$ for $\mathbf{H} \perp \mathbf{c}$ [Knebel 06, Knebel 11a, Park 06] and for $\mathbf{H}$ tilted by $\simeq 20°$ towards the direction $\mathbf{c}$ [Helm 20]. At ambient pressure, an increase of $T_N$ was also evidenced by heat-capacity measurements in low magnetic fields $\mathbf{H} \parallel \mathbf{a}$ (see Figure 4.25(b)) [Jiao 15]. In the low-dimensional hexagonal antiferromagnet BaNi$_2$V$_2$O$_8$ in a magnetic field $\mathbf{H} \perp \mathbf{c}$, a similar increase of $T_N$ was proposed to result from the field-induced effective transverse anisotropy [Knafo 07, Knafo 08]). Similarly, the field-stabilization of antiferromagnetism in CeRhIn$_5$ at ambient pressure, and its field-reentrance for $p \gtrsim p_c$ could be a consequence of a transverse magnetic anisotropy induced by the magnetic field.

Experiments on other weak XY antiferromagnets and modelings of their magnetic properties are needed to determine whether the behavior reported for CeRhIn$_5$ is representative, or not, of this class of materials.

## 4.3.2 Electronic properties

We have seen that, within a first approximation, the magnetic-field-temperature phase diagrams of heavy-fermion antiferromagnets can be explained by simple magnetic-polarization processes in relation with the magnetic anisotropy. However, more subtle effects generally need to be also considered. In Section 4.3.2.1, we show that intermediate antiferromagnetic phases, associated with a change of the magnetic structure, can be induced in fields smaller than the antiferromagnetic phase boundary $H_c$, i.e., before reaching a PPM regime. In Section 4.3.2.2, the enhancement of the effective masses are shown to indicate the presence of critical magnetic fluctuations at $H_c$. Section 4.3.2.3 presents dHvA measurements, which demonstrate that Fermi-surface reconstructions accompany the AF-to-PPM transition induced by a mag-netic field. Finally, Section 4.3.2.4 shows that the valence of a heavy-fermion antiferromagnet is weakly unaffected at $H_c$.

### 4.3.2.1 Intermediate phases

In heavy-fermion antiferromagnets, a magnetic field can induce an intermediate antiferromagnetic phase prior to the high-field polarized paramagnetic regime. In some of these systems, neutron diffraction showed that a change of the magnetic structure characterizes the intermediate magnetic phase. The case of $CeRhIn_5$ in a magnetic field $\mathbf{H} \parallel [110]$ was already presented in Section 4.3.1.5. Other examples are considered in this Section.

Figure 4.26 summarizes the properties of $Ce_{0.8}La_{0.2}Ru_2Si_2$ in a magnetic field $\mathbf{H} \parallel \mathbf{c}$. This antiferromagnet of Néel temperature $T_N = 5.8$ K lies in the vicinity of a quantum magnetic phase transition (see Figure 4.6(a) in Section 4.2.1, and Figure 4.10(c) in Section 4.2.2). Figure 4.26(a) shows the magnetic-field-variation of its low-temperature diffracted intensity $I(\mathbf{k} = 0)$ characteristic of ferromagnetic order, measured at the momentum transfer $Q = (1, 1, 0)$, and related to the magnetization $M$ by $I(\mathbf{k} = 0) \propto M^2$ [Jacoud 91]. It indicates two transitions at the fields $\mu_0 H_A = 0.8$ T and $\mu_0 H_B = 1.5$ T preceding the antiferromagnetic-phase-boundary critical field $\mu_0 H_c = 3.1$ T. These field-induced transitions are accompanied by changes of the antiferromagnetic structure, as shown by the neutron diffracted intensity at antiferromagnetic wavevectors $\mathbf{k} \neq 0$ shown at $T = 1.4$ K in Figure 4.26(b) and at $T = 4$ K in Figure 4.26(c) [Jacoud 91]. The temperature-magnetic-field phase diagram in Figure 4.26(d) indicates three different antiferromagnetic phases, separated by large hysteresis domains [Jacoud 91, Mignot 91a]. While antiferromagnetic phases I and II in low fields are ordered with the magnetic wavevector $\mathbf{k}_1 = (0.31, 0, 0)$, the antiferromagnetic phase IV, which develops in high fields at temperatures $T \lesssim T_N$, is ordered within a superposition of sine modulations of moments with the wavevectors $\mathbf{k}_1 = (0.31, 0, 0)$, $\mathbf{k}_2 = (0.31, 0.31, 0)$, and of a ferromagnetic component of wavevector $\mathbf{k} = 0$ corresponding to the polarization of the moments. The low-temperature and high-field phase III is ordered with the wavevector $\mathbf{k}_2^c = (1/3, 1/3, 0)$, which corresponds to a 'commensurate modification' of wavevector $\mathbf{k}_2$. A similar phase diagram was found for $CeRu_2(Si_{0.9}Ge_{0.1})_2$, which is also an antiferromagnet derived from $CeRu_2Si_2$ [Mignot 91b].

Figure 4.27 presents other examples of intermediate phases induced in heavy-fermion antiferromagnets in a magnetic field $H < H_c$.

- Figures 4.27 (a-b) show the low-temperature magnetization and electrical resistivity, and the magnetic-field-temperature phase diagram of the Ising antiferromagnet $CeNiGe_3$, of Néel temperature $T_N = 5.5$ K, in a magnetic field applied along its easy magnetic axis $\mathbf{a}$ [Mun 10]. Two first-order like transitions are observed at $\mu_0 H^* = 18.5$ T and at the antiferromagnetic phase boundary $\mu_0 H_c = 31$ T. An intermediate antiferromagnetic phase is stabilized in magnetic fields $H^* < H < H_c$. As shown in Figure 4.27 (c-d), very similar properties are observed in the Ising antiferromagnet $CeAuSb_2$ of Néel temperature $T_N = 6.6$ K in magnetic fields applied along it easy magnetic axis $\mathbf{c}$ [Zhao 16]. Its electrical resistivity shows two first-order transitions at $\mu_0 H^* = 2.8$ T and at the antiferromagnetic phase boundary $\mu_0 H_c = 5.6$ T, indicating the presence of an intermediate

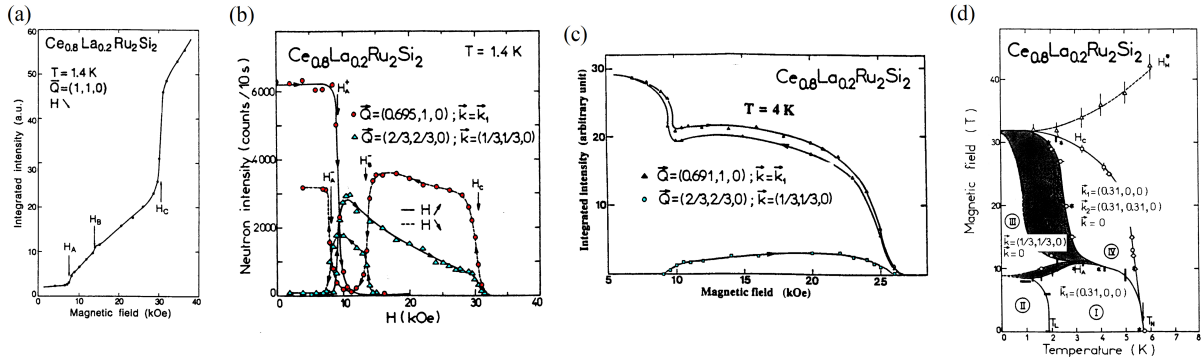

Figure 4.26: Magnetic-field-variation of elastic neutron diffraction intensity (a) at momentum transfer $\mathbf{Q} = (1, 1, 0)$ corresponding to wavevector $\mathbf{k} = (0, 0, 0)$ and (b) at momentum transfers $\mathbf{Q}_1 = (0.69, 1, 0)$ and $\mathbf{Q}_2^c = (2/3, 2/3, 0)$ corresponding to wavevectors $\mathbf{k}_1 = (0.31, 0, 0)$ and $\mathbf{k}_2^c = (1/3, 1/3, 0)$, respectively, at temperature $T = 1.4$ K, (c) at momentum transfers $\mathbf{Q}_1 = (0.69, 1, 0)$ and $\mathbf{Q}_2^c = (2/3, 2/3, 0)$ at temperature $T = 4$ K (from [Jacoud 91]), and (d) temperature magnetic phase diagram of $Ce_{0.8}La_{0.2}Ru_2Si_2$ in a magnetic field $\mathbf{H} \parallel \mathbf{c}$ (adapted from [Jacoud 91, Mignot 91a]).

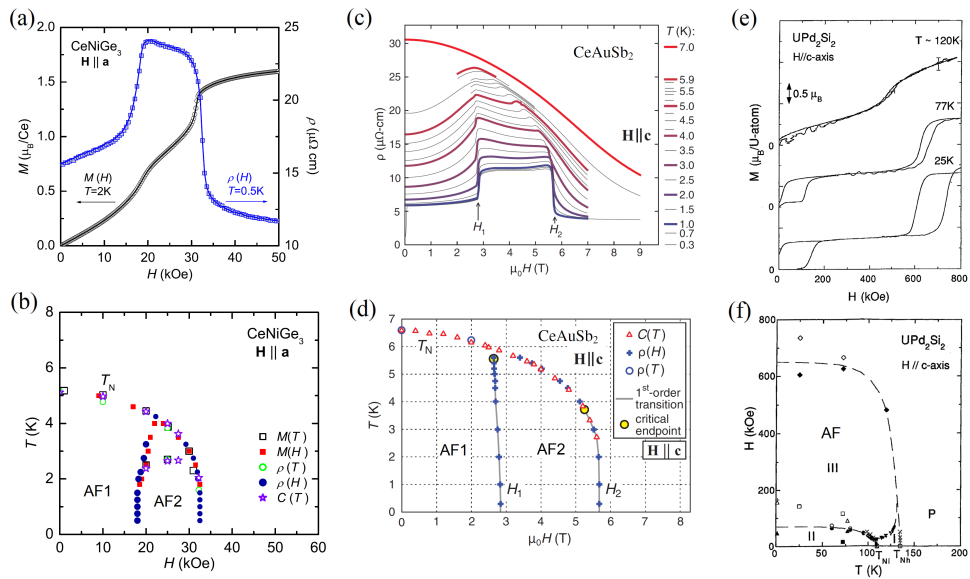

Figure 4.27: (a) Low-temperature magnetization and electrical resistivity versus magnetic field and (b) magnetic-field-temperature phase diagram of $CeNiGe_3$ in $\mathbf{H} \parallel \mathbf{a}$ (from [Mun 10]). (c) electrical resistivity versus magnetic field at different temperatures and (d) magnetic-field-temperature phase diagram of $CeAuSb_2$ in $\mathbf{H} \parallel \mathbf{c}$ (from [Zhao 16]). (e) magnetization versus magnetic field at different temperatures and (f) temperature-magnetic-field phase diagram of $UPd_2Si_2$ in $\mathbf{H} \parallel \mathbf{c}$ (from [Honma 98]). (1 T=$10^4$ Oe)

antiferromagnetic phase for $H^* < H < H_c$. The electrical resistivity and magnetization variations in these two compounds look similar than those reported for the weakly-Ising-anisotropic antiferromagnet YbNiSi$_3$, in a magnetic field applied along its easy axis **b** (see Figure 4.24 in Section 4.3.1.4). This could be the signature of a spin-flop transition at $H^*$ in the moderate-Ising antiferromagnets CeNiGe$_3$ and CeAuSb$_2$, the relatively small ratio $H_c/H^* \simeq 2$ resulting from a stronger anisotropy than in YbNiSi$_3$, for which $H_c/H^* \simeq 5$. However, CeNiGe$_3$ and CeAuSb$_2$ are characterized by an incommensurate magnetic structure at zero field, with the magnetic wavevectors **k** $= (0, 0.41, 1/2)$ [Ikeda 15] and **k** $= (0.135, 0.135, 1/2)$ [Yumnam 19], respectively. Alternatively to a spin-flop picture, and similarly to the Ce$_{0.8}$La$_{0.2}$Ru$_2$Si$_2$ and CeRu$_2$(Si$_{0.9}$Ge$_{0.1}$)$_2$ cases considered earlier in this Section, or to the CeRhIn$_5$ case considered in Section 4.3.1.5, the transition at $H^*$ in CeNiGe$_3$ and CeAuSb$_2$ may be driven by a modification of the magnetic structure. Neutron diffraction experiments are needed to establish the nature of the field-induced antiferromagnetic phases in these two systems.

- Figures 4.27 (e-f) present magnetization versus magnetic field curves at different temperature and the temperature-magnetic-field phase diagram, respectively, of the Ising antiferromagnet UPd$_2$Si$_2$ in a magnetic field **H** $\parallel$ **c** [Honma 98]. At zero-field, this system orders antiferromagnetically with the incommensurate wavevector **k**$_I = (0, 0, 0.73)$ at temperatures $T_N^l = 108$ K $< T < T_N = 135$ K and with the commensurate wavevector **k**$_{II} = (0, 0, 1)$ at temperatures $T < T_N^l$. At low-temperature, two hysteretic first-order metamagnetic transitions are induced at $\mu_0 H^* \simeq 0 - 20$ T and at the antiferromagnetic phase boundary $\mu_0 H_c \simeq 60 - 70$ T. The phase induced for $H > H^*$ corresponds to an antiferromagnetic order with the commensurate wavevector **k**$_{III} = (0, 0, 2/3)$ [Collins 93]. Interestingly, antiferromagnetic ordering with the same wavevector is established in the isostructural compound UNi$_2$Si$_2$ [Lin 91] at zero field, while an ordered phase with a wavevector **k** $= (2/3, 0, 0)$ is stabilized in U(Ru$_{0.96}$Rh$_{0.04}$)$_2$Si$_2$ in magnetic fields $\mu_0 H > 24$ T [Kuwahara 13] and in U(Ru$_{0.92}$Rh$_{0.08}$)$_2$Si$_2$ in magnetic fields $\mu_0 H > 21.6$ T [Prokeš 17b] (see also Chapter 6).

Heavy-fermion compounds are the place of magnetic interactions which induces magnetic fluctuations and long-range-ordered magnetic phases. In the parent paramagnetic compound CeRu$_2$Si$_2$, intersite magnetic fluctuations with the three main wavevectors **k**$_1$, **k**$_2$, and **k**$_3$ (see Figure 3.4 in Section 3.1.2) indicate the proximity to quantum phase transitions. While La- and Ge-dopings lead to the stabilization of long-range magnetic order with wavevector **k**$_1$, Rh-doping leads to the stabilization of long-range magnetic order with wavevector **k**$_3$ [Quezel 88, Haen 02, Mignot 91a, Watanabe 03]. We have further seen that the combination of doping and magnetic field permits to stabilized long-range magnetic order with **k**$_2$. More generally, the presence of intersite magnetic fluctuations, i.e., short-range order, with a given wavevector **k** in a quantum magnet can be considered as an indication for a nearby quantum magnetic phase transition, associated with long-range magnetic ordering with the same wavevector **k**. In many heavy-fermion magnets, an appropriate tuning (pressure, doping, or magnetic field) can lead to the increase of the magnetic correlations with **k**, ending with the stabilization of long-range magnetic order via a quantum phase transition.

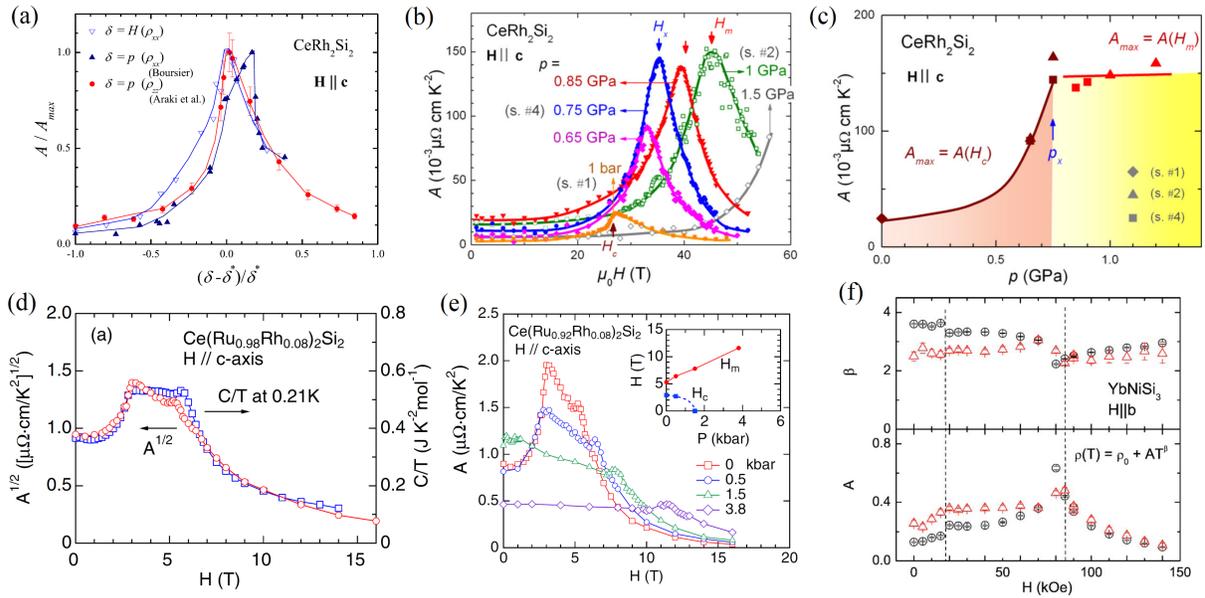

Figure 4.28: Comparison of variation of Fermi-liquid quadratic coefficient $A$ in a selection of heavy-fermion antiferromagnets in a magnetic field. (a) Comparison of magnetic-field variation, at ambient pressure, and pressure-variation, at zero-field, of $A$ in $CeRh_2Si_2$ in a magnetic field $\mathbf{H} \parallel \mathbf{c}$ (from [Knafo 10]). (b) Magnetic-field variation of $A$ at different pressures, and (c) pressure-dependence of the maximum value of $A(H)$, which is equal to $A(H_c)$ for $p < p_x = 0.75$ GPa and to $A(H_m)$ for $p > p_x$, of $CeRh_2Si_2$ in a magnetic field $\mathbf{H} \parallel \mathbf{c}$ (from [Knafo 17]). (d) Comparison of the magnetic-field variations of $\sqrt{A}$ and the heat-capacity Sommerfeld coefficient $\gamma =_{T \to 0} C_p/T$ and (e) magnetic-field variation of $A$ at different pressures of $Ce(Ru_{0.92}Rh_{0.08})_2Si_2$ in a magnetic field $\mathbf{H} \parallel \mathbf{c}$ (from [Aoki 12a]). The Inset of (e) presents the pressure-variations of the critical field $H_c$ and of the metamagnetic field $H_m$. (f) Magnetic-field-variation of the coefficients $A$ and $\beta$ extracted from fits to the electrical resistivity of $YbNiSi_3$, in a magnetic field $\mathbf{H} \parallel \mathbf{b}$, by $\rho = \rho_0 + AT^\beta$ (from [Bud'ko 07]).

### 4.3.2.2 Effective mass and magnetic fluctuations

We have seen in Sections 3.2.2 and 4.2.2 that the metamagnetic transition of heavy-fermion paramagnets, and the antiferromagnetic phase transition of heavy-fermion antiferromagnets can be the place of critical magnetic fluctuations, of ferromagnetic and antiferromagnetic nature, respectively. Here, a focus is made on the magnetic-field-induced phase transition of heavy-fermion antiferromagnets, for which an enhancement of the effective mass also indicate the presence of quantum critical magnetic fluctuations at $H_c$. It was emphasized that a non-Fermi-liquid behavior is induced at the critical field of the antiferromagnet $YbRh_2Si_2$ (see Figure 2.27 in Section 2.4.2.2) [Gegenwart 02, Custers 03]. Here, a focus is made on Fermi-liquid behaviors, which indicate an enhancement of the effective mass and, thus, of critical magnetic fluctuations under magnetic field. Three heavy-fermion antiferromagnets are considered.

- Figure 4.28(a) shows variations of the electrical-resistivity quadratic coefficient $A$ of the antiferromagnet $CeRh_2Si_2$ under pressure (at zero-field) and magnetic field applied along the easy magnetic axis $\mathbf{c}$ (at ambient pressure) [Knafo 10]. The same enhancement of $A$ by a factor $\simeq 10$ is induced at the critical pressure $p_c$ for $H = 0$, and at the critical field $H_c$ for $p = 1$ bar, indicating similar increases of the critical magnetic fluctuations at the pressure- and magnetic-field-boundaries of the antiferromagnetic phase (see Figure 4.7(c) in Section 4.2 and Figures 4.18(c) and 4.19(d-e) in Section 4.3.1.2). Figures 4.28(b-c) show that, under combined pressure and magnetic field, a much larger enhancement of the coefficient $A$, by a factor $\simeq 75$ in comparison with its zero-field and ambient pressure value, is observed on the $H_m$ line corresponding to the CPM $\rightarrow$ PPM transition [Knafo 17]. This enhancement indicates stronger critical magnetic fluctuations at the metamagnetic phase boundary of the CPM regime than at the boundaries $H_c$ and $p_c$ of the antiferromagnetic phase of $CeRh_2Si_2$.

- Figure 4.28(d) presents a comparison of the square root $\sqrt{A}$ of the electrical resistivity quadratic coefficient and of the heat-capacity Sommerfeld coefficient $\gamma$ of the antiferromagnet $Ce(Ru_{0.92}Rh_{0.08})_2Si_2$ in a magnetic field $\mathbf{H} \parallel \mathbf{c}$ (see Figure 4.20 in Section 4.3.1.2) [Aoki 12a]. Similar field-variations indicate that a Fermi-liquid description with an effective mass $m^* \sim \sqrt{A} \sim \gamma$ is appropriate. In this sample, $H_c$ and $H_m$ are decoupled at low temperature, and similar enhancements of $A$ and $\gamma$ are observed at these two transitions. Figure 4.28(e) further shows the evolution of $A$ versus $H$ at different pressures up to 3.8 kbar, emphasizing a reduction of the maximum values of $A$ observed at $H_c$, which decreases under pressure before vanishing at $p_c = 2$ kbar, and $H_m$, which increases under pressure.

- Figure 4.28(f) presents the results from a $\rho = \rho_0 + AT^\beta$ analysis of the electrical resistivity of the weakly-anisotropic antiferromagnet $YbNiSi_3$ in a magnetic field $\mathbf{H} \parallel \mathbf{b}$ [Bud'ko 07] (see Figure 4.24 in Section 4.3.1.4). Values of $\beta \gtrsim 2$ indicate that a Fermi-liquid quadratic description of the electrical resistivity is, within first approximation, appropriate. The spin-flop transition at $\mu_0 H_{sf} = 1.6$ T is accompanied by a small increase of $A$, which saturates at fields $H \gtrsim H_{sf}$, before passing through a maximum at $\mu_0 H_c = 8.5$ T. Here also, a Fermi-liquid description indicates the presence of critical magnetic fluctuations at the antiferromagnetic phase boundary $H_c$.

For the systems considered here, the variations of the electrical-resistivity quadratic coefficient $A$ and of the heat-capacity Sommerfeld coefficient $\gamma$ indicate an enhancement of the magnetic fluctuations at the antiferromagnetic boundary $H_c$ and at the boundary $H_m$ of the CPM regime (see Section 3.2.2). While critical ferromagnetic fluctuations have been evidenced at the metamagnetic field $H_m$ of the heavy-fermion paramagnet $CeRu_2Si_2$, the nature of the critical fluctuations at the boundary $H_c$ of a heavy-fermion antiferromagnetic have not been identified so far. Further effort is now needed to determine their nature and their relation with the enhancement of the effective mass, as determined from an analysis of heat-capacity and electrical-resistivity Fermi-liquid behaviors. For this purpose, inelastic neutron scattering experiments could be performed on an antiferromagnet under a magnetic field. The system

Ce(Ru$_{0.92}$Rh$_{0.08}$)$_2$Si$_2$, where $H_c$ and $H_m$ are decoupled at ambient pressure, could be an appropriate target. The characterization of the magnetic fluctuations at $H_c$ and $H_m$ may help understanding the different values of $A$ observed at these two characteristic field ($A(H_m) \gg A(H_c)$ in CeRh$_2$Si$_2$, $A(H_m) < A(H_c)$ in Ce(Ru$_{0.92}$Rh$_{0.08}$)$_2$Si$_2$).

### 4.3.2.3 Fermi surface

Here, the evolution of the Fermi-surface of antiferromagnets in a magnetic field is considered [Matsumoto 08, Matsumoto 10]. Figure 4.29(a) presents the three-dimensional magnetic-field-doping-temperature phase diagram of Ce$_{1-x}$La$_x$Ru$_2$Si$_2$, which is composed of an antiferromagnetic ground state stabilized for $x > x_c = 7.5$ % and a correlated paramagnetic regime stabilized for $x < x_c$ (see Section 4.2.2). A Kondo magnetic impurity limit is reached for $x \to 1$, for which $n_{4f} \to 0$, and the properties of pure LaRu$_2$Si$_2$ are considered as those of a non-$4f$-reference. In a magnetic field applied along the easy magnetic direction **c**, a transition from the AF ground state to a polarized paramagnetic regime is induced at $\mu_0 H_c$, which increases almost linearly with $(1-x)$ up to $\simeq 4$ T at $x_c$. A transition from the CPM ground state to a PPM regime is induced at $\mu_0 H_m$, which also increases almost linearly with $(1-x)$ up to $\simeq 7.8$ T at $x = 0$. Two sets of dHvA experiments are presented.

- The first one, whose results are shown in Figure 4.29(c), corresponds to experiments in a magnetic field **H** $\parallel$ **a**. Here, no transition is induced by the magnetic field, permitting to compare the properties of the AF and CPM ground states [Matsumoto 10]. For $x > 0.9$, i.e., for $n_{4f} < 0.1$, the frequencies and associated cyclotron masses extracted from the dHvA measurements are very similar to those of the non-$4f$-reference LaRu$_2$Si$_2$, indicating the validity of a picture with localized $4f$ electrons. For $x < 0.8$, the orbits $\delta$ (also labelled by $\beta_3$), $R_1$ and $R_2$ are not observed, indicating a different Fermi surface from that of the LaRu$_2$Si$_2$ reference. When $x$ becomes closer to 0, $n_f$ becomes closer to 1 and continuous modifications of the Fermi surface occur. The frequencies of the observed branches decrease continuously when $x$ is decreased, with a faster variation in the CPM regime for $x < x_c$. Their associated cyclotron masses increase first, pass through a maximum at $x_c$, and then decrease in the CPM regime. These modifications indicate that the $4f$ electrons participate to the Fermi surface and, thus, have an itinerant character.

- The second set of dHvA experiments, presented in Figure 4.29(d), was performed in magnetic fields **H** $\parallel$ **c**, with $H > H_N, H_c$. It compares the evolution of the Fermi surfaces of Ce$_{1-x}$La$_x$Ru$_2$Si$_2$ compounds in their PPM regime. For all La-contents $x$, three orbits of the branches $\beta_1$, $\beta_2$, and $\beta_3$ characteristic of the non-$4f$-reference LaRu$_2$Si$_2$ are observed, indicating similar Fermi surfaces. These three frequencies vary monotonously with $x$, with a faster rate when approaching $x_c$ and for $x < x_c$. The associated cyclotron mass also strongly varies but no maximum is observed at $x_c$, probably because there is no phase transition at $x_c$ for $H > H_N, H_c$. The interpretation of this second set of data is not trivial. On one hand, the similarity of all Fermi surfaces may indicate that all compounds in their PPM regimes correspond to a localized limit of $4f$ electrons. On the other hand, the variations of the dHvA frequencies and cyclotron masses may result from a contribution of the $4f$ electrons to the fermi surface.

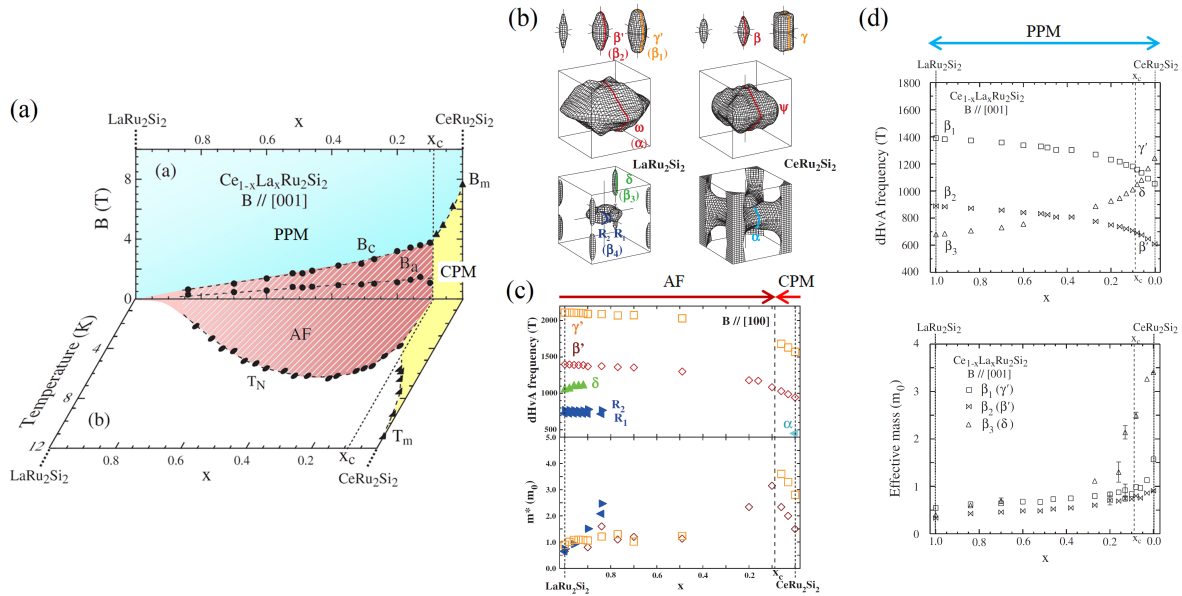

Figure 4.29: (a) Three-dimensional magnetic-field-doping-temperature phase diagram of $Ce_{1-x}La_xRu_2Si_2$ in a magnetic field **H** ∥ **c**. (b) Three-dimensional sketch of the Fermi surfaces calculated for $LaRu_2Si_2$ and $CeRu_2Si_2$, where the orbits of extremum surface perpendicular to the axis **a** are indicated. (c) Doping-$x$ variation of the dHvA frequencies and corresponding cyclotron masses, extracted for **H** ∥ **a**, i.e., with the AF → CPM phase boundary being crossed, and (d) Doping-$x$ variation of the dHvA frequencies and corresponding cyclotron masses, extracted for **H** ∥ **c**, with $H > H_c, H_m$ and a PPM ground state for all $x$ (adapted from [Matsumoto 08, Matsumoto 10]).

As shown here and in Sections 3.2.3 and 4.2.3, dHvA experiments demonstrate that the Fermi surface of heavy-fermion compounds strongly varies at their quantum magnetic phase transitions induced under pressure and magnetic field. They indicate that $4f$ electrons are more localized in the antiferromagnetic phase than in the correlated paramagnetic regime, and that a magnetic field leads to an increase of their localized character. However, extracting the degree of localization or itinerancy of the $4f$ electrons from dHvA experiments is not trivial. In the next Section, measurements of valence, which is an alternative indication of the degree of localization of the $4f$ electrons, will be considered.

#### 4.3.2.4 Valence

As shown in Section 4.2.4, XAS measurements indicate that heavy-fermion systems, and in particular those in an antiferromagnetic state, are very close to an integer-valence limit, corresponding to nearly-localized $4f$ electrons in Ce compounds (see for instance [Yamaoka 14]). In

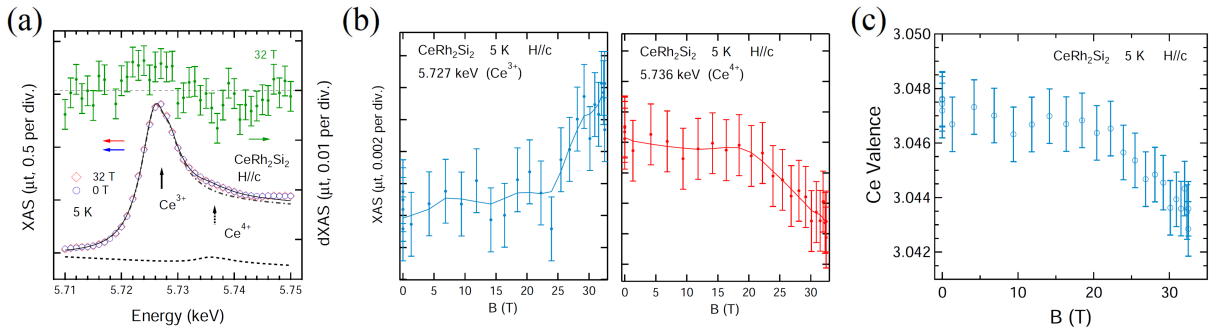

Figure 4.30: (a) x-ray absorption spectroscopy spectra measured at $T = 5$ K in a magnetic field $\mu_0 H = 0$ and 32 T, (b) magnetic-field dependence of the XAS intensity at the $Ce^{3+}$ and $Ce^{4+}$ positions, and (c) magnetic-field dependence of valence extracted from these low-temperature spectra on $CeRh_2Si_2$ in a magnetic field $\mathbf{H} \parallel \mathbf{c}$ (from [Matsuda 10]).

Figure 3.9 from Section 3.2.1, it was also shown that the valence of a heavy-fermion Ce-based paramagnet approaches its integer value and, thus, that the electron localization increases, in a magnetic field applied along the easy magnetic axis [Matsuda 12]. Here a focus is made on the case of a heavy-fermion antiferromagnet in a magnetic field along its easy magnetic axis

Figure 4.30 presents a XAS study of valence in the heavy-fermion antiferromagnet $CeRh_2Si_2$ through its strong first-order metamagnetic transition, observed in a magnetic field $\mu_0\mathbf{H} \parallel \mathbf{c}$ of 26 T [Matsuda 10] (see Figure 4.18 in Section 4.3.1.2). Figure 4.30(a) shows that a tiny difference can be seen from a comparison of two XAS spectra measured on both sides of the metamagnetic transition, at $\mu_0 H = 0$ and 32 T. Figure 4.30(b) presents the magnetic-field variation of the XAS intensity at two energies corresponding to the maxima of the integer valences $v = 3$ and 4, corresponding respectively to $Ce^{3+}$ and $Ce^{4+}$ configurations. While the intensities are almost field-independent for $\mu_0 H < 20$ T, the $Ce^{3+}$ weight increases and the $Ce^{4+}$ weight decreases for $\mu_0 H > 20$ T. Integration of these data permits to extract the valence variation with field plotted in Figure 4.30(c). It shows a constant valence for $\mu_0 H < 20$ T and a valence decrease $\Delta v \simeq -4 \cdot 10^{-3}$ in fields varying from 20 to 32 T. At all magnetic fields, the valence is close to the integer-valent value $v = 3$. The decrease of valence in a magnetic field indicates a progressive localization of the $f$-electrons.

The apparent contradiction between the conclusions from dHvA (see Sections 3.2.3, 4.2.3, and 4.3.2.3) and XAS experiments is a consequence of the dual itinerant-localized electronic character of heavy-fermion systems. While XAS experiments indicate that heavy-fermion systems, and more particularly the compounds in their AF and PPM phases, have nearly-localized $f$-electrons, dHvA experiments highlight the itinerant character of the $f$ electrons in the CPM regime, which is sufficient to induce significant deviations in comparison with a non-$4f$ Fermi surface. In addition, due to the magnetic-form factor, the magnetic properties of the $f$ electrons, i.e., long-range magnetic order and intersite magnetic fluctuations, observed by neutron scattering at non-zero momentum transfers are generally considered as resulting from localized $f$ electrons. We have seen that magnetic fluctuations of nearly-localized $f$-electrons drive the

Table 4.1: Néel temperature, critical magnetic field at the borderline of antiferromagnetism, and corresponding field-direction in various heavy-fermion antiferromagnets.

| Material | $T_N$ (K) | $H_c$ (T) | $\mathbf{H} \parallel$ | |
|---|---|---|---|---|
| $Ce_{1-x}La_xRu_2Si_2$ ($x = 10 \to 30 \to 80$ %) | $3.2 \to 6.3 \to 2$ | $3.6 \to 2.4 \to 0.6$ | **c** | ♯a |
| $CeRu_2(Si_{1-x}Ge_x)_2$ ($x = 10 \to 52$ %) | $6.6 \to 10.4$ | $3 \to 0.4$ | **c** | ♯b |
| $Ce(Ru_{0.92}Rh_{0.08})_2Si_2$ | 4 | 3 | **c** | ♯c |
| $CeCu_2Ge_2$ | 4 | 8 | $[\bar{1}10]$ | ♯d |
| $CeRh_2Si_2$ ($p = 0 \to 0.75 \to 0.85$ GPa) | $36 \to 25 \to 22$ | $26 \to 35.5 \to 24$ | **c** | ♯e |
| $CePd_2As_2$ | 14.7 | 0.95 | **c** | ♯f |
| $CeIn_3$ | 10 | 61 | (cubic) | ♯g |
| $CeIn_{2.75}Sn_{0.25}$ | 6.4 | 42 | (cubic) | ♯h |
| $CeRhIn_5$ | 3.8 | 52 | $[110]$ | ♯i |
| $CeAuSb_2$ | 6.6 | 5.6 | **c** | ♯j |
| $CeNiGe_3$ | 5.5 | 3.25 | **a** | ♯k |
| $YbRh_2Si_2$ | 0.07 | 0.06 | **c** | ♯l |
| $Yb_3Pt_4$ | 2.4 | 1.85 | $\perp$ **c** | ♯m |
| $YbNiSi_3$ | 5.1 | 8.3/9 | $\mathbf{b}/\perp \mathbf{b}$ | ♯n |
| $U_2Zn_{17}$ | 9.7 | 32 | $[11\bar{2}0]$ | ♯o |
| $U(Pd_{1-x}Ni_x)_2Al_3$ ($x = 0 \to 0.5 \to 1$) | $14.5 \to 16.5 \to 4.5$ | $18.5 \to 30 \to 78$ | $\perp$ **c** | ♯p |
| $UPb_3$ | 32 | 21 | **a** | ♯q |
| $U_2Rh_2Sn$ | 25 | 22.5 | **c** | ♯r |
| $URu_2Si_2$ ($p = 0.75 \to 3.9$ GPa) | $6 \to 26$ | $15 \to 48$ | **c** | ♯s |
| $UPt_2Si_2$ | 32 | 45/32 | **a/c** | ♯t |
| $UCo_2Si_2$ | 83 | 45 | **c** | ♯u |
| $UPd_2Si_2$ | 135 | 65 | **c** | ♯v |
| $USb_2$ | 202 | 63 ($T = 80$ K) | **c** | ♯w |
| $UNiAl$ | 19.3 | 11.5 | **c** | ♯x |
| $UNiGe$ | 42 | 25/10 | **b/c** | ♯y |
| $UIrGe$ | 16 | 21/14 | **b/c** | ♯z |
| $UPdIn$ | 20.8 | 16.3 | **c** | ♯aa |
| $U_2Rh_3Si_5$ | 25.7 | 14.1 | **b** | ♯ab |

♯a [Matsumoto 08, Matsumoto 10, Mignot 91a], ♯b [Matsumoto 11, Mignot 91a], ♯c [Aoki 12a], ♯d [Singh 11], ♯e [Knafo 10, Knafo 17],
♯f[Ajeesh17] ♯g [Ebihara 04], ♯h [Silhanek 06], ♯i [Takeuchi 01], ♯j [Zhao 16], ♯k [Mun 10], ♯l [Knebel 06], ♯m [Wu 11],
♯n [Avila 04, Grube 07], ♯o [Tateiwa 11], ♯p [Sakon 02, Mochidzuki 19], ♯q [Sugiyama 02], ♯r [Prokeš 17a],
♯s [Hassinger 08a, Aoki 09b, Knafo 20a], ♯t [Schulze Grachtrup 12], ♯u [Andreev 13], ♯v [Honma 98], ♯w [Stillwell 17], ♯x [Brück 94],
♯y [Havela 92], ♯z [Ramirez 87, Yoshii 06], ♯aa [Sugiura 90], ♯ab [Takeuchi 97].

Fermi-liquid and critical properties of heavy-fermion systems at their quantum magnetic phase transition. A picture based on nearly-localized $f$-electrons driving the magnetic properties, within an almost-integer value of valence, but whose small itinerant character may be sufficient to affect the Fermi surface, may possibly reconcile these different sets of complementary experiments. A description of both magnetic and Fermi surface properties within a unique model constitutes a challenge for theoreticians.

Figure 4.31: Critical antiferromagnetic field versus the Néel temperature for various heavy-fermion antiferromagnets (data from [Matsumoto 08, Matsumoto 10, Mignot 91a, Matsumoto 11, Mignot 91a, Aoki 12a, Singh 11, Knafo 10, Knafo 17, Ajeesh17, Ebihara 04, Silhanek 06, Takeuchi 01, Zhao 16, Mun 10, Knebel 06, Wu 11, Avila 04, Grube 07, Tateiwa 11, Sakon 02, Mochidzuki 19, Sugiyama 02, Prokeš 17a, Hassinger 08a, Aoki 09b, Knafo 20a, Schulze Grachtrup 12, Andreev 13, Honma 98, Stillwell 17, Brück 94, Havela 92, Ramirez 87, Yoshii 06, Sugiura 90, Takeuchi 97]).

## 4.4 Synthesis

Similarly to Section 3.3, where focus was given to heavy-fermion correlated paramagnets, a synthesis of the magnetic properties of a large number of heavy-fermion antiferromagnets is made here. Table 4.1 lists the values of the Néel temperature $T_N$ and of the critical field $H_c$ delimiting the antiferromagnetic ground state of these materials, and the field direction corresponding to the given values of $H_c$. Figure 4.31 presents a plot of $H_c$ versus $T_N$ of these materials, emphasizing the trend that large values of $T_N$ generally correspond to large values of $H_c$. However, the correspondence between these two quantities is not simple, as shown by the large scattering of the data in the graph, most of the data point collapsing in the wide area delimiting by the two lines $H_c = 7T_N$ and $H_c = 0.2T_N$. The non-proportionality of $H_c$ and $T_N$ appears more clearly when the properties of a given compound are tuned by pressure or chemical doping. It indicates that that a single parameter is not sufficient to relate $H_c$ and $T_N$ and, thus, to describe the antiferromagnetic phase of heavy-fermion systems. The presence of a long-range-ordered magnetic structure, sometimes of complex nature, generally combined with gapped and ungapped quasielastic excitations, may explain why multiple parameters are needed to relate $H_c$ and $T_N$ in heavy-fermion antiferromagnets.

This strongly contrasts with the conclusions about heavy-fermion correlated paramagnets, made by in Section 3.3 (Figure 3.14). The quasi-proportionality between the field and temperature boundaries $H_m$ and $T_\chi^{max}$, respectively, of the CPM regime, is compatible with the Fermi-liquid description of this regime driven by a single parameter, the effective mass $m^*$. The CPM regime is generally dominated by a broad quasielastic magnetic-fluctuations spectrum, whose energy linewidth can be identified as the energy scale driving the Fermi-liquid properties, in agreement with theoretical models of magnetic-fluctuations-driven Fermi-liquid behavior (see Section 2.4.3.2).

# Chapter 5

# Ferromagnetism

Heavy-fermion ferromagnets and nearly-ferromagnets are presented in this Chapter. Basic properties at a ferromagnetic phase transition are introduced in Section 5.1. The phenomena, including superconductivity, induced at a quantum ferromagnetic phase transition are presented in Section 5.2, and the effects of a magnetic field are considered in Section 5.3. In complement to this Chapter, a general introduction to ferromagnetism can be found in [Chikazumi 97], a review on the properties of itinerant-electron ferromagnets can be found in [Brando 16], and a review about U-based ferromagnetic superconductors can be found in [Aoki 19a].

## 5.1 Ferromagnetic ground state

The electronic properties of heavy-fermion ferromagnets are considered here. Section 5.1.1 presents signatures of ferromagnetic ordering in bulk thermodynamic and electrical-transport measurements. Section 5.1.2 presents experimental evidences of long-range magnetic order and magnetic fluctuations in these systems.

### 5.1.1 Bulk properties

A ferromagnetic state is characterized by a large magnetization in a magnetic field $H > H_c$, where the critical field $H_c$ is generally of a few hundreds of Oersted, and applied along an easy magnetic axis. Figure 5.1(a) presents the magnetization versus temperature measured on a selection of heavy-fermion U-based ferromagnets, in a magnetic field applied along their easy axis [Tateiwa 17]. The magnetization is enhanced at temperatures below the Curie temperature $T_C$ of these materials. Within first approximation, the magnetization of a ferromagnet is proportional to its order parameter, i.e., the ferromagnetic moment $\mu_{FM}$. $M$ is usually expressed in units of $\mu_B/$ magnetic atom. For the selection of materials considered here, the largest low-temperature magnetization $M = 1.75\ \mu_B/$ U is reached in the compound $UCu_2Ge_2$ with the largest $T_C \simeq 100$ K, while the smallest low-temperature magnetization $M = 0.4\ \mu_B/$ U is reached in the compound URhGe with the smallest $T_C = 9.5$ K. Reduced ferromagnetic

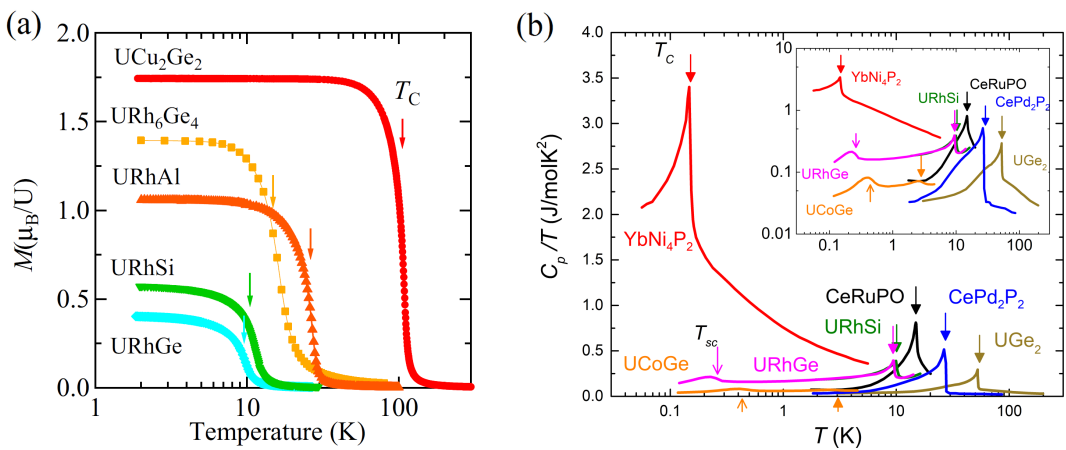

Figure 5.1: (a) Magnetization versus temperature in a magnetic field applied along the easy magnetic axis (from [Tateiwa 17]) and (b) heat capacity divided by temperature versus temperature for a selection of heavy-fermion ferromagnets (data from [Honda 03, Krellner 07, Aoki 11b, Troc 12, Steppke 13, Tran 14]). The non-electronic background has been subtracted to the heat capacity of UGe$_2$, CeRuPO, and CePd$_2$Si$_2$.

moment and Curie temperature generally indicate a nearby quantum ferromagnetic phase transition.

Figure 5.1(b) shows the temperature-variation of the heat capacity divided by temperature for a selection of heavy-fermion ferromagnets, for which $T_C$ varies from 150 mK (in YbCo$_2$Zn$_{20}$ [Torikachvili 07]) to 50 K (in UGe$_2$ [Troc 12]). In this graph, the non-electronic contribution to the heat capacity has been subtracted for UGe$_2$, CeRuPO, and CePd$_2$Si$_2$, where $T_C$ is high and phonons contribute to the heat capacity at temperatures near $T_C$. In these systems, a large heat-capacity precedes the establishment of ferromagnetism, indicating the development of magnetic fluctuations prior to long-range magnetic order. The integration of these curves generally leads to an electronic entropy variation $\Delta S \simeq R\ln 2$, as expected for low-temperature properties driven by an electronic doublet. At low temperatures, an enhancement of the Sommerfeld coefficient $\gamma =_{T\to 0} C_p/T$ indicates stronger quantum magnetic fluctuations in the compounds with the smaller $T_C$. Similarly to heavy-fermion antiferromagnets (see Figure 4.1(b) in Section 4.1.1), where the smallest Néel temperature was observed in an Yb-based compound (YRh$_2$Si$_2$ [Custers 03]), the smallest ferromagnetic ordering temperature is also observed in an Yb-based ferromagnetic compound (YbNi$_4$P$_2$ [Steppke 13]). In YbNi$_4$P$_2$, the low-temperature Sommerfeld coefficient reaches the highest value $\gamma \simeq 2$ J/molK$^2$ reported for a ferromagnet. In Section 3.1.1 (see Figure 3.1(b)), the heaviest Sommerfeld coefficient $\gamma \simeq 8$ J/molK$^2$ for a heavy-fermion paramagnet was also found for an Yb-based compound (YbCo$_2$Zn$_{20}$ [Torikachvili 07]). Similarly to many heavy-fermion antiferromagnets (see Section 4.2.1), a few U-based ferromagnets become superconducting near a quantum magnetic phase transition (see Section 5.2.1). Figure 5.1(b) shows the signature of the superconducting transitions at the temperatures $T_{sc} \simeq 250$ and 500 mK $< T_C$ in the heavy-fermion ferromagnets

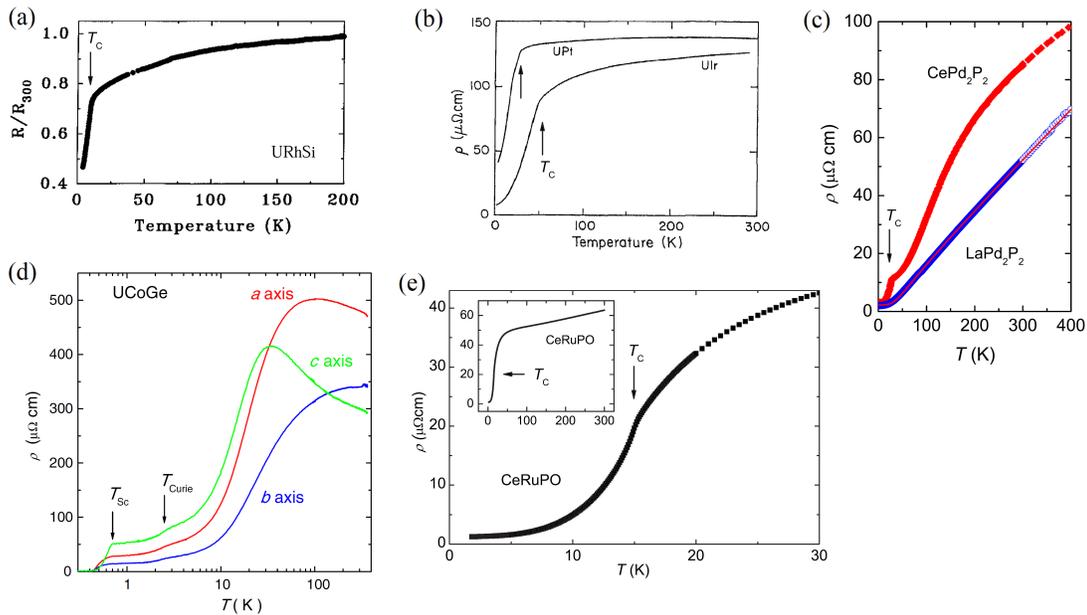

Figure 5.2: Electrical resistivity versus temperature of the ferromagnets (a) URhSi (from [Prokeš 96]), (b) UPt and UIr (from (from [Brändle 88]), (c) CePd$_2$Si$_2$ and its non-4$f$-reference LaPd$_2$Si$_2$ (from [Tran 14]), (d) UCoGe (from [Hattori 12]), and (e) CeRuPO (from [Krellner 07]).

URhGe and UCoGe, respectively [Aoki 11b].

In heavy-fermion antiferromagnets and nearly-antiferromagnets, a large high-temperature electrical resistivity indicates the presence of electronic correlations, presumably driven by magnetic fluctuations (see Figure 3.2(a) in Section 3.1 and Figures 4.6(c,e) in Section 4.2.1). Figure 5.2 shows that a large electronic contribution to the electrical resistivity is also observed in heavy-fermion ferromagnets at temperatures higher than the Curie temperature $T_C$ [Brändle 88, Prokeš 96, Krellner 07, Hattori 12, Tran 14]. This supports that intersite magnetic correlations precede the low-temperature onset of ferromagnetism. In UCoGe, the anisotropy of the electrical resistivity, which was measured with currents applied along the three main crystallographic directions a, b, and c, also indicates that anisotropic magnetic fluctuations develop at high temperatures [Hattori 12]. Similarly to antiferromagnets at temperatures below $T_N$ (see Figures 4.6(c,e) and 4.8(c,d) in Section 4.2.1), Figure 5.2 shows that ferromagnets at temperatures below $T_C$ are characterized by a fall of the electrical resistivity. In the next Section, signatures of long-range magnetic ordering and short-range magnetic fluctuations in heavy-fermion ferromagnets are considered.

## 5.1.2 Magnetic order and fluctuations

Magnetization measurements in fields $H \gtrsim H_c$ are often used to probe a ferromagnetic order parameter. However, the ferromagnetic phase transition at $T_C$ can be only observed for $H < H_c$. For $H > H_c$, a broad temperature crossover indicates a different groundstate, which corresponds to a polarized paramagnetic regime rather than a ferromagnetic phase (see Sections 5.3.1.1 and 5.3.1.2). Neutron diffraction permits to directly probe the order parameter of magnetically-ordered phases in zero-magnetic field. In a ferromagnet, magnetic Bragg peaks occur at momentum transfers $\mathbf{Q} = \tau$, where $\tau$ is a nuclear Bragg position, corresponding to the wavevector $\mathbf{k} = 0$. Neutron scattered intensity at a ferromagnetic peak can be expressed as $I(\mathbf{k} = 0) = I_0 + a\mu(\mathbf{k} = 0)^2$, where the moment $\mu(\mathbf{k} = 0) = \mu_{FM}$ is the ferromagnetic order parameter, $I_0$ is the intensity of the nuclear Bragg peak, and $a$ is a constant. Figure 5.3 presents two examples of neutron diffraction measurements, performed on the heavy-fermion ferromagnets URhGe [Aoki 01] and URhSi [Prokeš 03], indicating the onset of a ferromagnetic order at temperatures below their respective Curie temperature $T_C = 9.5$ and 10 K.

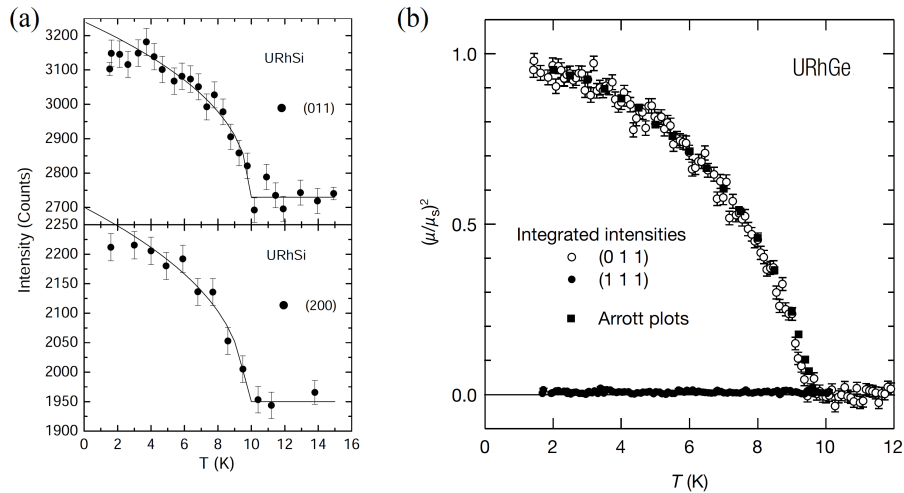

Figure 5.3: Neutron diffracted intensity at momentum transfers corresponding to the ferromagnetic wavevector $\mathbf{k} = 0$ (a) in URhSi (from [Prokeš 03]) and (b) in URhGe (from [Aoki 01]).

Similarly to heavy-fermion antiferromagnets, where critical fluctuations of the antiferromagnetic order parameter are peaked at the Néel temperature (see Figure 4.4(a,b) in Section 4.1.2), critical ferromagnetic fluctuations can be observed in the vicinity of the Curie temperature of ferromagnets. Figure 5.4 summarizes a study by inelastic neutron scattering of ferromagnetic fluctuations in the heavy-fermion ferromagnet $UGe_2$ [Huxley 03]. Due to strong nuclear Bragg peaks at the momentum transfers, as $\mathbf{Q} = (0, 0, 1)$ corresponding to the ferromagnetic wavevector $\mathbf{k} = 0$, and of the parasitic inelastic tails of these elastic peaks, ferromagnetic fluctuations were studied by Huxley *et al* [Huxley 03]. The energy spectra measured at $\mathbf{Q} = (0, 0, 1.04)$ shown in Figure 5.4(a) indicate an enhancement of the ferromagnetic fluctuations at the temperature $T = 54$ K $\simeq T_C = 52$ K. Figures 5.4(b-d) present $q$- and $T$-dependent plots of the relaxation rate $\Gamma(\mathbf{q})$ and of the static susceptibility $\chi'(\mathbf{q})$ extracted from a Lorentzian fit (see Equation 2.19 in Section 2.4.3.1) to the spectra measured on $UGe_2$. They show a sharp enhancement of $\chi'(\mathbf{q})$ and a sharp minimum of $\Gamma(\mathbf{q})$ for $\mathbf{q} \to 0$ and $T \to T_C$, indicating that the enhancement of ferromagnetic fluctuations at $T_C$ is accompanied by a vanishing of their char-acteristic energy scale. The product $\chi'(\mathbf{q})\Gamma(\mathbf{q})$ is found to be constant at temperatures $T \gtrsim T_C$, similarly to what was observed near $T_N$ in the antiferromagnets $Ce_{1-x}La_xRu_2Si_2$ (see Figure 4.11(c) in Section 4.2.2). However, a fall of $\chi'(\mathbf{q})\Gamma(\mathbf{q})$ occurs for $T < T_C$, which contrasts with the $T$-independent product reported for $T < T_N$ in $Ce_{1-x}La_xRu_2Si_2$

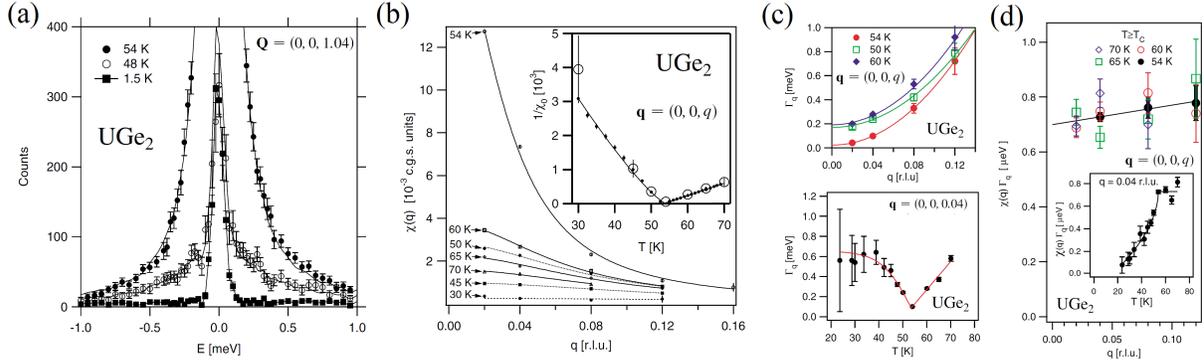

Figure 5.4: (a) Inelastic neutron scattering spectra measured at the momentum transfer $\mathbf{Q} = (0, 0, 1.04)$ at different temperatures, (b) wavevector-$q$-dependence of the static susceptibility $\chi'(\mathbf{q})$ extracted from the inelastic neutron scattering spectra at different temperatures (the Inset shows the variation of the inverse of the susceptibility $\chi'(\mathbf{q} \to 0) = \chi_0$ deduced from neutron measurements (large circles) of the susceptibility deduced from magnetization measurements (small circles), (c) wavevector-dependence of the relation rate $\Gamma(\mathbf{q})$ at different temperatures (top) and temperature-dependence of $\Gamma(\mathbf{q})$ at $\mathbf{q} = (0, 0, 0.04)$ (bottom), and (e) wavevector-dependence of $\chi'(\mathbf{q}\Gamma(\mathbf{q})$ at different temperatures (main panel) and temperature dependence of $\chi'(\mathbf{q}\Gamma(\mathbf{q})$ at $\mathbf{q} = (0, 0, 0.04)$ (inset), in the ferromagnet URhGe (from [Huxley 03]).

### 5.1.3 Critical magnetization

The bulk magnetic susceptibility $\chi$, or the magnetization $M = \chi H$, are natural probes to capture the critical ferromagnetic properties near a ferromagnetic phase transition. As shown by Equations 2.12 and 2.18 in Section 2.4.3.1, $\chi$ is related to an integration in energy of the ferromagnetic fluctuations spectra corresponding to wavevector $\mathbf{k} = 0$. Critical exponents related to ferromagnetic ordering can be extracted from magnetization and magnetic susceptibility measurements, but also from magnetostriction measurements. In this Section, the cases of two ferromagnets, UIr and URhGe, are considered in detail.

Figure 5.5(a) presents, within a log-log scale, the magnetization versus magnetic field measured on the ferromagnet UIr at different temperatures and in a magnetic field applied along its easy magnetic axis $[1, 0, \overline{1}]$ [Knafo 09b]. At a low temperature $T = 4.2$ K $\ll T_C = 45$ K, the magnetization saturates to a value $M_s = 0.5 \; \mu_B/$ U in magnetic fields higher than the critical field $\mu_0 H_c = 60$ mT. An upwards deviation of $M$ is also observed at magnetic fields $\mu_0 H > 10$ T, indicating the progressive quench of remaining magnetic fluctuations. At the Curie temperature $T_C$, a critical power-law behavior is observed in the magnetization, which

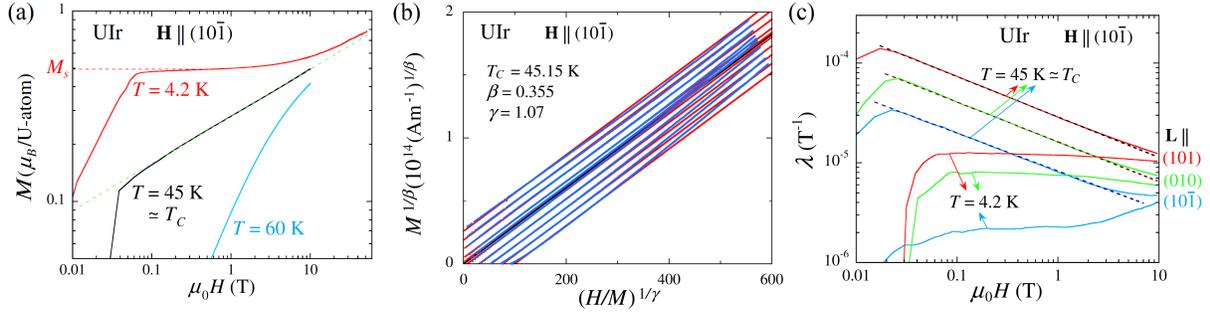

Figure 5.5: (a) Magnetization versus magnetic field at different temperatures, (b) Arrott plot of the magnetization data, and (c) magnetostriction coefficients measured along the directions $[1,0,1]$, $[0,1,0]$, and $[1,0,\overline{1}]$ of UIr in a magnetic field $\mathbf{H} \parallel [1,0,\overline{1}]$ (from [Knafo 09b]).

follows:

$$M \propto H^{1/\delta}, \tag{5.1}$$

with $\delta = 4.04 \pm 0.05$. The high-temperature magnetization, measured at $T = 60$ K, shows an upwards curvature tending asymptotically to the critical power-law in high-fields. The critical behavior in the magnetization at temperatures near $T_C$ can be analyzed using the Arrott equation of state [Arrott 67]:

$$M^{1/\beta} = c_1 \left( \frac{H}{M} \right)^{1/\gamma} - c_2 \left( T - T_C \right), \tag{5.2}$$

with the constants $c_1, c_2 > 0$. Figure 5.5(b) shows a plot of $M^{1/\beta}$ versus $(H/M)^{1/\gamma}$, from isothermal magnetization measurements made at temperatures close to $T_C$. This plot results from the best fit of Equation 5.2 from the data, which yields the critical exponents $\beta = 0.355 \pm 0.05$, and $\gamma = 1.07 \pm 0.1$. Deviations from the Arrott law are observed for $\mu_0 H \gtrsim 5$ T. At $T = T_C$, Equation 5.2 gives the scaling relation between the critical exponents:

$$\delta = \frac{\beta + \gamma}{\beta}. \tag{5.3}$$

Figure 5.5(c) presents the magnetic-field variation of the magnetostriction of UIr, defined by:

$$\lambda_i = \frac{1}{L_i} \frac{\partial L_i}{\partial (\mu_0 H)} = -\frac{\partial M}{\partial p_i}, \tag{5.4}$$

where $p_i$ is a uniaxial stress applied along $i$, and $L_i$ the sample length along $i$, measured with $\mathbf{L} \parallel [1,0,\overline{1}]$, $[1,0,1]$, and $[0,1,0]$, in a magnetic field $\mathbf{H} \parallel [1,0,\overline{1}]$. The right part of Equation 5.4 relates the magnetostriction coefficient to the uniaxial-pressure derivative of the magnetization. Anisotropic and field-independent low-temperature magnetostriction coefficients indicate anisotropic and constant values of $\partial M_s / \partial p_i$ in magnetic fields up to 5 T. The $H$-power law of $\lambda_i$ at $T_C$ can be derived by taking the $p_i$ derivative of Equation (5.2) and by assuming that $\partial T_C / \partial p_i$ is the relevant pressure derivative (i.e., $\partial c_1 / \partial p_i = 0$), which gives:

$$\lambda_i(H, T_C) = -A \frac{\partial T_C}{\partial p_i} H^{-1/\delta'},$$ (5.5)

where $A = c_1^{-\gamma/\delta'} c_2 \gamma / \delta$ is a positive constant and:

$$\delta' = \frac{\beta + \gamma}{1 - \beta}.$$ (5.6)

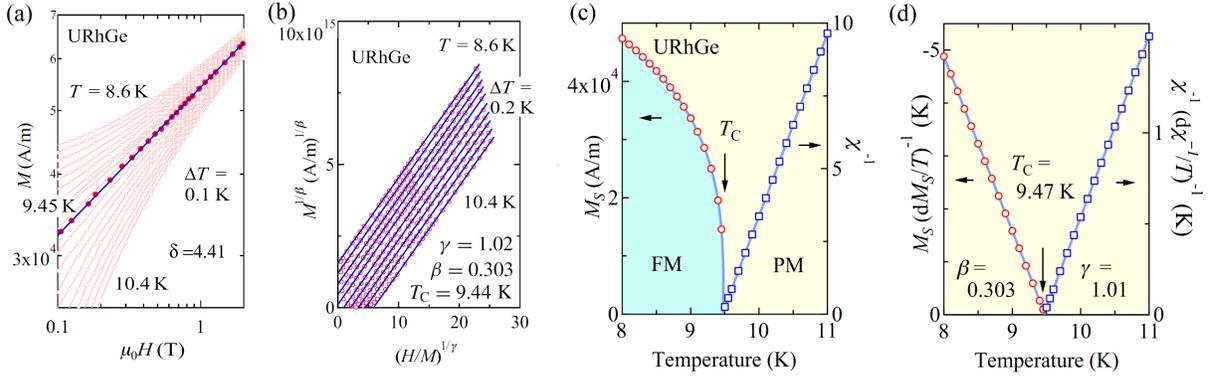

Figure 5.6: (a) Magnetization versus magnetic field at different temperatures, (b) Arrott plot of the magnetization data, (c) temperature-dependence of the spontaneous magnetization $M_s$ and of the inverse of the initial magnetic susceptibility $\chi^{-1}$ determined from the Arrott plot, and (d) Kouvel-Fisher plot for $M_s$ and $\chi^{-1}$ versus temperature, of URhGe in a magnetic field $\mathbf{H} \parallel \mathbf{c}$ (from [Tateiwa 14a]).

Using the values of $\beta$ and $\gamma$ determined from the Arrott method, the value $\delta' = 2.2 \pm 0.4$ is obtained using Equation 5.6, which agrees, within 15 %, with the exponents $\delta_i' \simeq 2.63$, 2.47, and $2.51 \pm 0.0.03$ of the power-law variations of $\lambda_i(H)$ at $T_C$ for $\mathbf{H} \parallel [1, 0, \bar{1}]$, $[1, 0, 1]$, and $[0, 1, 0]$, respectively. Equation 5.5 generalizes to non-mean-field exponents a formula determined by Belov for the particular case of mean-field behavior, where $\beta = 0.5$ and $\gamma = 1$ [Belov 56]. It also permits to extract the uniaxial pressure dependencies of the Curie temperature $\partial T_C / \partial p_i$ using ambient-pressure magnetization and magnetostriction measurements. This constitutes an alternative to the Ehrenfest relationship based on the combination of ambient- pressure specific-heat and thermal-expansion measurements (see [Knafo 09b, Sakarya 10]).

The study of UIr [Knafo 09b] was followed by the characterization of the critical magnetization of other U-based heavy-fermion ferromagnets. Figure 5.6 summarizes the investigation of the ferromagnetic criticality of URhGe in a magnetic field applied along its easy magnetic axis $\mathbf{c}$ [Tateiwa 14a]. While the magnetization at $T_C = 9.5$ K follows a $H$-power law with the critical exponent $\delta = 4.4$ (see Figure 5.6(a)), similar exponents $\gamma = 1.02$ and $\beta = 0.303$ than those found for UIr are extracted from an Arrott analysis of URhGe magnetization data (see Figure 5.6(b)). Alternatively to the Arrott method, the Kouvel-Fisher method also permits to extract the critical exponents [Kouvel 64]. This second method considers separately data with temperatures below and above $T_C$. For $T < T_C$, assuming that $c_1(H/M)^{1/\gamma} \ll c_2(T - T_C)$, $M^{1/\beta}$ and $M = M_s$, Equation 5.2 leads to $M_s = a(T_C - T)^{\beta}$, where $a$ is a positive constant, so that:

$$M_s \left( \frac{\partial M_s}{\partial T} \right)^{-1} = \frac{1}{\beta}(T - T_C). \tag{5.7}$$

Table 5.1: Critical exponents $\beta$, $\gamma$, $\delta$, and $\delta'$ extracted from the magnetization of U-based ferromagnets and expected for different universality classes [Collins 89].

| | $\beta$ | $\gamma$ | $\delta$ | $\delta'$ | |
|---|---|---|---|---|---|
| UIr | 0.355 | 1.07 | 4.01 | 2.20 | ♯a |
| UGe$_2$ | 0.33 | 1.0 | 4.16 | - | ♯b |
| URhGe | 0.30 | 1.0 | 4.4 | - | ♯b |
| URhAl | 0.29 | 1.47 | 6.08 | - | ♯c |
| UCo$_{0.98}$Os$_{0.02}$Al | 0.33 | 1.0 | 4.2 | - | ♯d |
| URhSi | 0.30 | 1.0 | 4.4 | - | ♯e |
| UTeS | 0.31 | 1.0 | 4.21 | - | ♯f |
| USeS | 0.30 | 1.0 | 4.34 | - | ♯f |
| 3D Heisenberg | 0.367 | 1.388 | 4.78 | 2.77 | ♯h |
| 3D XY | 0.345 | 1.316 | 4.81 | 2.54 | ♯h |
| 3D Ising | 0.326 | 1.238 | 4.80 | 2.32 | ♯h |
| 2D Ising | 0.125 | 1.75 | 15 | 2.14 | ♯h |
| Mean Field | 0.5 | 1 | 3 | 3 | ♯h |

♯a [Knafo 09b], ♯b [Tateiwa 14a], ♯c [Tateiwa 18], ♯d [Maeda 18],
♯e [Tateiwa 19a], ♯f[Tateiwa 19b], ♯h [Collins 89].

For $T > T_C$, assuming that $M^{1/\beta} \ll c_1(H/M)^{1/\gamma}$, $c_2(T - T_C)$ and $\chi = M/H$, Equation 5.2 leads to $\chi^{-1} = b(T - T_C)^{\gamma}$, where $b$ is a positive constant, so that:

$$\chi^{-1} \left( \frac{\partial \chi^{-1}}{\partial T} \right)^{-1} = \frac{1}{\gamma}(T - T_C). \tag{5.8}$$

While Figure 5.6(c) presents, from URhGe magnetization data, a plot of $M_s$ versus $T$ for $T < T_C$ and of $\chi^{-1}$ versus $T$ for $T > T_C$, Figure 5.6(d) shows Kouvel-Fisher plots of the same data with $M_s(\partial M_s/\partial T)^{-1}$ versus $T$ for $T < T_C$ and of $\chi^{-1}(\partial \chi^{-1}/\partial T)^{-1}$ versus $T$ for $T > T_C$, ending in similar critical exponents than those found with the Arrott method [Tateiwa 14a].

Table 5.1 summarizes the critical exponents extracted by the Arrott and Kouvel-Fisher methods for a large number of U-based ferromagnets, UGe$_2$, URhGe [Tateiwa 14a], URhAl [Tateiwa 18], UCo$_{0.98}$Os$_{0.02}$Al [Maeda 18], URhSi [Tateiwa 19a], UTeS, and USeS [Tateiwa 19b]. While none of these exponents, which were determined experimentally, correspond to those expected theoretically for standard universality classes (3D Heisenberg, 3D XY, 3D Ising, 2D Ising, and mean-field) [Collins 89], a surprising feature is that similar exponents $\beta \simeq 0.30 - 0.35$, $\gamma \simeq 1$, and $\delta \simeq 4 - 4.4$ are found to characterize UGe$_2$, URhGe, UCo$_{0.98}$Os$_{0.02}$Al,

URhSi, UTeS, and USeS, suggesting a possible common universality class for these ferromagnets. The case of URhAl is particular, since its critical exponents $\beta \simeq 0.29$, $\gamma \simeq 1.47$, and $\delta \simeq 6.1$ are quite different from the other U-compounds considered here. The investigation of a larger number of heavy-fermion ferromagnets is needed to identify possible universality classes for these compounds. The study of the critical exponents near a quantum ferromagnetic phase transition induced by chemical doping or pressure may also be of interest. These experiments appeal for theories beyond the standard models, to describe ferromagnetic criticality in heavy-fermion compounds.

## 5.2 Quantum ferromagnetic phase transitions

A quantum ferromagnetic phase transition, corresponding to the boundary of a ferromagnetic phase in the limit of zero-temperature, can be tuned by pressure or chemical doping in most of heavy-fermion ferromagnets. The quantum critical properties of heavy-fermion ferromagnets are considered here. Magnetic phase diagrams of critical ferromagnets are presented in Section 5.2.1. Signatures of enhanced magnetic fluctuations in the physical properties are presented in Section 5.2.2. Finally, the variation of valence at a ferromagnetic phase transition is considered in Section 5.2.3.

### 5.2.1 Bulk properties and phase diagrams

Different examples of ferromagnetic phase diagrams are presented. Quantum ferromagnetic-to-paramagnetic phase transitions are first introduced (Section 5.2.1.1). Then, the case of ferromagnetic superconductors is considered (Section 5.2.1.2). Finally, the phase diagrams of ferromagnets near an antiferromagnetic phase transition are presented, permitting to emphasize the competition between magnetic exchange interactions (Section 5.2.1.3).

#### 5.2.1.1 Ferromagnetic quantum phase transition

The ferromagnetic-to-paramagnetic quantum phase transition of a selection of heavy-fermion ferromagnets is presented in Figure 5.7:

- Figures 5.7 (a,b) show the magnetic susceptibility versus temperature of $YbCu_2Si_2$, measured in a magnetic field applied along the easy axis **c** and at different pressures, and its pressure-temperature phase diagram, respectively [Tateiwa 14b]. The large magnetic susceptibility and, thus, the large magnetization, developing at low temperature indicate that a ferromagnetic ground state is stabilized at pressures $p > p_c \simeq 8$ GPa.

- Figures 5.7 (c,d) show the magnetic susceptibility data (measured via the NMR Knight shift $\kappa$) and the magnetic phase diagram of $CeRu_{1-x}Fe_xPO$. Ferromagnetism is stabilized at dopings $x < x_c \simeq 0.85$ [Kitagawa 13]. Interestingly, a broad maximum in the susceptibility is observed in the paramagnetic compounds with doping $x > x_c$, suggesting the possible presence of antiferromagnetic correlations. This system is, indeed, also in the vicinity of an antiferromagnetic phase transition, which can be tuned under pressure (see Figure 5.9(d)).

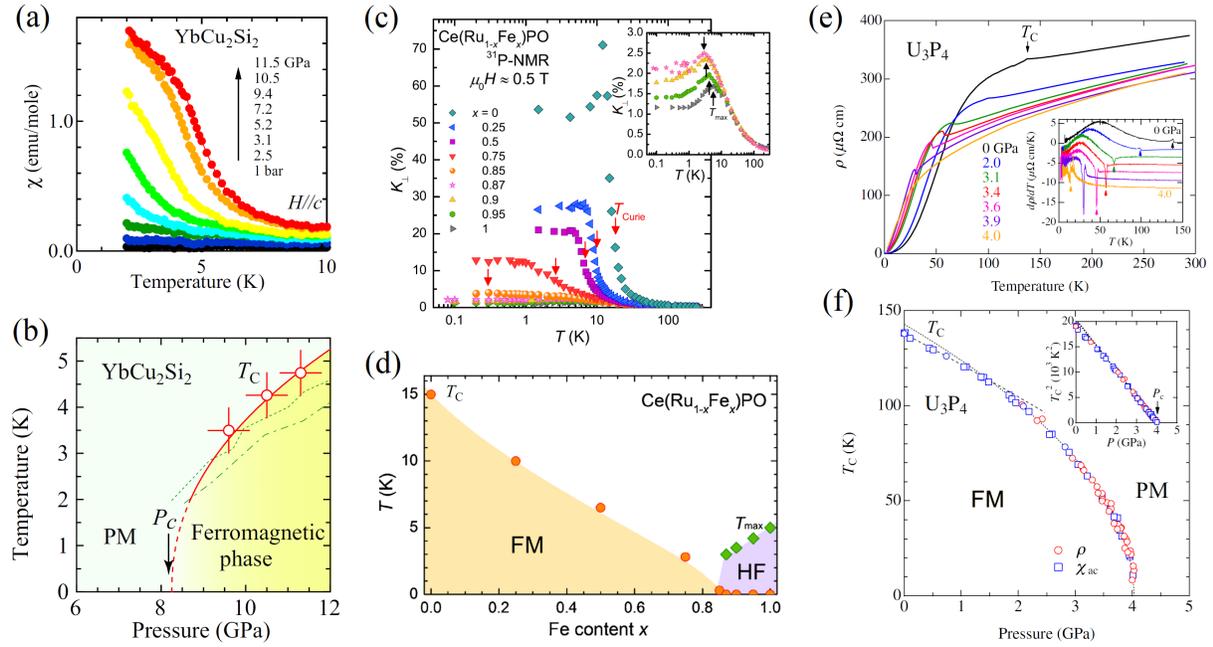

Figure 5.7: (a) Magnetic susceptibility versus temperature in a magnetic field **H** ∥ **c** and different pressures, and (b) pressure-temperature phase diagram of YbCu$_2$Si$_2$ (from [Tateiwa 14b]). (c) NMR Knight-shift versus temperature and (d) doping-temperature phase diagram of CeRu$_{1-x}$Fe$_x$PO (from [Kitagawa 13]). (e) Electrical resistivity versus temperature at different pressures and (f) pressure-temperature phase diagram of U$_3$P$_4$ (from [Araki 15]).

- Figures 5.7 (e,f) present electrical-resistivity data and the magnetic phase diagram of U$_3$P$_4$ under pressure [Araki 15]. This system is characterized by a large Curie temperature $T_C$ = 140 K at ambient pressure, and by a critical pressure $p_c$ = 4 GPa beyond which ferromagnetism is replaced by a paramagnetic ground state. The fall of $T_C$ is accompanied by a decrease of the electronic contribution to the electrical resistivity at high temperatures, and by its increase at low temperatures. Interestingly, the Curie temperatures varies as $T_c \sim \sqrt{p}$ over the broad pressure range $0 < p < 4$ GPa.

#### 5.2.1.2  Ferromagnetism and superconductivity

Superconductivity develops near a quantum ferromagnetic phase transition in the ferromagnets UGe$_2$, UIr, URhGe, and UCoGe. Their phase diagrams under pressure are displayed in Figure 5.8 [Pfleiderer 02, Akazawa 04, Aoki 19a]. While superconductivity appears at ambient pressure, below a superconducting temperature smaller than the ferromagnetic Curie temperature, i.e., $T_{sc} < T_C$, in URhGe [Aoki 01] and UCoGe [Huy 07], it is induced under pressure in UGe$_2$ [Saxena 00] and UIr [Akazawa 04].

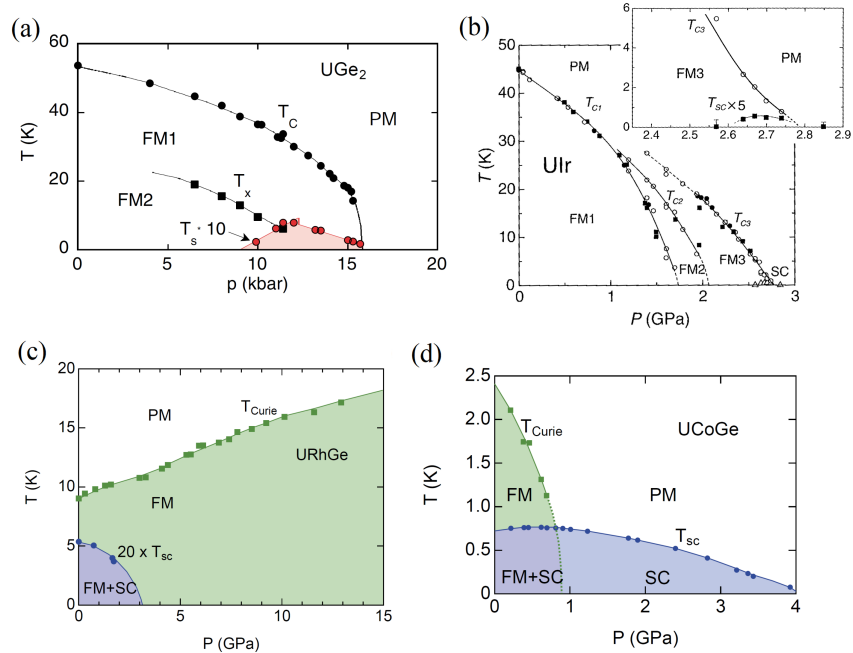

Figure 5.8: Pressure-temperature phase diagrams of (a) UGe$_2$ (from [Pfleiderer 02]), (b) UIr (from [Kobayashi 07]), (c) URhGe and (d) UCoGe (from [Aoki 19a]).

- Figure 5.8(a) shows that, in UGe$_2$, a transition from a low-pressure phase FM2 with large ferromagnetic moments to a high-pressure phase FM1 with small ferromagnetic moments is induced at low temperatures and pressures $p_x = 1.2$ GPa. It precedes the ferromagnetic phase transition at $p_c = 1.6$ GPa beyond which a paramagnetic regime is established [Huxley 01, Pfleiderer 02]. The temperature boundary $T_x$ between FM1 and FM2 increases with decreasing pressure, before reaching a critical end-point, beyond which it is replaced by a broad crossover at low pressure [Taufour 10]. The superconducting phase is observed only inside the ferromagnetic state, and the maximum superconducting temperature $T_{sc} = 800$ mK is observed at $p_x$. This suggests that the mechanism for superconductivity is related with the magnetic fluctuations induced at the transition FM2 → FM1. Figure 5.8(b) shows that three successive ferromagnetic phases are reported for UIr under pressure: FM1 at pressures $p < 1.5 - 2$ GPa, FM2 and FM3 at pressures 1.5-2 GPa $< p < p_c = 2.5$ GPa [Kobayashi 07]. Superconductivity in UIr is observed in the vicinity of the ferromagnetic phase transition at $p_c$, at which the superconducting temperature $T_{sc} = 140$ mK is maximum [Akazawa 04].

- Figures 5.8(c-d) show the pressure-temperature phase diagrams of the isostructural compounds URhGe and UCoGe [Aoki 19a]. In URhGe, $T_C$ increases while $T_{sc}$ decreases under pressure [Hardy 05], indicating the presence of a 'virtual' quantum ferromagnetic phase transition at a negative pressure. Oppositely, in UCoGe $T_C$ decreases under pressure, leading to a quantum ferromagnetic phase transition at $p_c \simeq 1$ GPa. Superconductivity is stable over a broad range of pressures, up to more than 2.5 GPa, with a maximum of the superconducting temperature $T_{sc} \simeq 600$ mK observed near $p_c$ [Hassinger 08b, Slooten 09].

A spin-triplet superconducting order parameter, with equal-spin pairing and robust against a large exchange field, has been proposed for these ferromagnetic superconductors [Aoki 19a]. In these systems, ferromagnetic fluctuations are suspected to be the 'glue' for the formation of the Cooper pairs. NMR experiments brought microscopic support for a triplet superconducting state in ferromagnetic superconductors, for which they further highlighted the role of magnetic fluctuations [Hattori 14, Manago 19a]. The situation in ferromagnetic superconductors contrasts with that of superconductors near a quantum antiferromagnetic phase transition, where antiferromagnetic fluctuations are suspected to drive the Cooper pairing mechanism, and for which a singlet superconducting order parameter is generally proposed. However, the presence of antiferromagnetic fluctuations cannot be excluded, in particular in ferromagnets near an antiferromagnetic phase transition (see Section 5.2.1.3).

### 5.2.1.3 Nearby antiferromagnetic phase transitions

Heavy-fermion magnets are the place of a subtle competition between magnetic interactions, either antiferromagnetic, with commensurate or incommensurate wavevectors $\mathbf{k} \neq 0$, or ferromagnetic, with the wavevector $\mathbf{k} = 0$. Figure 5.9 shows that, in several ferromagnetic compounds, this competition leads to an antiferromagnetic phase transition which can be revealed by tuning chemical doping or pressure.

- Figure 5.9(a) presents the temperature-dependence of neutron-diffracted intensities on $CeRu_2Ge_2$ at the momentum transfers $\mathbf{Q}_{FM} = (1, 1, 0)$, characteristic of the ferromagnetic wavevector $\mathbf{k} = 0$, and $\mathbf{Q}_1 = (0.7, 1, 0)$ characteristic of the antiferromagnetic wavevector $\mathbf{k}_1 = (0.3, 0, 0)$ [Raymond 99b]. A magnetic Bragg peak at $\mathbf{Q}_1$ indicates that long-range magnetic order with wavevector $\mathbf{k}_1$ is established in the phase AFM I at temperatures $T_C = 7.5$ K $< T < T_N = 8.3$ K. A transfer of magnetic spectral weight occurs at $T_C$, where the antiferromagnetic Bragg peak disappears and is replaced by a ferromagnetic Bragg peak at $\mathbf{Q}_{FM}$. Figure 5.9(b) shows that, under pressure, the ferromagnetic phase disappears and is replaced by a low-temperature antiferromagnetic phase AFM II, the phase AFM I remaining present at high temperatures [Wilhelm 99]. The pressure $p_c = 9$ GPa corresponds to a quantum antiferromagnetic phase transition, beyond which a paramagnetic ground state is established.

- Figure 5.9(c) presents an inter-uranium-atomic-distance-temperature phase diagram summarizing the properties of the $UCo_{1-x}Rh_xGe$ and $UIr_{1-x}Rh_xGe$ alloys [Pospíšil 18] (see also [Pospíšil 17, Pospíšil 20]). While UCoGe and URhGe are ferromagnetic, UIrGe is antiferromagnetic. A broad maximum in the magnetic susceptibility occurs at a temperature $T_\chi^{max} > T_{N,C}$ in UCoGe and UIrGe, indicating the onset of a high-temperature correlated paramagnetic regime in these systems, possibly controlled by antiferromagnetic correlations (see heavy-fermion paramagnets in Section 3.1.2). Starting from UCoGe, for which $T_\chi^{max} = 37.5$ K and $T_C = 3$ K, $T_\chi^{max}$ decreases and $T_C$ increases with Rh-doping. The decoupling of $T_\chi^{max}$ and $T_C$ vanishes with the merging of the two temperatures scales in $UCo_{0.6}Rh_{0.4}Ge$, where $T_C = 20$ K is maximum, before decreasing down to 9.5 K in URhGe. From URhGe, Ir-doping leads a quantum ferromagnetic-to-antiferromagnetic phase transition, at which the three temperature scales, $T_C$ for small Ir-doping, $T_\chi^{max}$ and $T_N$ for large Ir-doping, have their minimum values. $T_N$ and $T_\chi^{max}$ increase with Ir-doping up to 16.5 K and 30 K, respectively, in UIrGe. A quasi-proportionality between $T_\chi^{max}$ and $T_N$ indicates that, in the antiferromagnetic compounds, the magnetic fluctuations established at temperatures below $T_\chi^{max}$ are a precursor of the long-range antiferromagnetic order set at lower temperatures. A competition between different magnetic interactions is emphasized by this rich phase diagram composed of a high-temperature CPM regime in additions to the FM and AF phases.

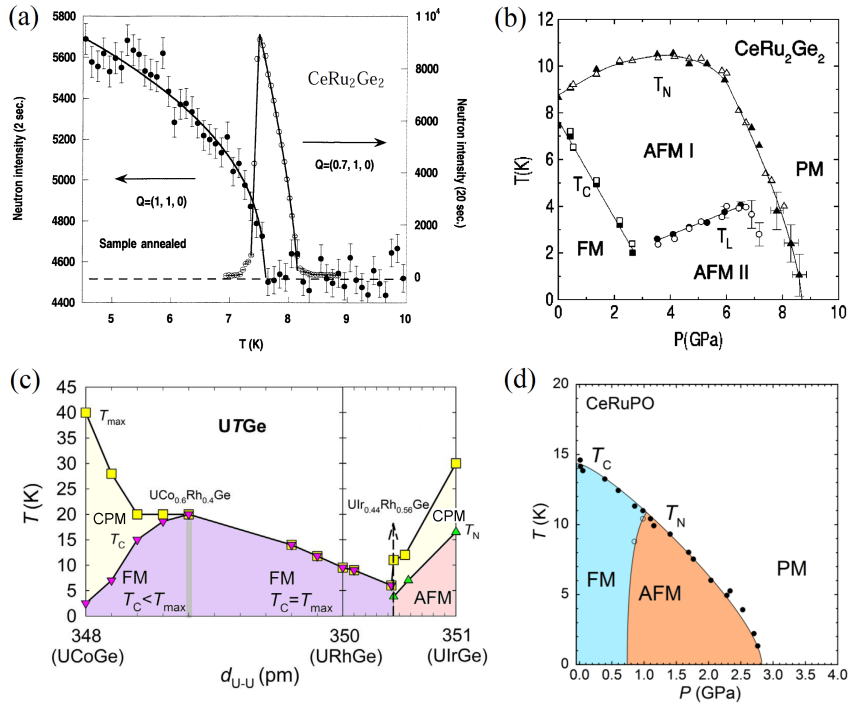

Figure 5.9: (a) Neutron diffracted intensity at the momentum transfers $\mathbf{Q} = (1, 1, 0)$ and $\mathbf{Q} = (0.7, 1, 0)$ versus temperature (from [Raymond 99b]) and (b) pressure-temperature phase diagram of $CeRu_2Ge_2$ (from [Wilhelm 99]), (c) inter-uranium-atomic distance - temperature phase diagram of $UCo_{1-x}Rh_xGe$ and $UIr_{1-x}Rh_xGe$ alloys (from [Pospíšil 18]), and (d) pressure-temperature phase diagram of CeRuPO (from [Kitagawa 14]).

- Figure 5.9(d) shows the pressure-temperature phase diagram of CeRuPO [Kitagawa 14]. In this compound, ferromagnetism is replaced under pressure by an antiferromagnetic phase, evidenced from NMR measurements. The ferromagnetic Curie temperature $T_C$ decreases under pressure and is replaced at $p^* = 1$ GPa by a Néel temperature $T_N$, which also decreases with pressure. A quantum antiferromagnetic phase transition is reached at $p_c = 2.8$ GPa, beyond which a paramagnetic ground state is established. The competition between the AF and FM interactions in CeRuPO may be related with the observation of a CPM regime, possibly driven by antiferromagnetic correlations, in the parent paramagnetic compound CeFePO (see Section 3.1.2).

## 5.2.2 Magnetic fluctuations

The investigation of heavy-fermion ferromagnets at their doping- or pressure-induced quantum phase transition permits to show signatures of critical magnetic fluctuations. Figure 5.10 presents a selection of such experimental studies.

- Figure 5.10(a) shows the electrical resistivity $\rho$ of $U_3P_4$ under pressure, plotted either versus $T^2$ (top left) or versus $T^{5/2}$ (top right), and the pressure-dependence of the coefficient $n$ extracted from a fit by $\rho = \rho_0 + AT^n$ to the electrical resistivity (bottom) [Araki 15]. A low-temperature Fermi-liquid behavior with $n = 2$ is observed at pressures $p < 3$ GPa, i.e., inside the ferromagnetic phase. Deviations from a Fermi-liquid regime are found at pressures near the critical pressure $p_c = 4$ GPa, and a non-Fermi-liquid regime is identified at $p_c$, where the coefficient $n = 5/3$ is minimum (see phase diagram in Figure 5.7(f)). Figure 5.10(b) presents the heat capacity versus temperature of $YbNi_4(P_{1-x}As_x)_2$ alloys, and the Inset of the Figure shows its doping-temperature phase diagram [Steppke 13]. The Sommerfeld coefficient $\gamma =_{T\to 0} C_p/T$ extrapolated to zero-temperature is maximum, indicating the enhancement of the magnetic fluctuations, at the critical pressure $x_c \simeq 0.08 - 0.09$, beyond which a paramagnetic ground state is stabilized. An exponential variation of the heat capacity $C_p/T \sim T^{-m}$, with $m \simeq 0.4$, is observed down to the lowest temperature at the concentration $x = 0.08$. This non-Fermi liquid behavior is the signature of diverging, or almost-diverging, critical magnetic fluctuations.

- The case of a ferromagnet near a quantum antiferromagnetic phase transition is considered in Figure 5.10(c), where the pressure-dependence of the electrical-resistivity quadratic coefficient $A$ of CeRuPO is presented [Kotegawa 13] (see pressure-temperature phase diagram in Figure 5.9(d)). A broad enhancement of $A$ is observed near the critical pressure $p_c = 2.8$ corresponding to the quantum antiferromagnetic-to-paramagnetic phase transition. However, $A$ remains small and without marked anomaly at the ferromagnetic-to-antiferromagnetic boundary $p^* = 1$ GPa. This indicates that the change of the magnetic structure at $p^*$ may not be accompanied by critical quantum magnetic fluctuations.

- NMR measurements of the relaxation rate constitute a microscopic probe to magnetic fluctuations in the limit of low energies $E \to 0$ and averaged over the reciprocal space (see Equation 2.20 in Section 2.4.3.1). However, they do not allow determining their wavevector-dependence nor their energy spectra. Figures 5.10(d-e) present the NMR relaxation rate, plotted as $S_\perp = 1/2T_1$ and $1/T_1T$ versus temperature, for $CeRu_{1-x}Fe_xPO$ alloys with different dopings [Kitagawa 12] and UCoGe under pressure [Manago 19b], respectively. The doping-temperature phase diagram of $CeRu_{1-x}Fe_xPO$ is shown in Figure 5.7(f) and the pressure-temperature phase diagram of UCoGe is shown in Figure 5.8(d) and recalled in the Inset of Figure 5.10(e). The NMR relaxation rate is enhanced at the Curie temperature $T_C$ of these two systems, and the fall of $T_C$ near the quantum ferromagnetic phase transition is accompanied by an enhancement of the relaxation rate in the zero-temperature limit, which indicates the presence of critical magnetic fluctua-tions at the quantum ferromagnetic phase transition. The Inset of Figure 5.10(d) shows a sharp enhancement of $S_\perp$ measured at the low-temperature $T = 200$ mK at the critical doping $x_c = 0.85$ of CeRu$_{1-x}$Fe$_x$PO. A fall of the relaxation rate is observed in the superconducting phase developing for $T < T_{sc}$ in UCoGe. The variation $1/T_1T \sim T^2$ was interpreted as the indication for a line-nodal superconducting gap by Manago *et al* [Manago 19b] (see also Section 2.5.3).

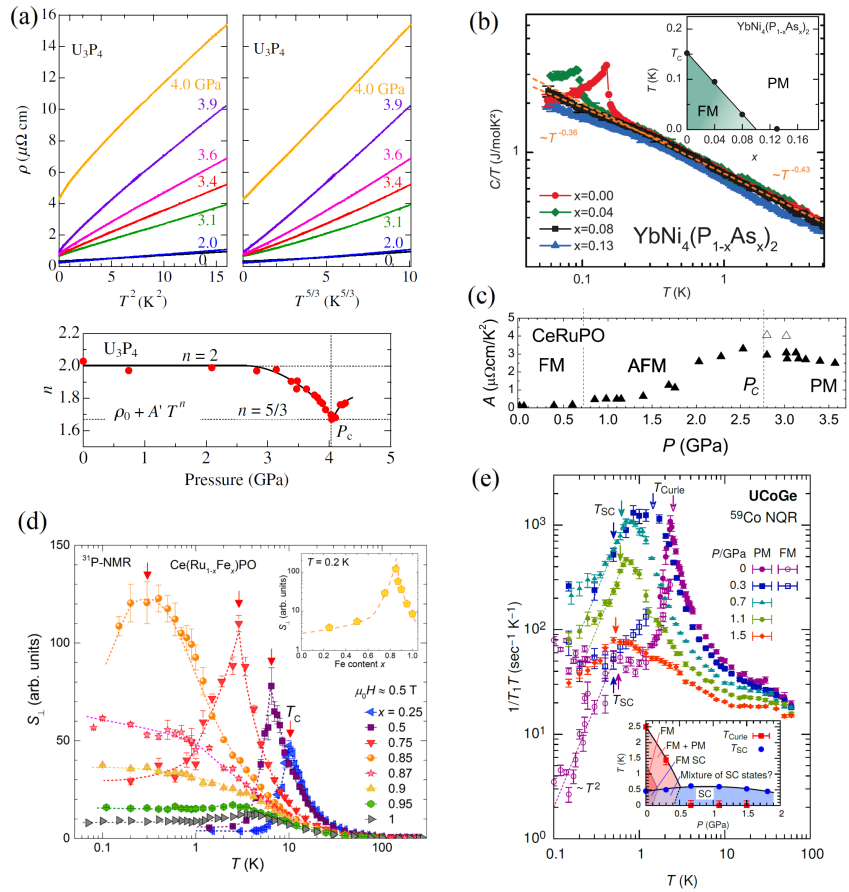

Figure 5.10: (a) Plot of electrical resistivity $\rho$ versus $T^2$ (top left) and versus $T^{5/2}$ (top right) at different pressures and pressure-dependence of the coefficient $n$ extracted from a fit by $\rho = \rho_0 + AT^n$ to the electrical resistivity (bottom) of $U_3P_4$ (from [Araki 15]). (b) Temperature variation of heat capacity divided by temperature of $YbNi_4(P_{1-x}As_x)_2$ alloys, whose doping-temperature phase diagram is indicated in the Inset (from [Steppke 13]). (c) Pressure-variation of the electrical-resistivity quadratic coefficient $A$ of CeRuPO (from [Kotegawa 13]). (d) Temperature-variation of the NMR relataxation rate, plotted as $S_\perp = 1/2T_1$, measured for $\mathbf{H} \parallel \mathbf{c}$ on $CeRu_{1-x}Fe_xPO$ alloys, with the doping variation of $S_\perp$ at $T = 0.2$ K in the Inset (from [Kitagawa 12]). (e) Temperature-dependence of the NMR relaxation rate $1/T_1T$ of UCoGe under pressure, and its pressure-temperature phase diagram in the Inset (from [Manago 19b]).

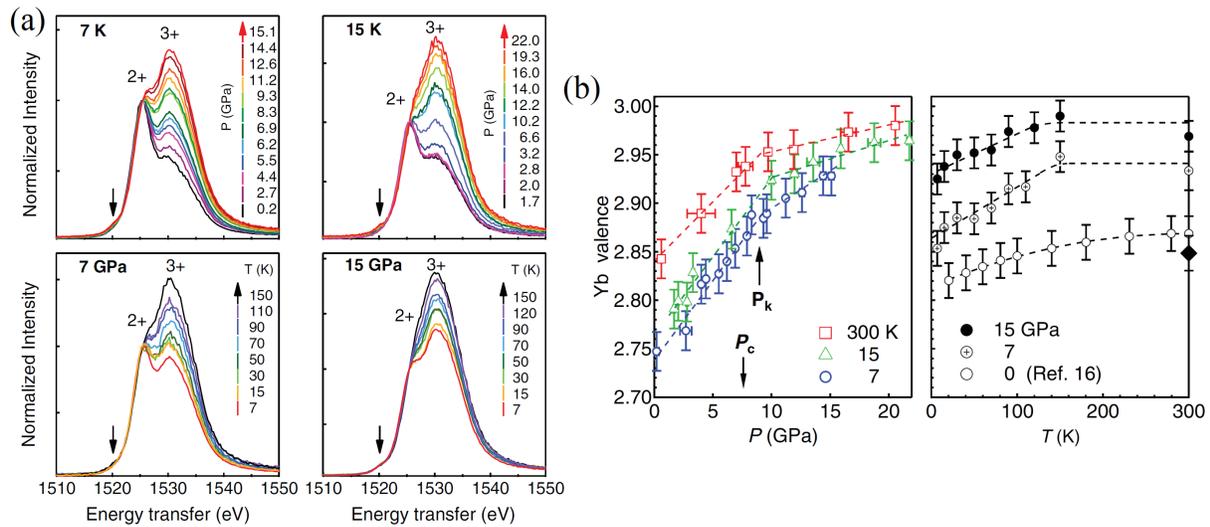

Figure 5.11: (a) RIXS spectra measured on YbCu$_2$Si$_2$ at different sets of temperatures and pressures and (b) temperature and pressure variation of the valence extracted from these spectra (from [Fernandez-Pañella 12]).

A microscopic understanding of the critical phenomena in play requests to determine the relationship between long-range magnetic orders and the magnetic fluctuations, implying first to identify them and to characterize their evolution in the electronic phase diagrams. Contrary to the case of a quantum antiferromagnetic phase transition, where fluctuations of the antiferromagnetic order parameter were shown to be peaked and, thus, critical [Knafo 09a], a microscopic study by inelastic neutron scattering of the wavevector-dependence of magnetic fluctuations at a quantum ferromagnetic phase transition is missing. Heavy-fermion ferromagnets sometimes lie near an antiferromagnetic phase transition, and a characterization of the magnetic fluctuations in the different parts, including the FM and AF phases and the CPM regime, of the phase diagram would be needed.

### 5.2.3 Valence

A study of valence in the heavy-fermion paramagnet YbCu$_2$Si$_2$ in the vicinity of its pressure-induced quantum ferromagnetic phase transition is summarized here [Fernandez-Pañella 12]. Figure 5.11(a) presents resonant x-ray scattering spectra measured at temperatures $T \geq 7$ K, i.e., at temperatures slightly higher than the Curie temperature $T_C$ of the pressure-induced ferromagnetic phase (see phase diagram in Figure 5.7(b)). The pressure- and temperature-variations of valence extracted from these spectra are summarized in Figure 5.11(b). They show that YbCu$_2$Si$_2$ at ambient pressure is in an intermediate-valent state, with a valence reaching its minimum value $v \simeq 2.75$ at low temperature. Valence increases under pressure, reaching a value $v \simeq 2.85 - 2.9$ in the vicinity of $p_c$. A kink characterizes the valence-variation in the vicinity of $p_c$, above which the valence becomes less pressure-dependent than at low pressure. This indicates that ferromagnetism is stabilized in YbCu$_2$Si$_2$ under pressure in a nearly-integer-valent regime. A similar valence variation was observed in the heavy-fermion paramagnet YbNi$_3$Ga$_9$ in the vicinity of a quantum phase transition, presumably of antiferromagnetic nature, induced under pressure (see Figure 4.15 in Section 4.2.4) [Matsubayashi 15]. In this intermediate-valent compound at ambient pressure, the pressure-induced magnetically-ordered phase was also found to develop in the vicinity of a nearly integer-valent state.

The studies of valence made on a few heavy-fermion compounds confirm that heavy-fermion physics, and the related quantum magnetic phase transitions, develop in systems in which the $f$ electrons are nearly-localized, i.e., where the valence is nearly-integer, in agreement with general considerations made in Section 2.2.3.

## 5.3 Magnetic-field-induced phenomena

The effects of a magnetic field applied on heavy-fermion ferromagnets are considered here. Section 5.3.1 presents the magnetic-moment reorientations observed in these systems under a magnetic field. Section 5.3.2 shows that superconductivity can be induced, or reinforced, by a magnetic field near the quantum metamagnetic transition of U-based ferromagnets. Sections 5.3.3 and 5.3.4 focus on the experimental signatures of the critical magnetic fluctuations and of possible Fermi-surface reconstructions, respectively, at these magnetic-field-induced transitions.

### 5.3.1 Magnetic-moment reorientations

The magnetic-moment reorientation processes occurring in heavy-fermion ferromagnets in a magnetic field are presented here. An introduction to basic effects of a magnetic field to ferromagnets is made in Section 5.3.1.1. The case of a magnetic field applied along an easy magnetic direction is considered in Section 5.3.1.2, while that of a magnetic field applied along a hard magnetic direction is considered in Section 5.3.1.2.

#### 5.3.1.1 Basic properties

Zero-field-cooled ferromagnets are composed of magnetic domains, in which the magnetic moments $\mu_{FM}$ are aligned along the same direction, 'up' in some domains, 'down' in the other domains. Figure 5.12(a) shows that, in a magnetic field applied along the easy magnetic direction, i.e., $\mathbf{H} \parallel \mu_{FM}$, a first-order transition accompanied by an hysteretic step-like variation of the magnetization occurs at a critical field $H_c$. Due to the hysteresis, a remanent magnetization $M_r$ is observed in 'hard' ferromagnets once the magnetic field is reduced from $H > H_c$ down to $H = 0$. In most of anisotropic materials, a magnetic field applied perpendicularly to the ferromagnetic moments, i.e., $\mathbf{H} \perp \mu_{FM}$, leads to a linear increase of the magnetization, which saturates at higher field values, generally without hysteresis. In some of them, meta-magnetism can also be induced by a transverse magnetic field (see Section 5.3.1.2). Figure 5.12(b) shows the typical zero-field-cooled magnetization-versus-temperature curves of a ferromagnet for different values of a magnetic field $\mathbf{H} \parallel \mu_{FM}$. For $H < H_c$, a sharp signature of the phase transition (a kink in $M(H)$ for a second-order phase transition) can be defined at the Curie temperature $T_C$ delimiting the low-temperature ferromagnetic phase. In magnetic fields $H > H_c$, there is no signature of phase transition and a broad crossover is observed at a temperature scale $T_{pol}$, which can be defined at the inflexion point of $M(H)$. While the ferromagnetic phase is delimited by the transitions at $T_C$ and $H_c$, the crossover at $T_{pol}$ indicates the progressive onset of a low-temperature polarized magnetic regime. In relation with the proximity of the ferromagnetic phase, ferromagnetic fluctuations, i.e., short-range ferromagnetic order, may be present in the PPM regime in fields near $H_c$. Similar ferromagnetic fluctuations were observed near the metamagnetic field $H_m$ in the PPM regime of the heavy-fermion paramagnet $CeRu_2Si_2$ (see Figure 3.12 in Section 5.3.3) [Sato 01, Flouquet 04]. Figure 5.12(c) shows the magnetic-field-temperature phase diagram of a ferromagnet in a field $\mathbf{H} \parallel \mu_{FM}$. It emphasizes the first-order transition at $H_c$ from a low-field ferromagnetic phase composed of magnetic domains and delimited by a sharp transition at $T_C$, to a high-field polarized regime, where all magnetic moments all aligned within the field direction, and delimited by a broad crossover at $T_{pol}$, which increases with $H$.

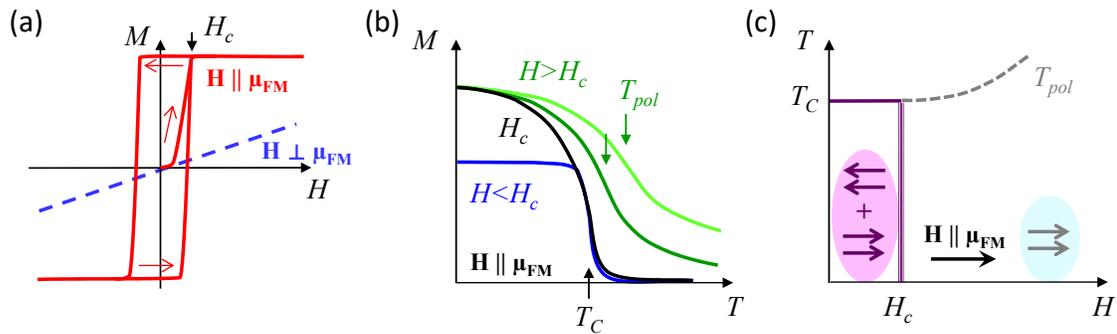

Figure 5.12: Schematic behavior expected for ferromagnets in a magnetic field. (a) Low-temperature magnetization versus magnetic field, in a magnetic field applied along easy and hard magnetic axes, (b) Magnetization versus temperature for different fields values, in a magnetic field applied along an easy magnetic axis and (c) magnetic-field-temperature phase diagram in a magnetic field applied along an easy magnetic axis.

The Curie temperature can be driven to zero under pressure $p$, or chemical doping $x$, leading to a quantum phase transition separating a ferromagnetic and a paramagnetic ground-states. In the following, the tuning parameter is noted $\delta$, with the convention that ferromagnetism is

(a)                                    (b)

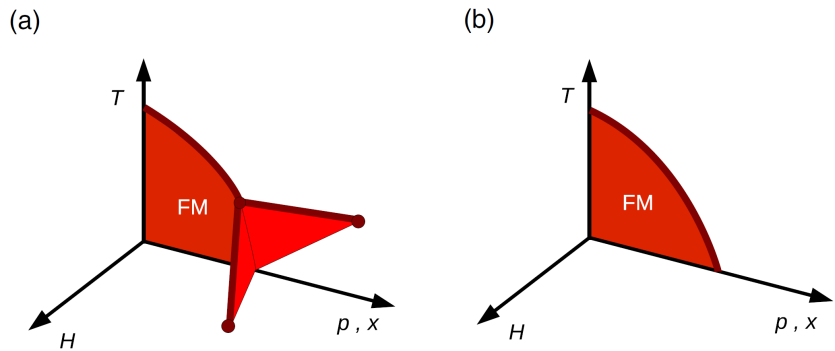

Figure 5.13: Schematic three-dimensional phase diagram expected for ferromagnets in a magnetic field combined with doping/pressure, (a) in the case of a discontinuous transition related with tricritical wings and (b) in the case of a continuous transition (from [Brando 16]).

stabilized for $\delta < \delta_c$. Figure 5.13 presents schematically the three-dimensional magnetic-field - pressure (or doping) - temperature phase diagrams representative of two classes of quantum ferromagnets in a magnetic field $\mathbf{H} \parallel \mu_{FM}$, for which a discontinuous or a continuous behavior of $T_C$ is reported at $\delta_c$ [Brando 16]. In these Figures, a critical field $H_c \to 0$ is considered and the ferromagnetic phase corresponds to a part of the $H = 0$ plane delimited by $T < T_C$ and $\delta < \delta_c$. Figure 5.13(a) presents the case of a discontinuous behavior of $T_C$ at $\delta_c$. $T_C$ drops from a non-zero value, corresponding to a tri-critical point, to a zero value at $\delta_c$. For $\delta \gtrsim \delta_c$, a magnetic field leads to a first-order transition to the polarized regime. This first-order transition is part of a wing-like plane, which ends into two quantum critical points at $(\pm H_{QCP}, \delta > \delta_c, T \to 0)$, also labeled as 'quantum critical endpoints' or 'quantum wing critical points' in the literature [Brando 16, Taufour 16]. Figure 5.13(b) presents the case of ferromagnets where $T_C$ decreases continuously when $\delta_c$ is approached, and for which no wing-like phase diagram is observed. Experimental realizations of discontinuous and continuous quantum critical ferromagnetic behaviors are presented in Sections 5.3.1.2 and 5.3.1.3, respectively.

#### 5.3.1.2   Field along an easy magnetic axis

In a magnetic field $\mathbf{H} \parallel \mu_{FM}$ applied along the easy-magnetic axis of a ferromagnet, the polarization of the magnetic moments generally occurs at a small critical field $\mu_0 H_c \ll 1$ T. These low-field effects are presented in Section 5.3.1.2.1. Close to a discontinuous critical point, where the Curie temperature suddenly collapses, wing-like phase diagrams can be observed. They are presented in Section 5.3.1.2.2.

#### 5.3.1.2.1   Polarization of ferromagnetic moments
Figure 5.14 presents hysteretic magnetization processes measured in three heavy-fermion ferromagnets in a magnetic field applied along their easy magnetic axis. It indicates small critical fields $\mu_0 H_c \simeq 50$ mT for UGe$_2$ in $\mathbf{H} \parallel \mathbf{a}$ (Figure 5.14(a))[Saxena 00], $\mu_0 H_c \simeq 5$ mT for UCoGe in $\mathbf{H} \parallel \mathbf{c}$ (Figure 5.14(b)) [Paulsen 12], and $\mu_0 H_c \simeq 100$ mT for CeRuPO in **H** ∥ **a** (Figure 5.14(c)) [Krellner 08]. For the three compounds, a large hysteresis leads to a strong remanent magnetization, whose value is similar to that of the saturated magnetization. Figure 5.14(b) shows different magnetization loops, for different maximum fields, emphasizing the additional superconducting diamagnetic signal (negative slope of M(H)) for small field-loops in UCoGe [Paulsen 12]. UGe$_2$ and URhGe have a strong Ising high-temperature magnetic anisotropy, whose easy magnetic axes, **a** and **c**, respectively, correspond to the direction of the moments $\mu_{FM}$ which order ferromagnetically at low temperature [Aoki 19a]. Oppositely, CeRuPO presents a weak XY magnetic anisotropy at high temperatures, but low-temperature ferromagnetic order is established within a Ising magnetic anisotropy, with $\mu_{FM}$ ∥ **c**, presumably due to a low-temperature anisotropic magnetic exchange [Krellner 08]. CeRuPO is not an isolated case and low-temperature magnetic ordering along a direction perpendicular to the high-temperature easy magnetic axis is also observed for other heavy-fermion ferromagnets [Hafner 19] and antiferromagnets (see the CeRhIn$_5$ [Bao 00, Takeuchi 01] and CePd$_2$Si$_2$ [Grier 84, van Dijk 00] cases in Figures 4.2(b-c) and 4.3(c-d), in Section 4.1.2).

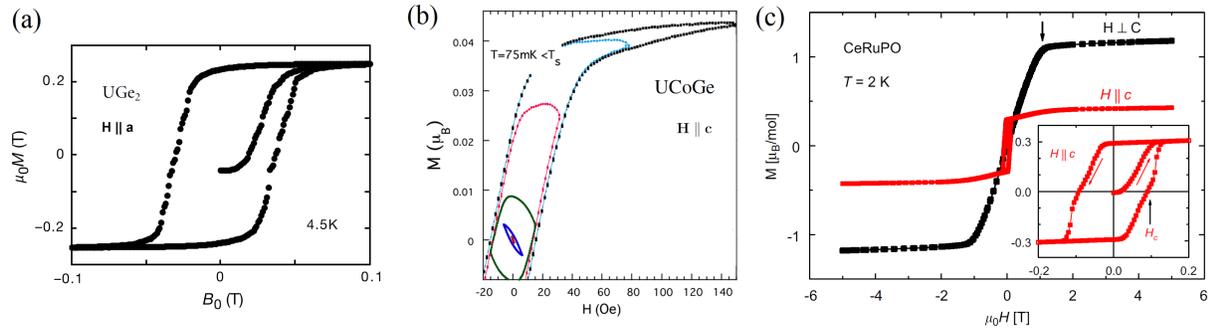

Figure 5.14: Hysteresis in the magnetization versus magnetic field of (a) UGe$_2$ at $T = 4.5$ K **H** ∥ **a** (from [Saxena 00]), (b) of UCoGe at $T = 75$ mK **H** ∥ **c** (from [Paulsen 12]), and of CeRuPO at $T = 2$ K **H** ∥ **c** (easy magnetic axis with hysteresis) and **H** ⊥ **c** (hard magnetic axis without hysteresis) (from [Krellner 08]).

Figure 5.15 focuses on thermodynamic signatures of the presence of a Curie temperature $T_C$ for $H < H_c$ and of its disappearance for $H > H_c$, for two examples of heavy-fermion ferromagnets, UIr and CeRuPO, in a magnetic field **H** ∥ $\mu_{FM}$. It shows that, for $H > H_c$, $T_C$ is replaced by a broad crossover at a temperature $T_{pol}$ characteristic of the field-polarized regime. Figure 5.15(a-c) present a set of magnetization data, plotted as $M$ versus $T$, $\chi = M/H$ versus $T$, and $|\partial\chi/\partial T|$ versus $T$, respectively, measured on UIr at different magnetic fields **H** ∥ $\mu_{FM}$ ∥ [10$\bar{1}$] [Knafo unp.]. Ferromagnetic ordering leads to a kink in $M(T)$, corresponding to a sharp maximum in $|\partial\chi/\partial T|$ at $T_C$. In increasing magnetic fields $H > H_c$, the maximum in $|\partial\chi/\partial T|$ becomes broader, and the associated temperature scale $T_{pol}$ increases. Figure 5.15(d) presents heat-capacity data, plotted as $C_p$ versus $T$, measured on UIr at zero field and different magnetic fields **H** ∥ [10$\bar{1}$] [Sakarya 10]. A step-like jump indicates the second-order nature of the ferromagnetic transition for $H = 0$. The transition disappears in magnetic fields $H > H_c$.

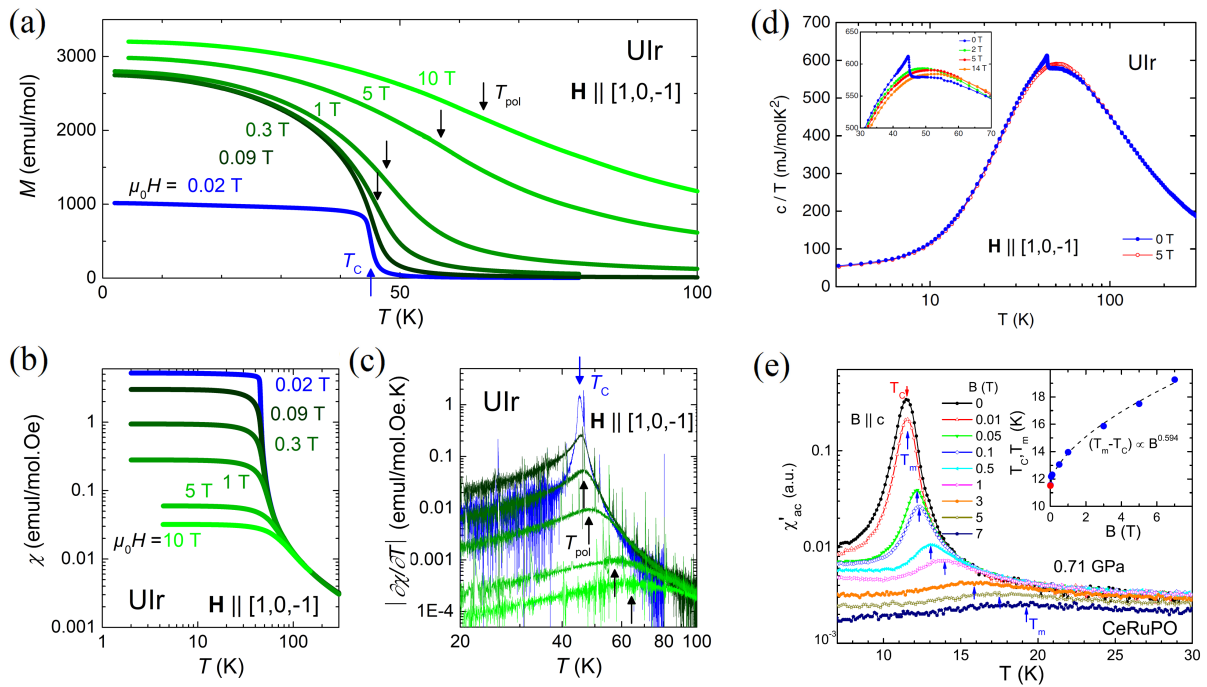

Figure 5.15: Temperature-variation of (a) the magnetization $M$, (b) the magnetic susceptibility $\chi = M/H$, (c) the absolute value of the temperature-derivative of $\chi$ (from [Knafo unp.]), and (d) the heat capacity divided by temperature of UIr at different magnetic-fields values with $\mathbf{H} \parallel [10\bar{1}]$ (from [Sakarya 10]). ac magnetic susceptibility $\chi'_{ac}$ versus temperature in CeRuPO at $p = 0.71$ GPa, at different magnetic-field values with $\mathbf{H} \parallel \mathbf{c}$ (from [Lengyel 15]).

Figure 5.15(e) presents ac-magnetic susceptibility $\chi'_{ac}$ versus temperature data measured on the ferromagnet CeRuPO with $\mathbf{H} \parallel \mu_{FM} \parallel \mathbf{c}$. In this set of data, the contribution from the static ferromagnetic order is not captured by $\chi'_{ac}$, which only probes the ferromagnetic fluctuations, i.e., the ferromagnetic short-range order. A sharp enhancement of $\chi'_{ac}$ at $T_C$ for $\mu_0 H \leq 0.01$ T is replaced by a broader maximum at a temperature, labeled by $T_m$ in the graph, characteristic of the polarized regime. The inset of Figure 5.15(e) further shows that a power-law behavior $(T_m - T_c) \sim H^{0.6}$ is observed.

#### 5.3.1.2.2 Wing-like phase diagrams

Experimental realizations of wing-like ferromagnetic phase diagrams are presented in Figure 5.16, where the isostructural Ising ferromagnet URhAl and nearly-ferromagnet UCoAl are considered.

- Figure 5.16(a) indicates that URhAl presents a discontinuous quantum ferromagnetic phase transition at the critical pressure $p_c = 5$ GPa, at which $T_C$ suddenly drops from 11 to 0 K [Shimizu 15]. Figure 5.16(b) shows that, for $p > p_c$, a magnetic field applied along the easy ferromagnetic direction $\mathbf{H} \parallel \mu_{FM} \parallel \mathbf{c}$ induces a first-order metamagnetic transition, from the zero-field paramagnetic phase to a high-field polarized mag-netic regime. $\mu_0 H_m$ grows linearly with $H$, from 0 T at $p_c$ to $\simeq$ 5 T at $p$ = 6 GPa. The metamagnetic boundary corresponds to a Wing-like plane in the three-dimensional magnetic-field-pressure-temperature phase diagram of URhAl shown in Figure 5.16(c).

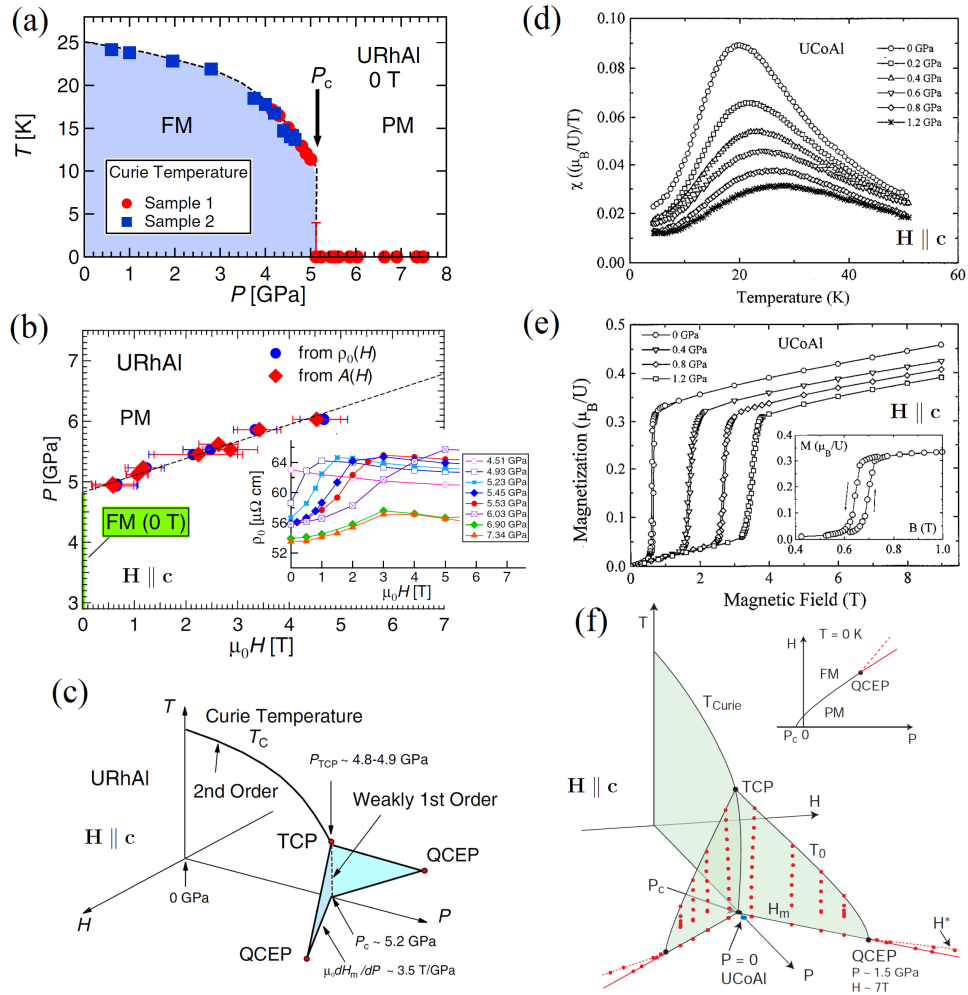

Figure 5.16: (a) Pressure-temperature phase diagram, (b) low-temperature pressure-magnetic field phase diagram, with the residual electrical resistivity versus magnetic field, at different pressures, in the Inset, and (c) three-dimensional magnetic field - pressure - temperature phase diagram of URhAl in a magnetic field **H** ∥ **c** (from [Shimizu 15]). (d) Magnetic susceptibility versus temperature at different pressures, (b) low-temperature magnetization versus magnetic field at different pressures (from [Mushnikov 99]), and (c) three-dimensional magnetic-field-pressure-temperature phase diagram of UCoAl in a magnetic field **H** ∥ **c** (from [Aoki 11c]).

• UCoAl is a paramagnet characterized, in a magnetic field applied along its easy magnetic axis **c**, by a broad maximum in the magnetic susceptibility at the temperature $T_\chi^{max}$ = 20 K and a sharp first-order metamagnetic transition at a relatively small mag-netic field $\mu_0 H_m = 0.7$ T [Mushnikov 99]. Figures 5.16(d-e) show that $T_\chi^{max}$ and $\mu_0 H_m$ both increase with pressure, up to 28 K and 3.5 T, respectively, at $p = 1.2$ GPa. Similarly to other heavy-fermion paramagnets (see Section 3.2), we can identify a correlated paramagnetic regime for $T < T_\chi^{max}$ and $H < H_m$. The three-dimensional magnetic-field-pressure-temperature phase diagram shown in Figure 5.16(f) indicates that the first-order character of the transition at $H_m$ is lost at a pressure of 1.5 GPa corresponding to a quantum critical endpoint [Aoki 11c]. For $p > 1.5$ GPa, $H_m$ continues to increase linearly with $p$, reaching 10.5 T at $p = 2.36$ GPa, and a broadening of the metamagnetic transition is observed.

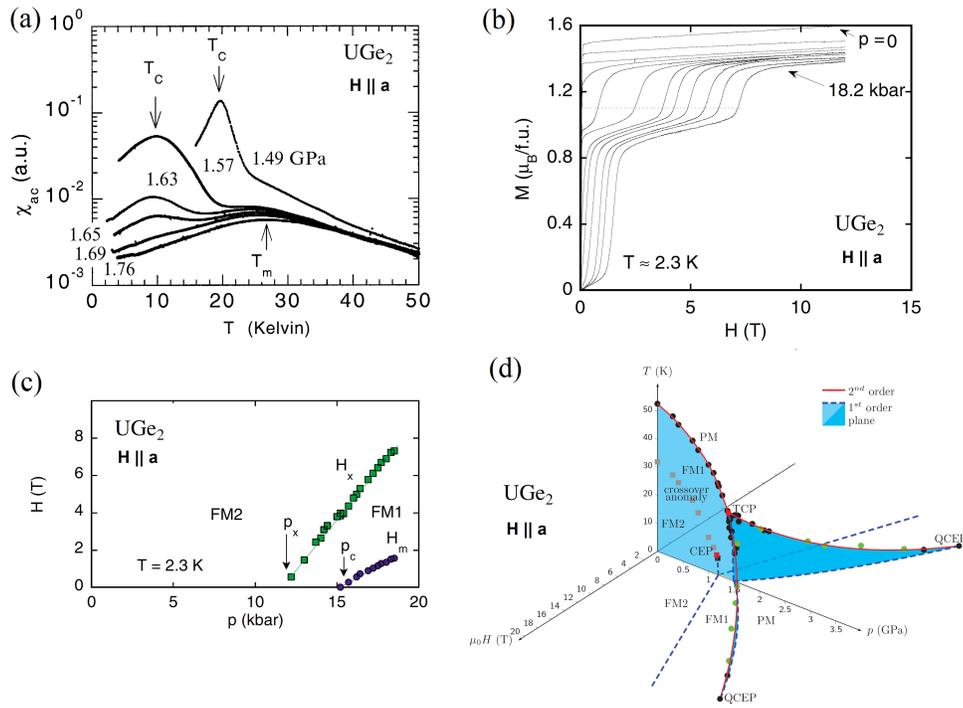

Figure 5.17: (a) ac magnetic susceptibility versus temperature (from [Huxley 00]), (b) low-temperature magnetization versus magnetic field measured at different pressures, (c) low-temperature pressure-magnetic-field phase diagram (from [Pfleiderer 02]), and (d) three-dimensional magnetic-field-pressure-temperature phase diagram of UGe$_2$ in a magnetic field **H ∥ a** (from [Taufour 11]).

In [Aoki 11c], it was proposed that UCoAl lies close to a discontinuous quantum ferromagnetic phase transition occurring at an effective 'negative' pressure $p_c^{eff} < 0$ (see Figure 5.16(f)). This proposition is consistent with the observation of a wing-like phase diagram in the isostructural compound URhAl (see Figure 5.16(c)) [Shimizu 15].

The ferromagnet UGe$_2$, of Curie temperature $T_C = 53$ K, is considered in Figure 5.17. This compound also presents a discontinuous quantum phase transition at $p_c \simeq 1.5$ GPa, beyond which a paramagnetic regime is stabilized. Figure 5.17(a) shows that, for $p > p_c$ and in a magnetic field applied along the easy magnetic axis, i.e., $\mathbf{H} \parallel \mu_{FM} \parallel \mathbf{a}$, a broad maximum in the magnetic susceptibility (here probed by an ac susceptibility measurement) at a temperature $T_\chi^{max} \simeq 25$ K [Huxley 00] indicates the onset of a correlated paramagnetic regime, similar to that observed in the nearly-ferromagnet UCoAl (see Figure 5.16(d)) or in heavy-fermion paramagnets (see Section 3.2). A particularity of UGe$_2$ is the presence of two metamagnetic transitions induced under pressure in a magnetic field $\mathbf{H} \parallel \mathbf{a}$, evidenced by sharp first-order jumps in the magnetization shown in Figure 5.17(b) [Pfleiderer 02]. Figure 5.17(c) shows that, for $p > p_x = 1.2$ GPa, a first metamagnetic transition occurs at $H_x$, which ends at the critical point $(p_x, H = 0)$ of the phase diagram [Pfleiderer 02]. $p_x$ separates a large-moment ferromagnetic phase at low pressure from a small-moment ferromagnetic phase at higher pressure. For $p > p_c$, a second metamagnetic transition occurs at the magnetic field $H_m$, which ends at the critical point $(p_c, H = 0)$. These two metamagnetic transitions are associated with a 'double-wing-like' three dimensional pressure-magnetic-field-temperature phase diagram, shown in Figure 5.17(d) [Taufour 11].

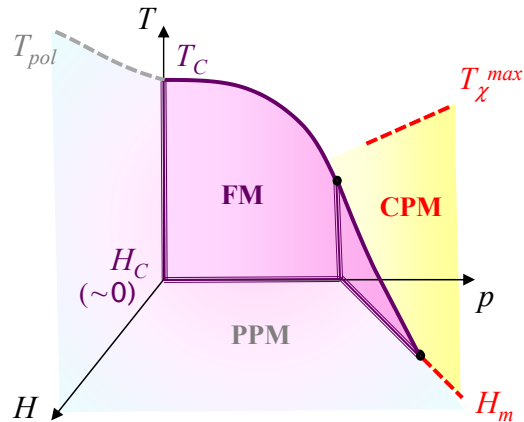

Figure 5.18: Alternative schematic three-dimensional magnetic-field-pressure-temperature phase diagram of ferromagnets with a discontinuous quantum phase transition.

Ferromagnets with a discontinuous quantum phase transition to a paramagnetic regime present the following properties, synthesized in the phase diagram shown in Figure 5.18:

- The high-field regime for $H > H_c$ is characterized by a broad temperature crossover marking the progressive onset of a polarized paramagnetic regime, similar to that observed at high fields in heavy-fermion antiferromagnets and nearly-antiferromagnets (see Sections 3.2 and 4.3).

- The high-pressure regime for $p > p_c$ is delimited by a broad maximum in the magnetic susceptibility at the temperature $T_\chi^{max}$ (observed in UCoAl and UGe$_2$) and by a metamagnetic transition at the magnetic field $H_m$. It can, thus, be identified as a correlated paramagnetic regime. Antiferromagnetic fluctuations can lead to a maximum of the magnetic susceptibility and were observed in the CPM regime of several heavy-fermion nearly-antiferromagnets (see Section 3.2). Inelastic neutron scattering experiments are needed to test if antiferromagnetic correlations may develop in the CPM regime of discontinuous quantum ferromagnets for $p > p_c$ as well.

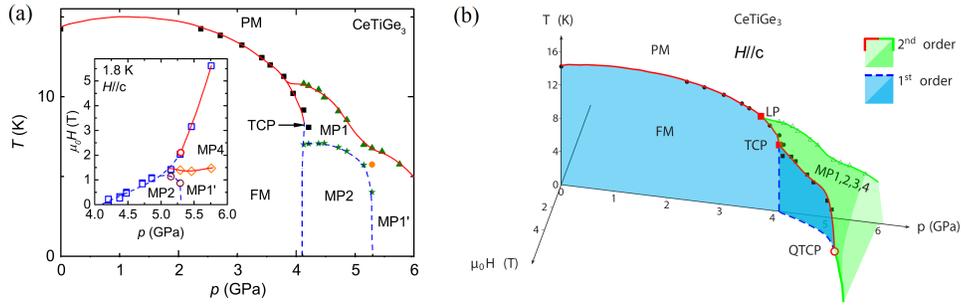

Figure 5.19: (a) zero-field pressure-temperature phase diagram, with the low-temperature pressure-magnetic-field phase diagram in the Inset, and (b) three-dimensional magnetic-field-pressure-temperature phase diagram of CeTiGe$_3$ in a magnetic field $\mathbf{H} \parallel \mathbf{c}$ (from [Kaluarachchi 18]).

- For most of heavy-fermion nearly-antiferromagnets, the ratio $R_{CPM} = T_\chi^{max}/H_m$ is nearly constant, varying in the range [0.5-2] (see Section 3.3). A deviation from this universal behavior is observed in the heavy-fermion nearly-ferromagnets UCoAl and UGe$_2$ under pressure, for which $H_m \to 0$ while $T_\chi^{max} \sim 25$ K at the critical pressure $p_c$. This deviation may be a consequence of the easiness to polarize a nearly-ferromagnet in a magnetic field $\mathbf{H} \parallel \mu_{FM}$.

In addition to paramagnet regimes with intersite magnetic fluctuations, antiferromagnetic phases can also exist near a quantum ferromagnetic phase transition. Figure 5.19 shows that a critical wing-like structure can be observed in ferromagnets driven to antiferromagnetism under pressure, as CeTiGe$_3$, under a magnetic field applied along the easy magnetic axis [Kaluarachchi 18]. The possibility of neighboring antiferromagnetic and ferromagnetic phases, and the related quantum critical phase diagrams has been considered theoretically by Belitz and Kirkpatrick [Belitz 17]. Multiple-wing structures, such that observed in UGe$_2$, or a wing-structure related with a quantum ferromagnetic-to-antiferromagnetic phase transition are also observed in other families of magnetic materials (see for instance the LaCrGe$_3$ ferromagnet [Kaluarachchi 17]).

### 5.3.1.3  Field along a hard magnetic axis

In a few ferromagnets, metamagnetism can be induced in higher magnetic field $\mathbf{H} \perp \mu_{FM}$ applied along a hard magnetic axis. Figure 5.20 presents magnetic susceptibility versus temperature, magnetization versus magnetic field, and electrical resistivity versus temperature for the isostructural orthorhombic heavy-fermion ferromagnets URhGe, UCoGe and URhSi, in magnetic fields applied along their main crystallographic directions $\mathbf{a}$, $\mathbf{b}$, and $\mathbf{c}$.

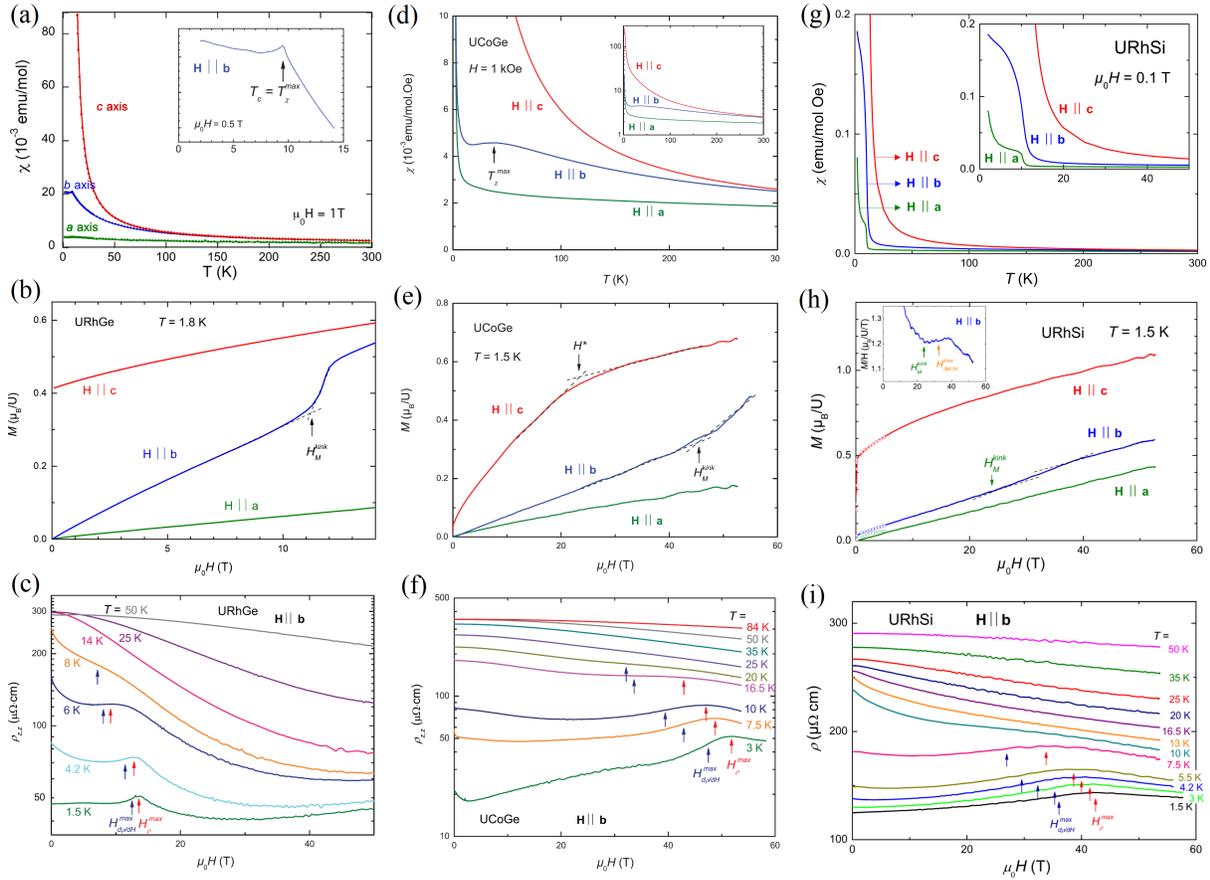

Figure 5.20: (a) Magnetic susceptibility versus temperature (from [Braithwaite 18]), (b) low-temperature magnetization versus magnetic field, for $\mathbf{H} \parallel \mathbf{a}, \mathbf{b}, \mathbf{c}$, and (c) electrical resistivity versus magnetic field for $\mathbf{H} \parallel \mathbf{b}$ at different temperatures (from [Knafo 12]) of URhGe. (d) Magnetic susceptibility versus temperature, (e) low-temperature magnetization versus magnetic field, for $\mathbf{H} \parallel \mathbf{a}, \mathbf{b}, \mathbf{c}$, and (f) electrical resistivity versus magnetic field for $\mathbf{H} \parallel \mathbf{b}$ at different temperatures of UCoGe (from [Knafo 12]). (g) Magnetic susceptibility versus temperature (from [Aoki unp.]), (h) low-temperature magnetization versus magnetic field, for $\mathbf{H} \parallel \mathbf{a}, \mathbf{b}, \mathbf{c}$, and (i) electrical resistivity versus magnetic field for $\mathbf{H} \parallel \mathbf{b}$ at different temperatures of URhSi (from [Knafo 19b]).

- The three samples have similar Ising uniaxial magnetic anisotropies, indicated by the hierarchy $\chi_c > \chi_b > \chi_a$ (Figure 5.20(a,d,g)) [Braithwaite 18, Knafo 12, Aoki unp.]. URhGe, UCoGe and URhSi order ferromagnetically for $T < T_C$, with $T_C = 9.5$, 2.7, and 10 K, respectively, and ferromagnetic moments $\mu_{FM} \parallel \mathbf{c}$ of amplitude of 0.4, 0.05, and 0.5 $\mu_B$/ U, respectively, revealed by the low-field magnetization in $\mathbf{H} \parallel \mathbf{c}$ (Figure 5.20(b,e,h)) [Knafo 12, Knafo 19b]. The small ferromagnetic moment in UCoGe indicates a nearby ferromagnetic phase transition (see Figure 5.8 in Section 5.2.1).

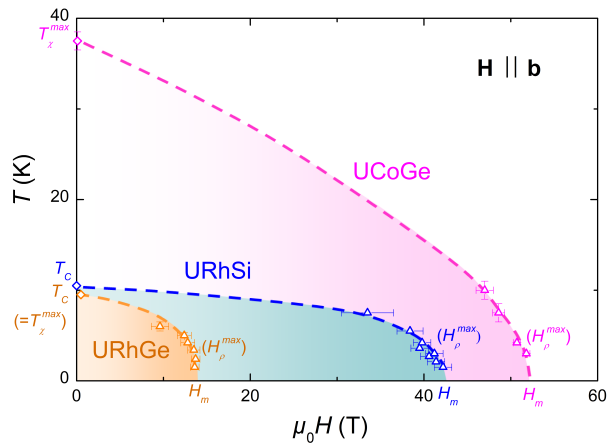

Figure 5.21: Magnetic-field-temperature phase diagrams of URhGe, UCoGe, and URhSi (from [Knafo 19b]).

- For the three compounds, signatures of metamagnetism are visible in the magnetization and electrical resistivity for **H** ∥ **b**. Sharp step-like jump in the magnetization and maximum in the electrical resistivity are observed at $\mu_0 H_m = 12$ T in URhGe (Figure 5.20(b,c)) [Knafo 12, Hardy 11]. Similar but broader anomalies are observed at $\mu_0 H_m \simeq 50$ T in UCoGe (Figure 5.20(e,f)) [Knafo 12], and much broader anomalies can be defined at $\mu_0 H_m \simeq 40$ T in URhSi (Figure 5.20(h,i)) [Knafo 19b]. The metamagnetic field and the sharpness of the associated anomalies are related with an anomaly in the magnetic susceptibility: a sharp kink and maximum in $\chi$ at $T_C = 9.5$ K in URhGe and a broad maximum in $\chi_b$ at $T_\chi^{max} = 37.5$ K in UCoGe. For URhSi, no clear anomaly can be observed in $\chi_b$. This may be related with the broad character of the metamagnetic crossover. The temperature-dependence of the electrical resistivity shown in Figure 5.20(c,f,i) indicates that the metamagnetic field decreases with increasing temperature, before vanishing at temperatures $T > T_C$ in URhGe and URhSi, and at temperatures $T > T_\chi^{max}$ in UCoGe.

The magnetic-field-temperature phase diagrams of the three compounds in a magnetic field **H** ∥ **b** are summarized in Figure 5.21. In URhGe, $H_c = H_m$ and $T_C = T_\chi^{max}$ are the boundaries of the ferromagnetic phase. In URhSi, $H_c = H_m$ is also verified but $T_\chi^{max}$ could not be defined. In UCoGe, a critical field $\mu_0 H_c \simeq 15$ T corresponding the fall of $T_C$, observed from electrical resistivity versus temperature data, was found to delimitate the ferromagnetic phase [Aoki 09a]. However, no signature of $H_c$ is visible in the magnetization and electrical resistivity versus magnetic field curves of UCoGe presented in Figures 5.20(e,f). The decoupling of $T_C$ and $T_\chi^{max}$ at low field is related with the decoupling of $H_c$ and $H_m$ at low temperature [Knafo 12]. A decoupling of $T_N$ and $T_\chi^{max}$ is also observed in heavy-fermion antiferromagnets near an antiferromagnetic phase transition (see Section 4.3.1.2) [Aoki 12a, Knafo 17].

The decoupling of $T_C$ and $T_\chi^{max}$ and its relation with metamagnetism was studied in UCo$_{1-x}$Rh$_x$Ge alloys in a magnetic field **H** ∥ **b** by Pospíšil *et al* [Pospíšil 20]. Figure 5.22(a) shows the mag-netization of $UCo_{1-x}Rh_xGe$ compounds measured at $T = 1.3$ K. It indicates the progressive increase of the metamagnetic field with Co-doping $x$. The doping-temperature and doping-magnetic-field phase diagrams are presented in Figures 5.22(b,c), respectively. For $x < 0.4$, the CPM regime is observed at temperatures $T_C < T < T_\chi^{max}$. The decoupling between $T_C$ and $T_\chi^{max}$ decreases when $x$ is increased, ending in a critical point at $x = 0.4$ separating the FM phase, the CPM and the high-temperature paramagnetic regimes. $T_C$ is maximum at the critical concentration. $H_m$ decreases with $x$ in a similar way than $T_\chi^{max}$, and the ratio $R_{CPM} = T_\chi^{max}/H_m = 0.75$ in $UCo_{1-x}Rh_xGe$ alloys [Pospíšil 20] is similar to the universal ratio observed in heavy-fermion paramagnets, where the role of antiferromagnetic fluctuations in the CPM regime was emphasized (see Section 3.3) [Aoki 13]. An inelastic-neutron-scattering study is needed to test whether antiferromagnetic fluctuations associated with a moment $\parallel \mathbf{b}$ may play a role in the CPM regime of UCoGe and $UCo_{1-x}Rh_xGe$ alloys (see also Figure 2.28 in Section 2.4.3.1).

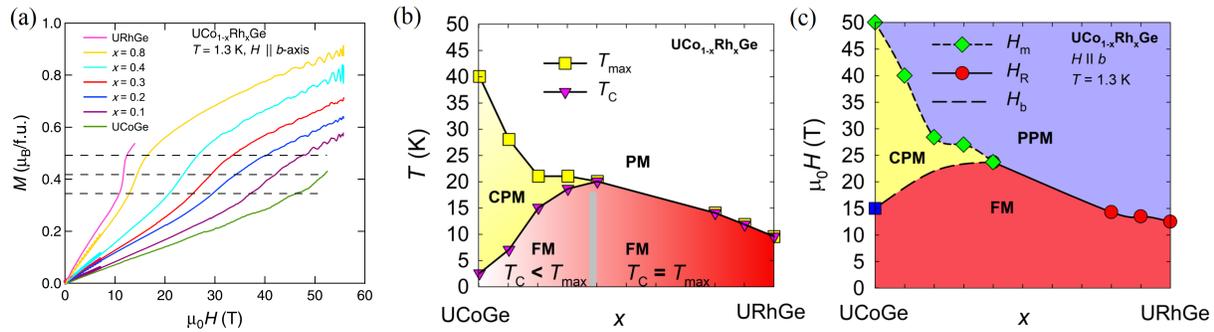

Figure 5.22: (a) Low-temperature magnetization versus magnetic field, (b) doping-temperature phase diagram, and (c) doping-magnetic-field phase diagram of $UCo_{1-x}Rh_xGe$ alloys in a magnetic field $\mathbf{H} \parallel \mathbf{b}$ (from [Pospíšil 20]).

### 5.3.2 Superconductivity

Superconductivity induced, or reinforced, by a magnetic field at a metamagnetic transition constitutes one of the most spectacular properties of heavy-fermion compounds. It was observed in the three ferromagnetic superconductors $UGe_2$, URhGe, and UCoGe, whose temperature-magnetic-field phase diagrams are shown in Figure 5.23.

- In $UGe_2$ at a pressure $p = 1.35$ GPa $> p_x$ in a magnetic field applied along the easy magnetic axis $\mathbf{a}$, an anomalous $S$-shape of $\mu_0 H_{c,2}$ indicates a reinforcement of superconductivity (see Figure 5.23(a-left panel)) [Huxley 01, Sheikin 01, Pfleiderer 02]. This reinforcement coincides with a metamagnetic transition occurring at $H_x$, at which the small magnetic moment from ferromagnetic phase FM1 are polarized (see Figure 5.17 in Section 5.3.1.2) [Pfleiderer 02].

- In URhGe at ambient pressure in a magnetic field applied along the hard magnetic axis

**b**, the low-field superconducting phase disappears at $\mu_0 H_{c,2} = 2$ T, and a high-field superconducting phase is induced in the vicinity of $\mu_0 H_m = 12$ T [Lévy 05, Aoki 19a] (see Figures 5.23(a-middle panel,b)). The maximum superconducting temperature of the high-field phase is twice higher than that of the zero-field superconducting phase. The metamagnetic transition field $H_m = H_c$ and the Curie temperature $T_C$ both delimitate the ferromagnetic phase (see also Figure 5.22(b-c) in Section 5.3.1.3).

- In UCoGe at ambient pressure and in a magnetic field applied along the hard magnetic axis **b**, an anomalous $S$-shape of $H_{c,2}$ indicates that superconductivity is reinforced around $\mu_0 H \simeq 15$ T [Aoki 09a, Aoki 19a] (see Figures 5.23(a-right panel,c)). A fall of $T_C$ was reported from electrical resistivity versus temperature data, corresponding to a critical field $\mu_0 H_c \simeq 15$ T coinciding with the field-reinforced superconductivity domain. However, no signature of metamagnetism was observed from field-dependent magnetization and electrical-resistivity measurements in this field range [Knafo 12]. Metamagnetism is observed at a much higher field $\mu_0 H_m = 50$ T, which delimitates, together with the temperature $T_\chi^{max} = 37.5$ K, a correlated paramagnetic regime surrounding the ferromagnetic phase (see also Figure 5.22(b-c) in Section 5.3.1.3) [Knafo 12].

The relationship between field-induced superconductivity and metamagnetism can be characterized by tuning the metamagnetic properties. Figures 5.24(a-c) summarize a set of experiments performed on URhGe, where pressure, uniaxial pressure, and a tilt of the magnetic field direction were adjusted. For comparison, Figure 5.24(d) shows that the effects of a tilt of the magnetic field on UCoGe.

- Figure 5.24(a) presents the magnetic-field-temperature phase diagrams of URhGe in **H** ∥ **b**, at different pressures up to 0.84 GPa [Miyake 09]. Pressure leads to a progressive collapse of the low-field superconducting phase, whose critical temperature $T_{sc}$ and field $H_{c,2}$ decrease under pressure (see Figure 5.8(c) in Section 5.2.1) [Hardy 05]. As well as $T_C$, $H_m$ increases under pressure, from 12 T at ambient pressure to $\simeq 15$ T at $p = 0.84$ GPa. Field-induced superconductivity is 'glued' to $H_m$ and is pushed to higher-field values. The increase of $H_m$ is accompanied by a decrease of the transition temperature $T_{sc}$ of the field-induced superconducting phase.

- Figure 5.24(b) presents temperature-magnetic-field- phase diagrams of URhGe in a magnetic field **H** ∥ **b**, at different uniaxial pressures $\sigma$ ∥ **b** up to 1.2 GPa [Braithwaite 19]. The effects of uniaxial pressure are opposed to those of hydrostatic pressure. An increase of the zero-field superconducting temperature and a decrease of the Curie temperature are observed. The initial slope of the low-field magnetization is related with a decrease of $H_m$. The field-induced superconducting phase is pushed down to lower field values, where it is enhanced up to almost 1 K for $\sigma = 1.2$ GPa, at a magnetic field $\mu_0 H = 4$ T.

- Figure 5.24(c) shows that, in URhGe, a magnetic field tilted away from **b** towards **c** leads to an enhancement of $H_m$. Its enhancement is accompanied by a weakening of the field-induced superconducting phase, which disappears at angles $\gtrsim 5°$ [Aoki 19a]. The Inset shows that, by considering separately the $H_b$ and $H_c$ components of the magnetic field along the directions **b** and **c**, respectively, a wing-like ferromagnetic phase diagram related with a discontinuous quantum ferromagnetic phase transition can be evidenced [Lévy 07].

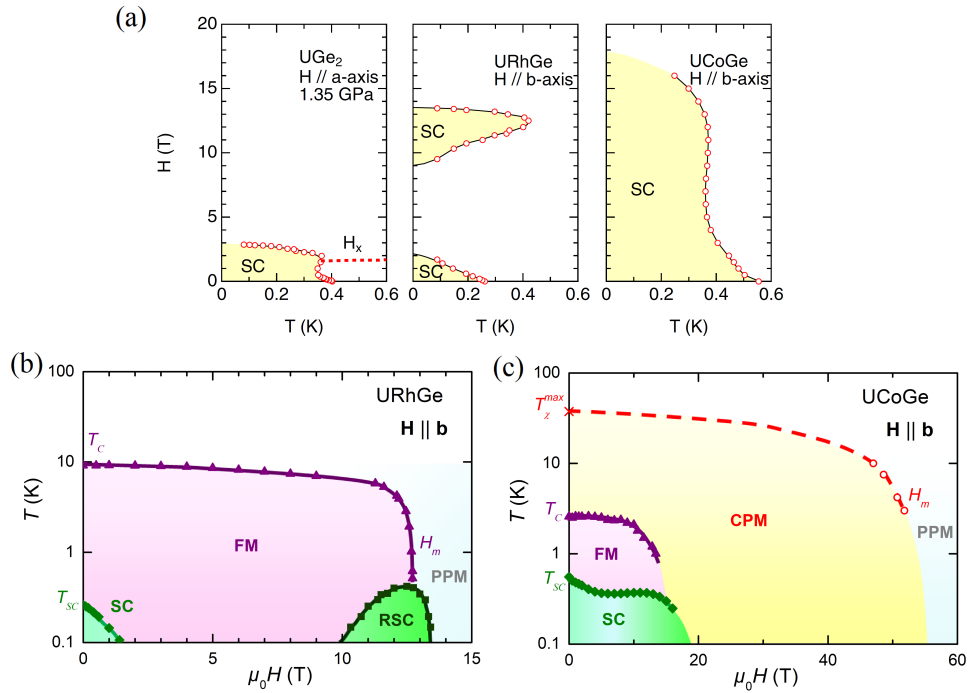

Figure 5.23: (a) Temperature-magnetic-field phase diagrams focusing on superconductivity in UGe$_2$ with **H** ∥ **a**, URhGe with **H** ∥ **b** and UCoGe with **H** ∥ **b** (from [Aoki 19a]). Magnetic-field-temperature phase diagrams emphasizing the interplay between magnetism and superconductivity in (b) URhGe (data from [Aoki 12b]) and (c) UCoGe with **H** ∥ **b** (data from [Aoki 09a, Knafo 12]).

- Figure 5.24(d) shows that, in UCoGe, a magnetic field tilted away from **b** towards **c** is accompanied by a sudden weakening of field-reinforced superconductivity [Aoki 19a]. The critical superconducting field, which reaches 20 T for **H** ∥ **b**, is strongly reduced, down to < 5 T at angles ≳ 5°.

High values of the superconducting critical fields are observed in U-based ferromagnetic superconductors, in particular for magnetic-field directions leading to a reinforcement or a reentrance of superconductivity. They are often considered as an indication for a spin-triplet mechanism of superconductivity, where no Pauli high-field limitation is expected [Aoki 19a]. The critical magnetic fluctuations at the metamagnetic transition, at $H_x$ in UGe$_2$ and $H_m$ in URhGe and UCoGe, are suspected to conduct to field-induced or reinforced superconductivity in these three ferromagnets. However, their microscopic characteristics, their feedback for superconductivity at zero field, and their evolution in a magnetic field need to be clarified for a quantitative description of the superconducting mechanism in the different phases.

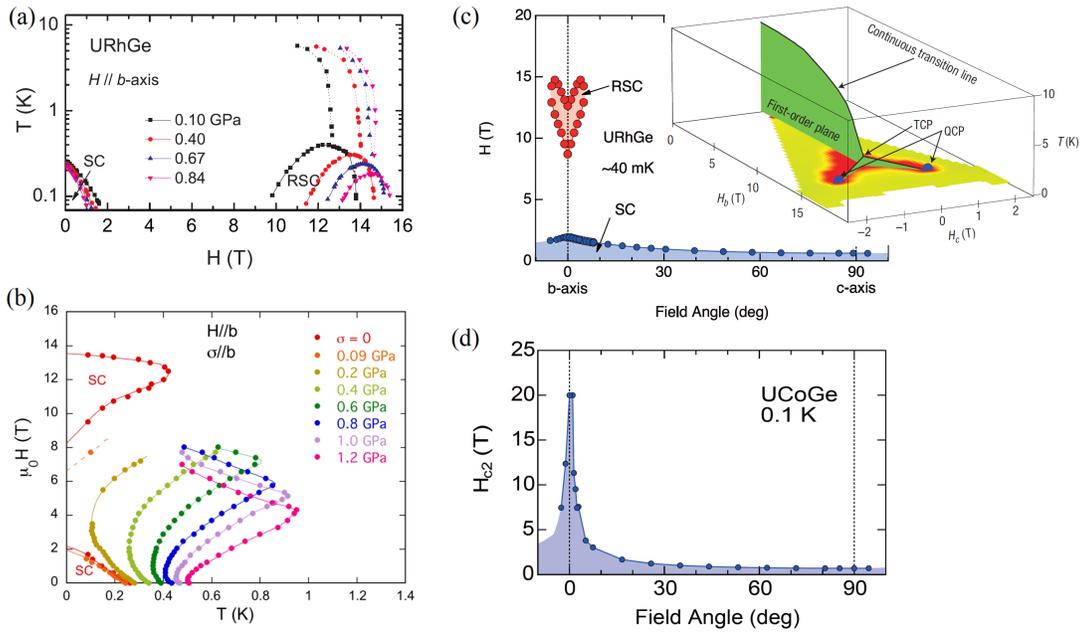

Figure 5.24: (a) Magnetic-field-temperature phase diagram at different pressures (from [Miyake 09]) and (b) temperature-magnetic-field phase diagram at different uniaxial pressures $\sigma \parallel \mathbf{b}$ (from [Braithwaite 19]) of URhGe in a magnetic field $\mathbf{H} \parallel \mathbf{b}$. (c) Low-temperature field-direction-dependent magnetic-field phase diagram of URhGe in a magnetic field varying from $\mathbf{H} \parallel \mathbf{b}$ to $\mathbf{H} \parallel \mathbf{c}$ (from [Aoki 19a]) and, in the Inset, corresponding three-dimensional magnetic-field-temperature phase diagram in magnetic fields $\mathbf{H} \perp \mathbf{a}$ (from [Lévy 07]). (d) Low-temperature field-direction-dependent magnetic-field phase diagram of UCoGe in a magnetic field varying from $\mathbf{H} \parallel \mathbf{b}$ to $\mathbf{H} \parallel \mathbf{c}$ (from [Aoki 19a])

## 5.3.3 Effective mass and magnetic fluctuations

Signatures of critical magnetic fluctuations induced at a metamagnetic transition, in a magnetic field applied along the easy-magnetic axis of nearly-ferromagnets (see Section 5.3.3.1) or along a hard magnetic axis of ferromagnets (see Section 5.3.3.2), are presented here. In the case of field-induced superconductivity, these critical magnetic fluctuations are suspected to drive an unconventional mechanism of superconductivity.

### 5.3.3.1 Easy magnetic axis

Figure 5.25 presents the variations of the Fermi-liquid coefficient $A$ extracted from electrical-resistivity measurements on UGe$_2$, URhAl and UCoAl under combined pressures and magnetic fields applied along the easy magnetic axis. Focus is given to their nearly-ferromagnetic regimes, from which wings in the three-dimensional phase diagram were reported (see Figures 5.17 and 5.16 in Section 5.3.1.2).

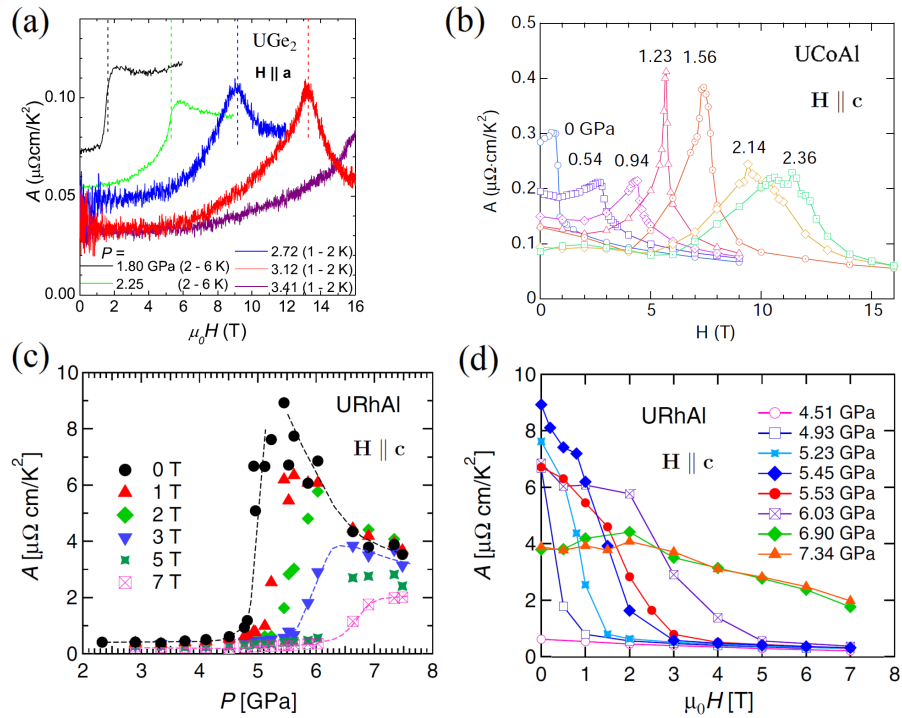

Figure 5.25: Variation of the Fermi-liquid quadratic coefficient $A$ of the electrical resistivity (a) of UGe$_2$ as function of a magnetic field $\mathbf{H} \parallel \mathbf{a}$, for different pressures (from [Kotegawa 11]), (b) of UCoAl as function of a magnetic field $\mathbf{H} \parallel \mathbf{c}$, for different pressures (from [Aoki 11c]), and of URhAl in a magnetic field $\mathbf{H} \parallel \mathbf{c}$, (c) as function of pressure a for different magnetic fields, and (d) as function of magnetic field, for different pressures (from [Shimizu 15]).

- In UGe$_2$ under pressures $p > p_c$, $A$ is enhanced at $H_m$ and similar values of $A(H_m)$ are extracted at the different pressures $1.8 \leq p \leq 3.4$ GPa (see Figure 5.25 (a)) [Kotegawa 11].

- In UCoAl, $A$ is maximum at $H_m$, but different values of $A(H_m)$ are extracted at the different pressures (see Figure 5.25 (b)) [Aoki 11c]. Sharp maxima of $A$ are identified at the quantum critical endpoint (see Figure 5.16(f) in Section 5.3.1.2).

- In URhAl, a maximum of $A$ is observed at the critical pressure $p_c$ for $H = 0$. For $p > p_c$, $A(H = 0)$ decreases with $p$, while $A(H)$ is constant for $H < H_m$, and decreases for $H > H_m$ (see Figures 5.25 (c-d)) [Shimizu 15].

For these three cases of metamagnetism in nearly-ferromagnets, $A$ reaches a maximum value at the metamagnetic field. This indicates an enhanced effective mass driven by quantum critical magnetic fluctuations, presumably of ferromagnetic nature (see the CeRu$_2$Si$_2$ case in Figure 3.12, Section 5.3.3 [Sato 01, Flouquet 04]). However, different variations of $A(H_m)$ are observed under pressure for the three nearly-ferromagnetic systems, suggesting a non-universal behavior. Similarly, different variations of $A(H_m)$ are observed at the metamagnetic transition of different nearly-antiferromagnets (see Section 4.3.2.2). This may result from subtle differences in the magnetic fluctuations spectra, and possibly in their wavevector dependence.

### 5.3.3.2 Hard magnetic axis

Figure 5.26 shows experimental signatures of critical magnetic fluctuations in the isostructural ferromagnets URhGe and URhSi in the vicinity of their metamagnetic transition, induced in a magnetic field applied along the hard magnetic axis $\mathbf{b}$.

- Figure 5.26(a) compares the field-variations of $A$ in URhGe and URhSi [Miyake 08, Knafo 19b]. The two systems are characterized by similar Curie temperatures $T_C \simeq 10$ K and ordered ferromagnetic moments $\mu_{FM} \simeq 0.5$ $\mu_B$/U. A sharp and intense peak is observed in URhGe at $\mu_0 H_m = 12$ T. A broader and lesser intense peak is observed in URhSi at $\mu_0 H_m = 40$ T. The critical magnetic fluctuations at $H_m$ are, thus, more intense in URhGe than in URhSi. Figure 5.26(b) shows that $A$ in URhGe at zero-field decreases under pressure, in agreement with a ferromagnetic groundstate stabilized under pressure and a quantum ferromagnetic phase transition at an effective negative pressure (see Figure 5.8(c) in Section 5.2.1) [Hardy 05]. For the different pressures, $A$ is maximum at $H_m$ and a decrease of $A(H_m)$ with increasing $p$ is observed. Under pressure, the decrease of $T_{sc}$ of the field-reentrant superconducting phase presumably results from less intense critical magnetic fluctuations, probed by the coefficient $A$, at $H_m$ (see Figure 5.24(a) in Section 5.3.2) [Miyake 09]. This confirms that a magnetically-mediated scenario of superconductivity may be relevant.

- Figure 5.26(c-d) presents different views on a set of NMR spin-spin relaxation rate $1/T_2$ measured on URh$_{0.9}$Co$_{0.1}$Ge under magnetic fields applied in the $(\mathbf{b}, \mathbf{c})$ plane [Tokunaga 15]. 10%-Co doping on URhGe enabled performing a $^{59}$Co NMR investigation of the magnetic fluctuations in a material whose magnetic properties are similar to those of pure-URhGe ($T_C$ increased by 1 K and $\mu_0 H_m$ increased by 1 T), but for which no superconductivity was observed. A sharp enhancement of the relaxation rate is observed at $H_m$ for $\mathbf{H} \parallel \mathbf{b}$. When the magnetic field is tilted from $\mathbf{b}$ to $\mathbf{c}$, $1/T_2$ remains maximum at $H_m$, which increases with the tilt (see the three-dimensional wing-like phase diagram shown in Figure 5.24(c), in Section 5.3.2) [Lévy 07, Aoki 19a]. $1/T_2(H_m)$ decreases with the tilt and is strongly reduced at the quantum critical endpoint. Longitudinal magnetic fluctuations, i.e., fluctuations $\mu \parallel \mathbf{b}$, may, thus, play a role for the development of field-induced superconductivity in pure URhGe.

Figure 5.27 presents a NMR study of the spin-lattice relaxation rate $1/T_1$ of UCoGe in various magnetic-fields directions [Hattori 12]. Knowing that $1/T_1$ probes transverse magnetic fluctuations, i.e., magnetic moments fluctuating perpendicular to the magnetic field, its enhancement measured for $\mathbf{H} \parallel \mathbf{a}, \mathbf{b}$ was interpreted as an indication for magnetic fluctuations $\mu \parallel \mathbf{c}$ (Figure 5.27(a)). Figure 5.27(b) shows the angular dependence of $1/T_1$ measured at different temperatures, in a magnetic field $\mu_0 H = 2$ T varying from $\mathbf{b}$ to $\mathbf{c}$. At $T = 20$ K, the angular variation of $1/T_1$ presents a broad maximum peaked for $\mathbf{H} \parallel \mathbf{b}$. At temperatures $T < 5$ K, it transforms into a sharp maximum, also centered on $\mathbf{H} \parallel \mathbf{b}$. Figure 5.27(c) shows that the amplitude of the magnetic fluctuations along $\mathbf{b}$, extracted from $1/T_1$, varies in a similar manner than $H_{c,2}$ in fields slightly tilted from $\mathbf{b}$ towards $\mathbf{c}$. This supports that magnetic fluctuations play a central role for the unconventional superconducting properties of UCoGe.

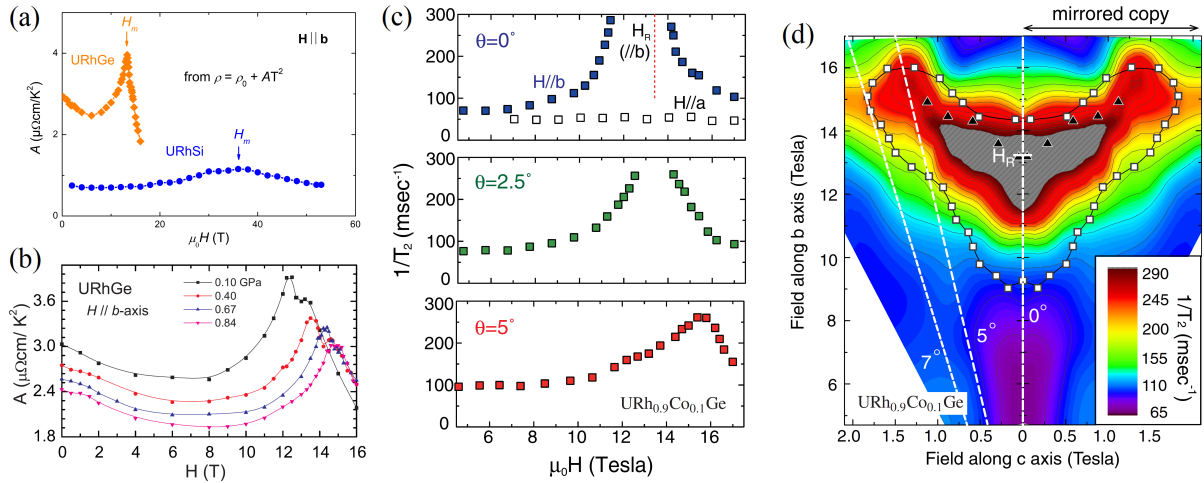

Figure 5.26: Variation of the Fermi-liquid quadratic coefficient $A$ of the electrical resistivity as function of a magnetic field $\mathbf{H} \parallel \mathbf{b}$ (a) of URhGe and URhSi (data from [Miyake 08, Knafo 19b]) and (b) of URhGe at different pressures (from [Miyake 09]). Variation of the NMR spin-spin relaxation rate $1/T_2$ of $URh_{0.9}Co_{0.1}Ge$ (c) versus magnetic field along the directions $\mathbf{H} \parallel \mathbf{a}$, $\mathbf{b}$, and two directions tilted by an angle $\theta = 2.5$ and $5°$, and (d) in the two-dimensional plane ($H_b$, $H_c$), corresponding to the two components of a magnetic field $\mathbf{H} = (0, H_b, H_c)$ (from [Tokunaga 15]).

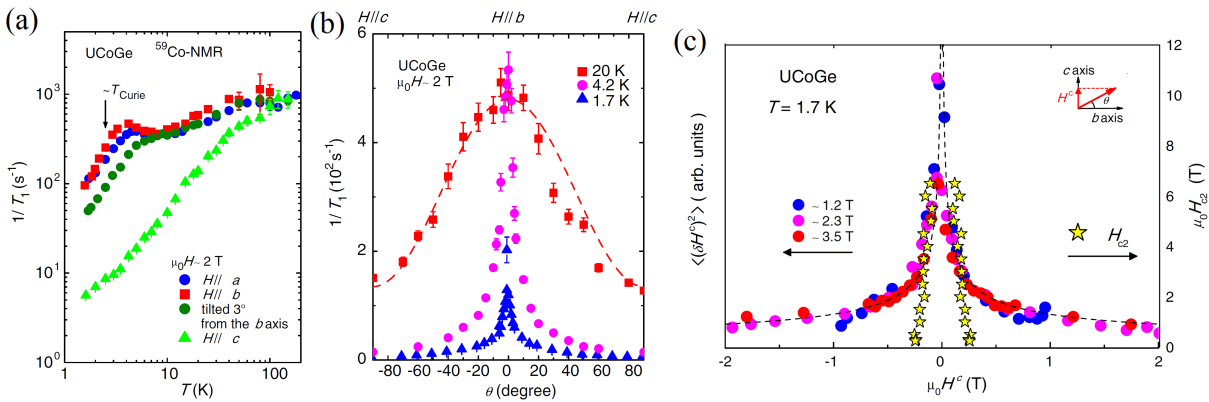

Figure 5.27: (a) Temperature variation of the NMR relaxation rate $1/T_1$ for a magnetic field applied the main crystallographic directions $\mathbf{a}$, $\mathbf{b}$, and $\mathbf{c}$, and a direction tilted by $3°$ from $\mathbf{b}$, (b) angular dependence of $1/T_1$, for a magnetic field tilted in the ($\mathbf{b}$, $\mathbf{c}$) plane, at different temperatures, and (c) amplitude of the magnetic fluctuations along the direction $\mathbf{c}$, extracted from the NMR relation rate $1/T_1$, measured in a field tilted from $\mathbf{b}$ towards $\mathbf{c}$ (from [Hattori 12]).

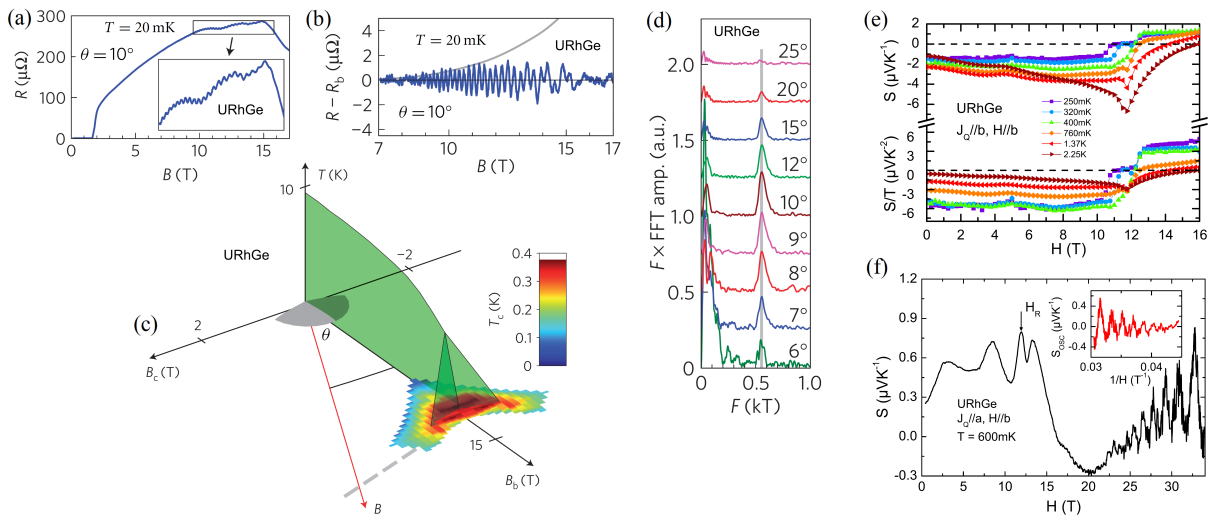

Figure 5.28: (a) Electrical resistance of URhGe versus temperature, with a zoom on the oscillating regime in the Inset, (b) extracted quantum oscillations in a magnetic field with an angle $\theta = 10°$, (c) three-dimensional $H_b$ - $H_c$ - temperature phase diagram including the definition of $\theta$, and (d) angular dependence of the Fourier transforms of the Subnikow-de Haas quantum oscillations of the resistivity (from [Yelland 11]). Seebeck thermoelectric-power coefficient (c) measured with a current $\mathbf{J} \parallel \mathbf{b}$ and plotted as $S$ and $S/T$ versus magnetic field, for different temperatures, and measured with a current $\mathbf{J} \parallel \mathbf{a}$ and plotted as $S$ versus magnetic field, at $T = 600$ mK, of URhGe in a magnetic field $\mathbf{H} \parallel \mathbf{b}$ (from [Gourgout 16]).

On one hand, from a NMR study of spin-spin relaxation rate $1/T_2$ of URh$_{0.9}$Co$_{0.1}$Ge, magnetic fluctuations parallel to the hard magnetic axis $\mathbf{b}$ were emphasized at the metamagnetic transition [Tokunaga 15]. These fluctuations were proposed to drive the field-reentrance of superconductivity in URhGe for $\mathbf{H} \parallel \mathbf{b}$. On the other hand, from a low-field NMR study of spin-lattice relaxation rate $1/T_1$, magnetic fluctuations parallel to the easy magnetic axis $\mathbf{c}$ were emphasized to drive the sharp angular dependence of the superconducting field of UCoGe in magnetic fields tilted from $\mathbf{H} \parallel \mathbf{b}$ [Hattori 12]. Further studies are needed to compare the evolution of the longitudinal and transverse components of the magnetic fluctuations in a high magnetic field. Inelastic neutron scattering are needed to determine whether, in addition to the ferromagnetic fluctuations, antiferromagnetic fluctuations are present in these ferromagnetic systems, and if they could play a role for superconductivity.

### 5.3.4   Fermi surface

While the Fermi surface of heavy-fermion antiferromagnets and nearly-antiferromagnets was investigated intensively (see Sections 3.2.3, 4.2.3, and 4.3.2.3), the Fermi surfaces of a few heavy-fermion ferromagnets were studied. Figure 5.28 summarizes investigations of the Fermi-surface of the ferromagnetic superconductor URhGe. In a magnetic field tilted from **b** towards **c**, Yelland *et al* observed quantum oscillations in the electrical resistance, whose Fourier transform leads to a peak at a frequency $F = 500$ T.this SdH frequency is almost independent of the tilt angle varying from 6 to 25° (see Figures 5.28(a-d)) [Yelland 11]. As shown in Figure 5.28(b), the quantum oscillations were found to disappear in magnetic fields higher than $\mu_0 H_m \simeq 15$ T, for a tilt angle of 10°. This was interpreted as a signature of a topological Fermi-surface transition at $H_m$ [Yelland 11]. Figure 5.28(e) shows thermoelectrical power $S$ measurements on URhGe in a magnetic field $\mathbf{H} \parallel \mathbf{b}$ [Gourgout 16]. An abrupt change of sign of $S$ at $H_m$ was also interpreted by Gourgout *et al* as the indication of a Fermi-surface modification. However, quantum oscillations were observed in the thermoelectric power in fields $H > H_m$ (see Figure 5.28(f)), and their Fourier transform led to a frequency $F \simeq 500$ T similar to that observed for $H < H_m$ in [Yelland 11]. Further sets of data are needed for a complete description of the Fermi surface of URhGe, and of its modification in a magnetic field. Beyond the URhGe case, the investigation of the Fermi-surface changes in other heavy-fermion ferromagnets, under pressure and magnetic field, may bring valuable information about their peculiar electronic properties.

# Chapter 6

# Magnetism and 'hidden-order' in $URu_2Si_2$

The paramagnet $URu_2Si_2$ occupies a special place in the family of heavy-fermion compounds. The nature of the 'hidden-order' (HO) phase developing in this material at temperatures below $T_0 = 17.5$ K constitutes a more than thirty-years-old unsolved mystery. Efforts were devoted to elucidate the order parameter of this phase, but none was successful to determine it unambiguously so far. Theories proposed to describe the HO within different approaches: localized, itinerant, or dual (localized/itinerant) $5f$-electrons, multipolar order, nematicity, spin liquid, 'hastatic' order, etc. [Fujimoto 11, Haule 10, Ikeda 12, Thomas 13, Chandra 13]. However, a difficulty is to evidence experimentally, in a direct and unambiguous manner, the order parameters proposed by these models.

$URu_2Si_2$ becomes superconducting below the temperature $T_{sc} = 1.5$ K, and rich magnetic properties can be revealed by tuning the system under pressure, chemical doping, and magnetic field. In this Chapter, the magnetic properties of $URu_2Si_2$ and their relations with the hidden-order phase are presented. The basic electronic properties of $URu_2Si_2$ are introduced in Section 6.1. Its quantum magnetic phase transitions induced under pressure or by chemical doping are presented in Section 6.2, and its high-magnetic-field properties are presented in Section 6.3. Complementary information about the hidden-order quest in $URu_2Si_2$ can be found in the reviews [Mydosh 11, Mydosh 20] devoted to this compound.

## 6.1 From high-temperature correlations to hidden order and superconductivity

We review here the properties of $URu_2Si_2$ at ambient pressure. The signatures in bulk properties of the correlated paramagnetic regime, the hidden-order phase, and superconductivity, which are successively stabilized when the temperature is decreased, are presented in Section 6.1.1. The evolution of the magnetic-fluctuations spectra is considered in Section 6.1.2, and the Fermi-surface properties are presented in Section 6.1.3. Finally, focus is made on the superconducting phase stabilized in $URu_2Si_2$ at low temperature in Section 6.1.4.

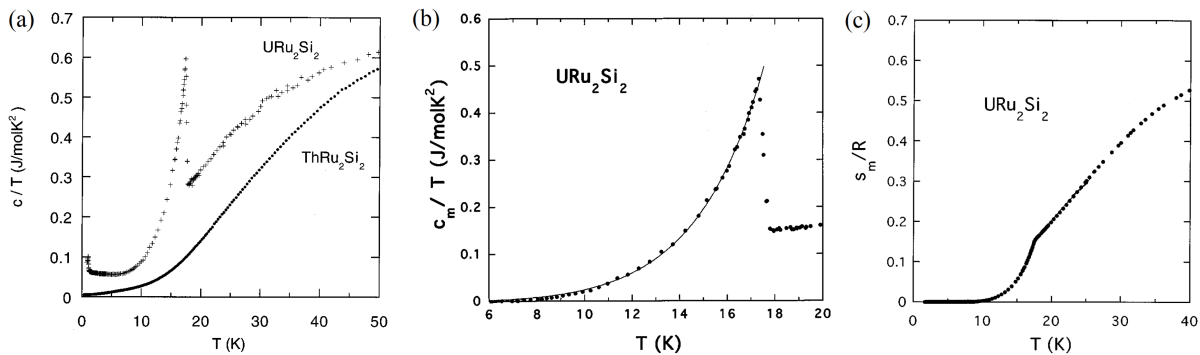

Figure 6.1: (a) Heat capacity $C_p$ versus temperature of URu$_2$Si$_2$ and ThRu$_2$Si$_2$, (b) magnetic heat capacity divided by temperature $C_p/T$ and (c) magnetic entropy expressed as $S_m/R$ versus temperature of URu$_2$Si$_2$ (from [van Dijk 97]).

### 6.1.1 Bulk properties

The transition at the temperature $T_0 = 17.5$ K to the hidden-order phase in URu$_2$Si$_2$ is associated with strong physical responses in various thermodynamic quantities. Figure 6.1(a) presents the temperature-variation of the heat capacity divided by temperature of URu$_2$Si$_2$ and of the non-$5f$-electron reference ThRu$_2$Si$_2$ [van Dijk 97] (see also [Palstra 85, Schlabitz 86]). The temperature-variations of the difference, which estimates the magnetic contribution to the heat capacity, and of the corresponding magnetic entropy are shown in Figures 6.1(b-c), respectively. A large variation of magnetic entropy $\Delta S_m \simeq 1/2R\ln 2$ in a temperature window up to 40 K indicates that the low-temperature electronic properties are those of a $5f$-electron-doublet groundstate, as in most of heavy-fermion compounds (see Sections 2.4.1.1, 4.1.1, and 5.1.1). The transition to the hidden-order phase leads to a sharp second-order step-like variation of $C_p/T$ at the temperature $T_0$ and is associated with an entropy variation $\Delta S_0 \simeq 0.15 \cdot R\ln 2$. A small antiferromagnetic moment, initially observed below $T_0$ by neutron diffraction [Broholm 87] but too small to explain this large entropy change, is now understood as an extrinsic property induced by crystal defects and distortions [Takagi 07, Niklowitz 10].

Figure 6.2 presents the temperature-dependence of anisotropic bulk properties of URu$_2$Si$_2$, in relation with the tetragonal structure of the compound.

- Magnetic susceptibility $\chi$ versus temperature curves measured in magnetic fields $\mathbf{H} \parallel \mathbf{a}, \mathbf{c}$ are shown in Figures 6.2(a-b) [Palstra 85, Sugiyama 99]. A strong Ising magnetic anisotropy is indicated by the hierarchy $\chi_c \gg \chi_a$. While $\chi_a$ is almost temperature-independent, $\chi_c$ shows a broad maximum peaked at the temperature $T_\chi^{max} = 55$ K. This maximum indicates the onset of a correlated paramagnetic regime associated with a Fermi-liquid behavior, as observed in other heavy-fermion magnets (see Sections 3.1.1, 4.3.1.2, and 5.2.1). At lower temperatures, the onset of the HO phase is accompanied by a kink in $\chi_c$ at $T_0$ and by a reduction of $\chi_c$ for $T < T_0$.

- Figures 6.2(c-d) emphasize the anisotropy of the electrical resistivity $\rho$ [Palstra 86]. $\rho_{xx}$

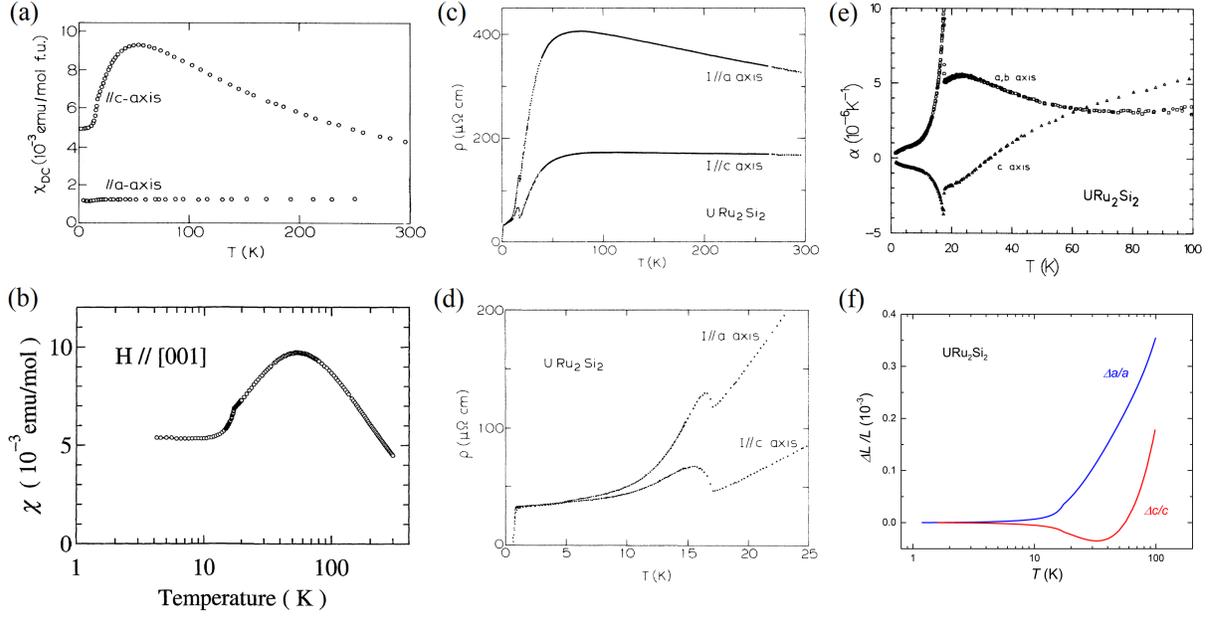

Figure 6.2: Temperature-dependence, in URu$_2$Si$_2$, of the magnetic susceptibility $\chi$ (a) for **H** ∥ **a**, **c** (from [Palstra 85]), (b) and for **H** ∥ **c** in a semi-logarithmic scale (from [Sugiyama 99]), of the electrical resistivity $\rho$ measured for **I** ∥ **a**, **c**, in temperatures (c) up to 300 K and (d) up to 25 K (from [Palstra 86]), (e) of the thermal expansion coefficient $\alpha_i$ (from [de Visser 86]) and (f) relative lengths $\Delta L/L$ for **L** ∥ **a**, **c** (adapted from [de Visser 86]).

measured with **I** ∥ **a** is twice larger than $\rho_{zz}$ measured with **I** ∥ **c**. A broad electronic contribution leads to a maximum in $\rho_{xx}$ at $T_\rho^{max} \simeq 75$ K and to a plateau in $\rho_{zz}$ in the temperature range 80-300 K. When the temperature is decreased, a step-like increase of the electrical resistivity, followed by a decrease at lower temperatures, is observed for the two current-directions at the transition temperature $T_0$ to the hidden-order phase. Zero-resistivity indicates the onset of superconductivity at temperatures below $T_{sc} = 1.5$ K.

- Figure 6.2(e) shows that the thermal-expansion coefficient $\alpha$ is also very anisotropic [de Visser 86]. Electronic correlations lead to broad Schottky-type anomalies at temperatures $T > T_0$: a positive anomaly in $\alpha_a$ extracted from the length **L** ∥ **a**, and a negative anomaly $\alpha_c$ extracted from **L** ∥ **c**. Sharp second-order step-like anomalies of same sign than the high-temperature anomalies are observed in $\alpha_a$ and $\alpha_c$ at the temperature $T_0$. As shown in Figure 6.2(f), relative length variations $\Delta a/a \simeq 3.5 \ 10^{-4}$ and $\Delta c/c \simeq 2 \ 10^{-4}$ in temperatures up to 100 K indicate that, as well as other heavy-fermion compounds, the valence of URu$_2$Si$_2$ is almost temperature-invariant, and that it presumably lies near an integer limit.

Beyond these reports of 'standard' thermodynamic and electrical-transport measurements, magnetic-torque measurements [Okazaki 11] followed by diffraction experiments [Tonegawa 14] were interpreted as the signatures of a nematic behavior in the HO phase, associated with a breaking of the four-fold rotational symmetry of the tetragonal lattice. However, this proposition for nematicity was contradicted by a series of diffraction [Choi 18, Tabata 14], NMR [Kambe 18], and thermodynamic [Wang 20] experiments. At the time of writing this review, the question of the symmetry of the hidden order remains open.

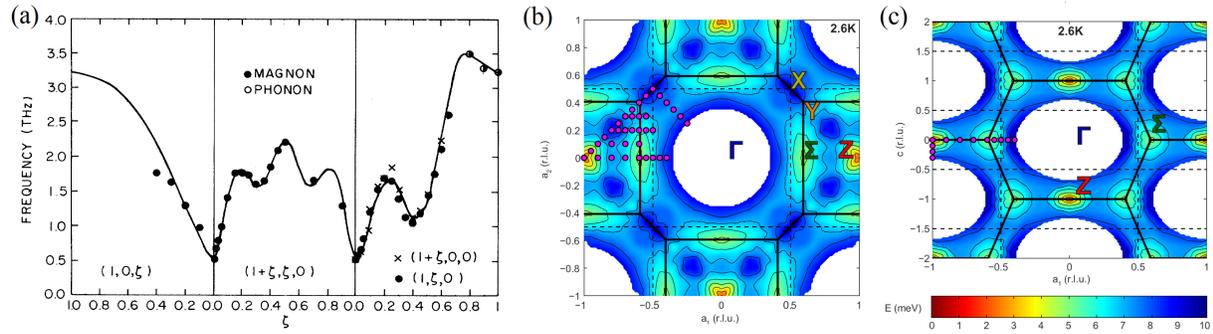

Figure 6.3: Dispersion of the gap in the magnetic fluctuations of URu$_2$Si$_2$ at low-temperature, as function of the magnetic wavevectors (a) along selected high-symmetry directions (from [Broholm 91]), and via color mappings of the gap value, (b) in the basal plane $\perp$ **c** and (c) in the plane (**a**, **c**) (from [Butch 15]).

### 6.1.2 Magnetic fluctuations

A peculiarity of URu$_2$Si$_2$ is the presence of dispersive gapped excitations spectra in its hidden-order phase. Figure 6.3(a) presents the variation of the gap in the low-energy magnetic excitations for different wavevectors along the high-symmetry lines $\mathbf{q} = (1, 0, \xi)$, $\mathbf{q} = (1 + \xi, \xi, 0)$, $\mathbf{q} = (1 + \xi, 0, 0)$ and $\mathbf{q} = (1, \xi, 0)$ of the reciprocal space [Broholm 91]. Two minimum gaps $E_0 = 1.6$ meV and $E_1 = 4.5$ meV are observed at the wavevectors $\mathbf{k}_0 = (0, 0, 1)$, which is equivalent to $(1, 0, 0)$, and $\mathbf{k}_1 = (0.6, 0, 0)$, respectively, where the neutron scattered intensity presents maxima [Bourdarot 03]. Color mappings of the gap variation in the planes $(1, 0, 0)$ and $(0, 0, 1)$ of the reciprocal space are presented in Figures 6.3(b-c) [Butch 15]. They emphasize that the minimum gaps, which correspond to the more intense low-energy magnetic excitations, mainly develop on high-symmetry points of the Brillouin-zone boundary.

The transition at $T_0$ is associated with strong modifications in the magnetic fluctuations spectra. Figures 6.4 and 6.5 show that the magnetic fluctuations at the wavevectors $\mathbf{k}_1$ and $\mathbf{k}_0$, respectively, are sharp and gapped in the hidden-order phase.

- Figures 6.4(a-b) present inelastic-neutron-scattering spectra measured on URu$_2$Si$_2$ at the wavevector $\mathbf{k}_1$, for a large set of temperatures from 1.6 to 150 K [Bourdarot 14]. A broad quasielastic, or nearly quasielastic, spectrum develops progressively at temperatures $T < 100$ K. At temperatures below the hidden-order transition temperature $T_0 = 17.5$ K, a gapping is induced and leads to a sharp inelastic peak. Figures 6.4(c-e) present the temperature-variation of the intensity (noted $A$ in the Figure), the relaxation rate $\Gamma_1$, and the gap $E_1$ (noted $\Delta$ in the Figure), respectively, extracted from a fit by a Lorentzian inelastic lineshape to the spectra [Broholm 91]. They show that the onset of the hidden-order phase at $T_0$ is accompanied by a small enhancement of the intensity of the fluctuations with wavevector $\mathbf{k}_1$, a sudden reduction of the associated relaxation rate $\Gamma_1$, and a sudden increase of the energy gap $E_1$.

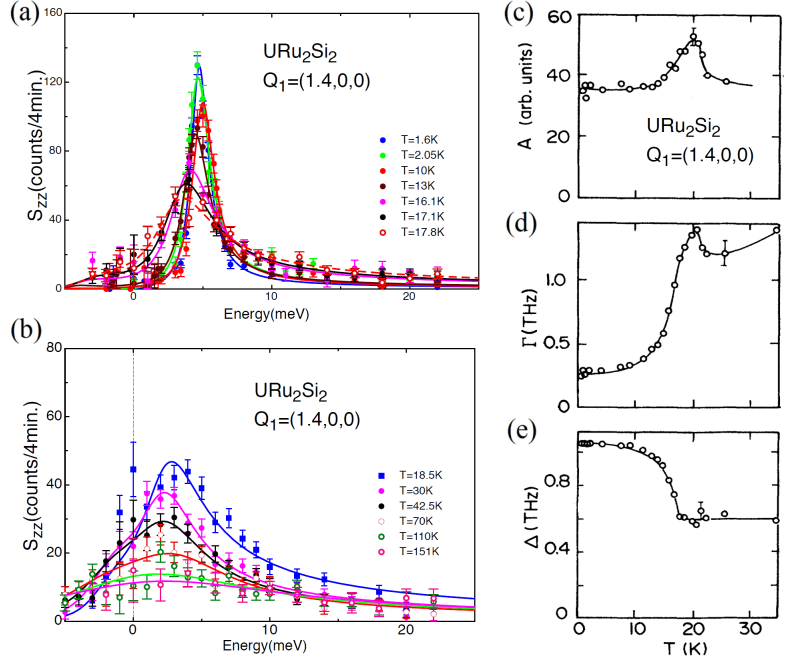

Figure 6.4: (a-b) Inelastic neutron scattering spectrum of URu$_2$Si$_2$ measured at different temperatures at the momentum transfer $\mathbf{Q}_1 = (1.4, 0, 0)$, corresponding to magnetic fluctuations at the wavevector $\mathbf{k}_1 = (0.6, 0, 0)$ (from [Bourdarot 14]). Temperature variation of (c) the amplitude $A = \chi'$, (d) the relaxation rate $\Gamma$, and (e) the gap $\Delta$, expressed in THz (1 THz = 4.1 meV) extracted from a fit by a Lorentzian function to the magnetic excitations spectrum at the momentum transfer $\mathbf{Q}_1 = (1, 0.4, 0)$, also corresponding to magnetic fluctuations at the wavevector $\mathbf{k}_1 = (0.6, 0, 0)$ (from [Broholm 91]).

- Figure 6.5(a) shows inelastic-neutron-scattering spectra measured on URu$_2$Si$_2$ at the wavevector $\mathbf{k}_0$ and temperatures from 1.5 to 27 K [Bourdarot 10a]. A quasielastic peak develops at temperatures $T \gtrsim T_0$ and is replaced by a sharp inelastic peak at temperatures $T < T_0$. Figures 6.5(b-d) present the temperature-variation of the gap $E_0$, the 'intensity' $\chi'_0$ and the relaxation rate $\Gamma_0$, respectively, extracted from a fit by a Lorentzian function for $T > T_0$, and from a fit by a harmonic-oscillator function at temperatures $T < T_0$, to these spectra. The transition to the hidden-order phase is accompanied by a sudden gapping of the magnetic excitations with wavevector $\mathbf{k}_0$, a significant increase of their intensity, and a strong reduction of the relaxation rate.

Intersite magnetic fluctuations with wavevector $\mathbf{k}_1$ develop at temperatures higher than $T_0$. In the hidden-order phase, they become sharply-gapped but their integrated intensity remains almost unaffected. On the contrary, intersite magnetic fluctuations with wavevector $\mathbf{k}_0$ are almost absent at high temperatures, they develop at temperatures near $T_0$ and their intensity, probed by $\chi'_0$ ($= \chi'(\mathbf{k}_0)$, also noted $= \chi(\mathbf{Q}_0, E = 0)$ in Figure 6.5(c) [Bourdarot 10a]), is strongly-enhanced in the hidden-order phase, where they also become gapped. Although $\chi'_0$ is the signature of an excited electronic state, its temperature-dependence is similar to that of the order parameter expected for a second-order phase transition. This indicates that the magnetic fluctuations with wavevector $\mathbf{k}_0$ may be a signature of the hidden-order phase. In Section 6.2.3, we will show that a periodicity with wavevector $\mathbf{k}_0$ has been proposed for the 'hidden' order parameter, following Fermi-surface studies of URu$_2$Si$_2$ under pressure.

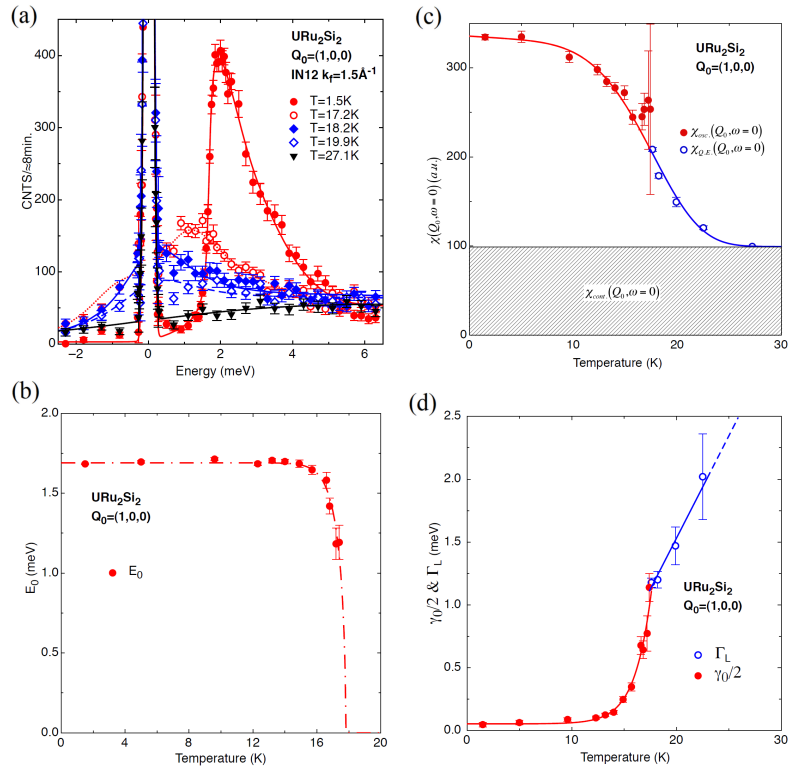

Figure 6.5: (a) Inelastic neutron scattering spectrum of URu$_2$Si$_2$ measured at different temperatures at the momentum transfer $\mathbf{Q}_1 = (1, 0, 0)$, corresponding to magnetic fluctuations at the wavevector $\mathbf{k}_0 = (0, 0, 1)$. Temperature variation of (b) the gap $E_0$, (c) the static magnetic susceptibility $\chi'$, and (d) the relaxation rate $\Gamma$ extracted from a fit, either by a harmonic-oscillator function at temperatures $T < T_0$ or by a Lorentzian function at temperatures $T > T_0$, to the magnetic excitations spectrum at the momentum transfer $\mathbf{Q}_0$ (from [Bourdarot 10a]).

Alternatively to inelastic neutron scattering, nuclear magnetic resonance allows studying microscopically the magnetic excitations. Figures 6.6(a-b) show the temperature-dependence of the NMR relaxation rate $1/T_1$ of URu$_2$Si$_2$ and of its non-5$f$-electron reference ThRu$_2$Si$_2$ in magnetic fields $\mathbf{H} \parallel \mathbf{c}$ and $\mathbf{H} \perp \mathbf{c}$ [Emi 15, Emi 17] (see also [Shirer 13]). Knowing that $1/T_1$ probes the magnetic fluctuations perpendicular to the magnetic field, its enhancement in URu$_2$Si$_2$ for $\mathbf{H} \perp \mathbf{c}$ is the signature of Ising-type magnetic fluctuations $\parallel \mathbf{c}$. At temperatures $T_0 < T < 50$ K, a Korringa $T$-linear variation of $1/T_1$ indicates a heavy-fermion behavior (see Section 2.4.3.2), in agreement with the nearly-saturated magnetic susceptibility observed in this regime (see Figure 6.2 in Section 6.1.1). At temperatures $T < T_0$, a downwards deviation of $1/T_1$ is compatible with the gapping of the magnetic fluctuations observed by inelastic neutron scattering (see Figures 6.4 and 6.5). Figure 6.6(c) shows spectra obtained by Raman scattering, which probes the electronic excitations at wavevector $\mathbf{k} = 0$ [Buhot 14]. Similarly to neutron scattering spectra measured at the antiferromagnetic wavevector $\mathbf{k}_0 = (0, 0, 1)$, Raman spectra indicate the presence in the hidden-order phase of a sharp inelastic peak, gapped with an energy $E_0 \simeq 1.7$ meV, at the ferromagnetic wavevector $\mathbf{k} = 0$.

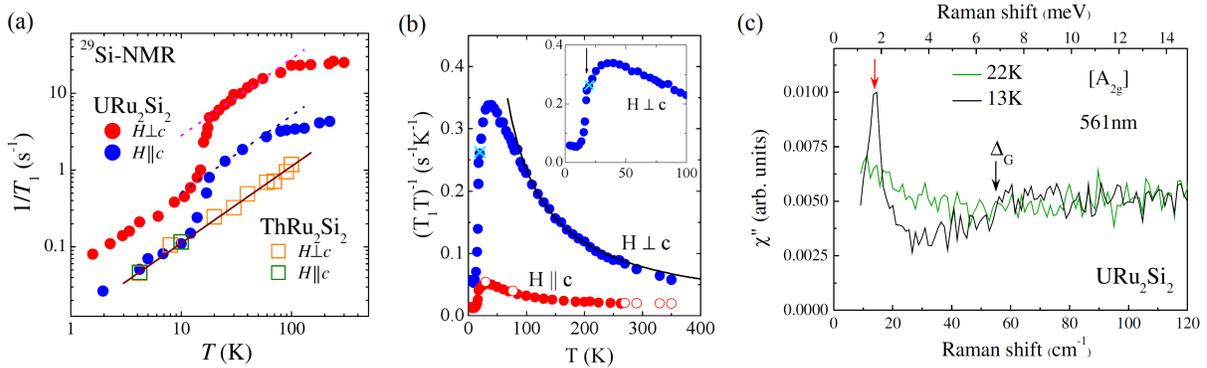

Figure 6.6: (a) NMR relaxation rate $1/T_1$ versus temperature of URu$_2$Si$_2$ and ThRu$_2$Si$_2$ in magnetic fields $\mathbf{H} \parallel \mathbf{c}$ and $\mathbf{H} \perp \mathbf{c}$ (from [Emi 15]). (b) NMR relaxation rate $1/T_1T$ versus temperature of URu$_2$Si$_2$ in magnetic fields $\mathbf{H} \parallel \mathbf{c}$ and $\mathbf{H} \perp \mathbf{c}$ (from [Emi 17]). (c) Raman spectra of URu$_2$Si$_2$ in $A_{2g}$ symmetry at the temperatures $T = 12$ and $22$ K (from [Buhot 14]).

### 6.1.3  Fermi surface

The Fermi surface of URu$_2$Si$_2$ in its zero- or low-field ground-state has been investigated by a large set of experiments, from de-Haas-van-Alphen and Shubnikov-de-Haas experiments [Bergemann 97, Ohkuni 99, Hassinger 10, Bastien 19], cyclotron-resonance [Tonegawa 13] and ARPES measurements [Santander-Syro 09, Yoshida 10, Yoshida 13, Chatterjee 13, Meng 13, Bareille 14], to thermodynamic and transport [Schoenes 87, LeR Dawson 89, Bel 04, Scheerer 12]. Figure 6.7 presents graphs extracted from a study of the Fermi surface by Shubnikov-de-Haas measurements [Bastien 19]. From quantum oscillations in the electrical resistivity (see Figure 6.7(a)), Fourier transforms lead to rich SdH spectra, as shown in Figure 6.7(b) for $\mathbf{H} \parallel \mathbf{c}$, where several orbits, $\eta$, $\gamma$, $\beta$, $\alpha$, and $\lambda$, can be evidenced. The angular dependence of the SdH frequencies of the different orbits, for magnetic fields rotating from directions [110] to [100], and from directions [100] to [001], is presented in Figure 6.7(c). The angular-dependencies of cyclotron masses extracted from the temperature-dependence of the spectra, are presented in Figures 6.7(d-f). They indicate that the Fermi surface of URu$_2$Si$_2$ is mainly three-dimensional.

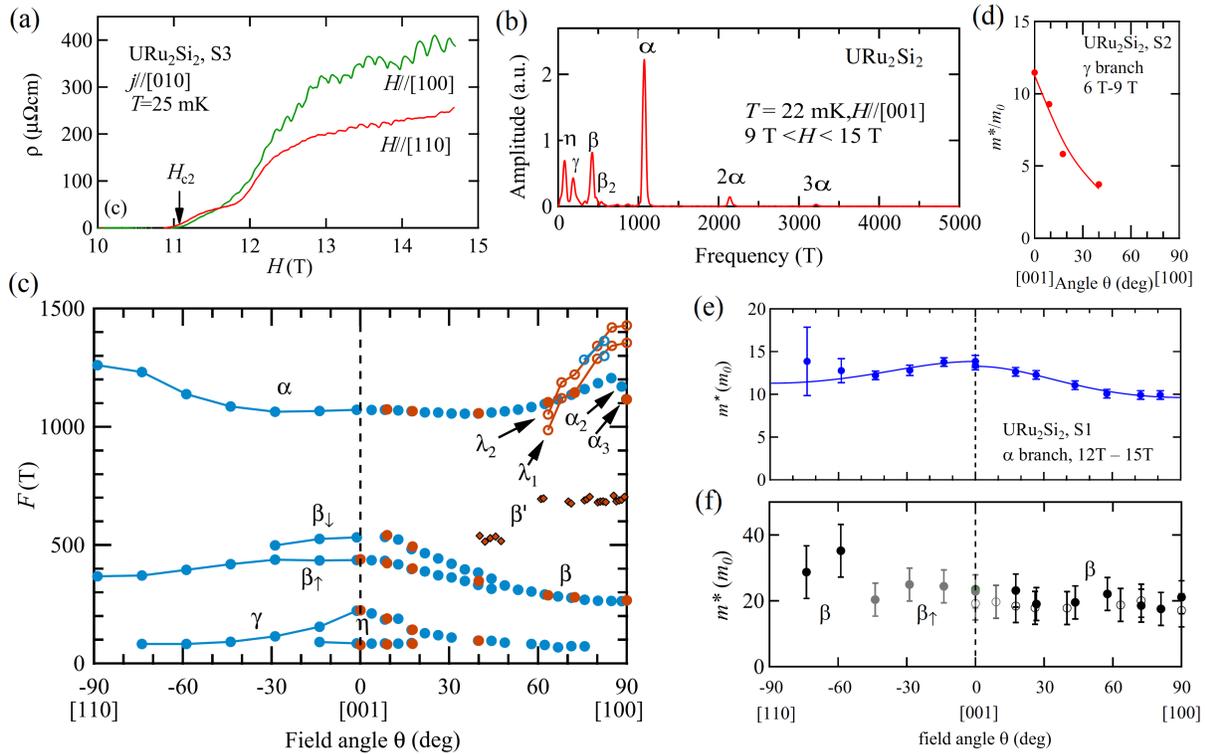

Figure 6.7: Investigation of the Fermi surface of URu$_2$Si$_2$ by Shubnikov-de Haas technique. (a) Low-temperature electrical resistivity presenting SdH quantum oscillations, for **H** ∥ [100] and [110], (b) Fourier transform of the SdH quantum oscillations measured for **H** ∥ [001], angular dependence of the (c) SdH frequencies and of the cyclotron masses extracted for (d) the $\gamma$ branch, (e) the $\alpha$ branch, and (f) the $\beta$ branch, for fields varying from [110] to [001], and from [001] to [100] (from [Bastien 19]).

Large masses $m^*_{c,\gamma} = (5-10)m_0$, $m^*_{c,\alpha} = (10-15)m_0$, and $m^*_{c,\beta} = (20-30)m_0$, where $m_0$ is the free-electron mass, are extracted for the orbits $\gamma$, $\alpha$, and $\beta$, respectively. They confirm that heavy-fermion properties in URu$_2$Si$_2$ are driven by itinerant electrons. A lighter mass $m^*_{c,\lambda} = (1.5-2)m_0$ is extracted for the measured orbits of the band $\lambda$ (see also [Scheerer 14]).

Figure 6.8 shows experimental data supporting that the Fermi surface of URu$_2$Si$_2$ is modified in its hidden-order phase.

- From a $H^2$ 'orbital' variation of the low-temperature electrical resistivity in a magnetic field $\mu_0\mathbf{H} \parallel \mathbf{c}$ up to 10 T (see Figure 6.8(a)), Kasahara *et al.* proposed that URu$_2$Si$_2$ is a compensated metal, i.e., a metal where similar densities of electrons and holes contribute to the Fermi surface [Kasahara 07]. Figure 6.8(b) further shows that, in a magnetic field $\mu_0\mathbf{H} \parallel \mathbf{a}$ up to 50 T, the large orbital field-increasing contribution to the electrical resistivity suddenly develops at temperatures $T < T_0$ [Scheerer 12]. This indicates an enhanced carrier mobility in the hidden-order phase.

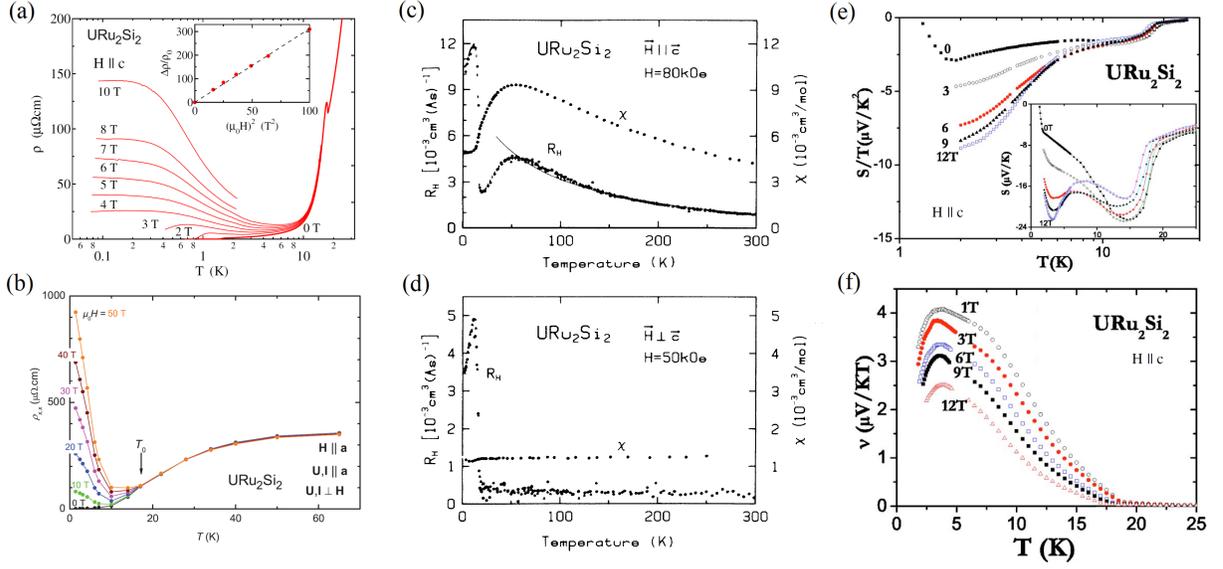

Figure 6.8: Signatures of Fermi-surface reconstruction in URu$_2$Si$_2$ at the temperature $T_0$ from thermodynamic and electrical transport measurements. Temperature variations of the electrical resistivity in magnetic fields (a) **H** ∥ **c** (from [Kasahara 07]) and (b) **H** ∥ **a** (from [Scheerer 12]), of the Hall coefficient $R_H$ and of the magnetic susceptibility in magnetic fields (c) **H** ∥ **c** and (d) **H** ∥ **a** (from [Schoenes 87]), of (e) the thermoelectric power divided by temperature $S/T$ and (f) of the Nernst coefficient $\nu$ in magnetic fields **H** ∥ **c** (from [Bel 04]).

- The temperature-dependence of the Hall-effect coefficient $R_H$ of URu$_2$Si$_2$, measured for **H** ∥ **c** and **H** ∥ **a**, is shown in Figures 6.8(c-d), respectively [Schoenes 87] (see also [LeR Dawson 89, Kasahara 07]) (see also [LeR Dawson 89]). A comparison with the temperature-dependence of the magnetic susceptibility $\chi$, also shown, indicates the presence of an anomalous contribution to the Hall coefficient. It can be written as $R_H = R_0 + a\chi R_S$, where $R_0$ is the normal Hall effect, $R_S$ is a temperature-independent extraordinary-Hall coefficient, and $a$ is a constant. A step-like increase of the normal Hall coefficient $R_0$ is observed at the transition temperature $T_0$ to the hidden-order state. This variation is compatible with a density of holes decreasing by a factor 10 in the hidden-order phase.

- Figure 6.8(e) shows the temperature-dependence of the thermoelectric power divided by temperature $S/T$, and of $S$ in the Inset, in magnetic fields $\mu_0$**H** ∥ **c** up to 12 T [Bel 04]. The negative sign of $S$ indicates that it is controlled by conduction electrons. A step-like decrease of $S$ is observed at $T_0$ and its comparison with heat-capacity data was understood as a signature of a lower electron density, and of an increased entropy per conduction electron, in the hidden-order phase [Bel 04]. Figure 6.8(f) shows the temperature-dependence of the Nernst coefficient $\eta$ in magnetic fields $\mu_0$**H** ∥ **c** up to 12 T [Bel 04]. Its large magnitude was understood as a confirmation of the large carrier mobility in the hidden-order phase [Bel 04].

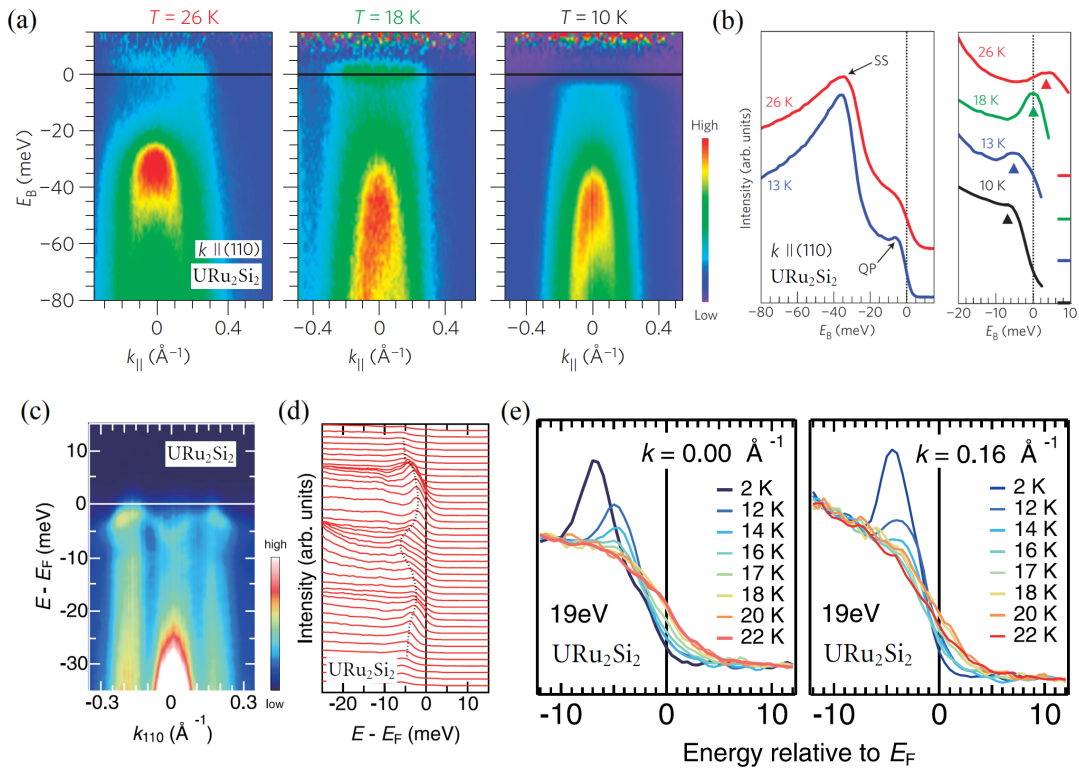

Figure 6.9: Signatures of Fermi surface reconstruction in URu$_2$Si$_2$ at the temperature $T_0$ from ARPES measurements. (a) Intensity mapping in $\mathbf{k}$, $E$-plane of ARPES spectra along the direction (110) measured at different temperatures, (b) corresponding ARPES spectra integrated within $\pm 0.1$ $^{-1}$ (left) and normalized by the Fermi-Dirac distribution (right) (from [Santander-Syro 09]). (c) Intensity mapping in $\mathbf{k}$, $E$-plane of ARPES spectra along the direction (110) measured at $T = 2$ K, (d) corresponding energy-distribution curves, and spectral cuts for (e) $k = 0$ and $k = 0.16$ $^{-1}$ obtained at different temperatures (from [Yoshida 13]).

Angle-resolved photo-emission spectroscopy (ARPES) experiments on URu$_2$Si$_2$ also permitted to show signatures of a Fermi-surface reconstruction at the transition temperature $T_0$ to the hidden-order phase. Figure 6.9(a-b) shows mapping and cuts of spectra measured along the direction (110) [Santander-Syro 09]. These spectra indicate a peculiar behavior of the Fermi surface at the temperature $T_0$, below which a quasiparticle peak shifts to energies below the Fermi energy $E_F$. From high-resolution ARPES measurements along the same direction (110), a dispersive energy distribution is revealed at temperatures $T < T_0$ (see Figures 6.9(c-d)) [Yoshida 13]. The temperature-dependence of these spectra shows that a sharp quasiparticle peak develops in the hidden-order phase. The corresponding energy of the peak is minimum and reaches a value $E = 4$ meV at the wavevector $k = 0.16$ Å$^{-1}$. In the future, efforts are needed to relate the spectra extracted from ARPES experiments to the wavevector-$\mathbf{k}$-dependent magnetic-excitation spectra extracted from neutron scattering. This may help understanding the dual itinerent-localized role played by $5f$ electrons in the formation of the hidden-order phase of URu$_2$Si$_2$.

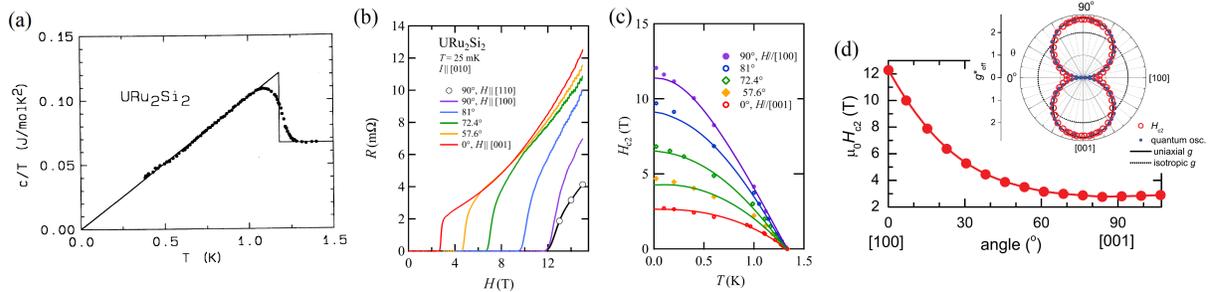

Figure 6.10: Characterization of superconductivity in URu$_2$Si$_2$. (a) Heat capacity divided by temperature $C_p/T$ versus temperature (from [van Dijk 95]). (b) Low-temperature electrical resistivity versus magnetic and (c) temperature-dependence of the superconducting critical field $H_{c,2}$ extracted from the electrical resistivity for different magnetic-field directions (from [Bastien 19]). (d) Angular dependence of the low-temperature superconducting critical field $H_{c,2}$ and, in the Inset, comparison in a polar plot of the angular dependence of $H_{c,2}$ and of the effective factor $g^*_{eff}$ deduced from quantum oscillations of the Fermi surface (from [Altarawneh 12]).

### 6.1.4 Superconductivity

URu$_2$Si$_2$ becomes superconducting at temperatures below T$_{sc} \simeq 1.3$ K [Palstra 85, Maple 86, Schlabitz 86]. Figure 6.10(a) shows that a second-order-like transition is visible in the heat-capacity [van Dijk 95]. The low-temperature critical superconducting field $\mu_0 H_{c,2}$ of URu$_2$Si$_2$ is strongly anisotropic, varying from 3 T for **H** ∥ **c** to 12 T for **H** ∥ **a** (see Figure 6.10(b-c)) [Bastien 19]. Within first approximation, this critical field can be described by spin-singlet Pauli-limitation assuming an anisotropic factor $g$, which was determined from the angular-dependence of quantum oscillations measurements (Figure 6.10(d)) [Altarawneh 12, Bastien 19].

URu$_2$Si$_2$ is currently considered as a realization of $d$-wave superconductivity. Figure 6.11(a) shows the temperature variation of the NMR Knight-shift $K$ measured for **H** ∥ **c** and **H** ⊥ **c** [Hattori 18]. A decrease of $K$ for **H** ∥ **c** was understood as the signature of spin-singlet superconductivity, with a chiral $d$-wave order parameter. This model also permitted to reproduce the temperature-dependence of the NMR relaxation rate $1/T_1$ in the superconducting state (see Figure 6.11(b)) [Hattori 18]. $1/T_1$ follows a Korringa $T$-linear behavior in the non-superconducting phase, either for $T > T_{sc}$ or for $H > H_{c,2}$. This indicates the presence of a Fermi-liquid regime driven by magnetic fluctuations, which are presumably responsible for the onset of unconventional superconductivity. Figures 6.11(c-d) further show that a tiny variation of the energy gap in the magnetic fluctuations with wavevector $\mathbf{k}_0$ is observed at temperatures $T < T_{sc}$ [Bourdarot 10b]. This observation emphasizes the subtle interplay between magnetism and superconductivity, but also with the hidden-order phase characterized by sharp inelastic excita-tions with wavevector $\mathbf{k}_0$.

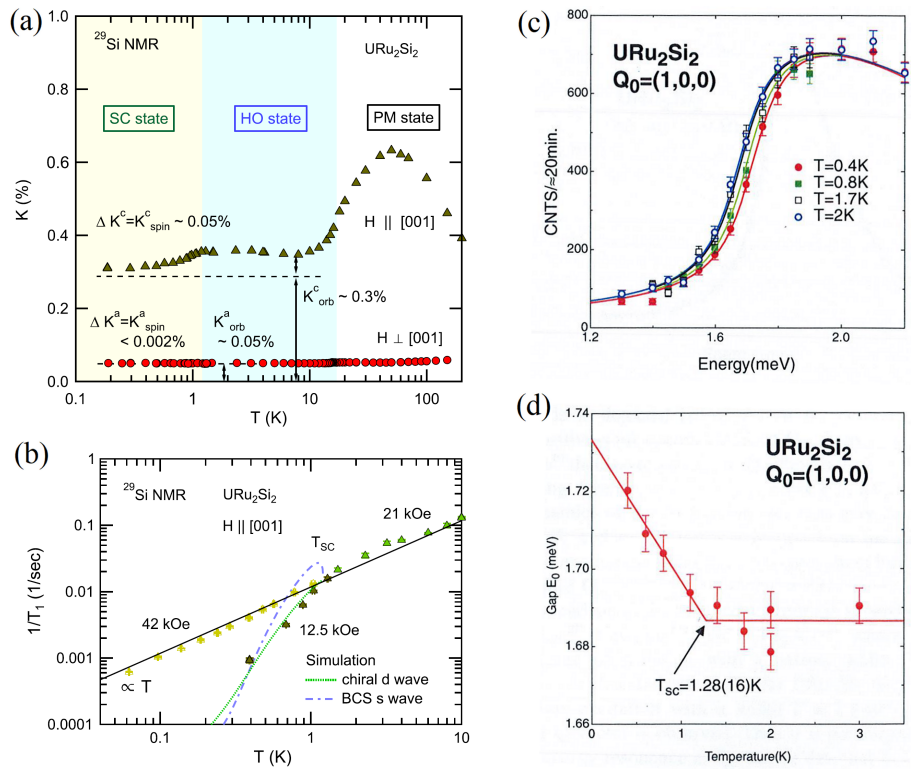

Figure 6.11: Magnetic properties of URu$_2$Si$_2$ in its superconducting phase. Temperature variation (a) of the NMR Knight-shift $K$ and (b) of the NMR relaxation rate $1/T_1$ for $\mathbf{H} \parallel \mathbf{c}$ and $\mathbf{H} \perp \mathbf{c}$ (from [Hattori 18]). (c) Inelastic neutron scattered intensity versus energy at the momentum transfer $\mathbf{Q}_0 = (1, 0, 0)$ and different temperatures, and (d) temperature-dependence of the associated energy gap (from [Bourdarot 10b]).

## 6.2 Quantum magnetic phase transitions

In early studies of URu$_2$Si$_2$ at ambient pressure, the transition at the temperature $T_0 = 17.5$ K was ascribed to the onset of antiferromagnetism, in relation with an antiferromagnetic moment $\mu_{AF} \simeq 0.02 - 0.04\ \mu_B$/U observed by neutron diffraction [Broholm 87, Broholm 91, Yokoyama 04, Bourdarot 05]. However, this antiferromagnetic moment is too small to explain the entropy change at the transition temperature $T_0$ (see Figure 6.1 in Section 6.1.1) [Palstra 85, Schlabitz 86, van Dijk 97]. It was later understood as an extrinsic property induced by crystal defects and distortions [Takagi 07, Niklowitz 10] due to a neighboring quantum antiferromagnetic phase transition.

The stabilization of long-range magnetic order in URu$_2$Si$_2$ under pressure or chemical doping is considered here. Bulk properties and electronic phase diagrams are presented in Section 6.2.1. Signatures of magnetic ordering and of modifications in the magnetic fluctuations spectra are presented in Section 6.2.2. Finally, Section 6.2.3 presents a Fermi-surface study of URu$_2$Si$_2$ through its pressure-induced quantum antiferromagnetic phase transition.

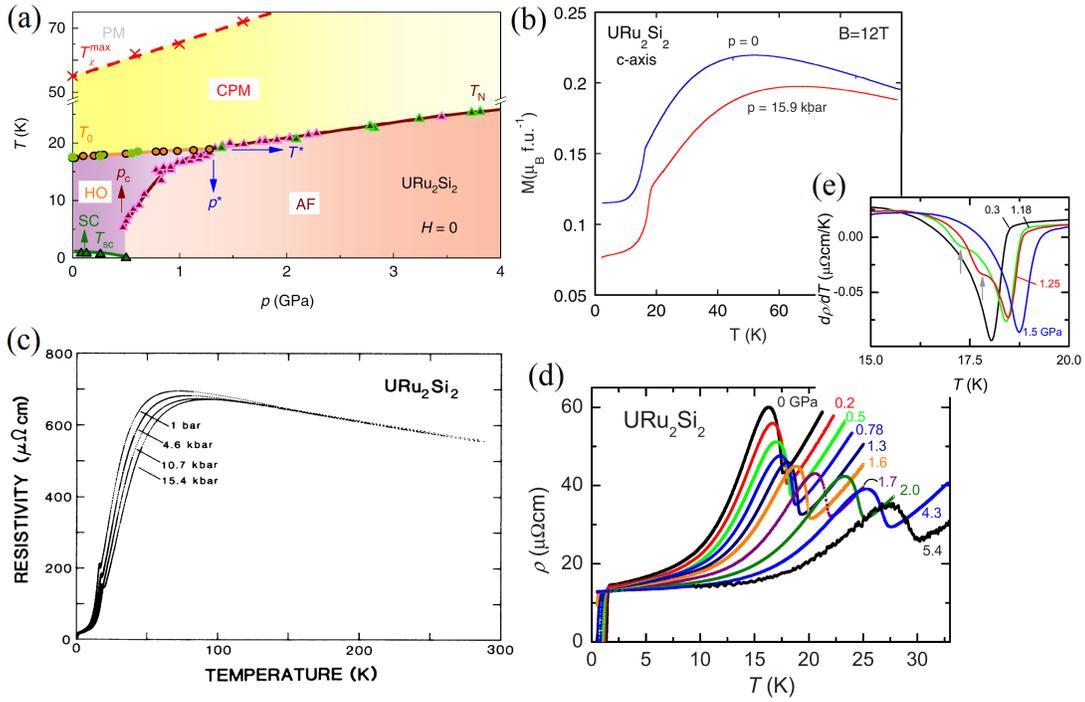

Figure 6.12: (a) Pressure-temperature phase diagram of URu$_2$Si$_2$ (from [Knafo 20a], constructed with data from [Pfleiderer 06, Hassinger 08a]). (b) Magnetization $M$ versus temperature, within a magnetic field $\mu_0\mathbf{H} \parallel \mathbf{c}$ of 12 T (from [Pfleiderer 06]), electrical resistivity $\rho$ versus temperatures (c) up to 300 K (from [McElfresh 87]), and (d) up to 35 K, and temperature-derivative of the electrical resistivity $\partial\rho/\partial T$ of URu$_2$Si$_2$ at different pressures (from [Hassinger 08a]).

## 6.2.1 Bulk properties and phase diagrams

The application of pressure leads to the destabilization of the hidden-order phase and to the appearance of an antiferromagnetic phase in URu$_2$Si$_2$. The magnetic-field-temperature phase diagram of URu$_2$Si$_2$, constructed from electrical resistivity data from [Hassinger 08a] and magnetic susceptibility data from [Pfleiderer 06], is shown in Figure 6.12(a) [Knafo 20a]. The correlated paramagnetic regime is stabilized under pressure, as indicated by the increase with $p$ of the temperature $T_\chi^{max}$, from 55 K at $p = 0$ to > 70 K at $p = 1.6$ GPa, where a broad maximum in the magnetic susceptibility is observed (see Figure 6.12(b)) [Pfleiderer 06]. A broad maximum in the electrical resistivity also indicates the onset of the CPM regime. The variation of its characteristic temperature, from 75 K at $p = 0$ to > 90 K at $p = 1.5$ GPa, confirms that the CPM regime is stabilized under pressure (see Figure 6.12(c)) [McElfresh 87]. Electrical-resistivity-versus-temperature data shown in Figure 6.12(d) indicate that $T_0$ also increases under pressure [Hassinger 08a]. At lower temperatures, superconductivity disappears at the critical pressure $p_c = 0.5$ GPa, beyond which an antiferromagnetic phase is stabilized at temperatures $T < T_N$ (see Figure 6.12(d)) [Hassinger 08a]. Under pressures $p_c < p < p^* = 1.3$ GPa, $T_N < T_0$ and the hidden-order and antiferromagnetic phases are stabilized successively when the temperature is decreased (see Figure 6.12(e)) [Hassinger 08a]. $T_N$ and $T_0$ increase under pressure and merge at the critical point $(p^*, T^* = 18$ K), beyond which $T_N$ continues to increases.

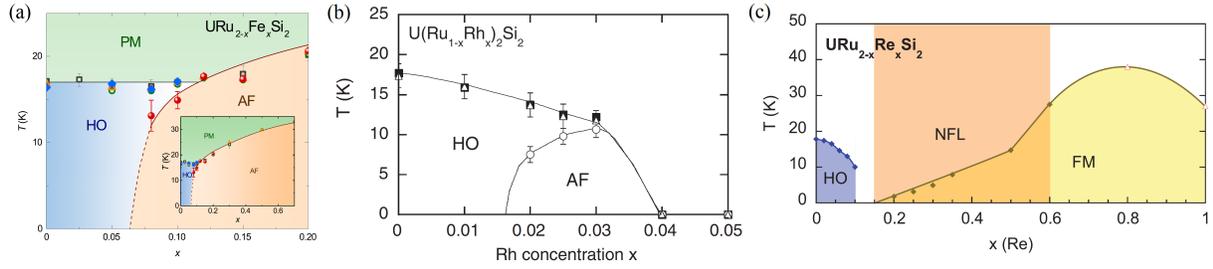

Figure 6.13: Doping-temperature phase diagrams of (a) $URu_{2-x}Fe_xSi_2$ (from [Ran 16]), (b) $U(Ru_{1-x}Rh_x)_2Si_2$ (from [Yokoyama 04]), and (c) $URu_{2-x}Re_xSi_2$ (from [Williams 12]).

Figure 6.13 shows that chemical doping can also lead to the destabilization of the hidden-order and, via a quantum magnetic phase transition, to the appearance of long-range magnetic ordering in $URu_2Si_2$.

- The doping-temperature phase diagram of Fe-doped $URu_{2-x}Fe_xSi_2$ alloys, presented in Figure 6.13(a) [Ran 16], shows striking similarities with that obtained under pressure (see Figure 6.12(a)), with the presence of a quantum phase transition at $x_c \simeq 0.07$, a critical point at $(x^* = 0.12, T^* = 18$ K), and the increase of $T_N$ under doping up to at least $x = 0.2$.

- Figure 6.13(b) shows that Rh-doped $U(Ru_{1-x}Rh_x)_2Si_2$ alloys are also characterized by a quantum antiferromagnetic phase transition, at $x_c \simeq 0.015$, from the HO phase to an AF phase [Yokoyama 04]. However, their phase diagram differs from those under pressure and Fe-doping: $T_0$ decreases with $x$, ending at a critical point $(x^* \simeq 0.3, T^* \simeq 10$ K) where it merges with $T_N$, and $T_N$ vanishes at a doping $x_{c,2} \lesssim 0.04$ near $x^*$.

- The phase diagram of Re-doped $URu_{2-x}Re_xSi_2$ alloys, shown in Figure 6.12(c), also shows a destabilization with doping of the hidden-order phase. $T_0$ decreases with $x$, before vanishing at a critical doping $x_c \gtrsim 0.1$ [Williams 12]. Interestingly, a ferromagnetic phase is stabilized for $x > x_c$.

The possibility to tune the paramagnet $URu_2Si_2$ towards antiferromagnetic and ferromagnetic phases indicates that, as well as many heavy-fermion compounds (see Figure 5.9 in Section 5.2.1 and Figure 5.19 in Section 5.3.1.2), it is subject to a competition between ferromagnetic and antiferromagnetic interactions. The presence of low-energy gapped magnetic excitations at the wavevectors $\mathbf{k}_0 = (0, 0, 1)$ and $\mathbf{k}_1 = (0.6, 0, 0)$, probed by inelastic neutron scattering (see Figures 6.3, 6.4, and 6.5 in Section 6.1.2) [Broholm 91], indicates the proximity of $URu_2Si_2$ to quantum antiferromagnetic phase transitions corresponding to the onset of long-range order with the same wavevectors. As well, the presence of a low-energy gapped magnetic excitation at the wavevectors $\mathbf{k} = 0$, probed by Raman scattering (see Figure 6.6(c) in Section 6.1.2) [Buhot 14], may be an indication for a nearby quantum ferromagnetic phase transition.

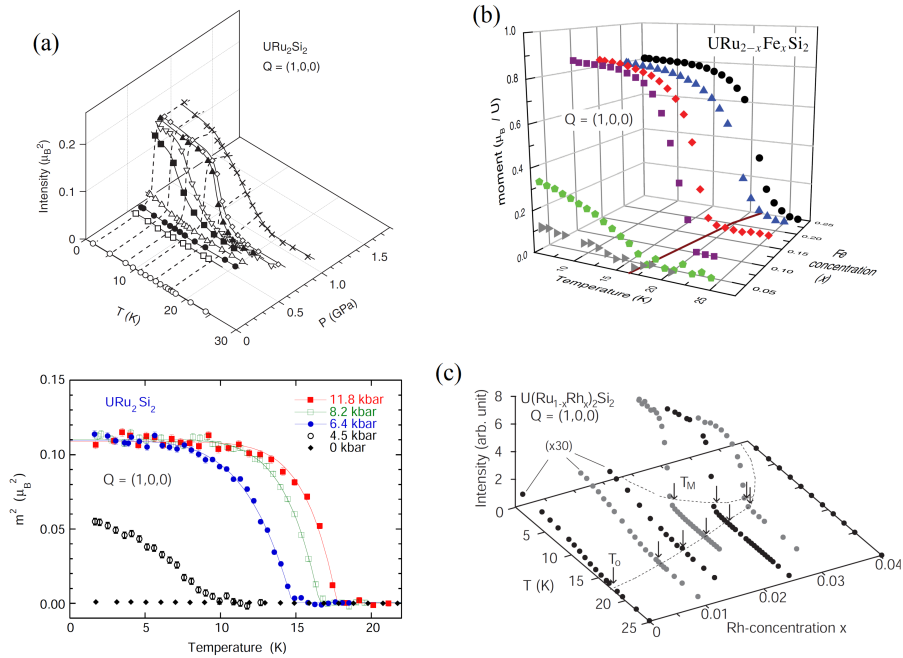

Figure 6.14: Neutron diffracted intensity at the momentum transfer $\mathbf{Q}_0 = (1, 0, 0)$ corresponding to the momentum transfer $\mathbf{k}_0 = (1, 0, 0)$, expressed in normalized units of magnetic moment $\mu$, or its square $\mu^2$, of (a) $URu_2Si_2$ under pressure (from [Bourdarot 05, Amitsuka 07]), (b) $URu_{2-x}Fe_xSi_2$ (from [Das 15]), and (c) $U(Ru_{1-x}Rh_x)_2Si_2$ (from [Yokoyama 07]).

### 6.2.2 Magnetic order and fluctuations

Figure 6.14 presents the neutron-diffracted-intensity-versus-temperature data measured at the momentum transfer $\mathbf{Q}_0 = (1, 0, 0)$ in $URu_2Si_2$ under pressure [Bourdarot 05, Amitsuka 07], or doped with Fe [Das 15] or Rh [Yokoyama 07]. A magnetic Bragg peak develops at $\mathbf{Q}_0$ for $p > p_c \simeq 0.4 - 0.5$ GPa in the pure compound, for $x > x_c \simeq 0.05$ in Fe-doped $URu_{2-x}Fe_xSi_2$ alloys, and for $x > x_c \simeq 0.015$ in Rh-doped $U(Ru_{1-x}Rh_x)_2Si_2$ alloys. The antiferromagnetic phases induced by pressure, Fe- and Rh-doping (see Figure 6.12(a) and Figures 6.13(a-b) in Section 6.2.1) are, thus, associated with the same order parameter, an antiferromagnetic moment $\mu_{AF} \parallel \mathbf{c}$ with the magnetic wavevector $\mathbf{k}_0 = (0, 0, 1)$. The antiferromagnetic moments reach different values, up to $\mu_{AF} = 0.3 - 0.4$ $\mu_B$/U in $URu_2Si_2$ under pres-sure [Bourdarot 05, Amitsuka 07], 0.8 $\mu_B$/U in URu$_{2-x}$Fe$_x$Si$_2$ [Das 15], and 0.25 $\mu_B$/U in U(Ru$_{1-x}$Rh$_x$)$_2$Si$_2$ [Yokoyama 07], possibly due to different sets of quantum-magnetic-fluctuations spectra.

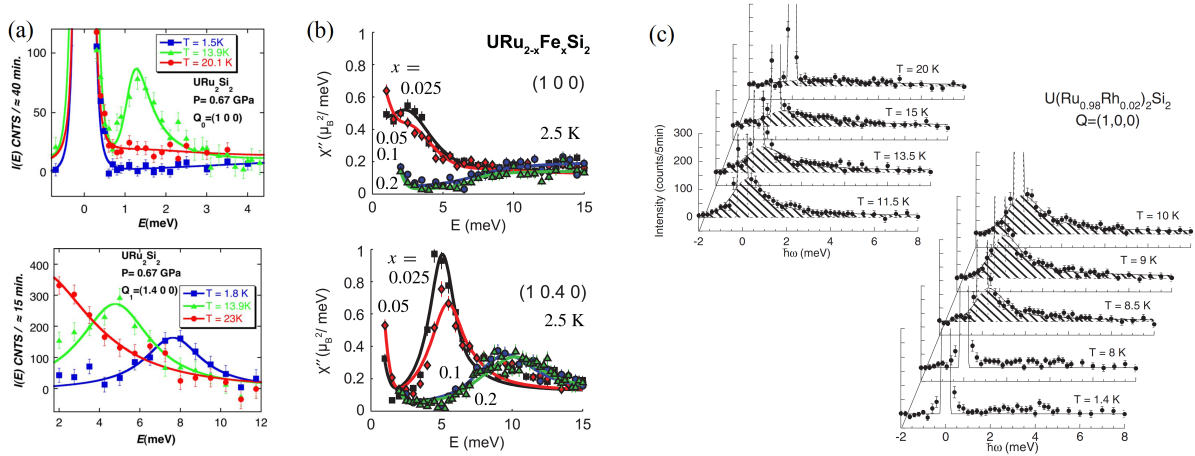

Figure 6.15: (a) Magnetic fluctuations spectra measured in URu$_2$Si$_2$ under a pressure $p = 0.67$ GPa, at the temperatures $T = 20 - 23$ K (paramagnetism), 13.9 K (hidden order) and 1.5-1.8 K (antiferromagnetism), at the momentum transfers $Q_0 = (1, 0, 0)$ (top) and $Q_1 = (1.4, 0, 0)$ (bottom) (from [Villaume 08]). (b) Magnetic fluctuations spectra measured in URu$_{2-x}$Fe$_x$Si$_2$, at the temperature $T = 2.5$ K and the momentum transfers $Q_0 = (1, 0, 0)$ (top) and $Q_1 = (1, 0.4, 0)$ (bottom), for the concentration $x = 0.025$ and 0.05 (hidden order) and $x = 0.1$ and 0.2 (antiferromagnetism) (from [Butch 16]). (c) Magnetic fluctuations spectra measured in U(Ru$_{0.98}$Rh$_{0.02}$)$_2$Si$_2$ at the momentum transfer $Q_0 = (1, 0, 0)$, at temperatures from 1.4 to 20 K (from [Yokoyama 04])

Figure 6.15 presents inelastic-neutron-scattering (INS) spectra measured on URu$_2$Si$_2$ under a pressure and on URu$_{2-x}$Fe$_x$Si$_2$ and U(Ru$_{1-x}$Rh$_x$)$_2$Si$_2$ alloys, at momentum transfers corresponding to the magnetic wavevectors $k_0 = (0, 0, 1)$ and $k_1 = (0.6, 0, 0)$. Measurements at the momentum transfers $Q_0 = (1, 0, 0)$ and $Q_1 = (1.4, 0, 0)$ (or equivalently $Q_1 = (1, 0.4, 0)$) correspond to magnetic fluctuations with wavevectors $k_0$ and $k_1$, respectively.

- Figure 6.15(a) shows INS spectra measured on URu$_2$Si$_2$ at the pressure $p_c < p < 0.67$ GPa $< p^*$, at three temperatures $T = 20 - 23$ K characteristic of the correlated paramagnetic regime, $T = 13.9$ K characteristic of the hidden-order phase, and $T = 1.5 - 1.8$ K characteristic of the antiferromagnetic phase [Villaume 08]. In the CPM regime, magnetic fluctuations with wavevector $k_1$ are intense, corresponding to a low-energy nearly-quasielastic spectrum, and magnetic fluctuations with wavevector $k_0$ are weak. Strong changes are induced in the AF and HO phases. In the HO phase, the magnetic-fluctuations spectra present sharp anomalies at $k_0$ with a gap $E_0 \simeq 1.3$ meV, and at $k_1$ with a gap $E_1 \simeq 5$ meV. In the AF phase, low-energy magnetic fluctuations at $k_0$ have almost vanished, while those at $k_1$ have been pushed towards higher energies, with a gap $E_1 \simeq 8$ meV.

- Figure 6.15(b) shows INS spectra measured on $URu_{2-x}Fe_xSi_2$ alloys at the temperature $T = 2.5$ K [Butch 16]. In the HO phase, i.e., for $x < x_c \simeq 0.05$, inelastic peaks are observed at $\mathbf{k}_0$ with a gap $E_0 \simeq 2.5$ meV, and at $\mathbf{k}_1$ with a gap $E_1 \simeq 5$ meV, similarly to the spectra reported for the pure compound (see Figures 6.3, 6.4, and 6.5 in Section 6.1.2) [Broholm 91]. In the AF phase, i.e., for $x > x_c \simeq 0.05$, the low-energy magnetic fluctuations with wavevector $\mathbf{k}_0$ have almost vanished (a broad signal remains in higher energies), while those with wavevector $\mathbf{k}_1$ have been reduced and pushed towards higher energies, with a gap $E_1 \simeq 8 - 10$ meV.

- Figure 6.15(c) shows INS spectra measured on $U(Ru_{0.98}Rh_{0.02})_2Si_2$ at different temperatures [Yokoyama 04]. In the HO phase of this compound, i.e., at temperatures $T_N \simeq 8.5 < T < T_0 \simeq 15$ K, the low-energy fluctuations with wavevector $\mathbf{k}_0$ are intense. On the contrary, they have almost vanished in the CPM regime, at temperatures $T > T_0$, and in the AF phase at temperatures $T < T_N$.

These sets of experiments on $URu_2Si_2$ under a pressure, or doped with Fe or Rh, indicate that low-energy magnetic fluctuations with wavevector $\mathbf{k}_0$ are enhanced in the HO phase, and that they vanish in the CPM regime (see also Figures 6.4 and 6.5 in Section 6.1.2 [Bourdarot 10a, Bourdarot 14, Broholm 91]) and AF phase. They confirm that the order parameter of the HO phase may be related with low-energy magnetic fluctuations with wavevector $\mathbf{k}_0$. They also indicate that the magnetic fluctuations with wavevector $\mathbf{k}_0$ can be considered as precursor of a neighboring antiferromagnetic phase, whose long-range order is modulated with $\mathbf{k}_0$ (see the prototypical $CeRu_2Si_2$ case; Figure 3.4 in Section 3.1.2, Figures 4.10 and 4.11 in Section 4.2.2).

### 6.2.3 Fermi surface

In Section 6.1.3, the Fermi surface of $URu_2Si_2$ was shown to be modified at onset of the hidden-order phase at the temperature $T_0 = 17.5$ K. Figure 6.16 presents studies by quantum oscillations techniques of the Fermi surface of $URu_2Si_2$ under pressure. de-Haas-van-Alphen quantum oscillations in the low-temperature magnetization of $URu_2Si_2$, in a magnetic field $\mathbf{H} \parallel \mathbf{a}$ and at a pressure $p = 0.5$ GPa, are shown in Figure 6.16(a-top) [Nakashima 03a]. Their FFT spectra, plotted at different pressures in Figure 6.16(a-bottom), present a peak corresponding to the orbit $\alpha$ (and several of its harmonics). They are quite similar in the low-pressure HO and high-pressure AF phases (see Figure 6.12(a) in Section 6.2.1). Shubnikov-de-Haas experiments in a magnetic field $\mathbf{H} \parallel \mathbf{c}$ are considered in Figures 6.16(b-c) [Hassinger 10] (see also [Nakashima 03b]). Figure 6.16(b) shows that similar FFT spectra are extracted from the SdH quantum oscillations measured in the HO phase at ambient pressure and in the AF phase at $p = 1.55$ GPa. The pressure-dependence of the SdH frequencies, presented in Figure 6.16(c-top), indicates subtle changes at the critical pressure $p_c$ (noted $p_x$ and estimated at $p_x \simeq 0.8$ GPa in [Hassinger 10]): a kink in the $p$-variation of $F_\alpha$, an increased splitting of $F_\beta$, and a step in the $p$-variation of $F_\gamma$. The cyclotron effective masses continuously decrease with increasing $p$, and no clear anomaly is observed at $p_c$ (see Figure 6.16(c-bottom)).

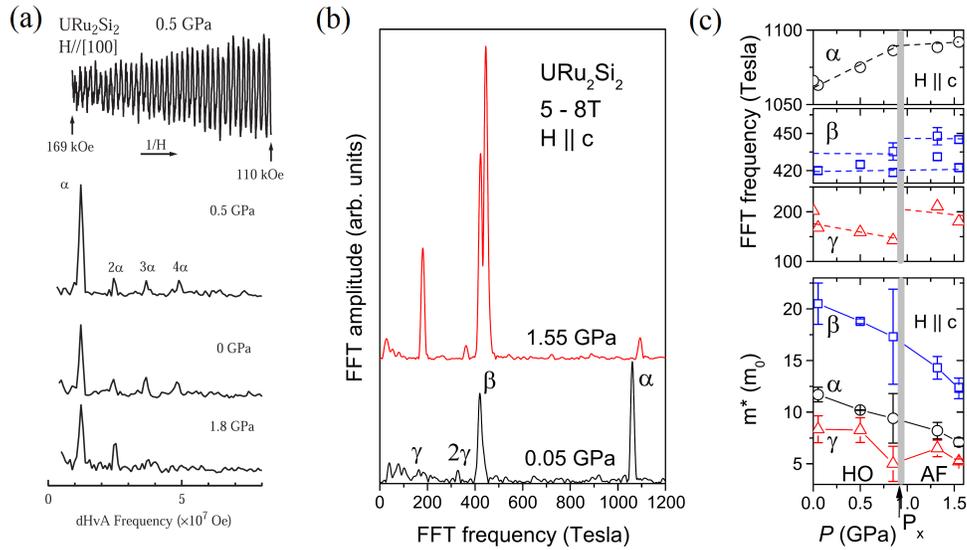

Figure 6.16: (a) de-Haas-van-Alphen quantum oscillations in the magnetization of URu$_2$Si$_2$ in a magnetic field **H** ∥ **a**, at $p = 0.5$ GPa (top), and FFT spectra extracted at different pressures (bottom) (from [Nakashima 03a]). (b) FFT spectra of Shubnikov-de-Haas quantum oscillations in the electrical resistivity of URu$_2$Si$_2$ in a magnetic field **H** ∥ **c**, at the pressures $p = 0.05$ and 1.55 GPa. (c) Pressure-dependence of the SdH frequencies of the orbits $\alpha$, $\beta$, and $\gamma$ (top) and of their cyclotron masses (bottom) (from [Hassinger 10]).

Knowing that the HO phase is the place of strong magnetic fluctuations with wavevector **k**$_0$ = (0, 0, 1), and that the AF phase is associated with antiferromagnetic long-range order with the same wavevector **k**$_0$, it has been proposed that the hidden-order parameter also has a periodicity with wavevector **k**$_0$ [Haule 09]. This proposition is compatible with the observation of similar Fermi surfaces in the HO and AF phases of URu$_2$Si$_2$ (see Figure 6.16) [Hassinger 10].

## 6.3 Magnetic-field-induced phenomena

The application of a magnetic field along the easy magnetic axis **c** of URu$_2$Si$_2$ permits to reveal rich electronic properties: the destabilization of the hidden-order phase, the appearance of a spin-density wave, a cascade of Fermi-surface reconstructions, etc. A review of these high-fields effects is given here. Section 6.3.1 presents bulk properties and the electronic phase diagram of URu$_2$Si$_2$ in a magnetic field **H** ∥ **c**. Field-induced spin-density-wave long-range order and modifications of the magnetic fluctuations are presented in Section 6.3.2. Field-induced cascades of Fermi-surface reconstructions are considered in Section 6.3.3. Finally, the properties of URu$_2$Si$_2$ in three-dimensional i) magnetic-field-pressure-temperature and ii) magnetic-field-chemical-doping-temperature phase diagrams are presented in Sections 6.3.4 and 6.3.5, respectively.

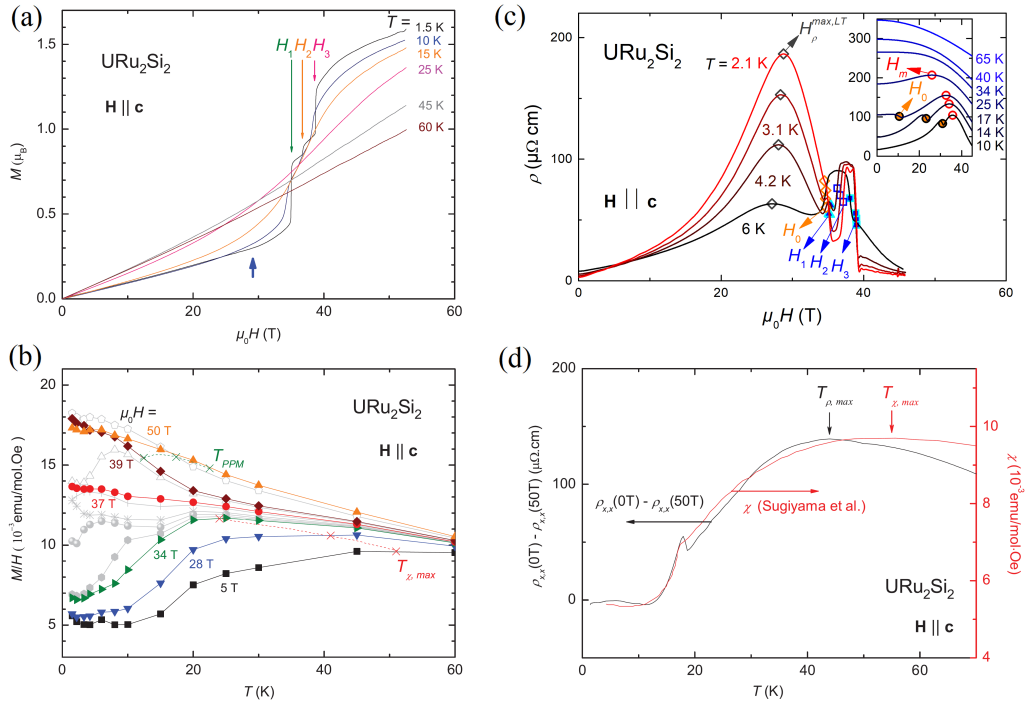

Figure 6.17: (a) Magnetization versus magnetic field, at different temperatures, (b) magnetic susceptibility $\chi = M/H$ versus temperature, at different magnetic fields (from [Scheerer 12]), (c) electrical resistivity $\rho$ versus magnetic field, at different temperatures (from [Knafo 20b]), and (d) comparison of $\rho(0\,\text{T}) - \rho(50\,\text{T})$ and $\chi(T)$ (from [Scheerer 12]) of URu$_2$Si$_2$ in a magnetic field $\mathbf{H} \parallel \mathbf{c}$.

### 6.3.1 Bulk properties and phase diagram

The first observation, by magnetization and electrical-resistivity measurements, of field-induced phase transitions in URu$_2$Si$_2$ under a magnetic field $\mathbf{H} \parallel \mathbf{c}$ [deBoer 86, de Visser 87b] has been followed by a plethora of experiments using a wide spectrum of techniques. In this Section, a selection of experiments describing the magnetic-field-temperature phase diagram of the system are presented.

Figure 6.17 presents the magnetization $M$ and electrical-resistivity $\rho$ measured on URu$_2$Si$_2$ at different temperatures and under a magnetic field $\mathbf{H} \parallel \mathbf{c}$. These curves show signatures of different electronic phases and regimes boundaries, which are described below:

- *SDW phase.* At low-temperature, first-order transitions are observed at $\mu_0 H_1 = 35$ K, $\mu_0 H_2 = 37.5$ T for increasing field and $36.3$ T for decreasing field, and $\mu_0 H_3 = 39$ T, leading to step-like anomalies in $M$ (Figure 6.17(a)) [Scheerer 12] and $\rho$ (Figure 6.17(c)) [Knafo 20b]. They delimitate a field-induced spin-density-wave phase established for $H_1 < H < H_3$ (see Section 6.3.2).

- *HO phase.* At low-temperature, the hidden-order phase is destabilized at a magnetic

field $\mu_0 H_0 \simeq 34.5$ K $\lesssim \mu_0 H_1$, which can be defined at a sharp maximum in $\partial \rho / \partial H$ (Figure 6.17(c)) [Knafo 20b]. A maximum of $\rho$ at $\mu_0 H_{\rho,max}^{LT} \simeq 29$ T, inside the hidden-order phase, is driven by Fermi-surface reconstructions (see Section 6.3.3) [Scheerer 12, Scheerer 14].

- *CPM regime.* Figure 6.17(b) shows that the maximum, at the temperature $T_\chi^{max}$, in the magnetic susceptibility $\chi = M/H$ vanishes in fields higher than $35-39$ T. This maximum is induced by a crossover into the CPM regime. At temperatures below 40 K, a maximum of $\rho_{x,x}$ is observed at the field $H_m$, which increases with decreasing temperatures. $H_m$ extrapolates to 37 T at low-temperatures, where it is masked by the anomalies at $H_0$, $H_1$, $H_2$, and $H_3$ (Figure 6.17(c)). In relation with this field-induced anomaly, a maximum in the zero-field resistivity is observed in the bare data at 70 K (see Figure 6.2 in Section 6.1.1). Knowing that a $T$-dependent, but $H$-independent, electron-phonon scattering contribution $\rho_{x,x}^{e-ph}$ adds to the purely electronic term $\rho_{x,x}^{e-e}$, Figure 6.17(d) presents a estimation of the electronic term by $\rho_{x,x}^{e-e}(T, 0T) = \rho_{x,x}(T, 0T) - \rho_{x,x}(T, 50T)$ and compares it with the low-field magnetic susceptibility $\chi$ (from [Sugiyama 90]) [Scheerer 12]. The similarity between the temperature-dependences of $\rho_{x,x}^{e-e}$ and $\chi$ indicates that the high-temperature maxima in $\rho$ and $\chi$ are driven by the onset of the CPM regime. The boundaries $T_\chi^{max}$ and $H_m$ of the CPM regime follow the universal law $R_{CPM} = T_\chi^{max}/\mu_0 H_m \simeq 1$ K/T observed for most of heavy-fermion paramagnets (see Figure 3.14 in Section 3.3). It indicates that $T_\chi^{max}$ and $H_m$ are controlled by a single parameter, which supports a Fermi-liquid description of the CPM regime (see Section 2.4.3). This parameter may be related with magnetic fluctuations with wavevector $\mathbf{k}_1$, which develop in the CPM regime, and whose energy scale, the relaxation rate $\Gamma_1 \simeq 4$ meV $\simeq 40$ K (see Section 6.2.2) [Broholm 91], is of same order than $T_\chi^{max}$.

- PPM regime: For $H > H_3$, the low-temperature magnetization nearly saturates to a value $M > 1.5$ $\mu_B$/U indicating the stabilization of a polarized paramagnetic regime (Figure 6.17(a)). The temperature scale $T_{PPM}$, characteristic of a crossover to the PPM regime, is defined at the inflexion point of the low-temperature increase of $\chi$ and is found to increase with $H$ (Figure 6.17(b)) [Scheerer 12].

Signatures in other bulk quantities of the phase transitions induced in URu$_2$Si$_2$ by a magnetic field $\mathbf{H} \parallel \mathbf{c}$ are presented in Figure 6.18. Heat-capacity measurements show that the second-order step-like anomaly at $T_0$ in zero-field is replaced by a first-order sharp peak in a field $\mu_0 H = 33.5$ T, where $T_0 = 5$ K (Figure 6.18(a)) [Jaime 02]. In higher fields, a second-order step and a first-order sharp peak are observed at temperatures of 7 and 5 K, respectively, which delimitate the field-induced SDW phase. Magnetostriction measurements show sharp first-order step-like variation of the relative lengths $\Delta a/a$ and $\Delta c/c$ at the transition fields $H_1$, $H_2$, and $H_3$ (Figure 6.18(b)) [Wartenbe 19]. In fields varying from 0 to 50 T, the amplitude of the variations $\Delta a/a = 3.7 \ 10^{-4}$ and $\Delta c/c = -1.6 \ 10^{-4}$ indicate that the valence is almost unaffected and might remain near an integer value. Sharp anomalies are also observed at the transition fields $H_1$, $H_2$, and $H_3$ in the elastic constants $C_{11}/10$, $C_{33}$, $C_{44}$, $C_{66}$ and $(C_{11} - C_{12})/2$ probed by ultrasound measurements, indicating a subtle relationship between the electronic and lattice

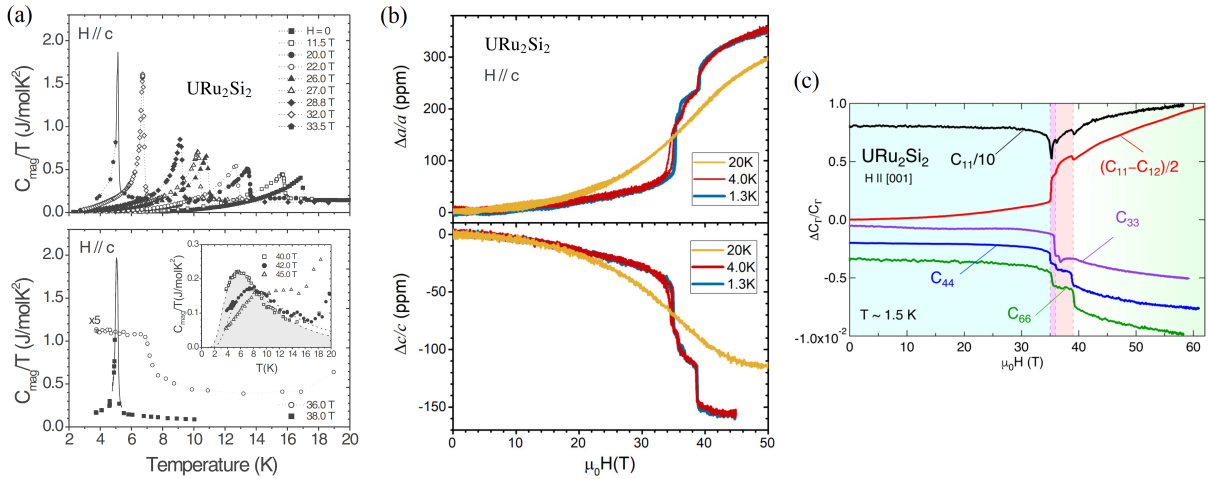

**Figure 6.18:** (a) Heat capacity divided by temperature $C_p/T$ versus temperature, at different magnetic fields (from [Jaime 02]), (b) relative length variations $\Delta a/a$ and $\Delta b/b$ versus magnetic field, at different temperatures (from [Wartenbe 19]), and (c) elastic constants $C_{11}/10$, $C_{33}$, $C_{44}$, $C_{66}$ and $(C_{11}-C_{12})/2$ versus magnetic field, at $T = 1.5$ K, (from [Yanagisawa 18]) of URu$_2$Si$_2$ in a magnetic field $\mathbf{H} \parallel \mathbf{c}$.

properties at the SDW phase boundaries (Figure 6.18(c)) [Yanagisawa 18].

The magnetic-field-temperature phase diagram of URu$_2$Si$_2$ in $\mathbf{H} \parallel \mathbf{c}$ is presented in Figure 6.19 [Scheerer 12, Knafo 20b]. It emphasizes that the field-induced destabilization of the high-temperature CPM regime precedes that of the HO phase, which both end with the stabilization of a low-temperature SDW phase in a field window [35 T $-$ 39 T]. The SDW phase is composed of at least two different sub-phases delimited by the critical field $H_2$, which presents a relatively large hysteresis. Different shapes of the boundaries of the SDW phase were reported from different experimental probes [Sugiyama 99, Jaime 02, Harrison 03, Kim 03a, Suslov 03, Kim 03b, Oh 07, Jo 07, Correa 12]. The proposition that the SDW phase lies beneath a dome, which is well-separated from the HO phase boundary, was made in [Mydosh 20]. It is consistent with the last set of electrical-resistivity data presented in Figure 6.17(c) [Knafo 20a, Knafo 20b].

### 6.3.2 Magnetic order and fluctuations

Figure 6.20 presents studies of the magnetic fluctuations in URu$_2$Si$_2$ under a magnetic field $\mathbf{H} \parallel \mathbf{c}$. In fields up to 17 T, i.e., well inside the HO phase, the magnetic-fluctuations spectra measured with a momentum transfer $\mathbf{Q}_0 = (1, 0, 0)$ corresponding to the wavevector $\mathbf{k}_0 = (0, 0, 1)$ are continuously modified, with an intensity decreasing and a gap increasing with field (Figure 6.20(a)) [Bourdarot 03]. On the contrary, the magnetic fluctuations measured with a momentum transfer $\mathbf{Q}_1 = (1.4, 0, 0)$ corresponding to the wavevector $\mathbf{k}_1 = (0.6, 0, 0)$ are almost field-independent (Figure 6.20(b)). Figure 6.20(c) shows the field-variation of the energy-gap extracted from these spectra: while the gap $E_0$ at wavevector $\mathbf{k}_0$ increases con-tinuously, from 1.6 meV at zero field to 2.5 meV at $\mu_0 H = 17$ T, the gap $E_1$ at wavevector $\mathbf{k}_1$ remains constant and equal to 4.5 meV [Bourdarot 03]. A linear extrapolation of the field-variations of the two gaps leads to their crossing at a magnetic field $\mu_0 H \simeq 35$ T near the low-temperature metamagnetic field $H_m$. This extrapolation is compatible with a loss of anti-ferromagnetic correlations, i.e., wavevector-$\mathbf{q}$ dependent magnetic fluctuations, at $H_m$. Figure 6.20(d) further shows a comparison of the field-variation of the coefficient $\sqrt{A}$, extracted from electrical-resistivity-versus-temperature measurements, with that of the specific-heat coefficient $\gamma$, extracted from magnetization-versus-temperature experiments [Levallois 09, Scheerer 12]. A Fermi-liquid behavior implying the relationships $\gamma \sim \sqrt{A}$ and $(\partial^2 M/\partial T^2)_H = (\partial \gamma/\partial \mu_0 H)_T$ has been assumed. An enhancement of $\gamma$ is observed in fields higher than 30 T, and its saturation indicates an enhanced effective mass and, thus, enhanced magnetic fluctuations in the critical field-window [$35 - 39$ T] coinciding with the SDW phase. By analogy with the CeRu$_2$Si$_2$ case (see Figure 3.12(d) in Section 3.2.2) [Sato 01, Flouquet 04], the field-induced critical magnetic fluctuations, peaked in the vicinity of $H_m$, may be ferromagnetic in URu$_2$Si$_2$ too.

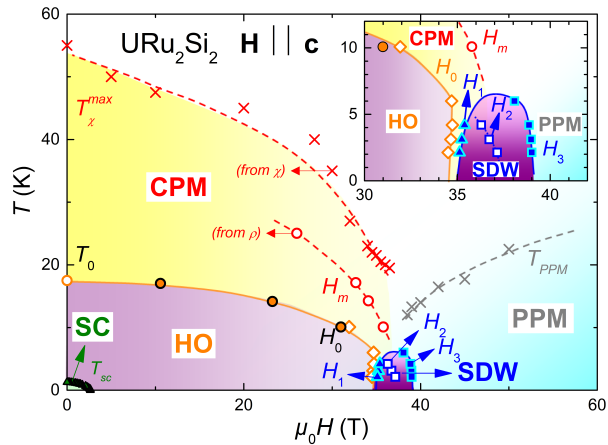

Figure 6.19: Magnetic-field temperature phase diagram of URu$_2$Si$_2$ in a magnetic field $\mathbf{H} \parallel \mathbf{c}$ (adapted from [Scheerer 12, Knafo 20b]).

The microscopic nature of the SDW phase induced by a magnetic field $\mathbf{H} \parallel \mathbf{c}$ in URu$_2$Si$_2$ was revealed by neutron diffraction [Knafo 16]. Figure 6.21(a) shows the magnetic-field dependence of the neutron-diffraction intensity measured at the momentum transfers $\mathbf{Q}_1 = (0.6, 0, 0)$ and $(1.6, 0, -1)$ corresponding to the wavevector $\mathbf{k}_1 = (0.6, 0, 0)$, at temperatures $T = 2$ and 18 K. While no intensity develops at $T = 18$ K, magnetic Bragg peaks are observed at $T = 2$ K in the field window [$35 - 39$ T]. As shown by scans in the reciprocal space in Figure 6.21(b), the field-induced Bragg peaks are centered on the wavevector $\mathbf{k}_1$. The phase stabilized for $H_1 < H < H_3$ is a spin-density-wave phase of wavevector $\mathbf{k}_1$. A small reduction of the diffracted intensity suggests that a subtle change of magnetic structure occurs at the intermediate field $H_2$. Figure 6.21(c) presents the three-dimensional pressure-magnetic-field-temperature phase diagram of URu$_2$Si$_2$, in which sketches of the magnetic structures are presented, with wavevector $\mathbf{k}_0$ for the pressure-induced AF phase, wavevector $\mathbf{k}_1$ for the field-induced SDW

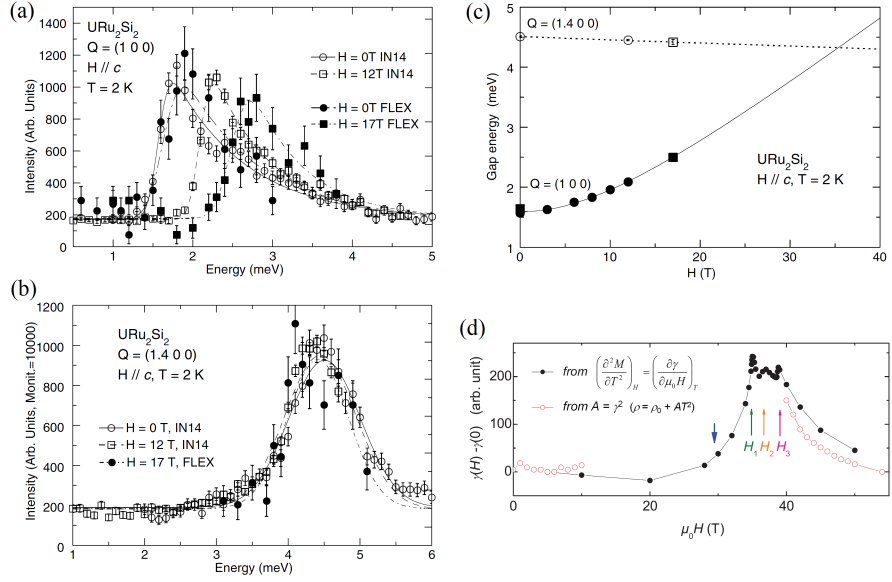

Figure 6.20: Inelastic neutron scattering spectra at the momentum transfers (a) $\mathbf{Q}_0 = (1, 0, 0)$ and (b) $\mathbf{Q}_1 = (1.4, 0, 0)$, in different magnetic fields up to 17 T and at the temperature $T = 2$ K, (c) magnetic-field-dependence of the excitation gaps at the two wavevectors (from [Bourdarot 03]), and field-variation of Sommerfeld coefficients extracted from a Fermi-liquid description of the magnetization and electrical resistivity (from [Levallois 09, Scheerer 12]), of URu$_2$Si$_2$ in a magnetic field $\mathbf{H} \parallel \mathbf{c}$.

phase and wavevector $\mathbf{k} = 0$ for the PPM regime.

The two magnetically-ordered phases of URu$_2$Si$_2$ are associated with a wavevector, $\mathbf{k}_0$ under pressure and $\mathbf{k}_1$ in a magnetic field, at which intense magnetic fluctuations with a minimum of the energy gap develop at zero-field and ambient pressure (see Figures 6.3, 6.4, and 6.5 in Section 6.1.2) [Broholm 91]. Intersite magnetic fluctuations in the paramagnetic hidden-order phase indicate, thus, the presence of neighboring quantum magnetic phase transitions at which long-range magnetic order with the same wavevectors can be established. Similarly, the precursor role of magnetic fluctuations can be emphasized in other heavy-fermion compounds. In the paramagnet CeRu$_2$Si$_2$, intersite magnetic fluctuations peaked at wavevectors $\mathbf{k}_1 = (0.31, 0, 0)$, $\mathbf{k}_2 = (0.31, 0.31, 0)$, and $\mathbf{k}_3 = (0, 0, 0.35)$ [Rossat-Mignod 88, Sato 99, Kadowaki 04] can be transformed into long-range order, with either $\mathbf{k}_1$, $\mathbf{k}_2$, or $\mathbf{k}_3$ wavevectors, by chemical doping (with La, Ge, or Rh) and/or a magnetic field [Quezel 88, Haen 02, Mignot 91a, Watanabe 03].

As in other heavy-fermion systems, a microscopic understanding of the mechanisms driving to magnetic exchange, which result in magnetic fluctuations or order with particular wavevectors, is lacking for URu$_2$Si$_2$. Fermi-surface-driven changes of RKKY magnetic-exchange interactions $J_{ex}(\mathbf{k})$ may lead to the appearance of long-range magnetic order, either antiferromagnetism with wavevector $\mathbf{k}_0$ or spin-density wave with wavevector $\mathbf{k}_1$, under pressure and magnetic field, respectively. Within this picture, pressure-enhanced exchange $J_{ex}(\mathbf{k}_0)$ would result in antiferromagnetic order with wavevector $\mathbf{k}_0$, while magnetic-field-enhanced exchange $J_{ex}(\mathbf{k}_1)$ would lead to a spin-density wave with wavevector $\mathbf{k}_1$. Understanding how these exchange parameters and the related magnetic excitations evolve under pressure and magnetic field could constitute a decisive step in the description of the electronic properties, including the HO phase, of URu₂Si₂.

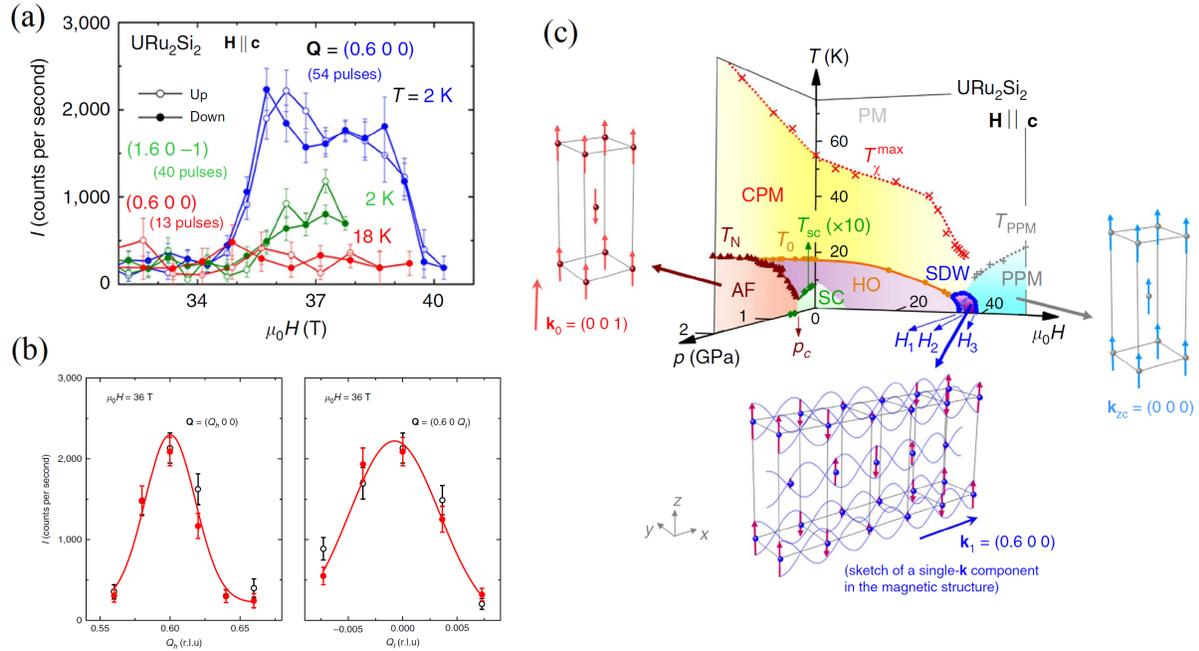

Figure 6.21: (a) Neutron diffracted intensity at the momentum transfer $\mathbf{Q}_1 = (0.6, 0, 0)$ and $(1.6, 0, -1)$ at $T = 2$ and 18 K, (b) $Q_h$ and $Q_l$ scans around the magnetic Bragg peak at $\mathbf{Q}_1 = (0.6, 0, 0)$, $\mu_0 H = 36$ T and $T = 2$ K, and (c) three-dimensional pressure-magnetic-field-temperature phase diagram of URu₂Si₂ in a magnetic field $\mathbf{H} \parallel \mathbf{c}$, with schemes representing the ordered magnetic structures (from [Knafo 16]).

### 6.3.3 Fermi surface

A cascade of Fermi-surface reconstructions is induced in URu₂Si₂ by a magnetic field $\mathbf{H} \parallel \mathbf{c}$. Figure 6.22 shows signatures of these Fermi-surfaces changes in bulk properties. The field-dependence of the low-temperature transverse electrical resistivity $\rho_{x,x}$ measured with an electrical current $\mathbf{I} \perp \mathbf{H}$ is shown for compounds of different qualities in Figure 6.22(a) [Levallois 09, Scheerer 12, Scheerer 14, Ran 17, Knafo 20a, Knafo 20b]. While $\rho_{x,x}$ is almost sample-independent for $H > H_1$, it is sample-dependent for $H < H_1$, i.e., in the HO phase. A field-enhancement of $\rho_{x,x}$ is induced by an orbital effect controlled by the cyclotron motion of conduction electrons. It indicates that URu₂Si₂ is a compensated metal with a high carrier mobility [Kasahara 07]. This enhancement is larger at lower temperature and for samples of higher residual-resistivity ratio, i.e., of higher quality. The orbital contribution is absent in lon-gitudinal transverse electrical resistivity $\rho_{z,z}$ measured with an electrical current **I** ∥ **H** (see Figure 6.22(b)). For all samples, a maximum of the orbital contribution to $\rho_{x,x}$ is observed at $\mu_0 H_{\rho,max}^{LT} \simeq 29$ T, i.e., inside the hidden-order phase. It indicates a Fermi-surface reconstruction associated with a reduction of the carrier mobility [Scheerer 12, Scheerer 14]. Figures 6.22(c-e) show that the Nernst coefficient $N$ and the thermoelectric power coefficient $S$, measured with a longitudinal or transverse configuration, are also strongly magnetic-field-dependent, suggesting Fermi-surface modifications. In particular, sign-changes of $S$ indicate modifications in the roles played by electron and holes on approaching metamagnetism [Pourret 13].

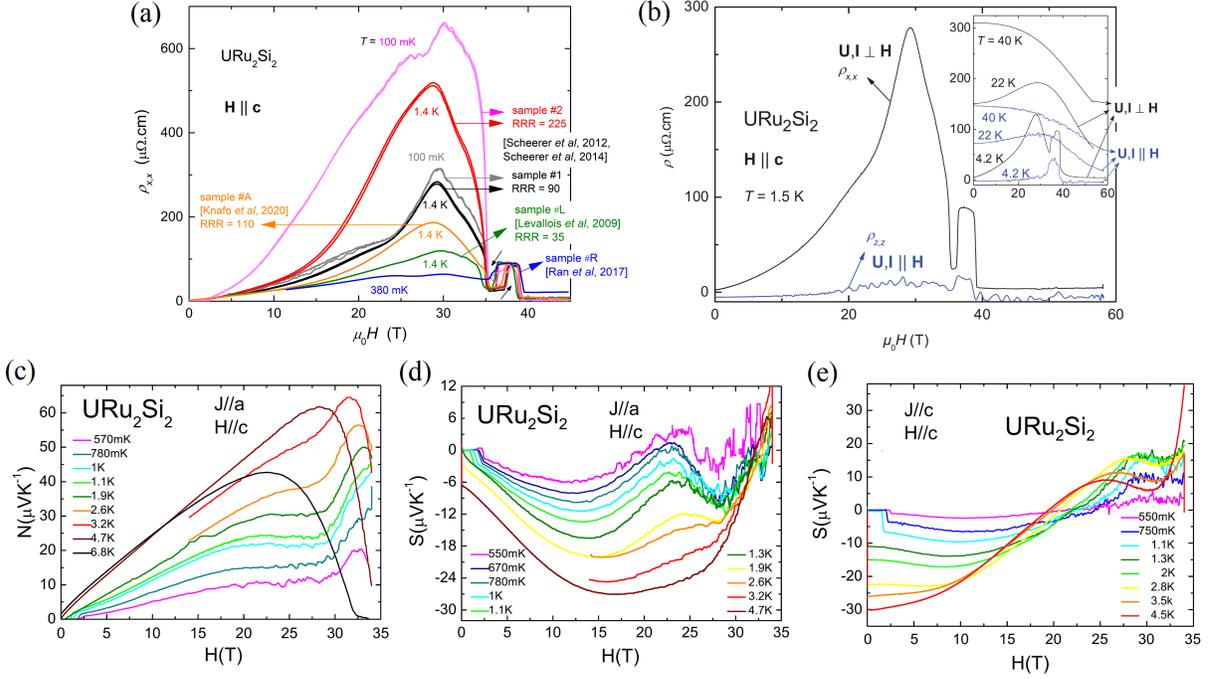

Figure 6.22: (a) Low-temperature electrical resistivity $\rho_{xx}$ versus magnetic field of different URu$_2$Si$_2$ samples, of various residual resistive ratios (RRR) in a magnetic field **H** ∥ **c** (from [Knafo 20b], includes data from [Levallois 09, Scheerer 12, Scheerer 14, Ran 17, Knafo 20a]). (b) Low-temperature electrical resistivities $\rho_{xx}$ and $\rho_{zz}$ of URu$_2$Si$_2$ versus magnetic field, measured with currents **I** ∥ **a** and **I** ∥ **c**, respectively, in a magnetic field **H** ∥ **c** [Scheerer 12]. Magnetic-field variation of (c) the Nernst signal $N$ and the thermoelectric power $S$ (d) with a current **I** ∥ **a** and (e) with a current **I** ∥ **c** of URu$_2$Si$_2$ in a magnetic field **H** ∥ **c** and at different temperatures (from [Pourret 13]).

Magnetic-field-induced Fermi-surface reconstructions in URu$_2$Si$_2$ were evidenced by Shubnikov-de-Haas experiments. Figure 6.23 presents the magnetic-field dependence of SdH quantum oscillations of the electrical resistivity in the field windows [15 − 35 T] (Figure 6.23(a)) [Scheerer 14], [36 − 37.5 T] (field-up, Figure 6.23(b-left)), [37 − 38.75 T] (field-down, Figure 6.23(b-right)) [Altarawneh 11], and [40 − 45 T] (Figure 6.23(c)) [Harrison 13]. The field-dependencies of the frequencies extracted from different SdH studies [Jo 07, Shishido 09, Altarawneh 11,

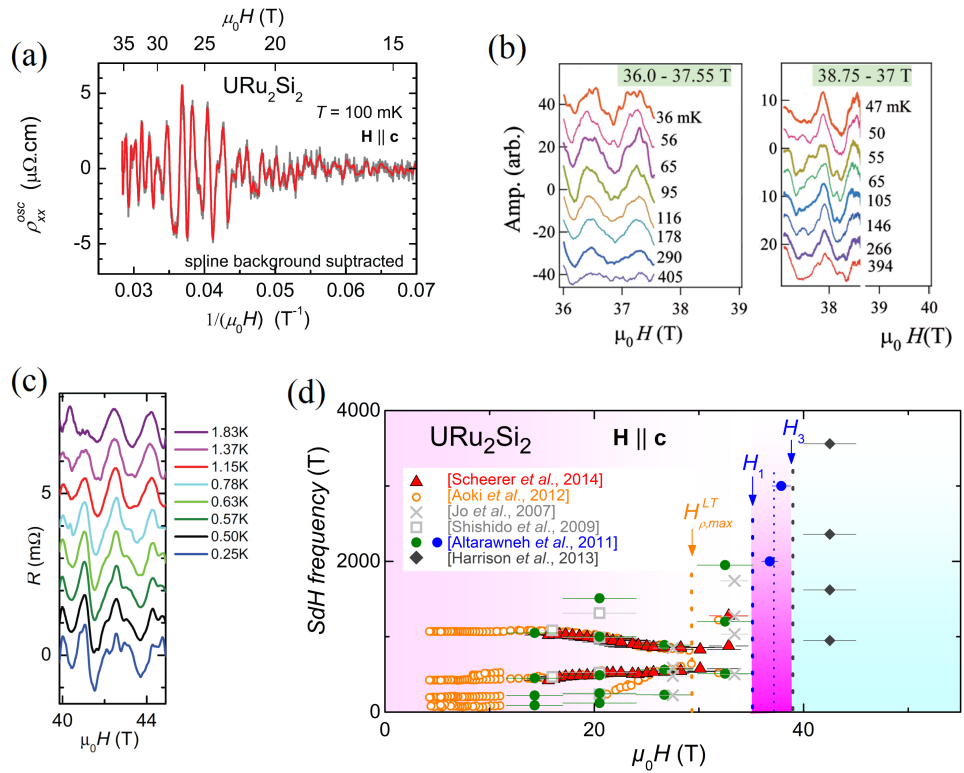

Figure 6.23: Shubnikov-de-Haas quantum oscillations of the electrical resistivity of URu$_2$Si$_2$ in different windows of magnetic field **H** ∥ **c**: (a) [15-35 T] (from [Scheerer 14]), (b) field-up of [36-37.5 T] (left) and field-down of [37-39 T] (from [Altarawneh 11]), and (c) [40-45 T] (from [Harrison 13]). (d) Magnetic-field variation of the SdH frequencies of URu$_2$Si$_2$ in a magnetic field **H** ∥ **c** (from [Knafo 18], includes data from [Jo 07, Shishido 09, Altarawneh 11, Aoki 12c, Harrison 13, Scheerer 14]).

Aoki 12c, Harrison 13, Scheerer 14] is synthesized in Figure 6.23(d) [Knafo 18]. Differences result from the choice of different field-windows for making the FFT spectra and from different noise-levels in the experiments. However, all set of data lead to similar qualitative conclusions:

- The SdH frequencies $F_\eta \simeq 90$ T, $F_\gamma \simeq 200$ T, $F_\beta \simeq 420$ T, and $F_\alpha \simeq 1060$ T and, thus, the Fermi surface of URu$_2$Si$_2$, are almost unchanged in magnetic fields up to 15 T [Aoki 12c].

- A progressive and continuous modification of the Fermi surface is visible in fields from 15 to 30 T, leading to a narrowing of the observed frequencies in a window between 500 and 1000 T [Shishido 09, Altarawneh 11, Aoki 12c, Scheerer 14].

- In fields higher than 30 T, which coincides with the maximum in the orbital contribution to the electrical resistivity, a sudden modification of the Fermi surface is reported, and higher frequencies, in a window from 500 to 2000 T, are observed [Jo 07, Altarawneh 11, Aoki 12c, Scheerer 14].

- Frequencies of 2000 T and 3000 T are extracted in the SDW phase, for $H_1 < H < H_2$ and $H_2 < H < H_3$, respectively [Altarawneh 11].

- A large spectrum of frequencies, varying from 500 to 3500 T, is observed in the PPM regime [Harrison 13].

The modifications of the Fermi surface of URu$_2$Si$_2$ in a magnetic field $\mathbf{H} \parallel \mathbf{c}$ support that conduction electrons contribute to the magnetic properties and play a role in the hidden-order phase. The Fermi-surface reconstructions inside the HO phase are related with a reduction of the carrier mobility and are a precursor of the field-destabilization of the HO. SdH frequencies tend to increase with increasing magnetic fields, which leads to the formation of larger Fermi-surfaces. From the observation of smaller effective masses and larger Fermi velocities in the PPM regime than in the HO phase, Harrison *et al* proposed that the $5f$ electron become localized in the PPM regime [Harrison 13]. A similar conclusion was obtained from Fermi-surface studies of the PPM regime in other heavy-fermion compounds (see Sections 3.2.3 and 4.3.2.3).

### 6.3.4 Combination of pressure and magnetic field

We have seen in Sections 6.2 and 6.3 that pressure and a magnetic field $\mathbf{H} \parallel \mathbf{c}$ both lead to the destabilization of the hidden-order phase in URu$_2$Si$_2$. However, they induce different magnetic phenomena. The application of pressure leads to an increase of $T_\chi^{max}$, indicative of the stabilization of the CPM regime, to the stabilization of an antiferromagnetic phase with wavevector $\mathbf{k}_0 = (0, 0, 1)$, but to no noticeable change of the Fermi surface. On the contrary, the application of a magnetic field $\mathbf{H} \parallel \mathbf{c}$ leads to a decrease of $T_\chi^{max}$, indicating the destabilization of the CPM regime, to the stabilization of a SDW phase corresponding to magnetic order with $\mathbf{k}_1 = (0.6, 0, 0)$ in a narrow field window, to a high-field polarized regime, and to a succession of Fermi-surface reconstructions. By presenting the effects of their combined application, this Section clarifies the respective roles of pressure and magnetic field on URu$_2$Si$_2$.

An investigation by Aoki *et al* by thermal expansion and neutron scattering of URu$_2$Si$_2$ under pressure combined with magnetic fields $\mu_0\mathbf{H} \parallel \mathbf{c}$ up to 15 T is summarized in Figure 6.24 [Aoki 09b]. Figure 6.24(a) shows a series of temperature-magnetic-field phase diagrams obtained at different pressures from thermal-expansion experiments. It indicates that, for $p > p_c$, a magnetic field can lead to an AF-to-HO phase transition, due to the faster field-reduction of $T_N$ in comparison with that of $T_0$. Field-induced reentrance of the HO phase is observed at the pressure $p = 0.92$ GPa $< p^*$, for which $T_N = 14$ K is decoupled from $T_0 = 18$ K at zero-field, and at the pressure $p = 1.39$ GPa $> p^*$, for which $T_N = T_0$ at zero-field. The field-induced decoupling of $T_N$ and $T_0$ is indicated by well-separated anomalies in the thermal-expansion data (Figure 6.24(d)). Neutron diffracted intensity at the magnetic Bragg peak with wavevector $\mathbf{k}_0$ vanishes at the antiferromagnetic boundary $H_c$, as shown at $p = 0.72$ GPa in Figure 6.24(b). INS measurements at the same pressure show that an inelastic peak, associated with an energy gap $E_0 = 2$ meV, develops at the wavevector $\mathbf{k}_0$ in the HO phase induced by $\mu_0 H = 15$ T $> H_c$ (Figure 6.24(c)). They confirm that gapped inelastic fluctuations with wavevector **k**$_0$ are a signature of the HO phase (see Figure 6.5 in Section 6.1.2 [Bourdarot 10a] and Figure 6.15 in Section 6.2.2 [Villaume 08]).

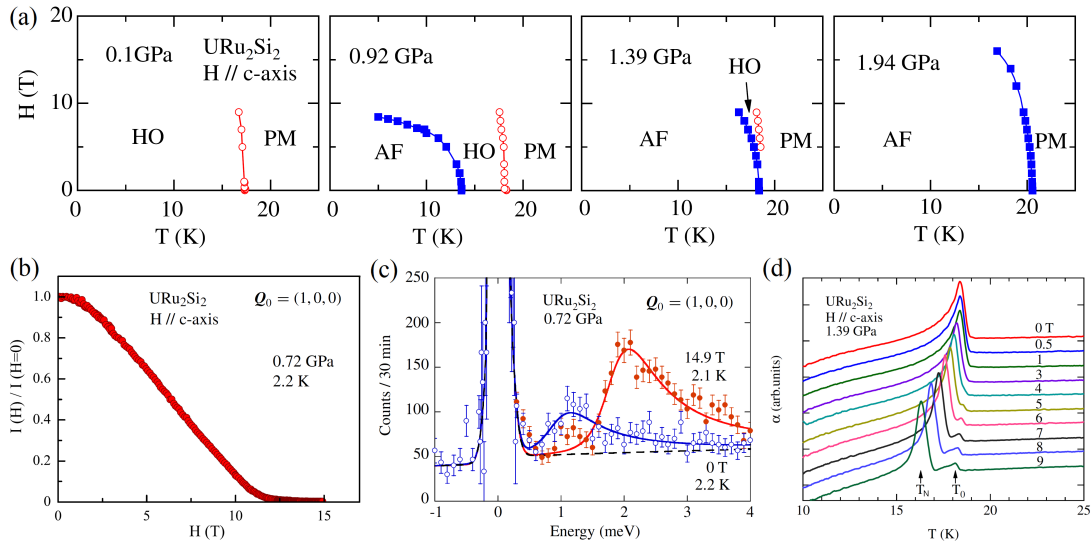

Figure 6.24: (a) Temperature-magnetic-field phase diagrams of URu$_2$Si$_2$ under pressure combined with a magnetic field **H** ∥ **c**. (b) Magnetic-field-dependence of neutron diffracted intensity at the momentum transfer **Q**$_0$ = (1, 0, 0) in URu$_2$Si$_2$ at $p = 0.72$ GPa and $T = 2.2$ K with a magnetic field **H** ∥ **c**. (c) Inelastic neutron scattering spectra at the momentum transfer **Q**$_0$ = (1, 0, 0) in URu$_2$Si$_2$ at $p = 0.72$ GPa and $T = 2.1$–2.2 K, for a magnetic field $\mu_0$**H** ∥ **c** of 0 and 14.9 T. Thermal expansion coefficient $\alpha$ versus temperature of URu$_2$Si$_2$ at $p = 1.39$ GPa for different values of a magnetic field **H** ∥ **c** (from [Aoki 09b]).

Figure 6.25 considers URu$_2$Si$_2$ under combined pressures up to 4 GPa and magnetic fields $\mu_0$**H** ∥ **c** up to 60 T [Knafo 20a]. Low-temperature electrical-resistivity-versus-magnetic-field curves are shown at different pressures in Figure 6.25(a), and the low-temperature pressure-magnetic-field phase diagram and three-dimensional pressure-magnetic-field-temperature phase diagram are shown in Figures 6.25(b,c), respectively. A large number of competing electronic states are observed: two magnetically ordered phases (antiferromagnetism and SDW) and two magnetic regimes (correlated and polarized paramagnetisms), in addition to the HO and superconducting phases. Their characteristic are detailed below:

- *SDW phase.* This low-temperature phase, observed between $\mu_0 H_1 = 35$ T and $\mu_0 H_3 = 39$ T at ambient pressure, moves towards higher fields and progressively shrinks under pressure [Jo 07, Knafo 20a]. Its trace is lost at pressures $p \gtrsim 2$ GPa.

- *HO phase.* This phase is observed at zero field for $p < p_c = 0.5$ GPa and for $H > H_c$, where $H_c$ is the AF boundary, under pressures $p_c < p < p^{**} = 3.25$ GPa. At low temperature and ambient pressure, its upper critical field $\mu_0 H_0 = 34.5$ T delimits the onset of the SDW phase and can be defined as a sharp maximum in $\partial \rho / \partial H$ [Knafo 20b].

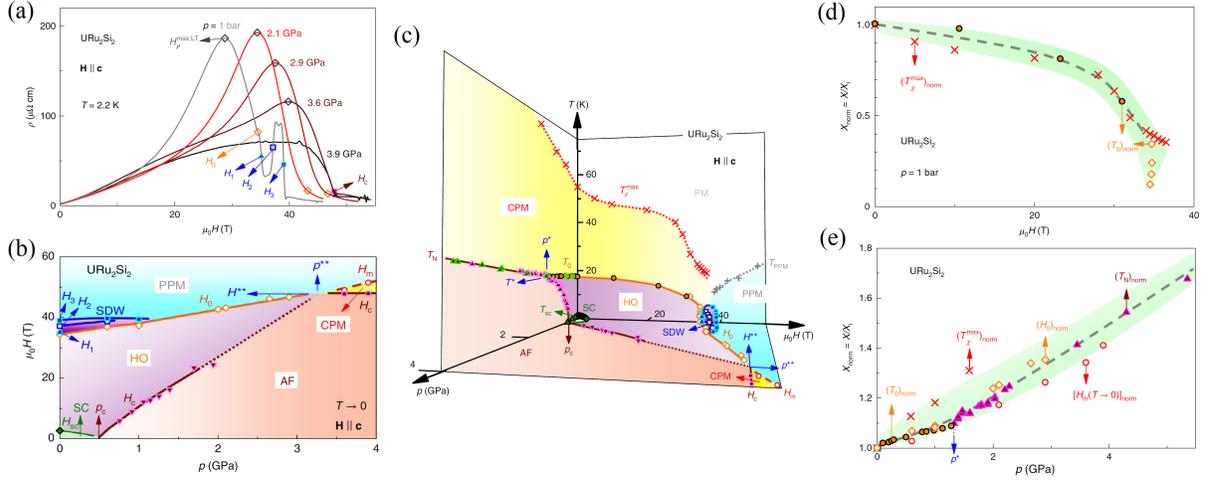

Figure 6.25: (a) Electrical resistivity $\rho$ versus magnetic field of URu$_2$Si$_2$ at different pressures combined with a magnetic field **H** ∥ **c** and $T = 2.2$ K. (b) Low-temperature pressure-magnetic-field phase diagram and (c) three-dimensional pressure-magnetic-field-temperature phase diagram of URu$_2$Si$_2$ in a magnetic field **H** ∥ **c**. Field and pressure variations of normalized quantities: plot of normalized values of $T_\chi^{max}$ and $T_0$ versus $H$, and (e) plot of normalized values of $T_\chi^{max}$ and $T_0$, $T_N$, $H_0$ and $H_m$ versus $p$ (from [Knafo 20a], includes data from [Pfleiderer 06, Hassinger 08a, Aoki 09b, Aoki 10, Scheerer 12]).

Under pressure, $H_0$ is defined as a sharp kink in $\rho$ characteristic of a transition to the SDW for $p \lesssim 1$ GPa, or as a broad kink in $\rho$ characteristic of a crossover to the PPM regime for $p \gtrsim 2$ GPa. $\mu_0 H_0$ increases almost linearly with $p$, reaching 48 T at $p^{**}$. The broad maximum in $\rho$ observed at $\mu_0 H_\rho^{max,LT} = 29$ T, i.e., inside the HO phase, at low temperature and ambient pressure collapses under pressure and disappears in the AF phase.

- *AF phase.* This phase is observed at zero field for $p > p_c = 0.5$ GPa. Its upper-field boundary $H_c$ at the onset of field-reentrant HO was observed under pressures up to 2 GPa by thermal expansion [Aoki 09b], but could not be observed by electrical resistivity, where it was presumably masked by the large orbital signal [Knafo 20a]. For $p > p^{**}$, $\mu_0 H_c$ saturates at 48 T and marks the onset of the CPM regime.

- *CPM regime.* It is observed at temperatures $T_{0,N} < T < T_\chi^{max}$ and under magnetic fields $H_{0,c} < H < H_m$. For $p > p^{**}$, the CPM regime can be identified at low temperature in a narrow field window $H_c < H < H_m$.

- *PPM regime.* It is ultimately established in high fields and corresponds to a weakly-correlated regime.

- *SC phase.* It is restricted to the low pressures $p < p_c$ and low fields $\mu_0 H < 3$ T.

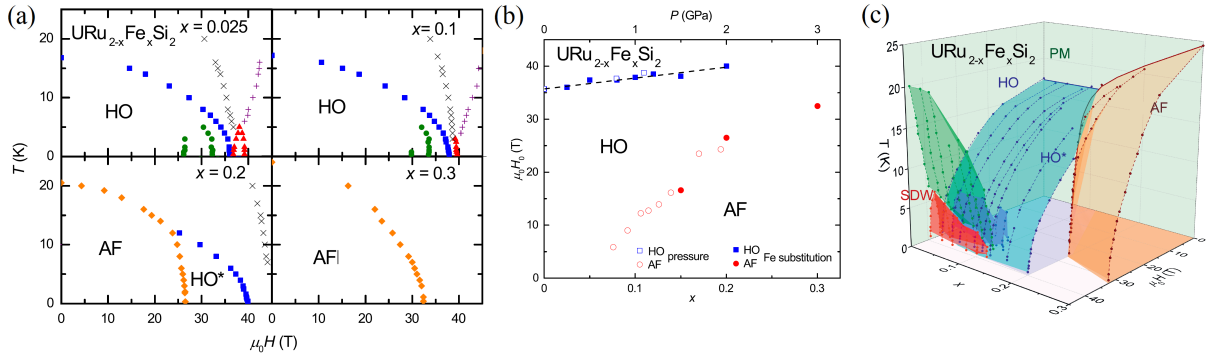

Figure 6.26: (a) Magnetic-field-temperature phase diagram for different values of Fe-doping $x$, (b) low-temperature doping-magnetic-field phase diagram, and (c) three-dimensional magnetic-field-doping-temperature phase diagram of URu$_{2-x}$Fe$_x$Si$_2$ in a magnetic field $\mathbf{H} \parallel \mathbf{c}$ (from [Ran 17]).

Beyond this apparent complexity, simple general trends are observed: the fall of the main energy scales under magnetic field and their increase under pressure [Knafo 20a]. Figure 6.25(d) shows that $T_\chi^{max}$ and $T_0$ decrease and are almost proportional under magnetic field. As well, Figure 6.25(e) shows that $T_\chi^{max}$, $T_0$, $T_N$ (for $p > p^*$), $H_0$, and $H_m$ increase and are almost proportional under pressure. This indicates that the pressure and magnetic-field variations of these quantities are controlled, within first approximation, by the variation of a single parameter $\delta$. In agreement with a Fermi-liquid picture, this parameter may be the energy scale $\Gamma_1$ of the magnetic fluctuations which are present in the CPM regime, and which controls its boundaries $T_\chi^{max}$ and $H_m$. However, not all of the boundaries of the 3D phase diagram of URu$_2$Si$_2$ are fully controlled by $\delta$. A single energy scale is not sufficient to explain why the HO is replaced by antiferromagnetism under pressure. The situation concerning the AF phase for $p < p^{**}$ is more subtle, since $T_N$ and $H_c$ are not proportional and, thus, not controlled by a single parameter. Further experimental and theoretical efforts are needed for a quantitative understanding of the three-dimensional phase diagram of URu$_2$Si$_2$. For this purpose, the knowledge of the evolution of the magnetic fluctuations spectra under pressure and magnetic field, and of the relation between their characteristic energies with the different field- and temperature-boundaries may be mandatory.

### 6.3.5 Combination of doping and magnetic field

In Section 6.2.1, chemical doping was presented as an alternative to pressure for driving URu$_2$Si$_2$ to a quantum antiferromagnetic phase transition. We consider here high-magnetic-field investigations performed on two families of doped URu$_2$Si$_2$ alloys:

- *URu$_{2-x}$Fe$_x$Si$_2$ alloys.* The doping-temperature phase diagram has striking similarities with the pressure-temperature phase diagram of URu$_2$Si$_2$ (see Figure 6.13(a)) [Ran 16]. This similarity was confirmed by an electrical-resistivity study of URu$_{2-x}$Fe$_x$Si$_2$ com-pounds in a magnetic field $\mathbf{H} \parallel \mathbf{c}$, for which phase diagrams are presented in Figure 6.26 [Ran 17]. Two-dimensional magnetic-field-temperature phase diagrams at different pressure and doping-magnetic field phase diagrams at low temperatures are shown in Figure 6.26(a,b), respectively. The three-dimensional doping-magnetic-field-temperature phase diagram is shown in Figure 6.26(c). They indicate a rapid collapse of the SDW phase, a field-induced AF-to-HO phase transition, and a pressure-induced-increase of the different field-scales. These features are very similar to those observed under pressure combined with magnetic field (see Figure 6.25 in Section 6.3.4).

- *U(Ru$_{1-x}$Rh$_x$)$_2$Si$_2$ alloys.* The doping-temperature phase diagram also shows an anti-ferromagnetic phase with long-range ordering with wavevector $\mathbf{k}_0$, induced for doping $x > x_c \simeq 0.015$. However, $T_0$ and $T_N$ both decrease with doping, ending in a para-magnetic ground-state for $x > x_{c,2} \simeq 0.04$. The effects of a magnetic field $\mathbf{H} \parallel \mathbf{c}$ applied on two Rh-doped alloys, of contents $x = 0.04$ and $0.08$, are considered in Figure 6.27. In these two compounds, the ground state is a CPM regime. Their magnetization show two first-order steps, at $\mu_0 H_1 = 26$ T and $\mu_0 H_2 = 37$ T in U(Ru$_{0.96}$Rh$_{0.04}$)$_2$Si$_2$ (Figure 6.27(a)) [Kim 04, Oh 07, Kuwahara 13] and at $\mu_0 H_1 = 22$ T and $\mu_0 H_2 = 38$ T in U(Ru$_{0.92}$Rh$_{0.08}$)$_2$Si$_2$ (Figure 6.27(b)) [Prokeš 17b]. The field-induced phase delimited by $H_1$ and $H_2$ corresponds to a dome preceding the onset of a PPM regime (Figure 6.27(c)) [Kim 04]. Neutron diffraction showed that an antiferromagnetic order associated magnetic moments with wavevector $\mathbf{k}_2 = (2/3, 0, 0)$ is established in magnetic fields $H_1 < H < H_2$ for the two compounds. Knowing that a ferromagnetic component of wavevector $\mathbf{k} = 0$ is induced by the field-polarization of the moments, an 'up-up-down' square magnetic structure was proposed for the field-induced antiferromagnetic phase of U(Ru$_{0.96}$Rh$_{0.04}$)$_2$Si$_2$ and U(Ru$_{0.92}$Rh$_{0.08}$)$_2$Si$_2$ (Figure 6.27(d)). This proposition, based on an antiferromagnetic component with wavevector $\mathbf{k}_2 = (2/3, 0, 0)$ twice more intense than the ferromagnetic component, is consistent with the observation of magnetization jumps $\Delta M/3$ at $H_1$ and $2\Delta M/3$ at $H_2$.

Since it leads to almost-equivalent effects than pressure, Fe-doping constitutes a precious alternative to pressure for the study of the electronic properties of URu$_2$Si$_2$. Fe-doped alloys may be considered for experiments using techniques which are not available, or difficult, under pressure (neutron diffraction in pulsed magnetic fields, inelastic neutron scattering, ARPES experiments, etc.). On the contrary, Rh-doping leads to different effects than pressure. However, the similarity between the magnetic wavevector $\mathbf{k}_2 = (2/3, 0, 0)$ of the AF phase stabilized by a magnetic field in the Rh-doped compounds, and the magnetic wavevector $\mathbf{k}_1 = (0.6, 0, 0)$ of the SDW phase stabilized by a magnetic field in pure URu$_2$Si$_2$, suggests a possible common origin. To reveal it, series of new experiments may be needed: i) high-field diffraction experiments in compounds with lower Rh-content, for which the ground-state is either the HO or the AF phase, ii) inelastic neutron scattering to determine the dispersive features in the magnetic excitation of these Rh-doped compounds, and their possible relationship with the field-induced magnetic phase. Indeed, the wavevector $\mathbf{k}_1 = (0.6, 0, 0)$ of the SDW phase induced by a magnetic field in URu$_2$Si$_2$ [Knafo 16] was found to coincide with that of strong inelastic magnetic fluctuations observed at zero-field [Broholm 91]. In parallel, high-field diffraction experiments and inelastic neutron scattering experiments may be performed on Fe-doped compounds too. Unveiling the relation between the field-induced phase and the excitations of the zero-field ground-states of these alloys may constitute a keystone for solving the hidden-order mystery in URu$_2$Si$_2$.

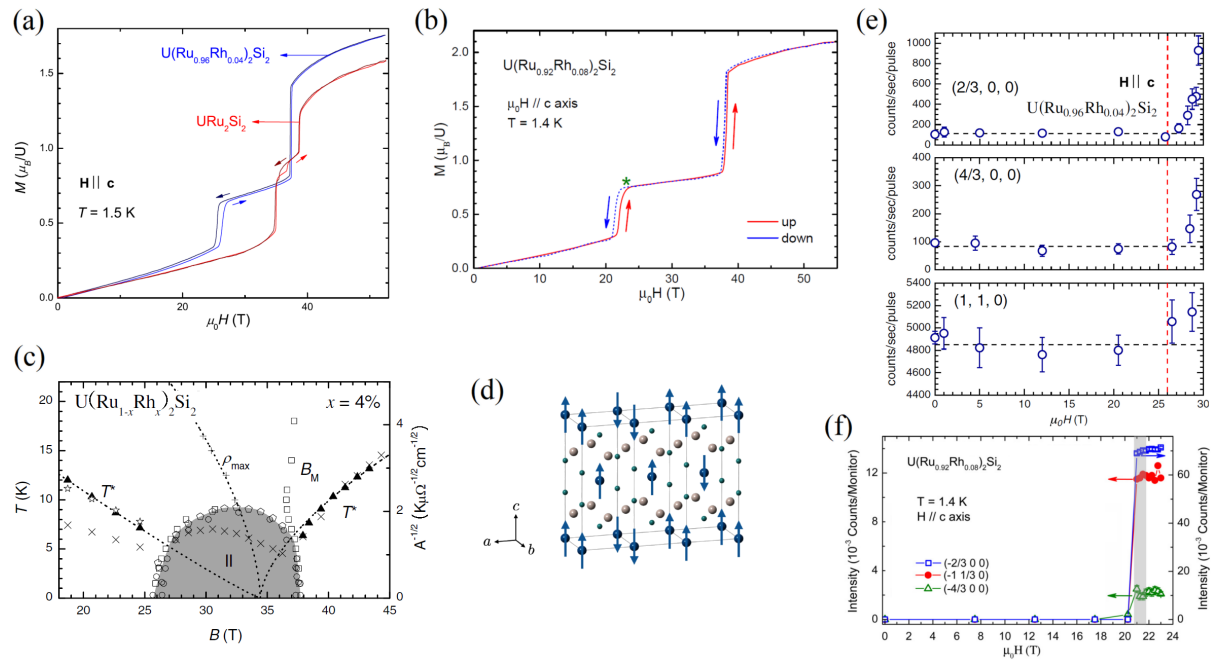

Figure 6.27: Magnetization versus magnetic field of (a) URu$_2$Si$_2$ and U(Ru$_{0.96}$Rh$_{0.04}$)$_2$Si$_2$ at $T = 1.5$ K (from [Kuwahara 13]) and (b) of U(Ru$_{0.92}$Rh$_{0.08}$)$_2$Si$_2$ at $T = 1.4$ K (from [Prokeš 17b]), in a magnetic field $\mathbf{H} \parallel \mathbf{c}$. (c) Magnetic-field-temperature phase diagram of U(Ru$_{0.96}$Rh$_{0.04}$)$_2$Si$_2$ in a magnetic field $\mathbf{H} \parallel \mathbf{c}$ (from [Kim 04]). (d) Scheme of the magnetic structure in the field-induced antiferromagnetic phase and (e) neutron diffracted intensity, at the momentum transfers $\mathbf{Q} = (2/3, 0, 0)$, $(4/3, 0, 0)$, and $(1, 1, 0)$, of U(Ru$_{0.96}$Rh$_{0.04}$)$_2$Si$_2$ in a magnetic field $\mathbf{H} \parallel \mathbf{c}$ (from [Kuwahara 13]). (f) Neutron diffracted intensity at the momentum transfers $\mathbf{Q} = (-2/3, 0, 0)$, $(-1, 1/3, 0)$, and $(4/3, 0, 0)$ of U(Ru$_{0.92}$Rh$_{0.08}$)$_2$Si$_2$ at $T = 1.4$ K in a magnetic field $\mathbf{H} \parallel \mathbf{c}$ (from [Prokeš 17b]).

# Chapter 7

# Magnetism and superconductivity in UTe$_2$

Superconductivity was discovered in 2019 in the paramagnet UTe$_2$ [Ran 19a, Aoki 19b]. In less than two years, a large number of experiments, by different techniques and within different external conditions, were performed. They evidenced multiple superconducting phases induced under pressure and magnetic field, in the vicinity of magnetic phase transitions or crossovers. A magnetically-ordered phase induced under pressure has not been characterized so far, and the question whether UTe$_2$ at ambient pressure is a nearly-ferromagnet or a nearly-antiferromagnet is still open. This Chapter summarizes the first experimental studies performed on the new compound UTe$_2$. Section 7.1 presents its properties at ambient pressure and low magnetic field. Focus is then given to its high-pressure properties, in Section 7.2, and to its high-field properties, in Section 7.3.

## 7.1  Ambient-pressure and low-magnetic-field properties

An introduction to the anisotropic magnetic properties of UTe$_2$, probed by bulk measurements, is made in Section 7.1.1. Magnetic-fluctuations studies are presented in Section 7.1.2. ARPES measurements of the Fermi surface are presented in Section 7.1.3. Finally, the properties of the low-field and ambient-pressure superconducting phase of UTe$_2$ are detailed in Section 7.1.4.

### 7.1.1  Bulk properties

UTe$_2$ is a paramagnet characterized by a strong Ising uniaxial magnetic anisotropy. Its magnetic susceptibility $\chi$, presented in Figure 7.1(a), shows a large low-temperature enhancement for a magnetic field $\mathbf{H}$ applied along the easy magnetic axis $\mathbf{a}$ [Ikeda 06]. A broad maximum is observed at $T_\chi^{max} = 35$ K for $\mathbf{H} \parallel \mathbf{b}$, which becomes the hardest magnetic axis at low temperatures. On the contrary, $\chi$ presents a low-temperature enhancement for $\mathbf{H} \parallel \mathbf{c}$. Figure 7.1(b) shows the NMR Knight-shift measured from the two non-equivalent Te sites, and whose temperature-dependence reproduces well that of the magnetic susceptibility, for the three field-directions $\mathbf{H} \parallel \mathbf{a}, \mathbf{b}, \mathbf{c}$ [Tokunaga 19]. The large value of $\chi$ at low-temperature, together with a scaling plot of $M/T$ versus $H/T^{1.5}$ of the magnetization data, were considered as an indication for the proximity of UTe$_2$ from a quantum ferromagnetic phase transition [Ran 19a]. However, Figure 7.1(c) presents a plot of $1/\chi$ versus $T$, from which negative Curie-Weiss temperatures $\theta_a = -60$ K, $\theta_b = -105$ K, and $\theta_c = -140$ K can be extracted from the high-temperature magnetic susceptibility for $\mathbf{H} \parallel \mathbf{a}$, $\mathbf{b}$, and $\mathbf{c}$, respectively [Knafo 2x]. These negative values are compatible with the presence of antiferromagnetic exchange. A plot of $\chi$ versus $T$ in a log-log scale, shown in the Inset of Figure 7.1(c), emphasizes that, for $\mathbf{H} \parallel \mathbf{a}$ and temperatures $20 \leq T \leq 300$ K, $\chi$ follows an anomalous $1/T^{0.75}$ variation rather than a 'standard' Curie-Weiss variation $1/(T - \theta)$. A downwards deviation from this high-temperature law, observed at $T < T^* \simeq 15$ K, is compatible with the onset of antiferromagnetic fluctuations (see Figure 2.29 in Section 2.4.3.1). A larger downwards deviation, observed for $\mathbf{H} \parallel \mathbf{b}$, leads to a maximum of $\chi$ at the temperature $T_\chi^{max}$. This maximum could also be the signature of antiferromagnetic fluctuations (see Section 3.3). Figure 7.1(d) shows that the electrical resistivity $\rho$, presented for the current directions $\mathbf{I} \parallel \mathbf{a}$, $\mathbf{b}$, is also anisotropic. A large electronic term is observed at temperatures $T < 200$ K, leading to broad maxima in $\rho_{x,x}$ at $T^{max}_{\rho_{x,x}} \simeq 70$ K, for $\mathbf{I} \parallel \mathbf{a}$, and in $\rho_{y,y}$ at $T^{max}_{\rho_{y,y}} \simeq 90$ K, for $\mathbf{I} \parallel \mathbf{b}$ [Ran 19a]. Similarly to other heavy-fermion compounds (see for instance URu$_2$Si$_2$; Figures 6.2(a-c) in Section 6.1.1 and Figure 6.17(d) in Section 6.3.1), the temperatures $T^{max}_{\chi}$, $T^{max}_{\rho_{x,x}}$, and $T^{max}_{\rho_{y,y}}$ can be ascribed to a crossover characteristic of the onset of a correlated paramagnetic regime.

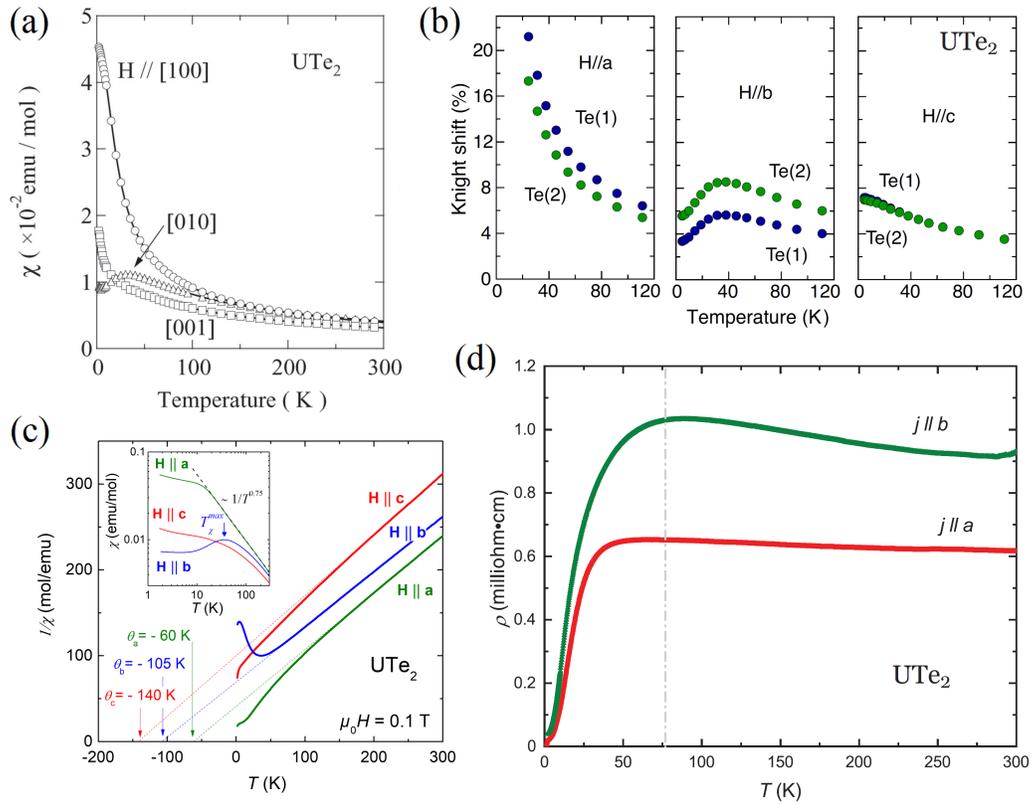

Figure 7.1: (a) Magnetic susceptibility $\chi$ versus temperature (from [Ikeda 06]), (b) NMR Knight-shift $K$ versus temperature (from [Tokunaga 19]), and (c) inverse magnetic susceptibility $1/\chi$ versus temperature (with a plot of $\chi$ versus $T$ in a log-log scale in the Inset) (from [Knafo 2x]) of UTe$_2$ in magnetic fields $\mathbf{H} \parallel \mathbf{a}$, $\mathbf{b}$, and $\mathbf{c}$. (d) Electrical resistivity of UTe$_2$ measured with currents $\mathbf{I} \parallel \mathbf{a}$ and $\mathbf{b}$ (from [Ran 19a]).

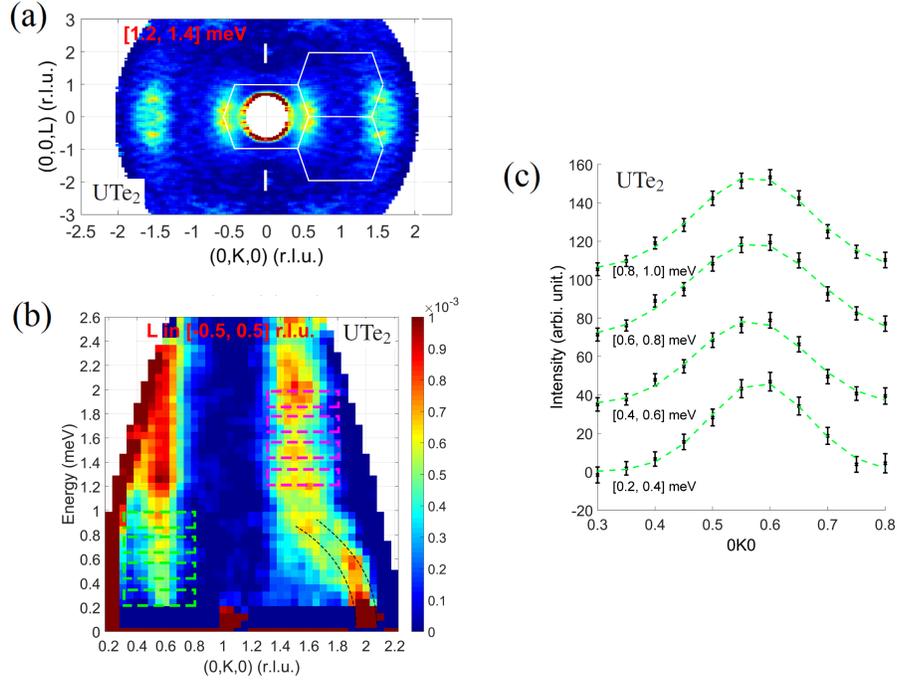

Figure 7.2: Studies of magnetic fluctuations in UTe$_2$ by neutron scattering. (a) Mapping of neutron scattered intensity in the scattering plane of wavevectors $(0, K, L)$, for energies transfers $1.2 \leq E \leq 1.4$ meV, (b) mapping of dynamical susceptibility $\chi''$ in the plane $(K, E)$, for $0.5 \leq L \leq 0.5$, and (c) neutron scattered intensity versus $K$ for different energy windows (from [Duan 2x]).

### 7.1.2 Magnetic fluctuations

A direct measurement by inelastic neutron scattering evidenced the presence of antiferromagnetic fluctuations in UTe$_2$ [Duan 2x]. Figures 7.2(a-b) show mappings of the magnetic-fluctuations intensity at wavevectors $(0, K, L)$, for energy transfers integrated in the window $1.2 \leq E \leq 1.4$ meV, and in the $(K, E)$ plane, for an integration over $-0.5 \leq L \leq 0.5$, respectively. The intensity versus wavevector-component-$K$ plots extracted from these spectra are shown in Figure 7.2(c) for different energy transfers. They indicate an enhancement of the magnetic fluctuations at the incommensurate wavevector $\mathbf{k} = (0, 0.57, 0)$. No signature of ferromagnetic fluctuations were evidenced in this work. A limit of this first INS experiment is the high mosaicity, by $\sim 15°$, of the arrangement of $> 60$ samples aligned together to reach a sufficient sample mass [Duan 2x]. Further neutron scattering experiments on a sample, or an arrangement of samples, with an improved mosaicity, are needed to confirm this first report of antiferromagnetic fluctuations, and to search for additional magnetic-fluctuation signals. Muon spin-relaxation/rotation experiments also evidenced signatures of magnetic fluctuations, whose nature was proposed to be ferromagnetic [Sundar 19].

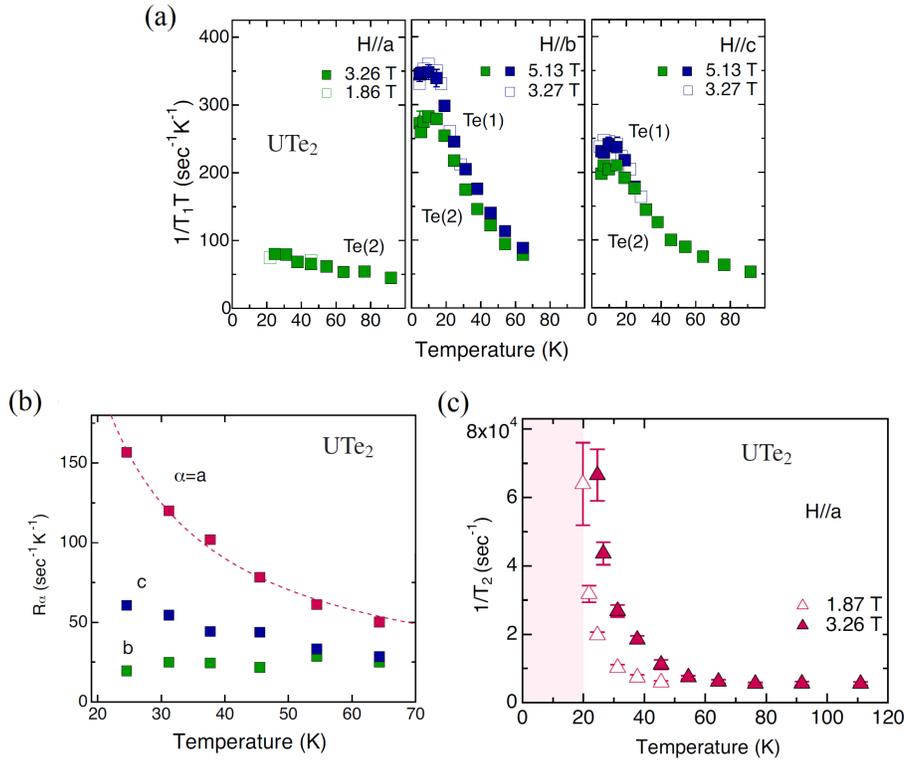

Figure 7.3: Studies of magnetic fluctuations in UTe$_2$ by NMR. Temperature-dependence (a) of the NMR spin-lattice relaxation rate $1/T_1T$ measured for $\mathbf{H} \parallel \mathbf{a}$, $\mathbf{b}$, and $\mathbf{c}$, (b) of the dynamical susceptibility components $R_a$, $R_b$, and $R_c$ deduced from $1/T_1T$, and (c) of the NMR spin-spin relaxation rate $1/T_2$ measured for $\mathbf{H} \parallel \mathbf{a}$ (from [Tokunaga 19]).

Complementarily to neutron and muon experiments, signatures of magnetic fluctuations were observed by NMR [Tokunaga 19]. The temperature-dependence of the spin-lattice relaxation rate $1/T_1T$ is shown in Figure 7.3(a) for magnetic fields $\mathbf{H} \parallel \mathbf{a}$, $\mathbf{b}$, and $\mathbf{c}$, and measured from the two Te-site signals. Knowing that $1/T_1T$ probes the magnetic fluctuations perpendicular to the magnetic-field direction, an anisotropic dynamical magnetic susceptibility, noted here as $R_i$, can be extracted from $(1/T_1T)_a < (1/T_1T)_c < (1/T_1T)_b$ (Figure 7.3(b)). At low temperature, the hierarchy $R_a > R_b$, $R_c$ is the signature of enhanced magnetic fluctuations parallel to the direction $\mathbf{a}$. This picture is confirmed by the enhancement of the spin-spin relaxation rate $1/T_2$, which probes magnetic fluctuations parallel to the magnetic-field direction, observed at low temperature for $\mathbf{H} \parallel \mathbf{a}$ (Figure 7.3(c)).

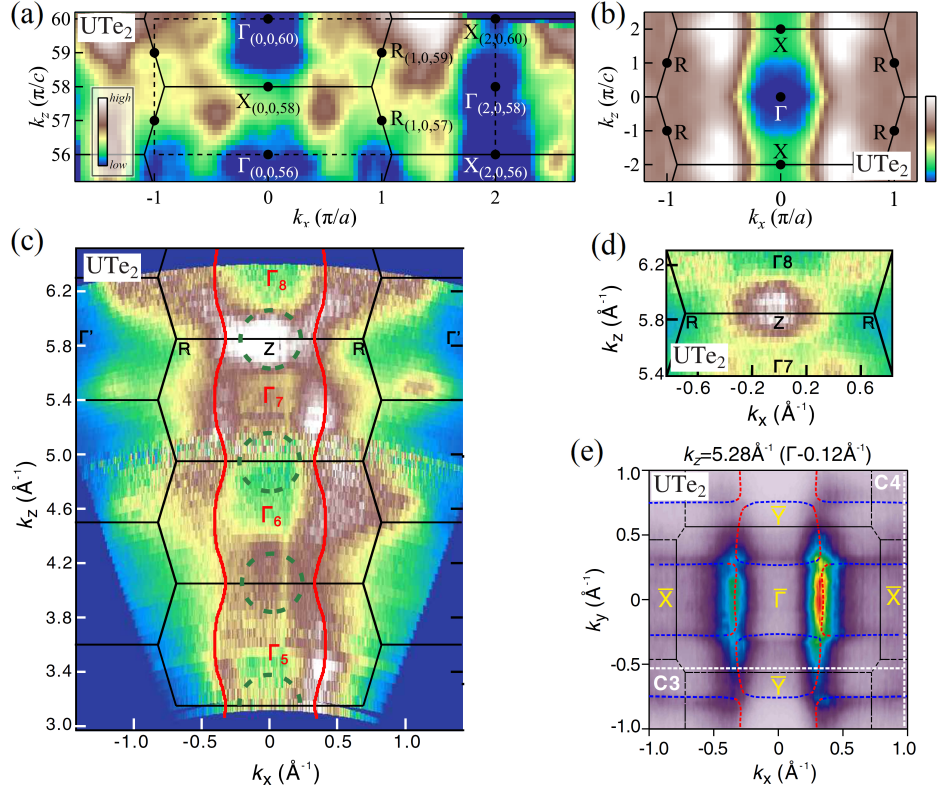

Figure 7.4: Mapping of the Fermi surface of UTe₂ by ARPES measurements, (a) in the $(k_x, k_z)$ plane, with (b) a zoom on the Brillouin zone, (from [Fujimori 19]), (c) in the $(k_x, k_z)$ plane, with (d) a zoom around the point $Z$ (labeled as X in (a,b)), and (e) in a $(k_x, k_y)$ plane (from [Miao 20]).

### 7.1.3 Fermi surface

The electronic structure and the Fermi surface of UTe₂ have been investigated by angle-resolved-photoemission spectroscopy [Fujimori 19, Miao 20]. Figures 7.4(a-b) present mappings of the Fermi surface of UTe₂ obtained by ARPES measurements in the $(k_x, k_z)$ plane of the reciprocal space [Fujimori 19]. In this work, a narrow nature of the bands in the vicinity of the Fermi energy and the itinerant character of $5f$ electrons were emphasized. Figures 7.4(c-e) present mappings of the Fermi surface in the $(k_x, k_z)$ and $(k_x, k_y)$ planes extracted from a second study by ARPES [Miao 20]. In this second work, Fermi-surface contributions from quasi-one-dimensional bands were identified (see for instance Figure 7.4(e)). Complementarily to ARPES experiments, SdH or dHvA investigations of quantum oscillations in the electrical resistivity or magnetization, respectively, are now necessarily for a finer characterization of the Fermi surface of UTe₂. Contrary to ARPES, quantum-oscillations techniques could also permit characterizing the evolution, and possible reconstructions, of the Fermi-surface of UTe₂ under pressure and magnetic field, in relation with the transitions to magnetically-ordered and superconducting phases.

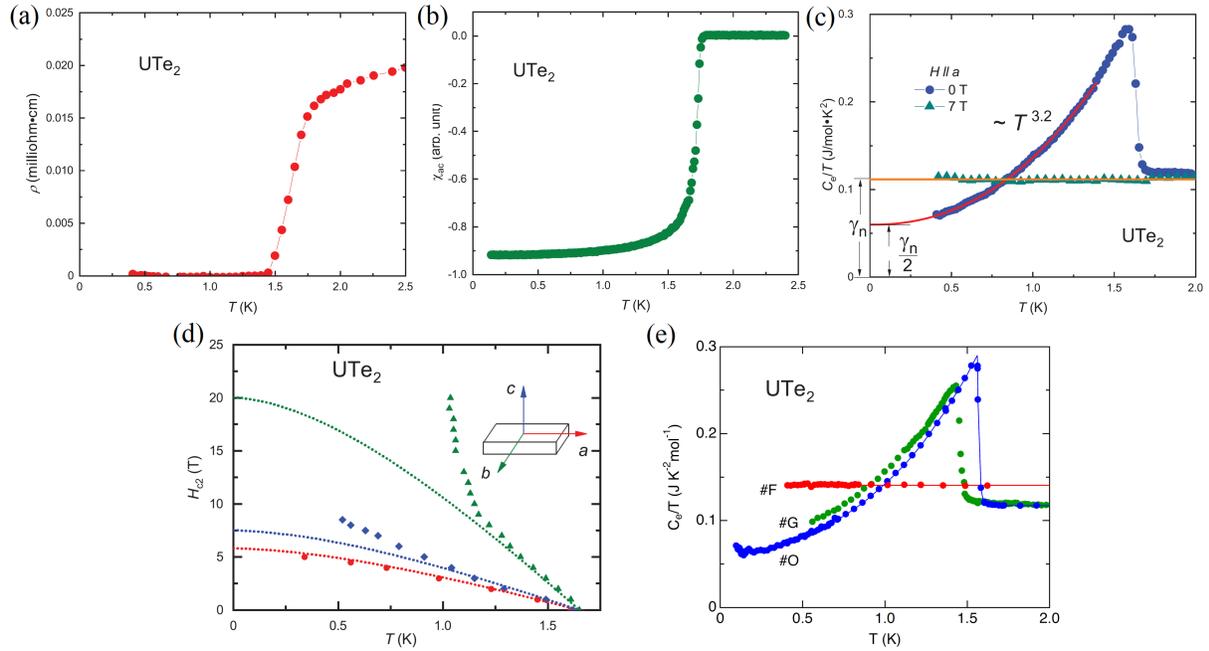

Figure 7.5: Signatures of superconductivity in UTe$_2$. Temperature dependence of (a) the electrical resistivity, (b) the ac susceptibility (field direction not specified), (c) the electronic heat capacity divided by temperature in a magnetic field $\mu_0 \mathbf{H} \parallel \mathbf{a}$ of 0 and 7 T, (d) and the superconducting field $H_{c,2}$ for $\mathbf{H} \parallel \mathbf{a}$, $\mathbf{b}$, and $\mathbf{c}$ (from [Ran 19a]). (e) Temperature dependence of the electronic heat capacity divided by temperature of three samples of different qualities (from [Aoki 19b]).

### 7.1.4 Superconductivity

Superconductivity was observed in UTe$_2$ at temperatures below a critical temperature $T_{sc}$ varying from 1.5 to 1.8 K, depending on the sample quality [Ran 19a, Aoki 19b, Cairns 20]. Figures 7.5(a-b) show the first evidences of superconductivity obtained by Ran *et al* from electrical-resistivity and ac-magnetic-susceptibility measurements [Ran 19a]. Figure 7.5(c) compares the temperature-dependence of the heat capacity divided by temperature $C_p/T$ measured at zero field and in a magnetic field $\mu_0 \mathbf{H} \parallel \mathbf{a}$ of 7 T [Ran 19a]. Under this magnetic field, superconductivity is absent and a heavy-fermion non-superconducting state characterized by a Sommerfeld coefficient $\gamma =_{T \to 0} C_p/T = 120$ mJ/molK$^2$ is revealed. The saturation of $C_p/T$ at temperatures $T > T_{sc}$ indicates that magnetic fluctuations driving a Fermi-liquid behavior are established prior to superconductivity. Figure 7.5(e) presents a plot of $C_p/T$ versus $T$ of UTe$_2$ samples of different qualities [Aoki 19b]. Sample $\sharp$F remains non-superconducting down to the lowest temperatures and presents a constant $C_p/T$ characteristic of a heavy Fermi liquid similar to that observed in a magnetic field $\mu_0 \mathbf{H} \parallel \mathbf{a}$ of 7 T (Figure 7.5(c)) [Ran 19a]. Samples $\sharp$G and

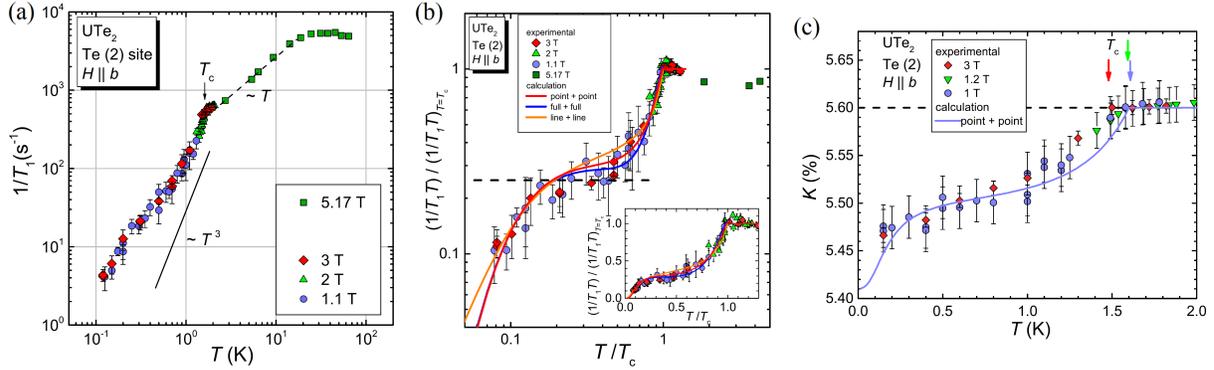

Figure 7.6: Temperature dependence of the NMR spin-lattice relaxation rate, plotted (a) as $1/T_1$ versus $T$, (b) as $1/T_1 T$ vs $T/T_{sc}$, and (c) of the NMR Knight-shift $K$ of UTe$_2$ in a magnetic field **H** ∥ **b**, at the onset of superconductivity (from [Nakamine 19]).

♯O are superconducting below the temperatures $T_{sc} = 1.45$ and $1.57$ K, respectively. A low-temperature extrapolation towards a smaller value of $C_p/T$ is observed for the sample of higher $T_{sc}$. Figure 7.5(d) shows that the critical superconducting field $H_{c,2}$ is strongly anisotropic [Ran 19a], with the hierarchy $H_{c,2}^a < H_{c,2}^c < H_{c,2}^b$ of the critical fields measured for **H** ∥ **a**, **c**, **b**, respectively. For these three directions of the magnetic field, $H_{c,2}$ exceeds the Pauli limit expected for a spin-singlet superconductor assuming a weak coupling limit and an isotropic factor $g = 2$, and a spin-triplet nature of superconductivity has been proposed [Ran 19a]. The proximity to a quantum ferromagnetic phase transition was also initially suspected (see Section 7.1.2), in agreement with a spin-triplet superconductivity possibly driven by ferromagnetic fluctuations [Ran 19a].

Figure 7.6 presents signatures of the superconducting phase observed from NMR experiments in a magnetic field **H** ∥ **b** [Nakamine 19]. Figure 7.6(a) shows that a Korringa law $1/T_1 \propto T$ characteristic of a Fermi-liquid regime is observed at temperatures $T < 20$ K, in agreement with the signatures of a Fermi liquid observed from heat-capacity measurements (Figure 7.5(c,e)) [Ran 19a, Aoki 19b]. The onset of superconductivity at $T_{sc}$ induces a downwards deviation from the Korringa law. A two-step-like increase of $1/T_1 T$ in the superconducting phase suggests the presence of at least two superconducting gaps (see Figure 7.6(b)). Fits to the data by two-gap models are shown in Figure 7.6(a) for different kinds of gaps (point-node, line-node, or full gaps). Figure 7.6(b) further shows that a decrease of the NMR Knight-shift $K$ by $\sim 0.1\%$ is observed for **H** ∥ **b**. This decrease was interpreted as an experimental evidence for spin-triplet superconductivity, and a fit to the data by a model with two node-gaps was proposed [Nakamine 19]. A spin-triplet superconducting order parameter with point-nodes was also proposed following studies by electrical and thermal transport, magnetic penetration depth, and heat-capacity measurements [Metz 19, Kittaka 20]. Following scanning-tunneling-spectroscopy experiments, UTe$_2$ was proposed to be a chiral superconductor [Jiao 20].

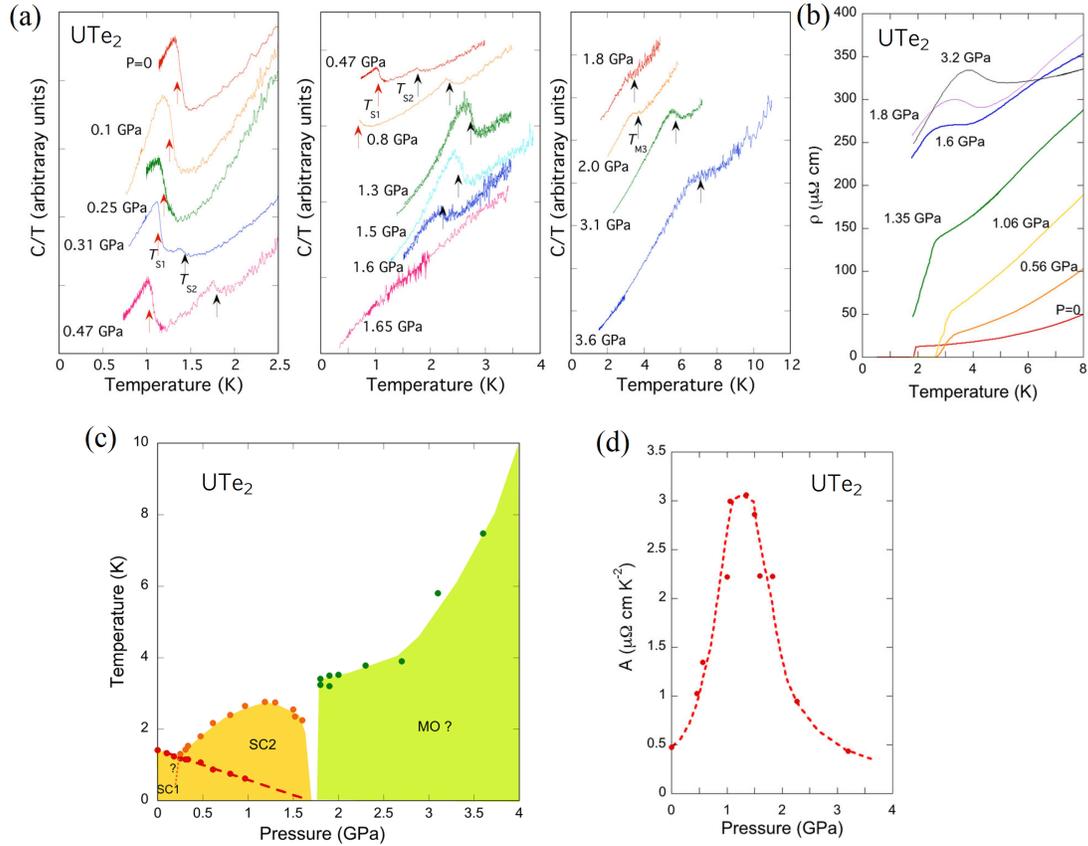

Figure 7.7: Temperature-dependence (a) of the heat capacity divided by temperature and (b) of the electrical resistivity of UTe$_2$ at different pressures. (c) Pressure-temperature phase diagram of UTe$_2$ and (d) pressure-dependence of the Fermi-liquid quadratic coefficient $A$ of the electrical resistivity (from [Braithwaite 19]).

## 7.2 Pressure-induced quantum magnetic phase transition

Pressure permits to tune the electronic properties of UTe$_2$: it enhances the temperature at the onset of superconductivity, which reaches a maximum at a critical pressure $p_c \simeq 1.75$ GPa. $p_c$ corresponds to a quantum magnetic phase transition, beyond which magnetic order is established and superconductivity disappears. The properties of UTe$_2$ under pressure are summarized in Figure 7.7 [Braithwaite 19] and described below:

- The combination of heat-capacity (Figure 7.7 (a)) and electrical-resistivity (Figure 7.7 (b)) measurements evidences the stabilization of a second superconducting phase, labeled SC2, under pressure $p$, in addition to the superconducting phase reported at ambient pressure, labeled SC1. The pressure-temperature phase diagram of UTe$_2$ is presented in Figure 7.7 (c). It shows that the superconducting temperature $T_{sc1}$ of phase SC1 decreases linearly with $p$ and extrapolates to 0 K for $p \rightarrow 1.5$ GPa. On the contrary, the superconducting temperature $T_{sc2}$ of phase SC2 increases with $p$, reaches a maxi-mum value of $\sim 2.8$ K at $p = 1.3$ GPa, before vanishing under pressures $p > p_c \sim 1.7$ GPa. At ambient pressure, a single superconducting transition was observed in several studies, where samples of different qualities were investigated [Ran 19a, Aoki 19b, Braithwaite 19, Cairns 20]. However, the observation of two nearby superconducting transitions at ambient pressure led to the proposition for a third superconducting phase in the pressure-temperature phase diagram [Thomas 20]. Future investigations are needed to determine whether the double transition observed by Thomas *et al* [Thomas 20] is an intrinsic ambient-pressure property or if it is induced by sample inhomogeneity or internal stress. The presence of a third superconducting phase at zero magnetic field has also been proposed in [Aoki 20] (see Figure 7.15(a)).

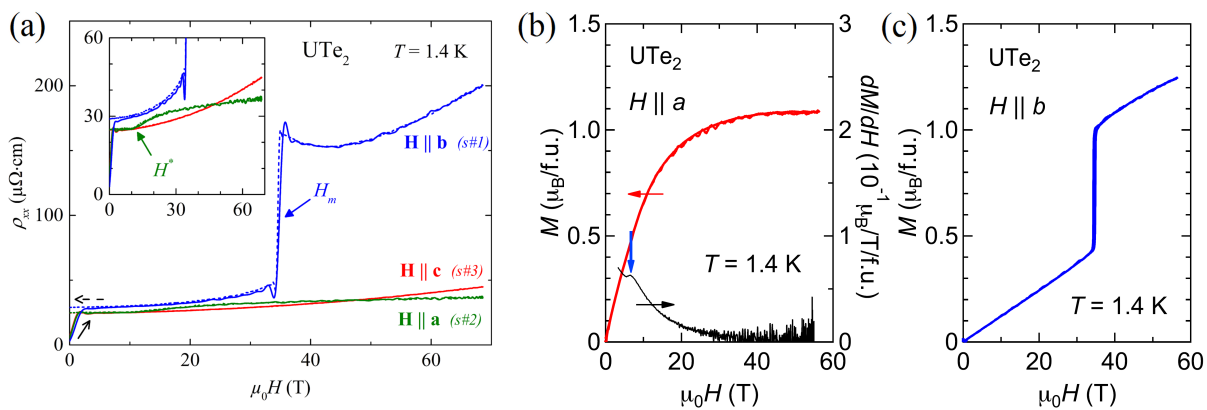

Figure 7.8: (a) Electrical resistivity versus magnetic field **H** ∥ **a**, **b**, and **c** (from [Knafo 19a]) and magnetization versus magnetic field (b) **H** ∥ **a** and (c) **H** ∥ **b** (from [Miyake 19]) of UTe$_2$ at $T = 1.4$ K.

- Signatures of a non-superconducting ordered phase were evidenced from heat-capacity and electrical-resistivity measurements on UTe$_2$ under pressures $p > p_c$ (Figures 7.7 (a-b)) [Braithwaite 19]. The high-pressure phase, presumably of magnetic nature, is delimited by a temperature, noted here $T_{MO}$, which is $> 3$ K and increases under pressure (Figure 7.7 (c)). Two high-pressure phases, instead of one as observed in [Braithwaite 19], were also reported in [Thomas 20]. Figure 7.7 (d) presents the pressure-variation of the quadratic coefficient $A$ extracted from a Fermi-liquid fit to the electrical resistivity. $A$ marks a sharp maximum at $p_c$, suggesting that quantum critical magnetic fluctuations accompany the onset of long-range magnetic order. Neutron diffraction experiments under pressure are mandatory to determine whether the pressure-induced phase(s) is(are) ferromagnetic, as proposed in [Ran 20, Lin 20], or antiferromagnetic, as proposed in [Aoki 20, Thomas 20].

## 7.3 Metamagnetism and superconductivity induced under a magnetic field

Multiple phenomena, from first-order metamagnetism to superconductivity, are induced in UTe$_2$ under a magnetic field. A review of these phenomena is given here. Section 7.3.1 presents bulk signatures of field-induced moment polarization processes. Section 7.3.2 focuses on the superconducting phases induced by a magnetic field in the vicinity of a metamagnetic transition. Section 7.3.3 presents the first experimental evidences for field-induced quantum critical magnetic fluctuations. Section 7.3.2 presents indications for Fermi-surface reconstructions induced in high magnetic fields. Finally, combined effects of magnetic fields and pressures are considered in Section 7.3.5.

### 7.3.1 Bulk properties and phase diagram

UTe$_2$ shows anisotropic properties in a magnetic field. Its electrical resistivity $\rho$ [Knafo 19a] and magnetization $M$ [Miyake 19] measured in magnetic fields $\mathbf{H} \parallel \mathbf{a}$, $\mathbf{b}$, and $\mathbf{c}$ and at a temperature $T = 1.4$ K, are presented in Figure 7.8. Different features are observed:

- For $\mathbf{H} \parallel \mathbf{b}$, large and sharp jumps of $\rho$ and $M$ indicate a first-order metamagnetic transition at $\mu_0 H_m = 35$ T induced (Figures 7.8(a,c)). A polarized paramagnetic regime is stabilized for $H > H_m$, where the magnetization reaches large values $M > 1$ $\mu_B$/U. The increase of $M$ in the PPM regime indicates the progressive quench of remaining magnetic fluctuations.

- For $\mathbf{H}$ applied along the easy magnetic axis $\mathbf{a}$, $M(H)$ increases with a large slope, which is progressively reduced for $\mu_0 H > 20$ T where the saturation at a value $M \simeq 1.1$ $\mu_B$/U is approached. A small variation of $\rho$ is reported in fields up to 70 T. A kink in $\rho$ at $\mu_0 H^* \simeq 10$ T indicates a possible phase transition or crossover [Knafo 19a]. A small anomaly in the magnetization, leading to a maximum in $\partial M / \partial H$, is also observed at $\simeq$ 6.5 T [Miyake 19], and anomalies at similar field values were observed in thermoelectric-power and Hall-effect measurements (see Figures 7.14(a-b) in Section 7.3.4) [Niu 20a]. These anomalies may result from Fermi-surface reconstructions.

- No transition nor crossover are observed in a magnetic field $\mathbf{H} \parallel \mathbf{c}$. The electrical resistivity, measured with a current $\mathbf{I} \perp \mathbf{H}$, follows a $H^2$ increase characteristic of an orbital contribution, which is driven by the field-induced cyclotron motion of the conduction electrons [Knafo 19a, Knafo 19c].

Figure 7.9 focuses on the properties of UTe$_2$ in a magnetic field $\mathbf{H} \parallel \mathbf{b}$. $\rho(H)$ [Knafo 19a] and $M(H)$ [Miyake 19] curves measured at different temperatures are presented in Figures 7.9(a-b). They show that the sharp first-order metamagnetic transition observed at $H_m$ and low temperature is replaced by a broad crossover at higher temperatures. Figure 7.9(c) shows the magnetic-field-temperature phase diagram constructed from the electrical-resistivity data. The metamagnetic transition is characterized by an hysteresis at low temperature. The step-like variation in $\rho$ and the hysteresis at $H_m$ are observed at temperatures $T < T_{CEP} = 7$ K. $T_{CEP}$ is the temperature of a critical endpoint, at which the first-order transition at $H_m$ terminates. For $T > T_{CEP}$, a crossover is characterized by a maximum in $\rho(H)$ at a magnetic field also noted by $H_m$. $H_m$ decreases with increasing temperature and vanishes at temperatures $T > 30$ K. This temperature scale almost coincides with $T_\chi^{max} = 35$ K, at which the magnetic susceptibility measured for **H ∥ b** is maximum (see Figure 7.1(a) in Section 7.1.1) [Ikeda 06]. $T_\chi^{max}$ and $H_m$ delimitate the same correlated paramagnetic regime, and the ratio $R_{CPM} = T_\chi^{max}/H_m = 1$ K/T in UTe₂ is similar to that observed in most of heavy-fermion paramagnets (See Section 3.3), indicating a conventional behavior. A peculiarity of UTe₂ is that its CPM regime is observed for a magnetic field applied along a hard magnetic axis, **b**, while in other paramagnets the CPM regime is generally observed for a magnetic field applied along the easy magnetic axis.

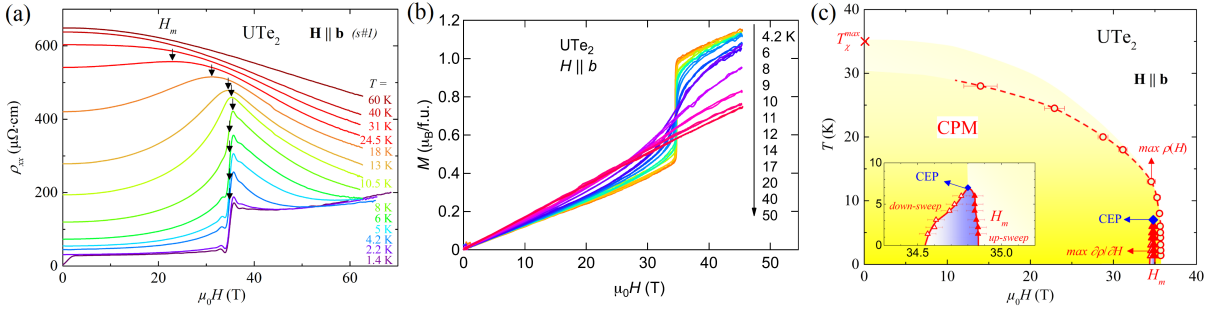

Figure 7.9: (a) Electrical resistivity versus magnetic field (from [Knafo 19a]) and (b) magnetization versus magnetic field of UTe₂ in a magnetic field **H ∥ b**, at different temperatures (from [Miyake 19]). (c) Magnetic-field-temperature phase diagram of UTe₂ in a magnetic field **H ∥ b** (from [Knafo 19a]).

### 7.3.2 Superconductivity

Figure 7.10 presents signatures of field-induced superconductivity observed in UTe₂ for **H ∥ b**. $\rho(H)$ curves measured at different temperatures are presented in Figure 7.10(a). A sequence zero-resistivity → non-zero-resistivity → zero-resistivity → non-zero-resistivity observed at temperatures 300 mK < $T$ < 800 mK indicates that superconductivity is induced by a magnetic field [Knebel 19] (see also [Ran 19b]). The magnetic-field-temperature phase diagram presented in Figure 7.10(b) shows that field-induced superconductivity occurs at temperatures below a critical temperature $T_{sc}$, which is maximum ($T_{sc}^{max} \simeq 1$ K) near $H_m$ [Knafo 2x]. Similarly to the URhGe case (see Section 5.3.2) [Lévy 05], field-induced superconductivity in UTe₂ is presumably driven by the critical magnetic fluctuations of a metamagnetic transition. A difference between these two systems is that URhGe is ferromagnetic while UTe₂ is paramagnetic, indicating that field-induced superconductivity can be established in compounds of different ground states.

The low-field superconducting phase of UTe₂ is labeled SC1. Its critical temperature $T_{sc}$ decreases with increasing field, and it extrapolates to a critical field $\mu_0 H_{c,2} \simeq 15$ T, masked by the appearance of the field-induced superconducting phase, labeled SC2. Electrical-resistivity measurements, as those presented here, cannot probe a transition between two superconducting phases and the presence of a phase transition between SC1 and SC2 at ambient pressure has not been experimentally directly proved so far. However, by continuity, the phase boundaries determined by heat-capacity under pressure combined with magnetic field (see Figure 7.16 in Section 7.3.5) support that a phase transition may separate SC1 and SC2 at ambient pressure. A striking feature of the phase diagram presented in Figure 7.10(b) is the sudden disappearance of the superconducting phase SC2 in the PPM regime stabilized for $H > H_m$. The upper field boundary of SC2 is almost temperature-independent and coincides with the prolongation of $H_m$ at low temperature. This indicates that the superconducting order parameter of phase SC2 is robust only in the low-internal-field CPM regime. On the contrary, the high internal magnetic field in the PPM regime constitutes an unfavorable condition leading to the disappearance of this superconducting phase.

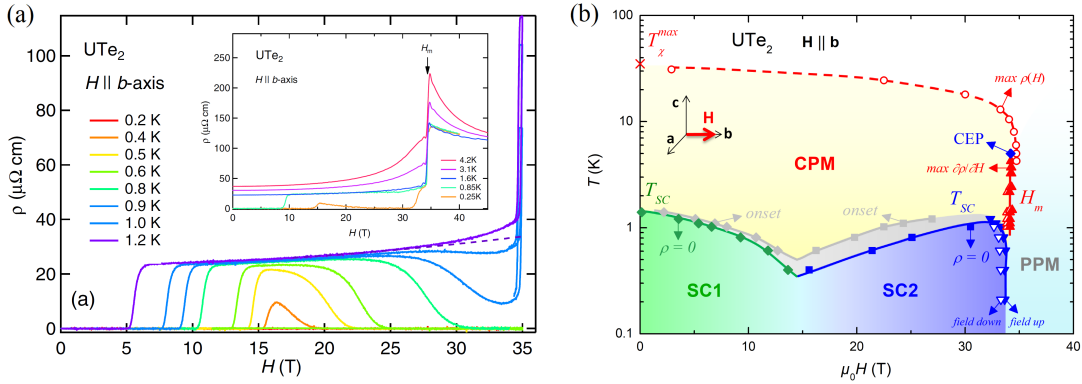

Figure 7.10: (a) Field-dependence of the electrical resistivity $\rho$ at different temperatures (from [Knebel 19]) and (b) magnetic-field-temperature phase diagram (from [Knafo 2x]) of UTe$_2$ in a magnetic field **H** ∥ **b**.

As well as its magnetic properties (see Sections 7.1.1 and 7.3.1), the superconducting properties of UTe$_2$ are anisotropic. Figure 7.11 presents signatures of the superconducting phases, extracted from low-temperature electrical-resistivity and TDO measurements, in a magnetic field applied along various directions. Figures 7.11(a-b) show that a tilt by a few degrees of **H**, from **b** towards **a** or **c**, leads to a strong reduction of the upper superconducting field $H_{c,2}$ [Knebel 19]. The plot of $H_{c,2}$ versus temperature shown in Figure 7.11(e) indicates that the field-induced superconducting phase SC2 vanishes in fields tilted by more than $\phi = 8°$ from **b** towards **a** [Knebel 19]. Similarly, SC2 vanishes in fields tilted by more than $\theta \simeq 15°$ from **b** towards **c**. Figure 7.11(c-d) shows that a third superconducting phase, labeled here as SC-PPM, is induced by a magnetic field of more than 40 T tilted from **b** towards **c**, with angles $25 < \theta < 40°$ [Ran 19b].

The low-temperature angle-magnetic-field phase diagram of UTe$_2$ in magnetic fields applied along variable directions from **b** to **a** and from **b** to **c** is presented in Figure 7.11(f) [Knebel 19, Ran 19b, Knafo 2x]. The low-field superconducting phase SC1 is characterized by critical fields $\mu_0 H_{c,2}^a = 6$ T, $\mu_0 H_{c,2}^b \simeq 15$ T, and $\mu_0 H_{c,2}^c = 10$ T measured for $\mathbf{H} \parallel \mathbf{a}$, $\mathbf{b}$, and $\mathbf{c}$, respectively. The metamagnetic field $H_m$ increases faster in fields tilted from $\mathbf{b}$ towards $\mathbf{a}$ than in fields tilted from $\mathbf{b}$ towards $\mathbf{c}$ [Ran 19b]. While the phase SC2 only develops in the CPM regime, in fields applied along directions close to $\mathbf{b}$, the phase SC-PPM only develops in the PPM regime, in fields applied near a direction tilted by $\theta \simeq 30°$ from $\mathbf{b}$ towards $\mathbf{c}$.

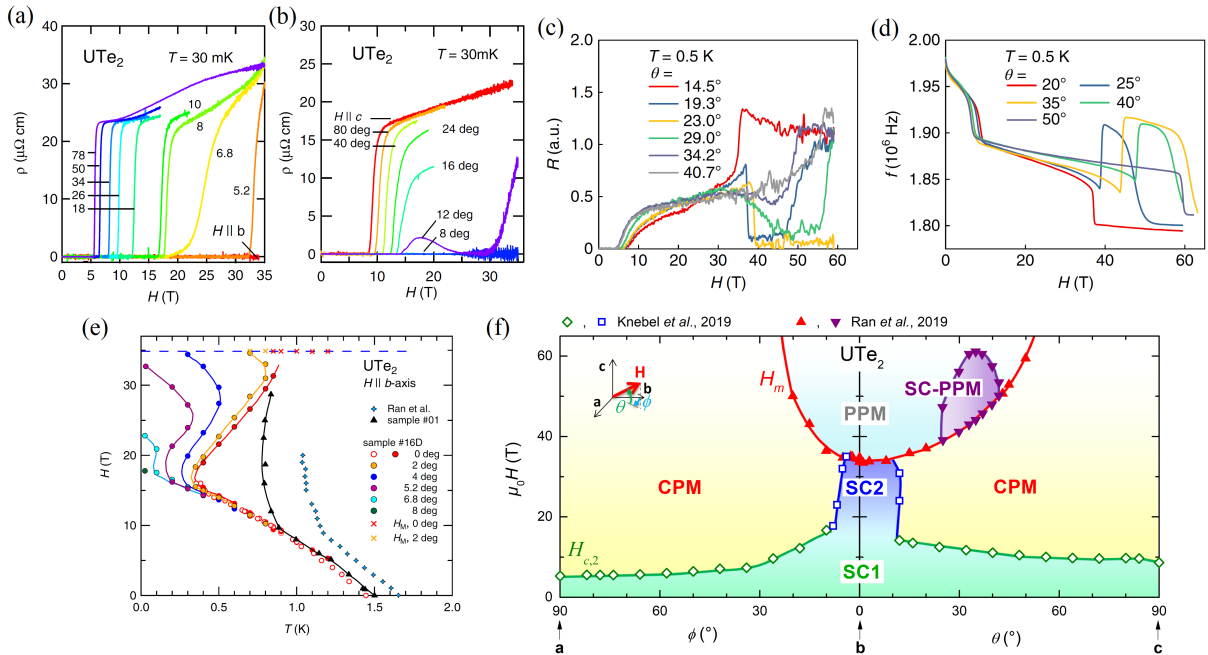

Figure 7.11: Field-dependence of the electrical resistivity (a) for different field-angles $\phi$ (from [Knebel 19]) and (b-c) $\theta$ (from [Knebel 19, Ran 19b]), and (d) of the TDO frequency for different field-angles $\theta$ (from [Ran 19b]). (e) Temperature-magnetic-field phase diagram of UTe$_2$ for different magnetic field directions, from $\mathbf{H} \parallel \mathbf{b}$ to $\mathbf{H}$ tilted towards $\mathbf{a}$ (from [Knebel 19]). (f) Low-temperature angle-magnetic-field phase diagram of UTe$_2$, in magnetic fields applied along variable directions from $\mathbf{b}$ to $\mathbf{a}$ (angle $\phi$) and from $\mathbf{b}$ to $\mathbf{c}$ (angle $\theta$) (from [Knafo 2x], includes data from [Knebel 19, Ran 19b]).

A focus on the superconducting phase SC-PPM induced in a magnetic field tilted by $\theta \simeq 27°$ from $\mathbf{b}$ to $\mathbf{c}$ is made in Figure 7.12. $\rho(H)$ curves are presented for different temperatures in Figure 7.12(a) and the magnetic-field-temperature phase diagram is presented in Figure 7.12(b). A decoupling between the low-field superconducting phase SC1, whose upper critical field reaches 10 T, and the high-field superconducting phase SC-PPM, which appears in fields higher than 45 T, is observed. The phase SC-PPM develops only in the PPM regime established in magnetic fields $\mu_0 H > \mu_0 H_m = 45$ T. Strikingly, the critical field at the set-up of a zero-resistivity state is almost temperature-independent and corresponds to the low-temperature extrapolation of $H_m$. The order parameter of the superconducting phase SC-PPM is robust only in the PPM regime, i.e., in a regime of high internal magnetic field. On the contrary, the low-internal-field CPM regime does not offer favorable conditions for the appearance of the phase SC-PPM.

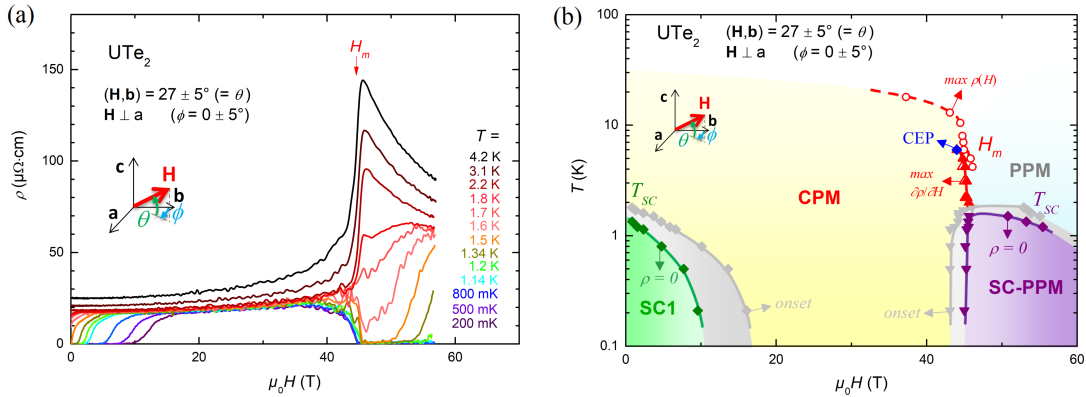

Figure 7.12: (a) Field-dependence of the electrical resistivity measured at different temperature and (b) magnetic-field-temperature phase diagram of UTe$_2$ in a magnetic field tilted from **b** to **c** by an angle $\theta \simeq 27°$ (from [Knafo 2x]).

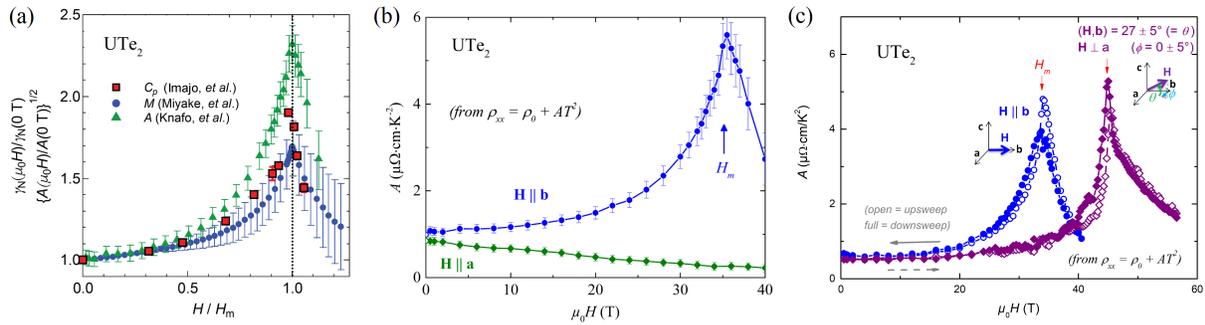

Figure 7.13: Magnetic-field-dependence (a) of the Fermi-liquid Sommerfeld coefficient $\gamma$ of UTe$_2$ in a magnetic field **H** ∥ **b**, estimated from heat-capacity, electrical-resistivity and magnetization experiments (from [Imajo 19], includes data from [Knafo 19a, Miyake 19]), and of the Fermi-liquid quadratic coefficient $A$ of the electrical resistivity of UTe$_2$ (b) in magnetic fields **H** ∥ **a** and **b** (from [Knafo 19a]) and (c) in magnetic fields **H** ∥ **b** and (from [Knafo 2x]).

### 7.3.3 Magnetic fluctuations

A Fermi-liquid description of UTe$_2$ permits to characterize its effective mass and, thus, the strength of its low-temperature magnetic fluctuations. The evolution of its Fermi-liquid properties in a magnetic field and their relation with field-induced superconductivity are considered here.

- Figure 7.13(a) presents a comparison of the Fermi-liquid Sommerfeld coefficient $\gamma$, directly estimated from heat-capacity measurements [Imajo 19], or indirectly estimated from electrical-resistivity (via the quadratic coefficient $A \sim \gamma^2$) [Knafo 19a] and magnetization experiments (via the relation $(\partial^2 M/\partial T^2)_H = (\partial \gamma/\partial \mu_0 H)_T$) [Miyake 19], of UTe$_2$ in a magnetic field **H** ∥ **b**. A maximum of $\gamma$, indicating the enhancement of crit-ical magnetic fluctuations, is observed at $H_m$. By analogy with the well-documented CeRu$_2$Si$_2$ case (see Figure 3.12(d) in Section 3.2.2) [Sato 01, Flouquet 04], critical ferromagnetic fluctuations may be responsible for the enhancement of $\gamma$ at $H_m$ in UTe$_2$. As in URhGe (see Section 5.3.2) [Lévy 05], the critical magnetic fluctuations at $H_m$ presumably drive the mechanism for field-induced superconductivity in UTe$_2$.

- Figure 7.13(b) shows that the Fermi-liquid electrical-resistivity coefficient $A$ decreases continuously in a magnetic field $\mathbf{H} \parallel \mathbf{a}$, which contrasts with the increase followed by a maximum observed for $\mathbf{H} \parallel \mathbf{b}$ [Knafo 19a]. This indicates that the anomalies observed at different field values in the magnetization, electrical resistivity and thermoelectric power for $\mathbf{H} \parallel \mathbf{a}$ (see Figure 7.9(a-b) in Section 7.3.1 and Figures 7.14(a-b) in Section 7.3.4) [Knafo 19a, Miyake 19, Niu 20a] are not associated with enhanced critical magnetic fluctuations.

- Figure 7.13(c) shows that similar enhancements of $A$ are observed at the metamagnetic fields $\mu_0 H_m = 35$ and 45 T, for $\mathbf{H} \parallel \mathbf{b}$ and $\mathbf{H}$ tilted by an angle $\theta \simeq 27°$ from $\mathbf{b}$ to $\mathbf{c}$, respectively [Knafo 2x]. Within an unconventional scenario of superconductivity, where magnetic fluctuations induce the superconducting pairing, one could expect that significant changes of the magnetic fluctuations strength drive the abrupt disappearance/appearance of superconductivity at $H_m$ for these two field-directions. However our results raise a serious hurdle to such picture since the field-driven enhancement of $A$ is very similar for $\mathbf{H} \parallel \mathbf{b}$ and $\mathbf{H}$ tilted by $\theta \simeq 27°$ from $\mathbf{b}$ to $\mathbf{c}$. Extra ingredients are, thus, needed to describe the field and angle domains of stability of these two field-induced superconducting phases.

### 7.3.4   Fermi surface

Figure 7.14 presents the field-dependence of electrical- and thermal-transport quantities measured on UTe$_2$ in magnetic fields $\mathbf{H} \parallel \mathbf{a}, \mathbf{b}$.

- Figure 7.14(a) shows a series of anomalies observed in the thermoelectric power $S$ for $\mathbf{H} \parallel \mathbf{a}$ [Niu 20a]. At $T = 600$ mK, a kink at $\mu_0 H = 5.5$ T is related with the destabilization of the superconducting phase and is followed by a succession of minima at $\mu_0 H_1 = 5.6$ T, $\mu_0 H_2 = 10.5$ T, and $\mu_0 H_3 = 21$ T. While the first anomaly disappears at temperatures $T \gtrsim T_{sc} = 1.6$ K, the anomalies at $H_1$, $H_2$, and $H_3$ broaden when the temperature is raised and their trace is lost at temperatures $T > 5$ K. Figure 7.14(b) further shows that kinks in the thermal conductivity $\kappa$ and in the electrical resistivity $\rho_{x,x}$, and a maximum in the Hall resistivity $\rho_{x,y}$ are observed at $H_1$. A maximum in the field-derivative of the magnetization $\partial M / \partial H$ is also observed at a field of 6.5 T $\simeq \mu_0 H_1$ (see Figure 7.9 in Section7.3.1) [Miyake 19]. A minimum in $\rho_{x,x}$ (initially identified as a kink in [Knafo 19a]) and a change of sign in $\rho_{x,y}$ are observed at $H_2$. Although their origin has not been clarified, these anomalies may be related to field-induced Fermi-surface reconstructions [Niu 20a].

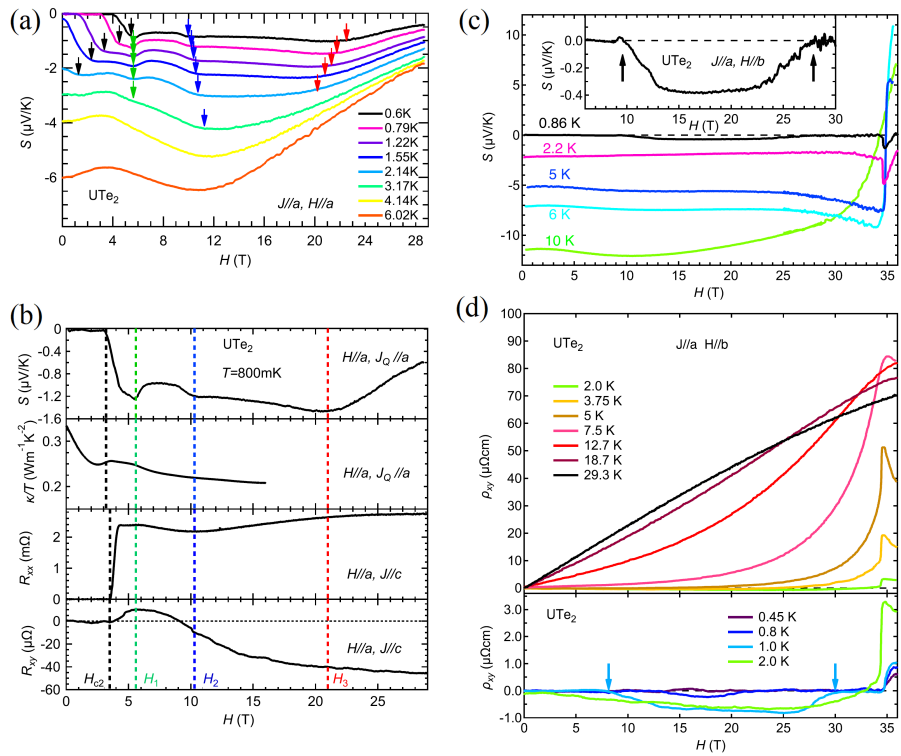

Figure 7.14: Field-dependence (a) of the thermoelectric power $S$ at different temperatures and (b) of the thermoelectric power $S$, the thermal conductivity $\kappa$, the electrical resistance $R_{x,x}$, and of the Hall resistance $R_{x,y}$ at $T = 800$ mK, of UTe$_2$ in a magnetic field $\mathbf{H} \parallel \mathbf{a}$ (from [Niu 20a]). Field-dependence (c) of the thermoelectric power $S$ and (d) of the Hall resistivity $\rho_{x,y}$ of UTe$_2$ at different temperatures in a magnetic field $\mathbf{H} \parallel \mathbf{b}$ (from [Niu 20b]).

- Figures 7.14(c-d) presents the magnetic-field-variations of the thermoelectric power $S$ and Hall resistivity $\rho_{x,y}$, respectively, of UTe$_2$ at different temperatures for $\mathbf{H} \parallel \mathbf{b}$ [Niu 20b]. At temperatures 800 mK $\leq T \leq 2$ K, the field-induced intermediate non-superconducting regime, between the phases SC1 and SC2, is associated with negative values of $S$ and $\rho_{x,y}$. At temperatures $T > 5$ K, $S$ is negative while $\rho_{x,y}$ is positive for $H < H_m$, and both $S$ and $\rho_{x,y}$ are positive for $H > H_m$. Knowing a hole contribution leads to positive terms in $S$ and $\rho_{x,y}$, while an electron contribution leads to negative terms in $S$ and $\rho_{x,y}$, these measurements indicate that the carrier properties and, thus, the Fermi surface, of UTe$_2$ are modified by a magnetic field $\mathbf{H} \parallel \mathbf{b}$.

These measurements are not definitive proofs and permit only to suspect the presence of Fermi-surface reconstructions in UTe$_2$ under magnetic fields. Direct measurements, by dHvA or SdH techniques, of the Fermi-surface are needed to confirm that the anomalies observed in electrical- and thermal-transport properties are induced by Fermi-surface reconstructions.

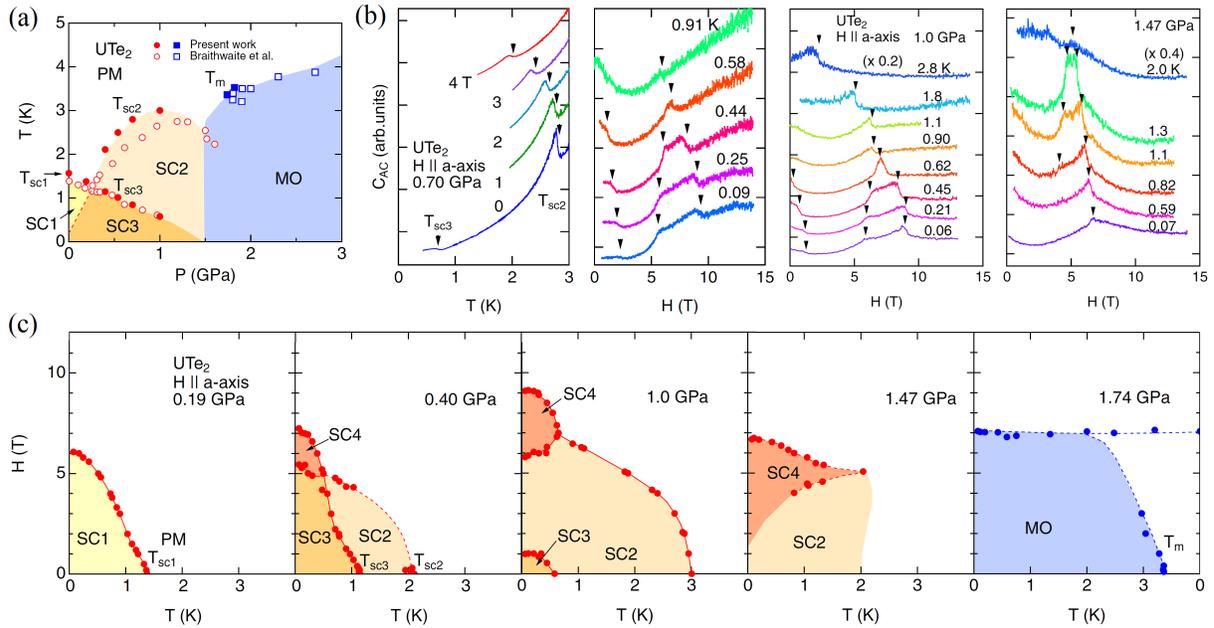

Figure 7.15: (a) Pressure-temperature phase diagram, (b) heat-capacity versus temperature and magnetic field and (c) magnetic-field-temperature phase diagrams of UTe$_2$ at different pressures, in a magnetic field **H ∥ a** (from [Aoki 20]).

### 7.3.5 Combination of pressure and magnetic field

Combined pressure and magnetic field applied on UTe$_2$ permit accessing a rich electronic phase diagram composed of multiple superconducting and magnetic phases. In this Section, we review series of experiments performed under magnetic fields applied either along the easy magnetic axis **a**, or along the hard magnetic axis **b**.

- Figure 7.15 shows that at least three, and possibly four - as proposed by Aoki *et al.*-, different superconducting phases, labeled SC1, SC2, SC3 and SC4, can be revealed in a magnetic field **H ∥ a** [Aoki 20]. Different anomalies are observed in heat-capacity measurements and permit to identify the boundaries between the superconducting phases. At a pressure $p = 1.74$ GPa, superconductivity has vanished and the magnetically-ordered phase is delimited by the temperature $T_{MO} = 3.4$ K and by the critical field $\mu_0 H_c = 7$ T.

- Figure 7.16 summarizes high-pressure experiments performed on UTe$_2$ in a magnetic field **H ∥ b**. Figure 7.16(a) shows that $H_m$ decreases under pressure and vanishes near the critical pressure $p_c \simeq 1.75$ GPa, beyond which a magnetically-ordered phase is stabilized [Knebel 20] (see Figure 7.7 in Section 7.2) [Braithwaite 19]. The temperature $T_\chi^{max}$ at the maximum of the magnetic susceptibility decreases under pressure in a similar manner than $H_m$. The pressure-independent ratio $R_{CPM} = T_\chi^{max}/H_m$ indicates that the CPM regime of UTe$_2$ under pressure behaves as that of conventional heavy-fermion paramagnets (see Section 3.3). Figure 7.16(b) shows that the fall of $T_\chi^{max}$ and $H_m$ drives the field-induced superconducting phase SC2 down to lower magnetic fields under pressure, ending in the loss of the S shape of the critical field [Knebel 20]. Electrical-resistivity and TDO-frequency versus magnetic field measurements are shown in Figure 7.16(c-d), respectively, for different pressures and at $T = 400$ mK [Lin 20]. A signature of the SC1 → SC2 transition is visible as a kink in the TDO-frequency measurements. At $p = 0.4$ GPa, the temperature-dependence of this anomaly, is shown in Figure 7.16(e). The low-temperature pressure-magnetic-field phase diagram presented in Figure 7.16(f), and the magnetic-field-temperature phase diagram shown for different pressures in Figures 7.16(g-i), indicate that the field-induced superconducting phase and the pressure-induced superconducting phase are the same phase, noted SC2 here (the labels SC1 and SC2 from [Lin 20] have been 'inverted' in the Figures presented here, to be consistent with the notations used in [Braithwaite 19, Aoki 20]).

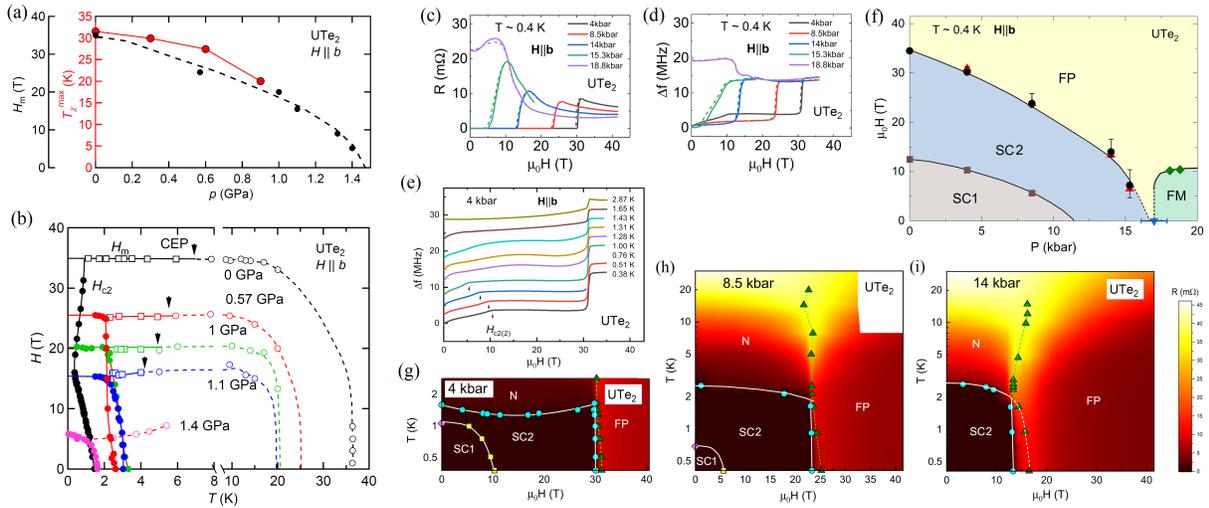

Figure 7.16: (a) Pressure-dependence of $H_m$ and $T_\chi^{max}$, and (b) temperature-magnetic-field phase diagram of UTe$_2$ at different pressures, in a magnetic field $\mathbf{H} \parallel \mathbf{b}$ (from [Knebel 20]). Field-dependence (c) of the electrical resistance and (d) of the TDO frequency at different pressures and $T = 400$ mK, and (e) of the TDO frequency at different temperatures and $p = 0.4$ GPa, (f) low-temperature pressure-magnetic-field phase diagram, magnetic-field-temperature phase diagram (g) at $p = 0.4$ GPa, (h) at $p = 0.85$ GPa, and (i) at $p = 1.4$ GPa of UTe$_2$ at different pressures, in a magnetic field $\mathbf{H} \parallel \mathbf{b}$ (adapted from [Lin 20]; labels SC1 and SC2 have been inverted to be consistent with the notations used in [Braithwaite 19, Aoki 20]).

In parallel to the works summarized above, field-induced superconductivity was also observed beyond the magnetically-ordered phase in a study where the direction of magnetic field was not precisely identified (direction tilted by $\simeq 30°$ from $\mathbf{a}$ towards the $(\mathbf{b}, \mathbf{c})$ plane) [Ran 20].

Pressure-induced modifications of the magnetic-fluctuations spectra may be related with the enhancement of $T_{sc}$ observed for the phase SC2 near the critical pressure $p_c$. Further investigations, by direct and indirect manners, of the magnetic fluctuations are needed to understand the nature of the superconducting phases induced in UTe$_2$. An experimental challenge will be to map out completely the electronic properties of UTe$_2$ for different directions of the magnetic field (up to more than 60 T) combined with pressure (up to at least 3 GPa), and by different sets of complementarily techniques (electrical resistivity, heat-capacity, magnetization, NMR etc.).